\documentclass[11pt]{article}
\usepackage{graphicx}
\usepackage{amssymb}
\usepackage{amsmath}
\usepackage{graphicx}
\usepackage{amstext}
\usepackage{esint}
\usepackage[utf8]{inputenc}
\usepackage{color}
\usepackage{float}
\usepackage{cite}
\usepackage{cite}
\usepackage{placeins}
\usepackage{titlesec}
\usepackage{breqn}
\usepackage{afterpage}

\titleformat*{\section}{\Huge\bfseries\centering}
\titleformat*{\subsection}{\LARGE\bfseries}
\titleformat*{\subsubsection}{\Large\bfseries}

 \numberwithin{equation}{section}

\def\beq{\begin{eqnarray}}    
\def\eeq{\end{eqnarray}}      

\newcommand{\rL}{\rho_\Lambda}

\newcommand{\dH}{\dot{H}}
\newcommand{\CC}{\Lambda}

\newcommand{\Om}{\Omega_m}
\newcommand{\Omo}{\Omega^0_{m}}

\newcommand{\OL}{\Omega_{\Lambda}}

\newcommand{\OLo}{\Omega^0_{\Lambda}}

\newcommand{\rco}{\rho^0_{c}}

\newcommand{\rmr}{\rho_m}

\newcommand{\rR}{\rho_r}

\newcommand{\rLo}{\rho^0_{\CC}}

\newcommand{\oD}{\omega_{\rm BD}}
\newcommand{\eBD}{\epsilon_{\rm BD}}
\newcommand{\Geff}{G_{\rm eff}}
\newcommand{\p}{\prime}
\newcommand{\pp}{\prime\prime}
\newcommand{\rvphi}{\rho_{\varphi}}
\newcommand{\dvphi}{\Delta{\varphi}}
\newcommand{\weff}{w_{\rm eff}}
\newcommand{\pvphi}{p_{\varphi}}

\newcommand{\rDE}{\rho_{\rm DE}}

\newcommand{\nueff}{\nu_{\rm eff}}
\newcommand{\nueffp}{\nu_{\rm eff}'}

\newcommand{\cH}{\mathcal{H}}

\newcommand{\Su}{[S1]\times}
\newcommand{\Sd}{[S2]\times}
\newcommand{\St}{[S3]\times}
\newcommand{\Sq}{[S4]\times} 

\newcommand{\xr}{x_r}
\newcommand{\xm}{x_m}
\newcommand{\xl}{x_\Lambda}
\newcommand{\xp}{x_\psi}

\begin{document}

\begin{center}

\thispagestyle{empty}
\vskip 0.6cm

\vspace{-0.2cm}
\begin{figure}[H]
 \centering
 \includegraphics[width=250pt]{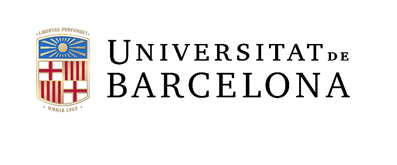}
\end{figure}
\vskip 2.5cm
{\huge {\bf IMPLICATIONS OF DYNAMICAL DARK ENERGY IN THE EXPANSION OF THE UNIVERSE AND THE STRUCTURE FORMATION}}\\
\vskip 3.2cm
{\Large Javier de Cruz P\'erez}
\vskip 0.6cm
{\Large Departament de Física Quàntica i Astrofísica}\\
\vskip 0.25cm
{\Large Universitat de Barcelona}\\
\vskip 2.5cm
{\Large April 2021}
\vskip 4cm

\end{center}

\newpage
\thispagestyle{empty}
\mbox{}
\newpage

\thispagestyle{empty}
\vskip 2cm
\vspace{-0.2cm}
\begin{figure}[H]
 \centering
 \includegraphics[width=250pt]{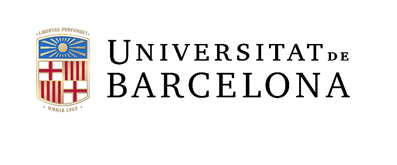}
\end{figure}
\vskip 0.5cm
\begin{center}
{\Large \bf Implications of Dynamical Dark Energy in the expansion of the Universe and the Structure Formation}
\vskip 1.0cm
by
\vskip 1.0cm
{\bf \Large Javier de Cruz P\'erez}
\end{center}
\vskip 0.5cm
A dissertation submitted in fulfillment of the requirements for the degree of {\it Doctor per la Universitat de Barcelona}. 
\newline
\newline
PhD program in Physics. Line of research: Particle Physics and Gravitation.
\newline
\newline
This thesis has been carried out under the supervision of
\vskip 1.5cm
\begin{center}
{\bf \Large Dr. Joan Sol\`a Peracaula}
\end{center}
\vskip 1.5cm
and under the tutoring of {\it Dr. Dom\`enec Espriu Climent}, both full professors of Theoretical Physics of the Department of Quantum Physics and Astrophysics of the University of Barcelona.

\vskip 1.5cm
\begin{flushright}
Barcelona, April 2021
\end{flushright}

\newpage
\thispagestyle{empty}
\mbox{}
\newpage

\thispagestyle{empty}
\vspace{5mm} 

\phantom{123}

\vspace{25mm} 

{\it A mis padres, a quienes les debo todo.}

\vfill

\phantom{123}

\newpage
\thispagestyle{empty}
\mbox{}
\newpage

\pagenumbering{arabic}
\begin{Huge}
\begin{center}
\textbf{Acknowledgments}
\end{center}
\end{Huge}  
This thesis would not have been possible without all those who have been by my side throughout these years. All the work done with enthusiasm during this period has helped me to learn many things not only about physics but also about life. It was definitely a wonderful time ! 
\newline
\newline
I would like to start by thanking my PhD supervisor Joan Sol\`a Peracaula, first for accepting me as his student and for being an exceptional guide. Thanks to him I have had the opportunity to work on fascinating topics and travel around the world meeting wonderful people. From him I have learned all the values that a physicist should have. I especially admire his passion for physics that has never faded and never will. Gr\`acies Joan, de tot cor, per tot el que has fet per mi i per tot el que m'has ensenyat. Mai ho oblidaré. 
\newline
\newline
I thank my collaborator and good friend Adri\`a G\'omez-Valent. He has been an example during the years in which this thesis has been carried out. His commitment when it comes to work has been encouraging to never give up, regardless of how difficult were the problems that we were facing. Gr\`acies Adri\`a per tot els bons moments que hem compartit i per tot el que he apr\`es de tu. 
\newline
\newline
Thanks to Cristian Moreno-Pulido, my collaborator and also a good friend. Ha estat un plaer poder treballar amb tu aquests \'ultims anys de doctorat. 
\newline
\newline
I would like to thank Professor Dom\`enec Espriu for accepting to be the tutor of my thesis. 
\newline
\newline
I want to express my gratitude to the Professors Nikolaos Mavromatos, Jordi Miralda and Diego Pav\'on for having accepted to be part of the PhD Thesis Committee. I also want to thank the Professors Antonio L\'opez Maroto, Enric Verdaguer and Jaume Garriga for having accepted to be supply members of the Committee. I really appreciate to count on you. 
\newline
\newline
I really thank the University of Barcelona where this thesis has been carried out and the Spanish government for funding my PhD research. I want to express my gratitude to all the great professors that taught me during my time there.
\newline
\newline
I would like to thank the Kansas State University for giving my the opportunity to carry out a 5-months stay in the Department of Physics. Many thanks to professor Bharat Ratra for accepting me as his student and for giving me the opportunity to meet great people: Narayan, Joe, Shulei, Tyler and Lado. 
\newline
\newline
I thank Professors Bharat Ratra and Chan-Gyung Park for their valuable collaboration and for agreeing to be the international referees of this thesis. I would like also to express my gratitude to Professor Federico Mescia for all the help in these years. 
\newline
\newline
Thank you very much to Professor Josep M. Pons for being an example of commitment and passion
for physics and mathematics. Gr\`acies per ensenyar-me a estimar i valorar com toca la doc\`encia. \newline
\newline
\newline
Thanks to all the wonderful people I have met thanks to studying at the University of Barcelona, Luciano, Alba, Dani, Carlos, Claudia, Adri\'an, Diego, Samira, Hao, Mar, Pere, Jordi, Carles ... We have shared great moments !
\newline
\newline
A mis querid\'isimos amigos Miguel, Gerard, Ernest, Juan Carlos, Joan, Dani, Toni, P\'ua, Ruiz, Caba, Mary, Jordi, Rafa, Christian, Marc, Mireia, Yuste ... Gracias por todos los momentos compartidos y por todos los que espero compartir en el futuro. 
\newline
\newline
A los buenos amigos que hice durante la carrera Iuri, Dario, Guim, Abel, Pablo, Mart\'i, Lucas, Gerard, Llu\'is, Pablo, Pau, Gerard, Eloi ... Agraeixo de tot cor a la meva gran amiga Catalina per tota l'ajuda durant aquests anys. 
\newline
\newline
A mi familia, por todo el aprecio y amor que me han dado y por estar siempre a mi lado, incluso en los momentos m\'as dif\'iciles.
\newline
\newline
A mis hermanos J\'ulia y V\'ictor a quines tanto quiero. No puedo imaginarme lo que hubiese sido crecer sin vosotros. 
\newline
\newline
A mi sobrina Arla, quien espero que en un futuro le eche un vistazo a estas p\'aginas con curiosidad.
\newline
\newline
A mis padres, a quienes va dedicada esta tesis. Me resulta imposible encontrar las palabras para agradeceros todo lo que habeis hecho por m\'i. Si hoy soy lo que soy es gracias a vosotros. Vuestro ejemplo y enseñanzas me servir\'an para enfrentarme a todo lo que la vida me depare. 
\newline
\newline
Thanks with all my heart to all !
\newpage

\begin{Huge}
\begin{center}
\textbf{Abstract}
\end{center}
\end{Huge}  
The high quality observations performed during the last two decades, have allowed to demonstrate, with high confidence range, that the Universe is in expansion and to be more precise in accelerated expansion. In order to explain the accelerated evolution the name of dark energy was coined. It refers to some mysterious form of diffuse energy presumably permeating all corners of the Universe as a whole. We may say that the canonical picture of our Universe defined in the framework of General Relativity, whose field equation were found by Einstein in 1917, is built upon the assumption that the observed acceleration is caused, in fact, by a rigid cosmological constant term denoted by $\Lambda$. Thanks to the aforementioned cosmological measurements, we have been able to pin down its value to an impressive level. Dark energy is not the only element, beyond the conventional baryons, photons and neutrinos, required by the observations since we also need large amounts of what is commonly call as dark matter. We call such an overall picture of the Universe the  ``concordance (or standard) cosmological model'' or simply $\Lambda$CDM.
\newline
\newline
Therefore, we attribute the observed accelerated expansion of the Universe to the existence of a repulsive force, exerted by the $\Lambda$ term, which works against the attractive gravitational force and tends to push the clusters of galaxies apart at a speed continuously increasing with the cosmic expansion. It is quite remarkable the fact that with so few ingredients (encoded in 6 degrees of freedom indeed) the concordance model has remained robust and unbeaten for a long time. It is consistent with a large body of very precise cosmological observations, like the cosmic microwave background, the baryonic acoustic oscillations or the large scale structure data. Just to illustrate what we have said, the $\Lambda$CDM is able to reproduce the anisotropies of the cosmic microwave background (the variations of the temperature field with respect to the mean value depending on the direction we are looking at) with an astonishing precision, even if they are of order $\mathcal{O}(10^{-5})$.
\newline
\newline
In spite of the success of the standard model of cosmology in explaining the current cosmological data, we still do not know what is the origin at a deep theoretical level, of its building block, namely the cosmological constant. We are still far from providing a satisfactory explanation at the level of fundamental physics, for instance, in the context of quantum field theory, quantum gravity or string theory. This fundamental failure is at the very heart of the cosmological constant problem, which basically expresses our complete inability to obtain even an approximation of the value of $\Lambda$, since each of the attempts made have failed in the prediction by many orders of magnitude. Such a dramatic discrepancy between the theoretical predictions and the observed value of $\Lambda$ together with the unsatisfactory theoretical picture, has motivated the cosmologist to seek for alternatives capable to predict, among other things, the cosmic acceleration beyond the cosmological constant term. One of the most popular models, for instance, has been the scalar fields models, which tries to explain the value of $\Lambda$ in a purely dynamical way. We can extend this idea to a more generic models for which the rigid cosmological term is replaced by some sort of  dynamical quantity which exhibits a mild time-evolution throughout or in part of the cosmic history. Even a more radical approach can be taken, namely, to assume that the General Relativity paradigm is not the correct one and that therefore we need to consider other possibilities. In regards to this, it was the model proposed by Brans and Dicke in 1961, the first proposal to extend General Relativity in order to accommodate possible variations of the Newtonian coupling. However we have to bear something in mind. As appealing as may they seem, none of the alternatives is free from the severity of the cosmological constant problem, since in one way or another we need to consider a fine tuning in order to adjust the small value of $\Lambda$. 
\newline
\newline
On the other hand, there is no doubt that we are leaving in a very promising epoch, to wit, the beginning of the era of precision cosmology. Due to the incredible advances in technology we are now able to explore the Cosmos in detail, even the most recondite corners in order to gather information. Let us just mention the most important advances in this field in recent years. On February 11th 2016 the first detection of the gravitational waves was announced by the LIGO and Virgo collaborations. Gravitational waves are one of the most important predictions of General Relativity and they are a disturbance of the space-time produced by the presence of accelerated masses. The event that allowed the detection of the gravitational waves was a merger of a pair of black holes. Precisely the second advance is related with black holes, which are also a prediction of Einstein's theory. On 10th April 2019 the Event Horizon Telescope published the first direct picture of a black hole and its vicinity. This couple of examples serves to illustrate how good are the improvements in the field of modern observations. Hopefully the incoming precise measurements will allow us in the near future to know more about the nature of dark energy/dark matter or even shed some light on other theoretical problems like the hierarchy of the neutrinos or the observed matter-antimatter asymmetry. 
\newline
\newline
In fact, these accurate measurements have revealed recently some tensions affecting the $\Lambda$CDM. In the first place, there is a worrisome tension between the disparate current values of the Hubble parameter $H_0$ obtained independently from measurements of the local Universe and the early Universe from the cosmic microwave background. Currently the $H_0$-tension has reached the $4\sigma$ level, and as cannot be otherwise, this fact has set the alarms bells ringing. In another vein, we have some trouble in the structure formation observations. The concordance model predicts more structure than is observed. This tension is in the level of 2-3$\sigma$ as is usually quantified through the parameter $\sigma_8$ (the current matter density root-mean-square fluctuations within spheres of radius $8h^{-1}$~Mpc, with $h\simeq 0.7$ the reduced Hubble constant). More specifically, the tension is in between measurements of the amplitude of the power spectrum of density perturbations inferred using CMB data against those directly measured by large scale structure formation on smaller scales. Whether all these tensions are the result of yet unknown systematic errors or really a signal for a underlying new physics is still unclear. There remains, however,  strongly the possibility that these discrepancies may just be the signal of a deviation from the $\Lambda$CDM model. 
\newline
\newline
In this thesis we focus on these problems of very practical nature. We study in detail a wide range of cosmological models, either well grounded in an underlying theory or based on a phenomenological approach, in order to contrast their predictions against the wealth of cosmological data and see if we can capture signs of new physics beyond the standard model. 
\newline
\newline
The layout of this thesis reads as follows. Chapter 1 is written in order to provide a general context that will allow to understand the results presented in the subsequent chapters. We start it by mentioning the most important events of the convulsed history of the cosmological constant, going from its birth, when Einstein himself introduced this term in the field equations to get an static Universe, to its actual status which, as we have mentioned before, turns out to be an essential part of the current standard model of cosmology. In this chapter we also deepen into the cosmological constant problem, formulating it in the context of quantum field theory in curved space-time. It is necessary to explain the main features of the $\Lambda$CDM model in order to be able to properly judge the performance of the other cosmological models studied in this thesis. This is the reason why also in this first chapter we provide the most important cosmological equations that characterize this model. We conclude this chapter by presenting some important alternatives to the $\Lambda$CDM, which can help to alleviate some of the tensions that affect the standard model.
\newline
Chapter 2 is dedicated to study the G-type and A-type models at the background and perturbation level. Both are particular cases of a more general class of cosmological models named Running Vacuum Models. These models are characterized by a time-evolving vacuum energy density, whose expression can be motivated from the renormalization group equation in the context of quantum field theory in curved space-time. In the case of the G-type models apart from the mild evolution of the vacuum energy density we also have a non-standard evolution of the Newtonian coupling $G(t)$ due to the expansion of the Universe. On the contrary, the evolution of $\rho_{\rm vac}$ in the A-type models is allowed due to an interaction with the ordinary matter. Compelling signs of dynamical vacuum energy are found and the $\sigma_8$-tension is loosened. 
\newline
As we have mentioned previously the scalar field models have been studied for a long time in order to see if they can handle some of the problems affecting the standard model of cosmology. In Chapter 3 we study in detail one of the most representative models characterized by the well-known Peebles \& Ratra potential. We compare its performance not only against the $\Lambda$CDM model but also against two dark energy parameterizations, to wit: the XCDM and the CPL. Once again promising signals of physics beyond the standard model are found. This is a very welcome fact since it is reassuring to find the same evidence, detected previously, in favor of dynamical dark energy, for the scalar field models, since they are different from the Running Vacuum Models in a fundamental way. 
\newline
Chapter 4 is dedicated to extend the analysis performed in Chapter 2, but this time we also study some phenomenological dynamical vacuum energy models. We update the observational data employed and we clearly identify where the signals in favour of a time-evolving vacuum come from. 
\newline
In Chapter 5 the $H_0$-tension is directly addressed in the context of the dynamical vacuum models, as well as, in the context of quasi-vacuum models, for which the equation of state parameter is not rigid but a mild evolving quantity. The main outcome that can be extracted is that while the aforementioned models are able to loosen the  $\sigma_8$-tension, they are too limited to ease the other tension affecting the $\Lambda$CDM, namely the $H_0$-tension. 
\newline
The results displayed in the chapters mentioned so far, have been obtained upon the consideration of an approximate treatment of the cosmic microwave background data through a compressed likelihood. This fact is amended in Chapter 6 where a full-likelihood for this data has been taken into account. The signs of dynamical dark energy found in Chapter 3 are reconfirmed.
\newline
Finally Chapter 7 is dedicated to the study of the Brans-Dicke model. As we have commented before this is the first extension of the General Relativity paradigm. The model incorporates an extra degree of freedom encoded in a scalar field, which in addition to the metric field is responsible for mediating the gravitational interaction. The performance of this model when dealing with, both, the $H_0$-tension and the $\sigma_8$-tension, is quite remarkable since it is able to alleviate them at a time. 
\newline
\newline
Additional and useful information is provided in the appendices. In the first of them, Appendix  \ref{Appendix_A}, a detailed explanation on how to deal with the different sign conventions is given. In Appendix \ref{Appendix_B} the way in which the perturbation equations are obtained, for the different cosmological models under study, is explained in the context of the Newtonian gauge. Appendices \ref{Appendix_C} and \ref{Appendix_D} are dedicated to provide some analytical and semi-analytical solutions for the Brans-Dicke model, for which it is not possible to find an analytical solution valid for the whole cosmic history. The last appendix, Appendix \ref{Appendix_E}, is focused on the explanation of the statistical methods employed to obtain the results displayed in the aforementioned chapters. 
\newpage

\begin{Large}
\textbf{List of publications}
\end{Large}  
\newline
\newline
This thesis is based on the following list of papers: 
\begin{enumerate}
\item {\it Hints of dynamical vacuum energy in the expanding Universe.}
\newline
J. Sol\`a Peracaula, A. G\'omez-Valent and J. de Cruz P\'erez
\newline
Astrophys. J. Lett. {\bf 811}, L14 (2015); [arXiv:1506.05793]
\item {\it First evidence of running cosmic vacuum: challenging the concordance model.}
\newline
J. Sol\`a Peracaula, A. G\'omez-Valent and J. de Cruz P\'erez
\newline
Astrophys. J. {\bf 836}, 43 (2017); [arXiv:1602.02103]
\item {\it Dynamical dark energy vs. $\Lambda = {\rm const.}$ in light of observations.}
\newline
J. Sol\`a Peracaula, J. de Cruz P\'erez and A. G\'omez-Valent 
\newline
Euro. Phys. Lett. {\bf 121}, 39001 (2018); [arXiv:1606.00450]
\item {\it Dynamical dark energy: scalar fields and running vacuum.}
\newline
J. Sol\`a Peracaula, A. G\'omez-Valent and J. de Cruz P\'erez
\newline
Mod. Phys. Lett. A{\bf 32}, 1750054 (2017); [arXiv:1610.08965]
\item {\it Possible signals of vacuum dynamics in the Universe.}
\newline
J. Sol\`a Peracaula, J. de Cruz P\'erez and A. G\'omez-Valent 
\newline
Mon. Not. Roy. Astron. Soc. {\bf 478}, 4357 (2018); [arXiv:1703.08218]
\item {\it The $H_0$ tension in light of vacuum dynamics in the Universe.}
\newline
J. Sol\`a Peracaula, A. G\'omez-Valent and J. de Cruz P\'erez
\newline
Phys. Lett. B. {\bf 774}, 317 (2017); [arXiv:1705.06723]
\item {\it Vacuum dynamics in the Universe versus a rigid $\Lambda={\rm const}$.}
\newline
J. Sol\`a Peracaula, A. G\'omez-Valent and J. de Cruz P\'erez
\newline
Int. J. Mod. Phys. A{\bf 32}, 1730014 (2017); [arXiv:1709.07451]
\item {\it Brans-Dicke cosmology mimicking running vacuum.}
\newline
J. de Cruz P\'erez and J. Sol\`a Peracaula
\newline
Mod. Phys. Lett. A{\bf 33}, 1850228 (2018); [arXiv:1809.03329]
\item {\it Signs of dynamical dark energy in current observations.}
\newline
J. Sol\`a Peracaula, A. G\'omez-Valent and J. de Cruz P\'erez
\newline
Phys. Dark. Univ. {\bf 25}, 100311 (2019); [arXiv:1811.03505]
\item {\it Brans-Dicke gravity with a cosmological constant smoothes out $\Lambda$CDM tensions.}
\newline
J. Sol\`a Peracaula, A. G\'omez-Valent, J. de Cruz P\'erez and C. Moreno-Pulido
\newline
Astrophys. J. Lett. {\bf 886}, L6 (2019); [arXiv:1909.02554]
\item {\it Brans-Dicke cosmology with a $\Lambda$-term: a possible solution to $\Lambda$CDM tensions.}
\newline
J. Sol\`a Peracaula, A. G\'omez-Valent, J. de Cruz P\'erez and C. Moreno-Pulido
\newline
Class. Quant. Grav. {\bf 37}, 245003 (2020); [arXiv:2006.04273]
\end{enumerate}

\newpage
\thispagestyle{empty}
\mbox{}
\newpage

\tableofcontents
\newpage
\newpage

\section{Introduction}\label{Introduction_chapter}

We deem that a proper introductory chapter is necessary in order to lay the foundations, that will allow to understand, the results displayed in the onward chapters of the dissertation. First of all in Sec. \ref{SubSec_1.1_Introduction_chapter}, we briefly summarize the fascinating history  of the cosmological constant (CC) and its possible connection with the vacuum energy density, which as we shall see, has been buried many times but one way or another has resurfaced \footnote{We want to make it clear, from the beginning, that this summary does not pretend to be exhaustive. Consequently only the most important events will be presented and regarding the different attempts of cosmological models only the main features will be provided. To know more about the history of cosmology see \cite{Kragh:2014jaa}}. Just after, in Sec. \ref{SubSec_1.2_Introduction_chapter}, we address the well-known problem of the CC formulating it in the context of quantum field theory (QFT). We will pay special attention to the Running Vacuum Models (RVM), which will be studied in depth throughout this thesis. To do so, we will have to deal with the concept of zero-point energy (ZPE), the ground state of a given quantum system. The building blocks of the standard model of cosmology, also known as $\Lambda$-Cold-Dark-Matter ($\Lambda$CDM), together with a brief explanation of the theory of inflation are presented in Sec. \ref{SubSec_1.3_Introduction_chapter}.  The precise cosmological measurements obtained in the recent years have revealed some tensions affecting the $\Lambda$CDM. We are going to comment on the most important ones. Boosted by the motivation of solving some of those tensions, a wide range of models, beyond the standard one, have been considered . In Sec. \ref{SubSec_1.4_Introduction_chapter}, we conclude the introductory chapter by presenting some of the most important alternatives to the $\Lambda$CDM. 
\subsection{Historical context of the cosmological constant}\label{SubSec_1.1_Introduction_chapter}

It is important to put into a historical context all the cosmological issues that we are going to talk about. Therefore let us start summarizing some of the most important facts related to the cosmological constant, a term that was introduced by Einstein himself in the field equations in 1917, and its possible connection with the vacuum energy in the context of QFT. We start explaining which was the motivation to introduce this constant in his field equations and how it affected the cosmology at that time. 
\newline
\newline
As it is well-known, since it is one of the most important events in the history of physics, in 1915 Albert Einstein \cite{Einstein:1915ca} found the field equations that relate the geometry of the space-time with the total amount of energy within it. They are written using tensors, in order to preserve the general covariance of the theory. What this means is that the physical laws remain unchanged under and arbitrary differential coordinate transformations. The resulting equations are non-linear and contain partial derivatives of the variables involved: 
\begin{equation}
R_{\mu\nu} -\frac{1}{2}R{g_{\mu\nu}} = 8\pi{G_{N}}T_{\mu\nu}, \footnote{Throughout this thesis we consider the natural units convention, namely: $\hbar = c=1$. However, regarding Newton's coupling $G_N$ we keep it explicitly in the equations, for convenience. When these conventions are not used it will be indicated.  }
\end{equation}
where $R_{\mu\nu}$ is the Ricci tensor, $R\equiv g^{\mu\nu}R_{\mu\nu}$ is the Ricci scalar, $G_N$ is the gravitational Newtonian coupling and finally $T_{\mu\nu}$ is the energy-momentum tensor, containing the contributions of all the species considered. The field equations are at the very heart of the theory of General Relativity (GR), which considers gravity as a geometric property of space, being its shape determined by the presence of some sort of energy.
\newline
\newline
Soon, it was realized that this set of equations was, indeed, the perfect tool to study the dynamics of the Universe once the content of matter is established. Of course this meant that cosmology was taken to a whole new level. At that time it was generally accepted the fact that the Universe was static and finite, so in order to fulfill these requirements, Einstein, in 1917 \cite{Einstein:1917}, introduced in his equations a constant called $\Lambda$\footnote{Actually he employed $\lambda$ to describe the cosmological constant term, however, taking into account that for the rest of the thesis we are going to denote this term with $\Lambda$ we stick with that from the beginning. In this section we employ the convention $(+,+,+)$, see Appendix \ref{Appendix_A}.}:
\begin{equation}\label{FieldEquations}
R_{\mu\nu} -\frac{1}{2}R{g_{\mu\nu}} +\Lambda{g_{\mu\nu}} = 8\pi{G_{N}}T_{\mu\nu}.     
\end{equation}
This is the origin of the cosmological constant. The addition of this constant term is perfectly consistent with the covariance principal of General Relativity, since the covariant derivative on both sides of \eqref{FieldEquations} leads to $\nabla^{\mu}{G_{\mu\nu}}=0$, where $G_{\mu\nu}\equiv R_{\mu\nu} -\frac{1}{2}R{g}_{\mu\nu} $ is defined as the Einstein tensor and $\nabla^{\mu}{\left(G_{N}T_{\mu\nu}\right)} = 0$. The former is the well-known Bianchi identity and it is satisfied purely due to the geometric properties of the tensors involved. The second equation is connected with the energy-momentum conservation, and if we assume that Newton's coupling $G_N$ is not a function neither of time nor space, it boils down to a simpler form, $\nabla^{\mu}{T_{\mu\nu}} = 0$. So, unless we consider some sort of interaction, we are left with %
\begin{equation}
\nabla_{\mu}\Lambda = \partial_{\mu}\Lambda = 0 \rightarrow \Lambda = {\rm const.}  
\end{equation}
To get a static Universe what is needed is an equilibrium between the gravitational attraction and the repulsion caused by the CC term. In a Universe full of dust (note that now we are considering the CC as a purely geometric term and not as another form of energy), with an energy density $\rho_m$, the equilibrium is achieved if $\Lambda = 4\pi{G_N}\rho_m$. This model is affected by stability problems and this was pointed out some years after by Lema\^\i tre \cite{Lemaitre:1927}.
\newline
\newline
Even if we consider that, as we have mentioned before, the general belief was that the Universe was static and finite, the introduction of the CC in the field equations was controversial from the beginning. The lack of a physical meaning, behind the CC term, provided by Einstein and the stability problems caused that the static Universe was far from being fully accepted by the cosmologists. Let us mention some relevant events of that period. Immediately after Einstein's publication the $\Lambda$ terms started to be deeply studied by the community and the first attempts either to connect the CC with something else or to remove it from the field equations were not far behind. The very first cosmological model after the introduction of the CC was presented by the Dutch physicist, Willem de Sitter in 1917 \cite{deSitter:1917} (actually just one month later after the born of the CC !). For the sake of convenience we will provide the features of this model later on. In 1918 the famous Austrian physicist, Erwin Schr\"odinger \cite{Schrodinger:1918}, suggested the possibility of identifying the CC with a negative pressure just by moving the term $\Lambda{g_{\mu\nu}}$ to the \textit{r.h.s.} of the field equations and defining $p_\Lambda \equiv -\Lambda/8\pi{G_N}$. However, Einstein refused this idea, as interesting as it may seem, because by doing this association we would go too deep into what he considered as a mere hypothesis. At that time, both of them were probably unaware of the fact that they were laying the basis of the debate, still alive, about whether dark energy exists or not. Efforts to obtain a cosmological model of the observed Universe, based on Einstein's field equations, continued, so that, in 1922 and 1924 the Russian physicist, Alexander Friedmann, came up with two fundamental articles for the development of modern cosmology \cite{Friedmann:1922} and  \cite{Friedmann:1924}. In them he studied different solutions of the GR field equations such as models with different values of the CC and spatial curvatures. In particular he considered the possibility of having a Universe with $\Lambda = 0$ and matter distributed in an homogeneous and isotropic way. \footnote{In agreement with the Cosmological Principle, which states that at large-scales matter in the Universe is distributed in a homogeneous and isotropic way. Although we do not have an experimental confirmation for this principle, most of the cosmological models are based on it. } Upon inserting these assumptions into Einstein's field equations he was able to calculate the rate at which the Universe would expand or contract, depending on the energy budget considered. The equations presented in this pair of articles are the ones that we are still using to study the evolution of the geometry of the Universe within a given model at the background level, as long as, we assume the same conditions of homogeneity and isotropy considered by Friedmann. 
\newline
\newline
As cannot be otherwise, the theoretical advances are closely linked to the experimental ones, and cosmology is not an exception. Between 1912 and 1915, Vesto Slipher, an American astronomer, observed the Doppler shift (a redshift to be more precise) in the spiral nebulae by measuring their radial velocities. What these observations mean, assuming they were correct, is that the nebulae were receding from us. Although the results were controversial, they can be considered as the first empirical hint of an expanding Universe and consequently the first experimental evidence against the static Universe model. In 1917, de Sitter, took into great consideration the measurements carried out by Slipher and presented a new cosmological model \cite{deSitter:1917}. In this early model the Universe is completely dominated by $\Lambda$, thus an eternal expansion is predicted due to the repulsive force exerted by the CC. It is the first dynamical cosmological model and it can perfectly accommodate the observed redshift of the spiral nebulae measured by Slipher. Despite the fact that the original model was not properly formulated, from the mathematical point of view, currently the model in which the energy budget is totally dominated by $\Lambda$ is known as the de Sitter model.
\newline
\newline
Einstein, aware of the growing evidence against the static model, not only from the theoretical point of view but also from the experimental perspective, began to wonder if it makes any sense to continue considering the cosmological constant term. In 1923 in a letter addressed to the German physicist and mathematician, Hermann Weyl, he expressed his doubts with the following words: \emph{If there is no quasi-static world, then away with the cosmological term.} As we are about to see things were going to get worse for the static model. 
\newline
\newline
In 1929 took place a breakthrough in cosmology. Of course we are talking about the article presented by the American astronomer Edwin Hubble \cite{Hubble:1929}, in which he displayed the results obtained after years of accurate observations. The main outcome that one can extract from his work is the linear relation found among the distance to the distant galaxies and the velocity with which they are moving away from us, namely: $v = H_0{d}$. We will see that this relation can indeed be derived, for small redshifts, from a model of an expanding Universe described by the Friedmann-Lema\^\i tre-Robertson-Walker (FLRW) metric. The term $H_0$ is known as the Hubble constant and its first measurement was $H_0 = 525 {\rm km/s/Mpc}$. Two years after, the results were confirmed by Hubble himself and Humason \cite{Hubble:1931}. As one can easily imagine, this revolutionary discovery forced to abandon definitively the static model, since it is not possible to accommodate the observed recession of the galaxies within that model. Actually, Hubble's work did not take, some of the cosmologist of the time, by surprise. Before Hubble presented his results, Lema\^\i tre, in 1927, published an article \cite{Lemaitre:1927} where he considered the possibility of a cosmological model in which appears $\Lambda > 0$ and a Universe filled with matter and in a state of expansion. The measured recession velocity of the extra galactic nebulae are just a consequence of the expansion of the Universe. He first published the article in a Belgium journal, with little impact, and this is the reason why the model began to be known after the publication of the Hubble relation. In 2018, with the aim of amending the injustice committed against Lema\^\i tre, it was decided that Hubble's law should actually be renamed as Hubble-Lema\^\i tre law. 
\newline
\newline
As expected the empirical relation between the recession velocity and the distance to the galaxies did no go unnoticed by most cosmologist at that time, who did their best to try to explain the ultimate cause of the expansion of the Universe. In 1930 de Sitter claimed that the element responsible for the expansion of the Universe cannot be other than the $\Lambda$ term. In an article called, \emph{Algemeen Handelsblad}, the Dutch physicist wrote: {\it What makes the Universe expands or swell up ? That is done by lambda. No other answer can be given.} As can be easily understood, in the light of the events that we have just mentioned, it was not a good time for the static model. The Hubble-Lema\^\i tre law was a serious blow to Einstein's static model, which we remember was already affected by theoretical problems. 
\newline
\newline
In 1931 Einstein definitely gave up the idea of the CC. The Soviet-American physicist George Gamow, wrote the following words \cite{Gamow:1970}: \emph{When I was discussing cosmological problems with Einstein, he remarked that the introduction of the cosmological term was the biggest blunder he ever made in his life.} We must say that if Einstein actually said that words, is a matter of debate, since there are no evidences, in his published works, that he made such a dramatic statement. What is clear is that due to the introduction of the $\Lambda$ term to get an static Universe Einstein missed the great opportunity to predict the phenomenon of the expansion of the Universe corroborated by Hubble's observations. In 1932, in collaboration with de Sitter, \cite{deSitterandEinstein:1932} Einstein studied a cosmological model with $\Lambda = 0$, $\Omega^0_k = 0$ (being that term the relative curvature density at present time) and the Universe filled completely with non-relativistic matter. This is the first time that Einstein studied a model, in which the idea of an expanding Universe is totally embraced. 
\newline
\newline
Even if Einstein himself abandoned the idea of the presence of the CC in the field equations, this did not mean at all that others did the same. Lema\^\i tre related, for the first time, the $\Lambda$ term with the notion of vacuum energy. However there were no mentions to a possible connection with the concept of ZPE. Unfortunately for Lema\^\i tre, his idea did not attract much attention at that time, since the conception that $\Lambda$ is a geometrical term, or in other words, has nothing to do with any form of energy was widely extended among the cosmologist. As a consequence there were  very few mentions to Lema\^\i tre's idea in the following years. Nevertheless, someone did pay attention to this promising connection. In 1933 it is possible to find a very interesting attempt to relate the cosmological constant term with the vacuum energy \cite{Bronstein:1931}. The Russian physicist, Malvi Bronstein, proposed a model which considers not only the CC term but also a possible time-evolution of it! Even if he actually did it only \emph{for the sake of generality}, it is quite remarkable that at that early stage of the modern cosmology the idea of a variable CC was considered. The model also includes an exchange between the vacuum energy and the ordinary matter. He related the evolution of $\Lambda$ with the dynamical character of the Universe allowing the principle of energy conservation to be violated, following Bohr's idea, suggested in \cite{Bohr:1932}. 
\newline
\newline
In the first decade after the end of the World War II those models with a non-zero CC were not well considered and the general opinion was that Einstein was right in taking out of the picture the $\Lambda$ term. The next attempts to revive the cosmological constant were due to the need to accommodate certain astrophysical measurements within a cosmological model. In the 1960s it was measured a peak for the number counts of quasar as a function of redshift at approximately $z\sim 2$ and in the first term it was thought that it could be explained thanks to a model with $\Lambda \neq 0$. Now there is an astrophysical explanation for this phenomenon. Going a bit further in time, in 1975, we can find another re-appearance in the context of astrophysics when James Gunn and Beatrice Tinsley studied how the deceleration parameter $q$, which is a dimensionless parameter that tell us about the acceleration of the expansion of the Universe, changes when a positive CC is considered. In the context of GR, when we have $q<0$ means that $\Lambda>0$. Their main conclusion was that the most favoured model was Lema\^\i tre's model with $H_0>80$km/s/Mpc. They stated that the ultimate nature of the $\Lambda$ terms was far from be completely understand but they suggested a possible connection with the QFT without entering in the details at all. 
\newline
\newline 
So far, we have mentioned some attempts to endow the CC with a physical meaning, but none of them deepen in the possible connection of $\Lambda$ with the quantum properties of the vacuum. We had to wait until the late 1960s in order to get this link properly formulated.
In the period of 1965-1968, thanks to the works of \'Erast Gliner \cite{Gliner:1966,Gliner:1970}, the connection between the CC and the Quantum Mechanics began to acquire importance. He worked with the possibility that the Universe started its expansion phase due to the fact that it was in a vacuum-like state, in a similar way that the theory of inflation, which was formulated some years later.
\newline
\newline
We have reached a critical point in the history of the cosmological constant. It is time to talk about the Russian physicist Yakov Zel'dovich and its formulation of what nowadays is known as {\it the cosmological constant problem} in the context of QFT \cite{Zeldovich:1967}. In one of his papers, \cite{Zeldovich:1968}, he stated: \emph{A clarification of the existence and magnitude of the cosmological constant will be of tremendous fundamental significance also for the theory of elementary particles.} He revived the $\Lambda$ term introduced by Einstein from the ashes once again. As we have already mentioned, the ZPE is the state of a quantum system with the lowest possible energy. The approach taken by Zel'dovich was to consider that the CC arises, precisely from the contribution of that ground state. We will say more about that later on, but the main idea behind the computation of the ZPE contribution is to consider that the quantum field is represented by an infinite collection of oscillators, each oscillating with its own mode or frequency. The total value is obtained upon integrating over all the modes. This type of integral gives an infinite result and something must be done in order to obtain a finite and physical result. Whether the ZPE contribution, emerged from the quantum fluctuations of the corresponding quantum field, is real or not is still a matter of debate today. The effect that we are going to talk about next, can be seen as an argument in favour of its existence and consequently can shed some light in the discussion. 
\newline
\newline
In 1948, took place an important event in the area of particle physics. That year the Dutch physicist Hendrik Casimir, predicted a phenomenon later known as the Casimir effect. What Casimir claimed was that there exists an attractive force created between two conducting and parallel plates even when the whole system can be considered to be in the vacuum and at zero temperature. The analytical expression found for the aforementioned force per unit area takes the expression:
\begin{equation}
F = -\frac{\pi^2}{240}\frac{\hbar{c}}{d^4}.    
\end{equation}
As can be noted this time we have explicitly written the constant $\hbar$ (the speed of light denoted by $c$ too) in order to underline the fact that the force is a pure quantum effect. With $d$ we refer to the distance between the plates and the minus sign explicitly says that the force is attractive, as it was mentioned before. The connection with QFT appears when one considers that because at zero temperature there are no real photons between the plates, thus the only thing that can produce the attraction among the plates is the vacuum itself, which in this case is the ground state of the quantum electrodynamics (QED). The Casimir effect received the first experimental support thanks to  Marcus Sparnaay \cite{Sparnaay:1958wg}. Nevertheless it is considered that the definitive confirmation of the existence of this phenomenon took place in 1997 \cite{Lamoreaux:1996wh}. Taking into account that the Casimir effect was predicted approximately 20 years ago than the moment when the connection between $\Lambda$ and the ground state of the QFT was established, it may come as a surprise that particle physicists were not involved earlier in the problem. Indeed it was Steven Weinberg, the American theoretical physicist, who wrote: \emph{Perhaps surprisingly it was a long time before particle physicists began to seriously  worry about this problem despite the demonstration in the Casimir effect of the reality of the ZPE.}
\newline
\newline 
Going back to Zel'dovich's contribution, he made clear his intention to bring back to life the cosmological term introduced in 1917. Referring to this issue he stated: \emph{The genius has been let out of the bottle and is no longer easy to force it back in ... In our opinion, a new field of activity arises, namely the determination of $\Lambda$.} As we have mentioned before if we try to solve the involved integrals in the computation of the ZPE inevitably we will meet infinities. A way out to this problem is to impose a cutoff $\Lambda_c$ which means that you are considering the theory of the quantum field in question only valid up to the value of the selected cutoff. By doing so, we are lead to a result like this $\rho_{\rm vac}\sim \Lambda^4_c$. Zel'dovich realized that for instance if the proton mass $m_p$, is the chosen value for the cutoff then $\rho_{\rm vac}\sim m^4_p\sim 1{\rm GeV}^4$. At the time of Zel'dovich was addressing the CC problem there was no experimental measurement for $\rho^0_\Lambda$, however the upper bound of the critical energy density $\rho^0_c \sim H^2_0/G_N \sim 10^{-47}{\rm GeV}^4$ was in force. In order to find a better agreement and due to the lack of ideas Zel'dovich tried to construct a dimensional consistent value for $\rho_{\rm vac}$, employing once again the proton mass. He ended up with the following estimation $\rho_{\rm vac}\sim G_N{m^6_p}\sim 10^{-38}{\rm GeV}^{4}$. The estimation can be even improved if instead of the proton mass is the pion mass $m_\pi \sim 0.1{\rm GeV}$ the one that enters in the computation. It is important to remain that these numbers are just numerical estimates and there is no theoretical support for them. It can be considered that at this point is when the CC problem starts.
\newline
\newline
In the period 1970-1990 the experimental verification of the electroweak (EW) theory, mainly developed by: Sheldon Glashow, Abdus Salam and Steven Weinberg \cite{Glashow:1961tr,Salam:1968rm,Weinberg:1967tq} together with the developments in theories of spontaneous symmetry breaking (SSB) drew the attention of some particle physicists (among other theoretical physicists) to the CC problem. In particular the intervention of Alan Guth, Andrei Linde, Andreas Albrecht, Paul Steinhardt and Alexei Starobinsky \cite{Starobinsky:1980te,Guth:1980zm,Linde:1981mu,Albrecht:1982wi} turned out to be crucial in the development of inflation. Later on, we will spent some time talking about the main characteristics of inflation, as well as, the handful of problems that it solves at once. Here, it is enough to say that inflation considers that the Universe underwent a period of accelerated expansion, triggered by some sort of source that exerts a negative pressure, the same effect that vacuum energy density produces. Consequently this was considered as an endorsement of the connection between the CC term and the particle physics. 
\newline
\newline
Another important discovery, that cannot be omitted, was the {\emph{accidental}} \footnote{Penzias and Wilson at the Bell Laboratory in New Jersey detected in 1965 a ``noise'' that seemed to come from all parts of the sky. They tried to eliminate it by all means, but they failed. It was then when they started to think that maybe that sound was something else than just noise. Indeed it was, they have discovered the cosmic microwave radiation. They received the Nobel Prize in Physics in 1978.} discovery, by the American radio astronomers Arno Penzias and Robert Wilson of an electromagnetic radiation measured in all directions of the sky and called cosmic microwave background (CMB) in 1965 \cite{Penzias:1965wn}. The origin of this relic radiation is in the early Universe and it is predicted by the Big Bang model. After the Big Bang the Universe consisted in a hot plasma of elementary particles, that formed complex structures as the Universe expanded. In particular when the Universe cooled enough electrons and protons could form neutral hydrogen atoms, this process is known as recombination. At that moment the photons were no longer scattered by the Thomson process and they traveled free. The CMB we observe at the present moment is composed of those same photons, only with lower temperature ($T\sim 3{\rm K}$), due to the expansion of the Universe. At first glance the CMB may appear as homogeneous and isotropic, but looking in more detail anisotropies of order $\Delta{T}/T\sim \mathcal{O}(10^{-5})$ can be appreciated. This data source has become one of the most important when testing cosmological models. 
\newline
\newline
In the period 1970-1990 many efforts were directed to discern whether or not we live in a Universe with $\Lambda =0$. Unfortunately, none of them were able to conclude neither if $\Lambda = 0$ nor $\Lambda \neq 0$ until 1998 came and the supernovae type Ia (SNIa) shed light on the matter. This year the Supernova Cosmology Project (SCP) \cite{Perlmutter:1998np} and the High-redshift Supernova Search Team (HSST) \cite{Riess:1998cb}, claimed, independently, the confirmation of the late-time cosmic acceleration, using both supernovae as a source of cosmological data. Supernovae are extremely luminous explosions and can be classified according to their characteristic spectrum. Both teams employed SNIa, taking advantage of the fact that for this kind of supernova, the absolute luminosity can be considered to be constant at the peak of brightness. Since the apparent luminosity can be measured, it is possible to determine the distance to the SNIa considered in each case. This is why they are usually called ``standard candles". 
The fact that the two teams, working independently, had come to the same conclusion, helped convince cosmologists of a model with $\Lambda\neq0$. 
\newline
\newline
Since then, the evidence in favor of the hypothesis $\Lambda\neq0$ has only grown. Due to the incredible amount of cosmological data, at our disposal, coming from different techniques: baryonic acoustic oscillations (BAO), large scale structure (LSS), strong lensing (SL), cosmic chronometers (CC), etc. The evidence in favour of $\Lambda\neq0$ has reached an astonishing level. For instance, according to the Planck 2018 TTTEEE+lowE+lensing results \cite{Aghanim:2018eyx}, within the $\Lambda$CDM, the obtained value for the relative density of the cosmological constant term is  $\Omega^0_\Lambda = 0.6847\pm 0.0073$, which implies that $\Omega^0_\Lambda > 0$ at $94\sigma$ c.l. ! 
\newline
\newline
Despite the tremendous success achieved in determining the value of the $\Lambda$ term and its corresponding relative energy density  $\Omega^0_\Lambda$, the job is far from be done. Remember, $\Lambda$ was introduced by Einstein to get an static Universe and no theoretical justification was provided. More than one hundred years has passed and we still have to provide an explanation for such a small value of the CC and explain from physical principles what is behind that mysterious term. As we have seen the history of the cosmological constant is full of ups and downs and surprises and we have no reason not to expect more of that in the incoming years.

\subsection{Vacuum energy density in QFT and the cosmological constant problem}\label{SubSec_1.2_Introduction_chapter}
It is time to properly introduce the old cosmological constant problem. As it was mentioned before, this problem was formulated, for the first time, by Zel'dovich \cite{Zeldovich:1967,Zeldovich:1968}, and basically consists in the tremendous mismatch between the computed value of the vacuum energy density, which from now on will be denoted as $\rho_{\rm vac}$, considering the different contributions from QFT, and the experimental measure $\rho^0_\Lambda$, which turns out to be of order $\sim\mathcal{O}(10^{-47}){\rm GeV}^{4}$ in natural units. In the  following we are going to deal with some of the contributions that, in principle, should be considered in the theoretical computation of the vacuum energy density. If we assume, that we are able to obtain all the contributions to get the total induced part we end up with the equality:
\begin{equation}\label{CC_equality_Introduction}
\rho_{\rm vac} = \frac{\Lambda}{8\pi{G_N}} + \rho_{\Lambda,\rm ind}.   
\end{equation}
So the general picture can be described as follows: the term $\rho_\Lambda \equiv \Lambda/8\pi{G_N}$, appearing in the Einstein-Hilbert action, is not yet the {\it physical} or {\it measurable} value of the vacuum energy density but a pure geometrical term. In order to get the value to which we have experimental access ($\rho_{\rm vac}$), we need to include in the term $\rho_{\Lambda,{\rm ind}}$ all the contributions coming from the QFT at any order of perturbation theory. Of course, the value of $\rho_{\rm vac}$ must be finite, since it is this quantity the one that must be equal to the experimental value $\rho^0_\Lambda$, meaning that if infinities arise in the computations of the induced contributions, they must be removed employing the renormalization techniques. We will provide more details about this general procedure later on. Hereafter we are going to compute some of the terms that we should include in the $\rho_{\Lambda,\rm ind}$ term. 
\newline
\newline
Let us start by looking at the part coming from the electroweak (EW) sector. This successful and amply tested QFT was formulated by Glashow, Weinberg and Salam \cite{Glashow:1961tr,Weinberg:1967tq,Salam:1968rm} and unifies in a beautiful way the electromagnetic interactions, mediated by photons, and the weak interactions, responsible for the nuclei decay. In order to be a renormalizable theory, two important ingredients are necessary: the local gauge invariance and the spontaneous symmetry breaking (SSB). The SSB mechanism is the only known way to generate fermion's masses while the gauge symmetry is preserved.
\newline
\newline
The implementation of this mechanism in the Standard Model of particle physics goes with a potential, which if we want to keep renormalizable it must not contain operators with dimension greater than four, in natural units. For the sake of convenience in this section we will use the sign convention $(-,+,+)$. See Appendix \ref{Appendix_A} to know the details. To study the SSB phenomena let us consider a simple potential for the Higgs boson with a real scalar field:
\begin{equation}\label{HiggsPotential}
{ {U}}^{\rm Class}_{\rm Higgs}(\Phi) = \frac{1}{2}m^2\Phi^2 + \frac{\lambda}{4!}\Phi^4  \quad (\lambda > 0).    
\end{equation}
Being the units of ${U}^{\rm Class}_{\rm Higgs}(\Phi)$ and $\Phi$, 4 and 1 respectively, in natural units.  As it was stated we are treating with the CC problem, as a consequence gravity cannot be omitted. Since we do not have a quantum theory for gravity, here we are going to address the problem considering the gravitational effects as if they were caused by an external field and then we quantize only the matter fields. Having said that, the potential, for the moment does not contain quantum effects, namely it is the Higgs potential at tree-level. Clearly the above potential is invariant under reflection transformation $\Phi \rightarrow -\Phi$, however this symmetry is broken when the scalar field acquires its ground state value, or in other words its vacuum expectation value $v\equiv \langle 0|\Phi|0\rangle $ \footnote{In order to check this the only thing that we need to do is introduce a tiny perturbation in the scalar field and make the replacement $\Phi = v + \delta\Phi$. The resulting potential is no longer invariant under a reflection transformation ${ U}^{\rm Class}_{\rm Higgs}(\delta\Phi) \neq {U}^{\rm Class}_{\rm Higgs}(-\delta\Phi)$.}. The total action for the system is composed by the usual Einstein-Hilbert (EH) action plus the action for the scalar field with potential ${U}^{\rm Class}_{\rm Higgs}(\Phi)$. For the sake of simplicity let us suppose that there are no other forms of matter. The total action, conveniently divided in two pieces, then takes the form:
\begin{equation}
S_{\rm tot} = -\frac{1}{16\pi{G_N}}\int d^{4}x \sqrt{-g}R + \tilde{S}[\Phi] 
\end{equation}
being 
\begin{equation}\label{ScalarFieldAction}
\tilde{S}[\Phi] = S[\Phi] -\int d^4x\sqrt{-g}\rho_\Lambda =  \int d^4x \sqrt{-g} \left( \frac{1}{2}g^{\mu\nu}\partial_\mu\Phi\partial_\nu\Phi - { U}^{\rm Class}_{\rm Higgs}(\Phi) -\rho_\Lambda \right).      
\end{equation}
Once we have the expression for the action it turns out very simple to obtain the associated energy-momentum tensor:
\begin{equation}
\tilde{T}^{\Phi}_{\mu\nu} = \frac{2}{\sqrt{-g}}\frac{\delta{\tilde{S}[\Phi]}}{\delta{g^{\mu\nu}}} = g_{\mu\nu}\rho_\Lambda + \partial_\mu\Phi\partial_\nu\Phi - g_{\mu\nu}\left(\frac{1}{2}g^{\alpha\beta}\partial_\alpha\Phi\partial_{\beta}\Phi - { U}^{\rm Class}_{\rm Higgs}(\Phi)\right)  \equiv g_{\mu\nu}\rho_\Lambda + T^{\Phi}_{\mu\nu}.  
\end{equation}
At this point, as previously mentioned, we stay at the classical level which means no quantum corrections and no bare parameters. Since we are interested in the vacuum contributions from the scalar field, or in other words, the contributions coming from the ground state, the kinetic term does not contribute. Consequently the vacuum expectation value of the energy-momentum tensor is given by:
\begin{equation}
\langle \tilde{T}^{\Phi}_{\mu\nu} \rangle = g_{\mu\nu}\rho_\Lambda + \langle {T}^{\Phi}_{\mu\nu} \rangle = (\rho_\Lambda + \langle {U}^{\rm Class}_{\rm Higgs}(\Phi) \rangle )g_{\mu\nu}.
\end{equation}
To get a non-null value for $\langle { U}^{\rm Class}_{\rm Higgs}(\Phi) \rangle$, we need the SSB mechanism to be in force, which means $m^2<0$, otherwise if $m^2>0$, $v= 0$ and consequently $\langle {U}^{\rm Class}_{\rm Higgs}(\Phi) \rangle=0$. At the classical level it is straightforward to find the ground state value of the scalar field once we have the potential. In this particular case it can be written in this way
\begin{equation}
\langle \Phi \rangle = v = \sqrt{\frac{-6m^2}{\lambda}}.
\end{equation}
This vacuum expectation value contributes to the induced vacuum energy due to the electroweak phase transition by the Higgs potential. That contribution can be computed if we first consider the following relation 
\begin{equation}
\frac{G_F}{\sqrt{2}} = \frac{g^2}{8M^2_W} = \frac{1}{2v^2}  
\end{equation}
between the Fermi scale $M_F = G^{-1/2}_F\simeq 293$ GeV and $M_W$, the mass of the bosons $W^{\pm}$. From the potential it is also possible to obtain the physical mass of the Higgs boson, determined by the oscillations of the field around the minimum, $M^2_H = \partial^2{U}^{\rm Class}_{\rm Higgs}(\Phi)/\partial^2\Phi|_{\Phi = v} = -2m^2 > 0$. With all of these ingredients we are finally able to compute the aforementioned EW contribution: 
\begin{equation}\label{EWcontribution}
\rho^{\rm EW}_{\Lambda, \rm ind} =   \langle { U}^{\rm Class}_{\rm Higgs}(\Phi = v) \rangle  = -\frac{3m^4}{2\lambda} = -\frac{1}{8\sqrt{2}}M^2_H{M^2_F} \sim \mathcal{O}(10^8){\rm GeV}^4. 
\end{equation}
The following comment is in order with respect to the above result: even if we stay at the classical level and we only look at the electroweak sector the predicted vacuum energy density is really far from the measured value which remember, turns out to be $\rho^0_{\Lambda}\sim 10^{-47}{\rm GeV}^4$, therefore:
\begin{equation}
\left|\frac{\rho^{\rm EW}_{\Lambda,\rm ind}}{\rho^0_{\Lambda}}\right|\sim \mathcal{O}(10^{55}).
\end{equation}
The Higgs boson has been detected in the Large Hadron Collider (LHC) \cite{Chatrchyan:2012ufa} and its mass is approximately $M_H\simeq 125$ GeV so \eqref{EWcontribution} is nor something hypothetical neither something that can be simply omitted. We should also take into account the contribution from quantum chromodynamics (QCD) the theory dealing with the strong interactions between quarks and gluons. Due to the phenomenon known as asymptotic freedom the behaviour of the characteristic coupling of QCD is opposite to that given in the electroweak sector in the sense that the lower is the energy the stronger is the coupling. This causes that the vacuum state, which is non-null because of the quark and gluon condensates, happens to be a non-perturbative state. We are not going to go into the details of the QCD ground state since it is a very complicate field, however we can get a rough estimation of the associated energy density just by considering the typical scale of the QCD, namely $\Lambda_{\rm QCD}\sim 0.2$ GeV and therefore $\rho^{\rm QCD}_{\Lambda,\rm ind}\sim \Lambda^4_{\rm QCD}\sim 10^{-3}{\rm GeV^{4}}$. As it can be seen even if this value is much smaller than the one obtained from the electroweak sector at classical level, $\rho^{\rm EW}_{\Lambda,\rm ind} = \langle { U}^{\rm Class}_{\rm Higgs}(\Phi = v) \rangle$,  is still much larger that the  measured one $\rho^{0}_{\Lambda}$. As if this were not enough we still need to account for the quantum corrections from the Higgs potential, as well as, the pure quantum effect from the ZPE coming from the different matter fields. So if we consider all the contributions that we have been talking about and we put them all  together, the following equality should hold:
\begin{equation}\label{CC_equality_Introduction_2}
\rho_{\rm vac} = \rho_\Lambda + \rho^{\rm QCD}_{\Lambda,\rm ind} + \rho^{\rm EW}_{\Lambda,\rm ind}  + \sum_{i =1} \hbar^i{{ U}^{(i)}_{\rm Higgs}} + \sum_{i =1} \hbar^i{{ V}^{(i)}_{\rm ZPE}}.   
\end{equation}
Remember that $\rho_\Lambda$ is the geometrical term that appears in the EH action and, as we just said $\rho^{\rm QCD}_{\Lambda,\rm ind}\sim 10^{-3}{\rm GeV}^4$ and $\rho^{\rm EW}_{\Lambda,\rm ind}=\langle{ U}^{\rm Class}_{\rm Higgs}(\Phi = v)\rangle\sim 10^{8}{\rm GeV}^4$. On the other hand the last two pieces of the \textit{r.h.s.} correspond to the quantum correction from the Higgs sector and the zero-point energy contribution from the different matter fields. We shall see a little more about those terms later on. The equality displayed in \eqref{CC_equality_Introduction_2} is nothing more than a slightly more extended version of \eqref{CC_equality_Introduction}, where we have specified a little bit more about the contributions that should be included in $\rho_{\Lambda,{\rm ind}}$. Therefore, on the right side of the above equality we have the QFT prediction of the physical value of the vacuum energy density to all orders of perturbation theory which in principle should be equal to the measured value of order $\sim \mathcal{O}(10^{-47}){\rm GeV^4}$ . Due to the already demonstrated presence of terms whose value exceed by many orders of magnitude the observational value, we are forced to choose the value of $\rho_{\Lambda}$ with a very high precision in order to get a value of the desired order. This of course sounds very unnatural. The mismatch is devastating even if we stick to the classical level and as one can imagine the situation does not get better if the quantum contributions come into play, since the value of $\rho_\Lambda$ must be changed, accordingly, as we consider higher orders of the perturbation expansion of the last two pieces of \eqref{CC_equality_Introduction_2}. One can naively think that we only have to worried about the first terms of the perturbative expansions, however this is not true at all. For instance, even if we consider the contributions of the 21$th$ order in perturbations theory for the Higgs potential we still have contributions comparable to the measured value for the vacuum energy density. See \cite{Sola:2013gha} for an extended discussion. All in all, we hope that what we have just explained serves to realize the severity of the cosmological constant problem, which is, without any shadow of doubt, one of the most difficult problems physics has ever had to deal with.
\subsubsection{Quantum corrections in Minkowski space-time}
So far, we have omitted the quantum effects coming from the Higgs potential. To amend this fact we have to deal with a powerful tool, the effective action. This object is tantamount to the action at the classical level in the sense that through the principle of least action it is possible to obtain the equations of motions for the ground state of the quantum field, which as we have mentioned before contains the quantum corrections up to the desired order in perturbation theory. As it is well known if someone wants to mess with the quantum correction should be ready to handle infinities and properly remove them by applying a renormalization scheme. Of course here we are not going to be free of that. Our ultimate goal is to apply this formalism to the potential \eqref{HiggsPotential}, however it turns out very easy to keep in the expressions a general potential $U(\Phi)$ and at the end obtain the result for the particular case of the Higgs potential. We are going to follow most of the notation employed in \cite{Peskin:1995ev}. In this section we explicitly write $\hbar$ (even if in most of the thesis we employ the natural units $\hbar =1$) to make the units clear at all time and because the perturbative expansion is going to be performed in powers of $\hbar$. The starting point is the generating functional of correlation functions: 
\begin{equation}
\mathcal{Z}[J] = e^{-\frac{i}{\hbar}E[J]} = \int \mathcal{D}\Phi e^{ \frac{i}{\hbar}\left[ S_m[\Phi] + \int d^4yJ(y)\Phi(y) \right] }\\.
\end{equation}
In the above expression we can see the term $S_m[\Phi]$, which is the classical matter action considering only the contribution of a scalar field, and $J$ which refers to an external source. 
What makes this object so useful is the fact that by simply taking functional derivatives we can get the correlation functions. It turns out convenient to also define the functional $E[J]$ which is the vacuum energy as a function of the external source. It is worth remembering that we want to compute the vacuum expectation value of the quantum field $\Phi$, to do so, it is necessary to compute the functional derivative of the energy functional with respect to the external source, namely:
\begin{equation}
\frac{\delta{E[J]}}{\delta{J(x)}} = -\frac{\int \mathcal{D}\Phi \Phi(x) e^{ \frac{i}{\hbar}\left[ S_m[\Phi] + \int d^4yJ(y)\Phi(y) \right] }}{\int \mathcal{D}\Phi e^{ \frac{i}{\hbar}\left[ S_m[\Phi] + \int d^4yJ(y)\Phi(y) \right] }} = -\langle \Omega |\Phi(x)| \Omega\rangle_J.    
\end{equation}
The last term in the above expression is the vacuum expectation value of the quantum field in the presence of an external source. Note the difference between the vacuum state $|\Omega\rangle $ when $J\neq 0$ and the one $|0\rangle$ when $J = 0$. In order to lighten the notation, from now on, we are going to use the following definition
\begin{equation}
\Phi_{\rm cl} \equiv \langle \Omega |\Phi(x)| \Omega\rangle_J.    
\end{equation}
Now we are ready to build up the effective action:
\begin{equation}
\Gamma[\Phi_{\rm cl}] \equiv -E[J] -\int d^4yJ(y)\Phi_{\rm cl}(y),     
\end{equation}
where as it can be appreciated it is a functional of $\Phi_{\rm cl}$. Once again through a functional derivative we can obtain the following relation
\begin{equation}
\frac{\delta\Gamma[\Phi_{\rm cl}]}{\delta\Phi_{\rm cl}} = -J(x), 
\end{equation}
which in the absence of the external sources it becomes simpler 
\begin{equation}\label{derivativeEA}
\frac{\delta\Gamma[\Phi_{\rm cl}]}{\delta\Phi_{\rm cl}} = 0.  
\end{equation}
This is the equation that we have been looking for in this section. As stated, when we consider the quantum corrections to the ground state the effective action replaces the classical action. Therefore, {\it if} (and sometimes this is a big {\it if}) we are able to compute the effective action and solve \eqref{derivativeEA} we would get the values of the stable quantum states of the theory in the absence of external sources and then the only thing left to do would be substitute those values in the effective action to know the induced contribution coming from the quantum corrections of the Higgs potential. Actually the induced contribution does not come from the effective action but from the effective potential. Fortunately, the effective action turns out to be an extensive quantity, namely it is proportional to the 4-dimensional volume, as a consequence:
\begin{equation}
\Gamma[\Phi_{\rm cl}] = -\int d^4xU_{\rm eff}(\Phi_{\rm cl})= -\mathcal{V}_{4}U_{\rm eff}(\Phi_{\rm cl}). \end{equation}
The above expression can be used as the definition of the effective potential, which remember is the extension of the classical potential when the quantum corrections must be taken into account. Quite often the effective action is not easy to compute even if we apply the perturbation theory. Here we stick at the first order beyond the tree level, and for this purpose, we are going to make use of the one loop approximation of the effective action computed for instance in \cite{Peskin:1995ev,Weinberg:1996kr,Parker:2009uva}
\begin{equation}
\Gamma[\Phi_{\rm cl}] = S^{\rm Ren}[\Phi_{\rm cl}] + \frac{i\hbar}{2}\ln{\rm det}\left[-\frac{\delta^2{S^{\rm Ren}}[\Phi]}{\delta\Phi(x)\delta\Phi(y)}\Bigg|_{\Phi = \Phi_{\rm cl}}\right] +\delta{S}[\Phi_{\rm cl}] + \mathcal{O}(\hbar^2),   
\end{equation}
where $S^{\rm Ren}[\Phi_{\rm cl}]$ is the classical action for the scalar field but containing only the renormalized parameters. On the other hand, $\delta{S}[\Phi_{\rm cl}]$ contains the counterterms that eventually will absorb the infinities that arise from the divergent integrals. 
\newline
\newline
One important thing that should be remembered when we stick to the renormalization program is that the parameters that appear in the action are {\it not} the physical ones, but the bare parameters, where the counterterms should be clearly separated from the renormalized parameters. Later on we shall see how we do that explicitly. In the case of the scalar field action the bare parameters are contained in the potential. Bearing this in mind we can write the action in flat space-time as: 
\begin{equation}
S_m[\Phi_{\rm cl}] = S^{\rm Ren}[\Phi_{\rm cl}] + \delta{S}[\Phi_{\rm cl}] = \int d^4x\left(\frac{1}{2}g^{\mu\nu}\partial_\mu\Phi_{\rm cl}\partial_\nu\Phi_{\rm cl} - U(\Phi_{\rm cl})\right)   
\end{equation}
where $U(\Phi_{\rm cl}) = U_{\rm Ren}(\Phi_{\rm cl}) + \delta{U(\Phi_{\rm cl})}$ and the last piece contains the counterterms. 
Applying the rules of the functional derivation to the classical action for the scalar field in flat space-time $S^{\rm Ren}[\Phi]$ we can work out the following result:
\begin{equation}
\frac{\delta^2{S^{\rm Ren}}[\Phi]}{\delta\Phi(x)\delta\Phi(y)}\Bigg|_{\Phi = \Phi_{\rm cl}} = -\left(\Box_x + \frac{\delta^{2}U_{\rm Ren}(\Phi_{\rm cl})}{\delta\Phi_{\rm cl}(y)\delta\Phi_{\rm cl}(x)})\right)\delta^{(4)}(x-y).    
\end{equation}
We have introduced the $\Box_x$ d'Alembert operator in addition to  $\delta^{(4)}(x-y)$, the four dimensional delta Dirac. For simplicity, from now on, we are going to use the following definition $U^{\prime\prime}_{\rm Ren}(\Phi_{\rm cl})\equiv\delta^{2}U_{\rm Ren}(\Phi_{\rm cl})/\delta\Phi_{\rm cl}(y)\delta\Phi_{\rm cl}(x)$. If we move to the momentum space we have:
\begin{equation}
\ln{\rm det}\left[-\frac{\delta^2{S^{\rm Ren}}[\Phi]}{\delta\Phi(x)\delta\Phi(y)}\Bigg|_{\Phi = \Phi_{\rm cl}}\right] = \frac{\mathcal{V}_4}{(2\pi)^4}\int d^4k \ln\left(-k^2 + U^{\prime\prime}_{\rm Ren}(\Phi_{\rm cl})\right). 
\end{equation}
As it can be easily verified this integral is divergent. In QFT in order to deal with this type of integrals it is necessary to apply some kind of regularization technique. For instance, the dimensional regularization, which is the one that we will be using, consists of applying the following change in the divergent integral 
\begin{equation}
\frac{1}{(2\pi)^4}\int d^4k \ln\left(-k^2 + U^{\prime\prime}_{\rm Ren}(\Phi_{\rm cl})\right)\rightarrow \frac{\mu^{4-d}}{(2\pi)^d}\int d^dk \ln\left(-k^2 + U^{\prime\prime}_{\rm Ren}(\Phi_{\rm cl})\right),     
\end{equation}
where $\mu$ is a parameter with dimensions of mass, which has been introduced in order to keep the correct dimensions. The term $d = 4-2\epsilon$ has been added in such a way that by doing the limit $\epsilon \rightarrow 0$ we will be able to distinguish the divergent parts from the non-divergent ones. These modifications allow us to split the integral into two pieces, one perfectly finite, nevertheless $\mu$-dependent, and the other one containing the infinities, which as we shall see are going to be removed thanks to the counterterms. Here we skip the details on how to solve the integral, but we refer the reader to the references \cite{Gomez-Valent:2017tkh,Peskin:1995ev} where all the finesse details can be found
\begin{equation}
\frac{\mu^{4-d}}{(2\pi)^d}\int d^4k \ln\left(-k^2 + U^{\prime\prime}_{\rm Ren}(\Phi_{\rm cl})\right) = -i\frac{(U^{\prime\prime}_{\rm Ren}(\Phi_{\rm cl}))^{2}}{32\pi^2}\left[\frac{1}{\epsilon} - \gamma_{\rm EM} + \frac{3}{2} + \ln\left(\frac{4\pi{\mu^2}}{U^{\prime\prime}_{\rm Ren}(\Phi_{\rm cl})}\right) + \mathcal{O}(\epsilon) \right].    
\end{equation}
In the above expression appears $\gamma_{\rm EM}\simeq 0.57721$, the Euler-Mascheroni constant, which comes from the expansion of the Gamma function as a power series of $\epsilon$. Now we are in position to plug all the pieces together and write down the one-loop approximation for the effective potential: 
\begin{equation}\label{general_effecive_potential}
U_{\rm eff}(\Phi_{\rm cl}) = U_{\rm Ren}(\Phi_{\rm cl})  - \frac{\hbar}{64\pi^2}(U^{\prime\prime}_{\rm Ren}(\Phi_{\rm cl}))^2\left[\frac{1}{\epsilon} - \gamma_{\rm EM} + \frac{3}{2} + \ln\left(\frac{4\pi{\mu^2}}{U^{\prime\prime}_{\rm Ren}(\Phi_{\rm cl})}\right) + \mathcal{O}(\epsilon) \right] + \delta{U}(\Phi_{\rm cl}) + \mathcal{O}(\hbar^2).   
\end{equation}
It is important to remark the presence of the terms $\sim 1/\epsilon$ whose contribution will be eliminated thanks to the counterterms, embodied in the piece $\delta{U}(\Phi_{\rm cl})$, leaving us with a finite expression for the effective potential. Let us go into the details of this procedure. 
\newline
\newline
At this stage it is crucial to properly introduce the bare parameters. As we previously mention, one may think that the parameters we wrote in \eqref{ScalarFieldAction} are directly the physical ones, or in other words, those parameters that we can measure. However this is not what renormalization tells us. The parameters in the Higgs action are the bare ones and can be expressed as:
\begin{equation}
\rho_\Lambda = \rho_\Lambda(\mu) + \delta\rho_\Lambda \quad     m^2 = m^2(\mu) + \delta{m^2} \quad \lambda = \lambda(\mu) + \delta\lambda.   
\end{equation}
It is important to highlight the dependence on the $\mu$ parameter in the renormalized parameters. As it has been mentioned before the counterterms, $\delta\rho_\Lambda$, $\delta{m^2}$ and $\delta\lambda$ are the responsible for removing the infinities that have arisen from the regularization procedure. Let us spoil the surprise at the end of the process by telling what will be the final result. The following equality must be hold
\begin{equation}
\rho_\Lambda + U_{\rm eff}(\Phi_{\rm cl}, m,\lambda;\mu) = \rho_\Lambda(\mu) + U^{\rm Ren}_{\rm eff}(\Phi_{\rm cl}(\mu),m(\mu),\lambda(\mu);\mu),     
\end{equation}
where in the {\it l.h.s.} we have counterterms and infinities, that of course cancel each other, while in the \textit{r.h.s.} we have only finite quantities but some of them are affected by an implicit dependence of the energy scale $\mu$. We will say more about that later on. Hereafter we denote by $U^{\rm Ren}_{\rm eff}$ the renormalized, i.e. free of infinities, effective potential. All in all, if we consider that the only possible contributions to the measurable vacuum energy density, $\rho_{\rm vac}$ are the ones coming from the Higgs sector, in terms of $U^{\rm Ren}_{\rm eff}$, then the final results will take the following form: 
\begin{equation}
\rho_{\rm vac} = \rho_\Lambda + \rho_{\Lambda,\rm ind} = \rho_\Lambda(\mu) + U^{\rm Ren}_{\rm eff}(\Phi_{\rm cl}(\mu), m(\mu),\lambda(\mu);\mu).    
\end{equation}
As expected our goal now is to obtain the expression for $U^{\rm Ren}_{\rm eff}(\Phi_{\rm cl}(\mu), m(\mu),\lambda(\mu);\mu)$. 
\newline
\newline
Substituting in \eqref{general_effecive_potential} the expression of the Higgs potential \eqref{HiggsPotential} evaluated at $\Phi = \Phi_{\rm cl}$ and considering what we have said about the bare parameters, the expression for physical vacuum energy density reads as follows:
\begin{align}
\rho_{\rm vac} &= \rho_\Lambda + \delta\rho_\Lambda + \frac{1}{2}(m^2+\delta{m^2})\Phi^2_{\rm cl} +  \frac{1}{4!}(\lambda+\delta{\lambda})\Phi^4_{\rm cl}\nonumber\\
&- \frac{\hbar}{64\pi^2}\left(m^4 + \lambda{m^2}\Phi^2_{\rm cl} + \frac{\lambda^2}{4}\Phi^4_{\rm cl}\right)\left[\frac{1}{\epsilon} - \gamma_{\rm EM} + \frac{3}{2} + \ln\left(\frac{4\pi{\mu^2}}{m^2 + (\lambda/2)\Phi^2_{\rm cl}}\right) + \mathcal{O}(\epsilon) \right] + \mathcal{O}(\hbar^2). 
\end{align}
Among the different options, in regards to the renormalization, we stick to the modified minimal subtraction one denoted as $\overline{\rm MS}$. In order to eliminate the infinities from the above expression the counterterms need to have the following form:
\begin{align}
\delta\rho_\Lambda &= \frac{\hbar{m^4}}{64\pi^2}\left(\frac{1}{\epsilon} - \gamma_{\rm EM} + \frac{3}{2}\right)\\
\delta{m^2} &= \frac{\hbar\lambda{m^2}}{32\pi^2}\left(\frac{1}{\epsilon} - \gamma_{\rm EM} + \frac{3}{2}\right)\\
\delta\lambda &= \frac{3\hbar\lambda^2}{32\pi^2}\left(\frac{1}{\epsilon} - \gamma_{\rm EM} + \frac{3}{2}\right).
\end{align}
Once we have subtracted the infinite terms the effective vacuum energy density boils down to 
\begin{align}\label{final_VED_Higgs}
\rho_{\rm vac} &= \rho_\Lambda + \frac{1}{2}m^2\Phi^2_{\rm cl} +  \frac{\lambda}{4!}\Phi^4_{\rm cl}\nonumber\\
&- \frac{\hbar}{64\pi^2}\left(m^4 + \lambda{m^2}\Phi^2_{\rm cl} + \frac{\lambda^2}{4}\Phi^4_{\rm cl}\right)\ln\left(\frac{4\pi{\mu^2}}{m^2 + (\lambda/2)\Phi^2_{\rm cl}}\right)  + \mathcal{O}(\hbar^2). 
\end{align}
This is the commonly known as the Coleman-Weinberg potential \cite{Coleman:1973jx}. All the parameters $\rho_\Lambda$, $m$, $\lambda$ as well as the scalar field $\Phi_{\rm cl}$ must be understood as explicit function of the $\mu$ parameter.
\newline
\newline
Before going any further, it is worth mentioning some characteristics of the expression of the effective potential. We take into account the quantum effects through a loop expansion in powers of the parameter $\hbar$, where the tree level is represented by the classical Higgs potential:
\begin{equation}
U^{\rm Ren}_{\rm eff}(\Phi_{\rm cl}) = U^{\rm Class}_{\rm Higgs}(\Phi_{\rm cl}) + \hbar{U^{(1)}_{\rm Higgs}} + \hbar^2{U^{(2)}_{\rm Higgs}} + \hbar^3{U^{(3)}_{\rm Higgs}} + \dots    
\end{equation}
For the purpose of this section it is enough to cut the series at the first order and the term $U^{(1)}_{\rm Higgs}$ can be identified with the term proportional to $\hbar$ displayed in \eqref{final_VED_Higgs}. As we explained in the last subsection even if we stay at the tree level the CC problem is difficult to assimilate but if we consider more and more terms, from the above expansion, the problem becomes catastrophic. 
\newline
It turns out that, the effective potential can be separated into two pieces, one consisting in loops with no external legs, the ZPE contribution denoted as $U_{\rm ZPE}$, and another one involving loops with external legs, that we call the scalar contribution $U_{\rm scalar}$. Therefore in a general way we have $U^{\rm Ren}_{\rm eff} = U_{\rm scalar} + U_{\rm ZPE}$, where:
\begin{align}
U_{\rm scalar} &= U^{\rm Class}_{\rm Higgs} + \hbar{U^{(1)}_{\rm scalar}} + \hbar^2{U^{(2)}_{\rm scalar}} + \hbar^3{U^{(3)}_{\rm scalar}} + \dots \\
U_{\rm ZPE} &= \hbar{U^{(1)}_{\rm ZPE}} + \hbar^2{U^{(2)}_{\rm ZPE}} + \hbar^3{U^{(3)}_{\rm ZPE}} + \dots
\end{align}
It should be noted that the ZPE contribution has no classical terms, which means that for $\hbar = 0$ we have $U_{\rm ZPE} = 0$. 
\newline
\newline
Going back to the expression of $\rho_{\rm vac}$ displayed in \eqref{final_VED_Higgs} it is important to realize that this quantity cannot depend on the arbitrary mass scale $\mu$, therefore we have to impose: 
\begin{equation}
\frac{d\rho_{\rm vac}}{d\mu} = \frac{d}{d\mu}\left[\rho_\Lambda(\mu) + U^{\rm Ren}_{\rm eff}(\Phi_{\rm cl}(\mu), m(\mu),\lambda(\mu);\mu) \right] = 0 .   
\end{equation}
Actually the above equality can be separated in two pieces due to the fact that the ZPE part is the only one that needs the term $\rho_\Lambda(\mu)$ in order to be finite and $\mu$-independent, whereas the $U_{\rm scalar}$ it already is by itself. So we end up we the following couple of differential equations:
\begin{align}\label{equation_for_beta_function}
\frac{d}{d\mu}\left[\rho_\Lambda(\mu) + U_{\rm ZPE}(m(\mu),\lambda(\mu);\mu) \right] &= 0 \\  
\frac{d}{d\mu}\left[U_{\rm scalar}(\Phi_{\rm cl}(\mu), m(\mu),\lambda(\mu);\mu) \right] &= 0.
\end{align}
From \eqref{equation_for_beta_function} we can obtain the $\beta$-function at first order defined as:
\begin{equation}
\beta^{(1)}_\Lambda\equiv \mu\frac{d\rho_\Lambda}{d\mu}.   
\end{equation}
This important function tells us how the $\rho_\Lambda(\mu)$ runs with the arbitrary scale $\mu$ because of the quantum effects considered in this section. At this stage it is possible to establish an analogy between the run of the vacuum energy density and the run of the couplings of QED or QCD (see \cite{Shapiro:2009dh} and references therein for an extended discussion), both having been experimentally confirmed. We would like to stress the fact that if we were able to compute the full effective action (or equivalently the full effective potential), it would be $\mu$-independent. However, since we do not have the full expression the presence of the arbitrary scale $\mu$ turns out to be completely necessary according to the renormalization group  method. We can summarize the situation by saying that, with an appropriate choice for $\mu$ and employing the renormalization group apparatus, we can get a good approximation of the behaviour of the vacuum energy density at some particular scales. Unlike QED and QCD, in cosmology the choice for $\mu$ is not obvious. We will talk more about that later on and we will propose some options to which the $\mu$ scale can be identified.  
\subsubsection{Running vacuum in QFT in curved space-time}
In this part we go a step further and we also take into account the effects of considering a curved space-time, in particular a space-time characterized by the FLRW metric. Unlike the previous section, here we do not base the computation on the $\overline{\rm MS}$ scheme where the counterterms play a fundamental role but on a full QFT calculus.  
We compute the renormalized vacuum energy-momentum tensor employing the adiabatic regularization prescription (ARP) \cite{Birrell:1982ix,Parker:2009uva}, and once we have its expression we will be able to compute the renormalized vacuum energy density. The convention adopted in this section is $(+,+,+)$, see Appendix \ref{Appendix_A} for the details.
\newline
\newline
For the sake of simplicity we are going to consider that the quantum contribution of the ZPE comes only from a non-minimally coupled scalar field, without self-interaction, in the FLRW conformal metric $ds^2 = a^2(\eta)\eta_{\mu\nu}dx^{\mu}dx^{\nu}$ with $\eta_{\mu\nu}$ = ${\rm diag}(-1,+1,+1,+1)$. Throughout this section we are going to employ mainly the conformal time, however in order to establish a connection with the running vacuum models (RVM's) it turns out convenient to also use, in some particular moments, the cosmic time $dt= ad\eta$ as a variable. When we work with the conformal time the Hubble function and its derivative take the following form $\mathcal{H} = da/ad\eta$ and $\mathcal{H}^{\prime} = d\mathcal{H}/d\eta$. On the other hand, when we use the cosmic time $H = da/adt$ and $\dot{H} = dH/dt$. Regardless of the variable used we write $R = 6/a^2\left(\mathcal{H}^{\prime} + \mathcal{H}^2\right) = 6\dot{H} + 12H^2$, when we talk about the Ricci scalar. The action for the non-minimally coupled scalar field reads: 
\begin{equation}
S[\Phi] = -\int d^4x\sqrt{-g}\left(\frac{1}{2}g^{\mu\nu}\partial_\mu\Phi\partial_\nu\Phi +\frac{1}{2}(m^2 + \xi{R})\Phi^2  \right), \end{equation}
where $\xi$ is the coupling parameter between the scalar field and the curvature. In general is needed for renormalization, however it is not for the present case since $\Phi$ is free as a quantum field. Keeping $\xi\neq 0$ will be of great help at the time of establishing the connection with the RVM's. By applying the principle of least action, {\it w.r.t.} the scalar field, we can obtain the Klein-Gordon equation:
\begin{equation}\label{Non_minimally_KG_equation}
(\Box -m^2 -\xi{R})\Phi = 0.     
\end{equation}
Where $\Box\Phi = g^{\mu\nu}\nabla_\mu\nabla_\nu\Phi = (-g)^{-1/2}\partial_\mu\left(\sqrt{-g}g^{\mu\nu}\partial_\nu\Phi\right)$. The next step would be to obtain the energy-momentum tensor, which can be written as:
\begin{align}\label{Non_minimall_EMT}
T^{\Phi}_{\mu\nu} = \frac{-2}{\sqrt{-g}}\frac{\delta{S[\Phi]}}{\delta{g}^{\mu\nu}} = (1-2\xi)\partial_\mu\Phi\partial_\nu\Phi + \left(2\xi -\frac{1}{2}\right)g_{\mu\nu}\partial^\alpha\Phi\partial_\alpha\Phi\\
-2\xi\Phi\nabla_\mu\nabla_\nu\Phi + 2\xi{g_{\mu\nu}\Phi\Box\Phi} + \xi{G_{\mu\nu}}\Phi^2 -\frac{1}{2}m^{2}g_{\mu\nu}\Phi^2.\nonumber
\end{align}
Being $G_{\mu\nu}$ the Einstein tensor. In order to consider the quantum effects we can expand the scalar field around its background value: 
\begin{equation}\label{Scalar_field_separation_Introduction}
\Phi(\eta,\vec{x}) = \bar{\Phi}(\eta)+\delta\Phi(\eta,\vec{x}), 
\end{equation}
where we have defined $\bar{\Phi}(\eta) \equiv \langle 0|\Phi(\eta,\vec{x})|0\rangle$ and considered $\langle 0|\delta\Phi(\eta,\vec{x})|0\rangle=0$. Both parts, $\bar{\Phi}$ and $\delta\Phi$, obey independently the Klein-Gordon equation. Due to the above separation we can know clearly separate the energy-momentum tensor in two pieces, the one associated to the classical part and the fluctuating one, whose 00-component can be associated with the ZPE. Taking this into consideration the total vacuum contribution reads as follows:
\begin{equation}\label{eq:Introduction_vacuum_tensor}
\langle T^{\rm vac}_{\mu\nu}\rangle = T^{\Lambda}_{\mu\nu} + \langle T^{\delta\Phi}_{\mu\nu}\rangle  = -\rho_{\Lambda}g_{\mu\nu}  + \langle T^{\delta\Phi}_{\mu\nu}\rangle.
\end{equation}
What we mean by this expression is that the effective or physical vacuum energy-momentum tensor is composed by the cosmological term plus the quantum fluctuations of the scalar field. Remember that $\rho_\Lambda$ is just the parameter that appears in the EH action, it is not yet the physical $\rho_{\rm vac}$ value to which we have experimental access. Now our goal is crystal clear, we have to compute $\langle T^{\delta\Phi}_{\mu\nu}\rangle$ and to do so it is necessary to quantize the part $\delta\Phi(\eta,\vec{x})$. If we Fourier-expand the perturbation in terms of the creation and annihilation operators (A$_{\vec{k}}$ and A$^{\dagger}_{\vec{k}}$ ) we obtain the following expression:
\begin{equation}\label{fluctuation_Fourier_expansion}
\delta\Phi(\eta,\vec{x})  = \frac{1}{a(2\pi)^{3/2}}\int d^3k\left[A_{\vec{k}}{e^{i\vec{k}\cdot \vec{x}}}h_k(\eta) + A^{\dagger}_{\vec{k}}e^{-i\vec{k}\cdot \vec{x}}h^{\ast}_k(\eta)\right].    
\end{equation}
Where we have employed the following notation $k\equiv |\vec{k}|$. The commutation relations for the operators are the usual ones
\begin{equation}\label{commutiation_relations}
[A_{\vec{k}},A^{\dagger}_{\vec{k}^{\prime}}] = \delta^{(3)}(\vec{k}-\vec{k}^{\prime}) \qquad [A_{\vec{k}},A_{\vec{k}^{\prime}}] = 0.     
\end{equation}
Where it is important to note that these operators are time-independent. Now if we substitute \eqref{fluctuation_Fourier_expansion} in \eqref{Non_minimally_KG_equation} considering \eqref{commutiation_relations} we are lead to the following linear differential equation:
\begin{equation}\label{DiffEq_frequency_modes}
h_k^{\prime\prime} + \Omega^2_k{h_k} = 0 \qquad \Omega^2_k(\eta) = \omega^2_k(m) + a^2\left(\xi -\frac{1}{6}\right)R. 
\end{equation}
with
\begin{equation}
\omega^2_k(m)\equiv k^2 + a^2m^2.    
\end{equation}
Due to the fact that $\Omega^2_k(\eta)$ is not a trivial function, there is no analytical solution to \eqref{DiffEq_frequency_modes}. By employing a recursive self-consistent iteration process we can get an approximate solution starting from the ansatz:
\begin{equation}\label{integral_ansatz}
h_k(\eta) = \frac{1}{\sqrt{2W_k(\eta)}}e^{i\int^{\eta} d\tilde{\eta}W_k(\tilde{\eta})}.    
\end{equation}
The Wronskian condition $h^{\prime}_{k}h^{*}_k - h_k{h^{*\prime}_{k}}=i$ must be followed. Now we have to deal with the following non-linear and inhomogeneous differential equation, obtained upon substituting \eqref{integral_ansatz} in \eqref{DiffEq_frequency_modes}:
\begin{equation}
\frac{1}{2}\left(\frac{W^{\prime\prime}_k}{W_k}\right) -\frac{3}{4}\left(\frac{W^{\prime}_k}{W_k}\right) + W^2_k = \Omega^2_k.
\end{equation}
In order to solve the above equation we are going to apply the Wentzel-Kramers-Brillouin (WKB) approximation, which is valid for large $k$ or equivalently short wavelength $\lambda$. In this context it turns out to be convenient to employ to notion of adiabatic vacuum \cite{Bunch:1980vc}, basically a state empty of high frequencies. We can organize the asymptotic series solution, obtained upon the application of the WKB method, in adiabatic orders. This constitutes the basis of the adiabatic regularization prescription. For more details about the use of this technique see \cite{Moreno-Pulido:2020anb}. In what follows we summarize the results presented in this reference. From now on we consider quantities of adiabatic order 0: $k^2$ and $a$, of adiabatic order 1: $a^{\prime}$ and $\mathcal{H}$, of adiabatic order 2: $a^{\prime\prime}$, $a^{\prime 2}$, $\mathcal{H}^{\prime}$ and $\mathcal{H}^2$. For each higher derivative that is considered, the corresponding order must be added in the adiabatic series. Taking into account the different order we have mentioned, the solution can be displayed as:
\begin{equation}
W_k = \omega^{(0)}_k + \omega^{(2)}_k + \omega^{(4)}_k +\dots  
\end{equation}
where the presence of only even terms is justified by the argument of general covariance. After a bit of algebra it is possible to obtain
\begin{eqnarray}\label{expanded_terms}
& \omega^{(0)}_k &\equiv \omega_k = \sqrt{k^2 + a^2{M^2}}\\
& \omega_k^{(2)}&= \frac{a^2 \Delta^2}{2\omega_k}+\frac{a^2 R}{2\omega_k}(\xi-1/6)-\frac{\omega_k^{\prime \prime}}{4\omega_k^2}+\frac{3\omega_k^{\prime 2}}{8\omega_k^3}\,,\\
&\omega_k^{(4)}&=-\frac{1}{2\omega_k}\left(\omega_k^{(2)}\right)^2+\frac{\omega_k^{(2)}\omega_k^{\prime \prime}}{4\omega_k^3}-\frac{\omega_k^{(2)\prime\prime}}{4\omega_k^2}-\frac{3\omega_k^{(2)}\omega_k^{\prime 2}}{4\omega_k^4}+\frac{3\omega_k^\prime \omega_k^{(2)\prime}}{4\omega_k^3}\,.
\end{eqnarray}
It should be noted that we are performing the expansions in terms of $1/\omega_k\sim 1/k$, i.e. a short wavelength expansion. This means that the ultraviolet divergent terms of the ARP are encapsulated in the first lower powers of $1/\omega_k$, while the higher adiabatic orders represent finite contributions. As it can be appreciated, we are working off-shell, which means we are using an arbitrary scale $M$ instead of the physical mass $m$. This fact allows us to relate the adiabatically renormalized theory at two different scales \cite{Ferreiro:2018oxx}. As we are about to see this turns out to be a key point. As expected upon fixing the value of $M$ to the physical value of the mass of the quantum field we recover the on-shell theory. We have made the definition $\Delta^2\equiv m^2 -M^2$ which is a quantity of adiabatic order 2. 
\newline
\newline
We have now all the necessary ingredients to compute the ZPE coming from the quantum fluctuations of the non-minimally coupled scalar field. Remember $\langle 0|\delta\Phi| 0\rangle =0$ so we can grab only the quadratic fluctuations. By doing so we can plug \eqref{fluctuation_Fourier_expansion} into \eqref{Non_minimall_EMT} and taking into account \eqref{commutiation_relations}, we can work out the following expression
\begin{equation}
\begin{split}
&\langle T_{00}^{\delta \Phi}\rangle=\frac{1}{4\pi^2 a^2}\int dk k^2 \left[ \left|h_k^\prime\right|^2+(\omega_k^2+a^2\Delta^2)\left|h_k\right|^2 \right.\\
&\left. +\left(\xi-\frac{1}{6}\right)\left(-6\mathcal{H}^2\left|h_k\right|^2+6\mathcal{H}\left(h_k^\prime h_k^{*}+h_k^{*\prime}h_k\right)\right)\right]\,,
\end{split}\label{T00}
\end{equation}
where we have integrated over solid angle and expressed the final equation in terms of $k=|\vec{k}|$. Using \eqref{expanded_terms} we can expand the different involved terms, appearing in the above expression, up to the 4$th$ adiabatic order:
\begin{equation}\label{eq:exphk2}
\begin{split}
&|h_k|^2=\frac{1}{2W_k}=\frac{1}{2\omega_k}-\frac{\omega_k^{(2)}}{2\omega_k^2}-\frac{\omega_k^{(4)}}{2\omega_k^2}+\frac{1}{2\omega_k}\left(\frac{\omega_k^{(2)}}{\omega_k}\right)^2+\dots\\
\end{split}
\end{equation}
\begin{equation}
\begin{split}\label{eq:exphkp2}
&|h_k^\prime|^2 =\frac{\left(W_k^\prime\right)^2}{8W_k^3}+\frac{W_k}{2}= \frac{\omega_k}{2}+\frac{\omega_k^{(2)}}{2}+\frac{\omega_k^{(4)}}{2}+\frac{1}{8\omega_k}\left(\frac{\omega_k^\prime}{\omega_k}\right)^2\left(1-3\frac{\omega_k^{(2)}}{\omega_k}\right)
+\frac{\omega_k^\prime \omega_k^{(2)\prime}}{4\omega_k^3}+\dots
\end{split}
\end{equation}
\begin{equation}
\begin{split}
h_k^\prime h_k^{*}+h_k^{*\prime}h_k &= -\frac{W_k^\prime}{2W_k^2}=-\frac{\omega_k^{\prime}}{2\omega_k^2}\left( 1-\frac{2\omega_k^{(2)}}{\omega_k}\right)-\frac{\omega_k^{(2)\prime}}{2\omega_k^2}+\dots\\
\end{split}
\end{equation}
to finally get the cumbersome expression:
\begin{equation}
\begin{split}
\langle T_{00}^{\delta \Phi} \rangle & =\frac{1}{8\pi^2 a^2}\int dk k^2 \left[ 2\omega_k+\frac{a^4M^4 \mathcal{H}^2}{4\omega_k^5}-\frac{a^4 M^4}{16 \omega_k^7}(2\mathcal{H}^{\prime\prime}\mathcal{H}-\mathcal{H}^{\prime 2}+8 \mathcal{H}^\prime \mathcal{H}^2+4\mathcal{H}^4)\right.\\
&+\frac{7a^6 M^6}{8 \omega_k^9}(\mathcal{H}^\prime \mathcal{H}^2+2\mathcal{H}^4) -\frac{105 a^8 M^8 \mathcal{H}^4}{64 \omega_k^{11}}\\
&+\left(\xi-\frac{1}{6}\right)\left(-\frac{6\mathcal{H}^2}{\omega_k}-\frac{6 a^2 M^2\mathcal{H}^2}{\omega_k^3}+\frac{a^2 M^2}{2\omega_k^5}(6\mathcal{H}^{\prime \prime}\mathcal{H}-3\mathcal{H}^{\prime 2}+12\mathcal{H}^\prime \mathcal{H}^2)\right. \\
& \left. -\frac{a^4 M^4}{8\omega_k^7}(120 \mathcal{H}^\prime \mathcal{H}^2 +210 \mathcal{H}^4)+\frac{105a^6 M^6 \mathcal{H}^4}{4\omega_k^9}\right)\\
&+\left. \left(\xi-\frac{1}{6}\right)^2\left(-\frac{1}{4\omega_k^3}(72\mathcal{H}^{\prime\prime}\mathcal{H}-36\mathcal{H}^{\prime 2}-108\mathcal{H}^4)+\frac{54a^2M^2}{\omega_k^5}(\mathcal{H}^\prime \mathcal{H}^2+\mathcal{H}^4) \right)
\right]\\
&+\frac{1}{8\pi^2 a^2} \int dk k^2 \left[  \frac{a^2\Delta^2}{\omega_k} -\frac{a^4 \Delta^4}{4\omega_k^3}+\frac{a^4 \mathcal{H}^2 M^2 \Delta^2}{2\omega_k^5}-\frac{5}{8}\frac{a^6\mathcal{H}^2 M^4\Delta^2}{\omega_k^7} \right.\\
& \left. +\left( \xi-\frac{1}{6} \right) \left(-\frac{3a^2\Delta^2 \mathcal{H}^2}{\omega_k^3}+\frac{9a^4 M^2 \Delta^2 \mathcal{H}^2}{\omega_k^5}\right)\right]+\dots \label{EMTFluctuations}
\end{split}
\end{equation}
As mentioned before see \cite{Moreno-Pulido:2020anb} to know all the details. We can separate \eqref{EMTFluctuations} into two different pieces, one containing the divergent parts and the other one having those parts that are perfectly finite
\begin{equation}
\langle T_{00}^{\delta \Phi}\rangle (M)= \langle T_{00}^{\delta \Phi}\rangle_{\rm Div}(M)+\langle T_{00}^{\delta \Phi}\rangle_{\rm Non-Div}(M). \label{DecompositionEMT}
\end{equation}
Regarding the last part it is possible to solve the integrals (see Appendix B of \cite{Moreno-Pulido:2020anb}) and write the expression up to the fourth order as: %
\begin{equation}\label{Non-DivergentPart}
\begin{split}
\langle T_{00}^{\delta \Phi}\rangle_{\rm Non-Div}(M) &=\frac{m^2 \mathcal{H}^2}{96\pi^2}-\frac{1}{960\pi^2 a^2}\left( 2\mathcal{H}^{\prime \prime}\mathcal{H}-\mathcal{H}^{\prime 2}-2\mathcal{H}^4\right)\\
&+\frac{1}{16\pi^2 a^2}\left(\xi-\frac{1}{6}\right) \left(2\mathcal{H}^{\prime \prime}\mathcal{H}-\mathcal{H}^{\prime 2}-3\mathcal{H}^4\right)+\frac{9}{4\pi^2 a^2}\left(\xi-\frac{1}{6}\right)^2 (\mathcal{H}^\prime \mathcal{H}^2 +\mathcal{H}^4)\\
&+ \left(\xi-\frac{1}{6}\right)\frac{3\Delta^2 \mathcal{H}^2}{8\pi^2}+\dots. 
\end{split}
\end{equation}
The divergent part is made up by the remaining  terms of \eqref{EMTFluctuations}. Here we cannot follow exactly the momentum subtraction scheme because the vacuum diagrams do not have external momentum. Instead, we renormalize the ZPE by subtracting the terms that appear up to the $4th$ adiabatic order at scale $M$, as a consequence the divergent parts will disappear \cite{Birrell:1982ix,Parker:2009uva}. So let us follow the procedure mentioned and compute the renormalized ZPE through: 
\begin{eqnarray}\label{EMTRenormalized}
\langle T_{00}^{\delta \Phi}\rangle_{\rm Ren}(M)&=&\langle T_{00}^{\delta \Phi}\rangle(m)-\langle T_{00}^{\delta \Phi}\rangle^{\rm (0-4)}(M)\nonumber\\
&=&\langle T_{00}^{\delta \Phi}\rangle_{\rm Div}(m)-\langle T_{00}^{\delta \Phi}\rangle_{\rm Div}(M)-\left(\xi-\frac{1}{6}\right)\frac{3\Delta^2 \cH^2}{8\pi^2}+\dots,
\end{eqnarray}
where the only remaining term from $\langle T^{\delta\Phi}_{00}\rangle_{\rm Non-Div}(m)- \langle T^{\delta\Phi}_{00}\rangle_{\rm Non-Div}(M)$ is the last piece of the above expression and the dots represent the contributions of higher adiabatic orders. After some considerable amount of algebra (see \cite{Moreno-Pulido:2020anb} in order to follow the details of the process) it is possible to solve the convergent integrals and get the following result: 
\begin{equation}
\begin{split}
&\langle T_{00}^{\delta \Phi}\rangle_{\rm Ren}(M)=\frac{a^2}{128\pi^2 }\left(-M^4+4m^2M^2-3m^4+2m^4 \ln \frac{m^2}{M^2}\right)\\
&-\left(\xi-\frac{1}{6}\right)\frac{3 \mathcal{H}^2 }{16 \pi^2 }\left(m^2-M^2-m^2\ln \frac{m^2}{M^2} \right)+\left(\xi-\frac{1}{6}\right)^2 \frac{9\left(2  \mathcal{H}^{\prime \prime} \mathcal{H}- \mathcal{H}^{\prime 2}- 3  \mathcal{H}^{4}\right)}{16\pi^2 a^2}\ln \frac{m^2}{M^2}+\dots
\end{split} \label{RenormalizedExplicit2}
\end{equation}
where $\langle T_{00}^{\delta \Phi}\rangle_{\rm Ren}(M)$, as expected from  \eqref{EMTRenormalized}, vanishes for $M = m$ up to 4$th$ adiabatic order. The reason why this happens is because we are subtracting terms computed up to the same adiabatic order what ensures the cancellation of all the divergences. Those terms beyond the $4th$ order are subleading and finite corrections, that satisfy the Appelquist-Carazzone decoupling theorem \cite{Appelquist:1974tg}, i.e. for large values of the physical mass $m$, of quantum fields, they are completely suppressed. While it is true that we have successfully removed the divergent terms we still have to deal with the problematic terms, $m^4$, $M^4$ and $m^2M^2$, or generally denoted as the quartic mass power terms, in the sense that make the contribution to the induced vacuum energy density huge. We are going to take care of that in the following. 
\newline
\newline
In order to make possible the renormalization program, in the context of QFT in curved space-time, we need to count on the high derivative (HD) terms in the classical effective action of vacuum. As a result, the field equations take the different form
\begin{equation}\label{Modified_EQ}
\frac{1}{8\pi{G_N(M)}}G_{\mu\nu} + \rho_\Lambda(M)g_{\mu\nu} + a_1(M)H^{(1)}_{\mu\nu} = T^{\bar{\Phi}}_{\mu\nu} + \langle{T^{\delta\Phi}_{\mu\nu}}\rangle_{\rm Ren}(M).    
\end{equation}
Note the explicit dependence of the different pieces in $M$. The term $H^{(1)}_{\mu\nu}$, which contains the high-derivative terms can be written in this way
\begin{equation}
H^{(1)}_{\mu\nu} = \frac{1}{\sqrt{-g}}\frac{\delta}{\delta{g^{\mu\nu}}}\int d^4x\sqrt{-g}R^2 = -2\nabla_\mu\nabla_\nu{R} + 2g_{\mu\nu}\Box{R} - \frac{1}{2}g_{\mu\nu}R^2 +2R{R_{\mu\nu}}  
\end{equation}
and its $00$-component in the FLRW metric reads
\begin{equation}
H^{(1)}_{00} = -18a^2\left(\dot{H}^2 -2\dot{H}\ddot{H}-6H^2\dot{H}\right).   
\end{equation}
As we mentioned at the beginning of this section, we want to relate the theory at different scales. Being that our goal, let us subtract from \eqref{Modified_EQ} the same equation but evaluated at the arbitrary scale $M_0$. We are left just with the following terms:
\begin{equation}
\langle T_{\mu \nu}^{\delta \Phi}\rangle_{\rm Ren}(M)- \langle T_{\mu \nu}^{\delta \Phi}\rangle_{\rm Ren}(M_0)=f_{G_N^{-1}}(m,M,M_0)  G_{\mu \nu}+f_{\rho_\CC}(m,M,M_0) g_{\mu \nu}+f_{a_1}(m,M,M_0) H^{(1)}_{\mu \nu}.  	\label{EinsteinDifferentScale}
\end{equation}
It turns out to be useful to introduce the following notation:
\begin{equation}
f_X(m,M,M_0)\equiv X(M)-X(M_0), 
\end{equation}
with $X= G^{-1}_N/8\pi, \rho_\Lambda, a_1$ (note we have omitted the term $1/8\pi$ in $f_{G^{-1}_N}$ ) respectively. If we now compute the $00$-component for the above expression and we compare it with the previously found expression for the renormalized energy-momentum tensor \eqref{RenormalizedExplicit2} we can get
\begin{align}
 f_{G^{-1}_N}&= \frac{1}{16\pi^2}\left(\xi -\frac{1}{6}\right)\left[M^2 - M^2_0 -m^2\ln\left(\frac{M}{M_0}\right)^2\right]\\
 f_{\rho_\Lambda}&= \frac{1}{128\pi^2}\left[M^4 - M^4_0 -4m^2(M^2-M^2_0) + 2m^4\ln\left(\frac{M}{M_0}\right)^2\right]\\
 f_{a_1}&= \frac{1}{32\pi^2}\left(\xi -\frac{1}{6}\right)^2\ln\left(\frac{M}{M_0}\right)^{2}.
\end{align}
We consider the vacuum state as that one satisfying:
\begin{equation}
\rho_{\rm vac}(M) = \rho_{\Lambda}(M) + \frac{\langle T^{\delta\Phi}_{00} \rangle_{\rm Ren}(M)}{a^2},    
\end{equation}
where it is worth to be mentioned we maintain the quantities generally evaluated at the scale $M$. Subtracting now the same expression but this time evaluated at $M=M_0$
\begin{equation}
\begin{split}
&\rho_{\rm vac}(M)-\rho_{\rm vac}(M_0)=\rho_{\Lambda}(M)-\rho_{\Lambda}(M_0)+\frac{\langle T_{00}^{\delta \Phi}\rangle_{\rm Ren}(M)- \langle T_{00}^{\delta \Phi}\rangle_{\rm Ren}(M_0)}{a^2}\nonumber\\
&=f_{\rho_\Lambda}(m,M,M_0)+\frac{f_{G_N^{-1}}(m,M,M_0)  G_{00}+f_{\rho_\CC}(m,M,M_0) g_{00}+f_{a_1}(m,M,M_0) H^{(1)}_{00}}{a^2}\nonumber\\
&=\frac{f_{G_N^{-1}}(m,M,M_0) }{a^2}  G_{00}+\frac{f_{a_1}(m,M,M_0)}{a^2} H^{(1)}_{00}\nonumber\\
&=\frac{3\cH^2}{a^2}\,f_{G_N^{-1}}(m,M,M_0)-\frac{18}{a^4}\left(\mathcal{H}^{\prime 2}-2\mathcal{H}^{\prime \prime}\mathcal{H}+3 \mathcal{H}^4 \right)\,f_{a_1}(m,M,M_0)\,,\phantom{\,\,\,}\label{RenormalizedVEa}
\end{split}
\end{equation}
where we have used the expressions for $G_{00}$ and $H^{(1)}_{00}$ to work out the final result. Using the FLRW metric it is possible to get the result:
\begin{equation}
\begin{split}
\rho_{\rm vac}(M)
=&\rho_{\rm vac}(M_0)+\frac{3}{16\pi^2}\left(\xi-\frac{1}{6}\right) H^2\left[M^2 - M_0^{2}-m^2\ln \left(\frac{M}{M_0}\right)^2\right]\\
&-\frac{9}{16\pi^2} \left( \xi-\frac{1}{6}\right)^2 \left(\dot{H}^2 - 2 H\ddot{H} - 6 H^2 \dot{H} \right)\ln \left(\frac{M}{M_0}\right)^{2}. \label{RenormalizedVE}
\end{split}
\end{equation}
As can be appreciated we have expressed the final result using the Hubble function and its derivative in terms of the cosmic time. It is important to remark that \eqref{RenormalizedVE} has no implication on the cosmological constant problem since we are not providing a value for $\Lambda$ obtained from first principles. We need the value of $\rho_{\rm vac}(M_0)$ in order to obtain the value of $\rho_{\rm vac}(M)$. However from the beginning the intention was not to solve the CC problem but to find out how the vacuum energy density changes its value when we move from one renormalized point to another. In other words, we are interested in the possible running of $\rho_{\rm vac}$, and this is precisely what \eqref{RenormalizedVE} provides. The following comment is in order. If we look at the expression \eqref{final_VED_Higgs} we can appreciate that the ZPE part (the one that remains when we set $\Phi_{\rm cl} =0$) contains a dependence on the energy scale parameter $\mu$, which, as we stated if it is properly related with the characteristic energy scale of the studied process we can obtain an accurate behaviour of the possible running of the vacuum energy density. This $\mu$-dependence is also present in the running of the couplings of the QED and QCD. However we should bear in mind that it is an artificial parameter emerged from the dimensional regularization process. Nevertheless it can be very useful if the right choice is made, as it is clear, for instance, from the experimentally measured running of the electric charge. On the other hand, \eqref{RenormalizedVE}, obtained in a curved space-time described by the FLRW metric, depends on the physicals scales $M$ and $M_0$, which have been present from the very beginning and they have not arisen from some sort of dimensional regularization technique. Then it is a full QFT calculation. We will use later on this expression in order to motivate the running vacuum models. 

\subsection{Standard model of cosmology}\label{SubSec_1.3_Introduction_chapter}

Beyond any doubt Einstein's discovery of General Relativity can be considered as one of the most important breakthrough in the history of theoretical physics. The field equations turned into the perfect tool, that we were chasing for so long, that allows us to create a testable theory about the evolution of the Universe. We would like to remark the word \textit{testable} because thanks to the incredible amount of data we have nowadays at our disposal, we can rule out those models that do not pass the strict experimental tests. Therefore, any serious attempt of cosmological model should explain the following observational facts:
\begin{itemize}
\item The accelerated expansion of the Universe. 
\item The detected cosmic microwave background (CMB). 
\item The observed structure formation. 
\end{itemize}
In this section we are going to present in detail the basic elements of the current standard model of cosmology, i.e. the $\Lambda$CDM, which is built upon the assumption of the existence not only of the dark matter but also dark energy. Let us start by talking about the left-hand side of Einstein's equations, the one that deals with the geometry of the Universe.  
\newline
\newline
An essential object we need, in order to be able to characterize the geometry and the causal structure of the Universe is the metric. It is described by a symmetric tensor field, $g_{\mu\nu}$, and allows us to convert observer dependent coordinate distance into physical distance. We can write the invariant line element as:
\begin{equation}
ds^2 = g_{\mu\nu}dx^\mu{dx^{\nu}}.     
\end{equation}
The next question must be, which is the metric able to describe our Universe ? Well, first let us establish the requirements that the metric has to meet. The Cosmological Principle (CP) postulates that if our Universe is observed at enough large scales is homogeneous and isotropic. This means that independently on which point or which direction we are looking at, we are going to observe exactly the same properties like density, temperature, structure formation $\dots$ What is more, we do know that our Universe is in accelerated expansion, so this feature must be also incorporated in the metric. At the beginning the symmetric tensor $g_{\mu\nu}$ starts with 10 {\it d.o.f.} but because of the symmetries aforementioned, the remaining {\it d.o.f.} can be embodied in one function of time, called scale factor $a(t)$ and one constant parameter $k$ whose role is to determine the geometry of the Universe ($k=0$, for a flat Universe, $k= -1$ for a closed Universe and finally $k=+1$ for an open Universe). All in all, the line element, in spherical coordinates, can be written in the following way:
\begin{equation}
ds^2 = -dt^2 + a^2(t)\left[ \frac{dr^2}{1 -k\frac{r^2}{r^2_0}} + r^2\left(d\theta^2 + {\rm sin}^2\theta{d\phi^2}\right)\right].   
\end{equation}
From now on we are only going to consider flat cosmological models, $k = 0$. For studies on curved models see \cite{Park:2018bwy,Ryan:2018aif,Park:2018fxx,Park:2018tgj,Ryan:2019uor,Park:2019emi,Khadka:2019njj,Kazantzidis:2018rnb,Cao:2020jgu,Khadka:2020vlh}. The coordinates $x^i = (r,\theta,\phi)$ are the comoving coordinates and they are related to the physical ones through the relation $x^i_{\rm phys} = a(t)x^i$. This means that if we place ourselves in the comoving system we do not feel the Hubble expansion, we only perceive peculiar velocities with respect to the Hubble flow.
\newline
\newline
Basically, we obtain the information from the Universe thanks to the light emitted by the distant objects. Due to the expansion of the Universe the wavelengths of the photons are stretched out, so, the wavelength of a photon $\lambda(t)$ emitted at time $t$ will be observed at time $t_0$ with a wavelength 
\begin{equation}
\lambda_{0} = \frac{a(t_0)}{a(t)}\lambda(t).    
\end{equation}
If we use the traditional definition of the redshift $z = (\lambda_0 - \lambda)/\lambda$, we are lead to the following relation
\begin{equation}\label{z_relation_a}
z = \frac{1}{a(t)} -1    
\end{equation}
where we have established $a(t_0) = 1$, as usual. The above relation connects the theoretical prediction $a(t)$, once a model is given, with the observed redshift $z$. As it was mentioned before in Section. \ref{SubSec_1.1_Introduction_chapter} Edwin Hubble came up with a relation between the speed with which the galaxies are moving away from us and the distance at which they are. This relation can also be derived from \eqref{z_relation_a}. If one expands the scale factor around its present value $a(t)= 1 + \frac{\dot{a}(t_0)}{a(t_0)}(t-t_0) + \mathcal{O}((t-t_0)^2)$ (the dots represent derivative {\it w.r.t.} the cosmic time) and uses the definition of the Hubble function at present time, namely: $H_0 = \dot{a}(t_0)/a(t_0)$ it is possible to obtain the aforementioned relation:
\begin{equation}
z\simeq H_0{d}    
\end{equation}
where $d$ is the distance to the galaxies (remember in our units convention $c=1$, so actually what we have is $d= c(t_0-t)$). Once we have defined the metric we can plug it into the Einstein equations and compute the cosmological equations that will determine the evolution of the Universe. First of all, we need to compute the Christoffel symbols. We assume they are unequivocally determined by the metric through the Levi-Civita connection, thus: 
\begin{equation}  
\Gamma^{\mu}_{\nu\lambda} = \frac{1}{2}{g^{\mu\alpha}}(   \partial_{\lambda}g_{\alpha\nu}  + \partial_{\nu}g_{\alpha\lambda} - \partial_{\alpha}g_{\nu\lambda}   ), 
\end{equation}
then it is possible to calculate the Ricci tensor 
\begin{equation}
R_{\mu\nu} = \partial_\alpha{\Gamma^{\alpha}_{\mu\nu}} - \partial_{\nu}\Gamma^{\alpha}_{\mu\alpha} + \Gamma^{\alpha}_{\mu\nu}\Gamma^{\beta}_{\alpha\beta} - \Gamma^{\alpha}_{\mu\beta}\Gamma^{\beta}_{\alpha\nu} 
\end{equation}
whose non-null components are
\begin{align}
R_{00}&= -3\left(H^2 + \dot{H}\right) \\
R_{ij}&= a^2\left(2\dot{H} + 3H^2\right)\delta_{ij}\\
R_{0i}&=R_{i0}= 0.
\end{align}
Once we have the components of $R_{\mu\nu}$, it is possible to get the expression of the Ricci scalar scalar by employing the definition:
\begin{equation}
R \equiv g^{\mu\nu}R_{\mu\nu} = 12H^2 + 6\dot{H}.    
\end{equation}
In order to know all the details about the results that we just listed see \cite{Amendola:2015ksp} and references therein. Having reached this point, the only thing left to do is to write down the  components of the Einstein tensor
\begin{align}
G_{00}&= 3H^2 \\
G_{ij}&= -a^2\left(2\dot{H} + 3H^2\right)\delta_{ij}\\
G_{0i}&=G_{i0}= 0,
\end{align}
which remember, can be obtained upon the consideration of the definition $G_{\mu\nu}\equiv R_{\mu\nu} -(1/2)Rg_{\mu\nu}$. 
\newline
\newline
\subsubsection{Cosmic inventory and cosmological equations}
So far, we have been busy with the geometrical part of the field equations but now it is time to pay attention to the energy budget. Einstein's equations tell us that if we want to know the geometry of the Universe, which in the FLRW metric boils down to know the exact form of the scale factor, we need to know what are the species that fill it. We can classify the content of matter in relativistic and non-relativistic particles, neutrinos, which transit from the relativistic (in the early Universe) to the non-relativistic regime (close to the present time) and finally the mysterious dark energy. For each of the species we are going to define the energy density denoted as $\rho_i$, the pressure $p_i$ and the corresponding Equation of State (EoS) parameter $w_i\equiv p_i/\rho_i$. For all the species, except for DE simply because we do not have enough information about it, we can get the expressions for the energy density and the pressure by working out the following integrals:
\begin{align}
\rho_i &= g_{*}\int \frac{d^3p}{(2\pi)^3}f(p,m,\mu)E(p,m)\label{general_rho_Introduction_chapter}\\
p_i &= g_{*}\int \frac{d^3p}{(2\pi)^3}f(p,m,\mu)\frac{p^2}{3E(p,m)}\label{general_p_Introduction_chapter},
\end{align}
where $g_{*}$ is the number of degrees of freedom, $E(p,m) = \sqrt{p^2 + m^2}$ is the relativistic expression for the energy (from now on we are going to simply write $E$) and $p = |\vec{p}|$. We also introduced the concept of phase space occupancy, which in equilibrium gives us the number of particles inside a given region of the phase space, at temperature $T$. Depending on whether we are dealing with fermions (half odd integer spin) or bosons (integer spin) we have to consider different distribution forms for $f(p,m,\mu)$:
\begin{equation}
f(p,m,\mu)=\left\lbrace\begin{array}{c} f_{\rm FD} = \frac{1}{e^{\frac{E-\mu}{T}} +1}\quad (\rm fermions)\\ f_{\rm BE} = \frac{1}{e^{{\frac{E-\mu}{T}}} -1}\quad (\rm bosons) \end{array}\right.    
\end{equation}
where $f_{\rm FD}$ is the Fermi-Dirac distribution and $f_{\rm BE}$ is known as the Bose-Einstein distribution. The parameter denoted as $\mu$ is the chemical potential, which gives us an idea of the necessary energy to change the number of particles. 
\newline
\newline
At the background level we consider that all the species behave as perfect fluids. The energy-momentum tensor for a general perfect fluid can be written in this way
\begin{equation}
T_{\mu\nu} = (\rho_i + p_i)U_{\mu}U_{\nu} + p_i{g_{\mu\nu}}  
\end{equation}
where $U_{\mu} = (-1,0,0,0)$ is the 4-velocity of the fluid in the comoving frame. It is important to note that in the FLRW metric, at the background level, everything is a function only of time. Of course the $\rho_i$ and $p_i$ of each specie are not an exception. Because of the Bianchi identity $\nabla^\mu{G_{\mu\nu}}=0$, we have as a consequence the conservation of the energy-momentum tensor $\nabla^\mu{T_{\mu\nu}}=0$, which leaves us with the local energy conservation equation
\begin{equation}
\sum_i \dot{\rho}_i + 3H(\rho_i + p_i) = 0     
\end{equation}
where the index $i$ runs for all the components. If all the species are self-conserved, namely at the background level they do not interact with each other and the EoS parameter $w_i$ remains constant, we can obtain the expressions for the energy densities and pressures
\begin{equation}\label{general_rho_and_p_scale_factor_Introduction_chapter}
\rho_i = \rho^0_i{a^{-3(1 +w_i)}}\quad   p_i = w_i\rho_i. 
\end{equation}
In order to know the corresponding EoS parameters for each of the components we need to use the expressions \eqref{general_rho_Introduction_chapter}-\eqref{general_p_Introduction_chapter} and our knowledge of their microscopic behaviour. However, as we are about to see, the case of dark energy is a little bit special, since the form of the equation of state parameter depends on the model considered. 
\newline
\newline
\begin{large}
{\bf Photons}
\end{large}
\newline
\newline
Let us go into the details of each component of the Universe and the first one will be the relativistic particles called photons. As it is well-known photons are massless bosons with chemical potential completely negligible. Therefore it is possible to find the exact solutions to the integrals \eqref{general_rho_Introduction_chapter}-\eqref{general_p_Introduction_chapter}, and express them in terms of the photon temperature $T_\gamma$
\begin{align}\label{rho_photons_temperature}
\rho_\gamma &= \frac{g_*}{30}\pi^2{T^4_\gamma}\\
p_\gamma &= w_\gamma{\rho_\gamma} = \frac{1}{3}\rho_\gamma.\label{p_photons_temperature} 
\end{align}
In the case of photons $g_*=2$ because we have to consider the two different spin states. The CMB spectra can be modeled in a very good approximation as a black body and the corresponding present temperature has been measured with a high degree of precision, being its value $T_{\gamma}(a=1) = 2.72548\pm 0.00057$K \cite{Fixsen:2009ug}. This fact allows us to compute the current fraction (with respect to the critical energy density) at present time. If we consider the relation between the CMB temperature and the scale factor $T_\gamma(a) = T_\gamma(a=1)/a$ we can find the expression 
\begin{equation}
\Omega_\gamma(a) = \frac{\rho_\gamma(a)}{\rho^0_{c}} \simeq 0.000024718h^{-2}a^{-4}.
\end{equation}
The dependence $\Omega_\gamma(a)\sim 1/a^4$ was expected since from \eqref{p_photons_temperature} we now that the EoS is $w_\gamma =1/3$. Plugging this value in \eqref{general_rho_and_p_scale_factor_Introduction_chapter} we obtain the before mentioned dependency. As we shall see later, the photons are not the only component that contributes to the total radiation energy density. 
\newline
\newline
\begin{large}
{\bf Baryons and Dark Matter}
\end{large}
\newline
\newline
In this part we are going to get the expressions for the energy density and pressure of baryons and dark matter (DM). Both of them fit into the category of non-relativistic particles and this implies that the temperature is always much smaller that the rest mass. Considering then $T\ll m$ (note that we do not add any subindex to the temperature or the mass) we can find the approximate expressions
\begin{align}
\rho_{\rm NR} &= g_{*}m^{5/2}\left(\frac{T}{2\pi}\right)^{3/2}e^{\frac{\mu -m}{T}}\left[ 1 + \frac{3}{2}\left(\frac{T}{m}\right) + \mathcal{O}\left(\frac{T}{m}\right)^2\right]\label{rho_NR_Introduction_chapter}\\
p_{\rm NR} &= g_{*}T^{5/2}\left(\frac{m}{2\pi}\right)^{3/2}e^{\frac{\mu -m}{T}}\left[ 1 - \frac{5}{2}\left(\frac{T}{m}\right) + \mathcal{O}\left(\frac{T}{m}\right)^2\right]\label{p_NR_Introduction_chapter}. 
\end{align}
They are valid for both bosons and fermions. From the above expressions we can obtain the approximate value of the EoS parameter
\begin{equation}
w_{\rm NR} = \frac{p_{\rm NR}}{\rho_{\rm NR}} = \frac{T}{m}\left[ 1 -4\left(\frac{T}{m}\right) + \mathcal{O}\left(\frac{T}{m}\right)^2\right].  
\end{equation}
As we can appreciate even the first term of the EoS parameter is suppressed by the ratio $T/m\ll 1$, therefore as a good approximation we can take $w_{\rm NR}\simeq 0$. Contrary to what happens with photons, the expressions \eqref{rho_NR_Introduction_chapter}-\eqref{p_NR_Introduction_chapter} are not just functions of temperature. What is more, we do not know the exact relation between the temperature of the non-relativistic species and the scale factor as we do for photons. As a consequence, unlike with photons we need to obtain from the cosmological observations \cite{Aghanim:2018oex} the present values for the relative energy densities and then make use of \eqref{general_rho_and_p_scale_factor_Introduction_chapter} in addition to $w_b = w_{dm} = 0$, to get the final expressions for the corresponding energy densities, which can be written as:
\begin{align}
\rho_b(a) &= \rho^0_b{a^{-3}} \\
\rho_{dm}(a) &=\rho^0_{dm}{a^{-3}}. 
\end{align}
The following comment about dark matter is in order: despite the fact that we have no direct experimental evidence of its existence we do have strong indirect evidences that makes its presence absolutely necessary. Just to mention a couple of examples it is not possible to explain the current CMB spectrum as well as the observed structure formation without DM. We also know that the total amount of DM cannot be exclusively composed by hot dark matter (HDM), since this is ruled out by observations, and this is why we only consider cold dark matter (CDM) ($T/m\ll 1$). There are some possible candidates for DM like: WIMPs (Weak Interacting Massive Particles), axions, supersymmetric particles and even primordial black holes, but here we are not going to mess with none of them. 
\newline
\newline
\newline
\begin{large}
{\bf Neutrinos}
\end{large}
\newline
\newline
From the Standard Model of particles we know that there are three different generations of neutrinos with different flavor, electron neutrino $\nu_e$, muon neutrino $\nu_\mu$ and finally tau neutrino $\nu_\tau$. They are fermions and contrary to the theoretical prediction of the SM, it seems that at least one of them is a massive particle. So far, it has not been possible to measure their individual masses, in case all three are massive. A quantum mechanical phenomenon, called neutrino oscillations \cite{Pontecorvo:1957cp,Pontecorvo:1967fh} in which a flavor violation takes place, predicts that neutrinos can be created with a given flavor but as the neutrino propagates there exists a non-null probability to observe that neutrino having another flavor. Neutrino oscillations are only sensitive to the difference in the square of the masses and despite the fact that we have an experimental confirmation \cite{deSalas:2017kay} of this important phenomenon we can not extract the individual masses of the three different neutrinos as we have mentioned before. Nonetheless, taking into account the experimental evidence we should include massive neutrinos in the cosmic inventory.
\newline
\newline
In the early Universe neutrinos were in thermal equilibrium with the cosmic plasma, consequently they were described by the Fermi-Dirac distribution with zero chemical potential. At some point of the cosmic history (we do not know yet when exactly) they decoupled from the cosmic plasma yet their distribution still was the Fermi-Dirac one. The neutrinos transit from the relativistic regime (in the early Universe when they were part of the cosmic plasma) to the non-relativistic one (at late times). It is not possible to find the analytical expressions for $\rho_\nu$ and $p_\nu$ for all the cosmic history, nevertheless we are going to compute their expressions in the relativistic and non-relativistic limit. 
\newline
\newline
\begin{large}
{\bf Relativistic Neutrinos}
\end{large}
\newline
\newline
Of course, they are characterized by having a mass much smaller than their temperature $m_\nu/T_\nu\ll 1$, and this allows us to compute the relativistic limit for the energy density and the corresponding pressure
\begin{align}
\rho^{\rm R}_\nu &= \frac{7\pi^2}{240}g_{*}T^4_\nu\left[1 + \frac{1}{35\pi^2}\left(\frac{m_\nu}{T_\nu}\right)^2 + \mathcal{O}\left(\frac{m_\nu}{T_\nu}\right)^4\right]\\
p^{\rm R}_\nu &= \frac{7\pi^2}{720}g_{*}T_\nu^4\left[1 - \frac{1}{35\pi^2}\left(\frac{m_\nu}{T_\nu}\right)^2 + \mathcal{O}\left(\frac{m_\nu}{T_\nu}\right)^4\right].
\end{align}
Regarding the equation of state it can be seen that neglecting terms of order $\mathcal{O}\left(m_\nu/T_\nu\right)^2$, we have $w^{\rm R}_\nu = p^{\rm R}_\nu/\rho^{\rm R}_\nu \simeq 1/3$. Like photons $\rho^{\rm R}_\nu \sim T^4_\nu$ and the EoS parameter takes the value $w^{\rm R}_\nu = 1/3$. This should not be surprising since as we stated at the beginning we are considering relativistic neutrinos. It turns out to be convenient to relate the neutrino temperature with the photon temperature and then express the total amount of radiation. Due to the entropy conservation in the $e^{+}e^{-}$ annihilation we can have the following relation $T_\nu = (4/11)^{1/3}T_\gamma$ \cite{Dodelson:2003ft}, which tells us that the temperature of the still hypothetical neutrino background is in fact smaller than the CMB one.
\newline
\newline
Putting all the ingredients together we end up with the expression of the total radiation contribution:
\begin{equation}\label{total_radiation_density}
\Omega_r(a) = \frac{\rho_\gamma(a) + \rho^{\rm R}_\nu}{\rho^0_{c}}  = \left(1 + N_{\rm eff}\frac{7}{8}\left(\frac{4}{11}\right)^{4/3}\right)\Omega_\gamma(a) \equiv \Omega^0_r{a^{-4}}
\end{equation}
where we have defined $\Omega^0_r = (1 + N_{\rm eff}(7/8)(4/11)^{4/3})\Omega^0_\gamma \simeq 0.000041816h^{-2}$ being $N_{\rm eff} $ the effective number of relativistic neutrinos. Depending on if we have 3,2,1 or 0 relativistic neutrinos $N_{\rm eff} = 3.046, 2.033, 1.020$ and $0.006$ respectively.
\newline
\newline
\begin{large}
{\bf Non-Relativistic Neutrinos}
\end{large}
\newline
\newline
The transition between the relativistic regime to the non-relativistic regime happens approximately when $T_\nu \sim m_\nu$. Now we are going to work out the expressions when $m_\nu \gg T_\nu$, that should be understood as an approximation in the before mention limit not as the exact solution. The energy density $\rho^{\rm NR}_\nu$ and $p^{\rm NR}_{\nu}$ take the following form:
\begin{align}
\rho^{\rm NR}_\nu &= \frac{3}{4}\frac{g_*}{\pi^2}\zeta(3)m_\nu{T^3_\nu}\left[1 + \frac{15}{2}\frac{\zeta(5)}{\zeta(3)}\left(\frac{T_\nu}{m_\nu}\right)^2 + \mathcal{O}\left(\frac{m_\nu}{T_\nu}\right)^4 \right]\\
p^{\rm NR}_\nu &= \frac{15}{4}\frac{g_*}{\pi^2}\zeta(5)\frac{T^5_\nu}{m_\nu}\left[1 - \frac{189}{4}\frac{\zeta(7)}{\zeta(5)}\left(\frac{T_\nu}{m_\nu}\right)^2 + \mathcal{O}\left(\frac{m_\nu}{T_\nu}\right)^4 \right],
\end{align}
where $\zeta(s) = \sum^{\infty}_{n=1}1/n^s $ is the Riemann zeta function and $\zeta(3)\simeq 1.20205$, $\zeta(5)\simeq 1.03692$ and $\zeta(7)\simeq 1.00834$. From the above expressions it can be clearly seen that $w^{\rm NR}_\nu = p^{\rm NR}_\nu/\rho^{\rm NR}_\nu \sim (T_\nu/m_\nu)^2 $ and taking into account that the ratio between the temperature and the mass is strongly suppressed we can take as a very good approximation $w^{\rm NR}_\nu\simeq 0$. We remark, once again, the need to solve numerically the expressions for neutrinos since we have no analytical formulas for the whole cosmic history. However, the expressions obtained in this section can be used to understand the behaviour of neutrinos in some particular limits.
\newline
\newline
The total non-relativistic contribution can be written as:
\begin{equation}
\Omega_m(a) = \frac{\rho_b(a) + \rho_{dm}(a) +\rho^{\rm NR}_\nu(a)}{\rho^0_c}    
\end{equation} 
which only scales like $\Omega_m(a)\sim a^{-3}$ when we consider as an approximation that all neutrinos are massless. If not, due to the non-analytical expression for the massive neutrinos we only can have its numerical value as a function of the scale factor. 
\newline
\newline
\begin{large}
{\bf Dark Energy}
\end{large}
\newline
\newline
Let us mention some of the most important features that have been attributed to DE. This mysterious form of energy, that permeates every corner of the Universe, has not been directly detected yet, nonetheless its existence is well supported on three different pillars:
\begin{itemize}
\item The observed change in the wavelength of the light incoming from the distant object, requires a Universe in accelerated expansion since approximately $z\sim 1$. Neither matter nor radiation are capable of producing such an effect, there has to be another component with negative pressure and this role can be played by DE. 
\item The current observations, point out to a Universe whose energy density is close to the critical one. Since the joint matter and radiation contribution is $\simeq \rho^0_{c}/3$ we need another component filling the remaining part. 
\item The observations of large-scale structure (BAO, structure formation and CMB) requires unavoidably the presence of DE. 
\end{itemize}
The first item comes from the SNIa observations, since thanks to the measured luminosity \cite{Riess:1998cb,Perlmutter:1998np} we had for the first time an indirect prove of the existence of DE. In order to have an accelerated expanding Universe the following condition must be fulfilled $\ddot{a} \sim -\sum_i (\rho_i + 3p_i) > 0 $ and this is only possible if there is some component, called generically dark energy, with EoS parameter $w_{\rm DE}<-1/3$. While this is the minimum requirement, we can consider different types of DE depending on the value of the EoS parameter. If we have $w_{\rm DE} = -1$ we can talk of vacuum energy. This type of DE has been already developed in previous sections and is going to be widely studied in this thesis. On the other hand if $w_{\rm DE}\lessapprox -1$ we have quintessence DE, while for $w_{\rm DE}\gtrapprox -1$ we have phantom DE. We want to stay open-minded and this is why we will study a wide range of cosmological models containing different assumptions about DE in order to see which ones are favoured by the cosmological data and which ones are not. 
\newline
\newline
Once we have computed the left-hand side of Einstein's equations, i.e. the part containing the geometrical terms, and we have given the details of the cosmic inventory we are finally in position to write, the cosmological equations for the standard model of cosmology known as $\Lambda$CDM, down: 
\begin{align}
&H^2 = \frac{8\pi{G_N}}{3}\left(\rho_\gamma(a) + \rho_b(a) + \rho_{dm}(a) + \rho_\nu(a) + \rho_{\rm vac}\right)\\
&3H^2 + 2\dot{H} = -8\pi{G_N}\left(p_\gamma(a) + p_\nu(a) + p_{\rm vac}\right).
\end{align}
It is important to note that in the case of the $\Lambda$CDM model, $\rho_{\rm vac} = \rho^0_\Lambda$ and $p_{\rm vac} = -\rho^0_\Lambda$ are constants since no evolution in time is considered within the standard model of cosmology. We denote the total neutrino contribution as $\rho_\nu \equiv \rho^{\rm R}_\nu + \rho^{\rm NR}_\nu$ while for the pressure we have $p_\nu \equiv p^{\rm R}_\nu + p^{\rm NR}_\nu$. We can easily work out the expression for the normalized Hubble function in terms of the relative energy densities $\Omega^0_i = \rho^0_i/\rho^0_c$
\begin{equation}
E^2(a) \equiv \frac{H^2(a)}{H^2_0} = \left(\Omega^0_b + \Omega^0_{dm}\right)a^{-3}  + \Omega^0_\gamma{a^{-4}}  + \frac{\rho_\nu(a)}{\rho^0_c} + \Omega^0_{\Lambda}
\end{equation}
This function is fundamental to characterize a cosmological model at the background level. 
\newline
\newline
Now that we have the expression of the Hubble function for the $\Lambda$CDM, as well as, the energy density for each component, we can compute two important moments in the cosmic history: the equality time between matter and radiation and the moment where the Universe transits from the decelerating regime to the accelerated one. For the sake of simplicity we will consider that all neutrinos are massless and of course relativistic, i.e. $N_{\rm eff} = 3.046$ in \eqref{total_radiation_density}.
\newline
The first of the aforementioned points is easy to find since all we need to do is equal the corresponding expressions for energy density of matter and radiation $\rho_r(z_{\rm eq}) = \rho_m(z_{\rm eq})$ and find the value  $z_{\rm eq}$ that meets the equality
\begin{equation}
z_{\rm eq} = \frac{\Omega^0_m}{\Omega^0_r} -1 \simeq 3405.  \end{equation}
Where we have used the values $\omega_b = 0.02237$ and $\omega_{dm} = 0.1200$ from the Planck 2018 TTTEEE+lowE+lensing \cite{Aghanim:2018eyx} results. Beyond this point, the non-relativistic matter becomes the dominant component of the Universe which means that the structure formation process can start since the pressure from the radiation component progressively decreases its preventive effect. At the recombination time, when baryons were released from photons, the structure formation process speeds up, and gives rise to the structure we can observe in the present time. 
\newline
\newline
Regarding the second point we need to employ the definition of the deceleration parameter commonly denoted as
\begin{equation}\label{deceleration_parameter}
q \equiv -\frac{\ddot{a}}{aH^2} = -1 -\frac{a}{2H^2}\frac{dH^2}{da}.    \end{equation}
Obviously the transition point has to happen much later than the period where the radiation was the dominant component, consequently we can neglect the contribution of $\Omega_r(a)$ to simplify the expression. In the $\Lambda$CDM, under the considerations that we mentioned, the deceleration parameter takes the form
\begin{equation}
q = \frac{1}{2}\frac{\Omega^0_m{a^{-3}} -2\Omega^0_\Lambda }{\Omega^0_\Lambda + \Omega^0_m{a^{-3}}}.  
\end{equation}
To find the transition point $z_{\rm tr}$ we only need to impose the condition $q = 0$, which is fulfilled by the following value of the redshift
\begin{equation}
z_{\rm tr} = \left(\frac{2\Omega^0_\Lambda}{\Omega^0_m}\right)^{1/3} -1    
\end{equation}
At this point, where the CC term is the dominant component, the Universe enters in a phase of accelerating expansion where galaxies are moving away from each other.
\newline
\newline
\subsubsection{Inflation}
It is worthwhile to dedicate a few words to explain inflation, which is a brief period of the cosmic history where an exponential expansion of the space takes place during the early Universe \cite{Starobinsky:1980te,Guth:1980zm,Linde:1981mu,Albrecht:1982wi}. It is important to make crystal-clear the fact that we have no direct experimental evidence for inflation, however, as it happens with DE, the indirect evidences supporting inflation are numerous and it turns out to be really complicate to explain the current Universe without taking it into account. So, hereafter we assume inflation as a part of the standard model of cosmology. This is the reason why a short explanation of the general features of the  mechanism and its main consequences are required. See \cite{Freese:1990rb,Boubekeur:2005zm,Kachru:2003sx,GarciaBellido:2001ky,Goncharov:1985yu,Stewart:1994ts,Dvali:1998pa,Burgess:2001vr,Cicoli:2008gp,Dobado:1994qp} to know more about different inflationary models.  
\newline
\newline
As we have already mentioned, the inflationary phase consists in a very rapid expansion of the space which happens after the Big Bang and before the radiation dominated epoch. As we have seen previously, in order to have an accelerating expanding Universe, the energy budget, must be dominated by a component with negative pressure. As expected there are several ways to implement inflation. Probably, one of the most studied types in the literature, are the so-called {\it inflaton} models where inflation is driven by a scalar field, called inflaton. Even if we cannot say that there is a standard model for inflation, here we are going to present the most important features of it within the context of inflaton models. To know more about alternatives to inflaton models see \cite{Sola:2015rra}. The associated energy density and pressure can be written in the usual way for a scalar field model:
\begin{align}
\rho_\Phi &= \frac{1}{2}\dot{\Phi}^2 + U(\Phi)\label{Inflation_rho_Introudction_chapter}\\
p_\Phi &= \frac{1}{2}\dot{\Phi}^2 - U(\Phi) \label{inflation_p_Introduction_chapter}
\end{align}
where $\Phi$ is a 1-dimensional field in natural units and $U(\Phi)$ is the potential considered in the different inflationary models. The dots represent derivative with respect to the cosmic time. As it can be appreciated from \eqref{inflation_p_Introduction_chapter} what it takes to get a negative pressure is a field configuration where the potential energy density dominates over the kinetic term. This can be realized if we consider the slow-roll approximation where the scalar field slowly evolves from the false vacuum of the potential towards the absolute minimum. Due to the slow evolution, the kinetic energy can be completely neglected and $\rho_\Phi$ can be well approximated as a constant. In this scenario, clearly the energy density associated to the scalar field quickly becomes the dominant component of the Universe which means we can write the Friedmann equation as:
\begin{equation}
\frac{1}{a}\frac{da}{dt} = \sqrt{\frac{8\pi{G_N}\rho_\Phi}{3}} \equiv H_\Lambda = {\rm const}.    
\end{equation}
The above differential equation can be easily solved to find the scale factor for the inflationary period
\begin{equation}\label{scale_factor_inflation_Introduction}
a(t) = a_{0}e^{H_\Lambda(t-t_{0})} \quad t>t_0
\end{equation}
where $a_{0}$ and $t_{0}$ are the scale factor and the cosmic time at the beginning of inflation respectively. As the scalar field goes down the hill of the potential, the potential energy density decreases whereas the kinetic energy increases. The end of inflation takes place when $\dot{\Phi}^2\simeq U(\Phi)$, i.e. the potential term is no longer the dominant one, and as a consequence, the pressure is not negative anymore being the ultimate consequence the reduction of the accelerated expansion of the Universe. Once this has happened the scalar field falls into the ground state and starts to oscillate around the minimum, giving rise to the process known as reheating. Unfortunately our knowledge about this mechanism is very poor but the main idea is to transfer the remaining energy, from the inflaton field, to the creation of standard model particles which filled the Universe. 
\newline
\newline
The natural question, having reached this point, is: why do we need an early period of expansion ? So far, we have only explained what is inflation and the main features of a toy model that allows us to implement it via a scalar field. It is time then to list the problems that inflation solves: %
\begin{itemize}
\item {\bf Horizon Problem:} As we do know now the CMB is highly isotropic. Regardless of the direction we look at in the sky we observe an spectra coming from an approximate black body with temperature $T_{\gamma}(a=1) = 2.72548\pm 0.00057$K \cite{Fixsen:2009ug} and tiny anisotropies of order $\Delta{T_\gamma}/T_\gamma\sim \mathcal{O}(10^{-5})$ \cite{Aghanim:2018eyx}. This fact points out that early on the CMB photons were in thermal equilibrium. After some moments of reflection a question pops up in our mind, how is this possible ? Should not photons coming from different directions have different temperature since they were never in causal contact ? What is more, we can extend these type of question to the structure formation observed in the Universe, how is possible that the Universe is isotropic and homogeneous ? This is known as the horizon problem. Inflation provides a beautiful solution to this problem since previously to the rapid expansion, caused by the inflation, the particles were in causal contact, reaching the thermal equilibrium, and it was after the end of inflation when the particles became disconnected from each other and evolve independently.
\item {\bf Flatness Problem}: From the Friedmann equation we can obtain the following relation valid for any value of the scale factor $1 -\Omega_{\rm tot}(a)= \Omega_k(a) = -k/\dot{a}^2$, where $k$ is the curvature parameter which determines the geometry of the Universe. If we derive the relation {\it w.r.t.} the cosmic time we are lead to the relation
\begin{equation}
\frac{d}{dt}\left(1 - \Omega_{\rm tot}(a)\right) = -2\frac{\ddot{a}}{\dot{a}}\left(1 - \Omega_{\rm tot}(a)\right). 
\end{equation}
For most of the cosmic history, until $z\sim 1$, the Universe has undergone a phase of decelerating expansion, which means that $\ddot{a}/\dot{a} = -|\ddot{a}/\dot{a}|$, consequently the quantity $\left(1 - \Omega_{\rm tot}(a)\right)$ grows with time. If now, we take the value from Planck 2018 TTTEEE+lowE+lensing+BAO \cite{Aghanim:2018eyx}, $\Omega^0_k = 0.0007\pm 0.0019$, we can infer that in order to get such a small value for $\Omega^0_k$ the value of $\Omega_k(a)$, at the inflation time, had to be really small. This is known as the flatness problem. Without inflation, we need to invoke a very precise fine-tuning in the initial conditions. However with inflation this situation is reached naturally due to the fact that, using \eqref{scale_factor_inflation_Introduction} we can get $1-\Omega_{\rm tot}(a)\sim e^{-2H_\Lambda(t-t_0)}$, which means that because of inflation $\Omega_{\rm tot}\simeq 1$ right after the end of inflation.
\item {\bf Exotic-relics problem:} Some exotic particles, such as the magnetic monopole, are predicted within a given model of grand unified theory, however they have never been observed. An accelerated expansion of the Universe, as the one predicted by inflation, could have diluted their density, making today almost impossible to observe them. 
\item {\bf Origin of the structure formation:} One of the most important advantages of inflation is the obtaining of a mechanism capable to explain the origin of the structure formation. The quantum fluctuations of the inflaton field, generated during inflation, are stretched up to macroscopic scales due to the exponential scale factor. It is generally claimed that these amplified fluctuations later served as the seed of the structures that we currently observe. 
\end{itemize}
\subsubsection{Problems and tensions of the $\Lambda$CDM model}\label{SubSec_1.4}
So far, the cosmological standard model, with the CC as a building block, turned out to be very successful in explaining the different phenomena observed. The explanation of the well-known baryonic acoustic oscillations (BAO) or the accurate prediction of the CMB spectra which contains anisotropies in the temperature $\Delta{T_\gamma}/T_\gamma\sim \mathcal{O}(10^{-5})$ are good examples yet not the only ones.
\newline
However, as we shall see in this section, the $\Lambda$CDM suffers from some problems and tensions, which may be interpreted either as systematic errors affecting the measurements or as a signal of new physics. Since we have not the ability to discern whether or not the observational data are affected by some sort of systematic errors we are going to focus on the theoretical problems and how they can be related with the possibility of having physics beyond the standard model. The first one of them has already been treated in extension, the cosmological constant problem, the tremendous mismatch between the predicted value for the vacuum energy density in the context of QFT, namely $\rho^{\rm QFT}_{\Lambda,{\rm ind}}$ and the observed value $\rho^0_\Lambda$ which turns out to be $\rho^{\rm QFT}_{\Lambda,{\rm ind}}/\rho^{0}_{\Lambda}\sim \mathcal{O}(10^{55})$ in natural units. Despite the great efforts made so far we have to admit we are still far from solving the problem. The available cosmological data can only be explained, in the context of GR, if there is some form of repulsive energy which causes the accelerated expansion of the Universe. The $\Lambda$CDM contemplates the simplest of all possibilities, namely a constant vacuum energy density whose value is equal to the observed one $\rho_{\rm vac} = \rho^0_\Lambda$ at any time in cosmic history. Even if the match in the comparison among the cosmological data and the theoretical predictions within the $\Lambda$CDM is astonishing, we still have some questions pending to deal with: which is the origin of the term $\rho^0_\Lambda$ ? Why precisely this tiny value ? Why this term remains constant while the other components evolve as the Universe expands ? The pressing need to answer these type of questions is one of the main reasons that drives the seek for other cosmological models beyond the $\Lambda$CDM. 
\newline
Another theoretical problem related with $\rho^0_\Lambda$ is the cosmic coincidence problem. It can be formulated simply with just one question: why, precisely now, $\rho^0_\Lambda/\rho^0_m\sim \mathcal{O}(1)$ ? While it is true that this fact can be just a coincidence, it is completely normal to ask ourselves, how on earth, if $\rho_m\sim a^{-3}$ while $\rho^0_\Lambda$ remains constant, they are of the same order in the present time ? As stated in \cite{Gomez-Valent:2017tkh} in order to consider that this fact is just a coincidence $\rho^0_\Lambda$ should not be linked with $\rho_m(a)$ and the equality of both energy densities could have happened at any moment of the cosmic history. It is not the main goal of this thesis to go into the details of this controversial problem, to know more about the issue \cite{Bianchi:2010uw} and references therein. 
\newline
The cosmological constant problem, as well as, the cosmic coincidence problem are theoretical problems that are rooted in our inability to explain the origin and the possible evolution of the $\rho_{\rm vac}$ term. Nevertheless, the $\Lambda$CDM is affected also by problems related with the apparent discrepancy between the obtained values for some parameters when different types of cosmological data are employed. Let us briefly summarize (they will be widely discussed in the next chapters) the main two problems of this type. 
\newline
The first of them is known as the $\sigma_8$-tension, and as the name already indicates, affects the value of the parameter $\sigma_8(z)$, the root-mean-square mass fluctuations on $R_8 = 8h^{-1}$Mpc scales
\begin{equation}
\sigma^2_8(z)=\frac{1}{2\pi^2}\int^{\infty}_{0}dk{k^2}P_m(k,z)W^2(kR_8),  
\end{equation}
where $P_m(k,z)$ is the matter power spectrum and $W(kR_8)$ is the top hat smoothing function. Recently, instead of quantifying the tension through the parameter $\sigma_8(0)$, another parameter has been employed  
\begin{equation}
S_8 = \sigma_8(0)\sqrt{\frac{\Omega^0_m}{0.3}} 
\end{equation}
which as can be appreciated also involves the value of the relative matter energy density at present time. 
The obtained value for the above parameter in the $\Lambda$CDM model, considering the data Planck 2018 TTTEEE+lowE+lensing \cite{Aghanim:2018eyx} is $S_8=0.832\pm 0.013$. On the other hand, the values obtained by different collaborations, using the weak gravitational lensing of galaxies, are DES \cite{Troxel:2017xyo} $S_8 = 0.777^{+0.036}_{-0.038}$, KiDS-450 \cite{Hildebrandt:2016iqg} $S_8=0.745\pm0.039$ and HSC \cite{Hikage:2018qbn} $S_8=0.780^{+0.030}_{-0.033}$ respectively. In order to properly quantify the tension among the obtained value by the Planck collaboration and the weak gravitational lensing measurements we use the following useful expression
\begin{equation}
T_X = \frac{|X_i - X_j|}{\sqrt{\sigma^2_{X_i} + \sigma^2_{X_j}}}
\end{equation}
where $X$ is the observable we are considering to account for the tension and $X_i,X_j$ are the measurements that are under comparison. Therefore, with respect to the Planck value, the tension for  the before mentioned collaborations takes the value, DES $T_{S_8} = 1.40\sigma$, KiDS $T_{S_8} = 2.12\sigma$ and HSC $T_{S_8} = 1.53\sigma$. Only in the KiDS case the tension is above $2\sigma$ so we can say that for the moment the tension is far from reaching a critical level. However it is important to keep an eye on the future measurements obtained from the weak gravitational lensing to see if the tension completely disappears or conversely, starts to increase. It is worthwhile to mention some models able to relieve the tension, some examples are: the running vacuum models \cite{Sola:2016jky,Sola:2016ecz,Sola:2017lxc}, models that include the $A_L$ lensing phenomenological parameter \cite{Aghanim:2018eyx,DiValentino:2015ola}, modified gravity models \cite{Ade:2015rim,DiValentino:2015bja} or a Brans-Dicke model with a cosmological constant \cite{Sola:2019jek}.
\newline
\newline
Without doubt, the most worrisome and consequently the most studied of the tensions is the $H_0$-tension, namely, the great discrepancy between the value from Planck 2018 TTTEEE+lowE+lensing \cite{Aghanim:2018eyx}  $H_0=67.36\pm0.54$ km/s/Mpc and the local measurements from the SH0ES collaboration \cite{Reid:2019tiq} $H_0=73.5\pm 1.4$ km/s/Mpc obtained with the cosmic distance ladder method using and improved calibration of the Cepheid period-luminosity relation. As can be easily checked the tension between the Planck value and the SH0ES one reaches the disturbing level of $T_{H_0} = 4.09\sigma$. The persistence in the discrepancies over the last few years turns out to be really puzzling since we are still far to conclude if the tension is rooted either in some sort of still unknown systematic or perhaps is the clearest sign we have right now of new physics. There are numerous attempts in the literature trying to solve this tension, here we just mention some of them: invoking an extra relativistic neutrino {\it d.o.f.} \cite{Aghanim:2018eyx,Jacques:2013xr,Carneiro:2018xwq,DiValentino:2015sam}, a coupled DE with an scalar field potential \cite{Gomez-Valent:2020mqn} and early dark energy models (EDE) \cite{Karwal:2016vyq, Banihashemi:2018oxo}. Another interesting possibility is the modification of the standard model at the recombination period since it has been proved that the value of $H_0$ is closely related with the comoving sound horizon $r_d$ \cite{Bernal:2016gxb,Chiang:2018xpn,Hart:2017ndk,Jedamzik:2020krr}. The aforementioned options are based on the idea to solve the tension within the GR paradigm, however the consideration of the Brans-Dicke theory opens a wide range of interesting possibilities to alleviate the tension \cite{Sola:2019jek,Sola:2020lba}.  
\newline
Other cosmological data can shed some light on the issue. The SPTPol survey \cite{Henning:2017nuy}, which also deals with CMB data, obtains a higher value for the Hubble constant $H_0 = 71.29\pm 2.12$ km/s/Mpc. The data extracted, for the H0LICOW team \cite{Wong:2019kwg}, from six gravitational lensed quasar point out also to a high value $H_0=73.3^{+1.8}_{-1.7}$km/s/Mpc, which is in tension with the value obtained, also from strong gravitational data, by the TDCOSMO+SLACS team \cite{Birrer:2020tax} $H_0 = 67.4^{+4.1}_{-3.2}$km/s/Mpc which clearly tells us that there are still a lot of stuff to understand. Finally the Gravitational Waves field turns out to be really promising and it is claimed that in the next decade it will provide accurate data \cite{Palmese:2020aof,Yu:2020vyy,Borhanian:2020vyr}. 
\newline
A final comment about the aforementioned $\sigma_8$-tension and $H_0$-tension is in order. We have to bear in mind that, while one tries to solve one of the tensions, should not make the other one worse. The $H_0$ and $\sigma_8$ are correlated so we have to be careful to not accommodate to much one of the values while we lost sight the other one. This is the only acceptable way to deal with these intriguing tensions. 
\subsection{Alternatives to the $\Lambda$CDM}\label{SubSec_1.4_Introduction_chapter}
In this section we present some of the models beyond the standard one that have been proposed over the years.  What all of them have in common, of course, is the fact that they deviate in one way or another from the theoretical framework where the $\Lambda$CDM is placed. As it is well-known the lack of a convincing explanation for the accelerated expansion of the Universe has motivated cosmologists to look for alternatives that do. Not only we are chasing for models able to provide a more physical mechanism for the expansion of the Universe but also we are interested in those models that can better handle the aforementioned tensions that affect the concordance model. No need to say that owing to the tremendous success of the $\Lambda$CDM, in explaining most of the cosmological observations, the alternative models should not depart to much from its theoretical predictions. For instance, as we shall see by considering a mild evolution of the rigid cosmological constant term we can fit the CMB data as well as the $\Lambda$CDM does but we can also improve the description of the large scale structure data.  In regards to the cosmological data, the increasing number of new and very precise measurements will allow us to put tight constraints over the models making possible to rule out those that do not pass the observational tests. In the meantime we should remind open-minded and consider all the possibilities. 
\subsubsection{Dark energy parameterizations}
The simplest possible extension of the $\Lambda$CDM model consists in endowing the DE density, $\rho_{\rm DE}$ with some sort of evolution over time and consider, at the same time, that it is self-conserved, i.e. $\dot{\rho}_{\rm DE} + 3H(1 + w_{\rm DE}(a))\rho_{\rm DE} = 0$. While it is possible to build models just by considering complicate functional forms for the EoS parameter, $w_{\rm DE}(a)$, here we want to keep things as simple as possible, this is why we only study two well-known different parameterizations, namely the XCDM \cite{Turner:1998ex} and the CPL \cite{Chevallier:2000qy,Linder:2002et,Linder:2004ng}. The first of them consists in considering $w_{\rm DE}(a) = w_0 = {\rm const}.$, nevertheless, this model differs from the standard one because $w_0\neq-1$, consequently the corresponding dark energy density can be written as $\rho_{\rm DE} = \rho^0_\Lambda{a^{-3(1 + w_0)}}$. Being $\rho^0_\Lambda$ the present measured value. The normalized Hubble function can be written as:
\begin{equation}
E^2(a) = (\Omega^0_b + \Omega^0_{dm})a^{-3} + \Omega^{0}_\gamma{a^{-4}} + \frac{\rho_\nu(a)}{\rho^0_c} + \Omega^0_\Lambda{a^{-3(1+w_0)}}.   
\end{equation}
For $w_0 = -1$ it boils down to that of the $\Lambda$CDM with rigid CC, as expected. On the other hand for $w_0 \gtrapprox -1$ the XCDM mimics quintessence, whereas for $w_0 \lessapprox -1$ it mimics phantom DE. 
\newline
A slightly more sophisticated DE parameterization is furnished by the CPL, in which  one assumes that the generic DE entity has a slowly varying EoS of the form
\begin{equation}
w_{\rm DE}(a) = w_0 + w_1(1-a) = w_0 + w_1\frac{z}{1+z}. 
\end{equation}
The CPL parameterization, in contrast to the XCDM one, gives room for a time evolution of the dark energy EoS owing to the presence of the additional parameter $w_1$, which satisfies $0<|w_1|\ll |w_0|$ with $w_0\gtrapprox-1$ or $w_0 \lessapprox-1$. The corresponding normalized Hubble function for the CPL can be easily computed: 
\begin{equation}
E^2(a) = (\Omega^0_b + \Omega^0_{dm})a^{-3} + \Omega^{0}_\gamma{a^{-4}} + \frac{\rho_\nu(a)}{\rho^0_c} + \Omega^0_\Lambda{a^{-3(1+w_0+w_1)}}e^{-3w_1(1-a)}.   
\end{equation}
Both the XCDM and the CPL parameterizations can be thought of as a kind of baseline frameworks to be refereed to in the study of dynamical DE. They can be used as a fiducial models to which we can compare other, more sophisticated, models for the dynamical DE. 

\subsubsection{Running vacuum models}
This kind of models is build upon the idea that the vacuum energy density should be a time depending quantity in cosmology. It is difficult to conceive an expanding Universe with a vacuum energy density that has remained immutable since the origin of time. Rather, a smoothly evolving energy density that inherits its time-dependence from cosmological variables, such as the Hubble rate $H(t)$, or the scale factor $a(t)$ is not only a qualitatively more plausible and intuitive idea, but it is also suggested by fundamental physics, in particular by QFT in curved space-time.
\newline
\newline
Remember that the renormalized ZPE contribution coming from the quantum fluctuations of a non-minimally coupled scalar field, in the FLRW metric, gives rise to the expression \eqref{RenormalizedVE}, which is useful to explore the running of the $\rho_{\rm vac}$, when we move from one scale to another. Once we insert the observational obtained value, at some particular scale, we are able to compute the value at another scale. Having said that, let us first relate the different values of the vacuum energy density at the present energy scale characterized by $M=H_0$ with the typical energy scale of most of the grand unified theories (GUT) $M_0 = M_X\sim\mathcal{O}(10^{16})$ GeV. We are interested in the description of the current Universe, consequently, those terms of order $\sim\mathcal{O}(H^4)$ (which include the terms $\dot{H}^2$, $H\ddot{H}$ and $\dot{H}H^2$) can be neglected. Considering what we just said, \eqref{RenormalizedVE}, takes the following form:   
\begin{equation}
\rho_{\rm vac}(M = H_0) \equiv \rho^0_\Lambda = \rho_{\rm vac}(M_0 = M_X) + \frac{3}{16\pi^2}\left(\xi -\frac{1}{6}\right)H^2_0\left[M^2_X + m^2\ln\left(\frac{H_0}{M_X}\right)^2\right].
\end{equation}
Where $\rho^0_\Lambda$ is the observational measurement at present time. The above expression can be simplified by defining the running parameter $\nu_{\rm eff}$: 
\begin{equation}\label{rho_Lambda_at_M_X}
\rho^0_{\Lambda}= \rho_{\rm vac}(M_0=M_X) + \frac{3\nu_{\rm eff}}{8\pi}H^2_0{M^2_{\rm pl}}
\end{equation}
where
\begin{equation}
\nu_{\rm eff} = \frac{1}{2\pi}\left(\frac{1}{6} -\xi\right)\frac{M^2_X}{M^2_{\rm pl}}\left(1 + \frac{m^2}{M^2_X}\ln\left(\frac{H_0}{M_X}\right)^2\right). 
\end{equation}
This dimensionless parameter is different from zero as long as $\xi \neq \frac{1}{6}$. Due to the presence of the ratio $M^2_X/M^2_{\rm pl}\ll 1$ we can always ensure that its value will be small. Once again we need to make use of the expression \eqref{RenormalizedVE}, but this time aiming to relate the value of the vacuum energy density at $M = H$ with the value already calculated \eqref{rho_Lambda_at_M_X} at $M_0 = M_X$. The results can be cast as follows:
\begin{equation}
\rho_{\rm vac}(H) = \rho^0_\Lambda - \frac{3\nu_{\rm eff}}{8\pi}H^2_0{M^2_{\rm pl}} + \frac{3\nu_{\rm eff}(H)}{8\pi}H^2{M^2_{\rm pl}}
\end{equation}
where we have employed the definition
\begin{equation}
\nu_{\rm eff}(H) = \frac{1}{2\pi}\left(\frac{1}{6} -\xi\right)\frac{M^2_X}{M^2_{\rm pl}}\left(1 + \frac{m^2}{M^2_X}\ln\left(\frac{H}{M_X}\right)^2\right).      
\end{equation}
As stated we are interested in the post-inflationary Universe, therefore the values that we are going to consider for $H(a)$ will not be much different from the present value $H_0$. This fact allows us to make the approximation $\nu_{\rm eff}(H)\simeq\nu_{\rm eff}$. All in all, the final expression can be written as: 
\begin{equation}\label{VED_ARP}
\rho_{\rm vac}(H)\simeq \rho^0_\Lambda + \frac{3\nu_{\rm eff}}{8\pi}(H^2-H^2_0)M^2_{\rm pl} = \rho^0_\Lambda + \frac{3\nu_{\rm eff}}{8\pi{G_N}}(H^2-H^2_0).
\end{equation}
For $\nu_{\rm eff} > 0$, the above expression can be conceived as if the vacuum decays into matter, since it value is larger in the past. Whereas for $\nu_{\rm eff} < 0$ is the other way around. See \cite{Chimento:2003iea,Olivares:2005tb,Salvatelli:2014zta,Yang:2018euj,Gomez-Valent:2020mqn,Sola:2017znb} to know more about interacting models. While it is true that the expression displayed in \eqref{VED_ARP} contains only the contribution of the ZPE coming from a scalar field, can serve as an ansatz to study in a phenomenological way the running vacuum models.
\newline
Actually, \eqref{VED_ARP} can be understood, as a particular case of a more general expression of the vacuum energy density considered in analogy to the renormalization group equation, that can be treated as an ansatz, but well motivated in the computations within a QFT \cite{Sola:2007sv,Sola:2013gha}:
\begin{equation}\label{rho_Lambda_derivative_RGE}
\frac{d\rho_{\rm vac}}{d\ln\mu^2} = \frac{1}{4\pi^2}\sum_i\left[ a_i{M^2_i}\mu^2 + b_i\mu^4 + c_i\frac{\mu^6}{M^2_i}+\dots\right]. 
\end{equation}
This equality describes in a very general way the possible running of the vacuum energy density triggered by the quantum effects coming from the different matter fields, either bosons or fermions. The coefficients $a_i, b_i, c_i \dots$ are dimensionless and $M_i$ represents the masses of the particles that contribute to the running of the vacuum. The energy scale $\mu$ is typically identified with $\sim aH^2+b\dot{H}$. Notice that $\rho_{\rm vac}(H)$ can only involve even powers of the Hubble rate $H$ or powers of the derivative $\dot{H}$, because of the covariance of the effective action. 
\newline
\newline
Integrating \eqref{rho_Lambda_derivative_RGE}, we can get an expression valid for the current Universe
\begin{equation}\label{RVM_energy_density}
\rho_{\rm vac}(H) = \frac{3}{8\pi{G_N}}\left(c_0 + \nu{H^2} + \alpha{\dot{H}}\right) + \mathcal{O}(H^4)    
\end{equation}
being $\alpha$ and $\nu$ dimensionless parameters and $c_0$ is a 2-dimensional constant in natural units. We emphasize that we always keep $c_0 \neq 0$ so as to ensure a smooth $\Lambda$CDM limit when the dimensionless coefficients are set to zero. These dimensionless coefficients can be computed in the context of QFT from the ratio squared of the different masses to the Planck mass and they can be interpreted as the $\beta$-function of the running of the vacuum energy density. The theoretical predictions establish $\alpha,\nu\sim 10^{-6}-10^{-3}$ \cite{Sola:2007sv}, however their values, at this moment, must be computed in a phenomenological way by constraining the models with the cosmological data and a proper statistical analysis. 
\newline
\newline
So far, we have justified the functional form of the vacuum energy density within the context of QFT, nevertheless, in order to allow the dynamics of $\rho_{\rm vac}(H)$, we are forced to introduce an extra ingredient with respect to the standard model of cosmology. In this thesis we are going to meanly consider three different scenarios derived from the local energy conservation equation: 
\begin{equation}\label{general_conservation_equation}
\frac{d}{dt}\left[G(t)\sum_N \rho_N\right] + 3G(t)H\sum_N\left(\rho_N + p_N\right) = 0    
\end{equation}
where, as usually, $N$ runs for the different components considered, e.g. baryons, dark matter, neutrinos, photons and vacuum. Let us generically denote by $w_{\rm DE}$ the equation of state parameter for the vacuum energy density, since as we are about to see in some cases we consider that it can mildly evolve with time.  The different possibilities that we study in different chapters of this thesis are: 
\newline
\newline
{\bf Scenario I:} $\dot{G} = 0$, $w_{\rm DE} = -1$ and $\sum_N \dot{\rho}_N + 3H(\rho_N + p_N) = 0$
\newline
\newline
In this first scenario the Newton coupling is treated as if it were a constant, $G = G_N$ and we consider dark energy as if it were vacuum, which means $\rho_{\rm vac} + p_{\rm vac} = 0$ in the local conservation equation. So, here we consider an exchange of energy between the ordinary matter and the vacuum component, which of course is not present in the $\Lambda$CDM model. 
\newline
\newline
{\bf Scenario II:} $\dot{G} \neq 0$, $w_{\rm DE} = -1$ and $\frac{d}{dt}(\rho_m + \rho_r) + 3H(\rho_m + p_m + \rho_r + p_r)  = 0$
\newline
\newline
Here the matter components $\rho_m = \rho_b + \rho_{dm} + \rho^{\rm NR}_{\nu}$ and $\rho_r = \rho_\gamma + \rho^{\rm R}_\nu$ follow exactly the same evolution laws than in the concordance model. In this case the vacuum does not interact with the ordinary matter but with the Newton coupling, which leaves us with the following equation: 
\begin{equation}
\frac{\dot{\rho}_{\rm vac}}{\rho_{\rm vac} + \rho_m + \rho_r } = -\frac{\dot{G}}{G}.    
\end{equation}
In this scenario not only the vacuum energy density inherits some dynamics, due to the expansion of the Universe, but also $G = G(t)$ does too. As cannot be otherwise, within that type of models, the predicted evolution for the Newton coupling is very slow being actually logarithmic $G(a)\simeq G_N\left(1 + \epsilon\ln(a)\right)$ with $|\epsilon|\ll 1$.
\newline
\newline
{\bf Scenario III:} $\dot{G} = 0$, $w_{\rm DE} \neq -1$ and $\sum_N \dot{\rho}_N + 3H(\rho_N + p_N) = 0$
\newline
\newline
On this last case we allow a tiny departure of the pure vacuum behaviour by considering a non-trivial expression for $w_{\rm DE}(t)$, while $\rho_{\rm vac}$ still has the functional form displayed in \eqref{RVM_energy_density} and interacts, like in Scenario I, with ordinary matter. 
\newline
\newline
Something that should not be omitted is the fact that it is also possible to accommodate the inflation mechanism within the RVM's. In order to produce the desired effects $\rho_{\rm vac}$ should acquire the following form:
\begin{equation}
\rho_{\rm vac}(H) = \frac{3}{8\pi G_N}\left( \bar{c}_0 + \bar{\nu}{H^2} + \bar{\alpha}\frac{H^4}{H^2_I}\right)    
\end{equation}
where $\bar{\nu}$ and $\bar{\alpha}$ are dimensionless parameters whereas the constants $H_I$ and $\bar{c}_0$ have dimension 1 and 2 respectively. Upon considering not only terms of order $\sim\mathcal{O}(H^2)$ but also those of order $\sim\mathcal{O}(H^4)$ it is possible to produce inflation in the period right before the radiation dominated era. Quite remarkable is the fact that terms like $\sim H^4$ can be generated in string-inspired mechanism, establishing a very intriguing connection with the RVM's \cite{Basilakos:2020qmu,Mavromatos:2020crd}. In order to know more details see \cite{Sola:2015rra} and references therein. Within this model we also find a scale factor of the form \eqref{scale_factor_inflation_Introduction}, as could not be otherwise, since we are interested in an expanding Universe completely dominated by a component with a negative pressure. 
\newline
\newline
In this thesis the RVM's are widely tested, putting them under the light of a large string of observational data, like the CMB spectrum, the BAO measurements or the important LSS observations (see \cite{Sola:2021txs} for an updated analysis). As expected the results depends of the dataset employed, but interestingly, as long as the triad CMB+BAO+LSS is included in the analysis an important conclusion can be extracted: the RVM's are favoured over the $\Lambda$CDM. 
\subsubsection{Quintessence scalar field models $\phi$CDM}
The literature is full of various attempts to explain, the CC and consequently, the current acceleration of the Universe employing scalar fields, see \cite{gibbons1985very,Abbott:1984qf,Ford:1987de,Weiss:1987xa,Peccei:1987mm,Wetterich:1994bg,Zlatev:1998tr,Chimento:2000kq,Chimento:2003iea,Tsujikawa:2013fta}  and references therein. The main idea behind this type of models is to describe DE in terms of some sort of scalar $\phi$ field with its corresponding potential $V(\phi)$, which can be the tail of a more complex effective potential in which inflation is also incorporated. We want to remark that, as the tittle of the section suggests, we are only interested in the quintessence scalar field models. The associated energy density $\rho_\phi$ evolves slowly with time until the scalar field becomes the dominant component, thus causing the accelerated expansion of the Universe we can see at the present time. The scalar field models possess a well-defined local Lagrangian description. The total action, including the EH term and the matter part, can be written, for a minimally coupled scalar field, in the following way:
\begin{equation}
S_{\rm tot} = S_{\rm EH} + S_\phi + S_m = \frac{M^2_{\rm pl}}{16\pi}\int d^4x\sqrt{-g}\left[R - \frac{1}{2}g^{\mu\nu}\partial_\mu\phi\partial_\nu\phi -V(\phi)\right]  + S_m. 
\end{equation}
Here $M_{\rm pl}=1/\sqrt{G_N}=1.22\times10^{19}{\rm GeV}$ is the Planck mass in natural units. As it can be appreciated in this convention the scalar field is taken to be dimensionless. Once we have the analytical expression for the scalar field action we can easily obtain the corresponding energy-momentum tensor thanks to:
\begin{equation}
T^{\phi}_{\mu\nu} = -\frac{2}{\sqrt{-g}}\frac{\delta{S_\phi}}{\delta{g^{\mu\nu}}} = \frac{M^2_{\rm pl}}{16\pi}\left[\partial_\mu\phi\partial_\nu\phi -g_{\mu\nu}\left(\frac{1}{2}g^{\alpha\beta}\partial_\alpha\phi\partial_\beta\phi + V(\phi)\right)\right]. 
\end{equation}
At this point we want to remark that we consider that the scalar field is an homogeneous quantity which means that at the background level is just a function of time. To know the details about the perturbation equations for the scalar field models see Appendix \ref{Appendix_B}. We can easily obtain the energy density and the pressure from the expression of the energy-momentum tensor:
\begin{equation}
\rho_\phi = T^{\phi}_{00} =  \frac{M^2_{\rm pl}}{16\pi}\left(\frac{\dot{\phi}^2}{2} + V(\phi)\right) \quad p_\phi = T^{\phi}_{ii} =  \frac{M^2_{\rm pl}}{16\pi}\left(\frac{\dot{\phi}^2}{2} - V(\phi)\right)
\end{equation}
The dots represent derivatives {\it w.r.t.} the cosmic time. Unlike the $\Lambda$CDM model the EoS parameter for the DE component is $w_\phi \neq -1$, this can be seen from
\begin{equation}
w_\phi = \frac{p_\phi}{\rho_\phi} = \frac{\dot{\phi}^2 -2V(\phi)}{\dot{\phi}^2 +2V(\phi)}. 
\end{equation}
In order to produce the late-time acceleration, the condition $w_\phi < -1/3$ must be fulfilled, which implies $\dot{\phi}/2\ll V(\phi)$. Taking this into account the EoS parameter can be approximated by $w_\phi\simeq -1 + \dot{\phi}^2/V(\phi)$. Due to the extra degree of freedom, embodied in the scalar field, it is possible to obtain the Klein-Gordon equation for a non-coupled model, upon functional derivative of $S_\phi$ with respect to the scalar field and considering the FLRW metric:
\begin{equation}
\ddot{\phi} + 3H\dot{\phi}  + \frac{\partial{V(\phi)}}{\partial\phi} = 0. 
\end{equation}
We must not fall into the error of thinking that the scalar field models are free from the old CC constant problem or equivalently free from some sort of fine-tuning problem. In order to have the proper shape the considered potential must adjust at least one free parameter to be able to produce the late-time accelerated expansion. To see a comprehensive list of possibilities for the potential check Table 1.1 of \cite{Gomez-Valent:2017tkh}. Also read Section 1.2.4 ({\it Scalar fields in the late-time Cosmology}) to get a summary of the different attempts based on the idea of scalar fields throughout the modern history of cosmology. 
\subsubsection{Brans-Dicke model with cosmological constant}\label{Introduction_BD_model}
The Brans-Dicke theory\,\cite{BransDicke1961} contains an additional gravitational degree of freedom as compared to GR, and therefore it genuinely departs from GR in a fundamental way. The new \textit{d.o.f.}  is represented by the scalar BD-field $\psi$, which is non-minimally coupled to curvature, $R$.
The BD-action reads as follows\,\,\footnote{We use natural units, therefore $\hbar=c=1$ and $G_N=1/M_{\rm pl}^2$, where $M_{\rm pl}\simeq 1.22\times 10^{19}$ GeV is the Planck mass. Regarding the sign convention we employ the $(+, +, +)$. See Appendix \ref{Appendix_A} for the details.}
\begin{eqnarray}
S_{\rm BD}=\int d^{4}x\sqrt{-g}\left[\frac{1}{16\pi}\left(R\psi-\frac{\oD}{\psi}g^{\mu\nu}\partial_{\nu}\psi\partial_{\mu}\psi\right)-\rL\right]+\int d^{4}x\sqrt{-g}\,{\cal L}_m(\Phi_i,g_{\mu\nu})\,. \label{eq:BDaction}
\end{eqnarray}
The (dimensionless) factor  in front of the kinetic term of $\psi$, i.e. $\oD$, will be referred to as the BD-parameter.
The last term of (\ref{eq:BDaction}) stands for the matter action $S_{m}$, which is constructed from the Lagrangian density of the matter fields, collectively denoted as $\Phi_i$. There is no potential for the BD-field $\psi$ in the original BD-theory, but we admit the presence of a CC term associated to $\rL$. The scalar field, $\psi$ has dimension $2$ in natural units (i.e. mass dimension squared), in contrast to the dimension $1$ of ordinary scalar fields. The effective value of $G$ at any time is thus given by $1/\psi$, and of course $\psi$ must be evolving very slowly with time. The field equations of motion ensue after performing variation with respect to both the metric and the scalar field $\psi$.  While the  first variation yields
\begin{equation}\label{eq:BDFieldEquation1}
\psi\,G_{\mu\nu}+\left(\Box\psi +\frac{\oD}{2\psi}\left(\nabla\psi\right)^2\right)\,g_{\mu\nu}-\nabla_{\mu}\nabla_{\nu}\psi-\frac{\oD}{\psi}\nabla_{\mu}\psi\nabla_{\nu}\psi=8\pi\left(\,T_{\mu\nu}-g_{\mu\nu}\rL\right)\,,
\end{equation}
the second variation gives the wave equation for $\psi$, which depends on the curvature scalar $R$. The latter can be eliminated upon tracing over the previous equation, what leads to a most compact result:
\begin{equation}\label{eq:BDFieldEquation2}
\Box\psi=\frac{8\pi}{2\oD+3}\,\left(T-4\rL\right)\,.
\end{equation}
Here we have assumed that both $\oD$ and $\rL$ are constants. To simplify the notation, we have written  $(\nabla\psi)^2\equiv g^{\mu\nu}\nabla_{\mu}\psi\nabla_{\nu}\psi$.
In the first field equation, $G_{\mu\nu}=R_{\mu\nu}-(1/2)Rg_{\mu\nu}$ is the Einstein tensor, and  on its \textit{r.h.s.} $T_{\mu \nu}=-(2/\sqrt{-g})\delta S_{m}/\delta g^{\mu\nu}$ is the energy-momentum tensor from matter. In the field equation for $\psi$,  $T\equiv T^{\mu}_{\mu}$ is the trace of the energy-momentum tensor for the matter part (relativistic and non-relativistic). The total energy-momentum tensor as written on the \textit{r.h.s.} of (\ref{eq:BDFieldEquation1})  is the sum of the matter and vacuum parts and adopts the perfect fluid form:
\begin{equation} \label{eq:EMT_BD_chapter}
\tilde{T}_{\mu\nu} =T_{\mu\nu}  -\rho_\CC g_{\mu\nu}=p\,g_{\mu\nu}+(\rho + p)U_\mu{U_\nu}\,,
\end{equation}
with $\rho \equiv \rho_m + \rho_r + \rho_\Lambda$ and $p \equiv p_m + p_r + p_\Lambda=p_r + p_\Lambda$,
where $p_m=0$, $p_r=(1/3)\rho_r$ and $p_\Lambda=-\rL$ stand for the pressures of dust matter (which includes baryons and DM), radiation and vacuum, respectively.
As in GR, we consider a possible CC term or constant vacuum energy density, $\rL$, in the BD-action (\ref{eq:BDaction}). The quantum matter fields usually induce an additional, and very large, contribution to $\rL$, this is of course the origin of the CC Problem.
Let us write down the field equations in the flat FLRW metric,  $ds^2=-dt^2 + a^2\delta_{ij}dx^idx^j$. Using the total density $\rho$ and pressure $p$ as indicated above, Eq.\,(\ref{eq:BDFieldEquation1}) renders the two independent equations
\begin{equation}\label{eq:Friedmannequation_BD_chapter}
3H^2 + 3H\frac{\dot{\psi}}{\psi} -\frac{\oD}{2}\left(\frac{\dot{\psi}}{\psi}\right)^2 = \frac{8\pi}{\psi}\rho
\end{equation}
\begin{equation}\label{eq:pressureequation_BD_chapter}
2\dot{H} + 3H^2 + \frac{\ddot{\psi}}{\psi} + 2H\frac{\dot{\psi}}{\psi} + \frac{\oD}{2}\left(\frac{\dot{\psi}}{\psi}\right)^2 = -\frac{8\pi}{\psi}p
\end{equation}
whereas (\ref{eq:BDFieldEquation2}) yields
\begin{equation}\label{eq:FieldeqPsi_BD_chapter}
\ddot{\psi} +3H\dot{\psi} = -\frac{8\pi}{2\oD +3}(3p-\rho)\,.
\end{equation}
Here dots indicate derivatives with respect to the cosmic time and $H=\dot{a}/a$ is the Hubble rate. For constant $\psi=1/G_N$, the first two equations reduce to the Friedmann and pressure equations of GR, and the third requires $\oD\to\infty$ for consistency. By combining the above equations we expect to find a local covariant conservation law, similar to GR.  This is because there is no interaction between matter and the BD field.  It can indeed be checked by explicit calculation. Although the calculation is more involved than in GR, the final result turns out to be the same:
\begin{equation}\label{eq:FullConservationLaw_BD_chapter}
\dot{\rho} + 3H(\rho + p)=\sum_{N} \dot{\rho}_N + 3H(\rho_N + p_N) = 0\,,
\end{equation}
where the sum is over all components, i.e. baryons, dark matter, radiation, neutrinos and vacuum. Here we take the point of view that all components are separately conserved in the main periods of the cosmic evolution.  In particular, the vacuum component obviously does not contribute in the sum since it is assumed to be constant and $\rho_\CC + p_\CC=0$ .
The expression (\ref{eq:FullConservationLaw_BD_chapter}) can also be obtained  upon lengthy but  straightforward computation of the covariant derivative on both sides of Eq.\,(\ref{eq:BDFieldEquation1}) and using the Bianchi identity satisfied by $G_{\mu\nu}$ and the field equation of motion for $\psi$.
\subsubsection{Modified gravity models}
So far we have considered models that assume the presence of dark energy (being a constant or a time-evolving quantity), however it is convenient to be open-minded and consider different alternatives present in the literature. See \cite{Amendola:2015ksp} in order to know about modified gravity models. Of course, we cannot simply omit the experimental evidence we have nowadays of how galaxies move away from each other in an accelerated way. Actually, in the previous section we have considered one of the first attempts of cosmological model not based on GR, namely the BD model, where gravity is not only mediated by the tensor $G_{\mu\nu}$ but also by the BD-field which is usually denoted as $\psi$. One can replace there the $\rho_\Lambda$ by a potential $U(\psi)$ and then remove completely the presence of a pure cosmological constant. We shall comment more about that in Chapter \ref{BD_gravity_chapter}. 
\newline
\newline
Of course it is possible to introduce a modification in the field equations without using a scalar field, and a classical example of that are the $f(R)$ models \cite{Capozziello:2002rd,Capozziello:2006uv,Brax:2008hh,Brookfield:2006mq,delaCruzDombriz:2008cp}, whose action take the general form:
\begin{equation}
S = \frac{1}{16\pi{G_N}}\int d^4x\sqrt{-g} f(R) + S_m(\Phi_i, g_{\mu\nu}).    
\end{equation}
As in other cases here $S_m$ denotes the matter action and $\Phi_i$ refer to the matter fields in general. Note that the EH action is just a particular case of the above action with $f(R) = R-2\Lambda$. It goes without saying that not all the forms for the function $f(R)$ are valid, since the considered model has to pass not only the cosmological test, but also the astrophysical ones in case no screening mechanism is considered. A brief comment on the frame where the models are considered turns out to be necessary. We say that we work in the Einstein frame when the Ricci scalar, appearing in the action, is not multiplied by a constant or replaced by a function of $R$, like in the case of $f(R)$ models, and it is also not multiplied by some sort of function of the scalar field, like in the BD model. If any of the above conditions are not fulfilled we say we are in the Jordan frame. Both frames are linked through a conformal transformation of the metric
\begin{equation}
\hat{g}_{\mu\nu} = \Omega^2{g_{\mu\nu}}    
\end{equation}
where the function $\Omega$ can vary depending on the model considered. There is no general agreement about which of the frames is the {\it physical} one, in case any of them are. In this thesis we adopt the convention to consider the physical frame as that one where the model is originally formulated. 
\newline
\newline
If we go back to the $f(R)$ models, in order to obtain the field equations, one has to choose between two different approaches. The first one is the metric formalism, where the connection between the Christoffel symbols $\Gamma^{\mu}_{\nu\lambda}$ and the metric $g_{\mu\nu}$ is the usual one. The other option is known as the Palatini formalism, which considers that both functions (the Christoffel symbols and the metric) must be considered as independent. Since for all the models studied in this thesis we assume the Levi-Civita connection, we stick to the former case. By considering the same process employed to obtain Einstein's field equations, vary the action with respect to $g_{\mu\nu}$ it is possible to obtain the following equations: 
\begin{equation}\label{eq:F_R_modified_field_equations}
F(R)R_{\mu\nu}  - \frac{1}{2}f(R)g_{\mu\nu} -\nabla_\mu\nabla_\nu{F(R)} + g_{\mu\nu}\Box{F(R)} = 8\pi{G_N}T_{\mu\nu}. 
\end{equation}
In the above expression appear $R_{\mu\nu}$ and $T_{\mu\nu}$, which are the usual Ricci tensor and energy-momentum tensor respectively, whereas the term $F(R) \equiv \partial{f}/\partial{R}$ has been introduced to simplify the notation. If we take the trace of \eqref{eq:F_R_modified_field_equations} we end up with the following relation:
\begin{equation}\label{eq:F_R_trace_equation}
3\Box{F(R)} + F(R)R -2f(R) = 8\pi{G_N}T    
\end{equation}
being $T$ the trace of the energy-momentum tensor. What \eqref{eq:F_R_trace_equation} means is that there is a propagating scalar degree of freedom due to the presence of the term $\Box{F(R)}$.
\newline
\newline
If the FLRW metric, $ds^2 = -dt^2 + a^2(t)\delta_{ij}dx^i{dx^j}$ is assumed, we can write the cosmological equations as:
\begin{align}
3H^2 &= 8\pi{G_N}\left(\rho_m + \rho_r + \rho_{\rm DE}\right) \\
2\dot{H} + 3H^2 &= -8\pi{G_N}\left(p_m + p_r + p_{\rm DE}\right)
\end{align}
where the terms $\rho_{\rm DE}$ and $p_{\rm DE}$ are defined as: 
\begin{align}
\rho_{\rm DE} &\equiv \frac{1}{8\pi{G_N}}\left(3H^2(1-F) + \frac{(FR-f)}{2} -3H\dot{F}\right)\\
p_{\rm DE} &\equiv \frac{1}{8\pi{G_N}}\left((2\dot{H} + 3H^2)(F-1) + 2H\dot{F} + \ddot{F} + \frac{(f-FR)}{2}\right). 
\end{align}
Upon considering the effective expressions for the energy density and the pressure it is possible to get the following local conservation equation $\dot{\rho}_{\rm DE} + 3H\left(\rho_{\rm DE} + p_{\rm DE}\right)=0$. If we continue with the analogy of dark energy we are allowed to define an EoS parameter given by the expression:
\begin{equation}
w_{\rm DE} = \frac{p_{\rm DE}}{\rho_{\rm DE}} = -1 + \frac{2\ddot{F} -2H\dot{F} -4\dot{H}F}{6H^2(1-F) + FR -f -6H\dot{F}}.     
\end{equation}
The above expression is useful to detect deviations from a pure GR model. 
\newpage
\thispagestyle{empty}
\mbox{}
\newpage

\section{First evidence of running cosmic vacuum: challenging the concordance model}\label{First_evidence_chapter}
We start our study of different models, with the capacity to provide a more general framework that the $\Lambda$CDM does, in order to accommodate a possible time evolution of the dark energy (DE) component, by studying the Running Vacuum Models (RVM's). These models were introduced in the previous chapter and are mainly characterized by having a vacuum energy density that evolves as the Universe expands. The particular functional form of $\rho_\Lambda = \rho_\Lambda(H)$ is motivated in the context of quantum field theory (QFT) in curved space-times. 
\newline
We would like to remark that our aim here is neither to solve the cosmological constant (CC) problem nor to solve the coincidence problem, but to study the aforementioned models from a phenomenological perspective. 
\newline
To do so, we will consider a large set of cosmological data SNIa+BAO+$H(z)$+LSS+BBN+CMB in order to test the possibility that the $\Lambda$-term and the associated vacuum energy density $\rho_\Lambda = \Lambda/(8\pi{G})$ could actually be a dynamical (``running") quantity whose rhythm of variation might be linked to the Universe expansion rate, $H(a)$. The idea is to check if this possibility helps to improve the description of the overall cosmological data as compared to the rigid assumption $\CC=$const. inherent to the concordance $\CC$CDM model. 
\newline
The plan of the chapter is as follows. In Section \ref{sec:RVMs_first_chapter} we describe the different types of running vacuum models (RVM's) that will be considered in this study. In Section \ref{sec:Fit_first_chapter} we fit these models to a large set of cosmological data on distant type Ia supernovae (SNIa), baryonic acoustic oscillations (BAO), the known values of the Hubble parameter at different redshift points, the large scale structure (LSS) formation data, the BBN bound on the Hubble rate, and, finally, the CMB distance priors from WMAP and Planck. We include also a fit of the data with the standard XCDM parameterization, which serves as a baseline for comparison. In Section \ref{sec:Discussion_first_chapter} we present a detailed discussion of our results, and finally in Section \ref{sec:Conclusions_first_chapter} we deliver our conclusions.


\begin{table}[t!]
\setcounter{table}{0}
\begin{center}
\resizebox{1\textwidth}{!}{
\begin{tabular}{| c | c |c | c | c | c | c | c | c | c |}
\multicolumn{1}{c}{Model} &  \multicolumn{1}{c}{$h$} &  \multicolumn{1}{c}{$\omega_b= \Omega_b h^2$} & \multicolumn{1}{c}{{\small$n_s$}}  &  \multicolumn{1}{c}{$\Omega_m$}&  \multicolumn{1}{c}{{\small$\nu_{\rm eff}$}}  & \multicolumn{1}{c}{$\omega$}  &
\multicolumn{1}{c}{$\chi^2_{\rm min}/dof$} & \multicolumn{1}{c}{$\Delta{\rm AIC}$} & \multicolumn{1}{c}{$\Delta{\rm BIC}$}\vspace{0.5mm}
\\\hline
$\Lambda$CDM  & $0.693\pm 0.003$ & $0.02255\pm 0.00013$ &$0.976\pm 0.003$& $0.294\pm 0.004$ & - & $-1$  & 90.44/85 & - & - \\
\hline
XCDM  & $0.670\pm 0.007$& $0.02264\pm0.00014 $&$0.977\pm0.004$& $0.312\pm0.007$ & - &$-0.916\pm0.021$  & 74.91/84 & 13.23 & 11.03 \\
\hline
A1  & $0.670\pm 0.006$& $0.02237\pm0.00014 $&$0.967\pm0.004$& $0.302\pm0.005$ &$0.00110\pm 0.00026 $ &  $-1$ & 71.22/84 & 16.92 & 14.72 \\
\hline
A2   & $0.674\pm 0.005$& $0.02232\pm0.00014 $&$0.965\pm0.004$& $0.303\pm0.005$ &$0.00150\pm 0.00035 $& $-1$  & 70.27/84 & 17.87 & 15.67\\
\hline
G1 & $0.670\pm 0.006$& $0.02236\pm0.00014 $&$0.967\pm0.004$& $0.302\pm0.005$ &$0.00114\pm 0.00027 $& $-1$  &  71.19/84 & 16.95 & 14.75\\
\hline
G2  & $0.670\pm 0.006$& $0.02234\pm0.00014 $&$0.966\pm0.004$& $0.303\pm0.005$ &$0.00136\pm 0.00032 $& $-1$  &  70.68/84 & 17.46 & 15.26\\
\hline
\end{tabular}}
\end{center}
\label{tableFit_first_chapter}
\begin{scriptsize}
\caption{\scriptsize The best-fit values for the $\CC$CDM, XCDM and the RVM's, including their statistical  significance ($\chi^2$-test and Akaike and Bayesian information criteria, AIC and BIC, see the text). The large and positive values of $\Delta$AIC and $\Delta$BIC strongly favor the dynamical DE options (RVM's and XCDM) against the $\CC$CDM (see text). We use $90$ data points in our fit, to wit: $31$ points from the JLA sample of SNIa, $11$ from BAO, $30$  from $H(z)$, $13$ from linear growth, $1$ from BBN, and $4$ from CMB (see S1-S7 in the text for references). In the XCDM model the EoS parameter $\omega$ is left free, whereas for the RVM's and $\CC$CDM is fixed at $-1$.  The specific RVM fitting parameter is $\nueff$, see Eq.\eqref{eq:xixip_first_chapter} and the text. For G1 and A1 models, $\nueff=\nu$. The remaining parameters are the standard ones ($h,\omega_b,n_s,\Omega_m$).  The quoted number of degrees of freedom ($dof$) is equal to the number of data points minus the number of independent fitting parameters ($5$ for the $\CC$CDM, $6$ for the RVM's and the XCDM. The normalization parameter M introduced in the SNIa sector of the analysis is also left free in the fit, cf. \cite{Betoule:2014frx}, but it is not listed in the table). For the CMB data we have used the marginalized mean values and standard deviation for the parameters of the compressed likelihood for Planck 2015 TT,TE,EE + lowP data from \cite{Huang:2015vpa}, which provide tighter constraints to the CMB distance priors than those presented in \cite{Ade:2015rim}}
\end{scriptsize}
\end{table}
\begin{table}[t!]
\begin{center}
\resizebox{1\textwidth}{!}{
\begin{tabular}{| c | c |c | c | c | c | c| c | c | c |}
\multicolumn{1}{c}{Model} &  \multicolumn{1}{c}{$h$} &  \multicolumn{1}{c}{$\omega_b= \Omega_b h^2$} & \multicolumn{1}{c}{{\small$n_s$}}  &  \multicolumn{1}{c}{$\Omega_m$}&  \multicolumn{1}{c}{{\small$\nu_{\rm eff}$}}  & \multicolumn{1}{c}{$\omega$}  &
\multicolumn{1}{c}{$\chi^2_{\rm min}/dof$} & \multicolumn{1}{c}{$\Delta{\rm AIC}$} & \multicolumn{1}{c}{$\Delta{\rm BIC}$}\vspace{0.5mm}
\\\hline
{\small $\Lambda$CDM} & $0.692\pm 0.004$ & $0.02254\pm 0.00013$ &$0.975\pm 0.004$& $0.295\pm 0.004$ & - & $-1$   & 86.11/78 & - & -\\
\hline
XCDM  &  $0.671\pm 0.007$& $0.02263\pm 0.00014 $&$0.976\pm 0.004$& $0.312\pm 0.007$& - & $-0.920\pm0.022$  & 73.01/77 & 10.78 & 8.67 \\
\hline
A1  & $0.670\pm 0.007$& $0.02238\pm0.00014 $&$0.967\pm0.004$& $0.302\pm0.005$ &$0.00110\pm 0.00028 $ & $-1$  & 69.40/77 & 14.39 & 12.27 \\
\hline
A2   & $0.674\pm 0.005$& $0.02233\pm0.00014 $&$0.966\pm0.004$& $0.302\pm0.005$ &$0.00152\pm 0.00037 $& $-1$  & 68.38/77 & 15.41 & 13.29\\
\hline
G1 & $0.671\pm 0.006$& $0.02237\pm0.00014 $&$0.967\pm0.004$& $0.302\pm0.005$ &$0.00115\pm 0.00029 $& $-1$  &  69.37/77 & 14.42 & 12.30\\
\hline
G2  & $0.670\pm 0.006$& $0.02235\pm0.00014 $&$0.966\pm0.004$& $0.302\pm0.005$ &$0.00138\pm 0.00034 $& $-1$  &  68.82/77 & 14.97 & 12.85\\
\hline
\end{tabular}}
\end{center}
\begin{scriptsize}
\caption{\scriptsize Same as in Table 1, but excluding from our analysis the BAO and LSS data from WiggleZ, see point S5) in the text.}
\end{scriptsize}
\label{tableFit2_first_chapter}
\end{table}


\subsection{Two basic types of RVM's}\label{sec:RVMs_first_chapter}
In an expanding Universe we may expect that the vacuum energy density and the gravitational coupling are functions of the cosmic time through the Hubble rate, thence $\rho_\Lambda=\rho_\Lambda(H(t))$ and $G=G(H(t))$. Adopting the canonical equation of state $p_\Lambda=-\rho_\Lambda(H)$ also for the dynamical vacuum,
the corresponding field equations in the
Friedmann-Lema\^\i tre-Robertson-Walker (FLRW) metric in flat space become formally identical to those
with strictly constant $G$ and $\CC$:
\begin{eqnarray}\label{eq:FriedmannEq_first_chapter}
&&3H^2=8\pi\,G(H)\,(\rho_m+\rR+\rho_\Lambda(H))\\
&&3H^2+2\dot{H}=-8\pi\,G(H)\,(p_r-\rho_\Lambda(H))\,. \label{eq:PressureEq_first_chapter}\,
\end{eqnarray}
The equations of state for the densities of relativistic ($\rho_r$) and dust matter ($\rho_m$) read
$p_r=(1/3)\rho_r$ and $p_m=0$, respectively.  Consider now the characteristic RVM structure
of the dynamical vacuum energy:
\begin{eqnarray}\label{eq:rhoL_first_chapter}
\rL(H;\nu,\alpha)=\frac{3}{8\pi G}\left(c_0+\nu H^2+\frac{2}{3}\alpha\,\dH\right)+{\cal O}(H^4)\,,\label{eq:rL_first_chapter}
\end{eqnarray}
where $G$ can be constant or a function $G=G(H;\nu,\alpha)$ depending on the particular model. The above expression is the form that has been suggested in the literature from the quantum corrections of QFT in curved space-time (cf. \,\cite{Sola:2013gha} and references therein). The terms  with higher powers of the Hubble rate have recently been used to describe inflation,
see e.g. \cite{Lima:2012mu,Lima:2014hia} and \cite{Sola:2015csa}, but these terms play no role at present and will be hereafter omitted. The coefficients $\nu$ and $\alpha$ have been defined dimensionless. They are responsible for the running of $\rho_\Lambda(H)$ and $G(H)$, and so for $\nu=\alpha=0$ we recover the $\CC$CDM, with $\rho_\Lambda$ and $G$ constants. The values of $\nu$ and $\alpha$ are naturally small in this context since they can be related to the $\beta$-functions of the running. An estimate in QFT indicates that they are of order $10^{-3}$ at most \cite{Sola:2007sv}, but here we will treat them as free parameters of the RVM and hence we shall determine them phenomenologically by fitting the model to observations. As previously indicated, a simple Lagrangian language for these models that is comparable to the scalar field DE description may not be possible, as suggested by attempts involving the anomaly-induced action\,\cite{Sola:2007sv,Sola:2013gha}.
\newline
\newline
Two types of RVM will be considered here: i) type-G models,  when matter is conserved and the running of $\rL(H)$ is compatible with the Bianchi identity at the expense of a (calculable) running of $G$; ii) type-$A$ models, in contrast, denote those with $G=$const. in which the running of $\rL$ must be accompanied with a (calculable) anomalous conservation law of matter. Both situations are described by the generalized local conservation equation $\nabla^{\mu}\left(G\,\tilde{T}_{\mu\nu}\right)=0$, where $\tilde{T}_{\mu\nu}=T_{\mu\nu}+\rL\,g_{\mu\nu}$ is the total energy-momentum tensor involving both matter and vacuum energy. In the FLRW metric, and summing over all energy components, we find
\begin{equation}\label{BianchiGeneral_first_chapter}
\frac{d}{dt}\,\left[G(\rho_m+\rho_r+\rL)\right]+3\,G\,H\,\sum_{i=m,r}(\rho_i+p_i)=0\,.
\end{equation}
If $G$ and $\rL$ are both constants, we recover the canonical  conservation law $\dot{\rho}_m+\dot{\rho}_r+3H\rho_m+4H\rho_r=0$ for the combined system of matter and radiation.
For type-G models Eq.\eqref{BianchiGeneral_first_chapter} boils down to $\dot{G}(\rho_m + \rho_r + \rho_\Lambda) + G\dot{\rho}_\Lambda =0$ since $\dot{\rho}_m + 3H\rho_m = 0$ and $\dot{\rho}_r + 4H\rho_r = 0$ for separated conservation of matter and radiation, as usually assumed.  Mixed type of RVM scenarios are possible, but will not be considered here.
\newline
We can solve analytically the type-G and type-A models by inserting equation \eqref{eq:rL_first_chapter} into \eqref{eq:FriedmannEq_first_chapter} and \eqref{eq:PressureEq_first_chapter}, or using one of the latter two and the corresponding conservation law \eqref{BianchiGeneral_first_chapter}. It is convenient to perform the  integration using the scale factor $a(t)$  rather than the cosmic time. For type-G models the full expression for the  Hubble function normalized to its current value, $E(a)=H(a)/H_0$, can be found to be
\begin{eqnarray}\label{eq:DifEqH_first_chapter} 
&&\left.E^2(a)\right|_{\rm type-G}=1+\left(\frac{\Omega_m}{\xi}+\frac{\Omega_r}{\xi^\prime}\right)\nonumber\\
&&\times\left[-1+a^{-4\xi^\prime}\left(\frac{a\xi^\prime+\xi\Omega_r/\Omega_m}{\xi^\prime+\xi\Omega_r/\Omega_m}\right)^{\frac{\xi^\prime}{1-\alpha}}\right]\,, 
\end{eqnarray}
where $\Omega_{i}=\rho_{i0}/\rho_{c0}$ are the current cosmological parameters for matter and radiation, and we have defined
\begin{equation}\label{eq:xixip_first_chapter}
\xi=\frac{1-\nu}{1-\alpha}\equiv 1-\nueff\,,\ \ \  \xi^\prime=\frac{1-\nu}{1-\frac{4}{3}\alpha}\equiv 1-\nueffp\,.    
\end{equation}
\begin{figure}[t!]
\setcounter{figure}{0}
\begin{center}
\includegraphics[angle=0,width=0.9\linewidth]{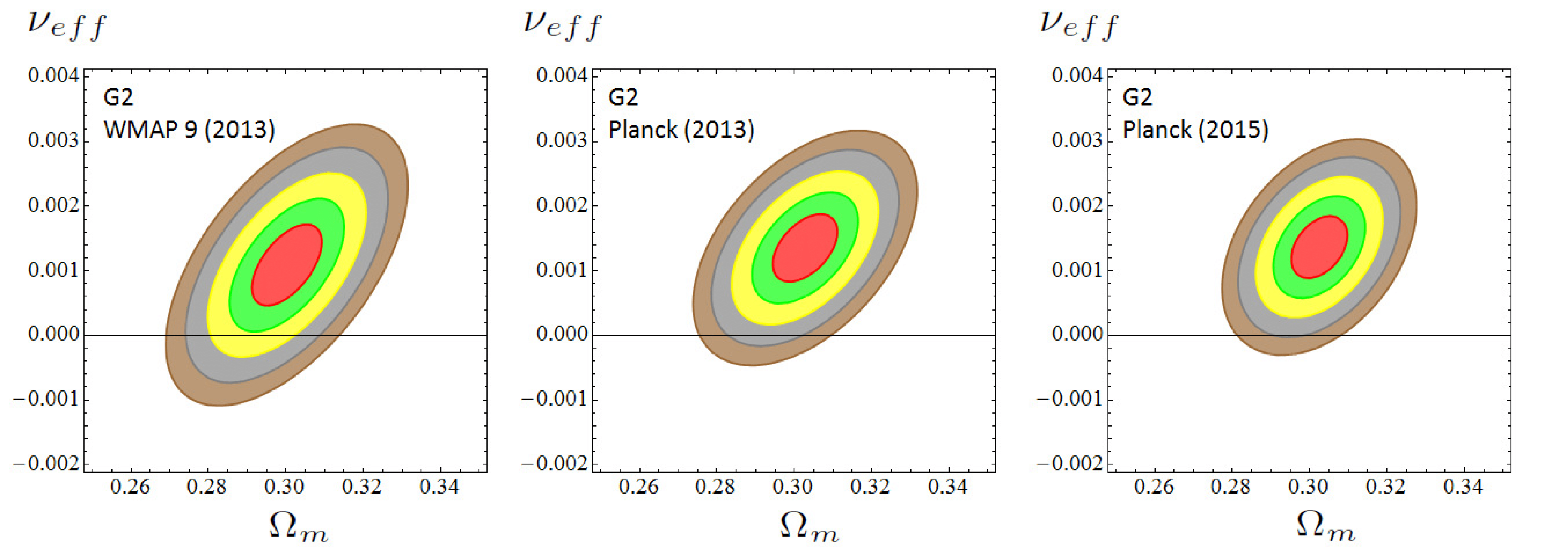}
\caption{\label{fig:G2Evolution_first_chapter}%
\scriptsize Likelihood contours in the $(\Omega_m,\nu_{\rm eff})$ plane for the values $-2\ln\mathcal{L}/\mathcal{L}_{max}=2.30$, $6.18, 11.81$, $19.33$, $27.65$ (corresponding to 1$\sigma$, 2$\sigma$, 3$\sigma$, 4$\sigma$ and 5$\sigma$ c.l.) after marginalizing over the rest of the fitting parameters indicated in Table 1. We display the progression of the contour plots obtained for model G2 using the 90 data points on SNIa+BAO+$H(z)$+LSS+BBN+CMB, as we evolve from the high precision CMB data from WMAP9, Planck 2013 and Planck 2015 -- see text, point S7). In the sequence, the prediction of the concordance  model ($\nueff=0$) appears increasingly more disfavored, at an exclusion c.l. that ranges from  $\sim 2\sigma$ (for WMAP9), $\sim 3.5\sigma$ (for Planck 2013) and up to  $4\sigma$ (for Planck 2015). Subsequent marginalization over $\Omega_m$ increases slightly the c.l. and renders the fitting values indicated in Table 1, which reach a statistical significance of $4.2\sigma$ for all the RVM's. Using numerical integration we can estimate that $\sim99.81\%$ of the area of the $4\sigma$ contour for Planck 2015 satisfies $\nu_{\rm eff}>0$. We also estimate that $\sim95.47\%$ of the $5\sigma$ region also satisfies $\nu_{\rm eff}>0$. The corresponding AIC and BIC criteria (cf. Table 1) consistently imply a very strong support to the RVM's against the $\CC$CDM.
}
\end{center}
\end{figure}
Note that $E(1)=1$, as it should. Moreover, for  $\xi,\xi^{\prime}\to 1$ (i.e. $|\nu,\alpha|\ll 1$)  $\nu_{\rm eff}\simeq \nu-\alpha$ and $\nu_{\rm eff}^\prime\simeq \nu-(4/3)\alpha$.
In the radiation-dominated epoch, the leading behaviour of Eq.\eqref{eq:DifEqH_first_chapter} is $\sim \Omega_r\,a^{-4\xi^{\prime}}$, while in the matter-dominated epoch is $\sim \Omega_m\,a^{-3\xi}$. Furthermore, for $\nu,\alpha\to 0$,  $E^2(a)\to 1+\Omega_m\,(a^{-3}-1)+\Omega_r(a^{-4}-1)$. This is the $\CC$CDM form, as expected in that limit. Note
that the following constraint applies among the parameters:
$c_0=H_0^2\left[\Omega_\Lambda-\nu+\alpha\left(\Omega_m+\frac43\,\Omega_r\right)\right]$,
as the vacuum energy density $\rho_\Lambda(H)$ must reproduce the current
value $\rho_\Lambda$ for $H=H_0$, using $\Omega_m+\Omega_r+\Omega_\Lambda=1$.  The explicit scale factor dependence of the vacuum energy density, i.e. $\rho_\Lambda=\rho_\Lambda(a)$,
ensues upon inserting \eqref{eq:DifEqH_first_chapter} into \eqref{eq:rL_first_chapter}. In addition, since the matter is conserved for type-G models, we can use the obtained expression for $\rho_\Lambda(a)$ to also infer the explicit form for $G=G(a)$ from \eqref{eq:FriedmannEq_first_chapter}.
We refrain from writing out these cumbersome expressions and we limit ourselves to quote some simplified forms. For instance, the expression for $\rho_\Lambda(a)$ when we can neglect the radiation contribution is simple enough:
\begin{equation}\label{eq:RhoLNR_first_chapter}
\rho_\Lambda(a)=\rho_{c0}\,
a^{-3}\left[a^{3\xi}+\frac{\Omega_m}{\xi}(1-\xi-a^{3\xi})\right]\,,
\end{equation}
where  $\rho_{c0}=3H_0^2/8\pi\,G_0$ is the current critical density and $G_0\equiv G(a=1)$ is the current value of the gravitational coupling. Quite obviously for $\xi=1$ we recover the $\CC$CDM form: $\rho_\Lambda=\rho_{c0}(1-\Omega_m)=\rho_{c0}\Omega_\Lambda=$const. As for the gravitational coupling, it
evolves logarithmically with the scale factor and hence changes very slowly\footnote{This is a welcome feature already expected in particular  realizations of type-G models in QFT in curved space-time\,\cite{Sola:2007sv,Sola:2013gha}. See also \cite{Grande:2011xf} }. It suffices to say that it behaves as
\begin{equation}\label{Gafunction_first_chapter}
G(a)=G_0\,a^{4(1-\xi^{\prime})}\,f(a)\simeq G_{0}(1+4\nueffp\,\ln\,a)\,f(a)\,,
\end{equation}
where $f(a)=f(a;\Omega_m,\Omega_r; \nu,\alpha)$ is a smooth function of the scale factor. We can dispense with the full expression here, but let us mention that $f(a)$ tends to one at present irrespective of the values of the various parameters $\Omega_m,\Omega_r,\nu,\alpha$ involved in it; and $f(a)\to1$ in the remote past ($a\to 0$) for $\nu,\alpha\to 0$ (i.e. $\xi,\xi^{\prime}\to 1$). As expected, $G(a)\to G_0$ for $a\to 1$, and $G(a)$ has a logarithmic evolution for $\nu_{\rm eff}^\prime\neq 0$.
Notice that the limit $a\to 0$ is  relevant for the BBN (Big Bang Nucleosynthesis) epoch and therefore $G(a)$ should not depart too much from $G_0$ according to the usual bounds on BBN. We shall carefully incorporate this restriction in our analysis of the RVM, see later on.
\newline
\newline
Next we quote the solution for type-A models. As indicated, in this case we have an anomalous matter conservation law. Integrating \eqref{BianchiGeneral_first_chapter} for $G=$const. and using \eqref{eq:rL_first_chapter} in it one finds $\rho_t(a)\equiv\rho_m(a)+\rho_r(a)=\rho_{m0}a^{-3\xi}+\rho_{r0}a^{-4\xi^{\prime}}$. We have assumed, as usual, that there is no exchange of energy between the relativistic and non-relativistic components. The standard expressions for matter and radiation energy densities are  recovered for $\xi,\xi^{\prime}\to 1$. The normalized Hubble function for type-A models is simpler than for type-G ones. The full expression including both matter and radiation reads:
\begin{equation}\label{eq:HubbleA_first_chapter}
\left.E^2(a)\right|_{\rm type-A}=1+\frac{\Omega_m}{\xi}\left(a^{-3\xi}-1\right)+\frac{\Omega_r}{\xi^\prime}\left(a^{-4\xi^\prime}-1\right)\,.
\end{equation}
From it and the found expression for $\rho_t(a)$ we can immediately derive the corresponding $\rho_\Lambda(a)$:
\begin{equation}\label{rLaTypaA_first_chapter}
\rho_\CC(a)=\rho_{\Lambda 0}+\rho_{m0}(\xi^{-1}-1)(a^{-3\xi}-1)+\rho_{r0}(\xi^{\prime-1}-1)(a^{-4\xi^\prime}-1)\,.
\end{equation}
Once more for $\nu,\alpha\to 0$ (i.e. $\xi,\xi^{\prime}\to 1$) we recover the $\CC$CDM case, as easily checked. In particular one finds $\rho_\Lambda\to\rho_{\Lambda{0}}=$const. in this limit.
\begin{figure}[t!]
\begin{center}
\centering
\includegraphics[angle=0,width=0.9\linewidth]{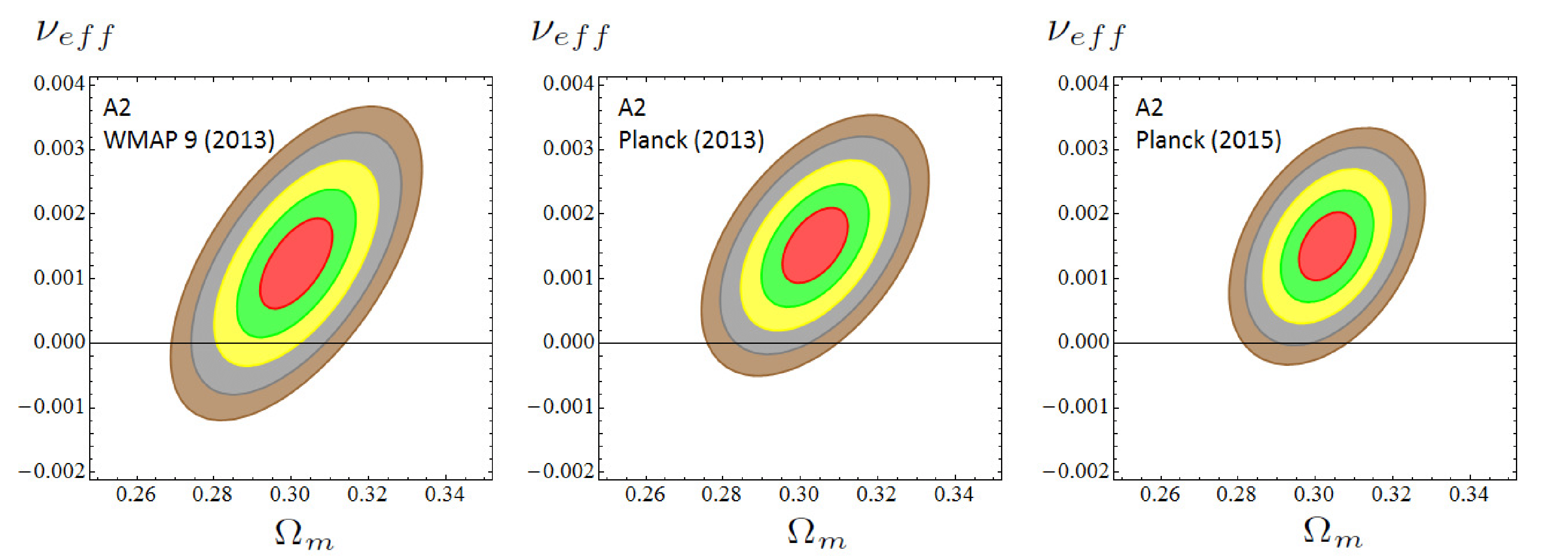}
\caption{\label{fig:A1Evolution_first_chapter}%
\scriptsize As in Fig.\,1, but for model A2. Again we see that the contours tend to migrate to the $\nueff>0$ half plane as we evolve from WMAP9 to Planck 2013 and Planck 2015 data. Using the same method as in Fig.\,1, we find that $\sim99.82\%$ of the area of the $4\sigma$ contour for Planck 2015 (and  $\sim95.49\%$ of the corresponding $5\sigma$ region) satisfies $\nu_{\rm eff}>0$.  The $\CC$CDM  becomes once more excluded at $\sim 4\sigma$ c.l.  ({cf. Table 1 for Planck 2015}).
}
\end{center}
\end{figure}
\begin{figure}[t!]
\begin{center}
\includegraphics[angle=0,width=0.65\linewidth]{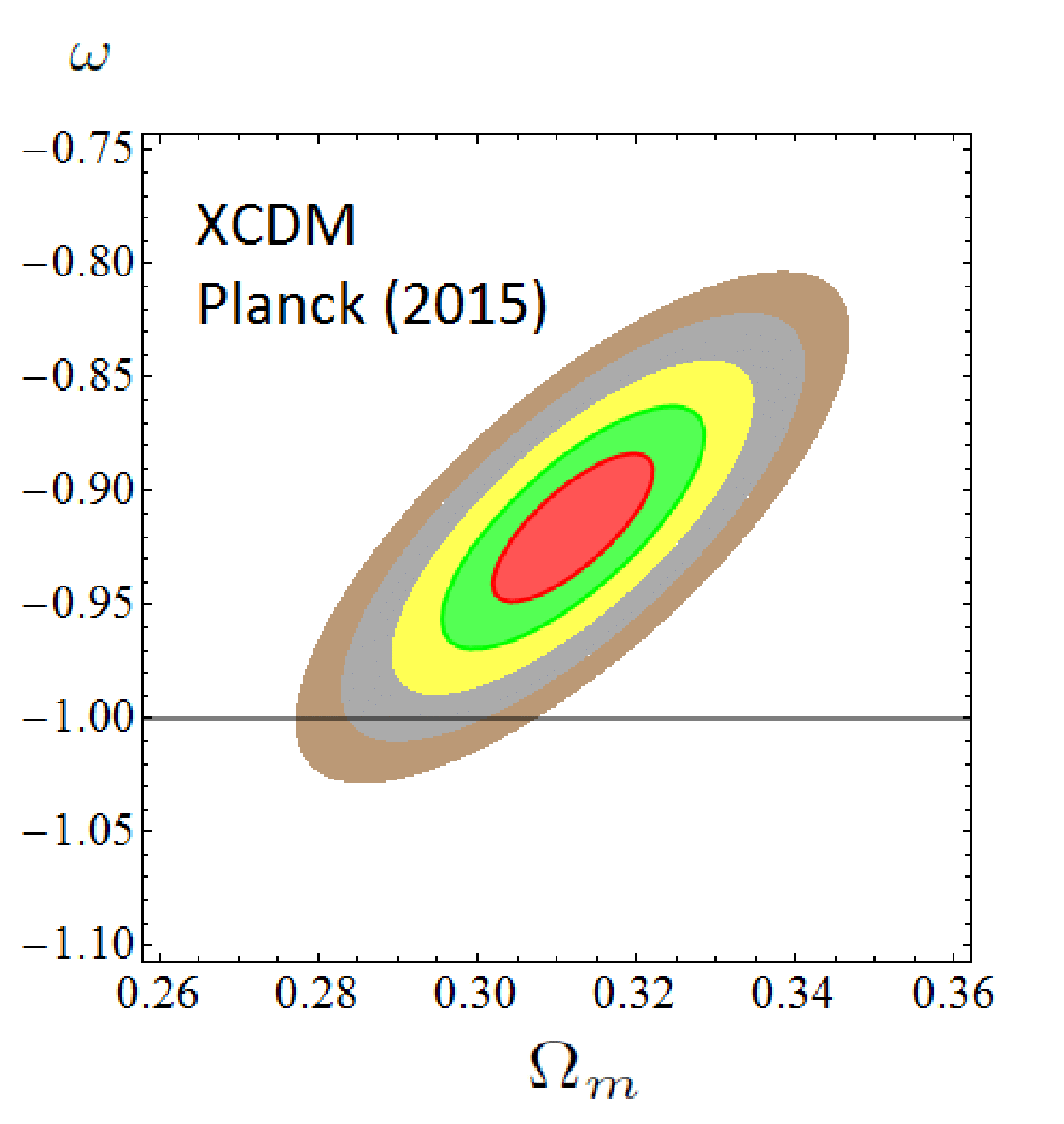}
\caption{\label{fig:XCDMEvolution_first_chapter}%
\scriptsize As in Fig.\,1 and 2, but for model XCDM and using Planck 2015 data. The $\CC$CDM is excluded at $\sim 4\sigma$ c.l.  ({cf. Table 1}).
}
\end{center}
\end{figure}
\subsection{Fitting the vacuum models to the data}\label{sec:Fit_first_chapter}
In order to better handle the possibilities offered by the type-G and type-A models as to their dependence on the two specific vacuum parameters $\nu,\alpha$, we shall refer to model G1 (resp. A1) when we address type-G (resp. type-A) models with $\alpha=0$ in Eq.\eqref{eq:rL_first_chapter}. In these cases $\nu_{\rm eff}=\nu$.  When, instead, $\alpha\neq 0$ we shall indicate them by G2 and A2, respectively. This classification scheme is used in Tables 1-2 and 5-7, and in Figs. 1-6. In the tables we are including also the XCDM (cf. Section \ref{sec:Discussion_first_chapter}) and the $\CC$CDM.
\newline
To this end, we fit the various models to the wealth of cosmological data compiled from distant type Ia supernovae (SNIa), baryonic acoustic oscillations (BAO), the known values of the Hubble parameter at different redshift points, $H(z_i)$, the large scale structure (LSS) formation data encoded in $f(z_i)\sigma_8(z_i)$, the BBN bound on the Hubble rate, and, finally, the CMB distance priors from WMAP and Planck, with the corresponding correlation matrices in all the indicated cases. Specifically, we have used $90$ data points (in some cases involving compressed data) from $7$ different sources S1-S7, to wit:
\begin{table}[t!]
\begin{center}
\begin{tabular}{| c | c | c |}
\multicolumn{1}{c}{$z$} &  \multicolumn{1}{c}{$H(z)$} & \multicolumn{1}{c}{{\small References}}
\\\hline
$0.07$ & $69.0\pm 19.6$ & \cite{Zhang:2012mp}
\\\hline
$0.09$ & $69.0\pm 12.0$ & \cite{Jimenez:2003iv}
\\\hline
$0.12$ & $68.6\pm 26.2$ & \cite{Zhang:2012mp}
\\\hline
$0.17$ & $83.0\pm 8.0$ & \cite{Simon:2004tf}
\\\hline
$0.1791$ & $75.0\pm 4.0$ & \cite{Moresco:2012jh}
\\\hline
$0.1993$ & $75.0\pm 5.0$ & \cite{Moresco:2012jh}
\\\hline
$0.2$ & $72.9\pm 29.6$ & \cite{Zhang:2012mp}
\\\hline
$0.27$ & $77.0\pm 14.0$ & \cite{Simon:2004tf}
\\\hline
$0.28$ & $88.8\pm 36.6$ & \cite{Zhang:2012mp}
\\\hline
$0.3519$ & $83.0\pm 14.0$ & \cite{Moresco:2012jh}
\\\hline
$0.3802$ & $83.0\pm 13.5$ & \cite{Moresco:2016mzx}
\\\hline
$0.4$ & $95.0\pm 17.0$ & \cite{Simon:2004tf}
\\\hline
$0.4004$ & $77.0\pm 10.2$ & \cite{Moresco:2016mzx}
\\\hline
$0.4247$ & $87.1\pm 11.2$ & \cite{Moresco:2016mzx}
\\\hline
$0.4497$ & $92.8\pm 12.9$ & \cite{Moresco:2016mzx}
\\\hline
$0.4783$ & $80.9\pm 9.0$ & \cite{Moresco:2016mzx}
\\\hline
$0.48$ & $97.0\pm 62.0$ & \cite{Stern:2009ep}
\\\hline
$0.5929$ & $104.0\pm 13.0$ & \cite{Moresco:2012jh}
\\\hline
$0.6797$ & $92.0\pm 8.0$ & \cite{Moresco:2012jh}
\\\hline
$0.7812$ & $105.0\pm 12.0$ & \cite{Moresco:2012jh}
\\\hline
$0.8754$ & $125.0\pm 17.0$ & \cite{Moresco:2012jh}
\\\hline
$0.88$ & $90.0\pm 40.0$ & \cite{Stern:2009ep}
\\\hline
$0.9$ & $117.0\pm 23.0$ & \cite{Simon:2004tf}
\\\hline
$1.037$ & $154.0\pm 20.0$ & \cite{Moresco:2012jh}
\\\hline
$1.3$ & $168.0\pm 17.0$ & \cite{Simon:2004tf}
\\\hline
$1.363$ & $160.0\pm 33.6$ & \cite{Moresco:2015cya}
\\\hline
$1.43$ & $177.0\pm 18.0$ & \cite{Simon:2004tf}
\\\hline
$1.53$ & $140.0\pm 14.0$ & \cite{Simon:2004tf}
\\\hline
$1.75$ & $202.0\pm 40.0$ & \cite{Simon:2004tf}
\\\hline
$1.965$ & $186.5\pm 50.4$ & \cite{Moresco:2015cya}
\\\hline
\end{tabular}
\end{center}
\label{compilationH_first_chapter}
\begin{scriptsize}
\caption{\scriptsize Current published values of $H(z)$ in units [km/s/Mpc] obtained using the differential-age technique (see the quoted references and point S4 in the text).}
\end{scriptsize}
\end{table}
\begin{table}[t!]
\begin{center}
\begin{tabular}{| c | c |c | c |}
\multicolumn{1}{c}{Survey} &  \multicolumn{1}{c}{$z$} &  \multicolumn{1}{c}{$f(z)\sigma_8(z)$} & \multicolumn{1}{c}{{\small References}}
\\\hline
6dFGS & $0.067$ & $0.423\pm 0.055$ & \cite{Beutler:2012px}
\\\hline
SDSS-DR7 & $0.10$ & $0.37\pm 0.13$ & \cite{Feix:2015dla}
\\\hline
GAMA & $0.18$ & $0.29\pm 0.10$ & \cite{Simpson:2015yfa}
\\ \cline{2-4}& $0.38$ & $0.44\pm0.06$ & \cite{Blake:2013nif}
\\\hline
DR12 BOSS & $0.32$ & $0.427\pm 0.052$  & \cite{Gil-Marin:2016wya}\\ \cline{2-3}
 & $0.57$ & $0.426\pm 0.023$ & \\\hline
 WiggleZ & $0.22$ & $0.42\pm 0.07$ & \cite{Blake:2011rj} \tabularnewline
\cline{2-3} & $0.41$ & $0.45\pm0.04$ & \tabularnewline
\cline{2-3} & $0.60$ & $0.43\pm0.04$ & \tabularnewline
\cline{2-3} & $0.78$ & $0.38\pm0.04$ &
\\\hline
2MTF & $0.02$ & $0.34\pm 0.04$ & \cite{Springob:2015pbs}
\\\hline
VIPERS & $0.7$ & $0.380\pm0.065$ & \cite{Granett:2015ppa}
\\\hline
VVDS & $0.77$ & $0.49\pm0.18$ & \cite{Guzzo:2008ac}\tabularnewline
 & & &\cite{Song:2008qt}
\\\hline
 \end{tabular}
\end{center}
\label{compilationLSS_first_chapter}
\begin{scriptsize}
\caption{\scriptsize Current published values of $f(z)\sigma_8(z)$. See the text, S5).}
\end{scriptsize}
\end{table}
\newline
\newline
{\bf S1)} The SNIa data points from the SDSS-II/SNLS3 Joint Light-curve Analysis (JLA) \cite{Betoule:2014frx}. We have used the $31$ binned distance modulus fitted to the JLA sample and the compressed form of the likelihood with the corresponding covariance matrix.
\newline
\newline
{\bf S2)} 5 points on the isotropic BAO estimator $r_s(z_d)/D_V(z_i)$: $z=0.106$ \cite{Beutler:2011hx}, $z=0.15$\, \cite{Ross:2014qpa}, $z_i=0.44, 0.6, 0.73$ \cite{Kazin:2014qga}, with the correlations between the last 3 points.
\newline
\newline
{\bf S3)} 6 data points on anisotropic BAO estimators: 4 of them on  $D_A(z_i)/r_s(z_d)$ and $H(z_i)r_s(z_d)$ at $z_i=0.32, 0.57$, for the LOWZ and CMASS samples, respectively. These data are taken from \cite{Gil-Marin:2016wya}, based on the Redshift-Space Distortions (RSD) measurements of the power spectrum combined with the bispectrum, and the BAO post-reconstruction analysis of the power spectrum (cf. Table 5 of that reference), including the correlations among these data encoded in the provided covariance matrices. We also use  2 data points  based on  $D_A(z_i)/r_s(z_d)$ and $D_H(z_i)/r_s(z_d)$ at $z=2.34$, from the combined LyaF analysis \cite{Delubac:2014aqe}. The correlation coefficient among these 2 points are taken from \cite{Aubourg:2014yra} (cf. Table II of that reference). We also take into account the correlations among the  BAO data and the corresponding $f\sigma_8$ data of \cite{Gil-Marin:2016wya} -- see S5) below and Table 4.
\newline
\newline
{\bf S4)} $30$ data points on $H(z_i)$ at different redshifts, listed in Table 3. We use only $H(z_i)$ values obtained by the so-called differential-age techniques applied to passively evolving galaxies. These values are  uncorrelated with the BAO data points. See also \cite{Farooq:2013dra,Sahni:2014ooa,Zheng:2016jlq} and \cite{Chen:2016uno}, where the authors make only use of Hubble parameter data in their analyses. We find, however, indispensable to take into account the remaining data sets to derive our conclusions on dynamical vacuum, specially the BAO, LSS and CMB observations. This fact can also be verified quite evidently in {Figures 5-6}, to which we shall turn our attention in Section \ref{sec:Discussion_first_chapter}.
\newline
\newline
{\bf S5)} $f(z)\sigma_8(z)$: 13 points. These are referred to in the text as LSS (large scale structure formation). {The actual fitting results shown in Table 1 make use of the LSS data listed in Table 4, in which we have carefully avoided possible correlations among them (see below). Let us mention that although we are aware of the existence of other LSS data points in the literature concerning some of the used redshift values in our Table 4 -- cf. e.g. \cite{Percival:2004fs,Turnbull:2011ty,Hudson:2012gt,Johnson:2014kaa}, we have explicitly checked that their inclusion or not in our numerical fits has no significant impact on our main result, that is to say, it does not affect the attained $\gtrsim4\sigma$ level of evidence in favor of the RVM's. This result is definitely secured in both cases, but we have naturally presented our final results sticking to the most updated data. }
\newline
\newline
The following observation is also in order. We have included both the WiggleZ and the CMASS data sets in our analysis. We are aware that there exists some overlap region between the CMASS and WiggleZ galaxy samples. But the two surveys have
been produced independently and the studies on the existing correlations among these observational results \cite{Beutler:2015tla,Marin:2015ula} show that the correlation is small. The overlap region of the CMASS and WiggleZ galaxy samples is actually not among the galaxies that the two surveys pick up, but between the region of the sky they explore. Moreover, despite almost all the WiggleZ region (5/6 parts of it) is inside the CMASS one, it only takes a very small fraction of the whole sky region covered by CMASS, since the latter is much larger than the WiggleZ one (see, e.g. Figure 1 in \cite{Beutler:2015tla}). In this paper, the authors are able to quantify the correlation degree among the BAO constraints in CMASS and WiggleZ, and they conclude that it is less than 4\%. Therefore, we find it justified to include the WiggleZ data in the main table of results of our analysis (Table 1), but we provide also the fitting results that are obtained when we remove the WiggleZ data points from the BAO and $f(z)\sigma_8(z)$ data sets (see Table 2). The difference is small and the central values of the fitting parameters and their uncertainties remain intact. Thus the statistical significance of Tables 1 and 2 is the same.
\newline
\newline
{\bf S6)} BBN:  we have imposed the average bound on the possible variation of the BBN speed-up factor, defined as the ratio of the expansion rate predicted in a given model versus that of the  $\CC$CDM model at the BBN epoch ($z\sim 10^9$). This amounts to the limit $|\Delta H^2/H_\Lambda^2|<10\%$ \,\cite{Uzan:2010pm}.
\newline
\newline
{\bf S7)} CMB distance priors:  $R$ (shift parameter) and $\ell_a$ (acoustic length) and their correlations  with $(\omega_b,n_s)$. For WMAP9 and Planck 2013 data we used the covariance matrix from the analysis of \cite{Wang:2013mha}, while for Planck 2015 data those of \cite{Huang:2015vpa}. Our fitting results for the last case are recorded in all our tables (except in Table 5 where we test our fit in the absence of CMB distance priors $R$ and $\ell_a$). We display the final contour plots for all the cases, see Figs. 1-2. Let us point out that in the case of the Planck 2015 data we have checked that very similar results ensue for all models if we use the alternative CMB covariance matrix from \cite{Ade:2015rim}. We have, however, chosen to explicitly present the case based on  \cite{Huang:2015vpa} since it uses the more complete compressed likelihood analysis for Planck 2015 TT,TE,EE + lowP data whereas \cite{Ade:2015rim} uses Planck 2015 TT+lowP data only.
\newline
Notice that G1 and A1 have one single vacuum parameter ($\nu$) whereas G2 and A2 have two ($\nu,\alpha$). There is nonetheless a natural alignment between $\nu$ and $\alpha$ for general type-G and A models, namely $\alpha=3\nu/4$, as this entails $\xi^{\prime}=1$ (i.e. $\nueffp=0$) in Eq.\eqref{eq:xixip_first_chapter}. Recall that for G2 models we have $G(a)\sim G_0\,a^{4(1-\xi^{\prime})}$ deep in the radiation epoch, cf. Eq.\eqref{Gafunction_first_chapter}, and therefore the condition $\xi^{\prime}=1$ warrants $G$ to take the same value as the current one, $G=G_0$,  at BBN. For model G1 this is not possible (for $\nu\neq 0$) and we adopt the aforementioned $|\Delta H^2/H_{\CC}^2|<10\%$ bound. We apply the same BBN restrictions to the A1 and A2 models, which have constant $G$. With this setting all the vacuum models contribute only with one single additional parameter as compared to the $\CC$CDM: $\nu$, for G1 and A1; and $\nueff=\nu-\alpha=\nu/4$, for G2 and A2.
\newline
\newline
For the statistical analysis, we define the joint likelihood function as the product of the likelihoods for all the data sets. Correspondingly, for Gaussian errors the total $\chi^2$ to be minimized reads:
\begin{equation}
\chi^2_{\rm tot}=\chi^2_{\rm SNIa}+\chi^2_{\rm BAO}+\chi^2_{H}+\chi^2_{f\sigma_8}+\chi^2_{\rm BBN}+\chi^2_{\rm CMB}\,.
\end{equation}
Each one of these terms is defined in the standard way, for some more details see e.g. \cite{Gomez-Valent:2014rxa}, although we should emphasize that here the correlation matrices have been included. The BAO part was split as indicated in S2) and S3) above. Also, in contrast to the previous analysis of \cite{Sola:2015wwa}, we did not use here the correlated $Omh^2(z_i,z_j)$ diagnostic for $H(z_i)$ data. Instead, we use
\begin{equation}\label{chi2H_first_chapter}
\chi^{2}_{\rm H}({\bf p})=\sum_{i=1}^{30} \left[ \frac{ H(z_{i},{\bf p})-H_{\rm obs}(z_{i})}
{\sigma_{H,i}} \right]^{2}\,.
\end{equation}
%
As for the linear structure formation data we have computed the density contrast $\delta_m=\delta\rho_m/\rho_m$ for each vacuum model by adapting the cosmic perturbations formalism for type-G and type-A vacuum models.
The matter perturbation, $\delta_m$, obeys a generalized equation which depends on the RVM type. For type-A models it reads (as a differential equation with respect to the cosmic time)
\begin{equation}\label{diffeqD_first_chapter}
\ddot{\delta}_m+\left(2H+\Psi\right)\,\dot{\delta}_m-\left(4\pi
G\rmr-2H\Psi-\dot{\Psi}\right)\,\delta_m=0\,,
\end{equation}
where $\Psi\equiv-\frac{\dot{\rho}_{\CC}}{\rmr}$. For $\rL=$const. we have $\Psi=0$ and Eq.\eqref{diffeqD_first_chapter} reduces to the $\CC$CDM form\,\footnote{For details on these equations, confer the comprehensive works \cite{Gomez-Valent:2014rxa}, \cite{Gomez-Valent:2014fda} and \cite{Gomez-Valent:2015pia}}. For type-G models the matter perturbation equation is explicitly given in \cite{Sola:2015wwa}. From here we can derive the weighted linear growth $f(z)\sigma_8(z)$, where $f(z)=d\ln{\delta_m}/d\ln{a}$ is the growth factor and $\sigma_8(z)$ is the rms mass fluctuation amplitude on scales of $R_8=8\,h^{-1}$ Mpc at redshift $z$. {It is computed from}
\begin{equation}
\begin{small}\sigma_8^2(z)=\frac{\delta_m^2(z)}{2\pi^2}\int_{0}^{\infty}k^2\,P(k,\vec{p})\,W^2(kR_8)\,dk\,,\label{s88generalNN_first_chapter}
\end{small}\end{equation}
with $W$ a top-hat smoothing function (see e.g. \cite{Gomez-Valent:2014rxa} for
details). The linear matter power spectrum reads $P(k,\vec{p})=P_0k^{n_s}T^2(k,\vec{p})$, where $\vec{p}=(h,\omega_b,n_s,\Omega_m,\nueff)$ is the fitting vector for the vacuum models we are analyzing (including the $\Lambda$CDM, for which $\nueff=0$ of course), and $T(k,\vec{p})$ is the transfer function, which we take from\,\cite{Bardeen:1985tr}, {upon introducing the baryon density effects through the modified shape parameter $\Gamma$\,\cite{Peacock:1993xg,Sugiyama:1994ed}. We have also explicitly checked that the use of the effective shape of the transfer function provided in \cite{Eisenstein:1997ik} does not produce any change in our results.}
\newline
{The expression \eqref{s88general_first_chapter} at $z=0$ allows us to write $\sigma_8(0)$ in terms of the power spectrum normalization factor $P_0$ and the primary parameters that enter our fit for each model (cf. Table 1). We fix $P_0$ from}
\begin{equation}
\begin{small}P_0=2\pi^2\frac{\sigma_{8,\Lambda}^2}{\delta^2_{m,\Lambda}(0)}\left[\int_0^\infty k^{2+n_{s,\Lambda}}T^2(k,\vec{p}_\Lambda)W^2(kR_{8,\Lambda})dk\right]^{-1}\,,\label{P0_first_chapter}
\end{small}\end{equation}
{in which we have introduced the vector of fiducial parameters $\vec{p}_\CC=(h_{\CC},\omega_{b,\CC},n_{s,\CC},\Omega_{m,\CC},0)$. This vector is defined in analogy with the fitting vector introduced before, but all its parameters are fixed and taken to be equal to those from the Planck 2015 TT,TE,EE+lowP+lensing analysis\,\cite{Ade:2015xua} with $\nueff=0$. The fiducial parameter $\sigma_{8,\Lambda}$ is also taken from the aforementioned Planck 2015 data.  However, $\delta_{m,\Lambda}(0)$ in \eqref{P0_first_chapter} is computable: it is the value of $\delta_m(z=0)$ obtained from solving the perturbation equation of the $\CC$CDM using the mentioned fiducial values of the other parameters. Finally, from $\sigma_8(z) = \sigma_8(0)\delta_m(z)/\delta_m(0)$ and plugging \eqref{P0_first_chapter} in \eqref{s88general_first_chapter} one finds:}
\begin{equation}
\begin{small}\sigma_{\rm 8}(z)=\sigma_{8, \Lambda}
\frac{\delta_m(z)}{\delta_{m,\CC}(0)}
\left[\frac{\int_{0}^{\infty} k^{2+n_s} T^{2}(k,\vec{p})
W^2(kR_{8}) dk} {\int_{0}^{\infty} k^{2+n_{s,\CC}} T^{2}(k,\vec{p}_\Lambda) W^2(kR_{8,\Lambda}) dk}
\right]^{1/2}\,.\label{s88general_first_chapter}
\end{small}\end{equation}
{Computing next the weighted linear growth rate  $f(z)\sigma_8(z)$ for each model under consideration, including the $\CC$CDM, all models become normalized to the same fiducial model defined above. The results for $f(z)\sigma_8(z)$ in the various cases are displayed in Fig.\,4 together with the LSS data measurements (cf. Table 4). We will further comment on these results in the next section.}
\begin{figure}[t!]
\begin{center}
\includegraphics[angle=0,width=0.9\linewidth]{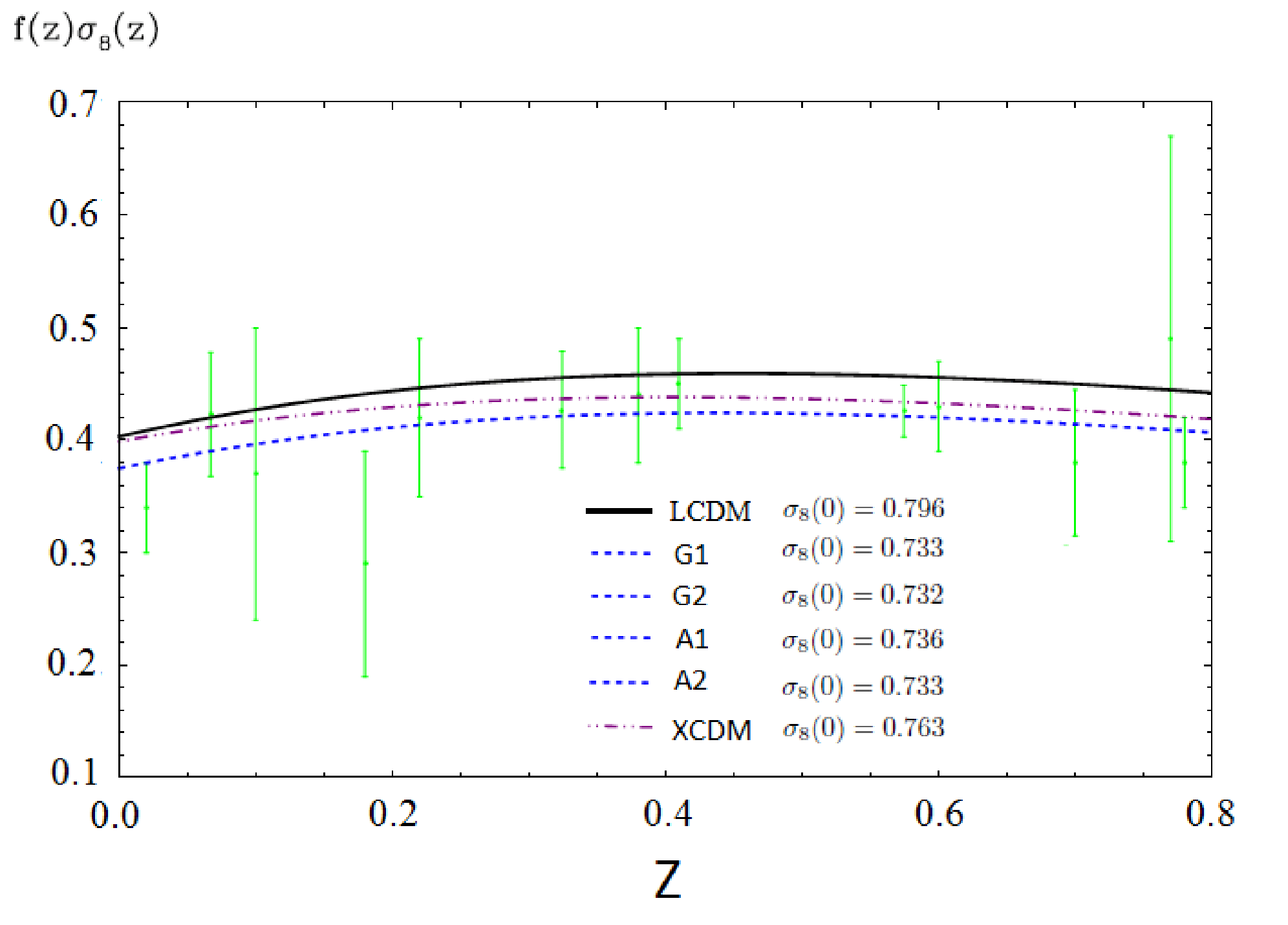}
\caption{\label{fig:fsigma8_first_chapter}%
\scriptsize {The $f(z)\sigma_8(z)$ data (Table 4) and the predicted curves by the RVM's, XCDM and the $\CC$CDM, using the best-fit values in Table 1. Shown are also the values of $\sigma_8(0)$ that we obtain for all the models. The theoretical prediction of all the RVM's are visually indistinguishable and they have been plotted using the same (blue) dashed curve.}
}
\end{center}
\end{figure}
\subsection{Discussion}\label{sec:Discussion_first_chapter}
Table 1 and Figures 1-2 present in a nutshell our main results. We observe that the effective vacuum parameter, $\nueff$, is neatly projected non null and positive for all the RVM's. The presence of this effect can be traced already in the old WMAP9 data (at $\sim 2\sigma$), but as we can see it becomes strengthened at $\sim 3.5\sigma$ c.l. with the  Planck 2013 data  and at $\sim4\sigma$ c.l.  with the Planck 2015 data -- see Figs. 1 and 2. For Planck 2015 data it attains up to $\gtrsim4.2\sigma$ c.l. for all the RVM's after marginalizing over the other fitting parameters.
%
\begin{table}[t!]
\begin{center}
\resizebox{1\textwidth}{!}{
\begin{tabular}{| c | c |c | c | c | c | c| c | c | c |}
\multicolumn{1}{c}{Model} &  \multicolumn{1}{c}{$h$} &  \multicolumn{1}{c}{$\omega_b= \Omega_b h^2$} & \multicolumn{1}{c}{{\small$n_s$}}  &  \multicolumn{1}{c}{$\Omega_m$}&  \multicolumn{1}{c}{{\small$\nu_{\rm eff}$}}  & \multicolumn{1}{c}{$\omega$} &
\multicolumn{1}{c}{$\chi^2_{\rm min}/dof$} & \multicolumn{1}{c}{$\Delta{\rm AIC}$} & \multicolumn{1}{c}{$\Delta{\rm BIC}$}\vspace{0.5mm}
\\\hline
$\Lambda$CDM  & $0.679\pm 0.005$ & $0.02241\pm 0.00017$ &$0.968\pm 0.005$& $0.291\pm 0.005$ & - & $-1$ & 68.42/83 & - & - \\
\hline
$X$CDM  & $0.673\pm 0.007$& $0.02241\pm 0.00017 $&$0.968\pm 0.005$& $0.299\pm 0.009$& - & $-0.958\pm0.038$  & 67.21/82  & -1.10 & -3.26 \\
\hline
A1  & $0.679\pm 0.010$& $0.02241\pm0.00017 $&$0.968\pm0.005$& $0.291\pm0.010$ &$-0.00001\pm 0.00079 $ & $-1$  & 68.42/82 & -2.31 & -4.47 \\
\hline
A2   & $0.676\pm 0.009$& $0.02241\pm0.00017 $&$0.968\pm0.005$& $0.295\pm0.014$ &$0.00047\pm 0.00139 $& $-1$  & 68.31/82 & -2.20 & -4.36 \\
\hline
G1 & $0.679\pm 0.009$& $0.02241\pm0.00017 $&$0.968\pm0.005$& $0.291\pm0.010$ &$0.00002\pm 0.00080 $& $-1$  &  68.42/82 & -2.31  &  -4.47\\
\hline
G2  & $0.678\pm 0.012$& $0.02241\pm0.00017 $&$0.968\pm0.005$& $0.291\pm0.013$ &$0.00006\pm 0.00123 $& $-1$  &  68.42/82 & -2.31  &  -4.47\\
\hline
\end{tabular}}
\end{center}
\label{tableFitRla_first_chapter}
\begin{scriptsize}
\caption{\scriptsize Same as in Table 1, but removing both the $R$-shift parameter and the acoustic length $l_a$ from our fitting analysis.}
\end{scriptsize}
\end{table}
\begin{table}[t!]
\begin{center}
\resizebox{1\textwidth}{!}{
\begin{tabular}{| c | c |c | c | c | c | c| c | c | c |}
\multicolumn{1}{c}{Model} &  \multicolumn{1}{c}{$h$} &  \multicolumn{1}{c}{$\omega_b= \Omega_b h^2$} & \multicolumn{1}{c}{{\small$n_s$}}  &  \multicolumn{1}{c}{$\Omega_m$}&  \multicolumn{1}{c}{{\small$\nu_{\rm eff}$}}  & \multicolumn{1}{c}{$\omega$}  &
\multicolumn{1}{c}{$\chi^2_{\rm min}/dof$} & \multicolumn{1}{c}{$\Delta{\rm AIC}$} & \multicolumn{1}{c}{$\Delta{\rm BIC}$}\vspace{0.5mm}
\\\hline
$\Lambda$CDM  & $0.685\pm 0.004$ & $0.02243\pm 0.00014$ &$0.969\pm 0.004$& $0.304\pm 0.005$ & - & $-1$  & 61.70/72 & - & - \\
\hline
XCDM  & $0.683\pm 0.009$ & $0.02245\pm 0.00015$ &$0.969\pm 0.004$& $0.306\pm 0.008$ & - & $-0.991\pm0.040$  & 61.65/71 & -2.30 & -4.29 \\
\hline
A1  & $0.685\pm 0.010$& $0.02243\pm0.00014 $&$0.969\pm0.004$& $0.304\pm0.005$ &$0.00003\pm 0.00062 $ & $-1$  & 61.70/71 & -2.36 & -4.34 \\
\hline
A2   & $0.684\pm 0.009$& $0.02242\pm0.00016 $&$0.969\pm0.005$& $0.304\pm0.005$ &$0.00010\pm 0.00095 $ & $-1$  & 61.69/71 & -2.35 & -4.33 \\
\hline
G1 & $0.685\pm 0.010$& $0.02243\pm0.00014 $&$0.969\pm0.004$& $0.304\pm0.005$ &$0.00003\pm 0.00065 $ & $-1$  & 61.70/71 & -2.36 & -4.34 \\
\hline
G2  & $0.685\pm 0.010$& $0.02242\pm0.00015 $&$0.969\pm0.004$& $0.304\pm0.005$ &$0.00006\pm 0.00082 $ & $-1$  & 61.70/71 & -2.36 & -4.34 \\
\hline
\end{tabular}}
\end{center}
\label{tableFitLSS_first_chapter}
\begin{scriptsize}
\caption{\scriptsize Same as in Table 1, but removing the points from the LSS data set from our analysis,  i.e. all the 13 points on $f\sigma_8$ .}
\end{scriptsize}
\end{table}
%
%
It is also interesting to gauge the dynamical character of the DE by performing a fit to the overall data in terms of the well-known XCDM parametrization, in which the DE is mimicked through the density $\rho_X(a)=\rho_{X0}\,a^{-3(1+\omega)}$ associated to some generic entity X, which acts as an ersatz for the $\CC$ term; $\rho_{X0}$ being the current energy density value of X and therefore equivalent to $\rho_{\CC 0}$, and  $\omega$ is the (constant) equation of state (EoS) parameter for X. The XCDM trivially boils down to the rigid $\CC$-term for $\omega=-1$, but by leaving $\omega$ free it proves a useful approach to roughly mimic a (non-interactive) DE scalar field with constant EoS. The corresponding fitting results are included in all our tables along with those for the RVM's and the $\CC$CDM. {In Table 1 (our main table) and in Fig. 3, we can see that the best fit value for $\omega$ in the XCDM is $\omega=-0.916\pm0.021$. Remarkably, it departs from $-1$ by precisely $4\sigma$.}
\newline
{Obviously, given the significance of the above result it is highly convenient to compare it with previous analyses of the XCDM reported by the Planck and BOSS collaborations. The Planck 2015 value for the EoS parameter of the XCDM reads $\omega = -1.019^{+0.075}_{-0.080}$  \cite{Ade:2015xua} and the BOSS one is $\omega = -0.97\pm 0.05$\,\cite{Aubourg:2014yra}. These results are perfectly compatible with our own result for $\omega$ shown in Table 1 for the XCDM, but in stark contrast to our result their errors are big enough as to be also fully compatible with the $\CC$CDM value $\omega=-1$. This is, however, not too surprising if we take into account that none of these analyses included LSS data in their fits, as explicitly indicated in their papers\,\footnote{Furthermore, at the time these analyses appeared they could not have used the important LSS and BAO results from \cite{Gil-Marin:2016wya}, i.e. those that we have incorporated as part of our current data set, not even the previous ones from\,\cite{Gil-Marin:2016wya}. The latter also carry a significant part of the dynamical DE signature we have found here, as we have checked.}. In the absence of LSS data we would find a similar situation. In fact, as our Table 6 clearly shows, the removal of the LSS data set in our fit induces a significant increase in the magnitude of the central value of the EoS parameter, as well as the corresponding error. This happens because the higher is $|\omega|$ the higher is the structure formation power predicted by the XCDM, and therefore the closer is such prediction with that of the $\CC$CDM (which is seen to predict too much power as compared to the data, see Fig. 4). In these conditions our analysis renders $\omega = -0.991\pm 0.040$, which is definitely closer to (and therefore compatible with) the central values obtained by Planck and BOSS teams. In addition, this result is now fully compatible with the $\CC$CDM, as in the Planck 2015 and BOSS cases, and all of them are unfavored by the LSS observations. This is consistent with the fact that both information criteria, $\Delta$AIC and $\Delta$BIC, become now slightly negative in Table 6, which reconfirms that if the LSS data are not used the $\CC$CDM performance is  comparable or even better than the other models. So in order to fit the  observed values of $f\sigma_8$, which are generally lower than the predicted ones by the $\CC$CDM, $|\omega|$ should decrease. This is exactly what happens for the XCDM, as well as for the RVM's, when the LSS data are included in our analysis (in combination with the other data, particularly with BAO and CMB data). It is apparent from Fig. 4 that the curves for these models are  then shifted below and hence adapt significantly better to the data points. Correspondingly, the quality of the fits increases dramatically, and this is also borne out by the large and positive values of  $\Delta$AIC and $\Delta$BIC, both above $10$ (cf. Table 1).}
\newline
{The above discussion explains why our analysis of the observations through the XCDM is sensitive to the dynamical feature of the DE, whereas the previous results in the literature  are not. It also shows that the size of the effect found with such a parametrization of the DE essentially concurs with the strength of the dynamical vacuum signature found for the RVM's using exactly the same data. This is remarkable, and it was not obvious a priori} since for some of our RVM's (specifically for A1 and A2) there is an interaction between vacuum and matter that triggers an anomalous conservation law, whereas for others (G1 and G2) we do not have such interaction (meaning that matter is conserved in them, thereby following the standard decay laws for relativistic and non-relativistic components). The interaction, when occurs, is however proportional to $\nueff$ and thus is small because the fitted value of $\nueff$ is small. This probably explains why the XCDM can succeed in nailing down the dynamical nature of the DE with a comparable performance. However not all dynamical vacuum models describe the data with the same efficiency, see e.g. \,\cite{Salvatelli:2014zta,Murgia:2016ccp,Li:2015vla}. In the XCDM case the departure from the $\CC$CDM takes the fashion of  ``effective quintessence'', whereas for the RVM's it appears as genuine vacuum dynamics. In all cases, however, we find unmistakable signs of DE physics beyond the $\CC$CDM (cf. Table 1), and this is a most important result of our work.
\newline
As we have discussed in Section \ref{sec:RVMs_first_chapter}, for models A1 and A2 there is an interaction between vacuum and matter. Such interaction is, of course, small because the fitted values of $\nueff$  are small, see Table 1. The obtained values are in the ballpark of $\nueff\sim {\cal O}(10^{-3})$ and therefore this is also the order of magnitude associated to the anomalous conservation law of matter. For example, for the non-relativistic component we have
\begin{equation}
\rho_m(a)=\rho_{m0}a^{-3\xi}=\rho_{m0}a^{-3(1-\nueff)}\,.
\end{equation}
This behaviour has been used in the works by \cite{Fritzsch:2012qc,Fritzsch:2015lua} as a possible explanation for the hints on the time variation of the fundamental constants, such as coupling constants and particle masses, frequently considered in the literature. The current observational values for such time variation are actually compatible with the fitted values we have found here. This is an intriguing subject that is currently of high interest in the field, see e.g.\,\cite{Uzan:2010pm} and \cite{Sola:2015xga}.
For models G1 and G2, instead, the role played by $\nueff$ and $\nueffp$ is different. It does not produce any anomaly in the traditional matter conservation law (since matter and radiation are conserved for type-G models), but now it impinges a small (logarithmic) time evolution on $G$ in the fashion sketched in Eq.\eqref{Gafunction_first_chapter}. Thus we find, once more, a possible description for the potential variation of the fundamental constants, in this case $G$, along the lines of the above cited works, see also \cite{Fritzsch:2016ewd}. There are, therefore, different phenomenological possibilities to test the RVM's considered here from various points of view.
\begin{table}[t!]
\begin{center}
\resizebox{1\textwidth}{!}{
\begin{tabular}{| c | c |c | c | c | c | c| c | c | c |}
\multicolumn{1}{c}{Model} &  \multicolumn{1}{c}{$h$} &  \multicolumn{1}{c}{$\omega_b= \Omega_b h^2$} & \multicolumn{1}{c}{{\small$n_s$}}  &  \multicolumn{1}{c}{$\Omega_m$}&  \multicolumn{1}{c}{{\small$\nu_{\rm eff}$}}  & \multicolumn{1}{c}{$\omega$}  &
\multicolumn{1}{c}{$\chi^2_{\rm min}/dof$} & \multicolumn{1}{c}{$\Delta{\rm AIC}$} & \multicolumn{1}{c}{$\Delta{\rm BIC}$}\vspace{0.5mm}
\\\hline
$\Lambda$CDM  & $0.693\pm 0.006$ & $0.02265\pm 0.00022$ &$0.976\pm 0.004$& $0.293\pm 0.007$ & - & $-1$  & 39.35/38 & - & - \\
\hline
XCDM  & $0.684\pm 0.010$ & $0.02272\pm 0.00023$ &$0.977\pm 0.005$& $0.300\pm 0.009$ & - & $-0.960\pm0.033$  & 37.89/37 & -1.25 & -2.30 \\
\hline
A1  & $0.681\pm 0.011$& $0.02254\pm0.00023 $&$0.972\pm0.005$& $0.297\pm0.008$ &$0.00057\pm 0.00043 $ & $-1$  & 37.54/37 & -0.90 & -1.95 \\
\hline
A2   & $0.684\pm 0.009$& $0.02252\pm0.00024 $&$0.971\pm0.005$& $0.297\pm0.008$ &$0.00074\pm 0.00057 $ & $-1$  & 37.59/37 & -0.95 & -2.00 \\
\hline
G1 & $0.681\pm 0.011$& $0.02254\pm0.00023 $&$0.972\pm0.005$& $0.297\pm0.008$ &$0.00059\pm 0.00045 $ & $-1$  & 37.54/37 & -0.90 & -1.95 \\
\hline
G2  & $0.682\pm 0.010$& $0.02253\pm0.00024 $&$0.971\pm0.005$& $0.297\pm0.008$ &$0.00067\pm 0.00052 $ & $-1$  & 37.61/37 & -0.97 & -2.02 \\
\hline
\end{tabular}}
\end{center}
\label{tableFitPlanck_first_chapter}
\begin{scriptsize}
\caption{\scriptsize Fitting results using the same data as in \cite{Ade:2015rim}}
\end{scriptsize}
\end{table}
%
%
%
We may reassess the quality fits obtained in this work from a different point of view. While the $\chi^2_{\rm min}$ value of the overall fit for any RVM and the XCDM is seen to be definitely smaller than the $\CC$CDM one, it proves very useful to reconfirm our conclusions with the help of the time-honored Akaike and Bayesian information criteria, AIC and BIC, see\,\cite{Akaike,Schwarz1978,Burnham2002}.
They read as follows:
\begin{equation}\label{eq:AICandBIC_first_chapter}
{\rm AIC}=\chi^2_{\rm min}+\frac{2nN}{N-n-1}\,,\ \ \ \ \ {\rm BIC}=\chi^2_{\rm min}+n\,\ln N\,.
\end{equation}
In both cases, $n$ is the number of independent fitting parameters and $N$ the number of data points used in the analysis.
To test the effectiveness of a dynamical DE model (versus the $\CC$CDM) for describing the overall data, we evaluate the pairwise differences $\Delta$AIC ($\Delta$BIC) with respect to the model that carries smaller value of AIC (BIC) -- in this case, the RVM's or the XCDM. The larger these differences the higher is the evidence against the
model with larger value of  AIC (BIC) -- the $\CC$CDM, in this case.
For $\Delta$AIC and/or $\Delta$BIC in the range $6-10$ one may claim ``strong evidence'' against such model; and, above 10, one speaks of ``very strong evidence''\,\cite{Akaike,Burnham2002}. The evidence ratio associated to rejection of the unfavored model is given by the ratio of Akaike weights, $e^{\Delta{\rm AIC}/2}$. Similarly, $e^{\Delta{\rm BIC}/2}$ estimates the so-called Bayes factor, which gives the ratio of marginal likelihoods between the two models\,\cite{Amendola:2015ksp}.
Table 1 reveals conspicuously that the $\CC$CDM appears very strongly disfavored (according to the above statistical standards) as compared to the running vacuum models. Specifically, $\Delta$AIC is in the range $17-18$ and $\Delta$BIC around $15$ for all the RVM's. These results are fully consistent and since both  $\Delta$AIC and  $\Delta$BIC are well above $10$ the verdict of the information criteria is conclusive. But there is another remarkable feature to single out at this point, namely the fact that the simple XCDM parametrization is now left behind as compared to the RVM's. While the corresponding XCDM values of $\Delta$AIC and $\Delta$BIC are also above $10$ (reconfirming the ability of the XCDM to improve the $\CC$CDM fit) they stay roughly $4$ points below the corresponding values for the RVM's. This is considered a significant difference from the point of view of the information criteria. Therefore, we conclude that the RVM's are significantly better than the XCDM in their ability to fit the data. In other words, the vacuum dynamics inherent to the RVM's seems to describe better the overall cosmological data than the effective quintessence behaviour suggested by the XCDM parametrization.
Being the ratio of Akaike weights and Bayes factor  much bigger for the RVM's than for the $\CC$CDM, the former appear definitely much more successful than the latter. The current analysis undoubtedly reinforces the conclusions of our previous study\,\cite{Sola:2015wwa}, with the advantage that the determination of the vacuum parameters is here much more precise and therefore at a higher significance level. Let us stand out some of the most important differences with respect to that work: 1) To start with, we have used now a larger and fully updated set of cosmological data; 2) The selected data set is uncorrelated and has been obtained from independent analysis in the literature, see points S1-S7) above and references therein; 3) We have taken into account all the known covariance matrices among the data; 4) In this work, $h$, $\omega_b$ and $n_s$ are not fixed a priori (as we did in the previous one), we have now allowed them to vary in the fitting process. This is, of course, not only a more standard procedure, but also a most advisable one in order to obtain unbiased results. The lack of consensus on the experimental value of $h$ is the main reason why we have preferred to use an uninformative flat prior  -- in the technical sense -- for this parameter. This should be more objective in these circumstances, rather than being subjectively elicited -- once more in the technical sense -- by any of these more or less fashionable camps for $h$ that one finds in the literature, \cite{Riess:2011yx,Chen:2011ab,Freedman:2012ny,Hinshaw:2012aka,Sievers:2013ica,Aubourg:2014yra,Ade:2015xua,Riess:2016jrr}, whose ultimate fate is unknown at present (compare, e.g. the value from \cite{Ade:2015xua} with the one from \cite{Riess:2016jrr}, which is $\sim 3\sigma$ larger than the former); 5) But the most salient feature perhaps, as compared to our previous study, is that we have introduced here a much more precise treatment of the CMB, in which we used not only the shift parameter, $R$, (which was the only CMB ingredient in our previous study) but the full data set indicated in S7) above, namely $R$ together with $\ell_a$ (acoustic length) and their correlations  with $(\omega_b,n_s)$.
%
%
\begin{figure}[t!]
\begin{center}
\includegraphics[angle=0,width=0.8\linewidth]{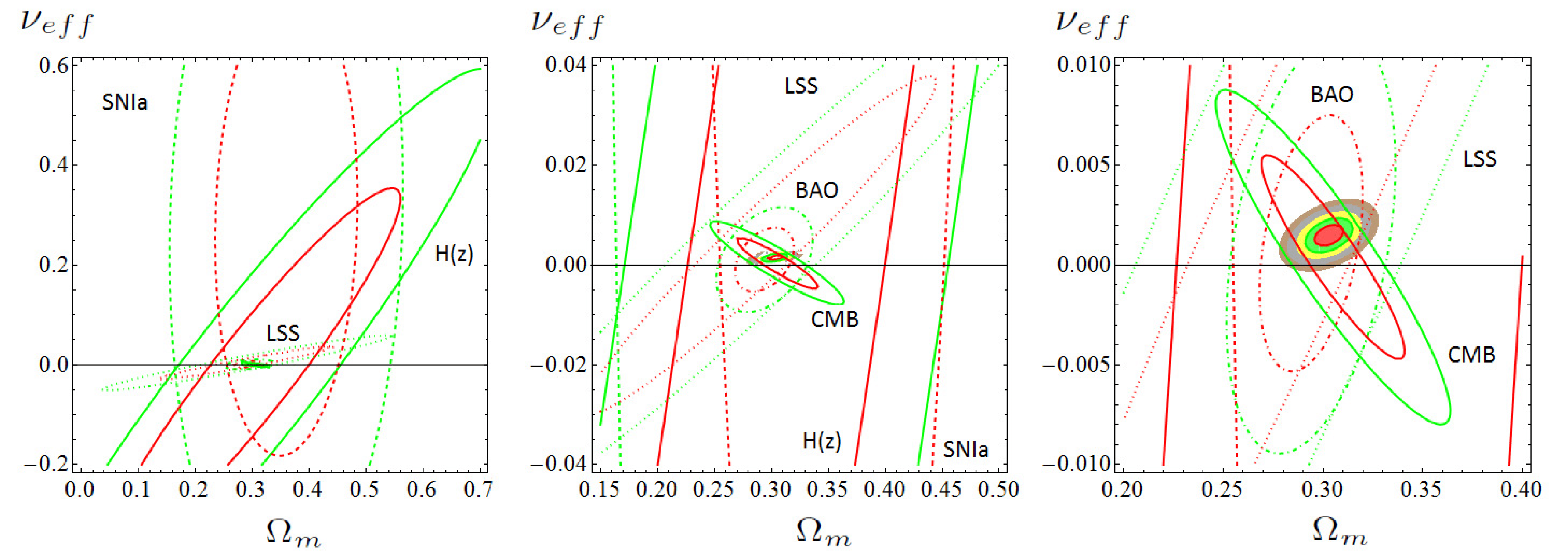}
\caption{\label{fig:A2reconstruction_first_chapter}%
\scriptsize Reconstruction of the contour lines for model A2, under  Planck 2015 CMB data (rightmost plot in Fig. 2) from the partial contour plots of the different SNIa+BAO+$H(z)$+LSS+BBN+CMB  data sources. The $1\sigma$ and $2\sigma$ contours are shown in all cases. For the reconstructed final contour lines we also plot the $3\sigma$, $4\sigma$ and $5\sigma$ regions.
}
\end{center}
\end{figure}
Altogether, this explains the substantially improved accuracy obtained in the current fitted values of the $\nueff$ parameter as compared to \cite{Sola:2015wwa}. In particular, in what concerns points 1-3) above  we should stress that for the present analysis we are using a much more complete and restrictive BAO data set. Thus, while in our previous work we only used 6  BAO data points based on the $A(z)$ estimator (cf. Table 3 of \cite{Blake:2011en}), here we are using a total of 11 BAO points (none of them based on $A(z)$, see S2-S3). These include the recent results from\,\cite{Gil-Marin:2016wya}, which narrow down the allowed parameter space in a more efficient way, not only because the BAO data set is larger but also owing to the fact that each of the data points is individually more precise and the known correlation matrices have been taken into account. Altogether, we are able to significantly reduce the error bars with respect to the ones we had obtained in our previous work. We have actually performed a practical test to verify what would be the impact on the fitting quality of our analysis if we would remove the acoustic length $l_a$ from the CMB part of our data and replace the current BAO data points by those used in \cite{Sola:2015wwa}. Notice that the CMB part is now left essentially with the $R$-shift parameter only, which was indeed the old situation. The result is that we recover the error bars' size shown in the previous papers, which are $\sim 4-5$ times larger than the current ones, i.e. of order $\mathcal{O}(10^{-3})$. We have also checked what would be the effect on our fit if we would remove both the data on the shift parameter and on the acoustic length;  or if we would remove only the data points on LSS. The results are presented in Tables 5 and 6, respectively. We observe that the  $\Delta$AIC and $\Delta$BIC values become $2-4$ points negative. This means that the full CMB and LSS data are individually very important for the quality of the fit and that without any of them the evidence of dynamical DE would be lost. If we would restore part of the CMB effect on the fit in Table 5 by including the $R$-shift parameter in the fitting procedure we can recover, approximately, the situation of our previous analysis, but not quite since the remaining data sources used now are more powerful.
\newline
It is also interesting to explore what would have been the result of our fits if we would not have used our rather complete SNIa+BAO+$H(z)$+LSS+BBN+CMB data set and had restricted ourselves to the much more limited one used by the Planck 2015 collaboration in the paper \cite{Ade:2015rim}. {The outcome is  presented in Table 7. In contrast to \cite{Ade:2015xua}, where no LSS (RSD) data were used, the former reference uses some BAO and LSS data, but their fit is rather limited in scope since they use only 4 BAO data points, 1 AP (Alcock-Paczynski parameter) data point, and one single LSS point, namely $f\sigma_8$ at $z=0.57$}, see details in that paper. In contradistinction to them, in our case we used 11 BAO and 13 LSS data points, some of them very recent and of high precision\,\cite{Gil-Marin:2016wya}.  From Table 7 it is seen that with only the data used in \cite{Ade:2015rim}  the fitting results for the RVM's are poor enough and cannot still detect clear traces of the vacuum dynamics. In particular, the $\Delta$AIC and $\Delta$BIC values in that table are moderately negative, showing that the $\Lambda$CDM does better with only these data. As stated before, not even the XCDM parametrization is able to detect any trace of dynamical DE with that limited data set, as the effective EoS is compatible with $\omega=-1$ at roughly $1\sigma$ ($\omega=-0.960\pm 0.033$). This should explain why the features that we are reporting here have been missed till now.
\newline
We complete our analysis by displaying in a graphical way the contributions from the different data sets to our final contour plots in {Figs. 1-3. We start analyzing the RVM's case.} For definiteness we concentrate on the rightmost plot for model A2 in Fig. 2, but we could do similarly for any other one {in Figs 1-2}. The result for model A2 is depicted in Fig. 5, where we can assess the detailed reconstruction of the final contours in  terms of the partial contours from the different SNIa+BAO+$H(z)$+LSS+BBN+CMB data sources. This reconstruction is presented through a series of three plots made at different magnifications. In the third plot of the sequence we can easily appraise that the BAO+LSS+CMB data subset plays a fundamental role in narrowing down the final physical region of the $(\Omega_m,\nueff)$ parameter space, in which all the remaining parameters have been marginalized over. This reconstruction also explains in very obvious visual terms why the conclusions that we are presenting here hinge to a large extent on considering the most sensitive components of the data. While CMB obviously is a high precision component in the fit, we demonstrate in our study (both numerically and graphically) that the maximum power of the fit is achieved when it is combined with the wealth of BAO and LSS data points currently available.
\begin{figure}[t!]
\begin{center}
\includegraphics[angle=0,width=0.52\linewidth]{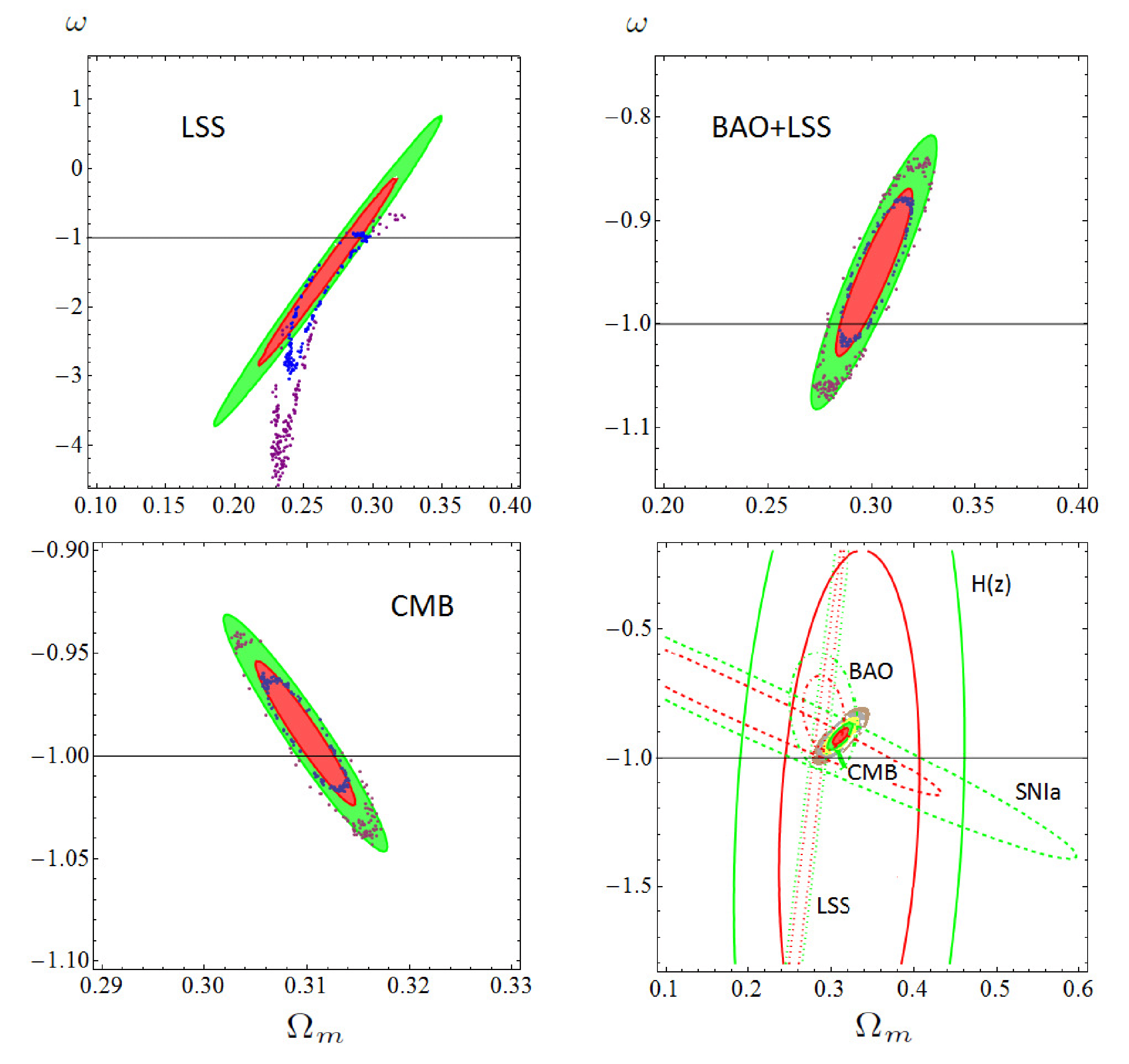}
\caption{\label{fig:XCDMreconstruction_first_chapter}%
{\scriptsize {\it Upper-left plot:} two-dimensional $\Omega_m-\omega$ contours at $1\sigma$ and $2\sigma$ c.l. for the XCDM, obtained with only the LSS data set. The dotted contours in blue and purple are the exact ones, whilst the red and green ellipses have been obtained using the Fisher's approximation. {\it Upper-right plot:} Same, but for the combination BAO+LSS. {\it Lower-left plot:} As in the upper plots, but for the CMB data. {\it Lower-right plot:} The Fisher's generated contours at $1\sigma$ and $2\sigma$ c.l. for all the data sets: SNIa (dotted lines), H(z) (solid lines), BAO (dot-dashed lines), LSS (dotted, very thin lines) and CMB (solid lines, tightly packed in a very small, segment-shaped, region at such scale of the plot). The exact, final, combined contours (from $1\sigma$ up to $5\sigma$) can be glimpsed in the small colored area around the center. See the text for further explanations and Fig. 3 for a detailed view.}
}
\end{center}
\end{figure}
{In Fig. 6 we show the corresponding decomposition of the data contours for the XCDM model as well. In the upper-left plot we display the two-dimensional contours at $1\sigma$ and $2\sigma$ c.l. in the $(\Omega_m,\omega)$ plane, found using only the LSS data set. The elliptical shapes are obtained upon applying the Fisher matrix formalism\,\cite{Amendola:2015ksp}, i.e. assuming that the two-dimensional distribution is normal (Gaussian) not only in the closer neighborhood of the best-fit values, but in all the parameter space. In order to obtain the dotted contours we have sampled the exact distribution making use of the Metropolis-Hastings Markov chain Monte Carlo algorithm \cite{Metropolis:1953am,Hastings:1970aa}. We find a significant deviation from the ideal perfectly Gaussian case. In the upper-right plot we do the same for the combination BAO+LSS. The continuous and dotted contours are both elliptical, which remarkably demonstrates the Gaussian behaviour of the combined BAO+LSS distribution. Needless to say the correlations among BAO and LSS data (whose covariance matrices are known) are responsible for that, i.e. they explain why the product of the non-normal distribution obtained from the LSS data and the Gaussian BAO one produces perfectly elliptical dotted contours for the exact BAO+LSS combination. Similarly, in the lower-left plot we compare the exact (dotted) and Fisher's generated (continuous) lines for the CMB data. Again, it is apparent that the distribution inferred from the CMB data in the $(\Omega_m,\omega)$ plane is a multivariate normal. Finally, in the lower-right plot we produce the contours at $1\sigma$ and $2\sigma$ c.l. for all the data sets in order to study the impact of each one of them. They have all been found using the Fisher approximation, just to sketch the basic properties of the various data sets, despite we know that the exact result deviates from this approximation and therefore their intersection is not the final answer. The final contours (up to $5\sigma$) obtained from the exact distributions can be seen in the small colored area around the center of the lower-right plot. The reason to plot it small at that scale is to give sufficient perspective to appreciate the contour lines of all the participating data. The final plot coincides, of course, with the one in Fig.3, where it can be appraised in full detail.}
\newline
{As it is clear from Fig. 6, the data on the $H(z)$ and SNIa observable are not crucial for distilling the final dynamical DE effect, as they have a very low constraining power. This was also so for the RVM case. Once more the final contours are basically the result of the combination of the crucial triplet of BAO+LSS+CMB data (upon taking due care in this case of the deviations from normality of the LSS-inferred distribution). The main conclusion is essentially the same as for the corresponding RVM analysis of combined contours in Fig. 5, except that in the latter there are no significant deviations from the normal distribution behaviour, as we have checked, and therefore all the contours in Fig. 5 can be accurately computed using the Fisher's matrix method. }
\newline
{The net outcome is that using either the XCDM or the RVM's the  signal in favor of the DE dynamics is clearly pinned down and in both cases it is the result of the combination of all the data sets used in our detailed analysis, although to a large extent it is generated from the crucial BAO+LSS+CMB combination of data sets. In the absence of any of them the signal would get weakened, but when the three data sets are taken together they have enough power to capture the signal of dynamical DE at the remarkable level of $\sim 4\sigma$.}
\subsection{Conclusions}\label{sec:Conclusions_first_chapter}
To conclude, the running vacuum models emerge as serious alternative candidates for the description of the current state of the Universe in accelerated expansion. These models have a close connection with the possible quantum effects on the effective action of QFT in curved space-time, cf. \cite{Sola:2013gha} and references therein. There were previous phenomenological studies that hinted in different degrees at the possibility that the RVM's could fit the data similarly as the $\CC$CDM, see e.g. the earlier works by \cite{Basilakos:2009wi,Grande:2011xf,Basilakos:2012ra,Basilakos:2014tha}, as well as the more recent ones by \cite{Gomez-Valent:2014rxa} and \cite{Gomez-Valent:2014fda}, including of course the study that precedes this work, \cite{Sola:2015wwa}.  However, to our knowledge there is no devoted work comparable in scope to the one presented here for the running vacuum models under consideration. The significantly enhanced level of dynamical DE evidence attained with them is unprecedented, to the best of our knowledge, all the more if we take into account the diversified amount of data used. Our study employed for the first time the largest updated SNIa+BAO+$H(z)$+LSS+BBN+CMB data set of cosmological observations available in the literature. Some of these data (specially the BAO+LSS+CMB part) play a crucial role in the overall fit and are substantially responsible for the main effects reported here. Furthermore, recently the BAO+LSS components have been enriched by more accurate contributions, which have helped to further enhance the signs of the vacuum dynamics. At the end of the day it has been possible to improve the significance of the dynamical hints from a confidence level of roughly $3\sigma$, as reported in our previous study \cite{Sola:2015wwa}, up to the $4.2\sigma$ achieved here. Overall, the signature of dynamical vacuum energy density seems to be rather firmly supported by the current cosmological observations. Already in terms of the generic XCDM parametrization we are able to exclude, for the first time, the absence of vacuum dynamics ($\CC$CDM) at  $4\sigma$ c.l., but such limit can be  even surpassed at the level of the RVM's and other related dynamical vacuum models, see the review \cite{Sola:2016zeg}.
\newline
{It may be quite appropriate to mention at this point of our analysis the very recent study of us \cite{Sola:2016hnq}, in which we have considered the well-known Peebles \& Ratra scalar field model with an inverse power law  potential $V(\phi)\propto \phi^{-{\alpha}}$ \cite{Peebles:1987ek}, where the power ${\alpha}$ here should, of course, not be confused with a previous use of $\alpha$ for model A2 in Section \ref{sec:RVMs_first_chapter}. In that study we consider the response of the Peebles \& Ratra model when fitted with the same data sets as those used in the current work. Even though there are  other recent tests of that model, see e.g. the works by \cite{Samushia:2009dd,Farooq:2013dra, Pourtsidou:2013nha,Pavlov:2013uqa,Arkhipova:2014dha,Pourtsidou:2016ico}, none of them used a comparably rich data set as the one we used here. This explains why the analysis of \cite{Sola:2016hnq} was able to  show that a non-trivial scalar field model, such as the Peebles \& Ratra model, is able to fit the observations at a level comparable to the models studied here. In fact, the central value of the ${\alpha}$ parameter of the potential is found to be non-zero at $\sim 4\sigma$ c.l., and the corresponding equation of state parameter $\omega$ deviates consistently from $-1$ also at the $4\sigma$ level. These remarkable features are only at reach when the crucial triplet of BAO+LSS+CMB data are at work in the fitting analysis of the various cosmological models. The net outcome of these investigations is that several models and parametrizations of the DE do resonate with the conclusion that there is a significant ($\sim 4\sigma$) effect sitting in the current wealth of cosmological data. The effect looks robust enough and can be unveiled using a variety of independent frameworks.}
Needless to say, compelling statistical evidence conventionally starts at $5\sigma$ c.l. and so we will have to wait for updated observations to see if such level of significance can eventually be attained. In the meanwhile the possible dynamical character of the cosmic vacuum, as suggested by the present study, is pretty high and gives hope for an eventual solution of the old cosmological constant problem, perhaps the toughest problem of fundamental physics.

\newpage

\newpage
\thispagestyle{empty}
\mbox{}
\newpage

\section{Dynamical dark energy: scalar fields and running vacuum}\label{Scalar_fiel_chapter}
In the previous chapter, a careful and complete analysis of the RVM's has been carried out. This models are built around the possible connection between the $\Lambda$-term and the vacuum energy. This assumption automatically leads us to the following equation of state $p_\Lambda = -\rho_\Lambda$. Nevertheless this is not the only possibility that can be considered as we shall see in this chapter.
\newline
Another very interesting option is the one in which dark energy (DE) is described in terms of a scalar field with some standard form for its potential. These kind of models, commonly denoted as $\phi$CDM models, have a local Lagrangian description. While it is true that the scalar field does not play an important role until the late time Universe, it can be part of a more general and complex potential acting in the inflationary period. 
\newline
Unlike for the RVM's models it is not possible to obtain an analytical expression for the Hubble function of the scalar field models. As a consequence we will work out some approximate solutions that will be used as an initial conditions to solve numerically the differential equations. For the sake of comparison, we present also the results obtained for the RVM as well as the results for two very well-known DE parameterizations, the XCDM and the CPL. The main upshot of the study is that a high evidence in favour of models with a dynamical DE is found.
\newline
This chapter is organized as follows: In Section \ref{sect:phiCDM_scalar_fields_chapter} we present the theoretical background of the $\phi$CDM model. We list the main cosmological equations and we write down their particular form when the Peebles \& Ratra potential $V(\phi)\sim \phi^{-\alpha}$ is chosen. We also find analytical solutions in the Matter Dominated Epoch (MDE) and in the Radiation Dominated Epoch (RDE). In sections \ref{sect:XCDMand CPL_scalar_fields_chapter} and \ref{sect:RVM_scalar_fields_chapter} for convenience we provide the background cosmological equations for the DE parameterizations before mentioned and for the RVM respectively. We extend the study from the background level to the perturbative level in Section \ref{sect:Structure_Formation_scalar_fields_chapter} whereas in Section \ref{Discussions_and_Conclusions_scalar_fields_chapter} we finally deliver our conclusions.


\subsection{$\phi$CDM with Peebles \& Ratra potential}\label{sect:phiCDM_scalar_fields_chapter}
Suppose that the dark energy is described in terms of some scalar field $\phi$ with a standard form for its potential $V(\phi)$, see below. We wish to compare its ability to describe the data with that of the $\CC$CDM, and also with other models of DE existing in the literature. The data used in our analysis will be the same one used in our previous studies\,\cite{Sola:2015wwa,Sola:2016jky,Sola:2016ecz} namely data on the distant supernovae (SNIa), the baryonic acoustic oscillations (BAO), the Hubble parameter at different redshifts, $H(z_i)$, the large scale structure formation data (LSS), and the cosmic microwave background (CMB). We denote this string of data as SNIa+BAO+$H(z)$+LSS+CMB.  Precise information on these data and corresponding observational references are given in the aforementioned papers and are also summarized in the caption of Table 1. The main results of our analysis are displayed in Tables 1-3 and Figures 1-5, which we will account in detail throughout our exposition.
\newline
\newline
We start by explaining our theoretical treatment of the $\phi$CDM model in order to optimally confront it with observations. The scalar field $\phi$ is taken to be dimensionless, being its energy density and pressure given by
\begin{equation}\label{eq:rhophi_scalar_fields_chapter}
\rho_\phi=\frac{M^2_{pl}}{16\pi}\left[\frac{\dot{\phi}^2}{2}+V(\phi)\right]\,,\ p_\phi=\frac{M^2_{pl}}{16\pi}\left[\frac{\dot{\phi}^2}{2}-V(\phi)\right]\,.
\end{equation}
Here $M_{pl}=1/\sqrt{G}=1.22\times 10^{19}$ GeV is the Planck mass, in natural units.
As a representative potential we adopt the original Peebles \& Ratra (PR) form\,\cite{Peebles:1987ek}:
\begin{equation}\label{eq:PRpotential_scalar_fields_chapter}
V(\phi)=\frac{1}{2}\kappa M_{pl}^2\phi^{-\alpha}\,,
\end{equation}
in which $\kappa$ and $\alpha$ are dimensionless parameters. These are to be determined in our fit to the overall cosmological data. The motivation for such potential is well described in the original paper \cite{Peebles:1987ek}. In a nutshell: such potential stands for the power-law tail of a more complete effective potential in which inflation is also comprised. We expect $\alpha$ to be positive and sufficiently small such that $V(\phi)$ can mimic an approximate CC that is decreasing slowly with time, in fact more slowly than the matter density. Furthermore, we must have $0<\kappa\ll 1$  such that $V(\phi)$ can be positive and of the order of the measured value $\rLo\sim 10^{-47}$ GeV$^4$. In the late Universe the tail of the mildly declining potential finally surfaces over the matter density (not far away in our past) and appears as an approximate CC which dominates since then. Recent studies have considered the PR-potential in the light of the cosmological data, see e.g. \cite{Pavlov:2013uqa,Arkhipova:2014dha,Farooq:2013dra,Pourtsidou:2013nha,Pourtsidou:2016ico}. Here we show that the asset of current observations indicates strong signs of dynamical DE which can be parametrized with such potential. In this way, we corroborate the unambiguous signs recently obtained with independent DE models\,\cite{Sola:2015wwa,Sola:2016jky,Sola:2016ecz} and with a similar level of confidence.
%
%
%
\begin{table}[t!]
\setcounter{table}{0}
\begin{center}
\begin{scriptsize}
\resizebox{1\textwidth}{!}{
\begin{tabular}{| c | c |c | c | c | c |}
\multicolumn{1}{c}{Model} &  \multicolumn{1}{c}{$\Omega_m$} &  \multicolumn{1}{c}{$\omega_b= \Omega_b h^2$} & \multicolumn{1}{c}{{\small$n_s$}}  &  \multicolumn{1}{c}{$h$} &
\multicolumn{1}{c}{$\chi^2_{\rm min}/dof$}
\\\hline
{$\Lambda$CDM} & $0.294\pm 0.004$ & $0.02255\pm 0.00013$ &$0.976\pm 0.003$& $0.693\pm 0.004$ & 90.44/85    \\
\hline
 \end{tabular}
 }
\caption{{\scriptsize The best-fit values for the $\CC$CDM parameters $(\Omega_m,\omega_b,n_s,h)$. We use a total of $89$ data points from SNIa+BAO+$H(z)$+LSS+CMB observables in our fit: namely $31$ points from the JLA sample of SNIa\,\cite{Betoule:2014frx}, $11$ from BAO\,\cite{Beutler:2011hx,Ross:2014qpa,Kazin:2014qga,Gil-Marin:2016wya,Delubac:2014aqe,Aubourg:2014yra}, $30$  from $H(z)$\,\cite{Zhang:2012mp,Jimenez:2003iv,Simon:2004tf,Moresco:2012jh,Moresco:2016mzx,Stern:2009ep,Moresco:2015cya}, $13$ from linear growth \cite{Gil-Marin:2016wya,Beutler:2012px,Feix:2015dla,Simpson:2015yfa,Blake:2013nif,Blake:2011rj,Springob:2015pbs,Granett:2015ppa,Guzzo:2008ac,Song:2008qt}, and $4$ from CMB\,\cite{Huang:2015vpa}. For a summarized description of these data, see \cite{Sola:2016jky}. The quoted number of degrees of freedom ($dof$) is equal to the number of data points minus the number of independent fitting parameters ($4$ for the $\CC$CDM). For the CMB data we have used the marginalized mean values and standard deviation for the parameters of the compressed likelihood for Planck 2015 TT,TE,EE + lowP data from \cite{Huang:2015vpa}. The parameter M in the SNIa sector\,\cite{Betoule:2014frx} was dealt with as a nuisance parameter and has been marginalized over analytically. The best-fit values and the associated uncertainties for each parameter in the table have been obtained by numerically marginalizing over the remaining parameters\,\cite{Amendola:2015ksp}}}
\end{scriptsize}
\end{center}
\label{tableFit1_scalar_fields_chapter}
\end{table}
The scalar field of the $\phi$CDM models satisfies the Klein-Gordon equation in the context of the Friedmann-Lema\^\i tre-Robertson-Walker (FLRW) metric: $\ddot{\phi}+3H\dot{\phi}+{dV}/{d\phi}=0$, where $H=\dot{a}/a$ is the Hubble function.
In some cases the corresponding solutions possess the property of having an attractor-like behaviour, in which a large family of solutions are drawn towards a common trajectory\,\cite{Ratra:1987rm,Zlatev:1998tr,Steinhardt:1999nw}. If there is a long period of convergence of all the family members to that common trajectory, the latter is called a ``tracker solution''\,\cite{Zlatev:1998tr,Steinhardt:1999nw}. When the tracking mechanism is at work, it funnels a large range of initial conditions into a common final state for a long time (or forever, if the convergence is strict). Not all potentials $V$ admit tracking solutions, only those fulfilling the ``tracker condition'' $\Gamma\equiv V\,V^{\prime\prime}/(V^\prime)^2>1$\,\cite{Zlatev:1998tr,Steinhardt:1999nw}, where $V^{\prime}=\partial V/\partial\phi$.  For the Peebles-Ratra potential \eqref{eq:PRpotential_scalar_fields_chapter}, one easily finds $\Gamma=1+1/\alpha$, so it satisfies such condition precisely for $\alpha>0$.
\newline
{It is frequently possible to seek power-law solutions, i.e. $\phi(t)=A\,t^p$,
for the periods when the energy density of the Universe is dominated by some conserved matter component} $\rho(a)=\rho_1\left({a_1}/{a}\right)^n$ (we may call these periods ``nth-epochs''). For instance,
$n=3$ for the matter-dominated epoch (MDE) and $n=4$ for the radiation-dominated epoch (RDE), with
$a_1$ the scale factor at some cosmic time $t_1$ when the corresponding component dominates. We define  $a=1$ as the current value. Solving Friedmann's equation in flat space,
$3H^2(a)=8\pi G\rho(a)$, we find  $H(t)={2}/(nt)$ as a function of the cosmic time in the nth-epoch. Substituting these relations in the  Klein-Gordon equation with the Peebles-Ratra potential \eqref{eq:PRpotential_scalar_fields_chapter} leads to
\begin{equation}
p=\frac{2}{\alpha+2}\,,\qquad A^{\alpha+2}=\frac{\alpha(\alpha+2)^2M_{pl}^2\kappa n}{4(6\alpha+12-n\alpha)}\,.
\end{equation}
From the power-law form we find the evolution of the scalar field with the cosmic time:
\begin{equation}\label{eq:initialphi_scalar_fields_chapter}
\phi(t)=\left[\frac{\alpha(\alpha+2)^2M_{pl}^2\kappa n}{4(6\alpha+12-n\alpha)}\right]^{1/(\alpha+2)}t^{2/(\alpha+2)}\,.
\end{equation}
In any of the nth-epochs the equation of state (EoS) of the scalar field remains stationary. A straightforward calculation from \eqref{eq:rhophi_scalar_fields_chapter}, \eqref{eq:PRpotential_scalar_fields_chapter} and \eqref{eq:initialphi_scalar_fields_chapter} leads to a very compact form for the EoS:
\begin{equation}\label{eq:EoSphi_scalar_fields_chapter}
w_\phi=\frac{p_{\phi}}{\rho_{\phi}}=-1+\frac{\alpha n}{3(2+\alpha)}\,.
\end{equation}
Since the matter EoS in the nth-epoch is given by $\omega_n=-1+n/3$, it is clear that \eqref{eq:EoSphi_scalar_fields_chapter} can be rewritten also as $w_\phi=(\alpha\omega_n-2)/(\alpha+2)$. This is precisely the form predicted by the tracker solutions\,\cite{Zlatev:1998tr,Steinhardt:1999nw}, in which the condition $w_\phi<\omega_n$ is also secured since $|\alpha|$ is expected small. In addition, $w_\phi$ remains constant in the RDE and MDE, but its value does \textit{not} depend on $\kappa$, only on $n$ (or $\omega_n$) and $\alpha$. The fitting analysis presented in Table 2 shows that $\alpha={\cal O}(0.1)>0$ and therefore $w_\phi\gtrsim-1$. It means that the scalar field behaves as quintessence in the pure RD and MD epochs (cf. the plateaus at constant values  $w_\phi\gtrsim-1$ in Fig.\,1). Notice that the behaviour of $w_\phi$ in the interpolating epochs, including the period near our time, is \textit{not} constant (in contrast to the XCDM, see next section) and requires numerical solution of the field equations. See also\, \cite{Podariu:1999ph} for related studies.
\newline
We can trade the cosmic time in \eqref{eq:initialphi_scalar_fields_chapter} for the scale factor. This is possible using $t^2=3/(2\pi G n^2\rho)$ (which follows from Friedmann's equation in the nth-epoch) and  $\rho(a)=\rho_{0}a^{-n}=\rho_{c0}\,\Omega\,a^{-n}$, where $\Omega=\Omega_m, \Omega_r$ are the present values of the cosmological density parameters for matter ($n=3$) or radiation ($n=4$) respectively, with $\rho_{c 0}=3H_0^2/(8\pi\,G)$  the current critical energy density. Notice that $\Omega_m=\Omega_{dm}+\Omega_b$ involves both dark matter and baryons. In this way we can determine $\phi$ as a function of the scale factor in the nth-epoch. For example, in the MDE we obtain
\begin{equation}\label{eq:Phia_scalar_fields_chapter}
\phi(a)=\left[\frac{\alpha(\alpha+2)^2\bar{\kappa}}{9\times 10^4\omega_m(\alpha+4)}\right]^{1/(\alpha+2)}a^{3/(\alpha+2)}\,.
\end{equation}
Here we have conventionally defined the reduced matter density parameter $\omega_m\equiv\Omega_m\,h^2$, in which the reduced Hubble constant $h$ is defined as usual from $H_0\equiv 100h\,\varsigma$, with $\varsigma\equiv 1 {\rm km/s/Mpc}=2.133\times10^{-44} {\rm GeV}$ (in natural units). Finally, for convenience we have introduced in \eqref{eq:Phia_scalar_fields_chapter} the dimensionless parameter $\bar{\kappa}$ through $\kappa\,M_{pl}^2\equiv \bar{\kappa}\,\varsigma^2$.
%
%
%
%
\begin{table}[t!]
\begin{center}
\begin{scriptsize}
\resizebox{1\textwidth}{!}{
\begin{tabular}{| c | c |c | c | c | c | c | c | c | c|c|}
\multicolumn{1}{c}{Model} &  \multicolumn{1}{c}{$\omega_m=\Omega_m h^2$} &  \multicolumn{1}{c}{$\omega_b=\Omega_b h^2$} & \multicolumn{1}{c}{{\small$n_s$}}  &  \multicolumn{1}{c}{$\alpha$} &  \multicolumn{1}{c}{$\bar{\kappa}$}&  \multicolumn{1}{c}{$\chi^2_{\rm min}/dof$} & \multicolumn{1}{c}{$\Delta{\rm AIC}$} & \multicolumn{1}{c}{$\Delta{\rm BIC}$}\vspace{0.5mm}
\\\hline
$\phi$CDM  &  $0.1403\pm 0.0008$& $0.02264\pm 0.00014 $&$0.977\pm 0.004$& $0.219\pm 0.057$  & $(32.5\pm1.1)\times 10^{3}$ &  74.85/84 & 13.34 & 11.10 \\
\hline
 \end{tabular}
 }
\caption{{\scriptsize The best-fit values for the parameter fitting vector \eqref{eq:vfittingPhiCDM_scalar_fields_chapter} of the $\phi$CDM model with Peebles \& Ratra potential (\ref{eq:PRpotential_scalar_fields_chapter}), including their statistical significance ($\chi^2$-test and Akaike and Bayesian information criteria, AIC and BIC, see the text). We use the same cosmological data set as in Table 1. The large and positive values of $\Delta$AIC and $\Delta$BIC strongly favor the $\phi$CDM model against the $\CC$CDM.  The specific $\phi$CDM fitting parameters are $\bar{\kappa}$ and $\alpha$. The remaining parameters  $(\omega_m,\omega_b,n_s)$ are standard (see text).  The number of independent fitting parameters is $5$, see Eq.\eqref{eq:vfittingPhiCDM_scalar_fields_chapter}-- one more than in the $\CC$CDM. Using the best-fit values and the overall covariance matrix derived from our fit, we obtain: $h=0.671\pm 0.006$ and $\Omega_m=0.311\pm 0.006$, which allows direct comparison with Table 1. We find $\sim 4\sigma$ evidence in favor of $\alpha>0$. Correspondingly the EoS of $\phi$ at present appears quintessence-like at $4\sigma$ confidence level: $w_\phi= -0.931\pm 0.017$.}}
 \end{scriptsize}
\end{center}
\label{tableFit2_scalar_fields_chapter}
\end{table}
Equation \eqref{eq:Phia_scalar_fields_chapter} is convenient since it is expressed in terms of the independent parameters that enter our fit, see below.  Let us note  that $\phi(a)$, together with its derivative $\phi^{\prime}(a)=d\phi(a)/da$, allow us to fix the initial conditions in the MDE (a similar expression can be obtained for the RDE). Once these conditions are settled analytically we have to solve numerically the Klein-Gordon equation, coupled to the cosmological equations, to obtain the exact solution. Such solution must, of course, be in accordance with \eqref{eq:Phia_scalar_fields_chapter} in the pure MDE. The exact EoS is also a function $w_\phi=w_\phi(a)$, which coincides with the constant value \eqref{eq:EoSphi_scalar_fields_chapter} in the corresponding nth-epoch, but interpolates nontrivially between them. At the same time it also interpolates between the MDE and the DE-dominated epoch in our recent past, in which the scalar field energy density surfaces above the non-relativistic matter density, i.e. $\rho_{\phi}(a)\gtrsim \rho_m(a)$, at a value of $a$ near the current one $a=1$. 
The plots for the deceleration parameter, $q=-\ddot{a}/aH^2$, and the scalar field EoS, $w_\phi(a)$, for the best fit parameters of Table 2 are shown in Fig. 1.  The transition point from deceleration to acceleration ($q=0$) is at $z_t=0.628$, {which is in good agreement with the values obtained in other works \cite{Farooq:2013hq,Farooq:2016zwm}}, and is also reasonably near the $\CC$CDM one ($z_t^{\CC{\rm CDM}}=0.687$) for the best fit values in Tables 1 and 2. The plots for $\phi(a)$ and the energy densities are displayed in Fig. 2. 
From equations  (2) and (6) we can see that in the early MDE the potential of the scalar field decays as $V\sim a^{-3\alpha/(2+\alpha)}\sim  a^{-3\alpha/2} $, where in the last step we used the fact that  $\alpha$ is small. Clearly the decaying behaviour of $V$ with the expansion is much softer than that of the matter density,  $\rho_m\sim a^{-3}$, and for this reason the DE density associated to the scalar field does not play any role until we approach the current time. This fact is apparent in Fig. 2 (right), where we numerically plot the dimensionless density parameters $\Omega_i(a)=\rho_i(a)/\rho_c(a)$ as a function of the scale factor, where $\rho_c(a)=3H^2(a)/(8\pi G)$ is the evolving critical density.
\newline
\newline
{As indicated above, the current value of the EoS} can only be known after numerically solving the equations for the best fit parameters in Table 2, with the result $w_\phi(z=0)= -0.931\pm 0.017$ (cf. Fig. 1). Such result lies clearly in the quintessence regime and with a significance of $4\sigma$. It is essentially consistent with the dynamical character of the DE derived from the non-vanishing value of $\alpha$ in Table 2.
\newline
In regard to the value of $h$, there is a significant tension between non-local measurements of $h$, e.g. \cite{Ade:2015xua,Aubourg:2014yra,Chen:2011ab,Sievers:2013ica,LHuillier:2016mtc,Bernal:2016gxb,Lukovic:2016ldd}, and local ones, e.g. \cite{Riess:2016jrr}. Some of these values can differ by $3\sigma$ or more. For the $\CC$CDM model we find $h=0.693\pm 0.004$ (cf. Table 1), which is in between the ones of \cite{Ade:2015xua} and \cite{Riess:2016jrr} and is compatible with the value presented in \cite{Hinshaw:2012aka}. For the $\phi$CDM, our best-fit value is $h=0.671\pm 0.006$ (cf. caption of Table 2), which differs by more than $3\sigma$ with respect to the $\CC$CDM one in our Table 1. Still, both remain perfectly consistent with the recent estimates of $h$ from Hubble parameter measurements at intermediate redshifts\,\cite{Chen:2016uno}. At the moment it is not possible to distinguish models on the sole basis of $H(z)$ measurements. Fortunately, the combined use of the different sorts of SNIa+BAO+$H(z)$+LSS+CMB data offers nowadays a real possibility to elucidate which models are phenomenologically preferred.
%
%
%
\begin{table}[t!]
\begin{center}
\begin{scriptsize}
\resizebox{1\textwidth}{!}{
\begin{tabular}{| c | c |c | c | c | c | c | c | c | c|c|}
\multicolumn{1}{c}{Model} &  \multicolumn{1}{c}{$\Omega_m$} &  \multicolumn{1}{c}{$\omega_b= \Omega_b h^2$} & \multicolumn{1}{c}{{\small$n_s$}}  &  \multicolumn{1}{c}{$h$} &  \multicolumn{1}{c}{$\nu$}&  \multicolumn{1}{c}{$w_0$} &  \multicolumn{1}{c}{$w_1$} &
\multicolumn{1}{c}{$\chi^2_{\rm min}/dof$} & \multicolumn{1}{c}{$\Delta{\rm AIC}$} & \multicolumn{1}{c}{$\Delta{\rm BIC}$}\vspace{0.5mm}
\\\hline
{\small XCDM} & $0.312\pm 0.007$ & $0.02264\pm 0.00014$ &$0.977\pm 0.004$& $0.670\pm 0.007$ & - & $-0.916\pm 0.021$ & - & 74.91/84 & 13.28 & 11.04\\
\hline
{\small CPL} & $0.311\pm 0.009$ & $0.02265\pm 0.00014$ &$0.977\pm 0.004$& $0.672\pm 0.009$ & - & $-0.937\pm 0.085$ & $0.064\pm 0.247$  & 74.85/83 & 11.04 & 6.61\\
\hline
{\small RVM} & $0.303\pm 0.005$ & $0.02231\pm 0.00015$ &$0.965\pm 0.004$& $0.676\pm 0.005$ & $0.00165\pm 0.00038$ & -1 & - & 70.32/84 & 17.87 & 15.63\\
\hline
 \end{tabular}
 }
\caption{{\scriptsize The best-fit values for the running vacuum model (RVM), together with the XCDM and CPL parametrizations, including also their statistical significance ($\chi^2$-test and Akaike and Bayesian information criteria, AIC and BIC) as compared to the $\CC$CDM (cf. Table 1). We use the same string of cosmological SNIa+BAO+$H(z)$+LSS+CMB data as in Tables 1 and 2. The specific fitting parameters for these models are $\nu,w_0,$ and $(w_0,w_1)$ for RVM, XCDM and CPL, respectively.  The remaining parameters  are standard. For the models RVM and XCDM the number of independent fitting parameters is $5$, exactly as in the $\phi$CDM. For the CPL parametrization there is one additional parameter ($w_1$). The large and positive values of $\Delta$AIC and $\Delta$BIC strongly favor the RVM and XCDM against the $\CC$CDM. The CPL is only moderately favored as compared to the $\CC$CDM and much less favored than the $\phi$CDM, RVM and XCDM.}}
 \end{scriptsize}
\end{center}
\label{tableFit3_scalar_fields_chapter}
\end{table}
%
Let us now describe the computational procedure that we have followed for the $\phi$CDM model. The initial conditions must be expressed in terms of the parameters that enter our fit. These are defined by means of the following $5$-dimensional fitting vector:
\begin{equation}\label{eq:vfittingPhiCDM_scalar_fields_chapter}
\vec{p}_{\phi{\rm CDM}}=(\omega_m,\omega_b,n_s,\alpha,\bar{\kappa})
\end{equation}
where $\omega_b\equiv\Omega_b\,h^2$ is the baryonic component and $n_s$ is the spectral index. These two parameters are specifically involved in the fitting of the CMB and LSS data ($\omega_b$ enters the fitting of the BAO data too), whereas the other three also enter the background analysis, see \cite{Sola:2015wwa,Sola:2016jky,Sola:2016ecz} and \cite{Gomez-Valent:2014rxa,Gomez-Valent:2014fda,Gomez-Valent:2015pia} for more details in the methodology. For the $\phi$CDM we have just one more fitting parameter than in the $\CC$CDM, i.e. $5$ instead of $4$ parameters (cf. Tables 1 and 2).
However, in contrast to the $\CC$CDM, for the $\phi$CDM we are fitting the combined parameter $\omega_m=\Omega_m h^2$ rather than $\Omega_m$ and $h$ separately.
The reason is that $h$ (and hence $H_0$) is not a direct fitting parameter in this case since the Hubble function values are determined from Friedmann's equation $3H^2=8\pi\,G(\rho_{\phi}+\rho_m)$, where $\rho_{\phi}$ is given in Eq.\eqref{eq:rhophi_scalar_fields_chapter} and $\rho_m=\rho_{c 0}\Omega_m a^{-3}=(3\times 10^4/8\pi G)\varsigma^2\,\omega_m\,a^{-3}$ is the conserved matter component. This is tantamount to saying that $h$ is eventually determined from the parameters of the potential and the reduced matter density $\omega_m$. For instance, in the MDE it is not difficult to show that
\begin{equation}\label{eq:barH2_scalar_fields_chapter}
\bar{H}^2(a)=\frac{\bar{\kappa}\,\phi^{-\alpha}(a)+1.2\times 10^5\,\omega_m\,a^{-3}}{12-a^2\phi^{\prime 2}(a)}\,,
\end{equation}
where we have defined the dimensionless $\bar{H}=H/\varsigma$, and used $\dot{\phi}=a\,H\,\phi^{\prime}(a)$. As we can see from \eqref{eq:barH2_scalar_fields_chapter}, the value of $h\equiv\bar{H}(a=1)/100$ is determined once the three parameters $(\omega_m,\alpha,\bar{\kappa})$ of the fitting vector \eqref{eq:vfittingPhiCDM_scalar_fields_chapter} are given, and then $\Omega_m=\omega_m/h^2$ becomes also determined.  Recall that $\phi(a)$ is obtained by solving numerically the Klein-Gordon equation under appropriate initial conditions (see below) which also depend on the above fitting parameters. As a differential equation in the scale factor, the Klein-Gordon equation reads
\begin{equation}\label{eq:KGa_scalar_fields_chapter}
\phi^{\prime\prime}+\phi^\prime\left(\frac{\bar{H}^\prime}{\bar{H}}+\frac{4}{a}\right)-\frac{\alpha}{2}\frac{\bar{\kappa}\phi^{-(\alpha+1)}}{(a\bar{H})^2}=0\,.
\end{equation}
It can be solved after inserting \eqref{eq:barH2_scalar_fields_chapter} in it, together with
\begin{equation}
\bar{H}^\prime=-\frac{3}{2a\bar{H}}\left(\frac{a^2\bar{H}^2\phi^{\prime 2}}{6}+10^4\,\omega_m a^{-3}\right)\,.
\end{equation}
The last formula is just a convenient rephrasing of the expression $\dot{H}=-4\pi\,G\left(\rho_m+p_m+\rho_{\phi}+p_{\phi}\right)$ upon writing it in the above set of variables. According to \eqref{eq:rhophi_scalar_fields_chapter}, the sum of density and pressure for $\phi$ reads $\rho_{\phi}+p_{\phi}=\dot{\phi}^2/(16\pi{G})=a^2 \bar{H}^2{\phi^{\prime}}^2 \varsigma^2/(16\pi{G})$, and  of course $p_m=0$ for the matter pressure after the RDE.
\newline
The initial conditions for solving \eqref{eq:KGa_scalar_fields_chapter} are fixed in the mentioned nth-epochs of the cosmic evolution. They are determined from the values of the fitting parameters in \eqref{eq:vfittingPhiCDM_scalar_fields_chapter}. For example if we set these conditions in the MDE they are defined from the expression of $\phi(a)$ in Eq.\eqref{eq:Phia_scalar_fields_chapter}, and its derivative $\phi^{\prime}(a)$, both taken at some point deep in the MDE, say at a redshift $z>100$, i.e. $a<1/100$.  The result does not depend on the particular choice in this redshift range provided we do not approach too much the decoupling epoch  ($z\simeq 1100$), where the radiation component starts to be appreciable. We have also iterated our calculation when we take the initial conditions deep in the RDE ($n=4$), in which the radiation component $\rho_r$ dominates. In this case $\omega_m=\Omega_m h^2$ is replaced by $\omega_r=\Omega_r h^2$, which is a function of the radiation temperature and the effective number of neutrino species, $N_{eff}$. We find the same results as with the initial conditions settled in the MDE. In both cases the fitting values do agree and are those indicated in Table 2. Let us also mention that when we start from the RDE we find that $\rho_{\phi}(a)\ll\rho_r(a)$ at (and around) the time of BBN (Big Bang Nucleosynthesis), where $a\sim 10^{-9}$, thus insuring that the primordial synthesis of the light elements remains unscathed.
\newline
Consistency with BBN is indeed a very important point that motivates the Peebles \& Ratra's inverse power potential $\phi$CDM, Eq.\eqref{eq:PRpotential_scalar_fields_chapter}, together with the existence of the attractor solution. Compared, say to the exponential potential, $V(\phi)=V_0\,e^{-\lambda\,\phi/M_{pl}}$, the latter is inconsistent with BBN (if $\lambda$ is too small) or cannot be important enough to cause accelerated expansion at the current time (if $\lambda$ is too large) \,\cite{Ratra:1987rm,Copeland:1997et}. This can be cured with a sum of two exponential with different values of $\lambda$\,\cite{Barreiro:1999zs}, but of course it is less motivated since involves more parameters. Thus, the PR-potential seems to have the minimal number of ingredients to successfully accomplish the job. In point of fact, it is what we have now verified at a rather significant confidence level in the light of the modern cosmological data.
\newline
Finally, let us mention that we have tested the robustness of our computational program by setting the initial conditions out of the tracker path and recovering the asymptotic attractor trajectory. This is of course a numerical check, which is nicely consistent with the fact that the Peebles \& Ratra potential satisfies the aforementioned tracker condition $\Gamma>1$. More details will be reported elsewhere.
\begin{figure}[t!]
\setcounter{figure}{0}
\begin{center}
\label{Parella2_scalar_fields_chapter}
\includegraphics[width=5.0in, height=2.32in]{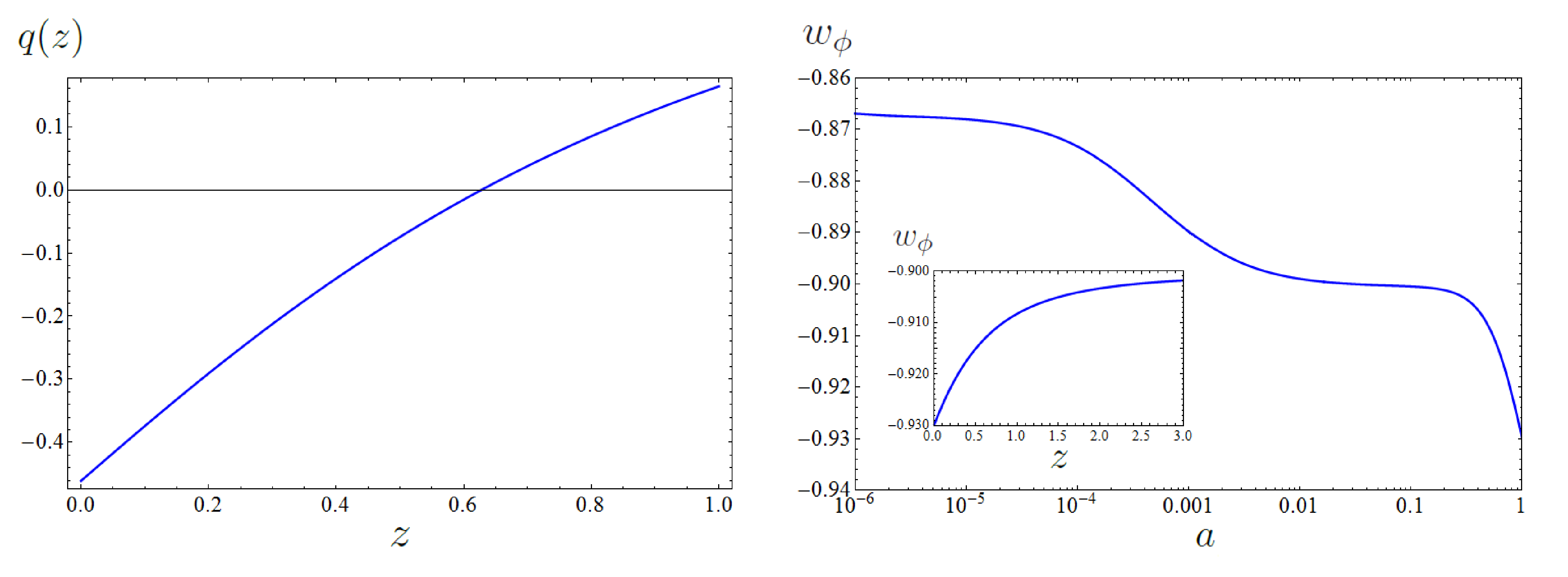}
\caption{\scriptsize
{\bf Left:} The deceleration parameter $q(z)$ for the recent Universe. The transition point where $q(z_t)=0$ is at $z_t=0.628$, for the best fit values of Table 2. {\bf Right:} The scalar field EoS parameter, $w_{\phi}(a)$, for the entire cosmic history after numerically solving the cosmological equations for the $\phi$CDM model with Peebles \& Ratra potential using the best-fit values of Table 2. The two plateaus from left to right correspond to the epochs of radiation and matter domination, respectively. The sloped stretch at the end, which is magnified in the inner plot in terms of  the redshift variable, corresponds to the recent epoch, in which the scalar field density (playing the role of DE) dominates. We find  $w_\phi(z=0)= -0.931\pm 0.017$.}
\end{center}
\end{figure}
%
\subsection{XCDM and CPL parametrizations}\label{sect:XCDMand CPL_scalar_fields_chapter}
The XCDM parametrization was first introduced in\,\cite{Turner:1998ex} as the simplest way to track a possible dynamics for the DE. Here one replaces the $\CC$-term with an unspecified dynamical entity $X$, whose energy density at present coincides with the current value of the vacuum energy density, i.e. $\rho_X^0=\rLo$. Its EoS reads $p_X=w_0\,\rho_X$, with $w_0=$const. The XCDM mimics the behaviour of a scalar field, whether quintessence ($w_0\gtrsim-1$) or phantom ($w_0\lesssim-1$), under the assumption that such field has an essentially constant EoS parameter around $-1$.  Since both matter and DE are self-conserved in the XCDM (i.e. they are not interacting), the energy densities as a function of the scale factor are given by $\rho_m(a)=\rho_m^0\,a^{-3}=\rho_{c 0}\Omega_m\,a^{-3}$ and $\rho_X(a)=\rho_X^0\,a^{-3(1+w_0)}=\rho_{c 0}(1-\Omega_m)\,a^{-3(1+w_0)}$.
Thus, the Hubble function in terms of the scale factor is given by
\begin{equation}\label{eq:HXCDM_scalar_fields_chapter}
H^2(a)=
H_0^2\left[\Omega_m\,a^{-3}+(1-\Omega_m)\,a^{-3(1+w_0)}\right]\,.
\end{equation}
A step further in the parametrization of the DE is the CPL prametrization\,\cite{Chevallier:2000qy,Linder:2002et,Linder:2004ng}, whose EoS for the DE is defined as follows:
\begin{equation}\label{eq:CPL_scalar_fields_chapter}
w=w_0+w_1\,(1-a)=w_0+w_1\,\frac{z}{1+z}\,,
\end{equation}
where $z$ is the cosmological redshift.
In contrast to the XCDM, the EoS of the CPL is not constant and is designed to have a well-defined asymptotic limit in the early Universe. The XCDM serves as a simple baseline to compare other models for the dynamical DE. The CPL further shapes the XCDM parametrization at the cost of an additional parameter ($w_1)$ that enables some cosmic evolution of the EoS. The Hubble function for the CPL in the MDE is readily found:
\begin{eqnarray}
\label{Hzzzquint_scalar_fields_chapter} H^2(z)&=&
H_0^2\,\left[\Omega_m\,(1+z)^3+(1-\Omega_m)
(1+z)^{3(1+w_0+w_1)}\,e^{-3\,w_1\,\frac{z}{1+z}}\right]
 \,.
\end{eqnarray}
%
\begin{figure}[t!]
\begin{center}
\label{Parella3_scalar_fields_chapter}
\includegraphics[width=5.0in, height=2.32in]{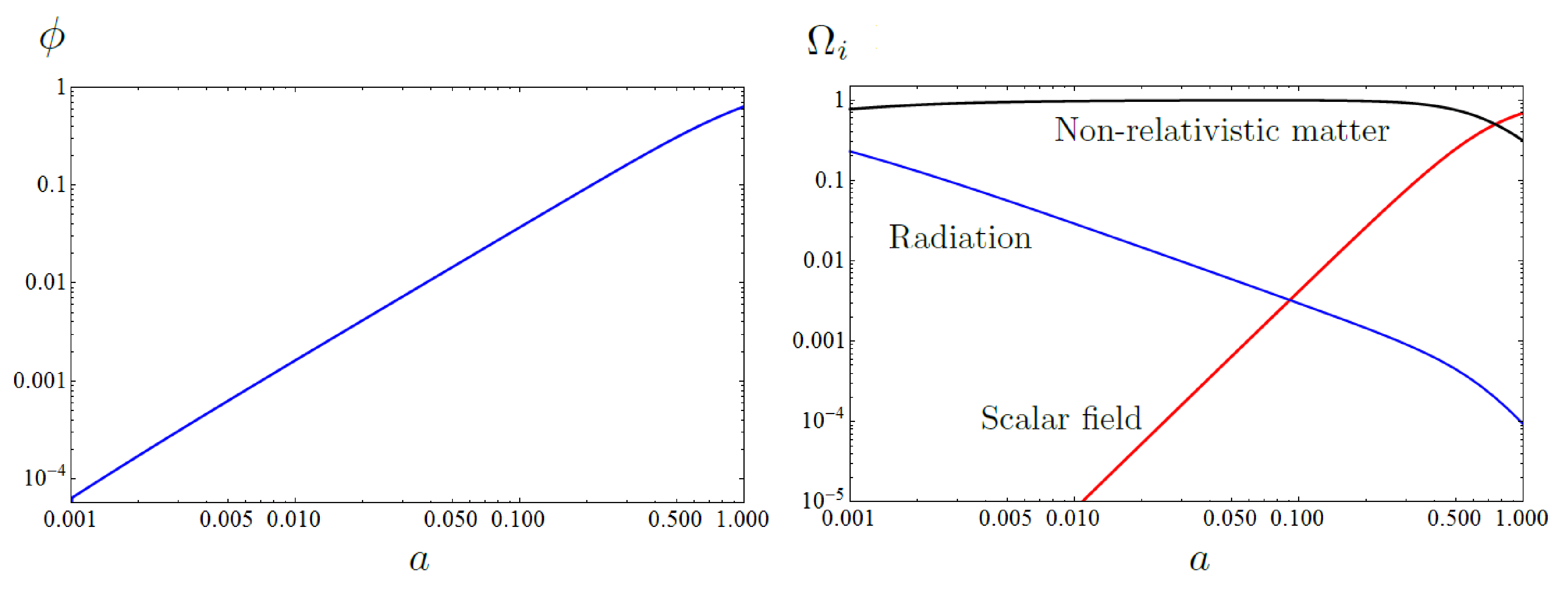}
\caption{\scriptsize Same as Fig.\,1, but for $\phi(a)$ and the density parameters $\Omega_i(a)$.  The crossing point between the scalar field density and the non-relativistic matter density lies very close to our time (viz. $a_c=0.751$, equivalently $z_c=0.332$), as it should. This is the point where the tail of the PR potential becomes visible and appears in the form of DE. The point $z_c$ lies nearer our time than the transition redshift from deceleration to acceleration, $z_t=0.628$ (cf. $q(z)$ in Fig. 1), similar to the $\CC$CDM.}
\end{center}
\end{figure}
It boils down to \eqref{eq:HXCDM_scalar_fields_chapter} for $w_1=0$, as expected. It is understood that for  the RDE  the term $\Omega_r (1+z)^4$ has to be added in the Hubble function. Such radiation term is already relevant for the analysis of the CMB data, and it is included in our analysis.
The fitting results for the XCDM and CPL parametrizations have been collected in the first two rows of Table 3.
Comparing with the $\phi$CDM model (cf. Table 2), we see that the XCDM parametrization also projects the effective quintessence option  at $4\sigma$ level, specifically $w_0=-0.916\pm0.021$.  The CPL parametrization, having one more parameter, does not reflect the same level of significance, but the corresponding AIC and BIC parameters (see below) remain relatively high as compared to the $\CC$CDM, therefore pointing also at clear signs of dynamical DE as the other models. Although the results obtained by the XCDM parametrization and the PR-model are fairly close (see Tables 2 and 3) and both EoS values  lie in the quintessence region, the fact that the EoS of the XCDM model is constant throughout the cosmic history makes it difficult to foresee if the XCDM can be used as a faithful representation of a given nontrivial $\phi$CDM model, such as the one we are considering here. The same happens for the extended CPL parametrization, even if in this case the EoS has some prescribed cosmic evolution. In actual fact, both the XCDM and CPL parametrizations are to a large extent arbitrary and incomplete representations of the dynamical DE.
\subsection{RVM: running vacuum}\label{sect:RVM_scalar_fields_chapter}
The last model whose fitting  results are reported in Table 3 is the running vacuum model (RVM). We  provide here the basic definition of it and some motivation -- see \,\cite{Sola:2013gha,Sola:2015rra,Sola:2016zeg} and references therein for details.  The RVM is a dynamical vacuum model, meaning that the corresponding EoS parameter is still $w=-1$ but the corresponding vacuum energy density is a``running'' one, i.e. it departs  (mildly) from the rigid assumption $\rL=$const. of the $\CC$CDM. Specifically, the form of $\rL$ reads as follows:
\begin{equation}\label{eq:RVMvacuumdadensity_scalar_fields_chapter}
\rho_\CC(H) = \frac{3}{8\pi{G}}\left(c_{0} + \nu{H^2}\right)\,.
\end{equation}
Here $c_0=H_0^2\left(1-\Omega_m-\nu\right)$ is fixed by the boundary condition $\rL(H_0)=\rLo=\rho_{c0}\,(1-\Omega_m)$. The dimensionless coefficient $\nu$ is expected to be very small, $|\nu|\ll1$, since the model must remain sufficiently close to the $\CC$CDM. The moderate dynamical evolution of $\rL(H)$ is possible at the expense of the slow decay rate of vacuum into dark matter (we assume that baryons and radiation are conserved\,\cite{Sola:2016ecz,Sola:2016zeg}).
\newline
In practice, the confrontation of the RVM with the data is performed by means of the following $5$-dimensional fitting vector:
\begin{equation}\label{eq:vfitting_scalar_fields_chapter}
\vec{p}_{\rm RVM}=(\Omega_m,\omega_b,n_s, h,
\nu)\,.
\end{equation}
The first four parameters are the standard ones as in the $\CC$CDM, while $\nu$ is the mentioned vacuum parameter for the RVM. Although it can be treated in a mere phenomenological fashion,
formally $\nu$ can be given a QFT meaning by linking it to the $\beta$-function of the running $\rL$\,\cite{Sola:2013gha,Sola:2015rra}. Theoretical estimates place its value in the ballpark of $\nu\sim 10^{-3}$ at most\,\cite{Sola:2007sv}, and this is precisely the order of magnitude pinned down for it in Table 3 from our overall fit to the data. The order of magnitude coincidence is reassuring.
Different realizations of the RVM are possible \cite{Sola:2015wwa,Sola:2016jky,Sola:2016ecz,Sola:2016zeg}, but here we limit ourselves to the simplest version. The corresponding Hubble function in the MDE reads:
\begin{equation}\label{eq:H2RVM_scalar_fields_chapter}
H^2(z)=H_0^2\,\left[1+\frac{\Omega_m}{1-\nu}\left((1+z)^{3(1-\nu)}-1\right)\right]\,.
\end{equation}
It depends on the basic fitting parameters $(\Omega_m, h,\nu)$, which are the counterpart of $(\omega_m,\alpha,\bar{\kappa})$ for the $\phi$CDM. The remaining two parameters are common and hence both for the RVM and the $\phi$CDM the total number of fitting parameter is five, see \eqref{eq:vfittingPhiCDM_scalar_fields_chapter} and \eqref{eq:vfitting_scalar_fields_chapter}.
Note that for $\nu=0$ we recover the $\CC$CDM case, as it should be expected.
%
%
%
%
\begin{figure}[t!]
\begin{center}
\label{Parella1_scalar_fields_chapter}
\includegraphics[width=3.5in, height=2.4in]{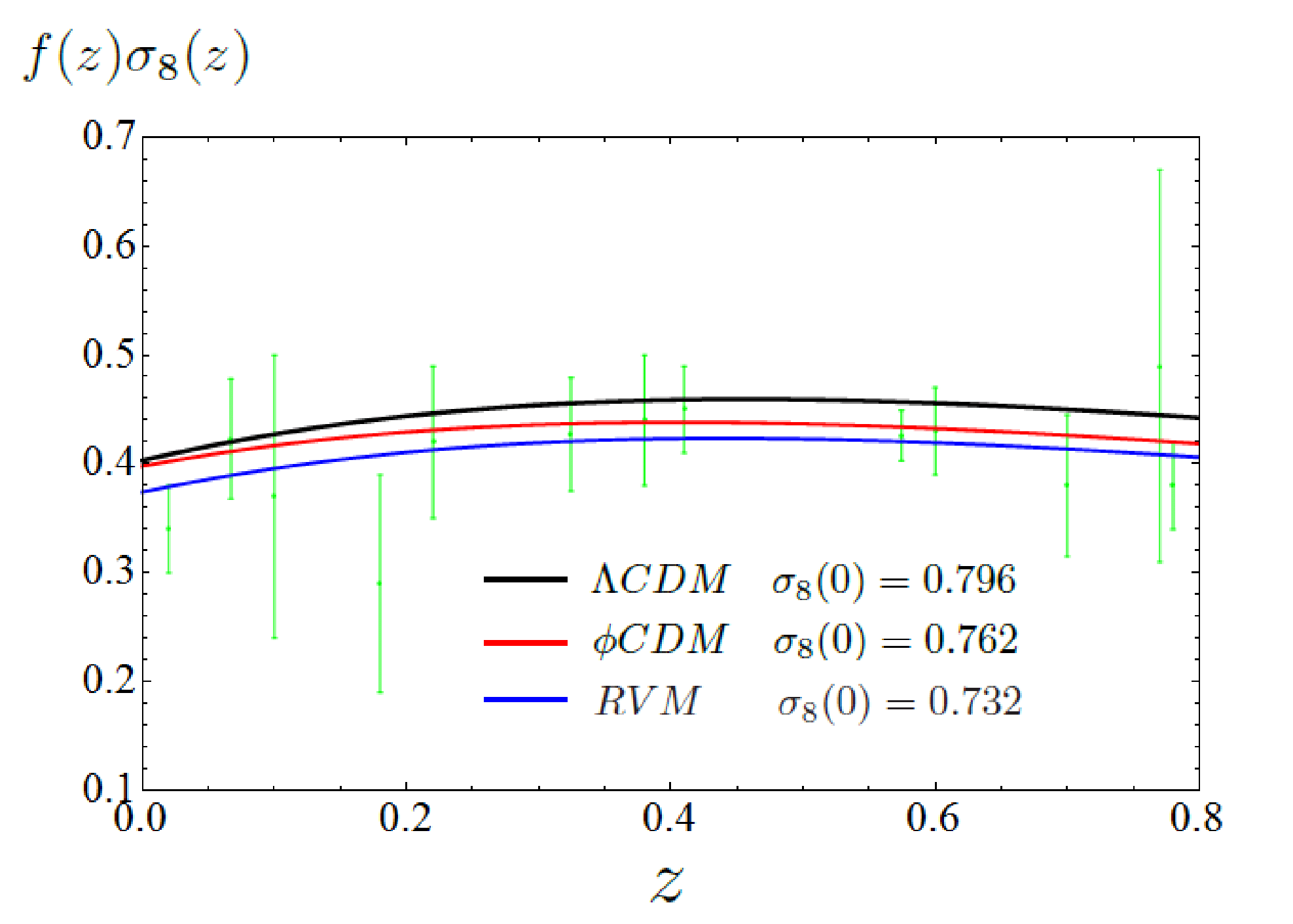}
\caption{\scriptsize The LSS data on the weighted linear growth rate, $f(z)\sigma_8(z)$, and the predicted curves by the various models, using the best-fit values in Tables 1-3. The XCDM and CPL lines are not shown since they are almost on top of the $\phi$CDM one. The values of $\sigma_8(0)$ that we obtain for the different models are also indicated.
}
\end{center}
\end{figure}
%
%
%
\subsection{Structure formation}\label{sect:Structure_Formation_scalar_fields_chapter}
A few observations on the analysis of structure formation are in order, as it plays a significantly role in the fitting results. On scales well below the horizon the scalar field perturbations are relativistic and hence can be ignored\,\cite{Peebles:1987ek}. As a result in the presence of non-interacting scalar fields the usual matter perturbation equation remains valid\,\cite{Amendola:2015ksp}. Thus, for the $\phi$CDM, XCDM and CPL models we compute the perturbations through the standard equation\,\cite{peebles:1993}
\begin{equation}\label{diffeqLCDM_scalar_fields_chapter}
\ddot{\delta}_m+2H\,\dot{\delta}_m-4\pi
G\rmr\,\delta_m=0\,,
\end{equation}
with, however, the Hubble function corresponding to each one of these models -- see the formulae in the previous sections.
\newline
For the RVM the situation is nevertheless different. In the presence of dynamical vacuum, the perturbation equation not only involves the modified Hubble function \eqref{eq:H2RVM_scalar_fields_chapter} but the equation itself becomes modified. The generalized perturbation equation reads\,\cite{Gomez-Valent:2014rxa,Gomez-Valent:2014fda,Gomez-Valent:2015pia}:
\begin{equation}\label{diffeqD_scalar_fields_chapter}
\ddot{\delta}_m+\left(2H+\Psi\right)\,\dot{\delta}_m-\left(4\pi
G\rmr-2H\Psi-\dot{\Psi}\right)\,\delta_m=0\,,
\end{equation}
where $\Psi\equiv -\dot{\rho}_{\Lambda}/{\rmr}$. As expected, for $\rL=$const. we have $\Psi=0$ and Eq.\eqref{diffeqD_scalar_fields_chapter} reduces to the standard one \eqref{diffeqD_scalar_fields_chapter}.
To solve the above perturbation equations we have to fix the initial conditions for $\delta_m$ and $\dot{\delta}_m$ for each model at high redshift, say at $z_i\sim100$ ($a_i\sim10^{-2}$), when non-relativistic matter dominates over the vacuum -- confer Ref.\,\cite{Gomez-Valent:2014rxa,Gomez-Valent:2014fda,Gomez-Valent:2015pia}.
\newline
Let us also note that, in all cases, we can neglect the DE perturbations at subhorizon scales. We have already mentioned above that this is justified for the $\phi$CDM. For the RVM it can be shown to be also the case, see \,\cite{Gomez-Valent:2014rxa,Gomez-Valent:2014fda,Gomez-Valent:2015pia}. The situation with the XCDM and CPL is not different, and once more the DE perturbations are negligible at scales below the horizon. A detailed study of this issue can be found e.g. in Ref.\,\cite{Grande:2011xf,Grande:2006nn}, in which the so-called $\CC$XCDM model is considered in detail at the perturbations level. In the absence of the (running) component $\CC$ of the DE, the $\CC$XCDM model reduces exactly to the XCDM as a particular case. One can see in that quantitative study that at subhorizon scales the DE perturbations become negligible no matter what is the  assumed value for the sound velocity of the DE perturbations (whether adiabatic or non-adiabatic).
\newline
The analysis of the linear LSS regime is conveniently implemented with the help of the weighted linear growth $f(z)\sigma_8(z)$, where $f(z)=d\ln{\delta_m}/d\ln{a}$ is the growth factor and $\sigma_8(z)$ is the rms mass fluctuation amplitude on scales of $R_8=8\,h^{-1}$ Mpc at redshift $z$. It is computed as follows:
\begin{equation}
\begin{small}\sigma_{\rm 8}(z)=\sigma_{8, \CC}
\frac{\delta_m(z)}{\delta^{\CC}_{m}(0)}
\sqrt{\frac{\int_{0}^{\infty} k^{n_s+2} T^{2}(\vec{p},k)
W^2(kR_{8}) dk} {\int_{0}^{\infty} k^{n_{s,\CC}+2} T^{2}(\vec{p}_\Lambda,k) W^2(kR_{8,\Lambda}) dk}}\,,\label{s88general_scalar_fields_chapter}
\end{small}\end{equation}
where $W$ is a top-hat smoothing function and $T(\vec{p},k)$ the transfer function (see e.g. \cite{Gomez-Valent:2014rxa,Gomez-Valent:2014fda,Gomez-Valent:2015pia} for details). Here $\vec{p}$ stands for the corresponding fitting vector for the various models, as indicated in the previous sections. In addition, we define a fiducial model, which we use in order to fix the normalization of the power spectrum. For that model we take the $\CC$CDM at fixed parameter values from the Planck 2015 TT,TE,EE+lowP+lensing analysis\,\cite{Ade:2015xua}. Such fiducial values are collected in the vector
$\vec{p}_\CC=(\Omega_{m,\CC},\omega_{b,\CC},n_{s,\CC},h_{\CC})$.
In Fig. 3 we display  $f(z)\sigma_8(z)$ for the various models using the fitted values of Tables 1-3 following this procedure.
%
\begin{figure}[t!]
\begin{center}
\label{Parella2b_scalar_fields_chapter}
\includegraphics[width=4.0in, height=2.2in]{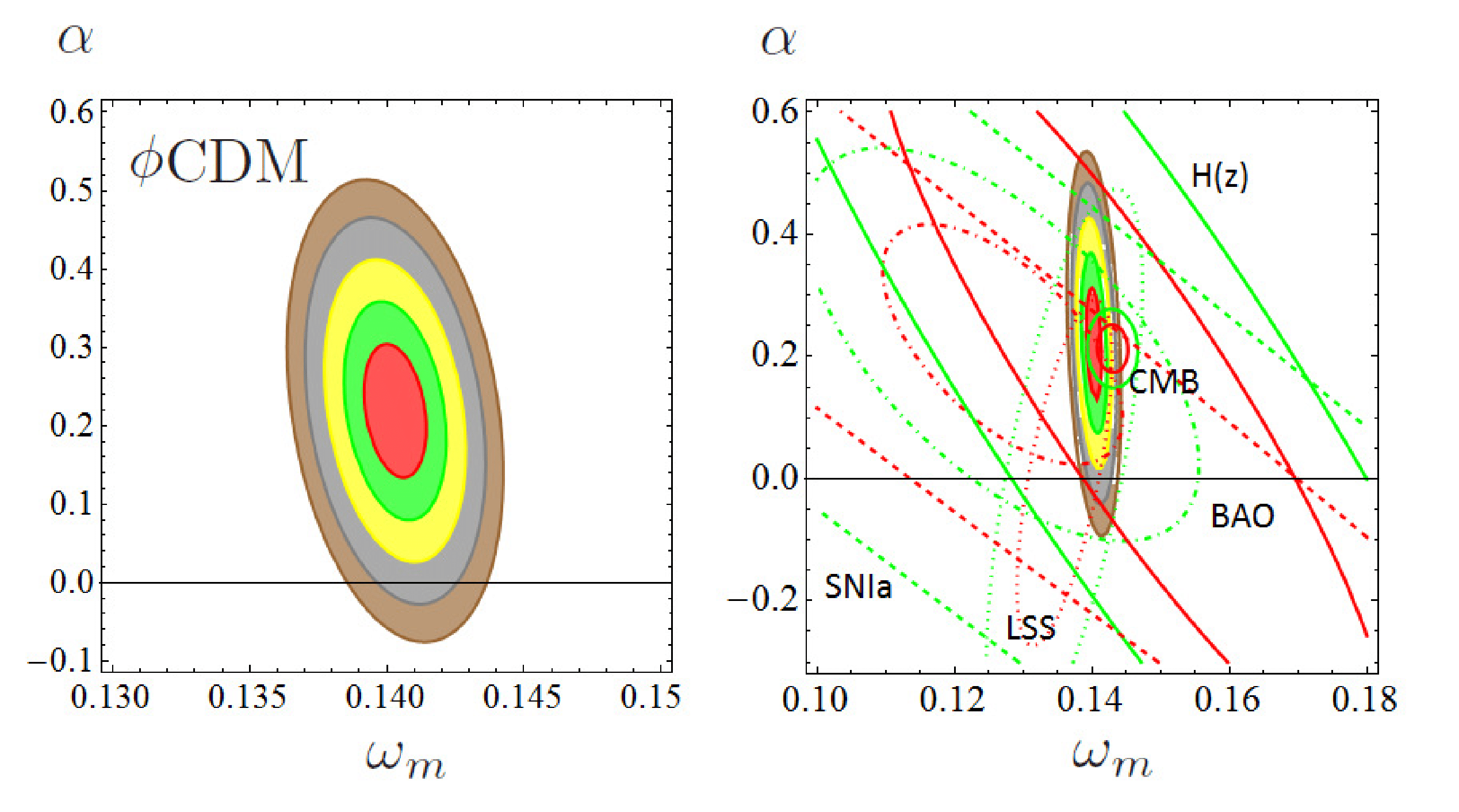}
\caption{\scriptsize {\bf Left}: Likelihood contours for the $\phi$CDM model in the ($\omega_m$,$\alpha$)-plane after marginalizing over the remaining parameters (cf. Table 2). The various contours correspond to 1$\sigma$, 2$\sigma$, 3$\sigma$, 4$\sigma$ and 5$\sigma$  c.l. The line $\alpha=0$ corresponds to the concordance $\CC$CDM model. The tracker consistency region $\alpha>0$ (see the text) is clearly preferred, and we see that it definitely points to dynamical DE at $\sim4\sigma$ confidence level. {\bf Right}: Reconstruction of the aforementioned contour lines from the partial contour plots of the different SNIa+BAO+$H(z)$+LSS+CMB data sources using Fisher's approach\,\cite{Amendola:2015ksp}. The $1\sigma$ and $2\sigma$ contours are shown in all cases, but for the reconstructed final contour lines we include the $3\sigma$, $4\sigma$ and $5\sigma$ regions as well. For the reconstruction plot we display a larger $\omega_m$-range to better appraise the impact of the various data sources.}
\end{center}
\end{figure}
\begin{figure}[t!]
\begin{center}
\label{Parella3b_scalar_fields_chapter}
\includegraphics[width=3.5in, height=2.11in]{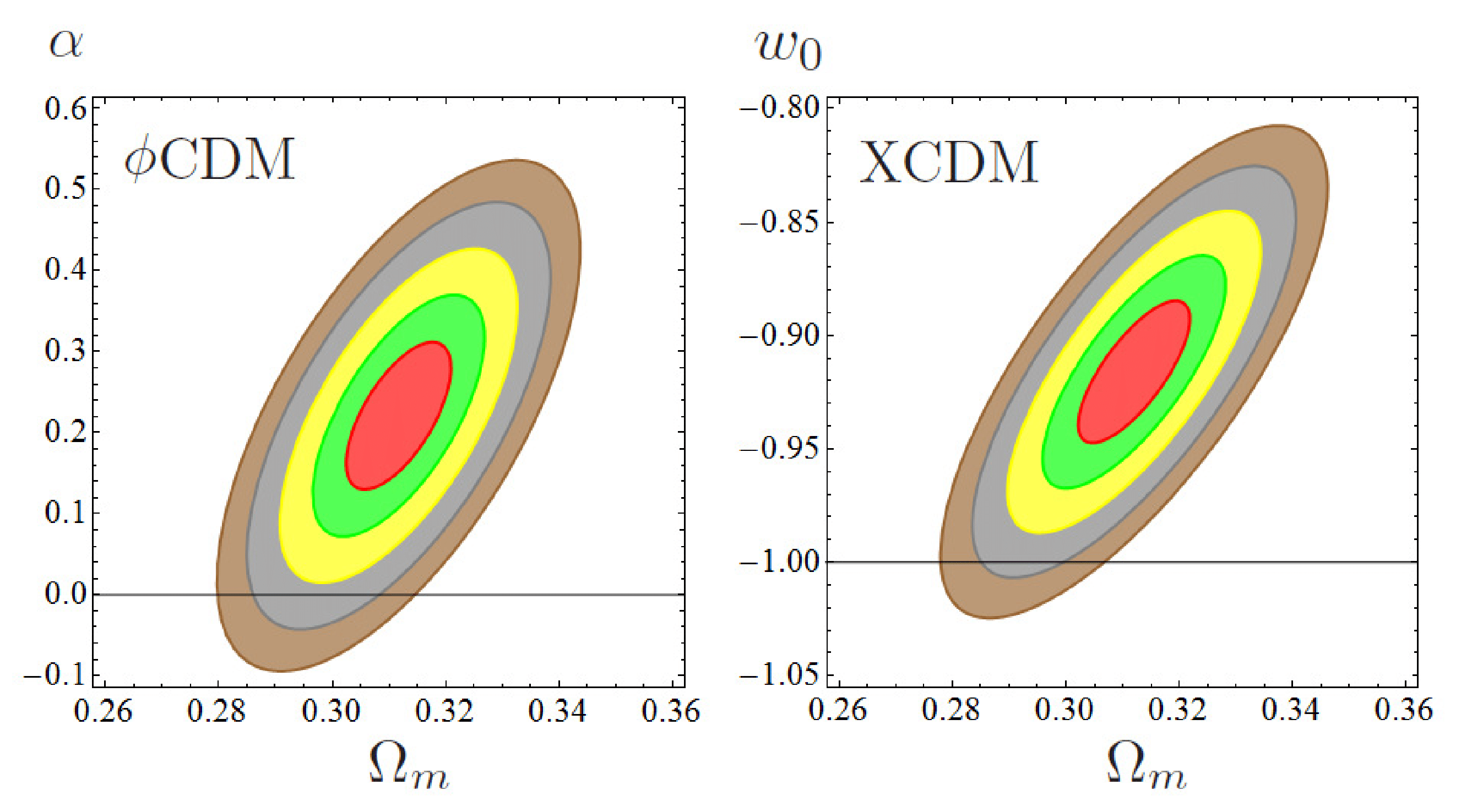}
\caption{\scriptsize Likelihood contours for the $\phi$CDM model (left) and the XCDM parametrization (right) in the relevant planes after marginalizing over the remaining parameters in each case (cf. Tables 2 and 3). The various contours correspond to 1$\sigma$, 2$\sigma$, 3$\sigma$, 4$\sigma$ and 5$\sigma$ c.l. The central values in both cases are $\sim4\sigma$ away from the $\CC$CDM, i.e. $\alpha=0$ and $w_0=-1$, respectively.}
\end{center}
\end{figure}
%
%
\subsection{Discussion and conclusions}\label{Discussions_and_Conclusions_scalar_fields_chapter}
The statistical analysis of the various models considered in this study is performed in terms of a joint likelihood function, which is the product of the likelihoods for each data source
and including the corresponding covariance matrices, following the standard procedure\,\cite{Amendola:2015ksp}. The contour plots for the $\phi$CDM and XCDM models are shown in Figures 4 and 5, where the the dynamical character of the DE is clearly demonstrated at $\sim 4\sigma$ c.l. More specifically, in the left plot of  Fig. 4 we display the final contour plots for $\phi$CDM in the plane $(\omega_m,\alpha)$ -- defined by two of the original parameters of our calculation, cf. Eq.\eqref{eq:vfittingPhiCDM_scalar_fields_chapter} --  together with the isolated contours of the different data sources (plot on the right). It can be seen that the joint triad of observable BAO+LSS+CMB conspire to significantly reduce the final allowed region of the ($\omega_m,\alpha$)-plane, while the constraints imposed by SNIa and $H(z)$ are much weaker. Finally, for the sake of convenience, in Fig. 5 we put forward the final $\phi$CDM and the XCDM contours in the more conventional $(\Omega_m,\alpha)$-plane. As for the RVM, see the contours in \cite{Sola:2016jky,Sola:2016ecz} and \cite{Sola:2016zeg}, where a dynamical DE effect $\gtrsim 4\sigma$  is recorded.
\newline
As noted previously, the three models $\phi$CDM, XCDM and RVM have the same number of parameters, namely 5, one more than the $\CC$CDM. The CPL, however, has 6 parameters. Cosmological models having a larger number of parameters have more freedom to accommodate observations. Thus, for a fairer comparison of the various nonstandard models with the concordance $\CC$CDM we have to invoke a suitable statistical procedure that penalizes the presence of extra parameters. Efficient criteria of this kind are available in the literature and they have been systematically used in different contexts to help making a selection of the best candidates among competing models describing the same data.  For a long time it has been known that the Akaike information criterion (AIC) and the Bayesian information criterion (BIC) are extremely valuable tools for a fair statistical analysis of this kind. These criteria are defined as follows\,\cite{Akaike,Schwarz1978,Burnham2002}:
\begin{equation}\label{eq:AICandBIC_scalar_fields_chapter}
{\rm AIC}=\chi^2_{\rm min}+\frac{2nN}{N-n-1}\,,\ \ \ \ \
{\rm BIC}=\chi^2_{\rm min}+n\,\ln N\,,
\end{equation}
where $n$ is the number of independent fitting parameters and $N$ the number of data points.
The larger are the differences $\Delta$AIC ($\Delta$BIC) with respect to the model that carries smaller value of AIC (BIC) the higher is the evidence against the model with larger value of  AIC (BIC) -- the $\CC$CDM in all the cases considered in Tables 2-3.
The rule applied to our case is the following\,\cite{Akaike,Schwarz1978,Burnham2002}: for $\Delta$AIC and $\Delta$BIC in the range $6-10$ we can speak of ``strong evidence'' against the $\CC$CDM, and hence in favor of the given nonstandard model. Above 10, one speaks of ``very strong evidence''. Notice that the Bayes factor is $e^{\Delta {\rm BIC}/2}$, and hence near 150 in such case.
\newline
A glance at Tables 2 and 3 tells us that for the models $\phi$CDM, XCDM and RVM, the values of $\Delta$AIC and $\Delta$BIC are both above 10. The CPL parametrization has only one of the two increments above 10, but the lowest one is above 6, therefore it is still fairly (but not so strongly) favored as the others. We conclude from the AIC and BIC criteria that the models $\phi$CDM, XCDM and RVM  are definitely selected over the $\CC$CDM as to their ability to optimally fit the large set of cosmological SNIa+BAO+$H(z)$+LSS+CMB data used in our analysis.  Although the most conspicuous model of those analyzed here appears to be the RVM {(cf. Tables 2 and 3)}, the scalar field model $\phi$CDM with Peebles \& Ratra potential also receives a strong favorable verdict from the AIC and BIC criteria. Furthermore, the fact that the generic XCDM and CPL parametrizations are also capable of detecting significant signs of evolving DE suggests that such dynamical signature is sitting in the data and is not privative of a particular model, although the level of sensitivity does indeed vary from one model to the other.
\newline
\newline
To summarize, the current cosmological data disfavors the hypothesis $\CC=$const. in a rather significant way. The presence of DE dynamics is confirmed by all four parametrizations considered here and with a strength that ranges between strong and very strong evidence, according to the Akaike and Bayesian information criteria. Furthermore, three of these parametrizations are able to attest such evidence at $\sim4\sigma$ c.l., and two of them ($\phi$CDM and RVM) are actually more than parametrizations since they are associated to specific theoretical frameworks.  The four approaches resonate in harmony with the conclusion that the DE is decreasing with the expansion, and therefore that it behaves effectively as quintessence.

\newpage
\thispagestyle{empty}
\mbox{}
\newpage


\section{Possible signals of vacuum dynamics in the Universe}\label{Possible_signals_chapter}
In this chapter we pursue with the study of the generic class of time-evolving vacuum models which can provide a better phenomenological account of the overall cosmological observations as compared to the $\Lambda$CDM. Among these models, the running vacuum model (RVM) appears to be the most favored one at a confidence level of $\sim 3\sigma$. Here we do not restrict the analysis of the dynamical vacuum models (DVM) to those that can only be motivated from the theoretical point of view since we also consider two pure phenomenological models. In this case, we limit the interactions, of the different species in game, to the dark sector, unlike what we did in previous chapters.
\newline
We also show that it is possible to extract fair signals of dynamical dark energy (DDE) by gauging the performance of the generic XCDM and CPL parameterizations. 
\newline
In all cases we confirm that the combined triad of modern observations on baryonic acoustic oscillations, large scale structure formation and the cosmic microwave background, provide the bulk of the signal sustaining a  possible dark energy dynamics. In the absence of any of these three crucial data sources, the DDE signal can not be perceived at a significant confidence level. Its possible existence could be a cure for some of the tensions existing in the $\Lambda$CDM when confronted to observations.
\newline
The guidelines of the chapter are as follows. In Sec.\,\ref{sect:DVMs_Possible_signals_chapter} we describe the dynamical vacuum models (DVMs). In Sec.\,\ref{sect:Fit_Possible_signals_chapter} we report on the set of cosmological data used, on distant type Ia supernovae (SNIa), baryonic acoustic oscillations (BAOs), the Hubble parameter values at different redshifts, the LSS data, and the cosmic microwave background (CMB) from Planck. In Sec.\,\ref{sect:perturbationsDVM_Possible_signals_chapter} we discuss aspects of structure formation with dynamical vacuum. The numerical analysis of the DVMs and a comparison with the standard XCDM and CPL parametrizations is the object of Sec.\,\ref{sect:numerical results_Possible_signals_chapter}. An ample discussion of the results along with a reanalysis under different conditions is developed in Sec.\,\ref{sect:discussion_Possible_signals_chapter}. Finally, in Sec.\,\ref{sect:conclusions_Possible_signals_chapter} we present our conclusions.
\subsection{Dynamical vacuum models}\label{sect:DVMs_Possible_signals_chapter}
{The gravitational field equations  are $G_{\mu\nu}=8\pi G\ \tilde{T}_{\mu\nu}$,
where $G_{\mu\nu}=R_{\mu \nu }-\frac{1}{2}g_{\mu \nu }R$ is the Einstein tensor and
$\tilde{T}_{\mu\nu}\equiv T_{\mu\nu}+g_{\mu\nu}\,\rL $ is the full energy-momentum tensor involving the effect of both matter and the vacuum energy density, with $\rL=\CC/(8\pi G)$.
The structure of $\tilde{T}_{\mu\nu}$ shows that the vacuum is dealt with as a perfect fluid carrying an equation of state (EoS)  $p_{\CC}=-\rho_{\CC}$. When the matter can also be treated as an ideal fluid and is distributed homogeneously and isotropically, as postulated by the Cosmological Principle, we can write
$\tilde{T}_{\mu\nu}= (\rho_{\Lambda}-p_{m})\,g_{\mu\nu}+(\rho_{m}+p_{m})U_{\mu}U_{\nu}$,
where $U_{\mu}$ is the bulk $4$-velocity of the cosmic fluid, $\rho_m$ is the proper energy density of matter and $p_m$ its isotropic pressure.}
We assume the standard cosmological framework grounded on the FLRW metric with flat three-dimensional slices: $ds^2=dt^2-a^2(t)\,d{\bf x}^2$, where $t$ is the cosmic time and $a(t)$ the scale factor. However, we admit that matter can be in interaction with vacuum, which is tantamount to saying that $\rL=\rL(\zeta)$ is a function of some cosmic variable evolving with time, $\zeta=\zeta(t)$. While this, of course, implies that $\dot{\rho}_{\CC}\equiv d\rL/dt\neq 0$ we assume that $\dot{G}=0$ in our study -- see \cite{Sola:2015wwa,Sola:2016jky} for studies including the option $\dot{G}\neq 0$. Such  vacuum dynamics is compatible with the Bianchi identity (see below) provided there is some energy exchange between vacuum and matter. It means that matter cannot be strictly conserved in these circumstances. The standard Friedmann and acceleration equations for the present Universe remain formally identical to the standard $\CC$CDM case:
\begin{eqnarray}
&&3H^2=8\pi\,G\sum_N\rho_N=8\pi\,G\,(\rho_m+\rho_r+\rho_\Lambda(\zeta))\label{eq:FriedmannEq_Possible_signals_chapter}\\
&&3H^2+2\dot{H}=-8\pi\,G\sum_Np_N=-8\pi\,G\,(p_r-\rho_\Lambda(\zeta))\label{eq:PressureEq_Possible_signals_chapter}\,.
\end{eqnarray}
Here $H=\dot{a}/a$ is the usual Hubble function,  $\rho_m=\rho_b+\rho_{dm}$ involves the pressureless contributions from baryons and cold DM, and $\rho_r$ is  the radiation density (with the usual EoS $p_r=\rho_r/3$).  We emphasize once more that in the above equations we stick to the EoS $p_{\CC}=-\rho_{\CC}$, although the vacuum is dynamical, $\rL(t)=\rL(\zeta(t))$, and its evolution is tied to the cosmic expansion. {The sums above run over all the components $N=dm,b,r,\CC$}.  In all of the dynamical vacuum models (DVMs) being considered here, the cosmic variable $\zeta$ is either the scale factor or can be expressed analytically in terms of it, $\zeta=\zeta(a)$, or equivalently in terms of the cosmological redshift, $z=a^{-1}-1$, in which we adopt the normalization $a=1$ at present.
{From the basic pair of equations \eqref{eq:FriedmannEq_Possible_signals_chapter}-\eqref{eq:PressureEq_Possible_signals_chapter}, a first integral of the system follows:

\begin{equation}
\begin{split}
\sum_{N} \dot{\rho}_N &+3\,H(\rho_N+p_N)= \label{BianchiGeneral_Possible_signals_chapter}\\
& \dot{\rho}_\CC +\dot{\rho}_{dm} + 3H\rho_{dm}+\dot{\rho}_b + 3H\rho_b+\dot{\rho}_r + 4H\rho_r =0\,.
\end{split}
\end{equation}
Such a first integral ensues also from the divergenceless property of the full energy-momentum tensor $\tilde{T}_{\mu\nu}$ in the FLRW metric, i.e. $\nabla^{\mu}\tilde{T}_{\mu\nu}=0$. The last property is a consequence of the Bianchi identity satisfied by the Einstein tensor, $\nabla^{\mu} G_{\mu\nu}=0$, and the assumed constancy of the Newtonian coupling $G$. It reflects the local conservation law of the compound system formed by matter and vacuum, and the consequent nonconservation of each of these components when taken separately.}
\newline
The concordance model assumes that matter and radiation are self-conserved after equality. It also assumes that baryons and CDM are separately conserved. Hence their respective energy densities satisfy $\dot{\rho}_b + 3H\rho_b=0$, $\dot{\rho}_r + 4H\rho_r=0$ and $\dot{\rho}_{dm} + 3H\rho_{dm}=0$.  In the presence of vacuum dynamics it is obvious that at least one of these equations cannot hold. Following our definite purpose to remain as close as possible to the $\CC$CDM, we shall assume that the first two of the mentioned conservation equations still hold good but that the last does not, meaning that the vacuum exchanges energy only with DM. The dilution laws for baryons and radiation as a function of the scale factor therefore take on the conventional $\CC$CDM forms:
\begin{equation}\label{eq:BaryonsRadiation_Possible_signals_chapter}
\rho_b(a) = \rho_{b0}\,a^{-3}, \ \ \ \ \ \ \ \rho_r(a)=\rho_{r0}\,a^{-4}\,,
\end{equation}
where $\rho_{b0}$ and $\rho_{r0}$ are the corresponding current values. In contrast, the density of DM is tied to the dynamics of the vacuum. {Taking into account the conserved components and  introducing the vacuum-dark matter interaction source, $Q$, we can write the interactive part of \eqref{BianchiGeneral_Possible_signals_chapter} into two coupled equations:}
\begin{equation}\label{eq:Qequations_Possible_signals_chapter}
\dot{\rho}_{dm}+3H\rho_{dm}=Q\,,\ \ \ \ \ \ \ \dot\rho_{\CC}=-{Q}\,.
\end{equation}
The solution of these equations will depend on the particular form assumed for $Q$, which determines the leakage rate of vacuum energy into dark matter or vice versa. Such a leakage must certainly be much smaller than the standard dilution rate of non-relativistic matter associated to the cosmic expansion (i.e. much smaller than $\sim a^{-3}$), as otherwise these anomalous effects would be too sharp at the present time. Therefore, we must have   $0<|Q|\ll\dot{\rho}_m$. The different DVMs will be characterized by different functions $Q_i$  ($i=1,2,..$).
\newline
Two possible phenomenological ansatzs  considered in the literature  are \cite{Salvatelli:2014zta,Murgia:2016ccp,Li:2015vla,Zhao:2016ecj}
\begin{eqnarray}
{\ \rm Model\ \ }Q_{dm}: \phantom{XX}Q_{dm}&=&3\nu_{dm}H\rho_{dm}\label{eq:PhenModelQdm_Possible_signals_chapter}\\
{\rm Model\ \ }Q_{\CC}:\phantom{XXx}Q_{\CC}&=&3\nu_{\CC}H\rho_{\CC}\,.\label{eq:PhenModelQL_Possible_signals_chapter}
\end{eqnarray}
The dimensionless parameters $\nu_{i}=(\nu_{dm},\nu_\CC)$ for each model  ($Q_{dm}$, $Q_\CC$) determine the strength of the dark-sector interaction in the sources $Q_i$ and enable the evolution of the vacuum energy density.  For $\nu_{i}>0$ the vacuum decays into DM (which is thermodynamically favorable \cite{Salim:1992mx,Lima:1995kd} whereas for $\nu_{i}<0$ is the other way around. This is also a relevant argument to judge the viability of these models, as only the first situation is compatible with the second law of thermodynamics.
There are many more choices for $Q$, see e.g. \cite{Bolotin:2013jpa,Costa:2016tpb}, but it will suffice to focus on these models and the RVM one defined in the next section to effectively assess the possible impact of the DVMs in the light of the modern observational data.
\subsubsection{The running vacuum model (RVM)}\label{sect:RVM_Possible_signals_chapter}
The last DVM under study is the so-called running vacuum model (RVM), which can be motivated in the context of QFT in curved space-time (cf. \cite{Sola:2013gha,Gomez-Valent:2014fda}, and references therein).
The model has some virtues and can be extended to afford an effective description of the cosmic evolution starting from inflation up to our days \cite{Sola:2013gha,Gomez-Valent:2014fda,Perico:2013mna,Lima:2012mu,Lima:2014hia,Lima:2015mca,Sola:2015csa}. For instance, in \cite{Sola:2015csa} it is suggested that the RVM could positively impinge on solving some of the fundamental cosmological problems, including the entropy problem. Intriguingly, the inherent tiny leakage of vacuum energy into matter within the RVM could also furnish an explanation for the possible slow time variation of the fundamental constants, an issue that has been examined in detail in \cite{Fritzsch:2012qc,Fritzsch:2015lua,Fritzsch:2016ewd}. See also the old work \cite{Terazawa:1981ga}. We shall not discuss here the implications for the early Universe, but only for the part of the cosmic history that is accessible to our measurements and can therefore be tested phenomenologically with the current data.
\newline
As advertised, for the specific RVM case the cosmic variable $\zeta$ in the field equations \eqref{eq:FriedmannEq_Possible_signals_chapter}-\eqref{eq:PressureEq_Possible_signals_chapter} can be identified with the Hubble rate $H$. The form of the RVM for the post-inflationary epoch and hence relevant for the current Universe reads as follows:
\begin{equation}\label{eq:RVMvacuumdadensity_Possible_signals_chapter}
\rho_\CC(H) = \frac{3}{8\pi{G}}\left(c_{0} + \nu{H^2}\right)\,.
\end{equation}
{Such structure  can be linked to a renormalization group (RG) equation for the vacuum energy density, in which the running scale $\mu$ of the RG is associated with the characteristic energy scale of the FLRW metric, i.e. $\mu=H$.
The additive constant $c_0=H_0^2\left(\Omega_\CC-\nu\right)$ appears because one integrates the RG equation satisfied by $\rL(H)$. It is fixed by the boundary condition $\rL(H_0)=\rLo$, where $\rLo$ and $H_0$ are the current values of these quantities; similarly $\OL=\rLo/\rco$  and  $\rco=3H_0^2/(8\pi G)$ are the values of the vacuum density parameter and the critical density today. The dimensionless coefficient $\nu$ encodes the dynamics of the vacuum at low energy and can be related with the $\beta$-function of the running of $\rL$.} Thus, we naturally expect $|\nu|\ll1$. An estimate of $\nu$ at one loop in QFT indicates that is of order $10^{-3}$ at most \cite{Sola:2007sv}, but here we will treat it as a free parameter. This means  we shall deal with the RVM on pure phenomenological grounds, hence fitting actually $\nu$ to the observational data (cf. Sec.\,\ref{sect:Fit_Possible_signals_chapter}).
\newline
In the RVM case, the source function $Q$ in \eqref{eq:Qequations_Possible_signals_chapter} is not just put by hand (as in the case of the DVMs introduced before). It is a calculable expression from \eqref{eq:RVMvacuumdadensity_Possible_signals_chapter}, using Friedmann's equation \eqref{eq:FriedmannEq_Possible_signals_chapter} and the local conservation laws \eqref{eq:BaryonsRadiation_Possible_signals_chapter}-\eqref{eq:Qequations_Possible_signals_chapter}. We find:
\begin{equation}\label{eq:QRVM_Possible_signals_chapter}
{\rm RVM:}\qquad Q=-\dot{\rho}_{\Lambda}=\nu\,H(3\rho_{m}+4\rho_r)\,,
\end{equation}
where we recall that $\rho_{m}=\rho_{dm}+\rho_{b}$, and that $\rho_b$ and $\rho_r$ are known functions of the scale factor -- see Eq.\eqref{eq:BaryonsRadiation_Possible_signals_chapter}. The remaining densities, $\rho_{dm}$ and $\rL$, must be determined upon further solving the model explicitly, see subsection\,\ref{sect:solvingDVM_Possible_signals_chapter}. If baryons and radiation would also possess a small interaction with vacuum and/or $G$ would evolve with time, the cosmological solutions would be different \cite{Gomez-Valent:2014rxa,Gomez-Valent:2014fda,Basilakos:2015wha,Sola:2015wwa,Sola:2016jky}. We can see from \eqref{eq:QRVM_Possible_signals_chapter} that the parameter $\nu$ plays a similar role as  $(\nu_{dm},\nu_\CC)$ for the more phenomenological models \eqref{eq:PhenModelQdm_Possible_signals_chapter} and \eqref{eq:PhenModelQL_Possible_signals_chapter}. The three of them will be collectively denoted $\nu_i$.
\subsubsection{Solving explicitly the dynamical vacuum models}\label{sect:solvingDVM_Possible_signals_chapter}
The matter and vacuum energy densities of the DVMs can be computed straightforwardly upon solving the coupled system of differential equations \eqref{eq:Qequations_Possible_signals_chapter}, given the previous explicit forms for the interacting source in each case and keeping in mind that, in the current framework, the baryon ($\rho_b$) and radiation ($\rho_r$) parts are separately conserved. After some calculations the equations for the DM energy densities $\rho_{dm}$ for each model (RVM, $Q_{dm}$, $Q_{\CC}$) can be solved in terms of the scale factor. Below we quote the final results for each case:
\begin{align}
&{\rm \textbf{RVM}}:\quad \rho_{dm}(a) = \rho_{dm0}\,a^{-3(1-\nu)}+ \rho_{b0}\left(a^{-3(1-\nu)} - a^{-3}\right) +
\frac{4\nu}{1 + 3\nu}\,{\rho_{r0}}\,\left(a^{-3(1-\nu)} - a^{-4}\right) \label{eq:rhoRVM_Possible_signals_chapter}\\
&{\rm \ \mathbf{Q_{dm}}}:\quad
\rho_{dm}(a) = \rho_{dm0}\,a^{-3(1-\nu_{dm})}\label{eq:rhoQdm_Possible_signals_chapter}\\
&{\rm \ \mathbf{Q_{\CC}}}:\quad
\rho_{dm}(a) =\rho_{dm0}\,a^{-3} + \frac{\nu_\CC}{1-\nu_\CC}\rLo\left(a^{-3\nu_\Lambda}-a^{-3}\right)\label{eq:rhoQL_Possible_signals_chapter}
\end{align}
%
\begin{small}
\begin{table}[t!]
\setcounter{table}{0}
\begin{center}
\begin{scriptsize}
\resizebox{1\textwidth}{!}{
\begin{tabular}{| c | c |c | c | c | c | c | c | c | c | c|}
\multicolumn{1}{c}{Model} &  \multicolumn{1}{c}{$h$} &    \multicolumn{1}{c}{$\Omega_m$}&  \multicolumn{1}{c}{{\small$\nu_i$}}  & \multicolumn{1}{c}{$w_0$} & \multicolumn{1}{c}{$w_1$} &\multicolumn{1}{c}{$\sigma_8(0)$} & \multicolumn{1}{c}{$\Delta{\rm AIC}$} & \multicolumn{1}{c}{$\Delta{\rm BIC}$}\vspace{0.5mm}
\\\hline
{$\CC$CDM} & $0.692\pm 0.004$ &  $0.296\pm 0.004$ & - & -1 & - & $0.801\pm0.009$ & - & -\\
\hline
XCDM  &  $0.672\pm 0.007$&  $0.311\pm 0.007$& - & $-0.923\pm0.023$ & - & $0.767\pm0.014$ & 8.55 & 6.31 \\
\hline
CPL  &  $0.673\pm 0.009$&  $0.310\pm 0.009$& - & $-0.944\pm0.089$ & $0.063\pm0.259$ & $0.767\pm0.015$ & 6.30 & 1.87 \\
\hline
RVM  & $0.677\pm 0.005$&  $0.303\pm 0.005$ & $0.00158\pm 0.00042$ & -1 & - & $0.736\pm0.019$ & 12.91 & 10.67 \\
\hline
$Q_{dm}$ &  $0.678\pm 0.005$&  $0.302\pm 0.005 $ & $0.00216\pm 0.00060 $ & -1 & - &  $0.740\pm0.018$  & 12.13 & 9.89 \\
\hline
$Q_\CC$  &  $0.691\pm 0.004$&  $0.298\pm 0.005$ & $0.00601\pm 0.00253$ & -1 & - &  $0.790\pm0.010$ & 3.41 & 1.17 \\
\hline \end{tabular}
 }
\end{scriptsize}
\end{center}
\caption{\scriptsize Best-fit values for the $\CC$CDM, XCDM, CPL and the three dynamical vacuum models (DVMs).   {The specific fitting parameters for each DVM are $\nu_{i}=\nu $ (RVM), $\nu_{dm}$($Q_{dm}$) and $\nu_{\CC}$($Q_{\CC}$), whilst for the XCDM and CPL are the EoS parameters $w_0$ and the pair ($w_0$,$w_1$), respectively. For the DVMs and the $\CC$CDM, we have $w_0=-1$ and $w_1=0$. The remaining parameters are as in the $\CC$CDM and are not shown. For convenience we reckon the values of $\sigma_8(0)$ for each model, which are not part of the fit but are computed from the fitted ones following the procedure indicated in Sec. \ref{linear_growth_Possible_signals_chapter} . The (positive) increments $\Delta$AIC and $\Delta$BIC (see the main text, Sec. \ref{subsect:AICandBIC_Possible_signals_chapter}) clearly favor the DDE options. The RVM and $Q_{dm}$ are particularly favored ($\sim 3.8\sigma$ c.l. and $3.6\sigma$, respectively). Our fit is performed over  a rich and fully updated SNIa+BAO+$H(z)$+LSS+CMB data set (cf. Sec. \ref{sect:Fit_Possible_signals_chapter})}.}
\label{tableFit1_Possible_signals_chapter}
\end{table}
\end{small}
In solving the differential equations \eqref{eq:Qequations_Possible_signals_chapter} we have traded the cosmic time variable for the scale factor using the chain rule $d/dt=aH d/da$. The corresponding vacuum energy densities can also be solved in the same variable, and yield:
\begin{align}
&{\rm \textbf{RVM}}:\quad \rho_\CC(a) = \rLo + \frac{\nu\,\rho_{m0}}{1-\nu}\left(a^{-3(1-\nu)}-1\right)+\frac{\nu{\rho_{r0}}}{1-\nu}\left(\frac{1-\nu}{1+3\nu}a^{-4} + \frac{4\nu}{1+3\nu}a^{-3(1-\nu)} -1\right)\label{eq:rhoVRVM_Possible_signals_chapter}\\
&{\rm \ \mathbf{Q_{dm}}}:\quad
\rho_\CC(a) = \rLo + \frac{\nu_{dm}\,\rho_{dm0}}{1-\nu_{dm}}\,\left(a^{-3(1-\nu_{dm})}-1\right)\label{eq:rhoVQdm_Possible_signals_chapter}\\
&{\rm \ \mathbf{Q_{\CC}}}:\quad
\rho_\CC(a) =\rLo\,{a^{-3\nu_\CC}}\label{eq:rhoVQL_Possible_signals_chapter}
\end{align}
One can easily check that for $a=1$ (i.e. at the present epoch) all of the energy densities \eqref{eq:rhoRVM_Possible_signals_chapter}-\eqref{eq:rhoVQL_Possible_signals_chapter} recover their respective current values $\rho_{N0}$ ($N=dm,\CC$). In addition,
for $\nu_{i}\to 0$ we retrieve for the three DM densities the usual $\CC$CDM expression $\rho_{dm}(a)=\rho_{dm 0}a^{-3}$, and the corresponding vacuum energy densities $\rL(a)$ boil down to the constant value $\rLo$ in that limit. The normalized Hubble rate $E\equiv H/H_0$ for each model can be easily obtained by plugging the above formulas, along with the radiation and baryon energy densities \eqref{eq:BaryonsRadiation_Possible_signals_chapter}, into  Friedmann's equation \eqref{eq:FriedmannEq_Possible_signals_chapter}. We find:
\begin{eqnarray}\label{eq:HRVM_Possible_signals_chapter}
{\rm \textbf{RVM}}:\quad E^2(a) &=& 1 + \frac{\Omega_m}{1-\nu}\left(a^{-3(1-\nu)}-1\right) \label{HRVM_Possible_signals_chapter}\\ \nonumber\\
& +& \frac{\Omega_r}{1-\nu}\left(\frac{1-\nu}{1+3\nu}a^{-4} + \frac{4\nu}{1+3\nu}a^{-3(1-\nu)} -1\right)\nonumber
\end{eqnarray}
\begin{eqnarray}\label{HQdm_Possible_signals_chapter}
{\rm \ \mathbf{Q_{dm}}}:\quad
E^2(a) &=& 1 + \Omega_b\left(a^{-3}-1\right) \\
&+& \frac{\Omega_{dm}}{1-\nu_{dm}}\left(a^{-3(1-\nu_{dm})}-1\right)
+ \Omega_r\left(a^{-4}-1\right)\nonumber
\end{eqnarray}
\begin{eqnarray}\label{HQL_Possible_signals_chapter}
{\rm \ \mathbf{Q_{\CC}}}:\quad
E^2(a) &=&\frac{a^{-3\nu_\CC}-\nu_\CC{a^{-3}}}{1-\nu_\CC}+\frac{\Omega_m}{1-\nu_\CC}\left(a^{-3}-a^{-3\nu_\CC}\right) \nonumber\\ &+&  \Omega_r\left(a^{-4} + \frac{\nu_\CC}{1-\nu_\CC}a^{-3} - \frac{a^{-3\nu_\CC}}{1-\nu_\CC}\right)
\end{eqnarray}
In the above expressions, we have used the cosmological parameters $\Omega_N=\rho_{N0}/\rco$ for each fluid component ($N=dm,b,r,\CC$), and defined $\Omega_m=\Omega_{dm}+\Omega_b$. Altogether, they satisfy the sum rule $\sum_N\Omega_N=1$. The normalization condition $E(1)=1$  in these formulas is apparent, meaning that the Hubble function for each model reduces to $H_0$ at present, as they should; and, of course, for $\nu_i\to 0$ we recover the $\CC$CDM form for $H$, as should be expected.
\begin{figure}[t!]
\setcounter{figure}{0}
\centering
\includegraphics[angle=0,width=1.03\linewidth]{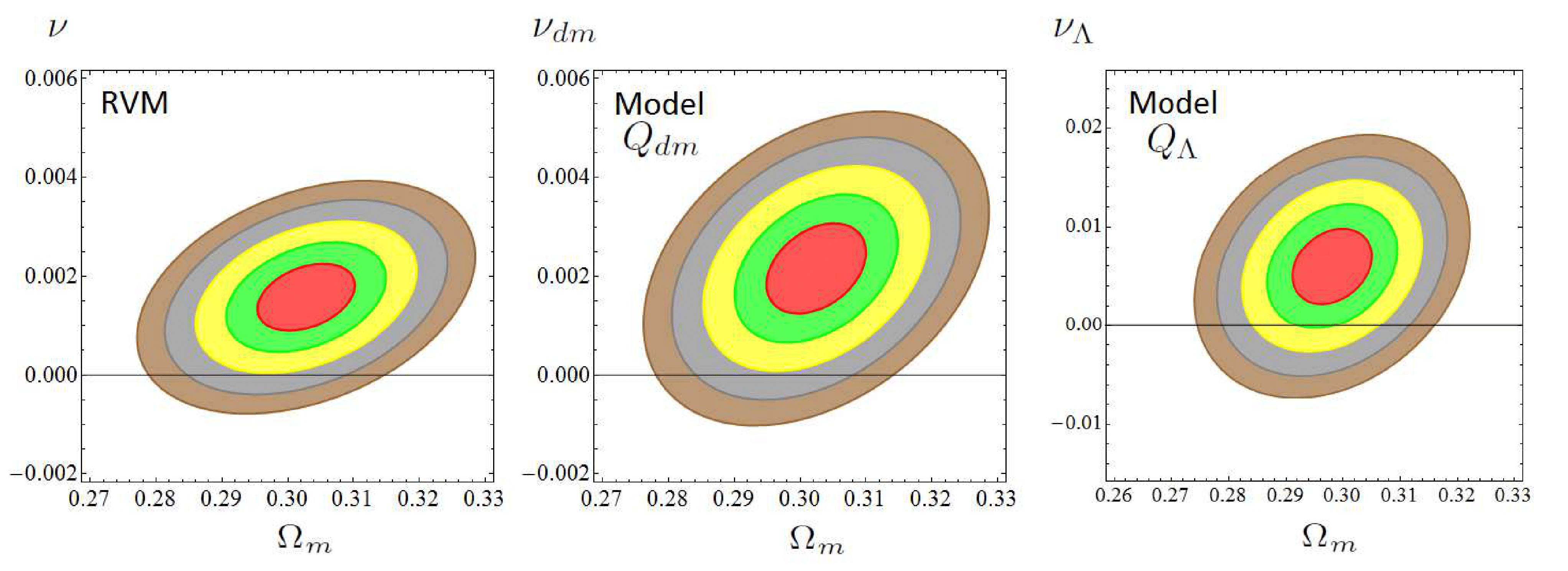}
\caption{\label{fig:G2Evolution_Possible_signals_chapter}%
\scriptsize Likelihood contours for the DVMs in the $(\Omega_m,\nu_i)$ plane for the values $-2\ln\mathcal{L}/\mathcal{L}_{max}=2.30$, $6.18, 11.81$, $19.33$, $27.65$ (corresponding to 1$\sigma$, 2$\sigma$, 3$\sigma$, 4$\sigma$ and 5$\sigma$ c.l.) after marginalizing over the rest of the fitting parameters. We estimate that for the RVM, $94.80\%$ (resp. $89.16\%$) of the 4$\sigma$ (resp. 5$\sigma$) area is in the $\nu>0$ region. For the Q$_{dm}$ we find that $95.24\%$ (resp. $89.62\%$) of the 4$\sigma$ (resp. 5$\sigma$) area is in the $\nu_{dm}>0$ region. Finally, for the Q$_{\CC}$ we estimate that $99.45\%$ (resp. $90.22\%$) of the $2\sigma$ (resp. $3\sigma$) area is in the $\nu_{\CC}>0$ region. Subsequent marginalization over $\Omega_m$ increases slightly the c.l. and renders the fitting values collected in Table 1. The $\CC$CDM ($\nu_i=0$) appears disfavored at $\sim 4\sigma$ c.l. in the RVM and $Q_{dm}$, and at $\sim 2.5\sigma$ c.l. for $Q_\CC$.}
\end{figure}
%
%
From the structure of equations \eqref{eq:rhoVRVM_Possible_signals_chapter} and \eqref{eq:HRVM_Possible_signals_chapter} we can readily see that the vacuum energy density for the RVM can be fully written as a function of a cosmic variable $\zeta$, which can be chosen to be not only the scale factor but the full Hubble function $\zeta=H$. The result is, of course, Eq.\eqref{eq:RVMvacuumdadensity_Possible_signals_chapter}. In contrast, for the $Q_{dm}$ and $Q_{\CC}$ models this is not possible, as it is clear on comparing equations \eqref{eq:rhoVQdm_Possible_signals_chapter}-\eqref{eq:rhoVQL_Possible_signals_chapter} and the corresponding ones \eqref{HQdm_Possible_signals_chapter}-\eqref{HQL_Possible_signals_chapter}. For these models $\rL$ can only be written as a function of the scale factor. Thus, the RVM happens to have the greatest level of symmetry since its origin is a RG equation in $H$ whose solution is precisely \eqref{eq:RVMvacuumdadensity_Possible_signals_chapter}.
\subsubsection{XCDM and CPL parametrizations}\label{sect:XCDMandCPL_Possible_signals_chapter}
Together with the DVMs , we fit also the same data through the simple XCDM parametrization of the dynamical DE, first introduced in \cite{Turner:1998ex}. Since both matter and DE are self-conserved (i.e., they are not interacting), the DE density as a function of the scale factor is simply given by $\rho_X(a)=\rho_{X0}\,a^{-3(1+w_0)}$, with $\rho_{X0}=\rho_{\CC 0}$, where $w_0$ is the (constant) EoS parameter of  the generic DE entity X in this parametrization. The normalized Hubble function is:
\begin{equation}\label{eq:HXCDM_Possible_signals_chapter}
E^2(a)=\Omega_m\,a^{-3}+\Omega_r\,a^{-4}+\OL\,a^{-3(1+w_0)}\,.
\end{equation}
For $w_0=-1$ it boils down to that of the $\CC$CDM with rigid CC term. The use of the XCDM parametrization throughout our analysis will be useful to roughly mimic a (noninteractive) DE scalar field with constant EoS. For $w_0\gtrsim-1$ the XCDM mimics quintessence, whereas for $w_0\lesssim-1$ it mimics phantom DE.
\newline
A slightly more sophisticated approximation to the behaviour of a noninteractive scalar field playing the role of dynamical DE is afforded by the CPL parametrization \cite{Chevallier:2000qy,Linder:2002et,Linder:2004ng}, in which one assumes that the generic DE entity $X$ has a slowly varying EoS of the form
\begin{equation}\label{eq:CPL_Possible_signals_chapter}
w_D=w_0+w_1\,(1-a)=w_0+w_1\,\frac{z}{1+z}\,.
\end{equation}
The CPL parametrization, in contrast to the XCDM one, makes allowance for a time evolution of the dark energy EoS owing to the presence of the additional parameter $w_1$, which satisfies $0<|w_1|\ll|w_0|$, with $w_0\gtrsim -1$ or $w_0\lesssim -1$. The expression \eqref{eq:CPL_Possible_signals_chapter} is seen to have a well-defined limit both in the early Universe ($a\to 0$, equivalently $z\to\infty$) and in the current one ($a=1$, or $z=0$).
The corresponding normalized Hubble function for the CPL can be easily found:
\begin{equation}
 E^2(a) = \Omega_m\,a^{-3}+ \Omega_r a^{-4}+\Omega_\CC
 a^{-3(1+w_0+w_1)}\,e^{-3\,w_1\,(1-a)}\,.
\label{eq:HCPL_Possible_signals_chapter}
\end{equation}
The XCDM and the CPL parametrizations can be conceived as a kind of baseline frameworks to be referred to in the study of dynamical DE.  We expect that part of the DDE effects departing from the $\CC$CDM should be captured by these parametrizations, either in the form of effective quintessence behaviour ($w\gtrsim -1$) or effective phantom behaviour ($w\lesssim-1$).
The XCDM, though,  is the most appropriate for a fairer comparison with the DVMs, all of which also have one single vacuum parameter $\nu_i$.
\subsection{Data sets and results}\label{sect:Fit_Possible_signals_chapter}
In this work, we fit the $\CC$CDM, XCDM, CPL and the three DVMs to the cosmological data from type Ia supernovae \cite{Betoule:2014frx}, BAOs \cite{Beutler:2011hx,Kazin:2014qga,Ross:2014qpa,Gil-Marin:2016wya,Delubac:2014aqe,Aubourg:2014yra}, the values of the Hubble parameter extracted from cosmic chronometers at various redshifts, $H(z_i)$ \cite{Zhang:2012mp,Jimenez:2003iv,Simon:2004tf,Moresco:2012jh,Moresco:2015cya,Moresco:2016mzx,Stern:2009ep}, the CMB data from Planck 2015 \cite{Ade:2015xua} and the most updated set of LSS formation data embodied in the quantity $f(z_i)\sigma_8(z_i)$,
{see the corresponding values and references in Table 2}. It turns out that the LSS data is very important for the DDE signal, and up to some updating performed here it has been previously described in more detail in \cite{Sola:2016jky}. We denote this string of cosmological data by SNIa+BAO+$H(z)$+LSS+CMB.
\newline
A guide to the presentation of our results is the following. The various fitting analyses and contour plots under different conditions (to be discussed in detail in the next sections) are displayed in four fitting tables, Tables 1 and 3-5, and in seven figures, Figs.\,1-7. {The main numerical results of our analysis are those recorded in Table 1. Let us mention in particular  Fig. 6, whose aim is to  identify what are the main data responsible for the DDE effect under study.  Bearing in mind the aforementioned significance of the LSS data, Fig. 7 is aimed to compare in a graphical way the impact of the $f(z)\sigma_8(z)$ and weak lensing data on our results.
The remaining tables and figures contain complementary information, which can be helpful for a more detailed picture of our rather comprehensive study.
Worth noticing are the results displayed in Table 5, which shows what would be the outcome of our analysis if we would restrict ourselves to the fitting data employed by the Planck 2015 collaboration \cite{Ade:2015xua}, where e.g. no LSS data were used and no DDE signal was reported.  Additional  details and considerations are furnished of course in the rest of our exposition}.
\begin{figure}[t!]
\centering
\includegraphics[width=15.0cm]{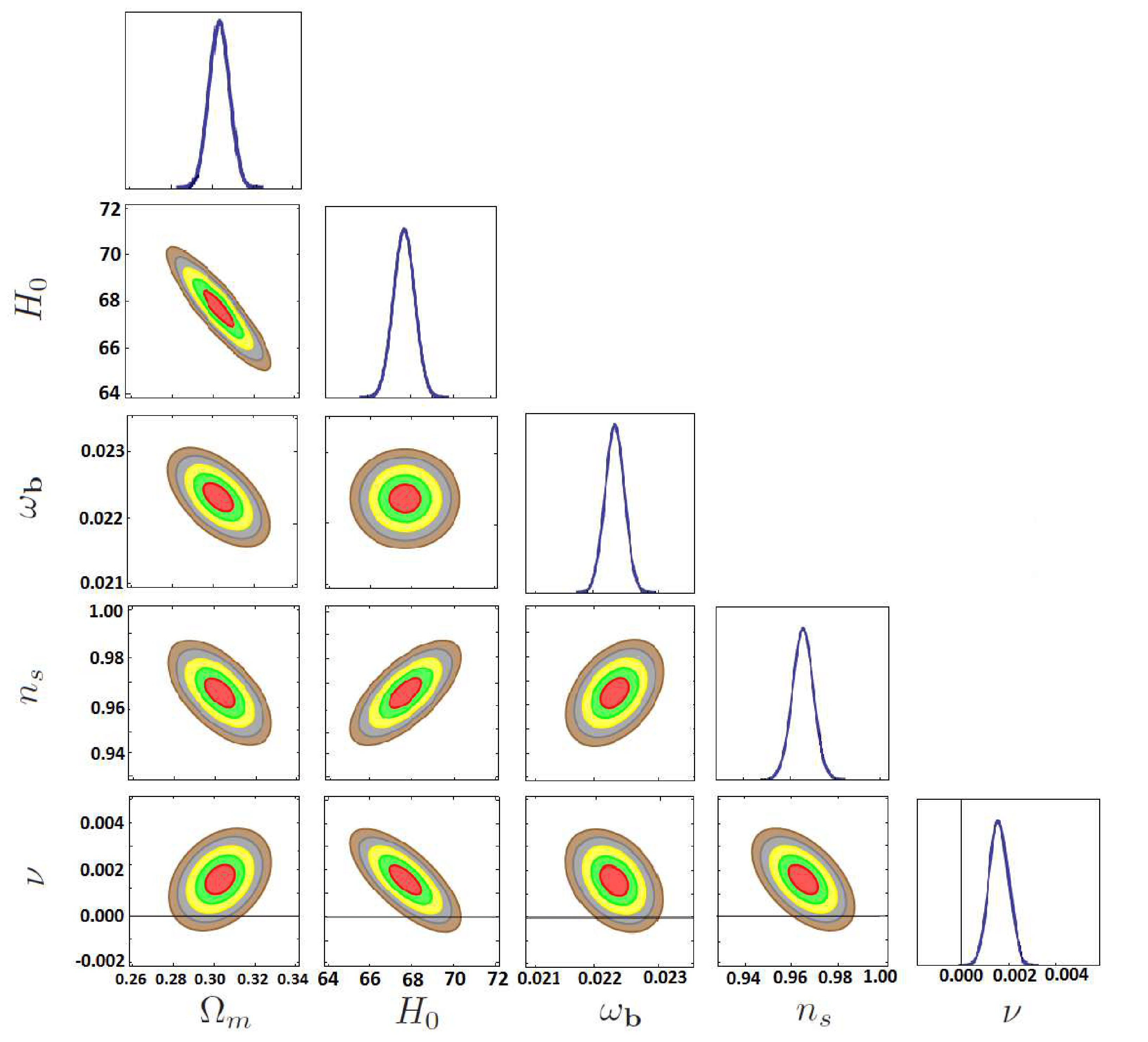}
\caption{\label{fig:MatrixPlot_Possible_signals_chapter}%
\scriptsize As in Fig.\,1, but projecting the fitting results of the RVM onto the different planes of the involved parameters ($H_0$ is expressed in Km/s/Mpc). The horizontal line $\nu=0$ in the plots of the last row corresponds to the $\CC$CDM.   It is apparent that it is significantly excluded at $\sim 4\sigma$ c.l. in all cases. The peak in the rightmost plot corresponds to the central value $\nu=0.00158$ indicated in Table 1.
}
\end{figure}
%
%
%
%
%
\begin{table}[t!]
\begin{center}
\resizebox{9cm}{!} {
\begin{tabular}{| c | c |c | c |}
\multicolumn{1}{c}{Survey} &  \multicolumn{1}{c}{$z$} &  \multicolumn{1}{c}{$f(z)\sigma_8(z)$} & \multicolumn{1}{c}{{\small References}}
\\\hline
6dFGS & $0.067$ & $0.423\pm 0.055$ & \cite{Beutler:2012px}
\\\hline
SDSS-DR7 & $0.10$ & $0.37\pm 0.13$ & \cite{Feix:2015dla}
\\\hline
GAMA & $0.18$ & $0.29\pm 0.10$ & \cite{Sugiyama:1994ed}
\\ & $0.38$ & $0.44\pm0.06$ & \cite{Blake:2013nif}
\\\hline
DR12 BOSS & $0.32$ & $0.427\pm 0.056$  & \cite{Gil-Marin:2016wya}\\
 & $0.57$ & $0.426\pm 0.029$ & \\\hline
WiggleZ & $0.22$ & $0.42\pm 0.07$ & \cite{Blake:2011rj} \tabularnewline
 & $0.41$ & $0.45\pm0.04$ & \tabularnewline
 & $0.60$ & $0.43\pm0.04$ & \tabularnewline
 & $0.78$ & $0.38\pm0.04$ &
\\\hline
2MTF & $0.02$ & $0.34\pm 0.04$ & \cite{Springob:2015pbs}
\\\hline
VIPERS & $0.7$ & $0.38^{+0.06}_{-0.07}$ &  \cite{Granett:2015ppa}
\\\hline
VVDS & $0.77$ & $0.49\pm0.18$ & \cite{Guzzo:2008ac,Song:2008qt}
\\\hline
 \end{tabular}
 }
\caption{\scriptsize Published values of $f(z)\sigma_8(z)$, referred to in the text as the LSS formation data.}
\end{center}
\end{table}
%
\subsection{Structure formation with dynamical vacuum}\label{sect:perturbationsDVM_Possible_signals_chapter}
Despite the theory of cosmological perturbations has been discussed at length in several specialized textbooks, see e.g. \cite{peebles:1993,Liddle:2000cg,Dodelson:2003ft}, the dynamical character of the vacuum produces some changes on the standard equations which are worth mentioning.  At deep subhorizon scales one can show that the matter density contrast $\delta_m=\delta\rho_m/\rho_m$ obeys the following differential equation (cf.\cite{Basilakos:2014tha,Gomez-Valent:2014rxa,Gomez-Valent:2017idt} for details):
\begin{equation}\label{diffeqD_Possible_signals_chapter}
\ddot{\delta}_m+\left(2H+\Psi\right)\,\dot{\delta}_m-\left(4\pi
G\rmr-2H\Psi-\dot{\Psi}\right)\,\delta_m=0\,,
\end{equation}
where $\Psi\equiv -\dot{\rho}_{\Lambda}/{\rmr}= Q/{\rmr}$, and the (vacuum-matter) interaction source $Q$ for each DVM is given in Sec.\,\ref{sect:DVMs_Possible_signals_chapter}. For $\rL=$const. and for the XCDM and CPL there is no such an interaction and Eq.\eqref{diffeqD_Possible_signals_chapter}\ reduces to
$\ddot{\delta}_m+2H\,\dot{\delta}_m-4\pi G\rmr\,\delta_m=0$, i.e. the $\CC$CDM form \cite{peebles:1993}.
We note that at the scales under consideration we are neglecting the perturbations of the vacuum energy density in front of the perturbations of the matter field. The justification for this has recently been analyzed in detail, cf.\cite{Gomez-Valent:2017idt}.
\newline
\newline
Let us briefly justify by two alternative methods the modified form \eqref{diffeqD_Possible_signals_chapter}, in which the variation of $\rho_\CC$ enters  through the Hubble function and the background quantity $\Psi$, but not through any perturbed quantity. We shall conveniently argue in the context of two well-known gauges, the synchronous gauge and the Newtonian conformal gauge, thus providing a twofold justification. In the synchronous gauge, vacuum perturbations $\delta\rL$ modify the momentum conservation equation for the matter particles in a way that we can easily get e.g. from the general formulae of \cite{Gomez-Valent:2017idt,Grande:2008re}, with the result
\begin{equation}\label{eq:ModifiedMomentum_Possible_signals_chapter}
\dot{v}_m+ \, H v_m=\frac{1}{a}\frac{\delta\rho_\Lambda}{\rho_m}-\Psi v_m\,,
\end{equation}
where  $\vec{v}=\vec{\nabla}v_m$ is the associated peculiar velocity, with potential $v_m$ (notice that this quantity has dimension of inverse energy in natural units). By setting $\delta\rho_\Lambda= a\,Q\,{v_m}=a\,\rho_m\,\Psi\,v_m$ the two terms on the \textit{r.h.s.} of \eqref{eq:ModifiedMomentum_Possible_signals_chapter} cancel each other and we recover the corresponding equation of the $\CC$CDM.
On the other hand, in the Newtonian or conformal gauge \cite{Mukhanov:1990me,Ma:1995ey} we find a similar situation. The analog of the previous equation is the modified Euler's equation in the presence of dynamical vacuum energy, 
\begin{equation}\label{eq:Euler_Possible_signals_chapter}
\frac{d}{d\eta}\left(\rho_mv_m\right)+4\mathcal{H}\rho_mv_m+\rho_m\phi-\delta\rho_\Lambda=0\,,
\end{equation}
where $\phi$ is the gravitational potential that appears explicitly in the Newtonian conformal gauge, and $\eta$ is the conformal time. Let an overhead circle denote a derivative with respect to the conformal time, $\mathring{f}=df/d\eta$ for any $f$. We define the quantities $\mathcal{H}=\mathring{a}/a=aH$ and $\bar{\Psi}=-\mathring{\rho}_{\Lambda}/\rho_m=a\Psi$, which are the analogues of $H$ and $\Psi$ in conformal time. Using the background local conservation equation \eqref{BianchiGeneral_Possible_signals_chapter} for the current epoch (neglecting therefore radiation) and rephrasing it in conformal time, i.e. $\mathring{\rho}_\Lambda+\mathring{\rho}_m+3\mathcal{H}\rho_m=0$, we can bring \eqref{eq:Euler_Possible_signals_chapter} to 
\begin{equation}\label{eq:ModifiedEuler_Possible_signals_chapter}
\mathring{v}_m+ \, \mathcal{H} v_m+\phi=\frac{\delta\rho_\Lambda}{\rho_m}-\bar{\Psi} v_m\,.
\end{equation}
Once more the usual fluid equation (in this case Euler's equation) is retrieved if we arrange that $\delta\rho_\Lambda=\rho_m\,\bar{\Psi}\,v_m=a\,\rho_m\,\Psi\,v_m$, as then the two terms on the \textit{r.h.s.} of \eqref{eq:ModifiedEuler_Possible_signals_chapter} cancel each other. Alternatively, one can use the covariant form $\nabla^{\mu} T_{\mu\nu}=Q_\nu$ for the local conservation law, with  the source 4-vector $Q_\nu= Q U_\nu$, where $U_\nu=(a,\vec{0})$ is the background matter 4-velocity in conformal time. By perturbing the covariant conservation equation one finds
\begin{equation}
\delta\left(\nabla^{\mu} T_{\mu\nu}\right)=\delta Q_\nu=\delta Q\,U_\nu+Q\delta U_\nu\,,
\end{equation}
where $\delta Q$ and $\delta U_\nu=a(\phi,-\vec{v})$ are the perturbations of the source function and the 4-velocity, respectively. Thus, we obtain
\begin{equation}
\delta\left(\nabla^{\mu} T_{\mu\nu}\right)=a(\delta Q+Q\phi,-Q\vec{v})\,.
\end{equation}
From the $\nu=j$ component of the above equation, we derive anew the usual Euler equation $\mathring{v}_m+ \, \mathcal{H} v_m+\phi=0$, which means that the relation $\delta\rho_\Lambda= a Q\,{v_m}=a\,\rho_m\,\Psi\,v_m$ is automatically fulfilled. So the analyses in the two gauges converge to the same final result for $\delta\rho_\Lambda$.
\newline
\newline
After we have found the condition that $\delta\rho_\Lambda$ must satisfy in each gauge so as to prevent that the vacuum modifies basic conservation laws of the matter fluid, one can readily show that any of the above equations \eqref{eq:ModifiedEuler_Possible_signals_chapter} or \eqref{eq:ModifiedEuler_Possible_signals_chapter} for each gauge (now with their \textit{r.h.s.} set to zero), in combination with the corresponding perturbed continuity equation and the perturbed $00$-component of Einstein's equations (giving Poisson's equation in the Newtonian approximation), leads to the desired matter perturbations equation \eqref{diffeqD_Possible_signals_chapter}, in accordance with the result previously derived by other means in Refs. \cite{Gomez-Valent:2014rxa,Basilakos:2014tha}. Altogether, the above considerations formulated in the context of different gauges allow us to consistently neglect the DE perturbations at scales down the horizon. This justifies the use of Eq.\eqref{diffeqD_Possible_signals_chapter} for the effective matter perturbations equation in our study of linear structure formation in the framework of the DVMs.  {See \cite{Gomez-Valent:2017idt} for an expanded exposition of these considerations.}
\newline
For later convenience let us also rewrite Eq.\eqref{diffeqD_Possible_signals_chapter} in terms of the scale factor variable rather than the cosmic time. Using $d/dt=aH\,d/da$ and denoting the differentiation  $d/da$ with a prime, we find:
\begin{equation}\label{diffeqDa_Possible_signals_chapter}
\delta^{\prime\prime}_m + \frac{A(a)}{a}\delta^\prime_m + \frac{B(a)}{a^2}\delta_m = 0\,,
\end{equation}
where the functions $A$ and $B$ of the scale factor are given by
\begin{align}
& A(a) = 3 + a\frac{H^{\prime}(a)}{H(a)} + \frac{\Psi(a)}{H(a)}\,,\label{deffA_Possible_signals_chapter}\\
& B(a) = - \frac{4\pi{G}\rho_m(a)}{H^2(a)} + \frac{2\Psi(a)}{H(a)} + a\frac{\Psi^{\prime}(a)}{H(a)}\label{deffB_Possible_signals_chapter}\,.
\end{align}
%
\begin{figure}[t!]
\centering
\includegraphics[width=0.7\linewidth]{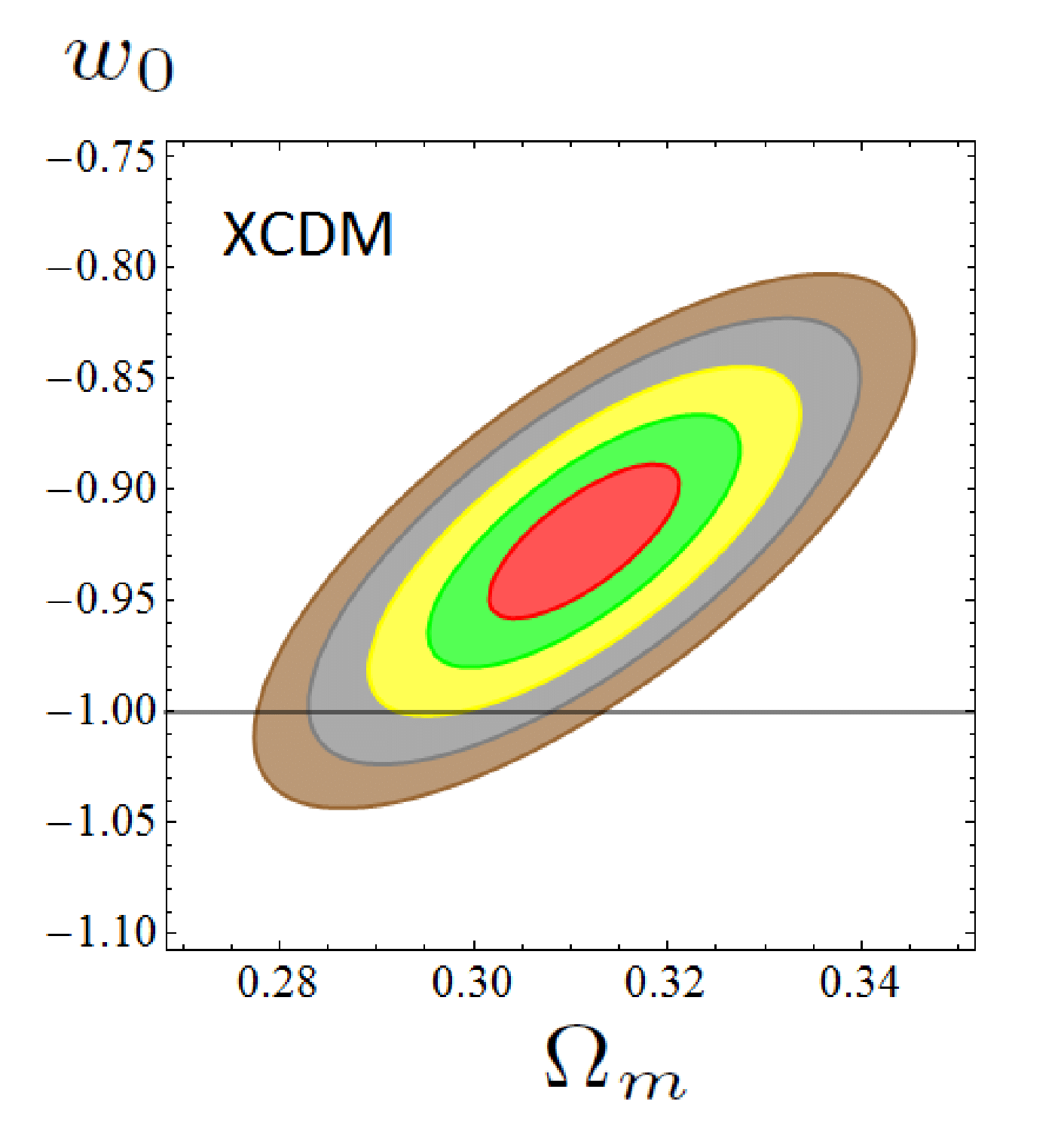}
\caption{\label{fig:XCDMEvolution_Possible_signals_chapter}%
\scriptsize As in Fig.\,1, but for model XCDM. The $\CC$CDM is excluded in this case at $\sim 3\sigma$ c.l. Marginalization over $\Omega_m$ increases the c.l. up to $3.35\sigma$ (cf. Table 1).
}
\end{figure}
%
%
\subsubsection{Initial conditions}\label{Initial_conditions_Possible_signals_chapter}
{In order to solve \eqref{diffeqDa_Possible_signals_chapter} we have to fix appropriate  initial conditions for $\delta_m(a)$ and $\delta^{\prime}_m(a)$ for each model at high redshift, say at $z_i\sim100$ ($a_i\sim10^{-2}$), when non-relativistic matter dominates both over the vacuum and the radiation contributions. In practice it  can be fixed at lower  redshifts, say of order $10$, where the subhorizon approximation is even more efficient \cite{Gomez-Valent:2018nib}, although the differences are small. For small values of the scale factor the normalized Hubble rate (squared) for each model, and the energy densities for the various components,  see equations \eqref{eq:rhoRVM_Possible_signals_chapter}-\eqref{HQL_Possible_signals_chapter}, can be significantly simplified.
As a result we obtain the leading form of the functions \eqref{deffA_Possible_signals_chapter}-\eqref{deffB_Possible_signals_chapter} for the different DVMs:}
\begin{eqnarray}
{\rm RVM}:\phantom{X} A &=& \frac{3}{2}(1+3\nu) \label{AFunction1_Possible_signals_chapter}\\
{\rm \ Q_{dm}}:\phantom{X} A &=& \frac{3}{2}\left(1 + \nu_{dm}\right) + 3\frac{\Omega_{dm}}{\Omega_m}\nu_{dm}+\mathcal{O}(\nu^2_{dm}) \nonumber\\ \label{AFunction2_Possible_signals_chapter}\\
{\rm \ Q_{\CC}}:\phantom{X} A &=& \frac{3}{2}\,,
\label{AFunction3_Possible_signals_chapter}
\end{eqnarray}
and
\begin{eqnarray}
{\rm RVM}:\phantom{X} B &=& -\frac{3}{2} +3\nu + \frac{9}{2}\nu^2  \label{BFunction1_Possible_signals_chapter}\\
{\rm \ Q_{dm}}:\phantom{X} B &=& -\frac{3}{2}\left(1-\nu_{dm} - \frac{\Omega_{dm}}{\Omega_m}\nu_{dm}\right)+\mathcal{O}(\nu^2_{dm}) \nonumber\\ \label{BFunction2_Possible_signals_chapter}\\
{\rm \ Q_{\CC}}: \phantom{X} B &=& -\frac{3}{2}\,.
\label{BFunction3_Possible_signals_chapter}
\end{eqnarray}
For $\nu_i\rightarrow 0$, we recover the $\CC$CDM behaviour $A \rightarrow \frac{3}{2}$ and $B \rightarrow -\frac{3}{2}$, as it should. This is already true for the $Q_{\CC}$ without imposing $\nu_{\CC}\rightarrow 0$, therefore its initial conditions are precisely the same as for the concordance model. Once the functions \eqref{deffA_Possible_signals_chapter}-\eqref{deffB_Possible_signals_chapter} take constant values (as it is the case here at the high redshifts where we fix the initial conditions), the differential equation \eqref{diffeqDa_Possible_signals_chapter} admits power-like solutions of the form $\delta_m(a_i)=a_i^{s}$. Of the two solutions, we are interested only in the growing mode solution, as this is the only one relevant for structure formation. For example, using \eqref{AFunction1_Possible_signals_chapter} and \eqref{BFunction1_Possible_signals_chapter} for the case of the RVM, the perturbations equation \eqref{diffeqDa_Possible_signals_chapter} becomes
\begin{equation}\label{diffeqDaRVM_Possible_signals_chapter}
\delta^{\prime\prime}_m + \frac{3}{2a}(1+3\nu)\delta^{\prime}_m - \left(\frac{3}{2}-3\nu-\frac{9}{2}\,\nu^2\right)\frac{\delta_m}{a^2} = 0\,.
\end{equation}
The power-law solution for the growing mode gives the result $\delta_m=a^{1-3\nu}$, which is exact even keeping the ${\cal O}(\nu^2)$ term. Nevertheless, as warned previously, in practice we can neglect all ${\cal O}(\nu_i^2)$ contributions despite we indicate their presence.  Repeating the same procedure for the other models, the power-law behaviour in each case for the growing mode solution $\delta_m\sim a^s$ is the following:
\begin{eqnarray}
{\rm RVM}:\phantom{X} s &=& 1-3\nu \label{svalues1_Possible_signals_chapter}\\
{\rm \ Q_{dm}}:\phantom{X} s &=& 1-\nu_{dm}\left(\frac{6\Omega_m + 9\Omega_{dm}}{5\Omega_m}\right) + \mathcal{O}(\nu^2_{dm})\nonumber\\ \label{svalues2_Possible_signals_chapter}\\
{\rm \ Q_{\CC}}:\phantom{X} s &=& 1\,.
\label{svalues3}
\end{eqnarray}
Imposing the above analytical results to fix the initial conditions, we are then able to solve numerically the full differential equation \eqref{diffeqDa_Possible_signals_chapter} from a high redshift $z_i\sim100$ ($a_i\sim10^{-2}$) up to our days. The result does not significantly depend on the precise value of $z_i$, provided it is in the matter-dominated epoch and well below the decoupling time ($z\sim 10^3$), where the radiation component starts to be nonnegligible.
\subsubsection{Linear growth and growth index}\label{linear_growth_Possible_signals_chapter}
The linear growth rate of clustering is an important (dimensionless)
indicator of structure formation \cite{peebles:1993}. It is defined as the logarithmic derivative of the linear growth factor $\delta_m(a)$ with respect to the log of the scale factor, $\ln a$. Therefore,
\begin{equation}\label{growingfactor_Possible_signals_chapter}
f(a)\equiv \frac{a}{\delta_m}\frac{d\delta_m}{d a}=\frac{d{\rm ln}\delta_m}{d{\rm ln}a}\,,
\end{equation}
where $\delta_m(a)$ is obtained from solving
the differential equation \eqref{diffeqDa_Possible_signals_chapter} for each model. The physical significance of $f(a)$ is that it determines the peculiar velocity flows \cite{peebles:1993}. In terms of the redshift variable, we have $f(z)=-(1+z)\,d{\ln\delta_m}/{dz}$, and thus the linear growth can also be used to determine the amplitude of the redshift distortions. This quantity has been analytically computed for the RVM in \cite{Basilakos:2015wha}. Here we shall take it into account for the study of the LSS data in our overall fit to the cosmological observations.
\newline
One usually expresses the linear growth rate of clustering
in terms of $\Omega_m(z)=\rho_m(z)/\rho_c(z)$, where $\rho_c(z)=3H^2(z)/(8\pi G)$ is the evolving critical density, as follows \cite{peebles:1993}:
\begin{equation}\label{eq:gammaIndex_Possible_signals_chapter}
f(z)\simeq \left[\Omega_{m}(z)\right]^{\gamma(z)}\,,
\end{equation}
where $\gamma$ is the so-called linear growth rate index. For the usual $\Lambda$CDM model, such an index is approximately given by $\gamma_{\CC} \simeq 6/11\simeq 0.545$. For models with a slowly varying  equation of state $w_D$ (i.e. approximately behaving as the XCDM, with $w_D\simeq w_0$) one finds the approximate formula $\gamma_D\simeq 3(w_D-1)/(6w_D-5)$ \cite{Wang:1998gt} for the asymptotic value when $\Omega_m\to 1$. Setting $w_D=-1+\epsilon$, it can be rewritten
\begin{equation}\label{eq:GammaIndex_Possible_signals_chapter}
\gamma_{D}\simeq \frac{6-3\epsilon}{11-6\epsilon}\simeq \frac{6}{11}\left( 1+\frac{1}{22}\,\epsilon\right)\,.
\end{equation}
Obviously, for $\epsilon\to 0$ (equivalently, $\omega_D\to -1$) one retrieves the $\CC$CDM case.
Since the current experimental error on the $\gamma$-index is of order $10\%$, it opens the possibility to discriminate cosmological models using such an index, see e.g. \cite{Pouri:2014nta}. In the case of the RVM and various models and frameworks, the function $\gamma(z)$ has been computed numerically in \cite{Gomez-Valent:2014rxa}. Under certain approximations, an analytical result can also be obtained for the asymptotic value \cite{Basilakos:2015vra}:
\begin{equation}\label{eq:GammaIndexRVM_Possible_signals_chapter}
\gamma_{\rm RVM}\simeq \frac{6+3\nu}{11-12\nu}\simeq \frac{6}{11}\left(1+\frac{35}{22}\,\nu\right)\,.
\end{equation}
This expression for the RVM is similar to \eqref{eq:gammaIndex_Possible_signals_chapter} for an approximate XCDM parametrization, and it reduces to the $\CC$CDM value for $\nu=0$, as it should.
\subsubsection{Weighted linear growth and power spectrum}\label{sect:fsigma8_Possible_signals_chapter}
A most convenient observable to assess the performance of our vacuum models in regard to structure formation is the combined quantity $f(z)\sigma_{8}(z)$, viz. the ordinary growth rate weighted with $\sigma_{8}(z)$, the rms total matter fluctuation
(baryons + CDM) on $R_8 = 8{h^{-1}}$ Mpc spheres at the given redshift $z$, computed in linear theory. It has long been recognized that this estimator is almost a model-independent way of expressing the observed growth history of the Universe, most noticeably it is found to be independent of the galaxy density bias \cite{Guzzo:2008ac,Song:2008qt}.
\newline
With the help of the above generalized matter perturbations equation \eqref{diffeqDa_Possible_signals_chapter} and the appropriate initial conditions, the analysis of the linear LSS regime is implemented on using  the weighted linear growth $f(z)\sigma_8(z)$. The variance of the smoothed linear density field on $R_8 = 8{h^{-1}}$ Mpc spheres at redshift $z$ is computed from
\begin{equation}
\sigma_8^2(z)=\delta_m^2(z)\int\frac{d^3k}{(2\pi)^3}\, P(k,\vec{p})\,\,W^2(kR_8)\,.
\label{s88generalNN_Possible_signals_chapter}
\end{equation}
Here ${P}(k,\vec{p})={P}_0\,k^{n_s}T^2(k)$ is the ordinary linear matter power spectrum (i.e. the coefficient of the two-point correlator of the linear perturbations), with $P_0$ a normalization factor, $n_s$ the spectral index and $T(k)$ the transfer function. Furthermore,
$W(kR_8)$ in the above formula is a top-hat smoothing function (see e.g. \cite{Gomez-Valent:2014rxa} for
details), which can be expressed in terms of the spherical Bessel function of order $1$, as follows:
\begin{equation}\label{eq:WBessel_Possible_signals_chapter}
W(kR_8)=3\,\frac{j_1(kR_8)}{kR_8}=\frac{3}{k^2R_8^2}\left(\frac{\sin{\left(kR_8\right)}}{kR_8}-\cos{\left(kR_8\right)}\right)\,.
\end{equation}
{Moreover, $\vec{p}$ is the  fitting vector with all the free parameters, including the specific vacuum parameters $\nu_i$ of the DVMs, or the EoS parameters $w_i$ for the XCDM/CPL parametrizations, as well as the standard parameters.}
\newline
{The power spectrum depends on all the components of the fitting vector. However, the dependence on the spectral index $n_s$ is  power-like, whereas the transfer function $T(k,\vec{q})$ depends in a  more complicated way on the rest of the fitting parameters (see below), and thus for convenience we collect them in the reduced fitting vector $\vec{q}$ not containing $n_s$.}
It is convenient to write the variance \eqref{s88generalNN_Possible_signals_chapter} in
terms of the dimensionless linear matter power spectrum, ${\cal P}(k,\vec{p})=\left(k^3/2\pi^2\right)\,P(k,\vec{p})$,
with
\begin{equation}\label{eq:PowerSpectrum_Possible_signals_chapter}
{\cal P}(k,\vec{p})={\cal P}_0k^{n_s+3}T^2(k,\vec{q})\,.
\end{equation}
The normalization factor ${\cal P}_0=P_0/2\pi^2$  will be determined in the next section in connection to the definition of the  fiducial model.
\newline
For the transfer function, we have adopted the usual BBKS form \cite{Bardeen:1985tr},
but we have checked that the use of the alternative
one by \cite{Eisenstein:1997ik} does not produce any significant change in our results. Recall that the wave number at equality, $k_{eq}$, enters the argument of the transfer function.
However,  $k_{eq}$ is a model-dependent quantity, which departs from the $\CC$CDM expression in those models in which matter and/or radiation are governed by an anomalous continuity equation, as e.g. in the DVMs. In point of fact $k_{eq}$ depends on all the parameters of the reduced fitting vector $\vec{q}$. For the concordance model, $k_{eq}$ has the simplest expression,
\begin{equation}\label{keqCCprev_Possible_signals_chapter}
k^\CC_{eq} = H_0\,\Omega_m\sqrt{\frac{2}{\Omega_r}}=\frac{\Omega_mh^2}{2997.9}\sqrt{\frac{2}{\omega_r}}\,\textrm{Mpc}^{-1}\,,
\end{equation}
where $\omega_r=\Omega_r h^2$. In the second equality we have used the relation $H_0^{-1}=2997.9 h^{-1}$ Mpc.
For the DVMs it  is not possible to find a formula as compact as the one above. Either the corresponding expression for $a_{eq}$ is quite involved, as in the RVM case,
\begin{equation}\label{eq:aeqRVM_Possible_signals_chapter}
\textrm{RVM}:\quad a_{eq}=\left[\frac{\Omega_r(1+7\nu)}{\Omega_m(1+3\nu)+4\nu\Omega_r}\right]^{\frac{1}{1+3\nu}}\,,
\end{equation}
or because $a_{eq}$ must be computed numerically, as for the models Q$_{dm}$ and Q$_\CC$. In all cases, for $\nu_i=0$ we retrieve the value of $a_{eq}$ in the $\CC$CDM.
\subsubsection{Fiducial model}\label{sect:FiducialModel_Possible_signals_chapter}
Inserting the dimensionless power spectrum \eqref{eq:PowerSpectrum_Possible_signals_chapter} into the variance \eqref{s88generalNN_Possible_signals_chapter} at $z=0$ allows us to write $\sigma_8(0)$ in terms of the power spectrum normalization factor ${\cal P}_0$ in \eqref{eq:PowerSpectrum_Possible_signals_chapter} and the primary parameters that enter our fit for each model. This is tantamount to saying that ${\cal P}_0$ can be fixed as follows:
\begin{equation}\label{P0_Possible_signals_chapter}
\begin{small}
{\cal P}_0=\frac{\sigma_{8,\Lambda}^2}{\delta^2_{m,\Lambda}}\left[\int_{0}^{\infty} k^{n_{s,\Lambda+3}}T^2(k,\vec{q}_{\Lambda})W^2(kR_{8,\Lambda})(dk/k)\right]^{-1}\,,
\end{small}
\end{equation}
%
\begin{figure}[t!]
\centering
\includegraphics[angle=0,width=0.7\linewidth]{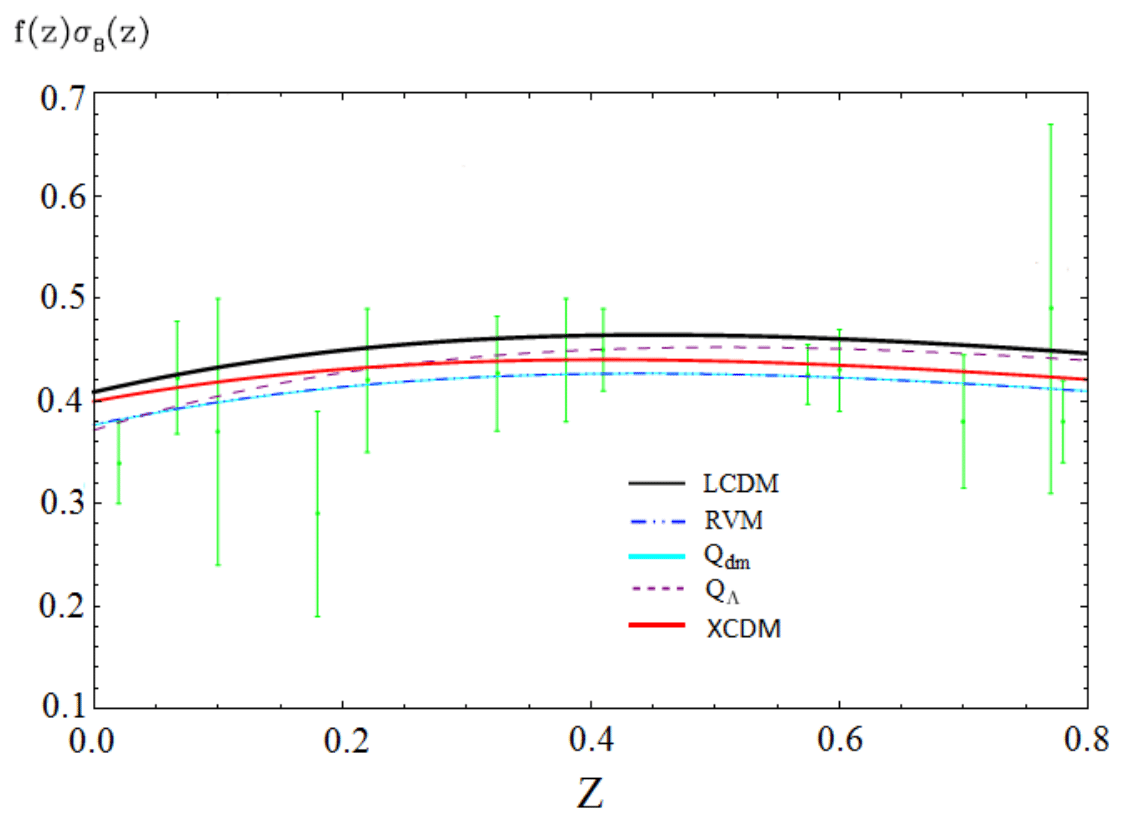}
\caption{\label{fig:fsigma8}%
\scriptsize The $f(z)\sigma_8(z)$ data (cf. Sec. \ref{sect:Fit_Possible_signals_chapter}) and the predicted curves by the $\CC$CDM, XCDM, and the DVMs, for the best-fit values in Table 1.}
\end{figure}
%
%
where the chosen values of the parameters in this expression define our fiducial model.  {The latter is characterized by the vectors of fiducial parameters $\vec{p}_\CC$
and  $\vec{q}_\CC$, defined in obvious analogy with the original fitting vectors but with all their parameters taken to be equal to those from the Planck 2015 TT,TE,EE+lowP+lensing analysis \cite{Ade:2015xua}, with $\nu_i=0$ for the DVMs and $w_0=-1, w_1=0$ for the XCDM/CPL parametrizations. The subindex $\CC$  carried by all the parameters denotes such a setting}.  In particular,  $\sigma_{8,\Lambda}\equiv\sigma_{8,\Lambda}(0)$ in \eqref{P0_Possible_signals_chapter} is also taken from the aforementioned Planck 2015 data.  However, $\delta_{m,\Lambda}\equiv\delta_{m,\Lambda}(0)$ in the same formula is computable: it is the value of $\delta_m(z=0)$ obtained from solving the perturbations equation of the $\CC$CDM, using the mentioned fiducial values of the other parameters. Finally, plugging the normalization factor \eqref{P0_Possible_signals_chapter} in \eqref{s88generalNN_Possible_signals_chapter} and using \eqref{eq:PowerSpectrum_Possible_signals_chapter} one finds:
\begin{equation}
\begin{small}\label{eq:sigma8normalized_Possible_signals_chapter}
\sigma_{8}(z)=\sigma_{8, \Lambda}
\frac{\delta_m(z)}{\delta_{m,\CC}}
\sqrt{\frac{\int_0^\infty k^{n_s+2} T^{2}(k,\vec{q})W^2(kR_{8})\, dk}{\int_0^\infty k^{n_{s,\CC}+2} T^{2}(k,\vec{q}_\Lambda) W^2(kR_{8,\Lambda})\, dk}}\,.
\end{small}
\end{equation}
%
For the fiducial $\CC$CDM, this expression just gives the scaling of $\sigma_{8,\CC}(z)$ with the redshift in the linear theory, that is to say, $\sigma_{8,\CC}(z)/\sigma_{8,\CC}=\delta_{m,\CC}(z)/\delta_{m,\CC}$. But for an arbitrary model, Eq. \eqref{eq:sigma8normalized_Possible_signals_chapter} normalizes the corresponding $\sigma_{8}(z)$ with respect to the fiducial value, $\sigma_{8, \Lambda}$. This includes, of course, our fitted $\CC$CDM, which is not the same as the fiducial $\CC$CDM. {So all fitted models are compared to the same fiducial model defined by the Planck 2015 results.} Similarly, upon computing with this method the weighted linear growth rate  $f(z)\sigma_8(z)$ for each model under consideration, (including the $\CC$CDM) the functions $f(z)\sigma_8(z)$ for all models become normalized to the same fiducial model. It is important to emphasize that one cannot adjust the power spectrum and the $f(z)\sigma_8(z)$ values independently. Therefore, we first normalize with Planck 2015 results, as above described, and from here we fit the models to the data, in which the LSS  component takes an essential part.
%
The connection of the normalization factor \eqref{P0_Possible_signals_chapter} with $A_s$ \cite{Ade:2015xua} can be easily found using standard formulae \cite{liddle_lyth_2000}. We find:
\begin{equation}\label{eq:PoAs_Possible_signals_chapter}
{\cal P}_0=\frac{4A_s}{25}\frac{k_{*}^{1-n_s}}{H_0^4 \Omega_m^2}\,,
\end{equation}
where $k_{*}=0.05$ Mpc$^{-1}$ is the pivot scale used by Planck. This follows from the fact that ${\cal P}_0$ is related to $\delta_H^2$ (the primordial amplitude of the gravitational potential) through ${\cal P}_0={\delta_H^2}/(H_0^{3+n_s} \Omega_m^2)$ and on the other hand we have  $\delta_H^2=(4/25) A_s (H_0/k_{*})^{n_s-1}$.
\newline
\newline
In Fig.\,4 we display the theoretical results for $f(z)\sigma_8(z)$ from the various models, side by side with the LSS data measurements, using the fitted values of Table 1. {The values that we find for $\sigma_8(0)$ for each model, with the corresponding uncertainties, are reckoned in Table 1}.
Inspection of Fig.\,4 shows that the DVMs provide a better description of the LSS data points as compared to the $\CC$CDM. {The XCDM parametrization takes an intermediate position, granting a better fit than the $\CC$CDM but a poorer one than the RVM and $Q_{dm}$}. One can see that it is necessary an overall reduction of  $\sim 8\%$ in the value of  $f(z)\sigma_8(z)$  with respect to the $\CC$CDM curve (the solid line on top of the others in that figure).  Once $\Omega_m $ is accurately fixed from the CMB data, the $\CC$CDM model does not have any further freedom to further adjust the low-$z$ LSS data.  This can be seen from Eq. \eqref{eq:PoAs_Possible_signals_chapter} and from the fact that the normalization amplitude of the power spectrum $A_s$ as given by Planck tolerates an error of order $2\%$ at most \cite{Ade:2015xua} and, therefore, such residual freedom cannot be invested to adjust the structure formation data, it is simply insufficient as we have checked. Thus, there seems to be no way at present to describe correctly  both the CMB and the LSS data within the  $\CC$CDM. This is of course at the root of the so-called  $\sigma_8$-tension, one of the important problems of the $\CC$CDM mentioned in the introduction -- see e.g. \cite{Macaulay:2013swa, Battye:2014qga,Basilakos:2016nyg,Basilakos:2017rgc} for additional discussion and references.
\newline
In contrast, the DVMs can provide a possible clue.  For example, for  the RVM case an analytical explanation has recently been provided  in  Refs. \cite{Gomez-Valent:2017idt,Gomez-Valent:2017tkh} showing why the dynamical vacuum can help in relaxing such tension. Recall that for $\nu=0$ the equality point between matter and radiation as given in Eq. \eqref{eq:aeqRVM_Possible_signals_chapter} boils down to the $\CC$CDM value. However, for $\nu\neq 0$ a nonnegligible contribution is obtained, despite the smallness of  $\nu$. Indeed, one can show that the $\nu$-effect causes a negative correction to the transfer function, which at linear order in $\nu$ is proportional to $6\nu\ln({\Omega_m}/{\Omega_r}) \simeq 50\,\nu$.  Since $\nu$ is fitted to be of order $\sim 10^{-3}$ in Table 1, it follows that the aforementioned negative correction can easily enhance the final effect up to near  $ 10\%$  level.  Upon a careful analysis of all the contributions, it eventually amounts to a  $\sim 8\%$ reduction of the weighted growth rate $f(z)\sigma_8(z)$ as compared to the $\CC$CDM value \cite{Gomez-Valent:2017idt,Gomez-Valent:2017tkh}. This is precisely the reduction with respect to the $\CC$CDM prediction that is necessary in order to provide a much better description of the LSS data, see Fig.\,4. Interestingly enough, {as a bonus one also obtains an excellent description of the current weak lensing data,  see  Sec.\,\ref{subsect:weaklensing_Possible_signals_chapter}.}
\newline
\newline
\subsection{Main numerical results}\label{sect:numerical results_Possible_signals_chapter}
For the statistical analysis, we define the joint likelihood function as the product of the likelihoods for all the data sets. Correspondingly, for Gaussian errors the total $\chi^2$ to be minimized reads:
\begin{equation}\label{chi2s_Possible_signals_chapter}
\chi^2_{\rm tot}=\chi^2_{\rm SNIa}+\chi^2_{\rm BAO}+\chi^2_{H}+\chi^2_{\rm LSS}+\chi^2_{\rm CMB}\,.
\end{equation}
Each one of these terms is defined in the standard way and they include the corresponding covariance matrices.
%
%
%
\begin{table}[!]
\begin{center}
\begin{scriptsize}
\begin{tabular}{| c | c |c | c | c | c | c |}
\multicolumn{1}{c}{Model} &  \multicolumn{1}{c}{$h$}  &  \multicolumn{1}{c}{$\Omega_m$}&  \multicolumn{1}{c}{{\small$\nu_i$}}  & \multicolumn{1}{c}{$w$} & \multicolumn{1}{c}{$\Delta{\rm AIC}$} & \multicolumn{1}{c}{$\Delta{\rm BIC}$}\vspace{0.5mm}
\\\hline
{$\CC$CDM} & $0.685\pm 0.004$ & $0.304\pm 0.005$ & - & -1 &  - & -\\
\hline
XCDM  &  $0.684\pm 0.009$&  $0.305\pm 0.007$& - & $-0.992\pm0.040$ &  -2.25 & -4.29 \\
\hline
RVM  & $0.684\pm 0.007$&  $0.304\pm 0.005$ & $0.00014\pm 0.00103$ & -1 &   -2.27 & -4.31 \\
\hline
$Q_{dm}$ &  $0.685\pm 0.007$&  $0.304\pm 0.005 $ & $0.00019\pm 0.00126 $ & -1 &   -2.27 & -4.31 \\
\hline
$Q_\CC$  &  $0.686\pm 0.004$&  $0.304\pm 0.005$ & $0.00090\pm 0.00330$ & -1 &   -2.21 & -4.25 \\
\hline
 \end{tabular}
\end{scriptsize}
\end{center}
\caption{\scriptsize Same as in Table 1, but removing the LSS data set from our fitting analysis.}
\label{tableFit4_Possible_signals_chapter}
\end{table}
%
%
Table 1 contains the main fitting results, whereas the other tables display complementary information.
We observe from Fig.\,1 that the vacuum parameters, $\nu$ and $\nu_{dm}$, are neatly projected non null and positive for the RVM and the Q$_{dm}$.  {In the particular case of the RVM,  Fig.\,2 displays in a nutshell our main results in all possible planes of the fitting parameter space.  Fig.\, 4, on the other hand, indicates that the XCDM is also sensitive to the DDE signal. In all cases the LSS data play an important role (cf. Fig.\, 4).  Focusing on the model that provides the best fit, namely the RVM,  Figs.\,5-6 reveal the clue to the main data sources responsible for the final results.} We will further comment on them in the next sections. Remarkably enough, the significance of this dynamical vacuum effect reaches up to about $\sim 3.8\sigma$ c.l. after marginalizing over the remaining parameters.
\subsubsection{Fitting the data with the XCDM and CPL parametrizations}\label{sect:XCDMandCPLnumerical_Possible_signals_chapter}
Here we further elaborate on the results we have found by exploring now the possible time evolution of the DE in terms of the well-known XCDM and CPL parametrizations (introduced in Sec.\,\ref{sect:XCDMandCPL_Possible_signals_chapter}). For the XCDM,  $w=w_0$ is the (constant) equation of state (EoS) parameter for X, whereas for the CPL there is also a dynamical component introduced by $w_1$, see Eq.\eqref{eq:CPL_Possible_signals_chapter}. The corresponding fitting results for the XCDM parametrization is included in all our tables, along with those for the DVMs and the $\CC$CDM. For the main Table 1, we also include the CPL fitting results. For example, reading off Table 1 we can see that the best-fit value for $w_0$ in the XCDM is
\begin{equation}\label{eq:woRVM_Possible_signals_chapter}
w_0=-0.923\pm0.023.
\end{equation}
It is worth noticing that this EoS value is far from being compatible with a rigid $\CC$-term. It actually departs from $-1$ by precisely $3.35\sigma$ c.l. In Fig.\, 3 we depict the contour plot for the XCDM in the $(\Omega_m,w_0)$ plane. Practically all of the $3\sigma$-region lies above the horizontal line at $w_0=-1$. Subsequent marginalization over $\Omega_m$ renders the result \eqref{eq:woRVM_Possible_signals_chapter}. Concerning the CPL, we can see from Table 1 that the errors on the fitting parameters are larger, specially on $w_1$, but it concurs with the XCDM that DE dynamics is also preferred (see also Sec.\,\ref{subsect:AICandBIC_Possible_signals_chapter}).
%
%
\begin{small}
\begin{table}[t!]
\begin{center}
\begin{scriptsize}
\begin{tabular}{| c | c |c | c | c | c | c | c | c | c | c|}
\multicolumn{1}{c}{Model} &  \multicolumn{1}{c}{$h$} &    \multicolumn{1}{c}{$\Omega_m$}&  \multicolumn{1}{c}{{\small$\nu_i$}}  & \multicolumn{1}{c}{$w$}  & \multicolumn{1}{c}{$\Delta{\rm AIC}$} & \multicolumn{1}{c}{$\Delta{\rm BIC}$}\vspace{0.5mm}
\\\hline
{$\CC$CDM} & $0.679\pm 0.005$ &  $0.291\pm 0.005$ & - & -1  &  - & -\\
\hline
XCDM  &  $0.674\pm 0.007$&  $0.298\pm 0.009$& - & $-0.960\pm0.038$   & -1.18 & -3.40 \\
\hline
RVM  & $0.677\pm 0.008$&  $0.296\pm 0.015$ & $0.00061\pm 0.00158$ & -1  & -2.10 & -4.32 \\
\hline
$Q_{dm}$ &  $0.677\pm 0.008$&  $0.296\pm 0.015 $ & $0.00086\pm 0.00228 $ & -1    & -2.10 & -4.32 \\
\hline
$Q_\CC$  &  $0.679\pm 0.005$&  $0.297\pm 0.013$ & $0.00463\pm 0.00922$ & -1   & -1.98 & -4.20 \\
\hline \end{tabular}

\end{scriptsize}
\end{center}
\caption{\scriptsize Same as in Table 1 but removing the CMB data set from our  fitting analysis.}
\label{tableFitextra_Possible_signals_chapter}
\end{table}
\end{small}
%
%
%
\begin{table}[t!]
\begin{center}
\begin{scriptsize}
\begin{tabular}{| c | c |c | c | c | c | c |}
\multicolumn{1}{c}{Model} &  \multicolumn{1}{c}{$h$} &    \multicolumn{1}{c}{$\Omega_m$}&  \multicolumn{1}{c}{{\small$\nu_i$}}  & \multicolumn{1}{c}{$w$} &\multicolumn{1}{c}{$\Delta{\rm AIC}$} & \multicolumn{1}{c}{$\Delta{\rm BIC}$}\vspace{0.5mm}
\\\hline
{$\CC$CDM} & $0.694\pm 0.005$ &  $0.293\pm 0.007$ & - & -1 &   - & -\\
\hline
XCDM  &  $0.684\pm 0.010$&  $0.299\pm 0.009$& - & $-0.961\pm0.033$ &   -1.20 & -2.39 \\
\hline
RVM  & $0.685\pm 0.009$&  $0.297\pm 0.008$ & $0.00080\pm 0.00062$ & -1 &   -0.88 & -2.07 \\
\hline
$Q_{dm}$ &  $0.686\pm 0.008$&  $0.297\pm 0.008 $ & $0.00108\pm 0.00088 $ & -1 &   -1.02 & -2.21\\
\hline
$Q_\CC$  &  $0.694\pm 0.006$&  $0.293\pm 0.007$ & $0.00167\pm 0.00471$ & -1 &  -2.45 & -3.64 \\
\hline
 \end{tabular}
\end{scriptsize}
\end{center}
\caption{\scriptsize As in Table 1, but using the same data set as the Planck Collaboration \cite{Ade:2015rim}. }
\label{tableFit6_Possible_signals_chapter}
\end{table}
%
%
{Remarkably, from the rich string of SNIa+BAO+$H(z)$+LSS+CMB data we find that even the simple XCDM parametrization is able to capture nontrivial signs of dynamical DE in the form of effective quintessence behaviour ($w_0\gtrsim -1$), at more than $3\sigma$ c.l.} Given the significance of this fact, it is convenient to compare it with well-known previous fitting analyses of the XCDM parametrization, such as the ones performed by the Planck and BOSS collaborations 2-3 years ago. The Planck 2015 value for the EoS parameter of the XCDM reads $w_0 = -1.019^{+0.075}_{-0.080}$ \cite{Ade:2015xua} and the BOSS one is $w_0 = -0.97\pm 0.05$ \cite{Aubourg:2014yra}. These results are perfectly compatible with our own fitting value for $w_0$ given in \eqref{eq:woRVM_Possible_signals_chapter}, but in stark contrast to it their errors are big enough as to be also fully compatible with the $\CC$CDM value $w_0=-1$. This is not too surprising if we bear in mind that none of these analyses included large scale structure formation data in their fits, as explicitly recognized in the text of their papers.
\newline
\newline
In the absence of the modern LSS data we would indeed find a very different situation to that in Table 1. As our Table 3 clearly shows, the removal of the LSS data set in our fit induces a significant increase in the magnitude of the central value of the EoS parameter for the XCDM, as well as of the corresponding error. This happens because the higher is $|w|$ the higher is the structure formation power predicted by the XCDM, and therefore the closer is such a prediction with that of the $\CC$CDM (which is seen to predict too much power as compared to the data, see Fig.\,4). Under these conditions our analysis renders $w = -0.992\pm 0.040$ (cf. Table 3), which is manifestly closer to (in fact consistent with) the aforementioned central values (and errors) obtained by Planck and BOSS teams. In addition, this result is now fully compatible with the $\CC$CDM, as in the Planck 2015 and BOSS cases, and all of them are unfavored by the LSS data.
\newline
From the foregoing observations it becomes clear that in order to improve the fit to the observed values of $f(z)\sigma_8(z)$, which generally appear lower-powered with respect to those predicted by the $\CC$CDM (cf. Fig.\,4), $|w|$ should decrease. This is just what happens in or fit for the XCDM, see Eq.\eqref{eq:woRVM_Possible_signals_chapter}.  At the level of the DVMs this translates into positive values of $\nu_i$, as these values cause the vacuum energy to be larger in our past; and, consequently, it introduces a time modulation of the growth suppression of matter. It is apparent from Fig.\,4 that the $f(z)\sigma_8(z)$ curves for the vacuum models are shifted downwards (they have less power than the $\CC$CDM) and hence adapt significantly better to the LSS data points.
%
%
\begin{figure}[t!]
\centering
\includegraphics[angle=0,width=1.03\linewidth]{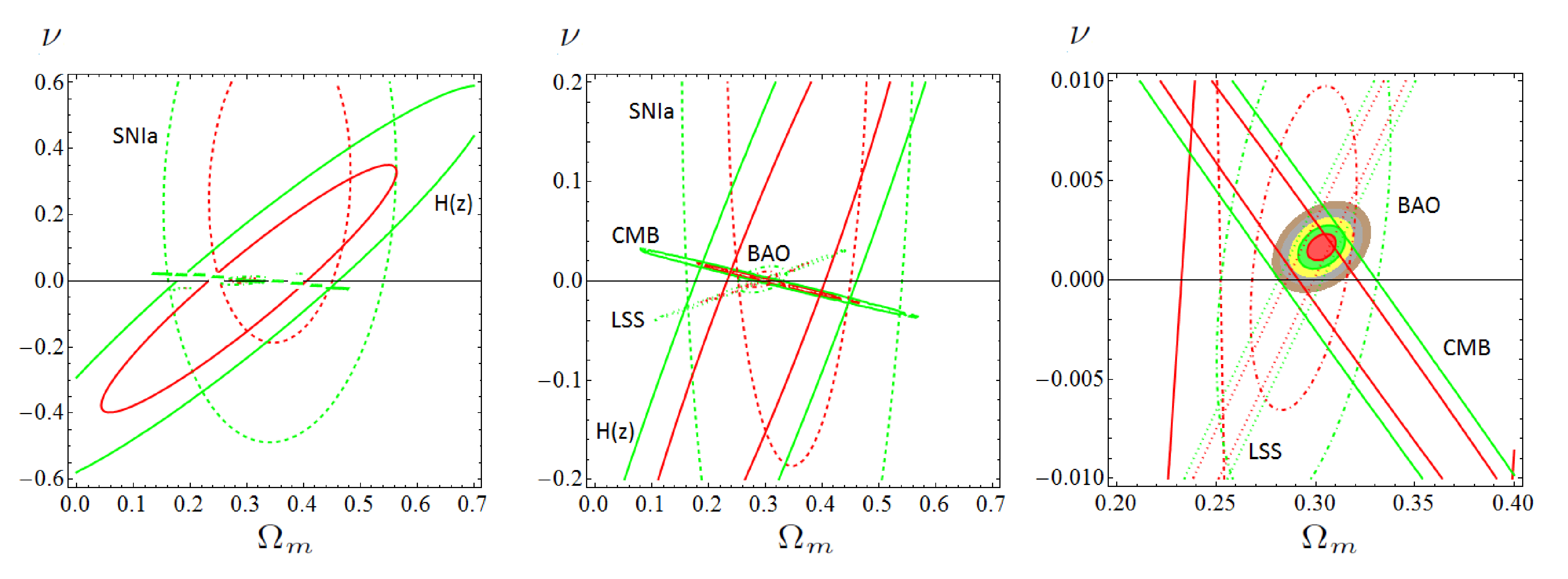}
\caption{\scriptsize Reconstruction of the contour lines for the RVM, from the partial contour plots of the different SNIa+BAO+$H(z)$+LSS+CMB data sources. The $1\sigma$ c.l. and $2\sigma$ c.l. contours are shown in all cases. For the reconstructed final contour lines we also plot the $3\sigma$, $4\sigma$ and $5\sigma$ confidence level regions.
}
\end{figure}


\subsubsection{Comparing the competing vacuum models through Akaike and Bayesian information criteria}\label{subsect:AICandBIC_Possible_signals_chapter}
We may judge the fit quality obtained for the different vacuum models in this work from a different perspective. Although the $\chi^2_{\rm min}$ value of the overall fits for the main DVMs (RVM and $Q_{dm}$) and XCDM appear to be definitely smaller than the $\CC$CDM one, it proves extremely useful to reassess the degree of success of each competing model by invoking the time-honored Akaike and Bayesian information criteria, denoted as AIC and BIC \cite{Akaike,Schwarz1978,KassRaftery1995}.
The Akaike information criterion is defined as follows:
\begin{equation}\label{eq:AIC_Possible_signals_chapter}
{\rm AIC}=\chi^2_{\rm min}+\frac{2nN}{N-n-1}\,,
\end{equation}
whereas the Bayesian information criterion reads
\begin{equation}\label{eq:BIC_Possible_signals_chapter}
{\rm BIC}=\chi^2_{\rm min}+n\,\ln N\,.
\end{equation}
In these formulas, $n$ is the number of independent fitting parameters and $N$ the number of data points. The added terms on $\chi^2_{\rm min}$ represent the penalty assigned by these information criteria to the models owing to the presence of additional parameters.
To test the degree of success of a dynamical DE model  (versus the $\CC$CDM) with the information criteria, we have to evaluate the pairwise differences $\Delta$AIC ($\Delta$BIC) between the AIC and BIC values of the $\CC$CDM with respect to the corresponding values of the models having a smaller value of these criteria -- in our case the DVMs, XCDM and CPL. The larger these (positive) differences are the higher is the evidence against the model with larger value of  AIC (BIC) -- i.e. the $\CC$CDM in the present case.
\newline
\newline
According to the standard usage, for $\Delta$AIC and/or $\Delta$BIC below 2 one judges that there is ``consistency'' between the two models under comparison; in the range $2-6$ there exists a ``positive evidence'' in favor of the model with smaller value of AIC and/or BIC; for values within $6-10$ one may claim ``strong evidence'' in favor of such a model; finally, above 10, one speaks of ``very strong evidence''. The evidence ratio associated to acceptance of the favored model and rejection of the unfavored model is given by the ratio of Akaike weights, $A\equiv e^{\Delta{\rm AIC}/2}$. Similarly, $B\equiv e^{\Delta{\rm BIC}/2}$ estimates the so-called Bayes factor, which gives the ratio of marginal likelihoods between the two models \cite{Amendola:2015ksp}.
{Table 1 reveals conspicuously that the $\CC$CDM appears disfavored when confronted to  the DDE models. The most favored one is the RVM, followed by the $Q_{dm}$ and next by the XCDM.  In the case of the CPL and  Q$_{\CC}$  the improvement is only mild.}
\newline
The AIC and BIC criteria can be thought of as a modern quantitative formulation of Occam's razor, in which the presence of extra parameters in a given model is conveniently penalized so as to achieve a fairer comparison with the model having less parameters.
\subsection{Discussion}\label{sect:discussion_Possible_signals_chapter}
In this section we consider in more detail some important aspects and applications of our analysis. In particular we identify which are the most important data sources which are responsible for the possible DDE signal and show that in the absence of any of these important ingredients the signal becomes weakened or completely inaccessible.
\newline
\subsubsection{Testing the impact of the different data sets in our analysis and comparing with Planck 2015}\label{sect:OtherDataSets_Possible_signals_chapter}
The current work follows the track of \cite{Sola:2015wwa} and is also firmly aligned with \cite{Sola:2016jky,Sola:2017lxc}. Although the models analyzed in \cite{Sola:2015wwa,Sola:2016jky} have some differences with respect to the ones treated here, the outcome of the analysis points to the very same direction, to wit: some DVMs  and the XCDM fit better the available data than the $\Lambda$CDM. But we want to emphasize some important aspects of the analysis carried out in this paper as compared to other analyses:
\begin{itemize}
\item We have used a large and fully updated set of cosmological  SNIa+BAO+$H(z)$+LSS+CMB observations. To our knowledge, this is one of the most complete and consistent data sets used in the literature, see \cite{Sola:2016jky} up to some updating introduced here, specially concerning the LSS data.
\item We have removed all data that would entail double counting and used the known covariance matrices in the literature. As an example, we have avoided to use Hubble parameter data extracted from BAO measurements, {and restricted only to those based on the differential age (i.e. the cosmic chronometers).}
\item We have duly taken into account all the known covariance matrices in the total $\chi^2$-function \eqref{chi2s_Possible_signals_chapter}, which means that we have accounted for all the known correlations among the data.  Not all data sets existing in the literature are fully consistent, sometimes they are affected from important correlations that have not been evaluated. We have discussed the consistency of the present data in \cite{Sola:2016jky}.
\end{itemize}
{We have conducted several practical tests in order to study the influence of different data sets in our fitting analysis. As previously mentioned, we have checked what is the impact on our results if we omit the use of the LSS data (cf. Table 3), but in our study we have also assessed what happens  if we disregard the  CMB data (cf. Table 4) while still keeping all the remaining observations. The purpose of this test is to illustrate once more the inherent $\sigma_8$-tension existing between the geometry data and the structure formation data.   In both cases, namely when we dispense with the LSS or the CMB data, we find that  for all the models under study the error bars for the fitted DDE parameters ($w_i, \nu_i$) become critically larger  (sometimes they increase a factor 2-4)  than those displayed in Table 1,  and as a consequence  these parameters become fully compatible with the $\CC$CDM values (in particular $\nu_i=0$ for the DVMs) within  $1\sigma$ c.l. or less, which is tantamount to saying that the DDE effect is washed out. At the same time, and in full accordance with the mentioned results, the $\Delta$AIC and $\Delta$BIC information criteria become negative, which means (according to our definition in Sec.\,\ref{subsect:AICandBIC_Possible_signals_chapter})  that none of these DDE models fits better the data than the $\CC$CDM under these particular conditions.  These facts provide incontestable evidence of the strong constraining power of the LSS as well as of the CMB data, whether taken individually or in combination, and of their capability for narrowing down the allowed region in the parameter space. In the absence of either one of them, the $\Lambda$CDM model is preferred  over the DDE models, but only at the expense of ignoring the CMB input, or the LSS data, both of which are of course of utmost importance.  Thus, the concordance model is now able to fit the LSS data better only because it became free from the tight CMB constraint on $\Omega_m$, which enforced the latter to acquire a larger value. Without such constraint, a lower $\Omega_m$ value can be chosen by the fitting procedure,  what in turn enhances the agreement with the $f(z)\sigma_8(z)$ data points.  We have indeed checked that the reduction of $\Omega_m$ in the $\CC$CDM  directly translates into an $8.6\%$ lowering of $\sigma_8(0)$ with respect to the value shown in Table 1 for this model, namely we find that  $\sigma_8(0)$
 changes from  $0.801\pm 0.009$ (as indicated in Table 1) to  $ 0.731\pm 0.019 $  when the CMB data are not used. Such substantial decrease tends to optimize the fit of the LSS data, but only at the expense of ruining the fit to the CMB when these data are restored. This is, of course, the very meaning of the $\sigma_8$-tension, which cannot be overcome at the moment within the $\CC$CDM. }
\newline
{In stark contrast with the situation in the  $\CC$CDM,  when the vacuum is allowed to acquire a mild dynamical component the $\sigma_8$-tension can be dramatically  loosened, see \cite{Gomez-Valent:2017idt,Gomez-Valent:2017tkh} for a detailed explanation. This can be seen immediately on comparing the current auxiliary tables 3 and 4 with the main Table 1. Recall that a positive vacuum energy suppresses the growth of structure formation and this is one of the reasons why the $\CC$CDM model is highly preferred to the CDM with $\CC=0$. Similarly, but at a finer and subtler  level of precision, a time modulation of the growth suppression through dynamical vacuum energy or in general DDE should further help in improving the adjustment of the LSS data.  In our case this is accomplished e.g. by the $\nu$-parameter of the RVM, which enables a dynamical modulation of the growth suppression through the $\sim \nu H^2$ component of the vacuum energy density -- cf. Eq.\eqref{eq:RVMvacuumdadensity_Possible_signals_chapter}. The presence of this extra degree of freedom allows the DVMs to better adjust the LSS data without perturbing the requirements from the CMB data (which can therefore preserve the standard $\Omega_m$ value obtained by  Planck 2015). The fact that this  readjustment of the LSS data by a dynamical component in the vacuum energy is possible is because the epoch of structure formation is very close to the epoch when the DE starts to dominate, which is far away from the epoch when the CMB was released,  and hence any new feature of the DE can play a significant role in the LSS formation epoch without disrupting the main features of the  CMB.  Let us recall at this point that the presence of the extra parameter from the DDE models under discussion is conveniently penalized by the Akaike and Bayesian information criteria in our analysis, and thus the DDE models appear to produce a better fit than the $\CC$CDM under perfectly fair conditions of statistical comparison between competing models describing the same data.}
\newline
{The conclusion of our analysis is clear: no signal of DDE can be found without the inclusion of the CMB data and/or the LSS data, even keeping the rest of observable within the fit. Both the LSS and CMB data are crucial ingredients to enable capturing the DDE effect, and the presence of BAO data just enhances it further.  This conclusion is  additionally confirmed by our study of the deconstruction and reconstruction of the RVM contour plots  in Figs. 5-6 and is discussed at length  in the next section. }

\begin{figure}[t!]
\centering
\includegraphics[angle=0,width=0.6\linewidth]{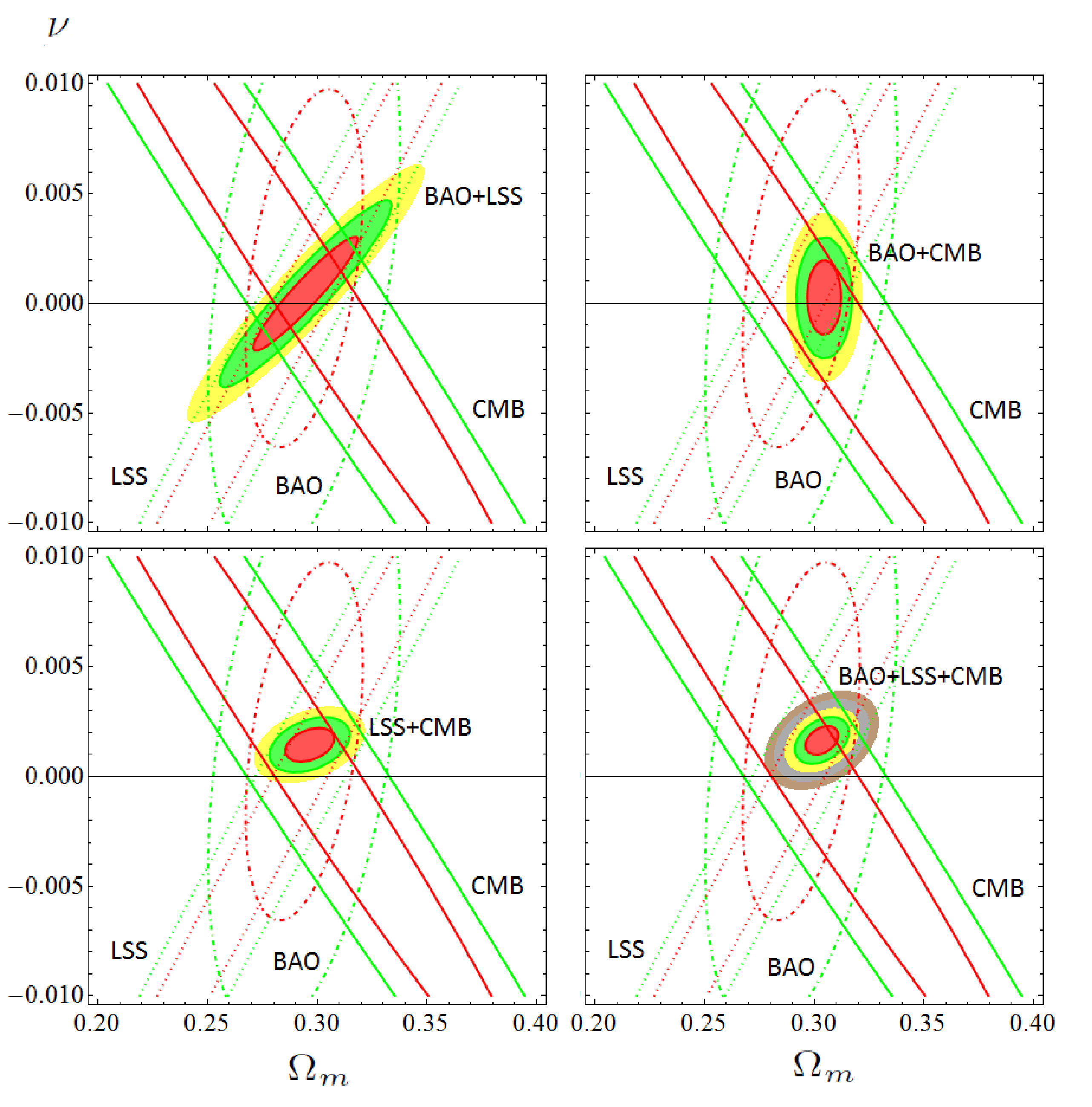}
\caption{\label{RecRVMtriad_Possible_signals_chapter}%
\scriptsize As in Fig.\,5, but considering the effect of only the BAO, LSS and CMB in all the possible combinations: BAO+LSS, BAO+CMB, LSS+CMB and BAO+LSS+CMB. As discussed in the text, it is only when such a triad of observable is combined that we can see a clear $\lesssim 4\sigma$ c.l. effect, which is comparable to intersecting the whole set of SNIa+BAO+$H(z)$+LSS+CMB data.}
\end{figure}
%
%
We close this section by answering a most natural question. Why the dynamical DE signal that we are glimpsing here escaped undetected from the fitting analyses of Planck 2015 ?  The answer can be obtained by repeating our fitting procedure and restricting ourselves to the much more limited data sets used by the Planck 2015 collaboration, more precisely in the papers \cite{Ade:2015xua,Ade:2015rim}. In contrast to \cite{Ade:2015xua}, where no LSS (RSD) data were used, in the case of \cite{Ade:2015rim} they used only some BAO and LSS data, but their fit is rather limited in scope. Specifically, they used only 4 BAO data points, 1 AP (Alcock-Paczynski parameter) data point, and one single LSS point, namely the value of $f(z)\sigma_8(z)$ at $z=0.57$-- see details in that paper. Using this same data we obtain the fitting results presented in our Table 5. They are perfectly compatible with the fitting results mentioned in Sec.\,\ref{sect:XCDMandCPLnumerical_Possible_signals_chapter} obtained by Planck 2015 and BOSS \cite{Aubourg:2014yra}, i.e. none of them carries evidence of dynamical DE, with only the data used by these collaborations two-three years ago.
\newline
In contradistinction to them, in our full analysis presented in Table 1  we used 11 BAO and 13 LSS data points, some of them available only from the recent literature and of high precision \cite{Gil-Marin:2016wya}. From Table 5 it is apparent that with only the data used in \cite{Ade:2015rim} the fitting results for the RVM are poor enough and cannot still detect clear traces of the vacuum dynamics. In fact, the vacuum parameters are compatible with zero at $1\sigma$ c.l. and the values of $\Delta$AIC and $\Delta$BIC in that table are moderately negative, showing that the DVMs do not fit better the data than the $\Lambda$CDM model with only such a limited input.  In fact, not even the XCDM parametrization is capable of detecting any trace of dynamical DE with that limited data set, as the effective EoS parameter is compatible with $w_0=-1$ at roughly $1\sigma$ c.l. ($w_0=-0.961\pm 0.033$).
\newline
The features that we are reporting here have remained hitherto unnoticed in the literature, except in \cite{Sola:2015wwa,Sola:2016jky,Sola:2016hnq,Sola:2017lxc,Sola:2017znb}, and in \cite{Zhao:2017cud}. In the last reference the authors have been able to find a significant $3.5\sigma$ c.l. effect on dynamical DE, presumably in a model-independent way and following a nonparametric procedure, see also \cite{Wang:2015wga}. The result of \cite{Zhao:2017cud} is well along the lines of the present work.
%
\begin{figure}[t!]
\centering
\includegraphics[angle=0,width=0.75\linewidth]{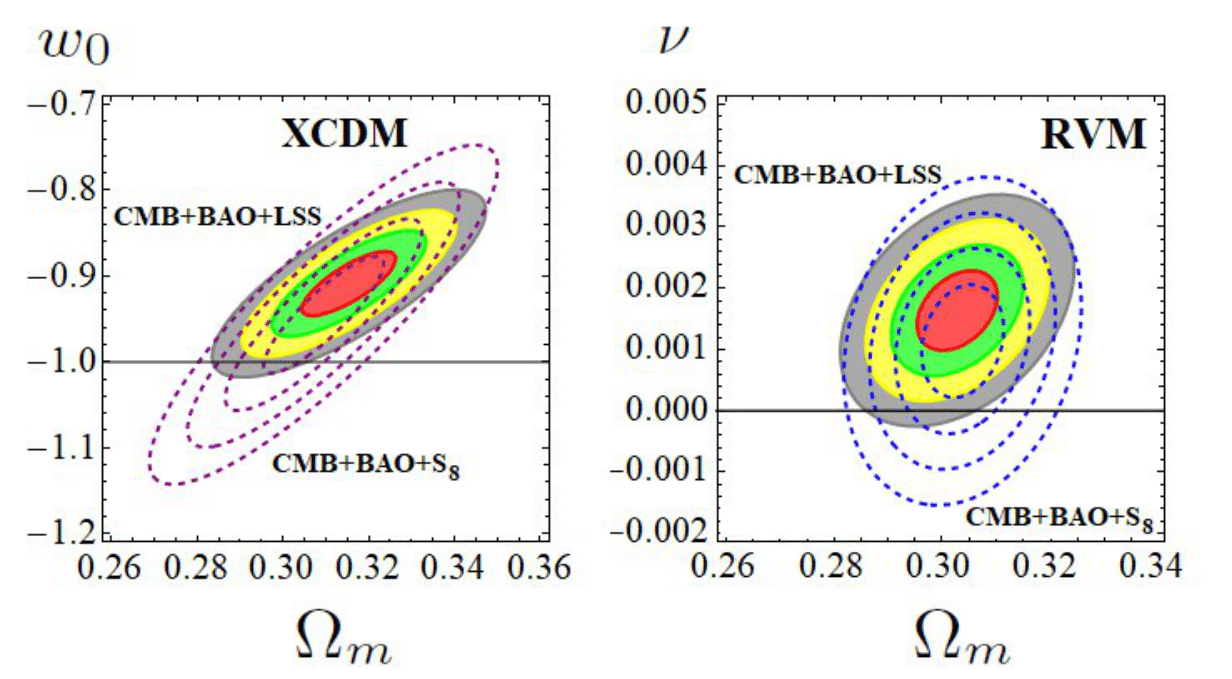}
\caption{\label{RecRVMtriad}%
\scriptsize Contour lines for the XCDM (left) and RVM (right) using the same CMB+BAO+LSS data as
in Table 1 (solid contours); and also when replacing the LSS data (i.e. the $f(z)\sigma_8(z)$ points) with the $S_8$ value obtained from the weak gravitational lensing data (Joudaki et al. 2018) (dashed lines).}
\end{figure}
%
%
\subsubsection{Deconstruction and reconstruction of the RVM contour plots}\label{subsect:deconstruction_Possible_signals_chapter}
We further complement our analysis by displaying in a graphical way the deconstructed contributions from the different data sets to our final contour plots in Fig.\,1, for the specific case of the RVM. One can do similarly for any of the models under consideration. The result is depicted in Fig.\,5, where we can assess the detailed deconstruction of the final contours in  terms of the partial contours from the different SNIa+BAO+$H(z)$+LSS+CMB data sources.
\newline
The deconstruction plot for the RVM case is dealt with in Fig.\,5, through a series of three plots made at different magnifications. In the third plot of the sequence we can immediately appraise that the BAO+LSS+CMB data subset plays a fundamental role in narrowing down the final physical region of the $(\Omega_m,\nu)$ parameter space, in which all the remaining parameters have been marginalized over. This deconstruction process also explains in very transparent visual terms why the conclusions that we are presenting here hinge to a large extent on some particularly sensitive components of the data. While there is no doubt that the CMB is a high precision component in the fit, our study demonstrates (both numerically and graphically) that the maximum power of the fit is achieved when it is combined with the wealth of BAO and LSS data points currently available.
\newline
To gauge the importance of the BAO+LSS+CMB combination more deeply, in Fig.\,6 we try to reconstruct the final RVM plot in Fig.\,1 (left) from only these three data sources. First we consider the overlapping regions obtained when we cross the pairs of data sources BAO+LSS, BAO+CMB, LSS+CMB and finally the trio BAO+LSS+CMB (in all cases excluding the SNIa and $H(z)$ data). One can see that neither the BAO+LSS nor the BAO+CMB crossings yield a definite sign for $\nu$. {This is consistent with the numerical results in Tables 3 and 4, where the removal of the LSS and the CMB data, respectively, renders rather poor fits with negative values of $\Delta$AIC and $\Delta$BIC.}
\newline
Remarkably, it is the LSS+CMB combination the one that carries a well-defined, positive, sign for $\nu$, as it is seen from the lower-left plot in Fig.\,6, where $\Delta$AIC and $\Delta$BIC are now both positive and above $6$ for the main DVMs (RVM and $Q_{dm}$), as we have checked. Finally, when we next intersect the pair LSS+CMB with the BAO data the signal peaks at  $3.8\sigma$ c.l., the final contours being now those shown in the lower-right plot of Fig.\,6.
The outcome of this exercise is clear. For the RVM case, we have checked that the final BAO+LSS+CMB plot in Fig.\,6 is essentially the same as the original RVM plot in Fig.\,1 (the leftmost one). In other words, the final RVM contour plot containing the information from all our five data sources can essentially be reconstructed with only the triad of leading observable BAO+LSS+CMB.
\subsubsection{Vacuum dynamics, structure formation and weak lensing data}\label{subsect:weaklensing_Possible_signals_chapter}
{Owing to the significant role played by the structure formation data in the extraction of the possible DDE signal we next  inquire into its impact when we use a different proxy to describe such data. Let us note that an account of the LSS observations does not only concern the $f(z)\sigma_8(z)$ data, but also the weak gravitational lensing constraints existing in the literature on the conventional quantity  $S_8\equiv \sigma_8(\Omega_m/0.3)^{0.5}$ \cite{Joudaki:2017zdt,Hildebrandt:2016iqg,Heymans:2013fya}. In Fig.\,7 we compare the respective results that we find for the XCDM (left) and the RVM (right)  when we use either the CMB+BAO+$f\sigma_8$ or the CMB+BAO+$S_8$  data sources.  For definiteness we use the recent study by \cite{Joudaki:2017zdt}, in which they carry
a combined analysis of cosmic shear tomography, galaxy- galaxy lensing tomography, and redshift-space multipole
power spectra using imaging data by the Kilo Degree Survey (KiDS-450) overlapping with the 2-degree Field Lensing Survey (2dFLenS) and the Baryon Oscillation Spectroscopic Survey (BOSS). They find $S_8 = 0.742 \pm 0.035$. Incidentally, this value is $2.6\sigma$ below the one provided by Planck's TT+lowP analysis [4]. Very similar results can be obtained using the weak gravitational lensing tomography data by KiDS-450 collaboration, $S_8=0.745\pm 0.039$ \cite{Hildebrandt:2016iqg}, and also by CFHTLenS,  $(\Omega_m/0.27)^{0.46}=0.770\pm 0.040$ \cite{Heymans:2013fya}.
In contrast, the result  $S_8=0.783^{+0.021}_{-0.025}$ provided by DES \cite{Abbott:2017wau} is more resonant with Planck, but due to its large uncertainty it is still fully compatible with \cite{Joudaki:2017zdt,Hildebrandt:2016iqg,Heymans:2013fya}.  From Fig. 7 we confirm (using both the XCDM and the RVM) that the contour lines computed from the data string CMB+BAO+$f\sigma_8$  are mostly contained within the contour lines from the alternative string CMB+BAO+$S_8$ and are shifted upwards. The former data set is therefore more precise and capable of resolving the DDE signal at a level of more than $3\sigma$, whereas with $S_8$ it barely surpasses
the $1\sigma$  c.l. within the RVM and even less with the XCDM, thus rendering essentially no DDE signal. The outcome of this additional test is that the use of the weak lensing data from $S_8$ as a replacement for the direct LSS measurements ($f\sigma_8$) is insufficient since it definitely weakens the evidence in favor of DDE.}
\subsection{Conclusions}\label{sect:conclusions_Possible_signals_chapter}
To conclude, in this work we aimed at testing cosmological physics beyond the standard or concordance $\CC$CDM model, which is built upon a rigid cosmological constant. We have presented a comprehensive study on the possibility that the global cosmological observations can be better described in terms of vacuum models equipped with a dynamical component that evolves with the cosmic expansion. This should be considered a natural possibility in the context of quantum field theory (QFT) in a curved background. Our task focused on three dynamical vacuum models (DVMs): the running vacuum model (RVM) along with two more phenomenological models, denoted $Q_{dm}$ and $Q_\CC$-- see Sec.\,\ref{sect:DVMs_Possible_signals_chapter}.
\newline
At the same time, we have compared the performance of these models with the general XCDM and CPL parametrizations. We have fitted all these models and parametrizations to the same set of cosmological data based on the observable SNIa+BAO+$H(z)$+LSS+CMB.
The remarkable outcome of this investigation is that in all the considered cases we find an improvement of the description of the cosmological data in comparison to the $\CC$CDM.
\newline
The ``deconstruction analysis'' of the contour plots in Sec.\,\ref{subsect:deconstruction_Possible_signals_chapter} has revealed which are the most decisive data ingredients responsible for the dynamical vacuum signal. We have identified that the BAO+LSS+CMB components play a momentous role in the overall fit, as they are responsible for the main effects uncovered here. The impact of the SNIa and $H(z)$ observable appears to be more moderate.  While the SNIa data were of course essential for the detection of a nonvanishing value of $\CC$, these data do not seem to have sufficient sensitivity (at present) for the next-to-leading step, which is to unveil the possible dynamics of $\CC$. The sensitivity for that seems to be reserved for the LSS, BAO and CMB data.
\newline
We have also found that the possible signs of DDE tend to favor an effective quintessence behaviour, in which the energy density decreases with the expansion. Whether or not the ultimate reason  for such a signal stems from the properties of the quantum vacuum or from some particular quintessence model, it is difficult to say at this point.
Quantitatively, the best fit is granted in terms of the RVM. The results are consistent with the traces of  DDE that can also be hinted at with the help of the XCDM and CPL  parametrizations.
\newline
In our work we have also clarified why previous fitting analyses based e.g. on the simple XCDM parametrization, such as the ones by the Planck 2015 \cite{Ade:2015xua,Ade:2015rim} and BOSS collaborations \cite{Aubourg:2014yra}, missed the DDE signature. Basically, the reason stems from not using a sufficiently rich sample of the most crucial data, namely BAO and LSS, some of which were unavailable a few years ago, and could not be subsequently combined with the CMB data.
\newline
More recently, signs of DDE at $\sim 3.5\sigma$ c.l. have been reported from non-parametric studies of the observational data on the DE, which aim at a model-independent result \cite{Zhao:2017cud}. The findings of their analysis are compatible with the ones we have reported here.
Needless to say, statistical evidence conventionally starts at $5\sigma$ c.l. and we will have to wait for updated observations to see if such a level of significance can be achieved in the future.

\newpage


\section{The $H_0$ tension in light of vacuum dynamics in the Universe}\label{H0_tension_chapter}
In this chapter we gauge how the Dynamical Vacuum Models (DVMs) deal with the $H_0$-tension, which comes from the huge discrepancy between the local measurement obtained by the SH0ES team \cite{Riess:2016jrr} $H_0 = 73.24\pm 1.74$ km/s/Mpc, denoted as $H^{\rm Riess}_0$ hereafter, and the one extracted from the Planck 2015 TT,TE,EE+lowP+lensing data \cite{Ade:2015xua}. 
\newline
It is important to remark that, as we pointed out in the introductory chapter, $H_0$ is the first ever parameter of modern cosmology. Having said that, we deem that a few lines about the comings and goings in regards to this parameters are deserved. 
\newline
The tension among the different measurements is inherit to its long and tortuous history. After Baade's revision (by a factor of one half \cite{Baade:1944zz}) of the exceedingly large value $\sim 500$ km/s/Mpc originally estimated by Hubble (which implied a Universe of barely two billion years only), the Hubble parameter was subsequently lowered to $75$ km/s/Mpc and finally to $H_0 = 55\pm 5$ km/s/Mpc where it remained for 20 years (until 1995) mainly under the influence of Sandage's devoted observations \cite{Sandage:1961zz}. Shortly after that period the first measurements of a non-vanishing positive, value of $\Lambda$ appeared \cite{Riess:1998cb,Perlmutter:1998np} and the typical range for $H_0$ moved upwards to $\sim 65$ km/s/Mpc. In the meantime, many different observational values of $H_0$ have piled up in the literature using different methods (see e.g. the median statistical analysis of $>500$ measurements considered in \cite{Chen:2011ab,Bethapudi:2017swc}).
\newline
As we stated before, there is still a dispute on the precise value of $H_0$, as it is clear from the results obtained by the Planck team, thanks to the observation of the CMB anisotropies and the value provided by the SH0ES team, which is based on the cosmic distance ladder method. We reexamine this tension, but not as an isolated conflict between these two particular sources of observations but rather in the light of the overall fit to the cosmological data SNIa+BAO+$H(z)$+LSS+CMB+$H_0$. This fact turns out to be crucial to properly study the $\sigma_8$-tension. 
\newline
This chapter is organized as follows: In Sec. \ref{sec:DVM_and_beyond_H0_tension_chatper} we provide the most relevant background formulas for the DVMs and also for what we have called the quasi-vacuum models ($w$DVM's) (see the next section for the details) under study, whereas in Sec. \ref{Sec:SF_H0_tension_chapter}, we do the same, but this time with the perturbation equations. In Sec. \ref{Sec:Discussion_H0_tension_chapter} we discuss and analyze the results obtained, which can be found in the corresponding tables and figures. Finally in Sec. \ref{Sec:Conclusions_H0_tension_chapter} we deliver the conclusions.  


\begin{table}[t!]
\setcounter{table}{0}
\begin{center}
\begin{scriptsize}
\resizebox{1\textwidth}{!}{
\begin{tabular}{ |c|c|c|c|c|c|c|c|c|c|}
\multicolumn{1}{c}{Model} &  \multicolumn{1}{c}{$H_0$(km/s/Mpc)} &  \multicolumn{1}{c}{$\omega_b$} & \multicolumn{1}{c}{{\small$n_s$}}  &  \multicolumn{1}{c}{$\Omega_m^0$} &\multicolumn{1}{c}{$\nu_i$} &\multicolumn{1}{c}{$w$} &\multicolumn{1}{c}{$\chi^2_{\rm min}/dof$} & \multicolumn{1}{c}{$\Delta{\rm AIC}$} & \multicolumn{1}{c}{$\Delta{\rm BIC}$}\vspace{0.5mm}
\\\hline
$\Lambda$CDM  & $68.83\pm 0.34$ & $0.02243\pm 0.00013$ &$0.973\pm 0.004$& $0.298\pm 0.004$ & - & -1  & 84.40/85 & - & - \\
\hline
XCDM  & $67.16\pm 0.67$& $0.02251\pm0.00013 $&$0.975\pm0.004$& $0.311\pm0.006$ & - &$-0.936\pm{0.023}$  & 76.80/84 & 5.35 & 3.11 \\
\hline
RVM  & $67.45\pm 0.48$& $0.02224\pm0.00014 $&$0.964\pm0.004$& $0.304\pm0.005$ &$0.00158\pm 0.00041 $ & -1  & 68.67/84 & 13.48 & 11.24 \\
\hline
$Q_{dm}$  & $67.53\pm 0.47$& $0.02222\pm0.00014 $&$0.964\pm0.004$& $0.304\pm0.005$ &$0.00218\pm 0.00058 $&-1  & 69.13/84 & 13.02 &10.78 \\
\hline
$Q_\Lambda$  & $68.84\pm 0.34$& $0.02220\pm0.00015 $&$0.964\pm0.005$& $0.299\pm0.004$ &$0.00673\pm 0.00236 $& -1  &  76.30/84 & 5.85 & 3.61\\
\hline
$w$RVM  & $67.08\pm 0.69$& $0.02228\pm0.00016 $&$0.966\pm0.005$& $0.307\pm0.007$ &$0.00140\pm 0.00048 $ & $-0.979\pm0.028$ & 68.15/83 & 11.70 & 7.27 \\
\hline
$w{Q_{dm}}$  & $67.04\pm 0.69$& $0.02228\pm0.00016 $&$0.966\pm0.005$& $0.308\pm0.007$ &$0.00189\pm 0.00066 $& $-0.973\pm 0.027$ & 68.22/83 & 11.63 & 7.20\\
\hline
$w{Q_\Lambda}$  & $67.11\pm 0.68$& $0.02227\pm0.00016 $&$0.965\pm0.005$& $0.313\pm0.006$ &$0.00708\pm 0.00241 $& $-0.933\pm0.022$ &   68.24/83 & 11.61 & 7.18\\
\hline
\end{tabular}}
\caption{{\scriptsize Best-fit values for the $\CC$CDM, XCDM, the three dynamical vacuum models (DVMs) and the three dynamical quasi-vacuum models ($w$DVMs), including their statistical significance ($\chi^2$-test and Akaike and Bayesian information criteria AIC and BIC).
For detailed description of the data and a full list of references, see \cite{Sola:2016jky} and \cite{Sola:2017jbl}. The quoted number of degrees of freedom ($dof$) is equal to the number of data points minus the number of independent fitting parameters ($4$ for the $\CC$CDM, 5 for the XCDM and the DVMs, and 6 for the $w$DVMs). For the CMB data we have used the marginalized mean values and {covariance matrix} for the parameters of the compressed likelihood for Planck 2015 TT,TE,EE + lowP+ lensing data from\,\cite{Wang:2015tua}. Each best-fit value and the associated uncertainties have been obtained by marginalizing over the remaining parameters.}}
\end{scriptsize}
\end{center}
\label{tableFit1_H0_tension_chapter}
\end{table}


%
\subsection{Dynamical vacuum models and beyond}\label{sec:DVM_and_beyond_H0_tension_chatper}
Let us consider a generic cosmological framework described by the spatially flat FLRW metric, in which matter is exchanging energy with a dynamical DE medium with a phenomenological equation of state (EoS) $p_{\CC}=w\rho_{\CC}$, where $w=-1+\epsilon$ (with $|\epsilon|\ll1$). Such medium is therefore of quasi-vacuum type, and for $w=-1$ (i.e. $\epsilon=0$) we precisely recover the genuine vacuum case. Owing, however, to the exchange of energy with matter, $\rL=\rL(\zeta)$ is in all cases a {\it dynamical} function that depends on a cosmic variable $\zeta=\zeta(t)$.  We will identify the nature of $\zeta(t)$ later on, but its presence clearly indicates that $\rL$ is no longer associated to a strictly rigid cosmological constant as in the $\CC$CDM. The Friedmann and acceleration equations read, however, formally identical to the standard case:
\begin{eqnarray}
&&3H^2=8\pi\,G\,(\rho_m+\rho_r+\rho_\Lambda(\zeta))\label{eq:FriedmannEq_H0_tension_chatper}\\
&&3H^2+2\dot{H}=-8\pi\,G\,(p_r + p_\Lambda(\zeta))\,.\label{eq:PressureEq_H0_tension_chatper}
\end{eqnarray}
Here $H=\dot{a}/a$ is the Hubble function, $a(t)$ the scale factor as a function of the cosmic time, $\rho_r$ is the energy density of the radiation component (with pressure $p_r=\rho_r/3$), and $\rho_m=\rho_b+\rho_{dm}$ involves the contributions from baryons and cold DM. The local conservation law associated to the above equations reads:
\begin{equation}\label{eq:GeneralCL_H0_tension_chatper}
\dot{\rho}_r + 4H\rho_r + \dot{\rho}_m + 3H\rho_m = Q\,,
\end{equation}
where
\begin{equation}\label{eq:Source_H0_tension_chatper}
Q = -\dot{\rho}_\CC - 3H(1+w)\rho_\CC\,.
\end{equation}
For $w=-1$ the last equation boils down to just $Q= -\dot{\rho}_\CC$, which is nonvanishing on account of $\rL(t)=\rL(\zeta(t))$.
\begin{table}[t!]
\begin{center}
\begin{scriptsize}
\resizebox{1\textwidth}{!}{
\begin{tabular}{ |c|c|c|c|c|c|c|c|c|c|}
\multicolumn{1}{c}{Model} &  \multicolumn{1}{c}{$H_0$(km/s/Mpc)} &  \multicolumn{1}{c}{$\omega_b$} & \multicolumn{1}{c}{{\small$n_s$}}  &  \multicolumn{1}{c}{$\Omega_m^0$} &\multicolumn{1}{c}{$\nu_i$} &\multicolumn{1}{c}{$w$} &\multicolumn{1}{c}{$\chi^2_{\rm min}/dof$} & \multicolumn{1}{c}{$\Delta{\rm AIC}$} & \multicolumn{1}{c}{$\Delta{\rm BIC}$}\vspace{0.5mm}
\\\hline
$\Lambda$CDM  & $68.99\pm 0.33$ & $0.02247\pm 0.00013$ &$0.974\pm 0.003$& $0.296\pm 0.004$ & - & -1  & 90.59/86 & - & - \\
\hline
XCDM  & $67.98\pm 0.64$& $0.02252\pm0.00013 $&$0.975\pm0.004$& $0.304\pm0.006$ & - &$-0.960\pm{0.023}$  & 87.38/85 & 0.97 & -1.29 \\
\hline
RVM  & $67.86\pm 0.47$& $0.02232\pm0.00014 $&$0.967\pm0.004$& $0.300\pm0.004$ &$0.00133\pm 0.00040 $ & -1  & 78.96/85 & 9.39 & 7.13 \\
\hline
$Q_{dm}$  & $67.92\pm 0.46$& $0.02230\pm0.00014 $&$0.966\pm0.004$& $0.300\pm0.004$ &$0.00185\pm 0.00057 $&-1  & 79.17/85 & 9.18 & 6.92 \\
\hline
$Q_\Lambda$  & $69.00\pm 0.34$& ${0.02224}\pm0.00016 $&$0.965\pm0.005$& $0.297\pm0.004$ &$0.00669\pm 0.00234 $& -1  &  82.48/85 & 5.87 & 3.61\\
\hline
$w$RVM  & $67.95\pm 0.66$& $0.02230\pm0.00015 $&$0.966\pm0.005$& $0.300\pm0.006$ &$0.00138\pm 0.00048 $ & $-1.005\pm0.028$ & 78.93/84 & 7.11 & 2.66 \\
\hline
$w{Q_{dm}}$  & $67.90\pm 0.66$& $0.02230\pm0.00016 $&$0.966\pm0.005$& $0.300\pm0.006$ &$0.00184\pm 0.00066 $& $-0.999\pm 0.028$ & 79.17/84 & 6.88 & 2.42\\
\hline
$w{Q_\Lambda}$  & $67.94\pm 0.65$& $0.02227\pm0.00016 $&$0.966\pm0.005$& $0.306\pm0.006$ &$0.00689\pm 0.00237 $& $-0.958\pm0.022$ &   78.98/84 & 7.07 & 2.61\\
\hline
\end{tabular}}

\caption{{\scriptsize The same as Table 1 but adding the $H_0^{\rm Riess}$ local measurement from Riess {\it et al.}\, \cite{Riess:2016jrr}}}
\end{scriptsize}
\end{center}
\label{tableFit2_H0_tension_chapter}
\end{table}
The simplest case is, of course, that of the concordance model, in which $\rL=\rLo=$const and $w=-1$, so that $Q=0$ trivially. However, for $w\neq -1$ we can also have $Q=0$ in a nontrivial situation, which follows from solving Eq.\,\eqref{eq:Source_H0_tension_chatper}. It corresponds to the XCDM parametrization\,\cite{Turner:1998ex}, in which the DE density is dynamical and self-conserved. It is easily found in terms of the scale factor:
\begin{equation}\label{eq:rhoXCDM_H0_tension_chapter}
\rL^{\rm XCDM}(a)=\rLo\,a^{-3(1+w)}=\rLo\,a^{-3\epsilon}\,,
\end{equation}
where $\rLo$ is the current value.
From \eqref{eq:GeneralCL_H0_tension_chatper} it then follows that the total matter component is also conserved. After equality it leads to separate conservation of cold matter and radiation.   In general, $Q$ can be a nonvanishing interaction source allowing energy exchange between matter and the quasi-vacuum medium under consideration; $Q$ can either be given by hand (e.g. through an {\it ad hoc} ansatz), or can be suggested by some specific theoretical framework. In any case the interaction source must satisfy $0<|Q|\ll\dot{\rho}_m$ since we do not wish to depart too much from the concordance model. Despite matter is exchanging energy with the vacuum or quasi-vacuum medium, we shall assume that radiation and baryons are separately self-conserved, i.e. $\dot{\rho}_r + 4H\rho_r =0$ and $\dot{\rho}_b + 3H\rho_b =0$, so that their energy densities evolve in the standard way: $\rho_r(a)=\rho_{r0}\,a^{-4}$ and $\rho_b(a) = \rho_{b0}\,a^{-3}$. The dynamics of $\rL$ can therefore be  associated to the exchange of energy exclusively with the DM (through the nonvanishing source $Q$) and/or with the possibility that the DE medium is not exactly the vacuum, $w\neq -1$, but close to it $|\epsilon|\ll 1$. Under these conditions, the coupled system of conservation equations \eqref{eq:GeneralCL_H0_tension_chatper}-\eqref{eq:Source_H0_tension_chatper} reduces to
\begin{eqnarray}
&&\dot{\rho}_{dm}+3H\rho_{dm}=Q\label{eq:Qequations1_H0_tension_chapter}\\
&&\dot{\rho}_\CC + 3H\epsilon\rho_\CC=-Q\,.\label{eq:Qequations2_H0_tension_chapter}
\end{eqnarray}
\begin{figure}[t!]
\setcounter{figure}{0}
\begin{center}
\label{FigLSS1_H0_tension_chapter}
\includegraphics[width=5.7in]{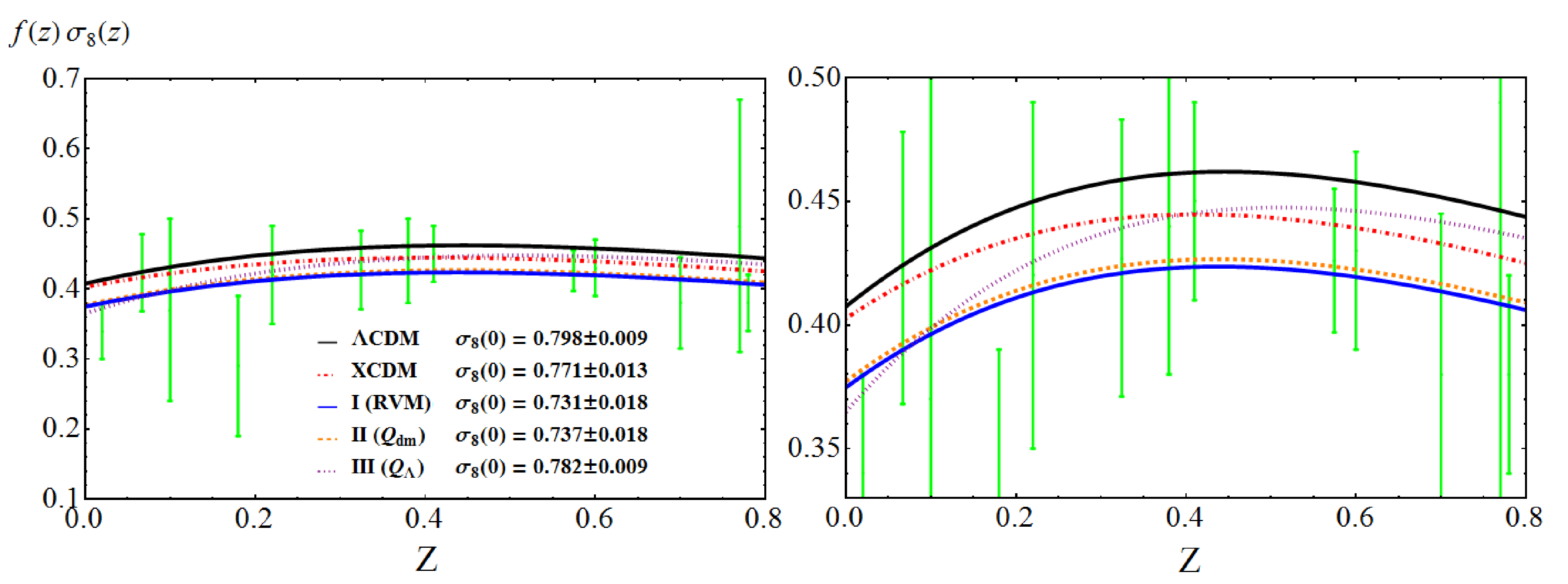}\ \ 
\caption{\scriptsize {\bf Left}: The LSS  structure formation data ($f(z)\sigma_8(z)$) versus the predicted curves by Models I, II and III, see equations \eqref{eq:QforModelRVM_H0_tension_chapter}-\eqref{eq:QforModelQL_H0_tension_chapter} for the case $w=-1$, i.e. the dynamical vacuum models (DVMs), using the best-fit values in Table 1. The XCDM curve is also shown. The values of $\sigma_8(0)$ that we obtain for the models are also indicated. {\bf Right}: Zoomed window of the plot on the left, which allows to better distinguish the various models.}
\end{center}
\end{figure}
In the following we shall for definiteness focus our study of the dynamical vacuum (and quasi-vacuum) models to the three interactive sources:
\begin{eqnarray}\label{eq:QforModelRVM_H0_tension_chapter}
&&{\rm Model\ I\ \ }(w{\rm RVM}):\ Q=\nu\,H(3\rho_{m}+4\rho_r)\label{eq:QforModelQdm_H0_tension_chapter}\\
&&{\rm Model\ II\ \ }(wQ_{dm}):\ Q_{dm}=3\nu_{dm}H\rho_{dm}\\
&&{\rm Model\ III\ \ }(wQ_{\CC}):\ Q_{\CC}=3\nu_{\CC}H\rho_{\CC}\,.\label{eq:QforModelQL_H0_tension_chapter}
\end{eqnarray}
Here $\nu_i=\nu, \nu_{dm},\nu_{\CC}$ are small dimensionless constants, $|\nu_i|\ll1$, which are determined from the overall fit to the data, see e.g. Tables 1 and 2.
The ordinal number names I, II and III will be used for short, but the three model names are preceded by $w$  to recall that, in the general case, the equation of state (EoS) is near the vacuum one (that is, $w=-1+\epsilon$). These dynamical quasi-vacuum models  are also denoted as $w$DVMs. In the particular case $w=-1$ (i.e. $\epsilon=0$) we recover the dynamical vacuum models (DVMs), which were previously studied in detail in\,\cite{Sola:2017jbl}, and in this case the names of the models will not be preceded by $w$.
\newline
\newline
In all of the above ($w$)DVMs, the cosmic variable $\zeta$ can be taken to be the scale factor, $\zeta=a(t)$, since they are all analytically solvable in terms of it, as we shall see in a moment. Model I with $w=-1$ is the running vacuum model (RVM), see\,\cite{Sola:2013gha,Sola:2015rra,Sola:2017jbl,Sola:2016zeg}. It is special in that the interaction source indicated in \eqref{eq:QforModelRVM_H0_tension_chapter} is not {\it ad hoc} but follows from an expression for the dynamical vacuum energy density, $\rL(\zeta)$, in which
$\zeta$ is not just the scale factor but the full Hubble rate: $\zeta=H(a)$.  The explicit RVM form reads
\begin{equation}\label{eq:RVMvacuumdadensity_H0_tension_chapter}
\rho_\CC(H) = \frac{3}{8\pi{G}}\left(c_{0} + \nu{H^2}\right)\,.
\end{equation}
\newline
The additive constant $c_0=H_0^2\left(\Omega^{0}_\Lambda-\nu\right)$ is fixed from the condition $\rL(H_0)=\rLo$, with $\Omega^{0}_\Lambda=1-\Omega^{0}_m-\Omega^{0}_r$. Combining the Friedmann and acceleration equations \eqref{eq:FriedmannEq_H0_tension_chatper}-\eqref{eq:PressureEq_H0_tension_chatper}, we find $\dot{H}=-(4\pi G/3)$
$\left(3\rho_m+4\rho_r+3\epsilon\rho_\CC\right)$,
and upon differentiating \eqref{eq:RVMvacuumdadensity_H0_tension_chapter} with respect to the cosmic time we are led to
$\dot{\rho}_\CC=-\nu\,H\left(3\rho_m+4\rho_r+3\epsilon\rho_\CC\right)$.
Thus, for $\epsilon=0$ (vacuum case) we indeed find $\dot{\rho}_\CC=-Q$ for $Q$ as in \eqref{eq:QforModelRVM_H0_tension_chapter}. However, for the quasi-vacuum case ($0<|\epsilon|\ll1$) Eq.\,\eqref{eq:Qequations2_H0_tension_chapter} does not hold if $\rL(H)$ adopts the form \eqref{eq:RVMvacuumdadensity_H0_tension_chapter}. This RVM form is in fact specific to the pure vacuum EoS ($w=-1$), and it can be motivated in QFT in curved space-time through a renormalization group equation for $\rL(H)$, what explains the RVM name\,\cite{Sola:2013gha}. In it, $\nu$ plays the role of the $\beta$-function coefficient for the running of $\rL$ with the Hubble rate. Thus, we naturally expect $|\nu|\ll1$ in QFT, see \cite{Sola:2013gha,Sola:2007sv}. Interestingly, the
RVM  form \eqref{eq:RVMvacuumdadensity_H0_tension_chapter} can actually be extended with higher powers of $H^n$ (typically $n=4$) to provide an effective description of the cosmic evolution from the inflationary Universe up to our days\,\cite{Lima:2012mu,Sola:2015rra}. Models II and III are purely phenomenological models instead, in which the interaction source $Q$ is introduced by hand, see e.g. Refs.\,\cite{Salvatelli:2014zta,Murgia:2016ccp,Li:2015vla,DiValentino:2017iww} and references therein.
%
%
The energy densities for the $w$DVMs can be computed straightforwardly. For simplicity, we shall quote here the leading parts only. The exact formulas containing the radiation terms are more cumbersome. In the numerical analysis we have included the full expressions. For the matter densities, we find:
\begin{eqnarray}
\rho^{\rm I}_{dm}(a) &=& \rho_{dm0}\,a^{-3(1-\nu)} + \rho_{b0}\left(a^{-3(1-\nu)} - a^{-3}\right) \nonumber \\
\rho^{\rm II}_{dm}(a) &=& \rho_{dm0}\,a^{-3(1-\nu_{dm})}
\label{eq:rhoms_H0_tension_chapter} \\
\rho^{\rm III}_{dm}(a) &=&\rho_{dm0}\,a^{-3} + \frac{\nu_\CC}{\nu_\CC + w}\rLo\left(a^{-3}-a^{-3(\epsilon + \nu_\CC)}\right)\nonumber\,,
\end{eqnarray}
and for the quasi-vacuum energy densities:
\begin{eqnarray}\label{eq:rhowLdensities_H0_tension_chapter}
\rho^{\rm I}_\CC(a) &=& \rLo{a^{-3\epsilon}} - \frac{\nu\,\rho_{m0}}{\nu + w}\left(a^{-3(1-\nu)}- a^{-3\epsilon}\right) \nonumber\\
\rho^{\rm II}_\CC(a)&=& \rLo{a^{-3\epsilon}} - \frac{\nu_{dm}\,\rho_{dm0}}{\nu_{dm} + w}\,\left(a^{-3(1-\nu_{dm})}- a^{-3\epsilon}\right)\\
\rho^{\rm III}_\CC(a) &=&\rLo\,{a^{-3(\epsilon + \nu_\CC)}}\,.\nonumber
\end{eqnarray}
Two specific dimensionless  parameters enter each formula, $\nu_{i}=(\nu,\nu_{dm},\nu_\CC)$  and $w=-1+\epsilon$. They are part of the fitting vector of free parameters for each model, as explained in detail in the caption of Table 1. For $\nu_{i}\to 0$ the models become noninteractive and they all reduce to the XCDM model case \eqref{eq:rhoXCDM_H0_tension_chapter}. For $w=-1$ we recover the DVMs results previously studied in \cite{Sola:2017jbl}.  Let us also note that for $\nu_{i}>0$ the vacuum decays into DM (which is thermodynamically favorable\,\cite{Sola:2017jbl}) whereas for $\nu_{i}<0$ is the other way around. Furthermore, when $w$ enters the fit, the effective behaviour of the $w$DVMs is quintessence-like for $w>-1$ (i.e. $\epsilon>0$) and phantom-like for $w<-1$ ($\epsilon<0$).
\newline
Given the energy densities \eqref{eq:rhoms_H0_tension_chapter}and \eqref{eq:rhowLdensities_H0_tension_chapter}, the Hubble function immediately follows. For example, for Model I:
\begin{equation}\label{eq:HubbewRVM_H0_tension_chapter}
H^2(a) = H_0^2\left[a^{-3\epsilon} + \frac{w}{w+\nu}\Omega^{0}_m\left(a^{-3(1-\nu)}- a^{-3\epsilon}\right)\right]\,.
\end{equation}
Similar formulas can be obtained for Models II and III. For $w=-1$ they all reduce to the DVM forms previously found in \cite{Sola:2017jbl}. And of course they all ultimately boil down to the $\CC$CDM form in the limit $(w, \nu_i)\to (-1, 0)$.

\begin{table}[t!]
\begin{center}
\begin{scriptsize}
\resizebox{1\textwidth}{!}{
\begin{tabular}{ |c|c|c|c|c|c|c|c|c|c|}
\multicolumn{1}{c}{Model} &  \multicolumn{1}{c}{$H_0$(km/s/Mpc)} &  \multicolumn{1}{c}{$\omega_b$} & \multicolumn{1}{c}{{\small$n_s$}}  &  \multicolumn{1}{c}{$\Omega_m^0$} &\multicolumn{1}{c}{$\nu_i$} &\multicolumn{1}{c}{$w$} &\multicolumn{1}{c}{$\chi^2_{\rm min}/dof$} & \multicolumn{1}{c}{$\Delta{\rm AIC}$} & \multicolumn{1}{c}{$\Delta{\rm BIC}$}\vspace{0.5mm}
\\\hline
$\Lambda$CDM  & $68.23\pm 0.38$ & $0.02234\pm 0.00013$ &$0.968\pm 0.004$& $0.306\pm 0.005$ & - & -1  & 13.85/11 & - & - \\
\hline
RVM  & $67.70\pm 0.69$& $0.02227\pm0.00016 $&$0.965\pm0.005$& $0.306\pm0.005$ &$0.0010\pm 0.0010 $ & -1  & 13.02/10 & -3.84 & -1.88 \\
\hline
$Q_\Lambda$  & $68.34\pm 0.40$& $0.02226\pm0.00016 $&$0.965\pm0.005$& $0.305\pm0.005$ &$0.0030\pm 0.0030 $& -1  &  12.91/10 & -3.73 & -1.77 \\
\hline
$w$RVM  & $66.34\pm 2.30$& $0.02228\pm0.00016 $&$0.966\pm0.005$& $0.313\pm0.012$ &$0.0017\pm 0.0016 $ & $-0.956\pm0.071$ & 12.65/9 & -9.30 & -4.22 \\
\hline
$w{Q_\Lambda}$  & $66.71\pm 1.77$& $0.02226\pm0.00016 $&$0.965\pm0.005$& $0.317\pm0.014$ &$0.0070\pm 0.0054 $& $-0.921\pm0.082$ &   12.06/9 & -8.71 & -3.63\\
\hline
$\Lambda$CDM* & $68.46\pm 0.37$ & $0.02239\pm 0.00013$ &$0.969\pm 0.004$& $0.303\pm 0.005$ & - & -1 & 21.76/12 & - & - \\
\hline
RVM*  & $68.48\pm 0.67$& $0.02240\pm0.00015 $&$0.969\pm0.005$& $0.303\pm0.005$ &$0.0000\pm 0.0010 $ & -1  & 21.76/11 & -4.36 & -2.77 \\
\hline
$Q_\Lambda$*  & $68.34\pm 0.39$& $0.02224\pm0.00016 $&$0.966\pm0.005$& $0.302\pm0.005$ &$0.0034\pm 0.0030 $& -1  &  20.45/11 & -3.05 & -1.46 \\
\hline
Ia ($w$RVM*)  & $70.95\pm 1.46$& $0.02231\pm0.00016 $&$0.967\pm0.005$& $0.290\pm0.008$ &$-0.0008\pm 0.0010 $ & $-1.094\pm0.050$ & 18.03/10 & -5.97 & -1.82 \\
\hline
IIIa ($w{Q_\Lambda}$*)  & $70.27\pm 1.33$& $0.02228\pm0.00016 $&$0.966\pm0.005$& $0.291\pm0.010$ &$-0.0006\pm 0.0042 $& $-1.086\pm0.065$ &   18.64/10 & -6.58  & -2.43 \\
\hline
\end{tabular}}

\caption{{\scriptsize Best-fit values for the $\CC$CDM and models RVM, Q$_{\Lambda}$, $w$RVM and $w$Q$_\Lambda$  by making use of the CMB+BAO data only. In contrast to Tables 1-2, we now fully dispense with the LSS data (see \cite{Sola:2016jky,Sola:2017jbl}) to test its effect. The starred/non-starred cases correspond respectively to adding or not the local value $H_0^{\rm Riess}$ from \cite{Riess:2016jrr} as data point in the fit. The AIC and BIC differences of the starred models are computed with respect to the $\Lambda$CDM*. We can see that under these conditions models tend to have $\Delta$AIC, $\Delta$BIC$<0$, including the last two starred scenarios, which are capable of significantly approaching $H_0^{\rm Riess}$}.}
\end{scriptsize}
\end{center}
\label{tableFit3_H0_tension_chapter}
\end{table}
\subsection{Structure formation: the role of the LSS data}\label{Sec:SF_H0_tension_chapter}
The analysis of structure formation plays a crucial role in comparing the various models. For the $\CC$CDM and XCDM  we use the standard perturbations equation\,\cite{peebles:1993}
\begin{equation}\label{diffeqLCDM_H0_tension_chapter}
\ddot{\delta}_m+2H\,\dot{\delta}_m-4\pi
G\rmr\,\delta_m=0\,,
\end{equation}
with, however, the Hubble function corresponding to each one of these models.
For the $w$DVMs, a step further is needed: the perturbations equation not only involves the modified Hubble function but the equation itself becomes modified. Trading the cosmic time for the scale factor and extending the analysis of \cite{Sola:2017jbl,Gomez-Valent:2014rxa,Gomez-Valent:2014fda} for the case  $w\neq -1$ ($\epsilon\neq 0$), we find
\begin{equation}\label{diffeqD_H0_tension_chapter}
\delta^{\prime\prime}_m + \frac{A(a)}{a}\delta_{m}^{\prime} + \frac{B(a)}{a^2}\delta_m = 0  \,,
\end{equation}
where the prime denotes differentiation with respect to the scale factor,
and the functions $A(a)$ and $B(a)$ are found to be as follows:
\begin{eqnarray}
&&A(a) = 3 + \frac{aH^{\prime}}{H} + \frac{\Psi}{H} - 3r\epsilon\label{eq:Afunction_H0_tension_chapter}\\
&&B(a) = -\frac{4\pi{G}\rho_m }{H^2} + 2\frac{\Psi}{H} + \frac{a\Psi^{\prime}}{H} -15r\epsilon - 9\epsilon^{2}r^{2} +3\epsilon(1+r)\frac{\Psi}{H} -3r\epsilon\frac{aH^{\prime}}{H}\,. \label{eq:Bfunction_H0_tension_chapter}
\end{eqnarray}
Here $r \equiv \rho_\Lambda/\rho_m$ and  $\Psi\equiv -\dot{\rho}_{\Lambda}/{\rmr}$. {For $\nu_i=0$ we have $\Psi=3Hr\epsilon$}, and after a straightforward calculation one can show that \eqref{diffeqD_H0_tension_chapter} can be brought to the standard form  Eq.\,\eqref{diffeqLCDM_H0_tension_chapter}.
%
%
%
\begin{figure}[t!]
\begin{center}
\label{FigLSS2_H0_tension_chapter}
\includegraphics[width=3.3in]{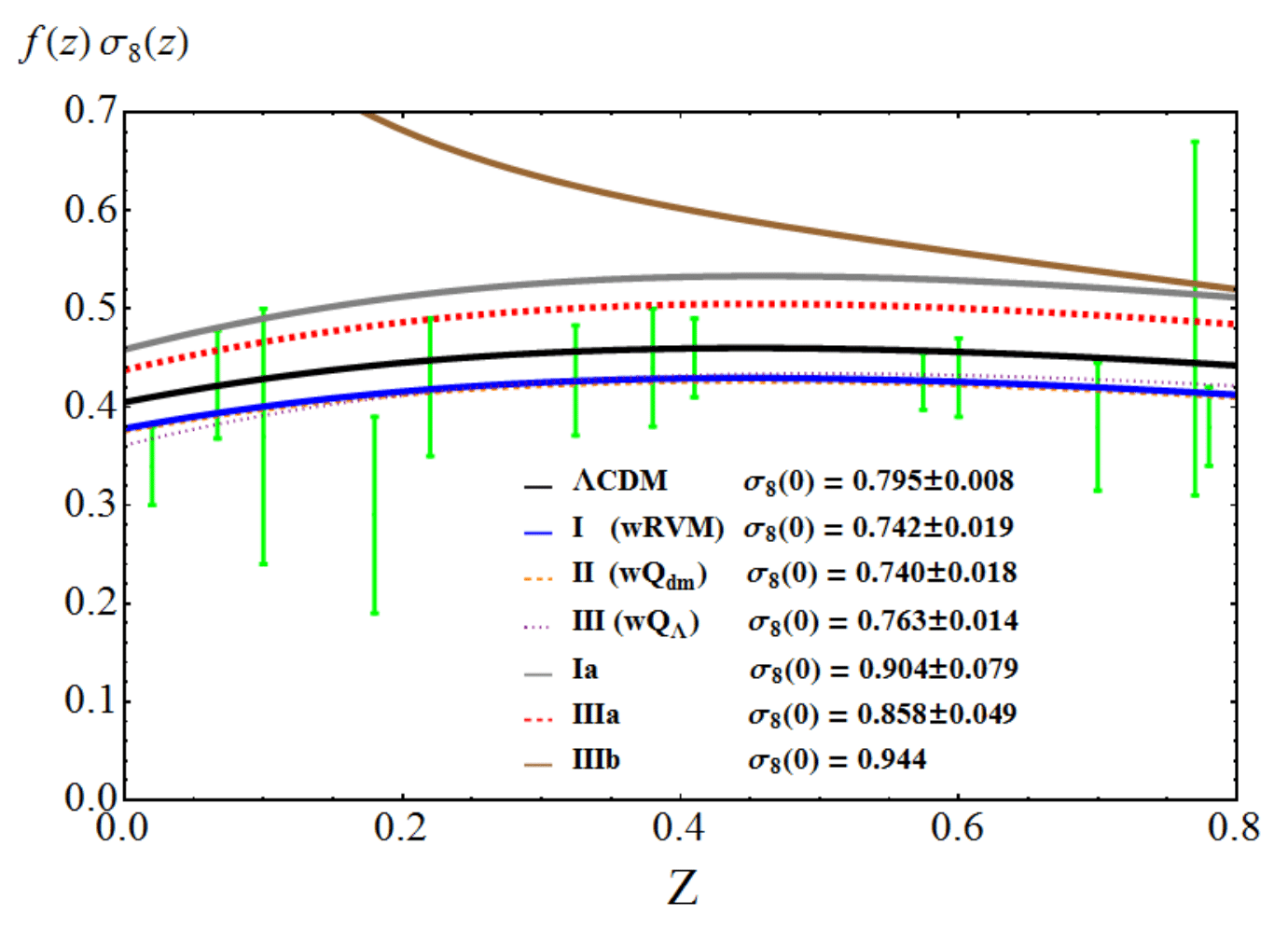}
\caption{\scriptsize The LSS  structure formation data ($f(z)\sigma_8(z)$) and the theoretical predictions for models \eqref{eq:QforModelRVM_H0_tension_chapter}-\eqref{eq:QforModelQL_H0_tension_chapter}, using the best-fit values in Tables 2 and 3. The curves for the cases Ia, IIIa correspond to special scenarios for Models I and III where the agreement with the  Riess {\it et al.} local value $H_0^{\rm Riess}$\,\cite{Riess:2016jrr} is better (cf. Table 3). The price, however, is that the concordance with the LSS data is now spoiled. Case IIIb is our theoretical prediction for the scenario proposed in \cite{DiValentino:2017iww}, aimed at optimally relaxing the tension with $H_0^{\rm Riess}$. Unfortunately, the last three scenarios lead to phantom-like DE and are in serious disagreement with the LSS data.
}
\end{center}
\end{figure}
%
%
%
\newline
\newline
To solve the above perturbations equations we have to fix the initial conditions on $\delta_m$ and ${\delta}^{\prime}_m$ for each model at high redshift, namely when non-relativistic matter dominates over radiation and DE, see\,\cite{Sola:2017jbl}.
Functions \eqref{eq:Afunction_H0_tension_chapter} and \eqref{eq:Bfunction_H0_tension_chapter} are then approximately constant and Eq.\,\eqref{diffeqD_H0_tension_chapter} admits power-law solutions $\delta_m(a) = a^{s}$. From explicit calculation we find that the values of $s$ for each model turn out to be:
\begin{eqnarray}\label{eq:svalues_H0_tension_chapter}
s^{\rm I} &=& 1 + \frac{3}{5}\nu\left(\frac{1}{w} -4\right) + \mathcal{O}(\nu^2)\nonumber  \\
s^{\rm II} &=& 1 -\frac{3}{5}\nu_{dm}\left(  1 + 3\frac{\Omega^{0}_{dm}}{\Omega^{0}_{m}} -\frac{1}{w}   \right) + \mathcal{O}({\nu_{dm}}^{2}) \\
s^{\rm III} &=&1\nonumber\,.
\end{eqnarray}
We can check that for $w=-1$ all of the above equations \eqref{diffeqD_H0_tension_chapter}-\eqref{eq:svalues_H0_tension_chapter} render the DVM results previously found in \cite{Sola:2017jbl}. The generalization that we have made to $w\neq -1$ ($\epsilon\neq 0$) has introduced several nontrivial extra terms in equations \eqref{eq:Afunction_H0_tension_chapter}-\eqref{eq:svalues_H0_tension_chapter}.
\newline
\newline
The analysis of the linear LSS regime is usually implemented with the help of the weighted linear growth $f(z)\sigma_8(z)$, where $f(z)=d\ln{\delta_m}/d\ln{a}$ is the growth factor and $\sigma_8(z)$ is the rms mass fluctuation on $R_8=8\,h^{-1}$ Mpc scales. It is computed as follows (see e.g. \cite{Sola:2017jbl,Sola:2016jky}):
\begin{equation}
\begin{small}\sigma_{\rm 8}(z)=\sigma_{8, \CC}
\frac{\delta_m(z)}{\delta^{\CC}_{m}(0)}
\sqrt{\frac{\int_{0}^{\infty} k^{n_s+2} T^{2}(\vec{p},k)
W^2(kR_{8}) dk} {\int_{0}^{\infty} k^{n_{s,\CC}+2} T^{2}(\vec{p}_\Lambda,k) W^2(kR_{8,\Lambda}) dk}}\,,\label{s88general_H0_tension_chapter}
\end{small}\end{equation}
where $W$ is a top-hat smoothing function and $T(\vec{p},k)$ the transfer function. The fitting parameters for each model are contained in $\vec{p}$.
Following the mentioned references, we have defined as fiducial model the $\CC$CDM at fixed parameter values from the Planck 2015 TT,TE,EE+lowP+lensing data\,\cite{Ade:2015xua}. These fiducial values are collected in  $\vec{p}_\CC$.
In Figs. 1-2 we display  $f(z)\sigma_8(z)$ for the various models using the fitted values of Tables 1-3. We remark that our BAO and LSS data include the bispectrum data points from \cite{Gil-Marin:2016wya} 
--see \cite{Sola:2017jbl} for a full-fledged explanation of our data sets.  In the next section, we discuss our results for the various models and assess their ability to improve the $\CC$CDM fit as well as their impact on the $H_0$ tension.


\begin{figure}[t!]
\begin{center}
\label{contours_H0_tension_chapter}
\includegraphics[width=5in]{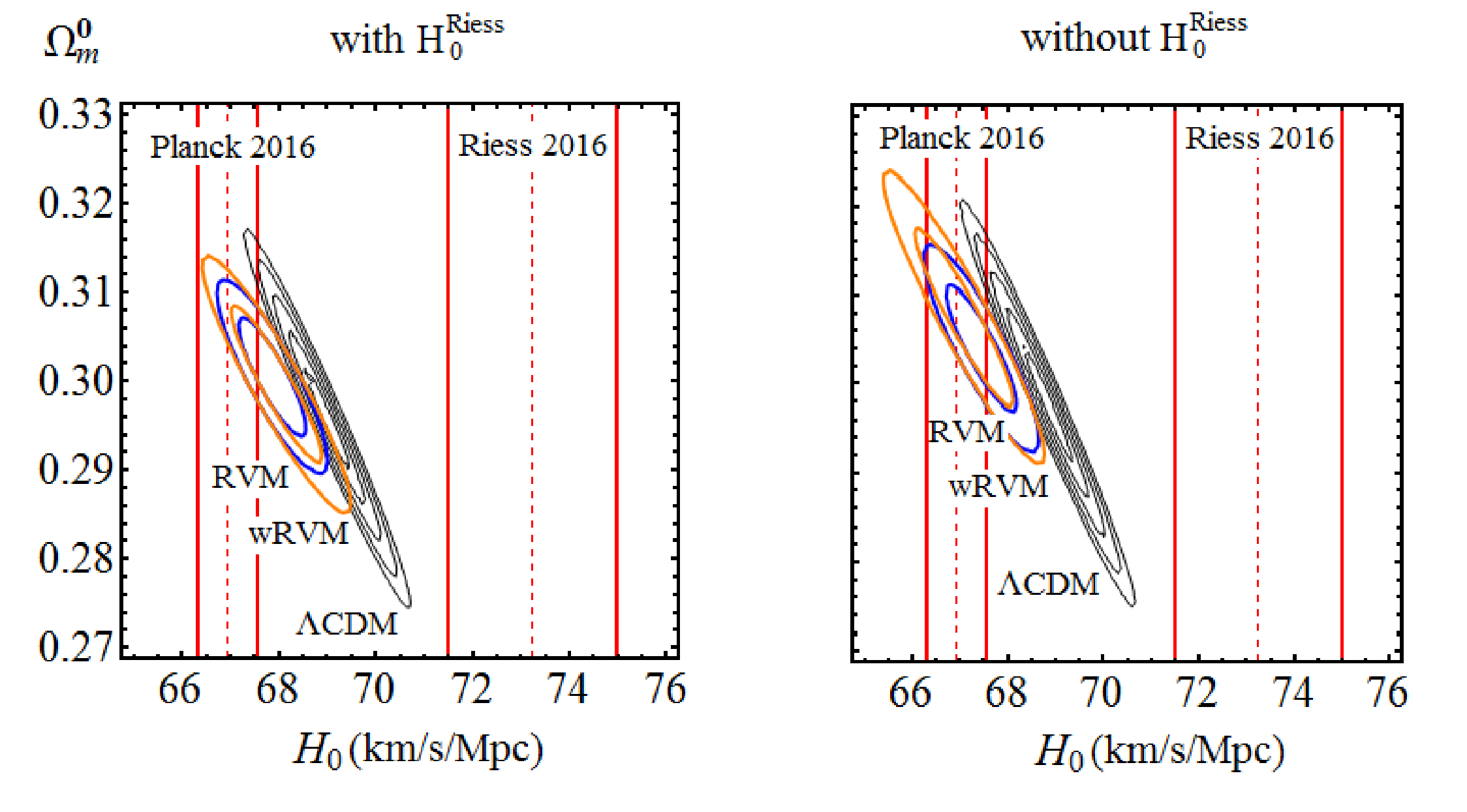}
\caption{\scriptsize Contour plots for the RVM (blue) and $w$RVM (orange) up to $2\sigma$, and for the $\CC$CDM (black) up to $5\sigma$ in the $(H_0,\Omega^0_m)$-plane. {Shown are the two relevant cases under study: the plot on the left is for when the local $H_0$ value of Riess {\it et al.}\,\cite{Riess:2016jrr} is included in the fit (cf. Table 2), and the plot on the right is for when that local value is {\it not} included (cf. Table 1). Any attempt at reaching the $H_0^{\rm Riess}$ neighborhood  enforces to pick too small values  $\Omega^0_m<0.27$ through extended contours that go beyond  $5\sigma$ c.l.} We also observe that the two ($w$)RVMs are much more compatible (already at $1\sigma$) with the $H_0^{\rm Planck}$ range than the $\CC$CDM. The latter, instead, requires some of the most external contours to reach the $H^{\rm Planck}_0$ $1\sigma$ region whether $H_0^{\rm Riess}$ is included or not in the fit. Thus, remarkably, in both cases when the full data string SNIa+BAO+$H(z)$+LSS+CMB enters the fit the $\CC$CDM has difficulties to overlap also with the $H_0^{\rm Planck}$ range at $1\sigma$, in contrast to the RVM and $w$RVM.
}
\end{center}
\end{figure}
\subsection{Discussion}\label{Sec:Discussion_H0_tension_chapter}
Following \cite{Sola:2017jbl}  the statistical analysis of the various models is performed in terms of a joint likelihood function, which is the product of the likelihoods for each data source
and includes the corresponding covariance matrices.
As indicated in the caption of Table 1,  the $\CC$CDM has $4$ parameters, whereas the XCDM and the DVMs have $5$, and finally any of the $w$DVMs has $6$. Thus, for a fairer comparison of the various nonstandard models with the concordance $\CC$CDM we have to invoke efficient criteria in which the presence of extra parameters in a given model is conveniently penalized so as to achieve a balanced comparison with the model having less parameters. The  Akaike information criterion (AIC) and the Bayesian information criterion (BIC) are known to be extremely valuable tools for a fair statistical analysis of this kind. They can be thought of as a modern quantitative formulation of Occam's razor. They read\,\cite{Akaike,Schwarz1978,KassRaftery1995}:
\begin{equation}\label{eq:AICandBIC_H0_tension_chapter}
{\rm AIC}=\chi^2_{\rm min}+\frac{2nN}{N-n-1}\,,\ \ \ \ \
{\rm BIC}=\chi^2_{\rm min}+n\,\ln N\,,
\end{equation}
where $n$ is the number of independent fitting parameters and $N$ the number of data points.
The bigger are the (positive) differences $\Delta$AIC and $\Delta$BIC with respect to the model having smaller values of AIC and BIC the higher is the evidence against the model with larger AIC and BIC. Take, for instance, Tables 1 and 2, where in all cases the less favored model is the $\CC$CDM (thus with larger AIC and BIC).
For $\Delta$AIC and $\Delta$BIC in the range $6-10$ one speaks of ``strong evidence'' against the $\CC$CDM, and hence in favor of the nonstandard models being considered. This is typically the situation for the RVM and $Q_{dm}$ vacuum models in Table 2 and for the three $w$DVMs in Table 1.
Neither the  XCDM nor the $Q_\CC$ vacuum model attain the ``strong evidence'' threshold in Tables 1 or 2. The XCDM parametrization, which is used as a baseline for comparison of the dynamical DE models, is nevertheless capable of detecting significant signs of dynamical DE, mainly in Table 1 (in which $H_0^{\rm Riess}$ is excluded), but not so in Table 2 (where $H_0^{\rm Riess}$ is included).  In contrast, model $Q_\CC$ does not change much from Table 1 to Table 2.
\newline
In actual fact, the vacuum model III ($Q_\CC$) tends to remain always fairly close to the $\CC$CDM. Its dynamics is weaker than that of the main DVMs (RVM and $Q_{dm}$). Being $|\nu_i|\ll 1$ for all the DVMs, the evolution of its vacuum energy density is approximately logarithmic: $\rL^{\rm III}\sim\rLo(1-3\nu_{\CC}\,\ln{a})$, as it follows from  \eqref{eq:rhowLdensities_H0_tension_chapter} with $\epsilon=0$. Thus, it is significantly milder in comparison to that of the main DVMs, for which $\rL^{\rm I, II}\sim \rLo\left[1+(\Omega^{0}_m/\Omega^{0}_\Lambda)\nu_i (a^{-3}-1)\right]$. The performance of $Q_\CC$ can only be slightly better than that of $\CC$CDM, a fact that may have not been noted in previous studies -- see \,\cite{Salvatelli:2014zta,DiValentino:2016hlg,Murgia:2016ccp,Li:2015vla,DiValentino:2017iww} and references therein.
\newline
\newline
According to the same jargon, when the differences $\Delta$AIC and $\Delta$BIC are both above 10 one speaks of ``very strong evidence'' against the unfavored model (the $\CC$CDM, in this case), wherefore in favor of the dynamical vacuum and quasi-vacuum models. It is certainly the case of the RVM and $Q_{dm}$ models in Table 1, which are singled out as being much better than the $\CC$CDM in their ability to describe the overall observations. From Table 1 we can see that the best-fit values of $\nu_i$ for these models are secured at a confidence level of $\sim 3.8\sigma$. These two models are indeed the most conspicuous ones in our entire analysis, and remain strongly favored even if $H_0^{\rm Riess}$\,\cite{Riess:2016jrr} is included (cf. Table 2). In the last case, the best-fit values of $\nu_i$ for the two models are still supported at a fairly large c.l. ($\sim 3.2\sigma$). {This shows that the overall fit to the data in terms of dynamical vacuum is a real option since the fit quality is not exceedingly perturbed in the presence of the data point $H_0^{\rm Riess}$. However, the optimal situation is really attained in the absence of that point, not only because the fit quality is then higher but also because that point remains out of the fit range whenever the large scale structure formation data (LSS) are included. {For this reason we tend to treat that input as an outlier -- see also \cite{Lin:2017bhs} for an alternative support to this possibility, which we comment later on}. In the following, we will argue that a truly consistent picture with all the data is only possible for $H_0$ in the vicinity of  $H_0^{\rm Planck}$ rather than in that of $H_0^{\rm Riess}$.}
\newline
\newline
{The conclusion is that the $H_0^{\rm Riess}$-$H_0^{\rm Planck}$  tension cannot be relaxed without unduly forcing the overall fit, which is highly sensitive to the LSS data. It goes without saying that one cannot have a prediction that matches both $H_0$ regions at the same time, so at some point new observations (or the discovery of some systematic in one of the experiments) will help to consolidate one of the two ranges of values and exclude definitely the other. At present no favorable fit can be obtained from the $\CC$CDM that is compatible with any of the two $H_0$ ranges. This is transparent from Figs. 3 and 4, in which the $\CC$CDM remains always in between the two regions. However, our work shows that a solution (with minimum cost) is possible in terms of vacuum dynamics. Such solution, which inevitably puts aside the $H_0^{\rm Riess}$ range, is however compatible with all the remaining data and tends to favor the Planck range of $H_0$ values. The DVMs can indeed provide an excellent fit to the overall cosmological observations and be fully compatible with both the $H_0^{\rm Planck}$ value and at the same time with the needed low values of the $\sigma_8(0)$  observable, these low values of $\sigma_8(0)$ being crucial to fit the structure formation data. Such strategy is only possible in the presence of vacuum dynamics, whilst it is impossible with a rigid $\CC$-term, i.e. is not available to the $\CC$CDM.}
In Fig.\,1 we confront the various models with the LSS data when the local measurement $H_0^{\rm Riess}$ is not included in our fit. The differences can be better appraised in the plot on the right, where we observe that the RVM and $Q_{dm}$ curves stay significantly lower than the $\CC$CDM one (hence matching better the data than the $\CC$CDM), whereas those of XCDM and $Q_\CC$ remain in between.
Concerning the $w$DVMs, namely the quasi-vacuum models in which an extra parameter is at play (the EoS parameter $w$), we observe a significant difference as compared to the DVMs (with vacuum EoS $w=-1$): they {\it all} provide a similarly good fit quality, clearly superior to that of the $\CC$CDM (cf. Tables 1 and 2) but in all cases below that of the main DVMs (RVM and $Q_{dm}$), whose performance is outstanding.
\newline
%
%
\begin{figure}[t!]
\begin{center}
\label{contours_H0_tension_chapter}
\includegraphics[width=5in]{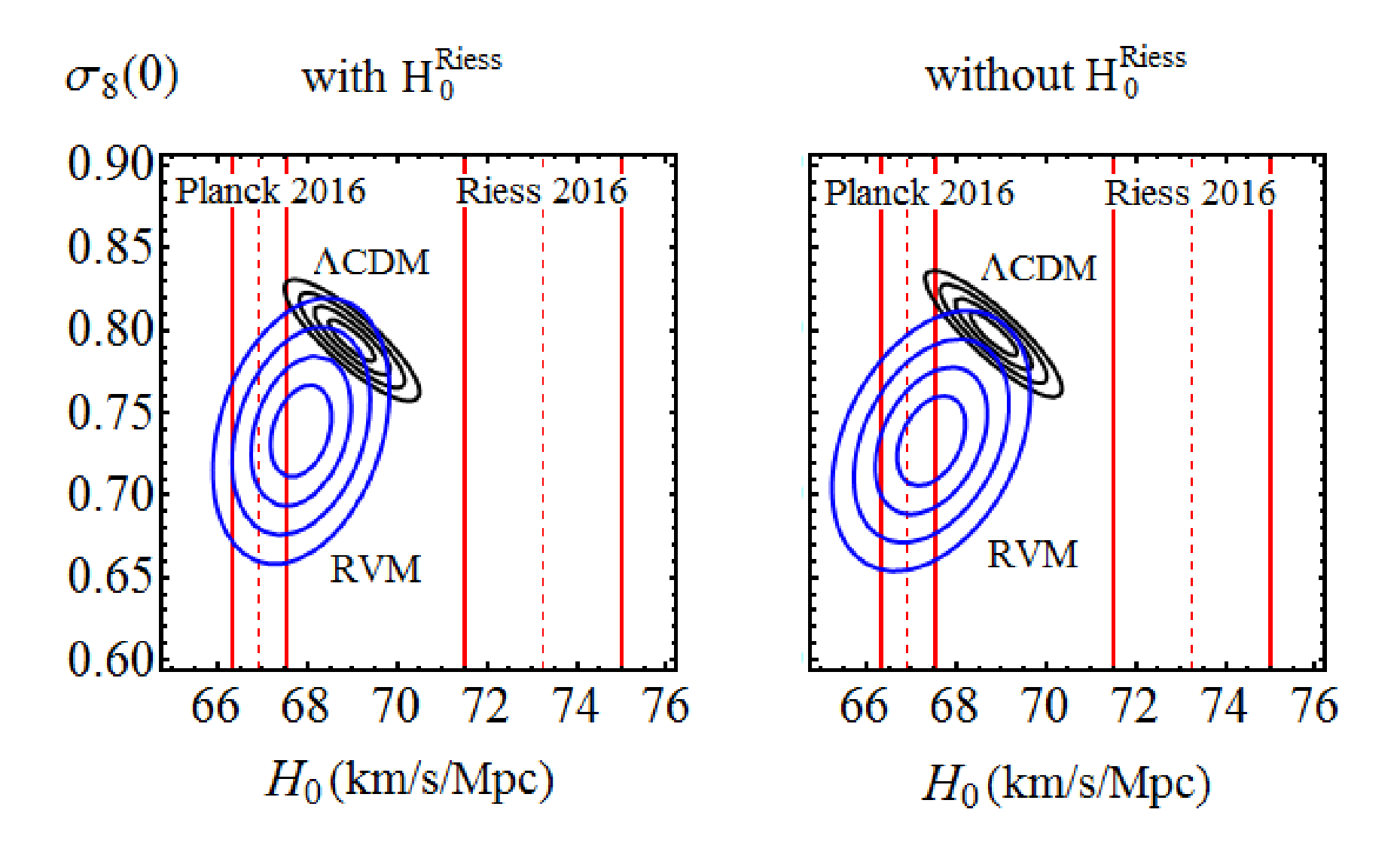}
\caption{\scriptsize {Contour lines for the $\CC$CDM (black) and RVM (blue) up to $4\sigma$ in the $(H_0,\sigma_8(0))$-plane. {As in Fig. 3, we present in the {\it left plot} the case when the local $H_0$ value of Riess {\it et al.}\,\cite{Riess:2016jrr} is included in the fit (cf. Table 2), whereas in the {\it right plot} the case when that local value is {\it not} included (cf. Table 1). Again, any attempt at reaching the $H_0^{\rm Riess}$ neighborhood  enforces to extend the contours beyond the $5\sigma$ c.l.}, which would lead to a too low value of $\Omega_m^{0}$ in both cases (cf. Fig. 3) and, in addition, would result in a too large value of $\sigma_8(0)$ for the RVM. Notice that $H_0$ and $\sigma_8(0)$ are positively correlated in the RVM (i.e. $H_0$ decreases when $\sigma_8(0)$ decreases), whilst they are anticorrelated in the $\CC$CDM ($H_0$ increases when $\sigma_8(0)$ decreases, and vice versa). It is precisely this opposite correlation feature with respect to the $\CC$CDM what allows the RVM to improve the LSS fit in the region where both $H_0$ and $\sigma_8(0)$ are smaller than the respective $\CC$CDM values (cf. Fig. 1). This explains why the Planck range for $H_0$ is clearly preferred by the RVM, as it allows a much better description of the LSS data.}}
\end{center}
\end{figure}
In Table 3, in an attempt to draw our fit nearer and nearer to $H_0^{\rm Riess}$\,\cite{Riess:2016jrr}, we test the effect of ignoring the LSS structure formation data, thus granting more freedom to the fit parameter space. We perform this test using the $\CC$CDM and models $(w)$RVM and $(w)Q_\CC$ (i.e. models I and III and testing both the vacuum and quasi-vacuum options), and we fit them to the CMB+BAO data alone. We can see that the fit values for $H_0$ increase in all starred scenarios (i.e. those involving the $H_0^{\rm Riess}$ data point in the fit), and specially for the cases Ia and IIIa in Table 3. Nonetheless, these lead to $\nu_i<0$ and $w<-1$ (and hence imply phantom-like DE); and, what is worse, the agreement with the LSS data is ruined (cf. Fig. 2) since the corresponding curves are shifted too high (beyond the $\CC$CDM one).  In the same figure we superimpose one more scenario, called IIIb, corresponding to a rather acute phantom behaviour ($w=-1.184\pm0.064$). The latter was recently explored in \cite{DiValentino:2017iww} so as to maximally relax the $H_0$ tension -- see also\,\cite{DiValentino:2016hlg}. Unfortunately, we find (see Fig.\,2) that the associated LSS curve is completely strayed since it fails to minimally describe the $f\sigma_8$ data (LSS).
In Fig. 3 we demonstrate in a very visual way that, in the context of the overall observations (i.e.  SNIa+BAO+$H(z)$+LSS+CMB), whether including or not including the data point $H_0^{\rm Riess}$ (cf. Tables 1 and 2), it becomes impossible to getting closer to  the local measurement $H_0^{\rm Riess}$ unless we go beyond the $5\sigma$ contours and end up with a too low value $\Omega^0_m<0.27$. These results are aligned with those of \cite{Zhai:2017vvt}, in which the authors are also unable to accommodate the $H_0^{\rm Riess}$ value when a string of SNIa+BAO+$H(z)$+LSS+CMB data (similar but {\it not} equal to the one used by us) is taken into account. Moreover, we observe in Fig.\,3 not only that both the RVM and  $w$RVM remain much closer to  $H^{\rm Planck}_0$ than to $H_0^{\rm Riess}$, but also that they are overlapping with the $H^{\rm Planck}_0$  range much better than the $\CC$CDM does. {The latter is seen to have serious difficulties in reaching the Planck range unless we use the most external regions of the elongated contours shown in Fig.\,3.}
\newline
\newline
Many other works in the literature have studied the existing $H_0$ tension. For instance, in \cite{Wang:2017yfu} the authors find $H_0 = 69.13\pm 2.34$ km/s/Mpc assuming the $\Lambda$CDM model. {Such result almost coincides with the central values of $H_0$ that we obtain in Tables 1 and 2 for the $\Lambda$CDM.  This fact, added to the larger uncertainties of the result, seems to relax the tension. Let us, however, notice that the value of \cite{Wang:2017yfu} has been obtained using BAO data only, what explains the larger uncertainty that they find. In our case, we have considered a much more complete data set, which includes CMB and LSS data as well. This is what has allowed us to better constrain $H_0$ with smaller errors and conclude that when a larger data set (SNIa+BAO+$H(z)$+LSS+CMB)  is used, the fitted value of the Hubble parameter for the $\Lambda$CDM is incompatible with the Planck best-fit value at about $4\sigma$ c.l. Thus, the $\Lambda$CDM model seems to be in conflict not only with the local HST estimation of $H_0$, but also with the Planck one!}
{Finally, in Figs. 4 and 5 we consider the contour plots (up to $4\sigma$ and $3\sigma$, respectively) in the $(H_0,\sigma_8(0))$-plane for different situations. Specifically, in the case of Fig.\,4 the plots on the left and on the right are in exact correspondence with the situations previously presented in the left and right plots of Fig.\,3, respectively\footnote{The $H_0^{\rm Planck}$ band indicated in Figs. 3-5 is that of \cite{Aghanim:2016yuo}, which has no significant differences with that of \cite{Ade:2015xua}}. As expected, the contours in the left plot of Fig.\,4 are slightly shifted (``attracted'') to the right (i.e. towards the $H_0^{\rm Riess}$ region) as compared to those in the right plot because in the former $H_0^{\rm Riess}$ was included as a data point in the fit, whereas $H_0^{\rm Riess}$ was not included in the latter. Therefore, in the last case the contours for the RVM are more centered in the $H_0^{\rm Planck}$ region and at the same time centered at relatively low values of $\sigma_8(0)\simeq0.73-0.74$, which are precisely those needed for a perfect matching with the experimental data points on structure formation (cf. Fig. 1). On the other hand, in the case of Fig. 5 the contour lines correspond to the fitting sets Ia, IIIa of Table 3 (in which BAO and CMB data, but \emph{no} LSS formation data, are involved). As can be seen, the contour lines in Fig.\,5 can attain the Riess 2016 region for $H_0$, but they are centered at rather high values ($\sim 0.9$) of the parameter $\sigma_8(0)$. These are clearly higher than the needed values $\sigma_8(0)\simeq 0.73-0.74$. This fact demonstrates once more that such option leads to a bad description of the structure formation data.
The isolated point in Fig.\,5 is even worst: it corresponds to the aforementioned theoretical prediction for the scenario IIIb proposed in \cite{DiValentino:2017iww}, in which the $H_0^{\rm Riess}$ region can be clearly attained but at the price of a serious disagreement with the LSS data. Here we can see, with pristine clarity, that such isolated point, despite it comfortably reaches the $H_0^{\rm Riess}$ region, it attains a value of  $\sigma_8(0)$ near $1$, thence completely strayed from the observations. This is, of course, the reason  why the upper curve in Fig.\,2 fails to describe essentially all points of the $f(z)\sigma_8(z)$ observable. So, as it turns, it is impossible to reach the $H_0^{\rm Riess}$ region without paying a high price, no matter what strategy is concocted to approach it in parameter space.}
\begin{figure}[t!]
\begin{center}
\label{contours_H0_tension_chapter}
\includegraphics[width=3in]{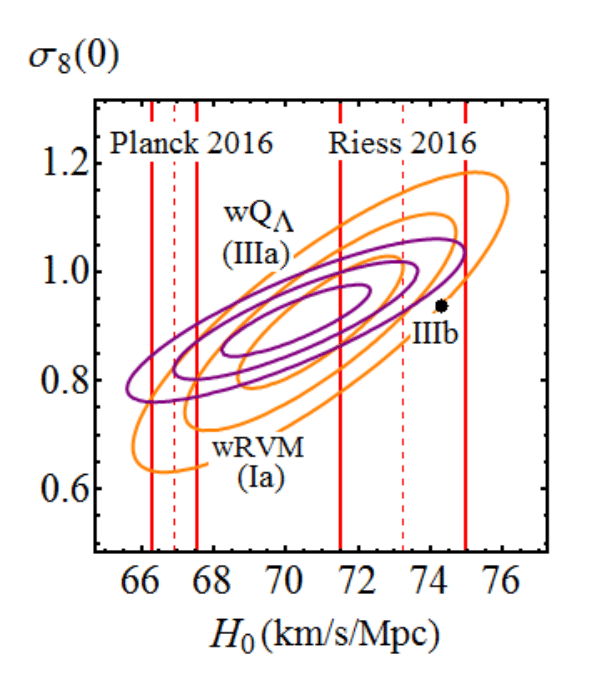}
\caption{\scriptsize {Contour lines for the models $w$RVM (Ia) and $w$Q$_\CC$ (IIIa) up to $3\sigma$ in the $(H_0,\sigma_8(0))$-plane, depicted in orange and purple, respectively, together with the isolated point (in black) extracted from the analysis of Ref. \cite{DiValentino:2017iww}, which we call IIIb. The cases Ia, IIIa and IIIb correspond to special scenarios with $w\neq -1$ for Models I and III in which the value $H_0^{\rm Riess}$  is included as a data point and then a suitable strategy is followed to optimize the fit agreement with such value. The strategy consists to exploit the freedom in $w$ and remove the LSS data from the fit analysis. The plot clearly shows that some agreement is indeed possible, but only if  $w$ takes on values in the phantom region ($w<-1$) (see text) and at the expense of an anomalous (too large) value of the parameter $\sigma_8(0)$, what seriously spoils the concordance with the LSS data, as can be seen in Fig. 2.
}}
\end{center}
\end{figure}
As indicated, we must still remain open to the possibility that the $H^{\rm Planck}_0$ and/or $H^{\rm Riess}_0$ measurements are affected by some kind of (unknown) systematic errors, although some of these possibilities may be on the way of  being ruled out by recent works. For instance, in \cite{Aylor:2017haa} the authors study the systematic errors in Planck's data by comparing them with the South Pole Telescope data. Their conclusion is that there is no evidence of systematic errors in Planck's results. If confirmed, the class of the $(w)$RVMs studied here would offer a viable solution to both the $H_0$ and $\sigma_8(0)$ existing tensions in the data, which are both unaccountable within the $\CC$CDM. Another interesting result is the ``blinded''  determination of $H_0$  from \cite{Zhang:2017aqn}, based on a reanalysis of the SNIa and Cepheid variables data from the older work by Riess et al. \cite{Riess:2011yx}. These authors find $H_0 = 72.5\pm 3.2$ km/s/Mpc, which should be compared with $H_0 = 73.8\pm 2.4$ km/s/Mpc\,\cite{Riess:2011yx}. Obviously, the tension with $H^{\rm Planck}_0$ diminished since the central value decreased and in addition the uncertainty has grown by $\sim 33\%$. We should now wait for a similar reanalysis to be made on the original sample used in \cite{Riess:2016jrr}, i.e. the one supporting the value  $H^{\rm Riess}_0$, as planned in \cite{Zhang:2017aqn}. { In \cite{Addison:2017fdm}  they show that by combining the latest BAO results with WMAP, Atacama Cosmology Telescope (ACT), or South Pole Telescope (SPT) CMB data produces values of $H_0$  that are $2.4-3.1\sigma$ lower than the distance ladder, independent of Planck. These authors conclude from their analysis  that it is not possible to explain the $H_0$  disagreement solely with a systematic error specific to the Planck data}. Let us mention other works, see e.g. \cite{Cardona:2016ems,Feeney:2017sgx}, in which a value closer to $H_0^{\rm Riess}$ is found and the tension is not so severely loosened; or the work\,\cite{Follin:2017ljs}, which excludes systematic bias or uncertainty in the Cepheid calibration step of the distance ladder measurement by\,\cite{Riess:2016jrr}. {Finally, we recall the aforementioned recent study \cite{Lin:2017bhs}, where the authors run a new (dis)cordance test to compare the constraints on $H_0$ from different methods and conclude that the local
measurement is an outlier compared to the others, what would favor a systematics-based explanation.}
Quite obviously, the search for a final solution to the $H_0$ tension is still work in progress.
\newline
\subsection{Conclusions}\label{Sec:Conclusions_H0_tension_chapter}
{The present updated analysis of the cosmological data SNIa+BAO+$H(z)$+LSS+CMB disfavors the hypothesis $\CC=$const. as compared to the dynamical vacuum models (DVMs). This is consistent with our most recent studies\,\cite{Sola:2015wwa,Sola:2016jky,Sola:2017jbl}. Our results suggest a dynamical DE effect near $3\sigma$ within the standard XCDM parametrization and near $4\sigma$ for the best DVMs.  Here we have extended these studies in order to encompass the class of quasi-vacuum models ($w$DVMs), where the equation of state parameter $w$ is near (but not exactly equal) to $-1$. The new degree of freedom $w$ can then be used to try to further improve the overall fit to the data. But it can also be used to check if values of $w$ different from $-1$ can relax the existing tension
between the two sets of  measurement of the $H_0$ parameter, namely those based:  i) on the CMB measurements by the Planck collaboration\,\cite{Ade:2015xua,Aghanim:2016yuo}, and ii) on the local measurement (distance ladder method) using Cepheid variables\,\cite{Riess:2016jrr}}
{Our study shows that the RVM with $w=-1$ remains as the preferred DVM for the optimal fit of the data. At the same time it favors the CMB measurements of $H_0$ over the local measurement.  Remarkably, we find that not only the CMB and BAO data, but also the LSS formation data (i.e. the known data on $f(z)\sigma_8(z)$ at different redshifts), are essential to support the CMB measurements of $H_0$ over the local one. We have checked that if the LSS data are not considered (while the BAO and CMB are kept), then there is a unique chance to try to accommodate the local measurement of $H_0$, but only at the expense of a phantom-like behaviour (i.e. for $w<-1$). In this region of the parameter space, however, we find that the agreement with the LSS formation data is manifestly lost, what suggests that the $w<-1$ option is ruled out. There is no other window in the parameter space where to accommodate the local $H_0$ value in our fit.  In contrast, when the LSS formation data are restored, the fit quality to the overall SNIa+BAO+$H(z)$+LSS+CMB observations improves dramatically and definitely favors the Planck range for $H_0$  as well as smaller values for $\sigma_8(0)$ as compared to the $\CC$CDM.}
{In short, our work suggests that signs of dynamical vacuum energy are encoded in the current cosmological observations. They  appear to be more in accordance with the lower values of $H_0$ obtained from the Planck (CMB) measurements than with the higher range of $H_0$ values obtained  from the present local (distance ladder) measurements, and provide smaller values of $\sigma_8(0)$ that are in better agreement with structure formation data as compared to the $\CC$CDM. We hope that with new and more accurate observations, as well as with more detailed analyses, it will be possible to assess the final impact of vacuum dynamics on the possible solution of the current tensions in the $\CC$CDM. }
\newpage
\newpage\phantom{123}
\newpage\phantom{123}

\section{Signs of dynamical dark energy in the current observations}\label{Signs_chapter}
In this chapter we assess the status of some dynamical dark energy (DDE) models, putting them in the light of a large body of cosmological observations. Their theoretical predictions are studied not only at the background level but also at the level of perturbations. In order to carry out a proper evaluation of the performance of the DDE models studied, we confront them with the standard model of cosmology, the $\Lambda$CDM, by computing the Bayes factor. This quantity involves some integrals which are far from be easy to solve. For this reason we use the numerical package called \texttt{MCEvidence}. We gauge the impact of the bispectrum in the LSS and BAO parts, and show that the subset of CMB+BAO+LSS observations may contain the bulk of the DDE signal. The chapter is organized in the following way: In Sec. \ref{sect:DDEmodels_signs_chapter} we present the different DDE models considered and we list the relevant background equations for each model. We would like to remark that while the perturbation equations are not provided in this chapter, all the modifications, in regards to this issue have been duly taken into account in the codes necessary to carry out the fits. In Sec. \ref{sect:Data_signs_chapter} we provide some details of  the cosmological data considered to test the models. In Sec. \ref{sect:SpectrumBispectrum_signs_chapter} we provide the necessary theoretical details to understand the difference between the spectrum and the bispectrum. We remark the importance of considering the bispectrum data in addition to the spectrum one to be able to capture the signs of the DDE. In Sec. \ref{sect:LinStructure_signs_chapter} we comment on the results displayed in the different tables and figures presented throughout this chapter. The aim of the next section, Sec. \ref{sect:Bayesian_evidence_signs_chapter}, is to explain why the Bayes factor is the perfect statistical tool to confront two different cosmological model and conclude which one is more favoured by the data. Finally in Sec. \ref{Sect:Conclusions_signs_chapter} we deliver our conclusions gathered after carefully study all the models under examination.

%
%
%
\renewcommand{\arraystretch}{1.1}
\begin{table}[t!]
\setcounter{table}{0}
\begin{center}
\resizebox{1\textwidth}{!}{

\begin{tabular}{|c  ||c | c | c || c | c |c | c |c|}
 \multicolumn{1}{c}{} & \multicolumn{3}{c}{DS1 with Spectrum (DS1/SP)} & \multicolumn{4}{c}{DS1 with Bispectrum (DS1/BSP)}
\\\hline
{\scriptsize Parameter} & {\scriptsize $\Lambda$CDM} & {\scriptsize XCDM} & {\scriptsize $\phi$CDM} & {\scriptsize $\Lambda$CDM} & {\scriptsize XCDM} & {\scriptsize CPL} & {\scriptsize $\phi$CDM}
\\\hline
{\scriptsize $H_0$ (km/s/Mpc)} & {\scriptsize $69.22^{+0.48}_{-0.49}$} & {\scriptsize $68.97^{+0.76}_{-0.79}$} & {\scriptsize $68.70^{+0.66}_{-0.61}$} & {\scriptsize $68.21^{+0.40}_{-0.38}$} & {\scriptsize $67.18^{+0.63}_{-0.68} $} & {{\scriptsize $67.17\pm 0.72$}} & {{\scriptsize $67.19^{+0.67}_{-0.64} $}}
\\\hline
$\omega_{cdm}$ & {\scriptsize $0.1155^{+0.0011}_{-0.0010}$} & {\scriptsize $0.1151^{+0.0013}_{-0.0014}$} & {\scriptsize $0.1147^{+0.0013}_{-0.0012}$} & {\scriptsize $0.1176^{+0.0008}_{-0.0009}$} & {\scriptsize $0.1161\pm 0.0012$} & {{\scriptsize $0.1161^{+0.0015}_{-0.0014}$}} & {\scriptsize $0.1160^{+0.0013}_{-0.0012}$}
\\\hline
$\omega_{b}$ & {{\scriptsize$0.02247^{+0.00020}_{-0.00019}$}} & {{\scriptsize$0.02251^{+0.00021}_{-0.00020}$}} & {{\scriptsize$0.02255^{+0.00020}_{-0.00022}$}} & {\scriptsize$ 0.02231^{+0.00019}_{-0.00018}$} & {{\scriptsize$0.02244^{+0.00021}_{-0.00020}$}} & {{\scriptsize$0.02244\pm 0.00021$}} &  {{\scriptsize $0.02245\pm 0.00021$}}
\\\hline
$\tau$ & {{\scriptsize$0.079^{+0.012}_{-0.013}$}} & {{\scriptsize$0.084^{+0.014}_{-0.017}$}} & {{\scriptsize$0.088\pm 0.016$}} & {{\scriptsize$0.059^{+0.012}_{-0.009}$}} & {{\scriptsize$0.074^{+0.013}_{-0.015}$}} & {{\scriptsize$0.074^{+0.016}_{-0.017}$}} &  {{\scriptsize$0.074^{+0.014}_{-0.016}$}}
\\\hline
$n_s$ & {{\scriptsize$0.9742\pm 0.0042$}} & {{\scriptsize$0.9752^{+0.0048}_{-0.0052}$}} & {{\scriptsize$0.9768\pm 0.0048$}} & {{\scriptsize$0.9678\pm 0.0038$}} & {{\scriptsize$0.9724\pm 0.0047$}} &  {{\scriptsize$0.9724^{+0.0049}_{-0.0052}$}} &  {{\scriptsize$0.9729^{+0.0044}_{-0.0048}$}}
\\\hline
$\sigma_8(0)$  & {{\scriptsize$0.813^{+0.008}_{-0.009}$}} & {{\scriptsize$0.811\pm 0.010$}} & {{\scriptsize$0.808\pm 0.010$}} & {\scriptsize $0.804^{+0.007}_{-0.008}$} & {{\scriptsize$0.795^{+0.010}_{-0.009}$}} & {{\scriptsize$0.795 \pm 0.010$}} &  {{\scriptsize$0.793\pm 0.009$}}
\\\hline
$w_0$ & {\scriptsize -1} & {{\scriptsize $-0.986^{+0.030}_{-0.029}$}} & - & {\scriptsize -1} &  {{\scriptsize$-0.945^{+0.029}_{-0.028}$}} &  {{\scriptsize$-0.934^{+0.067}_{-0.075}$}} & -
\\\hline
$w_1$ & - & - & - & - &  - & {{\scriptsize $-0.045^{+0.273}_{-0.204}$}} & -
\\\hline
$\left(\alpha,  10^{-3}\bar{\kappa}\right)$ & - & - & { \scriptsize $\left(<0.092, 37.3^{+1.9}_{-2.3}\right)$ } & - &  - & - & {\scriptsize $\left(0.150^{+0.070}_{-0.086}\right.$}, {\scriptsize $\left.33.5^{+1.1}_{-2.1}\right)$}
\\\hline
\end{tabular}}
\end{center}
\label{tableFit1_signs_chapter}
\caption{\scriptsize The mean fit values and $68.3\%$ confidence limits for  the considered models using dataset DS1, i.e.  all SNIa+$H(z)$+BAO+LSS+CMB data, with full Planck 2015 CMB likelihood.  In all cases a massive neutrino of  $0.06$ eV has been included. The first block involves BAO+LSS data using the matter (power) spectrum (SP) and is  labelled  DS1/SP. The second block includes both  spectrum and bispectrum, and is denoted DS1/BSP (see text). We display the fitting results for the relevant parameters, among them those that characterize the DDE models under discussion: the EoS parameter $w_0$ for XCDM, $w_0$ and $w_1$ for the CPL, the power $\alpha$ of the potential and $\bar{\kappa}\equiv\kappa [{\rm M_P/(km/s/Mpc)}]^2$ for $\phi$CDM, as well as the six conventional parameters:  the Hubble parameter $H_0$, $\omega_{cdm}=\Omega_{cdm} h^2$ and $\omega_{b}=\Omega_{b} h^2$ for cold dark matter and baryons, the reionization optical depth $\tau$, the spectral index $n_s$ of the primordial power-law power spectrum, and, for convenience, instead of the amplitudes $A_s$ of such spectrum we list the values of $\sigma_8(0)$.}
\end{table}
\subsection{DDE parametrizations and models}\label{sect:DDEmodels_signs_chapter}
We  consider two generic parametrizations of the DDE, together with a well-known  $\phi$CDM model, and confront them to a large and updated set of SNIa+$H(z)$+BAO+LSS+CMB observations.  Natural units are used hereafter, although we keep explicitly Newton's $G$, or equivalently the Planck mass:  $M_P = 1/\sqrt{G} = 1.2211\times 10^{19}$ GeV. Flat FLRW metric is  assumed throughout: $ds^2=-dt^2+a^2(t)(dx^2+dy^2+dz^2)$, where $a(t)$ is the scale factor as a function of the cosmic time.
\subsubsection{XCDM and CPL}
The first of the DDE parametrizations under study is the conventional XCDM\,\cite{Turner:1998ex}. In it both matter and DE are self-conserved (non-interacting) and the DE density is simply given by $\rho_{X}(a) = \rho_{X0}a^{-3(1+w_0)}$, where $\rho_{X0} = \rho_\Lambda$ is the current value and $w_0$ the (constant)  equation of state (EoS) parameter of the DE fluid.
For $w_0=-1$ we recover the $\Lambda$CDM model with a rigid CC. For $w_0 \gtrsim-1$ the XCDM mimics quintessence, whereas for $w_0 \lesssim -1$ it mimics phantom DE.
It is worth checking if a dynamical EoS for the DE can furnish a better description of the observational data. So we consider the well-known CPL parametrization\,\cite{Chevallier:2000qy,Linder:2002et}, which is characterized by the following  EoS:
%
\renewcommand{\arraystretch}{1.1}
\begin{table}[t!]
\begin{center}
\resizebox{1\textwidth}{!}{

\begin{tabular}{|c  ||c | c | c || c | c |c |}
 \multicolumn{1}{c}{} & \multicolumn{3}{c}{DS2 with Bispectrum  (DS2/BSP)} & \multicolumn{3}{c}{DS2/BSP with Planck 2018 (compressed likelihood)}
\\\hline
{\scriptsize Parameter} & {\scriptsize $\Lambda$CDM} & {\scriptsize XCDM}& {\scriptsize $\phi$CDM} & {\scriptsize $\Lambda$CDM} & {\scriptsize XCDM} & {\scriptsize $\phi$CDM}
\\\hline
{\scriptsize $H_0$ (km/s/Mpc)} & {\scriptsize $68.20^{+0.38}_{-0.41}$} & {\scriptsize $66.36^{+0.76}_{-0.86} $} & {\scriptsize $66.45\pm 0.74 $} & {\scriptsize $68.92\pm 0.31 $} & {\scriptsize $67.08\pm 0.73 $} & {\scriptsize $67.02\pm 0.70 $}
\\\hline
$\omega_{cdm}$ & {{\scriptsize$0.1176^{+0.0008}_{-0.0009}$}} & {{\scriptsize$0.1155^{+0.0014}_{-0.0012}$}} & {{\scriptsize$0.1154\pm 0.0013$}} & {{\scriptsize$0.1197\pm 0.0007$}} & {{\scriptsize$0.1191\pm 0.0008$}} & {{\scriptsize$0.1191\pm 0.0008$}}
\\\hline
$\omega_{b}$ & {{\scriptsize$0.02230\pm 0.00019$}} & {{\scriptsize$0.02247\pm 0.00021$}} & {{\scriptsize$0.02248^{+0.00020}_{-0.00021}$}} & {{\scriptsize$0.02254\pm 0.00013$}} & {{\scriptsize$0.02259\pm 0.00013$}} & {{\scriptsize$0.02259\pm 0.00013$}}
\\\hline
$\tau$ & {{\scriptsize$0.059\pm 0.010$}} & {{\scriptsize$0.083^{+0.014}_{-0.018}$}} & {{\scriptsize$0.083^{+0.013}_{-0.014}$}} & - & - & -
\\\hline
$n_s$ & {{\scriptsize$0.9681\pm 0.0039$}} & {{\scriptsize$0.9742^{+0.0048}_{-0.0053}$}} & {{\scriptsize$0.9746^{+0.0044}_{-0.0051}$}} & {{\scriptsize$0.9727\pm 0.0033$}} & {{\scriptsize$0.9734\pm 0.0032$}} & {{\scriptsize$0.9735\pm 0.0032$}}
\\\hline
$\sigma_8(0)$  & {{\scriptsize$0.805\pm 0.007$}} & {{\scriptsize$0.788\pm 0.010$}} & {{\scriptsize$0.789\pm 0.010$}} & {{\scriptsize$0.800\pm 0.008$}} & {{\scriptsize$0.774\pm 0.013$}} & {{\scriptsize$0.773\pm 0.012$}}
\\\hline
$w_0$ & {\scriptsize -1} & {{\scriptsize$-0.911^{+0.035}_{-0.034}$}} & - & {\scriptsize -1} & {{\scriptsize$-0.932\pm 0.025$}} &  -
\\\hline
$\left(\alpha,  10^{-3}\bar{\kappa}\right)$ & - & - & {{\scriptsize$\left(0.240^{+0.086}_{-0.102}\right.$}, {\scriptsize $\left.32.0^{+0.8}_{-1.4}\right)$}} & - & - &  {\scriptsize $\left(0.168\pm 0.068\right.$}, {\scriptsize $\left.32.7\pm 1.5\right)$}
\\\hline
\end{tabular}}
\end{center}
\label{tableFit1_signs_chapter}
\caption{\scriptsize As in Table 1, but using dataset DS2/BSP, which involves BAO+LSS+CMB data only. In the first block we use the full likelihood for Planck 2015, whereas in the second we use the compressed CMB likelihood for the more recent Planck 2018  data. See text for more details.}
\end{table}
%
\begin{equation}\label{eq:CPL_signs_chapter}
w(a) = w_0 + w_1(1-a) = w_0 + w_1\frac{z}{1+z}\,,
\end{equation}
where $z=a^{-1}-1$ is the cosmological redshift.
\newline
The corresponding Hubble rate $H=\dot{a}/a$ normalized with respect to the current value, $H_0=H(a=1)$, takes the form:
\begin{equation}\label{Hubble_function_DDE_PAR_signs_chapter}
E^2(a)\equiv\frac{H^2(a)}{H^2_0}=(\Omega_b+\Omega_{cdm}){a^{-3}} + \Omega_\gamma{a^{-4}}+ \frac{\rho_\nu(a)}{\rho_{c 0}} + \Omega_\Lambda{a^{-3(1+w_0+w_1)}}e^{-3w_1(1-a)}.
\end{equation}
Here $\Omega_i={\rho_{i 0}}/{\rho_{c 0}}$ are the current energy densities of baryons, cold dark matter, photons and CC/DE normalized with respect to the present critical density $\rho_{c 0}$.
The neutrino contribution,  $\rho_\nu(a)$ is more complicated since it contains a massive component, $\rho_{\nu, m}(a)$, apart from the massless ones. During the expansion of the Universe, the massive neutrino transits from a relativistic to a non-relativistic regime.  This process is nontrivial and has to be solved numerically. In \eqref{Hubble_function_DDE_PAR_signs_chapter}, for $w_1=0$ we recover the XCDM, and also setting   $w_0=-1$ we are back to the $\CC$CDM.
%
\begin{figure}
\begin{center}
\setcounter{figure}{0}
\label{FigContour_signs_chapter}
\includegraphics[width=4.3in, height=2.4in]{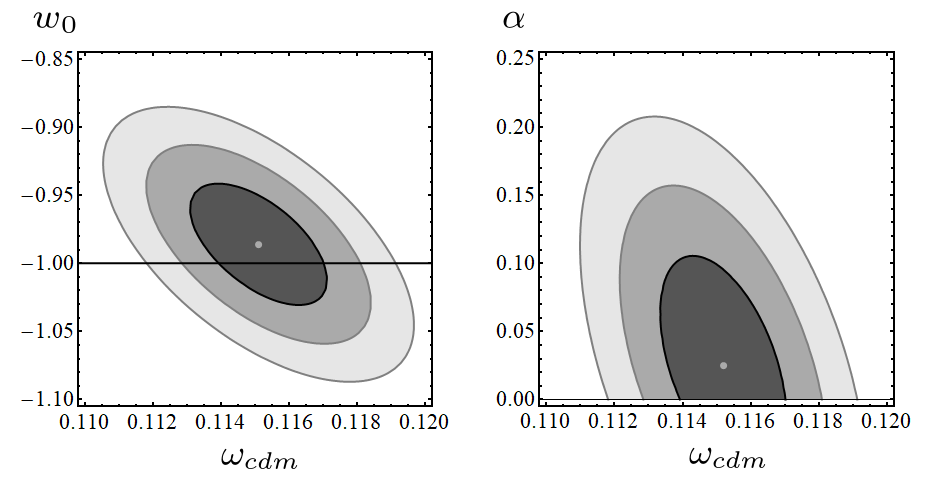}
\caption{\scriptsize {Likelihood contours for the XCDM  parametrization (left) and the considered $\phi$CDM model  (right) in the relevant planes $(\omega_{cdm},w_0)$ and $(\omega_{cdm},\alpha)$, respectively, after marginalizing over the remaining parameters. Dataset DS1/SP  is used in both cases (cf.  first block of Table 1). The various contours correspond to $1\sigma$, $2\sigma$ and $3\sigma$ c.l.}}
\end{center}
\end{figure}
%
%
\subsubsection{$\phi$CDM}
Let us now briefly summarize the theoretical framework of the $\phi$CDM, which has a well-defined local Lagrangian description. The DE is described here in terms of a scalar field, $\phi$, which we take dimensionless. Such field is minimally coupled to curvature ($R$) and the generic action is just the sum of the Einstein-Hilbert action $S_{\rm EH}$, the scalar field action $S_\phi$, and the matter action $S_m$:
\begin{equation}\label{eq:PhiCDMLagrangian_signs_chapter}
S=S_{\rm EH}+S_\phi+S_m=\frac{M_P^2}{16\pi}\int d^4 x \,\sqrt{-g}\,\left[R-\frac12\,g^{\mu\nu}\,\partial_{\mu}\,\phi\,\partial_\nu\phi- V(\phi)\right]+S_m\,.
\end{equation}
The energy density and pressure of $\phi$ follow from its energy-momentum tensor, $T^{\phi}_{\mu\nu}=\frac{-2}{\sqrt{-g}}\frac{\delta S_{\phi}}{\delta g^{\mu\nu}}$, and the fact that $\phi$ is a homogeneous scalar field which depends only on the cosmic time. Thus,
\begin{equation}\label{eq:densitypressure_signs_chapter}
\rho_\phi=T^{\phi}_{00} = \frac{M^2_P}{16\pi}\left( \frac{\dot{\phi}^2}{2} + V(\phi)\right) \quad p_\phi =T^{\phi}_{ii}= \frac{M^2_P}{16\pi}\left( \frac{\dot{\phi}^2}{2} - V(\phi)\right).
\end{equation}
Dots indicate derivatives with respect to the cosmic time, and notice that within our conventions $V(\phi)$ has dimension 2 in natural units.   
\newline
The field equation for $\phi$ is just the Klein-Gordon equation in curved space-time, $\Box\phi+\partial V/\partial\phi=0$, which for the FLRW metric leads to
\begin{equation}\label{eq:KleinGordon_signs_chapter}
\ddot\phi+3H\dot{\phi}+\frac{\partial V}{\partial\phi}=0\,.
\end{equation}
The corresponding Einstein's field equations read:
\begin{eqnarray}
&&3H^2=8\pi\,G\,(\rho_m+\rho_r+\rho_\phi)\label{eq:FriedmannEq_signs_chapter}\\
&&3H^2+2\dot{H}=-8\pi\,G\,(p_m+p_r+p_\phi)\label{eq:PressureEq_signs_chapter}\,.
\end{eqnarray}
Here $\rho_m=\rho_b+\rho_{cdm}+\rho_{\nu,m}$  involves the pressureless contributions from baryons and cold dark matter as well as the massive neutrino contribution. The latter evolves during the cosmic expansion from the relativistic regime (where $p_m=p_{\nu,m}\neq0$) to the non-relativistic one (where $p_{\nu,m}\simeq 0$).
On the other hand, $p_r=\rho_r/3$ is the purely relativistic part from photons and the massless neutrinos.
%
\begin{figure}
\begin{center}
\label{FigContour_signs_chapter}
\includegraphics[width=4.3in, height=2.4in]{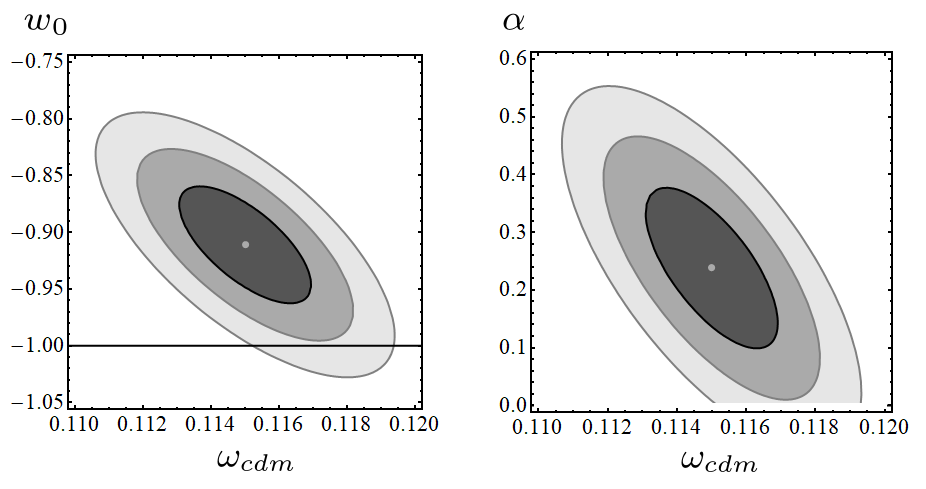}
\caption{\scriptsize{As in Fig. 1, but using dataset DS2/BSP. The various contours correspond, again, to $1\sigma$, $2\sigma$ and $3\sigma$ c.l.
The central values in both cases are shifted  $>2.5\sigma$ away from the $\CC$CDM, i.e. from $w_0 = -1$ and $\alpha=0$ in each case. Marginalization also over $\omega_{cdm}$ increases the c.l. up to $2.6\sigma$ and $2.9\sigma$, respectively  (cf.  first block of Table 2). }}
\end{center}
\end{figure}
%
As a representative potential for our analysis we borrow the traditional quintessence potential by Peebles and Ratra\,\cite{Peebles:1987ek,Ratra:1987rm},
\begin{equation}\label{eq:PRpotential_signs_chapter}
V(\phi) = \frac12\kappa M_P^2\phi^{-\alpha}\,,
\end{equation}
where $\kappa$ is a dimensionless parameter. In this context, the value of the CC at present ($t=t_0$) is given by $\rL=\rho_\phi(t_0)$. We expect $\kappa$  to be positive since we know that $\rho_\Lambda>0$ and the potential energy at present is dominant. In what follows, when referring to the $\phi$CDM model we will implicitly assume this form of the potential.  The power $\alpha$ in it should be positive as well, but sufficiently small so that $V(\phi)$ mimics an approximate CC slowly evolving with time.  In our case we take  $\alpha$  as a free parameter while $\kappa$ is a derived one, as discussed below.
\newline
The Klein-Gordon equation with the potential (\ref{eq:PRpotential_signs_chapter}) can be  written in terms of the scale factor as follows (prime stands for $d/da$, and we use $d/dt=a H d/da$):
\begin{equation}\label{eq:KGa_signs_chapter}
\phi^{\prime\prime}+\left(\frac{{H}^\prime}{{H}}+\frac{4}{a}\right) \phi^\prime-\frac{\alpha}{2}\frac{{\kappa{M^2_P}}\phi^{-(\alpha+1)}}{(a {H})^2}=0\,.
\end{equation}
When the cosmic evolution is characterized by the dominant energy density  $\rho(a)\propto {a}^{-n}$ ($n=3$ for the matter-dominated epoch and $n=4$ for radiation-dominated), i.e. when the DE density is negligible, one can search for power-law solutions of the Klein-Gordon equation, i.e. solutions of the form $\phi\propto a^p$.
However,  around the current time the effect of the potential is significant and the solution has to be computed numerically.
Equation (\ref{eq:KGa_signs_chapter}) is coupled to the cosmological equations under appropriate initial conditions. We can set these conditions at the radiation epoch, where we can neglect the DE. We find
\begin{equation}\label{eq:Initialphi_signs_chapter}
\phi(a) = \left[\frac{3\alpha(\alpha +2)^2 \kappa M_P^4}{64\pi (\alpha +6)}\ \left(\rho_r(a)+\rho_{\nu,m}\right(a))^{-1}\right]^{1/(\alpha +2)}\,.
\end{equation}
Notice that $\rho_{r,\rm tot}\equiv\rho_r+\rho_{\nu,m}$ is the total radiation energy density, which includes also the massive neutrino contribution. At the radiation epoch, however, $\rho_{r,\rm tot}$ behaves very approximately as $\sim a^{-4}$. Thus, Eq.(\ref{eq:Initialphi_signs_chapter}) together with $\phi^{\prime}(a)$ obtained from it can be used for the numerical solution of Eq.\,(\ref{eq:KGa_signs_chapter}) and of the EoS  of the scalar field at any subsequent epoch: $w_{\phi}(a)=p_{\phi}(a)/\rho_{\phi}(a)$.
\newline
\newline
To compare the theoretical predictions of the different models under study with the available observational data we have made use of the Einstein-Boltzmann code \texttt{CLASS} \cite{Blas:2011rf} in combination with the powerful Monte Carlo Markov chain (MCMC) sampler \texttt{MontePython}  \cite{Audren:2012wb}. In the particular instance of $\phi$CDM, we have conveniently modified \texttt{CLASS} such as to implement the shooting method, see \cite{StoerBulirsch1980} for a detailed exposition, what allowed us to consistently determine the value of $\kappa$ for each value of the free parameter $H_0$. As it is well-known, such numerical technique consists in replacing a boundary value problem with an initial value one through a large number of iterations of the initial conditions until finding the optimized solution. For the $\phi$CDM, the initial conditions are determined by (\ref{eq:Initialphi_signs_chapter}) and its time derivative, which are both explicitly dependent on $\kappa$.
%
%
\renewcommand{\arraystretch}{1.1}
\begin{table}[t!]
\begin{center}

\begin{tabular}{|c  ||c | c | c |}
 \multicolumn{1}{c}{} & \multicolumn{3}{c}{DS3 with Spectrum  (DS3/SP)}
\\\hline
{\scriptsize Parameter} & {\scriptsize $\Lambda$CDM} & {\scriptsize XCDM} & {\scriptsize $\phi$CDM}
\\\hline
{\scriptsize $H_0$ (km/s/Mpc)} & {\scriptsize $69.30^{+0.53}_{-0.52}$} & {\scriptsize $69.26^{+0.80}_{-0.79} $} & {\scriptsize $68.80^{+0.65}_{-0.58} $}
\\\hline
$\omega_{cdm}$ & {{\scriptsize$0.1154\pm 0.0011$}} & {{\scriptsize$0.1156^{+0.0014}_{-0.0015}$}} & {{\scriptsize$0.1147^{+0.0014}_{-0.0011}$}}
\\\hline
$\omega_{b}$ & {{\scriptsize$0.02249^{+0.00019}_{-0.00021}$}} & {{\scriptsize$0.02247\pm 0.00021$}} & {{\scriptsize$0.02254^{+0.00019}_{-0.00022}$}}
\\\hline
$\tau$ & {{\scriptsize$0.086^{+0.012}_{-0.014}$}} & {{\scriptsize$0.085^{+0.015}_{-0.016}$}} & {{\scriptsize$0.093^{+0.013}_{-0.014}$}}
\\\hline
$n_s$ & {{\scriptsize$0.9750^{+0.0045}_{-0.0044}$}} & {{\scriptsize$0.9746^{+0.0051}_{-0.0050}$}} & {{\scriptsize$0.9769^{+0.0044}_{-0.0049}$}}
\\\hline
$\sigma_8(0)$  & {{\scriptsize$0.819^{+0.009}_{-0.010}$}} & {{\scriptsize$0.818\pm 0.011$}} & {{\scriptsize$0.813\pm 0.010$}}
\\\hline
$w_0$ & {\scriptsize -1} & {{\scriptsize$-1.002^{+0.034}_{-0.035}$}} & -
\\\hline
$\left(\alpha,  10^{-3}\bar{\kappa}\right)$ & - & - & { \scriptsize $\left(<0.081, 37.6\pm 2.1\right)$ }
\\\hline
\end{tabular}
\end{center}
\label{tableFit1_signs_chapter}
\caption{\scriptsize As in Tables 1 and 2, but using dataset DS3/SP. The latter is similar to the one employed in the analyses of these models in \cite{Park:2018bwy,Park:2018fxx}, see the text for further details.
}
\end{table}
%
%
\subsection{Data}\label{sect:Data_signs_chapter}
The main fitting results of our analysis are presented in Tables 1 and 2. An additional Table 3 has also been introduced to illustrate the correct normalization of our results with other existing studies in the literature when we use comparable data. This issue will be further discussed throughout our presentation. To generate the fitting results displayed in Tables 1,  2 and 3 we have run the MCMC code \texttt{MontePython}, together with \texttt{CLASS}, over an updated data set SNIa+$H(z)$+BAO+LSS+CMB, consisting of: i) 6 effective points on the normalized Hubble rate (including the covariance matrix) from the Pantheon+MCT sample \cite{Scolnic:2017caz,Riess:2017lxs}, which includes 1063 SNIa. As it is explained in \cite{Riess:2017lxs}, the compression effectiveness of the information contained in such SNIa sample is extremely good; ii) 31 data from $H(z)$ from cosmic chronometers \cite{Jimenez:2003iv,Simon:2004tf,Stern:2009ep,Moresco:2012jh,Zhang:2012mp,Moresco:2015cya,Moresco:2016mzx,Ratsimbazafy:2017vga}; iii) 16 effective BAO points \cite{Kazin:2014qga,Gil-Marin:2016wya, Bourboux:2017cbm,Carter:2018vce,Gil-Marin:2018cgo}; iv) 19 effective points from LSS, specifically 18 points from the observable $f(z)\sigma_8(z)$  (obtained from redshift-space distortions -- RSD)\,\cite{Guzzo:2008ac,Song:2008qt,Blake:2011rj,Beutler:2012px,Blake:2013nif,Simpson:2015yfa,Okumura:2015lvp,Gil-Marin:2016wya,Howlett:2017asq,Shi:2017qpr,Gil-Marin:2018cgo,Mohammad:2018mdy} and one effective point from the weak lensing observable $S_8\equiv \sigma_8(0)\left({\Omega_m}/{0.3}\right)^{0.5}$ \cite{Hildebrandt:2016iqg}; v) Finally we make use of the full CMB likelihood from Planck 2015 TT+lowP+lensing\,\cite{Ade:2015xua}. It is important to emphasize that we take into account all the known correlations among data.  Owing to their special significance in this kind of analysis, we have collected the set of BAO and LSS  data points used in this chapter in the Tables 4 and 5, respectively.
\newline
The total data set just described will be referred to as DS1.  This is our baseline scenario since it involves a large sample of all sorts of cosmological data and will be used to compare with the possible additional effects that emerge when we enrich its structure, as we shall comment in a moment. For a more detailed discussion of the data involved in DS1, see \cite{Sola:2017znb,Sola:2015wwa,Sola:2017jbl}.
In this study, however, we wish to isolate also the effect from the triad of  BAO+LSS+CMB data. These ingredients may be particularly sensitive to the DDE, as shown in the previous references. Such subset of DS1 will be called  DS2 and contains the same data as DS1 except SNIa+$H(z)$. In the next section we define further specifications of these two datasets with special properties.
\begin{table}[t!]
\begin{center}
\resizebox{10cm}{!}{
\begin{tabular}{| c | c |c | c |c|c|}
\multicolumn{1}{c}{Survey} &  \multicolumn{1}{c}{$z$} &  \multicolumn{1}{c}{Observable} &\multicolumn{1}{c}{Measurement} & \multicolumn{1}{c}{{\small References}} & \multicolumn{1}{c}{{\small Data set}}
\\\hline
6dFGS+SDSS MGS & $0.122$ & $D_V(r_d/r_{d,fid})$[Mpc] & $539\pm17$[Mpc] &\cite{Carter:2018vce} & SP/BSP
\\\hline

 WiggleZ & $0.44$ & $D_V(r_d/r_{d,fid})$[Mpc] & $1716.4\pm 83.1$[Mpc] &\cite{Kazin:2014qga} & SP/BSP \tabularnewline
\cline{2-4} & $0.60$ & $D_V(r_d/r_{d,fid})$[Mpc] & $2220.8\pm 100.6$[Mpc]& &\tabularnewline
\cline{2-4} & $0.73$ & $D_V(r_d/r_{d,fid})$[Mpc] &$2516.1\pm 86.1$[Mpc] & &
\\\hline

DR12 BOSS (BSP) & $0.32$ & $Hr_d/(10^{3}km/s)$ & $11.549\pm0.385$   &\cite{Gil-Marin:2016wya} & BSP \\ \cline{3-4}
 &  & $D_A/r_d$ & $6.5986\pm0.1337$ & &\tabularnewline \cline{3-4}
 \cline{2-2}& $0.57$ & $Hr_d/(10^{3}km/s)$  & $14.021\pm0.225$ & &\\ \cline{3-4}
 &  & $D_A/r_d$ & $9.3869\pm0.1030$ & &\\\hline

DR12 BOSS (SP) & $0.38$ & $D_M(r_d/r_{d,fid})$[Mpc] & $1518\pm22$   &\cite{Alam:2016hwk} & SP \\ \cline{3-4}
 &  & $H(r_{d,fid}/r_d)$[km/s/Mpc] & $81.5\pm1.9$ & &\tabularnewline \cline{3-4}
 \cline{2-2}& $0.51$ & $D_M(r_d/r_{d,fid})$[Mpc] & $1977\pm27$ & &\\ \cline{3-4}
 &  & $H(r_{d,fid}/r_d)$[km/s/Mpc] & $90.4\pm1.9$ & &\\ \cline{3-4}
 \cline{2-2}& $0.61$ & $D_M(r_d/r_{d,fid})$[Mpc]  & $2283\pm32$ & &\\ \cline{3-4}
 &  & $H(r_{d,fid}/r_d)$[km/s/Mpc] & $97.3\pm2.1$ & &\\\hline

eBOSS & $1.19$ & $Hr_d/(10^{3}km/s)$ & $19.6782\pm1.5866$   &\cite{Gil-Marin:2018cgo}& SP/BSP \\ \cline{3-4}
 &  & $D_A/r_d$ & $12.6621\pm0.9876$ & &\tabularnewline \cline{3-4}
 \cline{2-2}& $1.50$ & $Hr_d/(10^{3}km/s)$  & $19.8637\pm2.7187$ & &\\ \cline{3-4}
 &  & $D_A/r_d$ & $12.4349\pm1.0429$ & &\\ \cline{3-4}
 \cline{2-2}& $1.83$ & $Hr_d/(10^{3}km/s)$  & $26.7928\pm3.5632$ & &\\ \cline{3-4}
 &  & $D_A/r_d$ & $13.1305\pm1.0465$ & &\\\hline

Ly$\alpha$-forest & $2.40$ & $D_H/r_d$ & $8.94\pm0.22$   &\cite{Bourboux:2017cbm} & SP/BSP
\\ \cline{3-4} &  & $D_M/r_d$ & $36.6\pm1.2$ & &\\\hline

\end{tabular}}
\caption{\scriptsize Published values of BAO data {used in the main analyses. In the last column we specify the data sets in which these points have been employed. We label the latter with SP if they are used in DS1/SP; with BSP if they are used in DS1/BSP and DS2/BSP; and finally with SP/BSP if they are used in the three data sets, DS1/SP, DS1/BSP, and DS2/BSP. See the quoted references, and the text in Sec. \ref{sect:Data_signs_chapter}}}
\end{center}
\end{table}
%
%
%
%
%
\begin{table}[t!]
\begin{center}
\resizebox{7cm}{!}{
\begin{tabular}{| c | c |c | c |c |}
\multicolumn{1}{c}{Survey} &  \multicolumn{1}{c}{$z$} &  \multicolumn{1}{c}{$f(z)\sigma_8(z)$} & \multicolumn{1}{c}{{\small References}} & \multicolumn{1}{c}{{\small Data set}}
\\\hline
2MTF & $0$ & $0.505\pm 0.084$ & \cite{Howlett:2017asq} & SP/BSP
\\\hline
6dFGS & $0.067$ & $0.423\pm 0.055$ & \cite{Beutler:2012px} & SP/BSP
\\\hline
SDSS-DR7 & $0.10$ & $0.376\pm 0.038$ & \cite{Shi:2017qpr} & SP/BSP
\\\hline
GAMA & $0.18$ & $0.29\pm 0.10$ & \cite{Simpson:2015yfa} & SP/BSP
\\ \cline{2-4}& $0.38$ & $0.44\pm0.06$ & \cite{Blake:2013nif} &
\\\hline
DR12 BOSS (BSP) & $0.32$ & $0.427\pm 0.056$  & \cite{Gil-Marin:2016wya} &  BSP\\ \cline{2-3}
 & $0.57$ & $0.426\pm 0.029$ & & \\\hline
DR12 BOSS (SP) & $0.38$ & $0.497\pm 0.045$ & \cite{Alam:2016hwk}&  SP\tabularnewline
\cline{2-3} & $0.51$ & $0.458\pm0.038$ & &\tabularnewline
\cline{2-3} & $0.61$ & $0.436\pm0.034$ & &

\\\hline
 WiggleZ & $0.22$ & $0.42\pm 0.07$ & \cite{Blake:2011rj} & SP/BSP\tabularnewline
\cline{2-3} & $0.41$ & $0.45\pm0.04$ & &\tabularnewline
\cline{2-3} & $0.60$ & $0.43\pm0.04$ & &\tabularnewline
\cline{2-3} & $0.78$ & $0.38\pm0.04$ & &
\\\hline
VIPERS & $0.60$ & $0.49\pm 0.12$ & \cite{Mohammad:2018mdy}& SP/BSP
\\ \cline{2-3}& $0.86$ & $0.46\pm0.09$ & &
\\\hline
VVDS & $0.77$ & $0.49\pm0.18$ & \cite{Guzzo:2008ac},\cite{Song:2008qt}& SP/BSP
\\\hline
FastSound & $1.36$ & $0.482\pm0.116$ & \cite{Okumura:2015lvp}& SP/BSP
\\\hline
eBOSS & $1.19$ & $0.4736\pm 0.0992$ & \cite{Gil-Marin:2018cgo}& SP/BSP \tabularnewline
\cline{2-3} & $1.50$ & $0.3436\pm0.1104$ & &\tabularnewline
\cline{2-3} & $1.83$ & $0.4998\pm0.1111$ & &

\\\hline
 \end{tabular}}
\caption{\scriptsize{As in Table 4, but for the published values of $f(z)\sigma_8(z)$. See the quoted references, and text in Sec. \ref{sect:Data_signs_chapter}}}
\end{center}
\end{table}
%
%
%
\subsection{Spectrum versus bispectrum}\label{sect:SpectrumBispectrum_signs_chapter}
The usual analyses of structure formation data in the literature are performed in terms of the matter power spectrum  $P({\bf k})$, referred to here simply as spectrum (SP). As we know, the latter is defined in terms of the two-point correlator of the density field $D({\bf k})$ in Fourier space, namely $\langle D({\bf k})\,D({\bf k}')\rangle=\delta({\bf k}+{\bf k}') P({\bf k})$, in which $\delta$ is the Dirac delta of momenta. For a purely Gaussian distribution, any higher order correlator of even order decomposes into sums of products of two-point functions, in a manner very similar to Wick's theorem in QFT. At the same time, all correlators of odd order vanish. This ceases to be true for non-Gaussian distributions, and the first nonvanishing correlator is then the bispectrum $B({\bf k}_1,{\bf k}_2,{\bf k}_3)$, which is formally connected to the three-point function
\begin{equation}\label{eq:bispectrum_signs_chapter}
\langle D({\bf k}_1)\,D({\bf k}_2)\, D({\bf k}_3)\rangle= \delta({\bf k}_1+{\bf k}_2+{\bf k}_3)B({\bf k}_1,{\bf k}_2,{\bf k}_3)\,.
\end{equation}
The Dirac $\delta$ selects in this case all the triangular configurations. Let us note that even if the primeval spectrum would be purely Gaussian, gravity makes fluctuations evolve non-Gaussian. Such deviations with respect to a normal distribution may be due both to the evolution of gravitational instabilities that are amplified from the initial perturbations, or even from some intrinsic non-Gaussianity of the primordial spectrum.  For example, certain implementations of inflation (typically multifield inflation models) unavoidably lead to a certain degree of non-Gaussianity\,\cite{liddle_lyth_2000}.  Therefore,  in practice  the bispectrum is expected to be a non-zero parameter in real cosmology, even if starting from perfect Gaussianity, which is in no way an absolute condition to be preserved.  On the other hand, such departure  should, of course,  be small.  But the dynamics of the DE is also expected to be small, so  there is a possible naturalness relationship between the two.
\newline
The bispectrum (BSP) has been described in many places in the literature, see e.g.\,\cite{liddle_lyth_2000,Amendola:2015ksp}  and references therein.  The physical importance of including the bispectrum cannot be overemphasized as it furnishes important complementary information that goes beyond the spectrum. If fluctuations in the structure formation were strictly Gaussian, their full statistical description would be contained in the two-point correlation function  $\langle D({\bf k})\,D({\bf k}')\rangle$ since, as already mentioned,  all  higher order correlators of even order can be expanded in terms of products of two-point functions.  In such a case the formal bispectrum defined above would identically vanish.  However,  there is no a priori reason for that to happen, and in general this is not what we expect if we take into consideration the reasons mentioned above. The crucial question is:  how to test the real situation in practice ?  While the above definitions are the formal ones, operationally (in other words, at the practical level of galaxy counting) one must resort to use SP and BSP estimators of empirical nature.  For the power spectrum estimator one may use $\langle F_2({\bf k}_1)F_2({\bf k}_2)\rangle$, where $F_2({\bf q})$ is the Fourier transform of an appropriately defined weighted field of density fluctuations, that is to say, one formulated in terms of the number density of galaxies\,\cite{Gil-Marin:2016wya}.
\newline
Similarly, a bispectrum estimator $\langle F_3({\bf k}_1)F_3({\bf k}_2) F_3({\bf k}_3)\rangle$ can be defined from the angle-average of closed triangles defined by the $\bf k$-modes, ${\bf k}_1,\,{\bf k}_2,\,{\bf k}_3$, where $F_3({\bf q})$ is the Fourier transform of the corresponding weighted field of density fluctuations defined  in terms of the number density of galaxies. It can be conveniently written as
\begin{equation}
 \langle F_3({\bf k}_1)F_3({\bf k}_2) F_3({\bf k}_3)\rangle=\frac{k_f^3}{V_{123}}\int d^3{\bf r}\, \mathcal{D}_{\mathcal{S}_1}({\bf r}) \mathcal{D}_{\mathcal{S}_2}({\bf r}) \mathcal{D}_{\mathcal{S}_3}({\bf r})\,,
 \label{eq:bis2_signs_chapter}
\end{equation}
i.e. through an expression involving a separate product of Fourier integrals
\begin{equation}
 \mathcal{D}_{\mathcal{S}_j}({\bf r})\equiv \int_{\mathcal{S}_j} d{\bf q}_j\, F_3({\bf q}_j)e^{i{\bf q}_j\cdot{\bf r}}\,.
\end{equation}
Here $k_f$ is the fundamental frequency, $k_f=2\pi/L_{\rm box}$, $L_{\rm box}$ the size of the box in which the galaxies are embedded and
\begin{equation}
 V_{123}\equiv\int_{\mathcal{S}_1} d{\bf q}_1\, \int_{\mathcal{S}_2} d{\bf q}_2\, \int_{\mathcal{S}_3} d{\bf q}_3\, \delta({\bf q}_1+{\bf q}_2+{\bf q}_3)
\end{equation}
is the number of fundamental triangles inside the shell defined by $\mathcal{S}_1$, $\mathcal{S}_2$ and $\mathcal{S}_3$, with $\mathcal{S}_i$ the region of the $k$-modes contained in a $k$-bin, $\Delta k$, around $k_i$. The Dirac $\delta$ insures that only closed triangles are included -- see the mentioned references for more details. The measurement of the bispectrum estimator $\langle F_3({\bf k}_1)F_3({\bf k}_2) F_3({\bf k}_3)\rangle$  is essential to be sensitive to possible higher order effects associated to non-Gaussianities in the distribution of galaxies. This task is what has been done in the important work\,\cite{Gil-Marin:2016wya}.
\newline
\newline
Here we wish to dwell on the impact of the bispectrum as a potential tracer of the DDE. Observationally, the data on BAO+ LSS (more specifically, the $f\sigma_8$ part of LSS) including both the spectrum (SP) and bispectrum (BSP) are taken from \cite{Gil-Marin:2016wya}, together with the correlations among these data encoded in the provided covariance matrices. The same data  including SP but no BSP has been considered in \cite{Alam:2016hwk}. In this chapter, we analyze the full dataset DS1 with spectrum only (dubbed DS1/SP) and also the same data when we include both SP and BSP (denoted DS1/BSP for short). In addition, we test the DDE sensitivity of the special subset DS2, which involves both SP+BSP components (scenario DS2/BSP). The contrast of results between the ``bispectrumless'' scenarios (i.e. the pure SP ones) and those including the matter bispectrum component as well (i.e. the SP+BSP ones)   will be made apparent in our study, specially through our devoted discussions in sections \ref{sect:LinStructure_signs_chapter} and \ref{sect:Bayesian_evidence_signs_chapter}.
%
\begin{figure}[!t]
\begin{center}
\label{LCDMtriangular_signs_chapter}
\includegraphics[width=5.5in, height=4.7in]{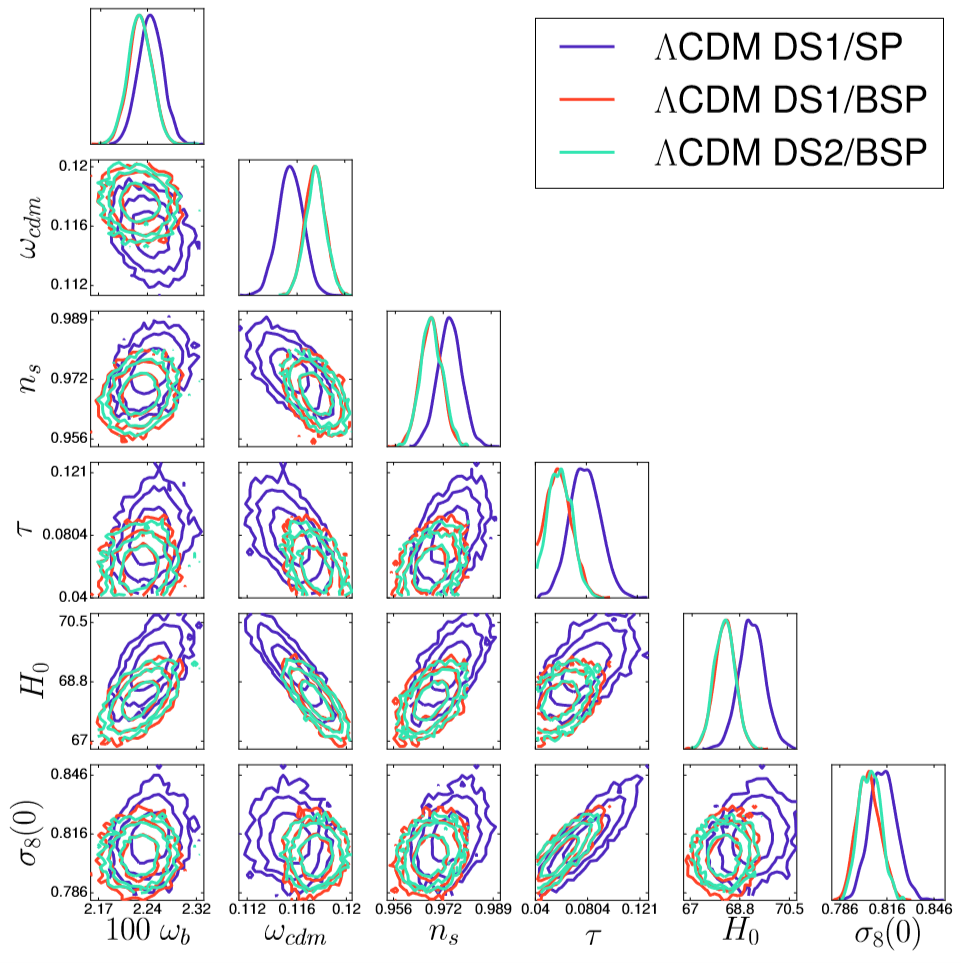}
\caption{\scriptsize Triangular matrix containing all the possible combinations of two-dimensional marginalized distributions for the various $\Lambda$CDM fitting parameters (at $1\sigma$, $2\sigma$ and $3\sigma$ c.l.), together with the corresponding one-dimensional marginalized likelihoods for each parameter. $H_0$ is expressed in km/s/Mpc. We present the results for the main data sets used in this chapter: DS1/SP, DS1/BSP, and DS2/BSP (cf. Tables 1 and 2 for the numerical fitting results, and the text for further details).}
\end{center}
\end{figure}
%
The numerical fitting results that we have obtained for the DDE models under examination  and the $\CC$CDM are displayed  in Tables 1 and 2. While we use Planck 2015 CMB data with full likelihood in almost all the cases, in the second block of Table 2 we report on the preliminary results obtained from the recent Planck 2018 CMB data under compressed likelihood  \cite{Aghanim:2018eyx,Chen:2018dbv}.  Let us note that the full likelihood for  Planck 2018 CMB data is not public yet.   We discuss the  results that we have obtained and their possible implications in the next two sections.
\newline
As indicated above, an additional fitting table has  been generated (Table 3) in order to illustrate the fact that the results presented here are consistent with other studies presented in the literature, namely for the case when we restrict our analysis to the baseline bispectrumless scenario introduced in this chacpter (i.e.  DS1/SP).  We substantiate this fact by introducing an alternative bispectrumless  scenario, which we call  DS3/SP.  The latter is essentially coincident with the one used in the  investigations of the same models recently presented by the authors of \cite{Park:2018bwy,Park:2018fxx}, although there exist some mild differences between our DS3/SP scenario and the one considered by these authors, to wit: (i) In the supernovae sector there appear two differences:  first, we use the compressed Pantheon+MCT data\,\cite{Scolnic:2017caz}  mentioned in the previous section, which in expanded form includes more than one thousand supernovae,  whereas they use the full (uncompressed) list of supernova data from the Pantheon compilation, but in their  list -- and here lies the second difference--  they do not include the 15 high-redshift SNIa ($z>1$) from the CANDELS and CLASH Multy-Cycle Treasury (MCT) programs. However these differences should not be significant at all, and we refer the reader to the detailed discussion in Ref.\,\cite{Riess:2017lxs} to justify why is so. In fact, in the latter  reference it is demonstrated  that the constraints derived from the full Pantheon+MCT compilation of supernovae of Type Ia are essentially indistinguishable from those derived from the corresponding compressed likelihood (see Fig. 3 of that reference); (ii)  Concerning the BOSS Ly$\alpha$-forest data, they use the exact distributions, whereas we use the Gaussian approximations. A direct comparison of the values of the fitting parameters reported in \cite{Park:2018bwy,Park:2018fxx} with those presented in our Table 3 confirms a high degree of compatibility, being the discrepancies in all cases  a small fraction of  $ 1\sigma$. For instance, for the XCDM we obtain $w_0=-1.002^{+0.034}_{-0.035}$, and the authors of \cite{Park:2018bwy} find $w_0=-0.994\pm 0.033$, which differs by less than  $0.17\sigma$  from the former, taking the errors in quadrature.  The difference with respect to the  value of $w_0$  corresponding to the DS1/SP dataset in Table 1 is also small:  $0.18\sigma$. On these grounds we judge that the approach adopted in our analysis is fully consistent and  that the tiny differences which may appear between the exact and compressed treatment of the SNIa and the exact and Gaussian Ly$\alpha$-forest likelihoods cannot be held responsible for any significant change in the main conclusions, as to the dynamical nature of the DE. Put another way, the preliminary signal of DDE that we find in this work cannot be attributed to the tiny differences among the bispectrumless  data sets existing in the literature, whether our DS1/SP, DS3/SP,  the exact one used e.g. by \cite{Park:2018bwy,Park:2018fxx} or any other consistent dataset employed by different authors.  All these bispectrumless scenarios are statistically equivalent since they involve a sufficiently complete and uncorrelated set of data from the various SNIa+$H(z)$+BAO+LSS+CMB sources  and therefore they are mutually consistent within a fraction of $ 1\sigma$ errors.  We must conclude that the emerging indications of DDE that we have encountered should rather originate from the peculiarities of the bispectrum component, which seems to be particularly sensitive to the dynamical features presumably sitting in the DE.
%
%
\begin{figure}
\begin{center}
\label{XCDMtriangular_signs_chapter}
\includegraphics[width=5.5in, height=4.7in]{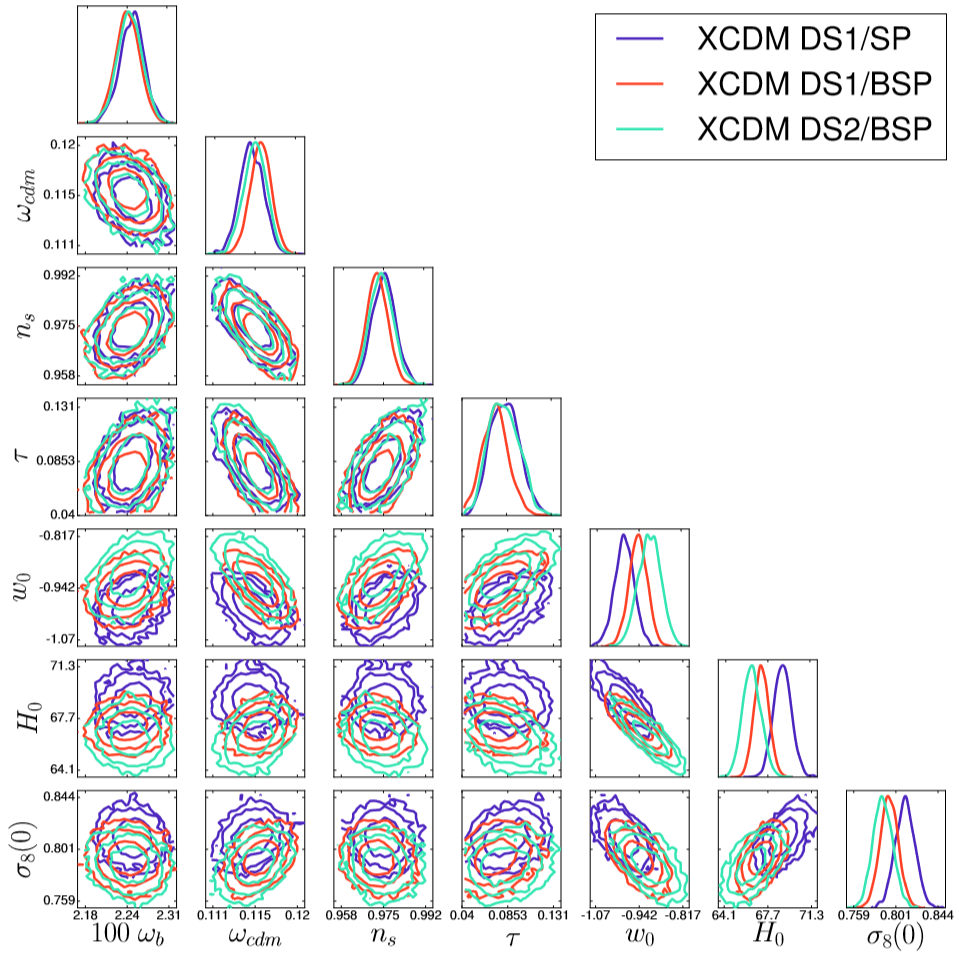}
\caption{\scriptsize  As in Fig. 3, but for the XCDM. Here we also show the two-dimensional contours and one-dimensional distribution for the EoS parameter $w_0$.}
\end{center}
\end{figure}
%
%
\subsection{Confronting DDE to observations}\label{sect:LinStructure_signs_chapter}
As we can see from Tables 1 and 2, the comparison of the $\CC$CDM with the DDE models points to a different sensibility of the data sets used to the dynamics of the DE.   If we start focusing on the results from the  bispectrumless scenario DS1/SP -- cf.  left block of Table 1 -- there is no evidence that the DDE models perform better than the $\CC$CDM. The XCDM, for instance,  yields a weak signal which is compatible with $w_0=-1$ (i.e. a rigid CC). This is consistent e.g. with the analysis of  \cite{Park:2018bwy}.  The $\phi$CDM model remains also inconclusive under the same data. The upper bound of  $\alpha<0.092$ at $1\sigma$  ($0.178$ at 2$\sigma$) recorded in that table  is consistent, too,  with the recent studies of\,\cite{Park:2018fxx}.  The same conclusion is  derived upon inspection of Table 3, based on the alternative bispectrumless  scenario DS3/SP, whose fitting results are indeed  consistent with those collected for our DS1/SP as well as with those from the mentioned references \cite{Park:2018bwy} and \cite{Park:2018fxx}.
%
%
\begin{figure}
\begin{center}
\label{phiCDMtriangular_signs_chapter}
\includegraphics[width=5.5in, height=4.7in]{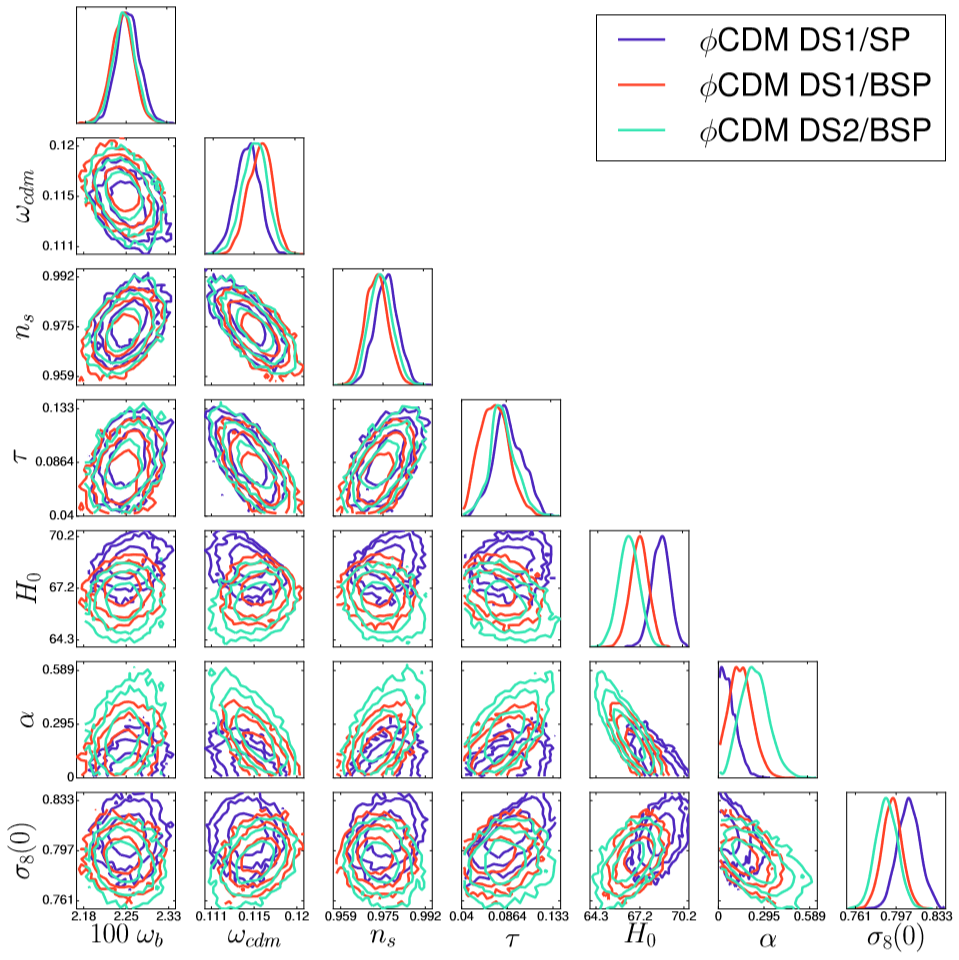}
\caption{\scriptsize As in Fig. 3, but for the $\phi$CDM model. We also include the two-dimensional contours and one-dimensional distribution for the parameter $\alpha$ of the potential (\ref{eq:PRpotential_signs_chapter}).}
\end{center}
\end{figure}
%
Notwithstanding, both the XCDM parametrization and the $\phi$CDM model fare significantly better than the $\CC$CDM
 if we consider the dataset DS1/BSP, i.e. upon including the bispectrum component of the BAO+LSS data.  The corresponding DDE signature is  about $2\sigma$ c.l. However, it is further enhanced within the restricted DS2/BSP dataset, where the XCDM and the $\phi$CDM reach in between  $2.5-3\sigma$ c.l. (cf. left  block of Table 2).  As for the CPL parametrization, Eq.\, (\ref{eq:CPL_signs_chapter}), we  record it explicitly in Table 1 for the DS1/BSP case only.  One can check that even in this case the errors in the EoS parameters are still too big to capture any clear sign of DE dynamics, owing to the additional parameter present in this model. Specifically, we find $w_0= -0.934^{+0.067}_{-0.075}$ and $w_1 = -0.045 ^{+0.273}_{-0.204}$,  hence fully compatible with a rigid CC ($w_0=-1, w_1=0$).
\newline
\newline
In Fig. 1 we show the contour plots  in the $(\omega_{cdm},w_0)$ and $(\omega_{cdm},\alpha)$ planes for the XCDM and $\phi$CDM models, respectively,  at different confidence levels, which are obtained with the (bispectrumless) DS1/SP data set.  It is obvious from these contours that the models do not exhibit a clear preference for  DDE.  The contour plots for the XCDM (on the left side of Fig. 1)  appear located roughly $50\%$ up and  $50\%$  down with respect to the cosmological constant divide $w_0=-1$, and therefore we cannot appraise any marked preference for a deviation into the quintessence ($w_0\gtrsim-1$) or the phantom region ($w_0\lesssim-1$). Similarly,  the contour plots for the parameter $\alpha$ of the  $\phi$CDM model (indicated on the right side of Fig. 1)  show that $\alpha$ is consistent with zero at $1\sigma$, which means that the potential  (\ref{eq:PRpotential_signs_chapter}) is perfectly compatible with a cosmological constant.
%
\begin{figure}
\begin{center}
\label{XCDMPlanck2018_signs_chapter}
\includegraphics[width=4.6in, height=2.7in]{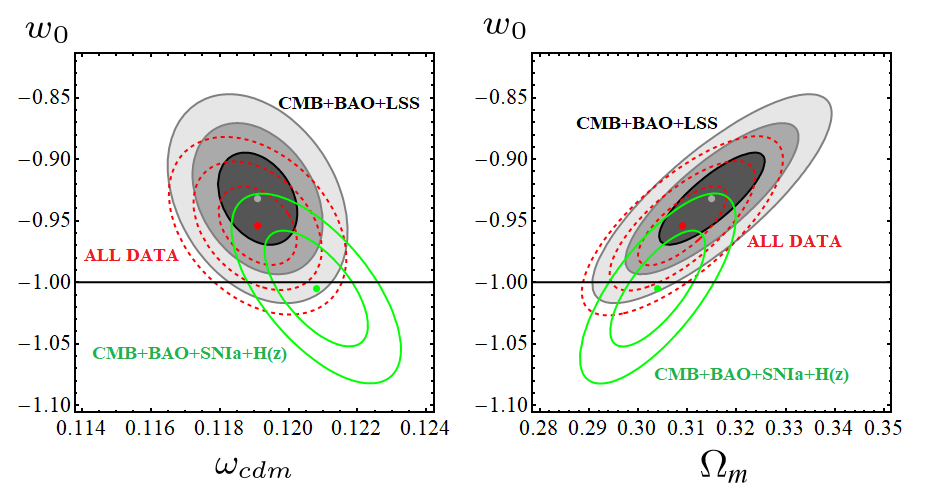}
\caption{\scriptsize Contour lines for the XCDM in the planes $(\omega_{cdm},w_0)$ and $(\Omega_m,w_0)$ obtained with different data set combinations and using the compressed Planck 2018 data \cite{Aghanim:2018eyx,Chen:2018dbv} instead of the full Planck 2015 likelihood \cite{Ade:2015xua}. More concretely we show the results obtained with the DS2/BSP data set (CMB+BAO+LSS, in grayish shades, see the second block of Table 2), the DS1/BSP (ALL DATA, dashed red contours), and the DS1/BSP but without LSS (CMB+BAO+SNIa+$H(z)$, solid green contours). The solid points indicate the location of the best-fit values in each case. When the LSS data points with bispectrum are considered in combination with the CMB and BAO constraints the DDE signal exceeds the $2\sigma$ c.l., regardless we use the Planck 2015 or 2018 information. See the text for further details.}
\end{center}
\end{figure}
%
%
In Fig. 2, however,  the situation has changed in a rather conspicuous way.  Once more  we display the corresponding  contour plots for the XCDM and $\phi$CDM models, but now they are obtained in the presence of the bispectrum, specifically we use in this case the DS2/BSP data set.  While we could use the entire  DS1/BSP set, the former subset (made exclusively  on BAO+LSS+CMB)  is slightly  more sensitive and we choose it to illustrate which data  tend to optimize the DDE signal.  After all not all data is expected to be equally sensitive, and this is something that has already been noted for other DDE models\,\cite{Sola:2017jbl}.  Coming back to Fig. 2 we can see that, in stark contrast with Fig. 1, there is a marked preference for the contours of the XCDM to shift upwards into the quintessence region.  Specifically,  the EoS parameter $w_0$  lies now more than $2.5\sigma$ away from $-1$  in the quintessence domain  $w_0\gtrsim-1$.  On the other hand, in the $\phi$CDM case (shown in the right plot of the same figure)  we consistently find $\alpha>0$ at a similar (actually slightly higher) c.l.
%
\begin{figure}
\begin{center}
\label{phiCDMPlanck2018_signs_chapter}
\includegraphics[width=4.6in, height=2.7in]{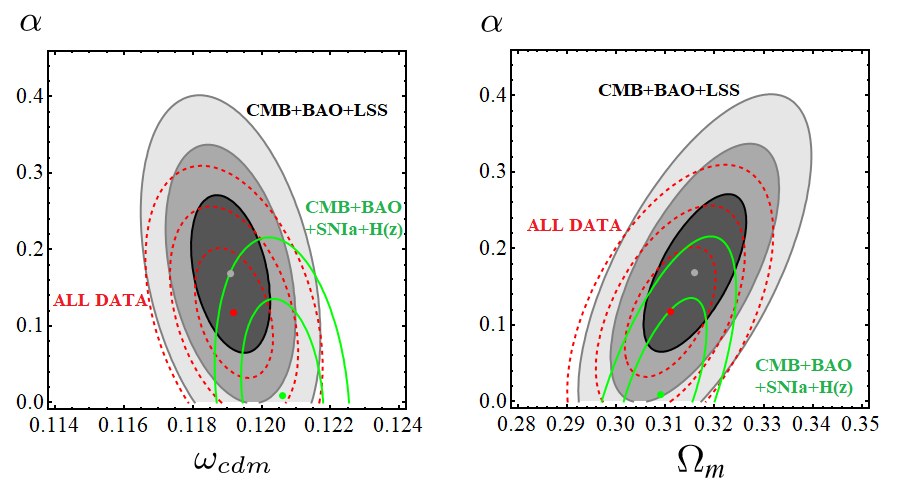}
\caption{\scriptsize The same as in Fig. 6, but for the $\phi$CDM model and in the planes $(\omega_{cdm},\alpha)$ and $(\Omega_m,\alpha)$.}
\end{center}
\end{figure}
%
Figures 3-5 are also very illustrative. They contain the confidence contours in all the relevant planes of the parameter space up to $3\sigma$ c.l., together with the marginalized one-dimensional posterior distributions for the various fitting parameters. In Fig. 3 we present the results obtained with the DS1/SP, DS1/BSP and DS2/BSP scenarios in the context of the $\Lambda$CDM. Figs. 4 and 5 contain the analogous plots for the XCDM and $\phi$CDM, respectively. These figures allow us to appreciate in a very straightforward way what are the shifts in the confidence regions caused by the various changes of our cosmological data sets. Fig. 3 shows, for instance, that in the $\Lambda$CDM the central values of all the parameters that enter the fit get shifted away from their DS1/SP values  when the bispectrum information is also taken into account. These shifts are specially important for the parameters $\omega_{cdm}$, $n_s$, $\tau$ and $H_0$, which in the DS1/BSP scenario are shifted $+1.48\sigma$, $-1.13\sigma$, $-1.23\sigma$ and $-1.62\sigma$ away with respect to the DS1/SP case. In contrast, $\omega_b$ and $\sigma_8(0)$ are lesser modified ($-0.60\sigma$ and $-0.79\sigma$, respectively). The removal of the SNIa+$H(z)$ data points from the DS1/BSP data set (which gives rise to the DS2/BSP one) does not produce any additional significant change in the $\Lambda$CDM, as it can be easily checked by comparing the DS1/BSP and DS2/BSP contours and the corresponding one-dimensional distributions of Fig. 3. In the DDE models the effect on the parameters introduced by the bispectrum signal is quite different. To begin with, the values of $\omega_{cdm}$, $\omega_b$, $n_s$ and $\tau$ remain very stable and completely compatible at $<1\sigma$ regardless of the data set under consideration. Moreover, the values of these parameters are also compatible with those of the $\Lambda$CDM in the DS1/SP scenario. This means that the bispectrum does not force in the XCDM and $\phi$CDM any important shift of the central values of $\omega_{cdm}$, $\omega_b$, $n_s$ and $\tau$ with respect to the typical values found in the $\Lambda$CDM when we only consider the matter power spectrum part of the LSS+BAO data sector. Conversely, when we move from the DS1/SP scenario to the DS1/BSP and, finally, to the DS2/BSP, we find a progressive and non-negligible displacement of the peaks of the one-dimensional distributions for $H_0$ and $\sigma_8(0)$ towards lower values, which is accompanied by a departure of the DDE parameters of the XCDM ($w_0$) and $\phi$CDM ($\alpha$) from the $\Lambda$CDM values ($-1$ and $0$, respectively). The global decrease of $\sigma_8(0)$ from DS1/SP to DS2/BSP  is  $ -1.6\sigma$ ($-1.3\sigma$) for XCDM ($\phi$CDM), and the one of $H_0$ is  $-2.3\sigma$ for both models. This fact allows to reduce the well-known $\sigma_8$-tension \cite{Macaulay:2013swa} in a more efficient way in the context of the DDE models. However, the low values of the Hubble parameter derived in the XCDM and $\phi$CDM from our fitting analyses keep the statistical tension with the local distance ladder determinations of \cite{Riess:2016jrr,Riess:2018uxu} high, reaching the latter the $\sim 3.8\sigma$ level. Nevertheless, it is important to remark that the values of $H_0$ obtained by us in the current study are  perfectly consistent with those obtained by some authors that apply model-independent reconstruction techniques and low-redshift data from SNIa, BAO and $H(z)$ data points from cosmic chronometers, see e.g. \cite{Yu:2017iju,Gomez-Valent:2018hwc,Feeney:2018mkj,Haridasu:2018gqm,Macaulay:2018fxi}. In the DDE scenarios analyzed in this chapter, the $\sigma_8$-tension is directly linked with the $H_0$-tension. We can only loosen the former at the expense of keeping the latter (cf. the contour lines in the $H_0$-$\sigma_8(0)$ plane of Figs. 4 and 5). This is consistent with our analysis of \cite{Sola:2017znb}, in which we  studied other DDE models pointing to the same conclusion. Future data might be able to elucidate the ultimate origin of the $H_0$-tension. If the true value of $H_0$ lies in the Planck-preferred region of $H_0\sim 66-68$ km/s/Mpc, then the DDE models can offer an efficient way of automatically relieving the $\sigma_8$-tension.
%
\begin{figure}
\begin{center}
\label{FigLSS2_signs_chapter}
\includegraphics[width=4.5in, height=3in]{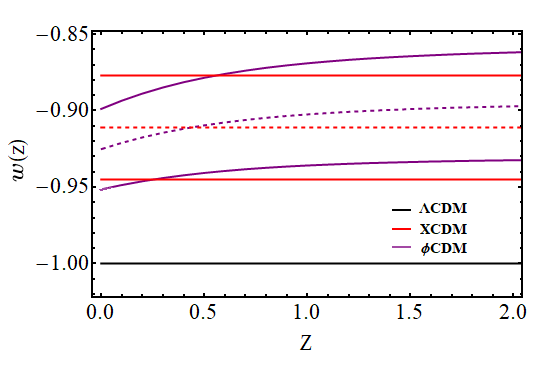}
\caption{\scriptsize  The EoS for the considered DE models
within the corresponding $1\sigma$ bands and under scenario DS2/BSP. For the XCDM the EoS
is constant and points to quintessence at  $\sim 2.6\sigma$ c.l (cf. Table 2). For the $\phi$CDM the EoS evolves with time and
is computed through a Monte Carlo analysis. The
current value is $w(z=0) = - 0.925\pm 0.026$ and favors once more
quintessence at  $\sim 2.9\sigma$ c.l.}
\end{center}
\end{figure}
%
From a more prospective point of view, in Figs. 6 and 7 we perform a preliminary exploration of  the corresponding results using  Planck 2018 data. The results for the same DDE models are obtained under different data set combinations, but on this occasion making use of the Planck 2018 compressed CMB data, given the mentioned fact that the full likelihood for Planck 2018 is not publicly available yet. These plots make, once more, palpable the importance of the bispectrum  for the study of the DDE.  Indeed,  it is only when the bispectrum is included in the analysis, together with the CMB and BAO data sets,  that a non-negligible signal in favor of the dynamical nature of the DE clearly pops out.
\newline
In Fig. 8 we plot the EoS of the various models in terms of the redshift near our time and  within the $1\sigma$ error bands for the XCDM and $\phi$CDM,  again for the DS2/BSP case. These bands  have been computed from a Monte Carlo  sampling of the $w(z)$ distributions.  The behaviour of the curves shows that the quintessence-like behaviour is sustained until the present epoch. For the $\phi$CDM we find  $w(z=0) = - 0.925\pm 0.026$, thus implying a DDE signal at  $\sim 2.9\sigma$ c.l., which is consistent with the XCDM.
\newline
\newline
Finally, in Fig. 9 we compute the matter power spectrum and the temperature anisotropies for the DDE models using Planck 2015 data, and also display the percentage differences with respect to the $\CC$CDM. As can be seen, the differences of the CMB anisotropies between the DDE models and the $\CC$CDM remain safely small,  of order $\sim1\%$ at most for the entire range.  The $3-5\%$ suppression of the matter power spectrum $P(k)$ of the DDE models in the relevant range of wave numbers $k$ with respect to the one of the concordance model is what gives rise to lower values of $\sigma_8(0)$ (cf. Tables 1-3).
%
\begin{figure}
\begin{center}
\label{FigLSS3_signs_chapter}
\includegraphics[width=6in, height=3.5in]{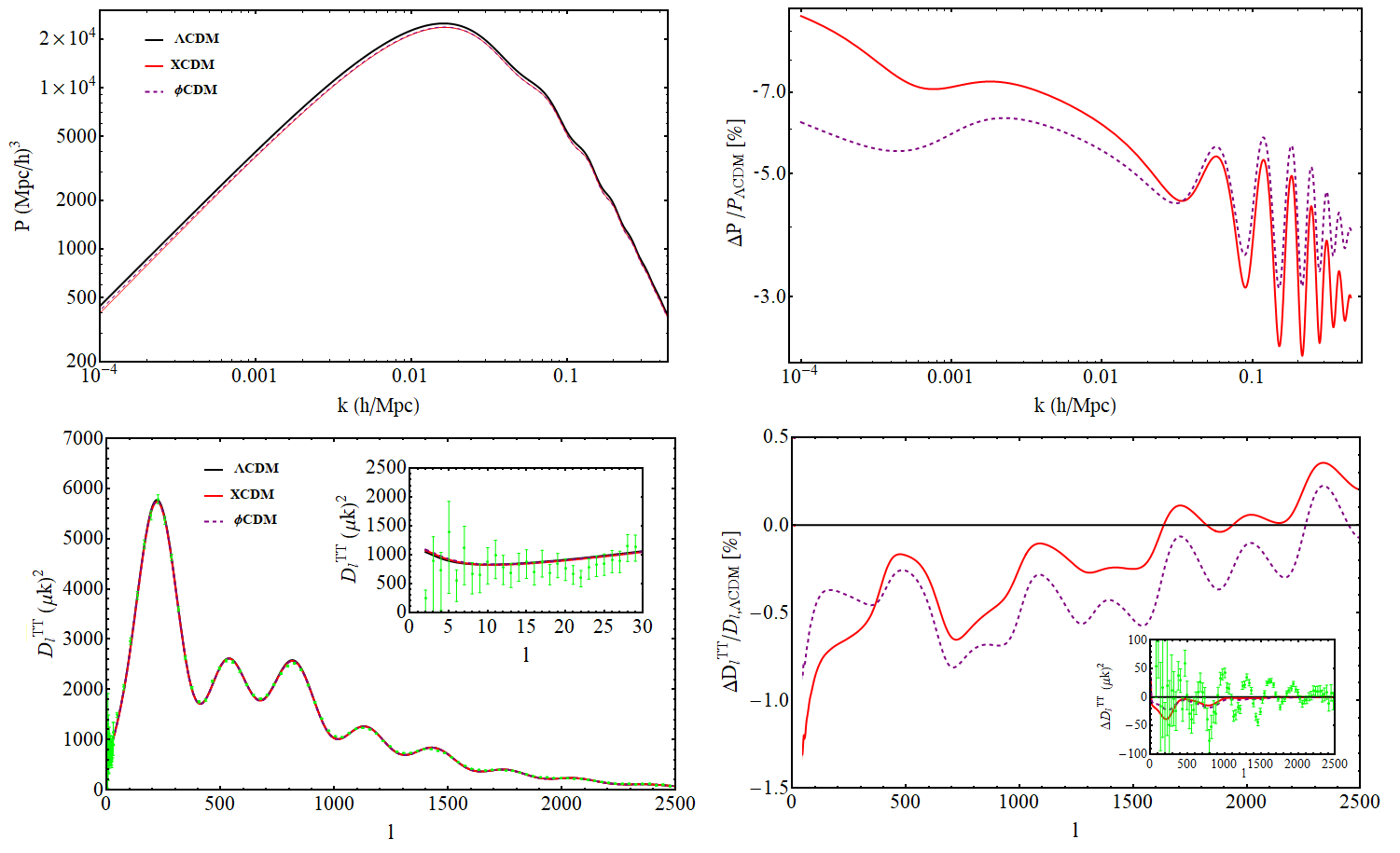}
\end{center}
\caption{\scriptsize (Upper plots) The power spectrum for the DDE models under consideration side by side with the relative (percentage) differences with respect to that of the $\CC$CDM.  We have used the best-fit values of the parameters of each model;  (Bottom plots) As before, but considering the CMB temperature anisotropies for Planck 2015 data. The inner plots show additional details including the data points with the corresponding errors in part or the whole available angular range.}
\end{figure}


%



%
\subsection{Bayesian evidence of DDE}\label{sect:Bayesian_evidence_signs_chapter}
The main results of this work are synthesized in Tables 1 and 2, and in Figs. 1-10.  Here we wish to quantify the significance of these results from the statistical point of view.  This is done in detail in Fig. 10.  In what follows we explain the meaning of this figure whereas the list of the most important formulas regarding the Bayesian evidence can be found int Appendix \ref{Appendix_E}. As remarked before, in the absence of BSP data the DDE signs are weak and we find consistency with previous studies. However, when we include BSP data and focus on the LSS+BAO+CMB observable (i.e. scenario DS2/BSP) the situation changes significantly.  Both models XCDM and $\phi$CDM  then consistently point to a $2.5-3\sigma$ effect.  Specifically,  the evolving EoS of the $\phi$CDM takes a value at present which lies $\sim 3\sigma$ away from the $\CC$CDM  ($w=-1$) into the quintessence region $w\gtrsim-1$ (cf. Fig. 8). To appreciate the significance of these results we compute the Bayesian evidence, based on obtaining the posterior probability function for the fitting parameters. The details of the procedure, as well as the notation employed, can be obtained in Sec. \ref{sec:Model_Selection_Appendix_E} of Appendix \ref{Appendix_E}.
\newline
\newline
%
\begin{table}[t!]
\begin{center}
\begin{tabular}{cc}
\hline
$\Delta {\rm BIC}=2\ln B_{ij}$ &$\ \ \ \ \ \ $ Bayesian evidence of model ${M}_i$ versus $M_j$ $\ \ $ \\ \hline
$0<\Delta {\rm BIC} < 2$ & {\rm weak evidence (consistency between both models)} \\
$2<\Delta {\rm BIC}< 6$ & {\rm  positive evidence} \\
$6<\Delta {\rm BIC}< 10$ & {\rm strong evidence} \\
$\Delta {\rm BIC} \geq 10$ & {\rm very strong evidence} \\
$\Delta {\rm BIC} < 0$ & {\rm counter-evidence against model $M_i$} \\
\hline
\end{tabular}
\caption{\scriptsize Conventional ranges of values of $\Delta {\rm BIC} $  used to judge the observational support for a given model $M_i$ with respect to the reference one $M_j$. See \cite{KassRaftery1995,Burnham2002} for more details. }
\label{tab:Delta BIC_signs_chapter}
\end{center}
\end{table}
%
%
%
\begin{figure}[!]
\begin{center}
\label{FigLSS2_signs_chapter}
\includegraphics[width=4.1in, height=3.1in]{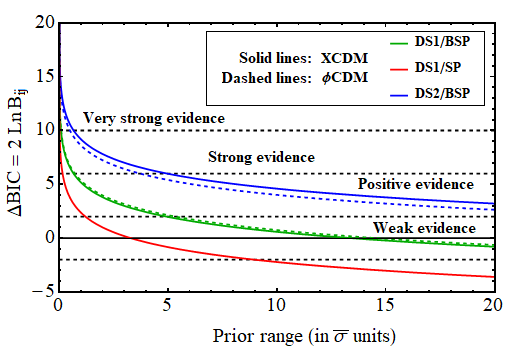}
\caption{\scriptsize  The full Bayesian evidence curves for the various models as compared to the $\CC$CDM as a function of the (flat) prior range  (in $\bar{\sigma}$ units, see text). The curves are computed using the exact evidence formula, Eq.\,(\ref{eq:DeltaBICExact_signs_chapter}), with the data indicated in Table 1 and the first block of Table 2 (i.e. with full Planck 2015 CMB likelihood). The marked evidence ranges conform with the conventional definitions of Bayesian evidence shown in Table 6.}
\end{center}
\end{figure}
%
%
%
%
In Fig. 10 we compute the full Bayesian evidence curves as a function of  the prior range.  We assume flat priors for the parameters in each model (we have no reason to assume otherwise). Thus $p(\theta|M)$ (the marginal likelihood) is constant and cancels out for all  common $\theta$ in the two models, rendering $B_{ij}$ (the Bayes factor) a pure ratio of marginal likelihoods. We define the unit of measure of the prior range, $\bar{\sigma}$, as half the sum of the upper and lower fitting errors of the DDE parameters in Tables 1 and 2. In Fig. 10 we rate the level of support of the DDE models as compared to the $\CC$CDM for each value of the prior range, following the standard definitions collected in Table 6\,\cite{KassRaftery1995,Burnham2002}.
\newline
As indicated, we utilized the code \texttt{MCEvidence}\,\cite{Heavens:2017afc} \footnote{We thank Y. Fantaye for helpful advice in the use of the numerical package \texttt{MCEvidence}} to compute the exact $\Delta  {\rm BIC}$ values in Fig. 10. We have actually compared these results with those obtained for the evidences computed with Gaussian (Fisher) likelihoods for the parameters and we have found that they are qualitatively similar but with non-negligible numerical differences. These were expected from the mild departures of Gaussianity from the exact distributions. The full evidence curves in Fig. 10  reconfirm the mentioned result that with the bispectrumless dataset DS1/SP  the evidence is weak. But at the same time we can recognize once more the impact of the bispectrum in the scenarios enriched with this component. Thus, it is remarkable to find a long tail of sustained positive Bayesian evidence both for the XCDM and $\phi$CDM within the scenario DS2/BSP, which corroborates the higher sensitivity of the BAO+LSS+CMB data to DDE.  The fact that with the full Planck 2015 data and with traditional models and parametrizations of the DE we can reach such significant level of evidence is encouraging. Even more reassuring is the fact that with the preliminary Planck 2018 data we can reach similar levels of evidence (cf. Table 2, second block, and also Figs. 6 and 7), which will have to be refined when the full Planck 2018 likelihood will be made public.
\newline
\newline
Last, but not least, let us emphasize one final important point of our analysis. {Our results with  BSP data show that the DDE values of $\sigma_8(0)$  (specially within the DS2/BSP scenario) are smaller than  the (bispectrumless)   $\CC$CDM value  quoted in Table 1  by roughly $2\sigma$,  thus essentially mitigating  the $\sigma_8$-tension\,\cite{Macaulay:2013swa}. As remarked in Sec. \ref{sect:LinStructure_signs_chapter} (related to our discussion  on Fig.\,3),  the $\CC$CDM value of  $\sigma_8(0)$  does also diminish with  BSP,  but it is insufficient to alleviate the $\sigma_8$-tension. The latter is  defined of course by taking as a reference the bispectrumless  $\CC$CDM result since the higher order corrections to the spectrum were never considered in previous analyses of the $\sigma_8$-tension. This is the reason  why we have compared with that value  in order to judge the performance of the DDE models in connection to such tension. } We have checked also the influence of massive neutrinos.  We find that a similar effect is possible within the $\CC$CDM (under  DS1/BSP)  for a sum of neutrinos masses of $\sum  m_\nu=  0.195\pm  0.076$ eV, leading to $\sigma_8(0)= 0.785\pm 0.014$.  Recall, however, that in all our fitting tables we have included  a light massive neutrino of  $0.06$ eV, which is the standard  assumption adopted in most analyses\,\cite{Ade:2015xua} so as to preserve  the so-called normal hierarchy of neutrino masses (as required from neutrino oscillation experiments).  A solution solely based on invoking a  substantially massive neutrino scenario is thus ruled out since it clashes with the usual constraints on neutrino oscillations; specifically,  it requires unnatural fine-tuning among the neutrino masses. The DDE option is, therefore, more natural.  There may be, of course, other factors that could have an influence on this analysis, such as e.g. the possible concurrence of spatial curvature effects. When curvature is allowed to be a free parameter the constraints on DE
dynamics may weaken since we then have an additional parameter.  We have already seen a weakening of DDE evidence in the case of the CPL as compared to the XCDM, owing to the presence of one more parameter in the former case as compared to the latter.  The effects of curvature in this kind of analyses  have been studied for the main models under consideration  in\,\cite{Park:2017xbl,Park:2018bwy,Park:2018fxx,Ooba:2018dzf,Park:2018tgj}, and previously in\,\cite{Ooba:2017npx,Farooq:2013dra}. For instance, in \cite{Park:2017xbl} (see also references therein)  the authors show that in the non-flat $\Lambda$CDM model the spatial curvature also allows a decrease of the central value of $\sigma_8(0)$ when only the TT+lowP+lensing Planck 2015 CMB data is employed in the fitting analysis. When a more complete data set including SNIa+$H(z)$+BAO+LSS is considered, though, such decrease is less significant. The same pattern is found e.g. for the $\phi$CDM in \cite{Park:2018fxx}. It is nonetheless important to point out that the effect of the bispectrum signal in the BAO+LSS data sector has not been studied in the context of a spatially curved Universe yet. It would be interesting to carefully assess it. Here, however, we have chosen to present our main results by placing  emphasis on the impact of the bispectrum on the study of the DDE within the simplest possible scenario, which is to assume spatially flat FLRW metric. In order to better isolate the effect, we have avoided additional considerations, except for revisiting the impossibility that massive neutrinos alone can solve these problems  in a natural way.  Owing to the considerable controversy in the literature on these matters we believe that by focusing the message we gain in clarity.  The study of additional effects can be, of course,  very interesting for future investigations, but the main focus of the current study is the impact of the bispectrum on the DDE.
%
%
\subsection{Conclusions}\label{Sect:Conclusions_signs_chapter}
In this chapter, we have tested the performance of cosmological physics beyond the standard or concordance $\CC$CDM model, which is built upon a rigid cosmological constant. We have shown  that the global cosmological observations can be better described in terms of models equipped with a dynamical DE component that evolves with the cosmic expansion.  Our task has focused on three dynamical dark energy (DDE) models: the  general XCDM and CPL parametrizations as well as a traditional scalar field model ($\phi$CDM), namely the original Peebles \& Ratra model based on the potential (\ref{eq:PRpotential_signs_chapter}). We have fitted them (in conjunction with the $\CC$CDM) to the same set of cosmological data based on the observables SNIa+BAO+$H(z)$+LSS+CMB. Apart from the global fit involving all the data, we have also tested the effect of separating the expansion history data (SNIa+$H(z)$) from the features of CMB and the large scale structure formation data (BAO+LSS, frequently interwoven), where LSS includes both the RSD and weak lensing measurements.  We have found that the expansion history data are not particularly sensitive to the dynamical effects of the DE, whereas the BAO+LSS+CMB are more sensitive.  Furthermore, we have evaluated for the first time the impact of the bispectrum component in the matter correlation function on the dynamics of the DE .  We have done this by including BAO+LSS data that involve both the conventional power spectrum and the bispectrum. The outcome is that when the bispectrum component is not included our results are in perfect agreement with previous studies of other authors, meaning that in this case we do not find clear signs of dynamical DE. In contrast, when we  activate the bispectrum component in the BAO+LSS sector (along with the corresponding covariance matrices) we can achieve a significant DDE signal at a confidence level of $2.5-3\,\sigma $ for the XCDM and the $\phi$CDM. We conclude that the bispectrum can be  a very useful tool to track possible dynamical features of the DE and their influence in the formation of structures in the linear regime.
\newline
A surplus of our analysis is that we have also found noticeably lower values of $\sigma_8(0)$ in the presence of the bispectrum, see the third from last row of Table 2. For a long time it has been known that there is an unexplained  `$\sigma_8$-tension' in the framework of the $\CC$CDM, which is revealed through the fact that the $\Lambda$CDM tends to provide higher values of $\sigma_8(0)$ than those obtained from RSD measurements.  Dynamical DE models, therefore, seem to provide a possible alleviation of such tension, specially when we consider the combined CMB+BAO+LSS measurements and with the inclusion of the bispectrum.
\newline
Finally, let us remark that although  we have used the full Planck 2015 CMB likelihood in our work, we have also advanced the preliminary results involving the recent Planck 2018  data under compressed CMB likelihood, still awaiting for the public appearance of the full Planck 2018 likelihood. The prospective results that we have obtained do consistently keep on favoring DDE versus a rigid cosmological constant.
\newline
To summarize, our study shows that it is possible to reach significant signs of DDE with the current data, provided we use the bispectrum in combination with the power spectrum.  The main practical conclusion that we can draw from our results is quite remarkable: the potential of the bispectrum for being sensitive to dynamical DE effects is perhaps more important than it was suspected until now. As it turns out, its more conventional application as a tool to estimate the bias between the observed galaxy spectrum and the underlying matter power spectrum may now be significantly enlarged, for the bispectrum (as the first higher-order correction to the power spectrum) might finally reveal itself as an excellent tracer of dynamical DE effects, and ultimately of the  ``fine structure'' of the DE.
\newpage

\section{Brans-Dicke cosmology with a $\Lambda$-term a possible solutions to $\Lambda$CDM tensions}\label{BD_gravity_chapter}
In this chapter we focus our efforts on study the well-known gravity model of Brans and Dicke with a cosmological constant (CC). As it will be seen throughout this chapter, this alternative to Einstein's General Relativity is characterized by having a scalar field, replacing the constant Newton's coupling $G_N$, which mediates the gravitational interaction together with the metric field. 
\newline
We wish to stick firmly to the $\Lambda$-term as the simplest provisional explanation for  the cosmic acceleration. But we want to do it on the grounds of  what we may call the BD-$\CC$CDM model, i.e. a cosmological framework still encompassing all the phenomenological ingredients of the concordance $\CC$CDM  model, in particular a strict cosmological constant term $\CC$ and dark matter along with baryons, but now all of them ruled over by a different gravitational paradigm: BD-gravity\,\cite{BransDicke1961, brans1962mach, dicke1962physical}, instead of GR.  Our intention is to combine the dynamical character of $G$ in the context of BD-gravity with a cosmological term $\Lambda$, aiming at finding a form of (effective) dynamical dark energy (DDE) capable to overcome the $H_0$-tension and the $\sigma_8$-tension, both, having been discussed in previous chapters.  Not all forms of DDE can make it. In this chapter we will discuss the fact that the simultaneous solution/alleviation of the two tensions requires a very particular form of DDE which is  ``aligned'' with the kind of observable to be adjusted.
\newline
{We show that the $H_0$-tension  can be significantly reduced in this context; and, interestingly enough, this can be achieved without detriment of the  $\sigma_8$-tension, hence fulfilling what we call, the golden rule.} We also find that an effective signature for dynamical DE may ultimately emerge from the BD-$\CC$CDM dynamics, and such signature is able to mimic quintessence at a confidence level in between $3-4\sigma$ level, depending on the datasets used. If a successful phenomenological description of the data and the loosening of the tensions could be reconfirmed with the advent of new and more precise data and new analyses in the future, it would be tantalizing to suggest the possibility that the underlying fundamental theory of gravity might actually be BD rather than GR. But there is still a long way to follow, of course.
\newline
The layout of this chapter reads as follows. In Section \ref{sec:BDgravity_BD_article} we introduce the basics of the Brans-Dicke (BD) model. {In particular, we discuss the introduction of the cosmological term in this theory and the notion of effective gravitational coupling.} Section \ref{sec:preview_BD_article} is a preview of the whole chapter, and in this sense it is a very important section where the reader can get a road map of the basic results of the chapter and above all an explanation of why BD-cosmology with a CC can be a natural and efficient solution to the $H_0$ and $\sigma_8$ tensions.  Section \ref{sec:EffectiveEoS_BD_article} shows how to parametrize  BD-cosmology as a departure from GR-cosmology. We show that BD with a CC appears as GR with an effective equation of state (EoS) for the dark energy which behaves quintessence-like at a high confidence level.  The remaining of the chapter presents the technical details and the numerical analysis, as well as complementary discussions. Thus, Section \ref{sec:StructureFormation_BD_article} discusses the perturbations equations for BD-gravity in the linear regime (leaving for Appendix \ref{Appendix_B} the technical details); Section \ref{sec:Mach_BD_article} defines four scenarios for the BD-cosmology in the light of  Mach's Principle; Section \ref{sec:MethodData_BD_article} carefully describes the data used from the various cosmological sources; Section \ref{sec:NumericalAnalysis_BD_article} presents the numerical analysis and results. {In Section \ref{sec:Discussion_BD_article} we perform a detailed discussion of the obtained results and we include a variety of extended considerations, in particular we assess the impact of massive neutrinos in the BD-$\CC$CDM framework. Finally, in Section \,\ref{sec:Conclusions_BD_article} we deliver our conclusions}. 
\subsection{BD-$\CC$CDM: Brans-Dicke gravity with a cosmological constant}\label{sec:BDgravity_BD_article}
Since the appearance of GR, more than one hundred years ago, many alternative theories of gravity have arisen, see e.g. \cite{Sotiriou:2008rp,Capozziello:2011et,Clifton:2011jh,Will:2014kxa} and references therein. The most important one, however, was proposed by Brans and Dicke almost sixty years ago\,\cite{BransDicke1961}. This theoretical framework  contains an additional gravitational  \textit{d.o.f.}  as compared to GR, and as a consequence it is different from GR in a fundamental way, see the previously cited reviews. The new \textit{d.o.f.}  is a scalar field, $\psi$, coupled to the Ricci scalar of curvature, $R$.  BD-gravity is indeed the first historical attempt to extended GR to accommodate variations in the Newtonian coupling $G$. A generalization of it has led to a wide panoply of scalar-tensor theories since long ago\,\cite{Bergmann:1968ve, Nordtvedt:1970uv, Wagoner:1970vr, Horndeski:1974wa}.  The theory is also characterized by  a   (dimensionless) constant parameter, $\oD$, in front of the kinetic term of $\psi$.
\subsubsection{Action and field equations}
In our study we will consider the original BD-action extended with a cosmological constant density term, $\rL$, as it is essential to mimic the conventional $\CC$CDM model based on GR and reproduce its main successes. In this way we obtain what we have called the `BD-$\CC$CDM model' in the introduction, i.e. the version of the  $\CC$CDM within the BD paradigm.
The BD action and the corresponding field equations can be found in Sec. \ref{Introduction_BD_model}. Throughout this chapter we consider the sign convention (+,+,+), see Appendix \ref{Appendix_A} for the details.
Hereafter,  for convenience, we will use a dimensionless BD-field, $\varphi$, and the inverse of the BD-parameter, according to the following definitions:
\begin{equation}\label{eq:definitions_BD_article}
\varphi(t) \equiv G_N\psi(t)\,,\qquad  \qquad\eBD \equiv \frac{1}{\oD}\,.
\end{equation}
In this expression, $G_N = 1/m_{\rm Pl}^2$, being $m_{\rm Pl}$ the Planck mass; $G_N$ gives the local value of the gravitational coupling, e.g. obtained from Cavendish-like  (torsion balance) experiments. Note that a nonvanishing value of $\eBD$ entails a deviation from GR. Being $\varphi(t)$  a dimensionless quantity,  we can recover GR by enforcing the simultaneous limits   $\eBD \rightarrow 0$ \textit{and} $\varphi\to 1 $.  We emphasize that it is not enough to set $\eBD \rightarrow 0$.  In this partial limit, we can only insure that $\varphi$ (and $\psi$, of course) does not evolve, but it does not fix its constant value. As we will see later on,  this feature can be important in our analysis. Using \eqref{eq:definitions_BD_article}, we can see that \eqref{eq:BDaction} can be rewritten
\begin{eqnarray}
S_{\rm BD}=\int d^{4}x\sqrt{-g}\left[\frac{1}{16\pi G_N}\left(R\varphi-\frac{\oD}{\varphi}g^{\mu\nu}\partial_{\nu}\varphi\partial_{\mu}\varphi - 2\Lambda\right)\right]+\int d^{4}x\sqrt{-g}\,{\cal L}_m(\chi_i,g_{\mu\nu})\,, \label{eq:BDactionvarphi_BD_article}
\end{eqnarray}
where $\Lambda$ is the cosmological constant, which is related with the associated vacuum energy density as $\rL=\CC/(8\pi G_N)$. The cosmological equations, written in terms of $\varphi$ and $\epsilon_{\rm BD}$ in the FLRW metric read as follows:
\begin{align}
3H^2 + 3H\frac{\dot{\varphi}}{\varphi} - \frac{1}{2\epsilon_{\rm BD}}\left(\frac{\dot{\varphi}}{\varphi}\right)^2 = \frac{8\pi{G_N}}{\varphi}\rho\label{eq:Friedmann_equation_varphi_BD_article} \\
2\dot{H} +3H^2 + \frac{\ddot{\varphi}}{\varphi} + 2H\frac{\dot{\varphi}}{\varphi} + \frac{1}{2\epsilon_{\rm BD}}\left(\frac{\dot{\varphi}}{\varphi}\right)^2 = -\frac{8\pi{G_N}}{\varphi}p \label{eq:Pressure_equation_varphi_BD_article} \\
\ddot{\varphi} + 3H\dot{\varphi} = \frac{8\pi{G_N}\epsilon_{\rm BD}}{2 + 3\epsilon_{\rm BD}}(\rho - 3p)\label{eq:Wave_equation_varphi_BD_article}
\end{align}
where the dots represent derivatives {\it w.r.t.} the cosmic time and {$\rho \equiv \rho_m + \rho_\gamma+\rho_\nu + \rho_\Lambda$ and $p \equiv p_m + p_\gamma + p_\nu + p_\Lambda$. The matter part $\rho_m \equiv \rho_b + \rho_{cdm}$, contains the pressureless contribution from baryons and cold dark matter. Photons are of course relativistic, so $p_\gamma=\rho_\gamma/3$. The functions $\rho_\nu$ and $p_\nu$ include the effect of massive and massless neutrinos, and therefore must be computed numerically. Since no interaction is considered between the BD scalar field and matter the local energy conservation equation takes the usual form}
\begin{equation}\label{eq:local_conservation_equation_BD_article}
\dot{\rho} + 3H(\rho+p) = \sum_N \dot{\rho}_N + 3H(\rho_N + p_N) = 0\end{equation} 
where the sum is over all components.
\subsubsection{Cosmological constant and vacuum energy in BD theory}
{There are several ways to introduce the cosmological constant in the BD framework, so  a few comments are in order at this point, see e.g. the comprehensive exposition\,\cite{Fujii:2003pa} and the works \cite{mathiazhagan1984inflationary, La1989, Weinberg1989, barrow1990extended} in the context of inflation. We can sum up the situation by mentioning  essentially three ways.
In one of them, the BD-action \eqref{eq:BDaction} is obtained upon promoting $G_N$ in the Einstein-Hilbert (EH) action with the $\rL$ term into a dynamical scalar field $1/\psi$,  adding the corresponding kinetic term and  keeping $\rho_\Lambda={\rm const.}$ The CC term $\Lambda$ is then related with the vacuum energy density through $\rho_\Lambda=\Lambda/(8\pi G_N )$. The EoS of the vacuum fluid is defined as $p_\Lambda=-\rho_\Lambda$. With these definitions, which we adopt throughout this chapter, the CC is not directly coupled to the BD-field. The latter, therefore, has a trivial (constant) potential, and the late-time acceleration source behaves as in the GR-$\Lambda$CDM model.} {Alternatively, one could also adopt the EH action with CC and promote $G_N$ to a dynamical scalar, adding the corresponding kinetic term as before, but now keeping $\Lambda={\rm const.}$ instead of $\rho_\Lambda={\rm const}$. In this case, $\Lambda$ is linearly coupled to $\psi$. The potential energy density for the scalar field takes the form $\rho_\Lambda(\psi)= \psi \Lambda/(8\pi)$, so it is time-evolving. The coupling between the cosmological constant and the BD-field modifies the equations of motion. Thus, it may also alter the physics with respect to the option \eqref{eq:BDaction}, at least when the dynamics of the scalar field is not negligible \cite{Nariai:1969vh,endo1977cosmological, Uehara1982, lorenz1984exact, barrow1990extended, romero1992brans, Tretyakova2012}.} {A third possibility, of course, is to consider more general potentials, but they do not have a direct interpretation of a  CC term as in the  original GR action with a cosmological term, see e.g. \cite{esposito2001scalar,Alsing:2011er, Ozer:2017oik,Faraoni:2004pi}.  Let us briefly explain why. If one starts from a general potential for the BD-field in the action, say some  arbitrary function of the BD-field $U(\psi)$ (not carrying along any additive constant)  in place of the constant term $\rL$ in \eqref{eq:BDaction}, then it is not so straightforward to  generate a CC term and still remain in a pure BD framework. For if one assumes that $\psi$ develops a vacuum expectation value\,\footnote{{We may assume, for the sake of the argument, that it is a classical ground state, since the BD-field is supposed to be part of the external gravitational field. Recall that we do not assume gravity  being a quantized theory here but just a set of background fields, in this case composed of the metric components $g_{\mu\nu}$ and $\psi$ (or equivalently $\varphi$)}.}, then the theory (when written in terms of the  fluctuating field around the ground state) would split into a non-minimal term coupled to curvature and a conventional EH term, so it would not be a pure BD theory.} {This point of view is perfectly possible and has been considered by other authors, on which we shall comment in some detail in Sec.\,\ref{sec:Discussion_BD_article}. However, proceeding in this way would lead us astray from our scientific leitmotif in this work (which is, of course, to remain fully within the BD paradigm).  Yet there is another option which preserves our BD philosophy, which is to perform a conformal transformation to the Einstein frame, where one can define a strict CC term.  Then we could impose that the effective potential in that frame, $V$,  is a constant. However, once this is done, the original potential in the (Jordan) BD frame, $U$, would no longer be constant since it would be proportional to $V$ times $\varphi^2$, with $\varphi$ defined as in \eqref{eq:definitions_BD_article}.  Conversely, one may assume $U=\rL={\rm const}$. in the Jordan frame (as we actually did) and then the vacuum energy density will appear mildly time-evolving in the Einstein frame (for sufficiently large $\omega_{\rm BD}$, of course).  This last option is actually  another way to understand why the vacuum energy density can be perceived as a slightly time-evolving quantity when the BD theory is viewed from the GR standpoint. It is also the reason behind the fact that the BD-$\CC$CDM framework mimics the so-called running vacuum model (RVM), see  Appendix \ref{sec:RVMconnection_BD_article} and references therein  for details.} {As indicated, in this work we opt for considering the definition provided in \eqref{eq:BDaction}, as we wish  to preserve the exact canonical form of the late-time acceleration source that is employed in the GR-$\Lambda$CDM model. At the same time we exploit the connection of the BD framework with the RVM and its well-known successful phenomenological properties, see e.g.\,\cite{Sola:2016ecz , Sola:2017jbl , Gomez-Valent:2018nib ,Gomez-Valent:2017idt,Sola:2017znb , sola2017first , Sola:2016hnq , Sola:2015wwa }.  In addition, in Sec.\,\ref{sec:EffectiveEoS_BD_article} we  consider a direct parametrization of the departures of BD-$\Lambda$CDM  from  GR-$\Lambda$CDM. }


\subsubsection{Effective gravitational strength}
From \eqref{eq:BDactionvarphi_BD_article} it follows that the quantity
\begin{equation}\label{eq:Gvarphi}
G(\varphi)=\frac{G_N}{\varphi}
\end{equation}
constitutes the effective gravitational coupling at the level of the BD-action. We will argue that $G(\varphi) $ is larger than $G_N$ because $\varphi<1$ (as it will follow from our analysis).  The gravitational field between two tests masses, however, is \textit{not} yet $G(\varphi) $ but the quantity  $G_{\rm eff}(\varphi)$  computed below.
\newline
\newline
{Let us remark that if one would like to rewrite the BD action in terms of a canonically normalized scalar field $\phi$ (of dimension $1$) having  a non-minimal coupling to curvature of the form  $\frac12\xi\phi^2 R$, it would suffice to redefine the BD-field as  $\psi=8\pi \xi\,\phi^2$   with $\xi=1/(4\oD)=\epsilon_{\rm BD}/4$, and then the scalar part of the action takes on the usual form}
\begin{eqnarray}
S_{\rm BD}=\int d^{4}x\sqrt{-g}\left(\frac12\,\xi\phi^2 R-\frac12 g^{\mu\nu}\partial_{\nu}\phi\partial_{\mu}\phi-\rL\right)+\int d^{4}x\sqrt{-g}\,{\cal L}_m(\chi_i,g_{\mu\nu})\,. \label{eq:BDaction2}
\end{eqnarray}
This alternative expression allows us to immediately connect with the usual parametrized post-Newtonian (PN) parameters, which restrict the deviation of the scalar-tensor theories of gravity with respect to GR\,\cite{Boisseau:2000pr, Clifton:2011jh,Will:2014kxa}.  Indeed, if we start from the generic scalar-tensor  action
\begin{eqnarray}
S_{\rm }=\int d^{4}x\sqrt{-g}\left(\frac12\, F(\phi) R-\frac12 g^{\mu\nu}\partial_{\nu}\phi\partial_{\mu}\phi-V(\phi)\right)+\int d^{4}x\sqrt{-g}\,{\cal L}_m(\chi_i,g_{\mu\nu})\,, \label{eq:BDactionST}
\end{eqnarray}
we can easily recognize from \eqref{eq:BDaction2} that $F(\phi)=\xi\phi^2$, and that the potential $V(\phi)$ is just replaced by the  CC density $\rL$.  {In this way we can easily apply the well-known formulae of the PN formalism. We find the following values for the main PN parameters $\gamma^{\rm PN}$ and $\beta^{\rm PN}$ in our case (both being equal to $1$ in strict GR)}:
\begin{equation}\label{eq:gammaPPN}
1-\gamma^{\rm PN} =\frac{F^{\prime}(\phi)^2}{F(\phi)+2F^{\prime}(\phi)^2}= \frac{4\xi}{1+ 8\xi}=\frac{\eBD}{1+ 2\eBD}\simeq \eBD +   {\cal O}(\eBD^2)
\end{equation}
and
\begin{equation}\label{eq:betaPPN}
 1-\beta^{\rm PN}=-\frac14\, \frac{F(\phi) F^{\prime}(\phi)}{2F(\phi)+3F^{\prime}(\phi)^2}\,\frac{d \gamma^{\rm PN}}{d\phi}  =0\,,
\end{equation}
where {the primes refer here to derivatives with respect to $\phi$}. We are neglecting terms of $ {\cal O}(\eBD^2)$  and, in the second expression,  we use the fact that  $d\gamma_{\rm PN}/d\phi=0$ since $\oD=$ const. (hence $\eBD=$ const. too) in our case.  {Therefore, in BD-gravity, $\gamma^{\rm PN}$ deviates from $1$  an amount given precisely by $\eBD$ (in linear order), whereas $\beta^{\rm PN}$ undergoes no deviation at all. Furthermore, the effective gravitational strength between two test masses in the context of the scalar-tensor framework
\eqref{eq:BDactionST} is well-known\,\cite{Boisseau:2000pr,Will:2014kxa,Clifton:2011jh}. In our case it  leads to the following result:}
\begin{equation}\label{eq:LocalGN}
G_{\rm eff}(\varphi) =\frac{1}{8\pi F(\phi)}\frac{2F(\phi) + 4F^{\prime}(\phi)^2}{2F(\phi)+3F^{\prime}(\phi)^2}=\frac{1}{8\pi\xi\phi^2} \frac{1+8\xi}{1+6\xi}= G(\varphi) \frac{2+4\eBD}{2+3\eBD}\,,
\end{equation}
where $G(\varphi)$ is the effective gravitational coupling in the BD action, as indicated in Eq. \eqref{eq:Gvarphi}.
Expanding linearly in $\eBD$, we find
\begin{equation}\label{eq:LocalGN2}
G_{\rm eff}(\varphi) =G(\varphi)\left[1+\frac12\,\eBD + {\cal O}(\eBD^2)\right] \,.
\end{equation}
{We confirm from the above two equations that the physical gravitational field undergone by two tests masses is not just determined by the effective  $G(\varphi)$ of the action but by  $G(\varphi)$ times a correction factor which depends on $\eBD$ and is larger (smaller) than $G(\varphi)$  for  $\eBD>0\,   (\eBD<0 )$.}
\newline
\newline
{From the exact formula \eqref{eq:LocalGN} we realize that if the local gravitational constraint ought to hold strictly, i.e. $G_{\rm eff}\to G_N$,  such formula would obviously enforce}
\begin{equation}\label{eq:LC}
\varphi=\frac{2+4\eBD}{2+3\eBD}\,.
\end{equation}
Due to Eq.\,\eqref{eq:gammaPPN}, the bound obtained from the Cassini mission\,\cite{Bertotti:2003rm}, $\gamma^{\rm PN}-1=(2.1\pm 2.3)\times 10^{-5}$, translates directly into a constraint on  $\eBD\simeq (-2.1\pm 2.3)\times 10^{-5}$ (in linear order), what implies $\oD\gtrsim 10^4$ for the BD-parameter. Thus, if considered together with the assumption $\varphi\simeq 1$ we would be left with very little margin for departures of $\Geff$ from $G_N$. However,  as previously  indicated,  we will not apply these local astrophysical bounds in most of our analysis since  we will assume that $\eBD$ may not be constrained in the cosmological domain and that the cosmological value of the gravitational coupling $G(\varphi)$  is different from $G_N$  owing to some possible variation of the BD-field $\varphi$ at the cosmological level. This can still be compatible with the local astrophysical constraints provided that we assume the existence of a screening mechanism in the local range which `renormalizes' the value of $\omega_{\rm BD}$ and makes it appear much higher than its `bare' value (the latter being accessible only at the cosmological scales, where matter is much more diluted and uninfluential) --- cf. Sec.\,\ref{sec:Mach_BD_article} for details on the various possible BD scenarios.  We know that this possibility remains open and hence it must be explored\,\cite{Avilez:2013dxa}, see also \cite{Amendola:2015ksp,Clifton:2011jh}  and references therein.
\newline
\newline
Henceforth we shall stick to the original BD-form (\ref{eq:BDaction}) of the action  since the field $\psi$ (or, alternatively, its dimensionless companion $\varphi$)  is directly related to the effective gravitational coupling and the  $\oD$ parameter can be ascribed as being part of the kinetic term of $\psi$.  In contrast, the form (\ref{eq:BDaction2}) involves both $\phi$ and $\xi=1/(4\oD)$ in the definition of the gravitational strength.
\subsection{Why  BD-$\CC$CDM alleviates at a time the $H_0$ and $\sigma_8$ tensions ?  Detailed preview and main results  of our analysis.}\label{sec:preview_BD_article}
The field $\varphi$ and the parameter $\epsilon_{\rm BD}$ defined in the previous section, Eq.\,\eqref{eq:definitions_BD_article}, are the fundamental new ingredients of BD-gravity as compared to GR  in the context of our analysis.  Any departure of $\varphi$ from one and/or of  $\eBD$ from zero should reveal an extra effect of the BD-$\CC$CDM model as compared to the  conventional GR-$\CC$CDM one.   We devote this section to study the influence of  $\varphi$ and $\eBD$ on the various observable we use in this work to constrain the BD model.  This preliminary presentation will serve as a preview of the results presented in the rest of the chapter and will allow us to anticipate why BD-$\Lambda$CDM is able to alleviate so efficiently both of the $H_0$ and $\sigma_8$ tensions that are found in the context of the traditional  GR-$\Lambda$CDM framework.
\newline
Interestingly, many Horndeski theories \cite{Horndeski:1974wa} reduce in practice to BD at cosmological scales \cite{Avilez:2013dxa}, so the ability of  BD-$\Lambda$CDM to describe the wealth of current observations can also be extended to other, more general, scalar-tensor theories of gravity. Hence, it is crucial to clearly identify the reasons why BD-$\Lambda$CDM leads to such an improvement in the description of the data. Only later on (cf. Sec. \ref{sec:NumericalAnalysis_BD_article}) we will fit in detail the overall  data to the BD-$\Lambda$CDM model and will display the final numerical results. Here, in contrast, we will endeavour  to show why BD-gravity has specific clues to the solution which are not available to GR.
\newline
We can show this in two steps. First, we analyze what happens when we set $\eBD=0$  at fixed values of $\varphi$ different from $1$.  From Eq.\,\eqref{eq:LocalGN} we can see that this scenario means to stick to the standard GR picture, but assuming that the effective Newtonian coupling can act at cosmological scales with a (constant) value $G_{\rm eff}=G(\varphi)$  different from the local one $G_N$.   In a second stage,  we  study the effect of the time dependence of $\varphi$  (triggered by a nonvanishing value of $\eBD$), i.e. we will exploit the departure of $G_{\rm eff}$  in  Eq.\,\eqref{eq:LocalGN} from $G_N$ caused by $\eBD\neq0$ \textit{and} a variable $G(\varphi)$.  It will become clear  from this two-step procedure why BD-gravity has the double ability to reduce the two $\CC$CDM tensions in an harmonic way. On the one hand, a value of $\varphi$ below $1$ in the late Universe increases the value of $G_{\rm eff}$  and hence of $H_0$, so it should be  able to significantly  reduce the $H_0$-tension;  and on the other hand {the dynamics of $\varphi$ triggered by a finite (but negative) value of $\eBD$ helps to suppress the structure formation processes in the Universe, since it enhances the Hubble friction and also leads to a decrease of the Poisson term in the perturbations equation.} {The upshot is that the $\sigma_8$-tension becomes reduced as well.} {Let us note that the lack of use of LSS data may lead to a different conclusion, in particular to $\eBD>0$, see e.g. \cite{Yadav:2019fjx}. This reference, in addition, uses only an approximate treatment of the CMB data through distance priors.}
\subsubsection{Role of $\varphi$, and the $H_0$-tension}\label{sec:rolesvarphiH0_BD_article}
Let us start, then, by studying how the observable change with $\varphi$ when $\eBD=0$,  for fixed values of the current energy densities. In the context of BD-gravity, as well as in GR,  if the energy densities are fixed at present we can fully determine their cosmological evolution, since we consider all the species are self-conserved. In BD-gravity, with $\eBD=0$, the Hubble function takes the  form
\begin{equation}\label{eq:H1}
H^2(a) = \frac{8\pi G_N}{3\varphi}\rho(a)\,,
\end{equation}
where  $\varphi={\rm const}$. We have just removed the time derivatives of the scalar field in the Friedmann equation \eqref{eq:Friedmann_equation_varphi_BD_article}. We define $H_0$ from the value of the previous expression at $a=1$ (current value). Recall from the previous section that $\rho$ is the sum of all the energy density contributions, namely $\rho \equiv \rho_m + \rho_\gamma + \rho_\nu + \rho_\Lambda$. Therefore,
\begin{equation}\label{eq:H1b}
H_0^2 = \frac{8\pi G_N}{3\varphi}\rho^0=\frac{8\pi G_N}{3\varphi}\left( \rho_m^0+\rho_\gamma^0+ \rho_\nu^{0}+\rL\right)\,.
\end{equation}
$\rho^0=\rho(a=1)$ is the total energy density at present, $\rho_\gamma^0$ is the corresponding density of photons and $\rho_\nu^{0}$ that of neutrinos, and finally $\rL$ is the original cosmological constant density in the BD-action \eqref{eq:BDaction}.
Using \eqref{eq:H1b} it proves now useful  to rewrite \eqref{eq:H1} in the alternative way:
\begin{equation}\label{eq:H2}
H^2(a)= H_0^2\left[1+\tilde{\Omega}_{m}(a^{-3}-1)+\tilde{\Omega}_\gamma(a^{-4}-1)+\tilde{\Omega}_\nu(a)-\tilde{\Omega}_\nu\right]\,,
\end{equation}
where we use the modified cosmological parameters (more appropriate for the BD theory):
\begin{equation}\label{eq:tildeOmegues}
\tilde{\Omega}_i\equiv\frac{\rho^{0}_i}{\rho^{0}}=\frac{\Omega_i}{\varphi}\,,\ \ \ \ \ \ \ \ \  \rho^{0}=\frac{3H_0^2}{8\pi G_N}\,\varphi=\rco\varphi\,.
\end{equation}
The tilde is to distinguish the modified $\tilde{\Omega}_i$ from the usual cosmological parameters $\Omega_i=\rho_i^{0}/\rco$ employed in the GR-$\Lambda$CDM model.  In addition,  $\tilde{\Omega}_{m}=\tilde{\Omega}_{cdm}+\tilde{\Omega}_b$ is the sum of the contributions from CDM and baryons; and $\tilde{\Omega}_\gamma$ and $\tilde{\Omega}_\nu$ are the current values for the photons and neutrinos. For convenience, we also define $\tilde{\Omega}_r=\tilde{\Omega}_\gamma+\tilde{\Omega}_\nu$.  We remark that he current total energy density $ \rho^{0}$ is related to the usual critical density $\rco=3H_0^2/(8\pi G_N)$  through a factor of $\varphi$, as indicated above.
The modified parameters obviously satisfy the canonical sum rule
\begin{equation}\label{eq:SumRuleBD}
\tilde{\Omega}_m+\tilde{\Omega}_r+\tilde{\Omega}_\CC=1\,.
\end{equation}
The form of \eqref{eq:H2} is completely analogous to the one found in GR-$\Lambda$CDM since in BD-$\CC$CDM the $\tilde{\Omega}_i$'s represent the fraction of energy carried by the various species in the current Universe, as the $\Omega_i$'s do in GR, so from the physical point of view the $\tilde{\Omega}_i$'s in BD and the $\Omega_i$'s in GR contain the same information\footnote{For $|\eBD|\neq 0$ and small, the $\tilde{\Omega}_i$ parameters defined in Eq.\,\eqref{eq:tildeOmegues} receive a correction of ${\cal O}(\eBD)$, see Sec.\,\ref{sec:eBDands8_BD_article}.}. $H_0$ represents in both cases the current value of the Hubble function. Nevertheless, there is a very important (although maybe subtle) difference, namely: in BD-$\Lambda$CDM there does not exist a  one-to-one correspondence between $H_0$ and $\rho^{0}$. In contradistinction to GR-$\Lambda$CDM, in the BD version of the concordance model the value of $\varphi$ can modulate $H_0$ for a given amount of the total (critical) energy density. In other words, given a concrete value of $H_0$ there is a $100\%$ degeneracy between $\varphi$ and $\rho^{0}$. This degeneracy is broken by the data, of course. The question we want to address is precisely whether there is still room for an increase of $H_0$ with respect to the value found in the GR-$\Lambda$CDM model once the aforementioned degeneracy is broken by observations. We will see that this is actually the case by analyzing what is the impact that $\varphi$ has on each observable considered in our analyses.
\newline
\newline
{\bf Supernovae data}
\newline
\newline
\noindent
In the case  of Type Ia  Supernovae data (SNIa)  we fit observational points on their apparent magnitudes $m(z)= M +5\log_{10}[D_L(z)/10{\rm pc}]$, where $M$ is the absolute magnitude of the SNIa and $D_L(z)$ is the luminosity distance. In a spatially flat Universe the latter reads,
\begin{equation}
D_L(z)=c(1+z)\int^z_{0}\frac{dz^\p}{H(z^\p)}\,,
\end{equation}
where we have momentarily kept $c$ explicitly for the sake of better understanding. Considering \eqref{eq:H2}, we can easily see that if we only use SNIa data, there is a full degeneracy between $M$ and $H_0$ in the computation of the apparent magnitudes, so it is not possible to obtain information about $\varphi$, since we cannot disentangle it from the absolute magnitude. As in GR-$\Lambda$CDM, we can only get constraints on the current fraction of matter energy in the Universe, i.e. $\tilde{\Omega}_m$.
\newline
\newline
{\bf Baryon acoustic oscillations}
\newline
\newline
\noindent
The constraints obtained from the analysis in real or Fourier space of the baryon acoustic oscillations (BAO) are usually provided by galaxy surveys in terms of $D_A(z)/r_s$ and $r_s H(z)$, or in some cases by a combination of these two quantities when it is not possible to disentangle the line-of-sight and transverse information, through the so-called dilation scale $D_V$ (cf. Sec. \ref{sec:MethodData_BD_article}),
\begin{equation}
\frac{D_V(z)}{r_s}=\frac{1}{r_s}\left[D_M^2(z)\frac{cz}{H(z)}\right]^{1/3}\,,
\end{equation}
$D_M=(1+z)D_{A}(z)$ being the comoving angular diameter distance, $D_A(z)=D_L(z)/(1+z)^2$ the proper angular diameter distance, and
\begin{equation}\label{eq:rs}
r_s=\int_{z_d}^{\infty}\frac{c_s(z)}{H(z)}\,dz
\end{equation}
the comoving sound horizon at the baryon drag epoch $z_d$. In the above equation, $ c_s(z)$ is  the sound speed of the photo-baryon plasma, which depends on the ratio $\rho_b^{0}/\rho^{0}_\gamma$. The current temperature of photons (and hence also its associated current energy density $\rho^{0}_\gamma$) is already known with high precision thanks to the accurate measurement of the CMB monopole \cite{Fixsen:2009ug}. Because of \eqref{eq:H1} it is obvious that we cannot extract information on $\varphi$ from BAO data when $\eBD=0$, since it cancels exactly in the ratio $D_A(z)/r_s$ and the product $r_s H(z)$. Thus, BAO data provide constraints on $\tilde{\Omega}_m$ and $\rho_b^{0}$, but not on $\varphi$.
\newline
\newline
{\bf Redshift-space distortions (RSD)}
\newline
\newline
\noindent
{The  LSS observable $f(z)\sigma_8(z)$, which is essentially determined from RSD measurements,  is of paramount importance to study the formation of structures in the Universe.  In BD-gravity, the equation of matter field perturbations is different from that of GR and  is studied in detail in Sec.\,\ref{sec:StructureFormation_BD_article}.  Here we wish to make some considerations which will help to have a rapid overview of why BD-gravity with a cosmological constant can also help to improve the description of structure formation as compared to GR. The exact differential equation for the linear density contrast of the matter perturbations,  $\delta_m=\delta\rho_m/\rho_m$,  in this context can be computed  at deep subhorizon scales and is given by (cf. Sec. \ref{sec:StructureFormation_BD_article}):}
%
\begin{equation}\label{eq:ExactPerturScaleFactor}
\delta_m^{\prime\prime}+\left(\frac{3}{a}+\frac{H^\prime(a)}{H(a)}\right)\delta_m^\prime-\frac{4\pi \Geff(a)}{H^2(a)}\frac{\rho_m(a)}{a^2}\,\delta_m=0\,,
\end{equation}
{where for the sake of convenience we express it in terms of the scale factor and hence prime denotes here  derivative {\it w.r.t.}  such variable:  $()^\prime \equiv d()/d a$. In the above equation, $\Geff(a)$ is the effective gravitational coupling \eqref{eq:LocalGN} with $\varphi=\varphi(a)$ expressed in terms of the scale factor:}
\begin{equation}\label{eq:LocalGNa}
\Geff(a) = G(\varphi(a)) \frac{2+4\eBD}{2+3\eBD}\,.
\end{equation}
It crucially controls the  Poisson term of the perturbations equation, i.e. the last term in \eqref{eq:ExactPerturScaleFactor}.
As we can see, it is $\Geff(\varphi)$ and not just $G(\varphi)$ the coupling involved in the structure formation data since it is $\Geff(\varphi)$ the gravitational coupling felt by the test masses in BD-gravity.
It is obvious that the above Eq.\eqref{eq:ExactPerturScaleFactor} boils down to the GR form for $\eBD=0$ and $\varphi=1$.
\newline
\newline
With the help of the above equations  we  wish first to assess the bearing that $\varphi$ can have on the LSS observable $f(z)\sigma_8(z)$ at  fixed values of the current energy densities and for $\eBD=0$. Recall that when  $\eBD=0$ the BD-field cannot evolve at all, so it just remains fixed at some value.  For this consideration, we therefore set $\psi=$const. in equations \eqref{eq:Friedmannequation_BD_chapter} and \eqref{eq:pressureequation_BD_chapter} and of course the BD-theory becomes just a GR version with an effective coupling $\Geff=G(\varphi)=$const.  which nevertheless need not be identical to  $G_N$.  In these conditions, it is easy to verify that Eq.\,\eqref{eq:ExactPerturScaleFactor} adopts the following simpler form, in which $\Geff$ drops from the final expression:
\begin{equation}\label{eq:DC}
\delta_m^{\prime\prime}+\frac{3}{2a} \left(1-w(a)\right) \delta_m^\prime -\frac{3}{2a^2}\frac{\rho_m(a)}{\rho(a)}\delta_m=0\,.
\end{equation}
In the above expression, $w(a)=p(a)/\rho(a)$ is the equation of state (EoS) of the total cosmological fluid, hence  $\rho(a)$ and $p(a)$ stand respectively for the total density and pressure of the fluids involved (cf. Sec. \ref{sec:BDgravity_BD_article}). In particular, during the epoch of structure formation the matter particles contribute a negligible contribution to the pressure and the dominant component is that of the  cosmological term: $p(a)\simeq p_\CC=-\rL$.
\newline
It is important to realize that $\varphi=$ const. does not play any role in \eqref{eq:DC}. This means that its constant value, whatever it may be,  does not affect the evolution of the density contrast, which is only determined by the fraction of matter, $\rho_m(a)/\rho(a)$,  and the  EoS of the total cosmological fluid.  The equation that rules the evolution of the density contrast is exactly the same as in the GR-$\Lambda$CDM model. Matter inhomogeneities grow in the same way regardless of the constant value $\Geff$ that we consider. Matter tends to clump more efficiently for larger values of the gravitational strength, of course, but the Hubble friction also grows in this case, since such an increase in $\Geff$ also makes the Universe to expand faster.
Surprisingly, if $\eBD=0$, i.e. if $\Geff=$const., both effects compensate each other. Thus, the BD growth rate $f(a)=a\delta_m^\prime(a)/\delta_m(a)$ does not change  {\it w.r.t.} the GR scenario either. But what happens with $\sigma_8(z)$ ? It is computed through the following expression:
\begin{equation}\label{eq:sigma8}
\sigma_8^2(z)=\frac{1}{2\pi^2}\int_0^\infty dk\,k^2\,P(k,z)\,W^2(kR_8)\,,
\end{equation}
in which  $P(k,z)$ is the power spectrum of matter fluctuations and $W(kR_8)$ is the top hat smoothing function in Fourier space, with $R_8=8h^{-1}$ Mpc. Even for $\eBD=0$ one would naively expect \eqref{eq:sigma8} to be sensitive to the value of $\varphi$, since some relevant features of the power spectrum clearly are. For instance, the scale associated to the matter-radiation equality reads
\begin{equation}
k_{eq}=a_{eq}H(a_{eq})=H_0\tilde{\Omega}_m\sqrt{\frac{2}{\tilde{\Omega}_r}}\,,
\end{equation}
and $H_0\propto\varphi^{-1/2}$, so the peak of $P(k,z)$ is shifted when we change $\varphi$. Also the window function itself depends on $H_0$ through $R_8$. Despite this, the integral \eqref{eq:sigma8} does not depend on the Hubble function for fixed energy densities at present, and hence neither on $\varphi$. To see this, let us  first rescale the wave number  as follows  $k=\bar{k}h$, and we obtain
\begin{equation}\label{eq:sigma8v2}
\sigma_8^2(z)=\frac{1}{2\pi^2}\int_0^\infty d\bar{k}\,\bar{k}^2\,\underbrace{P(k=\bar{k}h,z)\,h^3}_{\equiv \bar{P}(\bar{k},z)}\,W^2(\bar{k}\cdot 8\,{\rm Mpc^{-1}})\,.
\end{equation}
The only dependence on $h$ is now contained in $\bar{P}(\bar{k},z)$. We can write $P(k,z)=P_0k^{n_s}T^2(k/k_{eq})\delta^2_m(z)$, with $T(k/k_{eq})$ being the transfer function and
\begin{equation}
P_0=A_s\frac{8\pi^2}{25}\frac{k_*^{1-n_s}}{(\tilde{\Omega}_m H_0^2)^2}\,,
\end{equation}
with $A_s$ and $n_s$ being the amplitude and spectral index of the dimensionless primordial power spectrum, respectively, and $k_*$ the corresponding pivot scale. The last relation can be found using standard formulae, see e.g. \cite{GorbunovRubakovBook,Amendola:2015ksp}. Taking into account all these expressions we obtain
\begin{equation}
\bar{P}(\bar{k},z)=A_s\frac{8\pi^2}{25}\frac{\bar{k}^{1-n_s}_*\bar{k}^{n_s}}{(10^4\varsigma^2\tilde{\Omega}_m)^2} T^2(\bar{k}/\bar{k}_{eq})\delta^2_m(z)\,,
\end{equation}
where we have used $H_0=100\, h\, \varsigma$  with $\varsigma\equiv 1\,{\rm km/s/Mpc}=2.1332\times 10^{-44}$ GeV (in natural units). We see that all factors of $h$ cancel out. Now it is obvious that $\sigma_8(z)$ is not sensitive to the value of $\varphi$. We have explicitly checked this with our own modified version of the Einstein-Boltzmann system solver \texttt{CLASS} \cite{Blas:2011rf}, in which we have implemented the BD-$\Lambda$CDM model (see Sec. \ref{sec:MethodData_BD_article} for details). The product $f(z)\sigma_8(z)$ does not depend on $\varphi$ when $\eBD=0$, so RSD data cannot constrain $\varphi$ either.
\newline
\newline
\newline
{\bf Strong Lensing time delay distances, distance ladder determination of $H_0$, and cosmic chronometers}
\newline
\newline
\noindent
In this chapter, we will use the Strong Lensing time delay angular diameter distances provided by the H0LICOW collaboration \cite{Wong:2019kwg}, see Sec. \ref{sec:MethodData_BD_article}. Contrary to SNIa and BAO data, these distances are not relative, but absolute. This allows us to extract information not only on $\Om$, but on $H_0$ too.  Furthermore,  the data on $H(z)$ obtained from cosmic chronometers (CCH) give us information about these two parameters as well. Cosmic chronometers have been recently employed in the reconstruction of the expansion history of the Universe using Gaussian Processes and the so-called Weighted Function Regression method \cite{Yu:2017iju,Gomez-Valent:2018hwc,Haridasu:2018gqm}, which do not rely on any particular cosmological model. The extrapolated values of the Hubble parameter found in these analyses are closer to the best-fit GR-$\Lambda$CDM value reported by Planck \cite{Aghanim:2018eyx}, around $H_0\sim (67.5-69.5)$ km/s/Mpc, but they are still compatible within  $\sim 1\sigma$ c.l. with the local determination obtained with the distance ladder technique \cite{Riess:2018uxu,Riess:2019cxk,Reid:2019tiq} and the Strong Lensing time delay measurements by H0LICOW \cite{Wong:2019kwg}. The statistical weight of the CCH data is not as high as the one obtained from these two probes, so when combined with the latter, the resulting value for $H_0$ is still in very strong tension with Planck \cite{Gomez-Valent:2018hwc,Haridasu:2018gqm}. As mentioned before, $H_0^2\propto \rho^{0}/\varphi$ when $\eBD=0$. Thus, we can alleviate in principle the $H_0$-tension by keeping the same values of the current energy densities of all the species as in the best-fit GR-$\Lambda$CDM model reported by Planck \cite{Aghanim:2018eyx}, lowering the value of $\varphi$ down at cosmological scales, below $1$, and assuming some sort of screening mechanism acting on high enough density regions that allows us to evade the solar system constraints and keep unmodified the stellar physics needed to rely on CCH, SNIa, the H0LICOW data, and the local distance ladder measurement of $H_0$. By doing this we do not modify at all the SNIa, BAO and RSD observable {\it w.r.t.} the GR-$\Lambda$CDM, and we automatically improve the description of the H0LICOW data and the local determination of $H_0$, which are the observable that prefer higher values of the Hubble parameter.  Let us also mention that the fact that $\varphi<1$ throughout the cosmic history (which means $G>G_N$)  allows to have a larger value of $H$ (for similar values of the density parameters) at any time as compared to the GR-$\CC$CDM and hence a smaller value of the sound horizon distance $r_s$, Eq.\,\eqref{eq:rs}, what makes the model to keep a good  fit to the BAO data. This is confirmed by the numerical analysis presented in Tables 3-6 and 10 as compared to the conventional $\CC$CDM values, see Tables 3-5 and 7.  While the claim existing in the literature  that models which predict smaller values of $r_s$ are the preferred ones for solving the $H_0$-tension is probably correct,  we should point out that this sole fact is no guarantee of success, as one still needs in general a compensation mechanism at low energies which prevents $\sigma_8$ from increasing and hence worsening such tension. In the BD-$\CC$CDM such compensation mechanism is provided by a  negative value of $\eBD$ (as we will show later), and for this reason the two tensions can be smoothed at the same time in an harmonic way.
\newline
\newline
Overall, as we have seen from the above discussion,  according to the (long) string of supernovae, baryonic acoustic oscillations, cosmic chronometers, Strong Lensing and local Hubble parameter data (SNIa+BAO+RSD+CCH+H0LICOW+$H_0$) it is possible to loosen the $H_0$-tension, and this is already very remarkable, but we still have to see whether this is compatible with the very precise measurements of the photon temperature fluctuations in the CMB map or not. More specifically, we have to check whether it is possible to describe these anisotropies while keeping the current energy densities close to the best-fit GR-$\Lambda$CDM model from Planck, compatible with $\varphi<1$.
\newline
\newline
\begin{figure}[t!]
\setcounter{figure}{0}
\begin{center}
\label{Cls-varphi_BD_article}
\includegraphics[width=6in, height=4in]{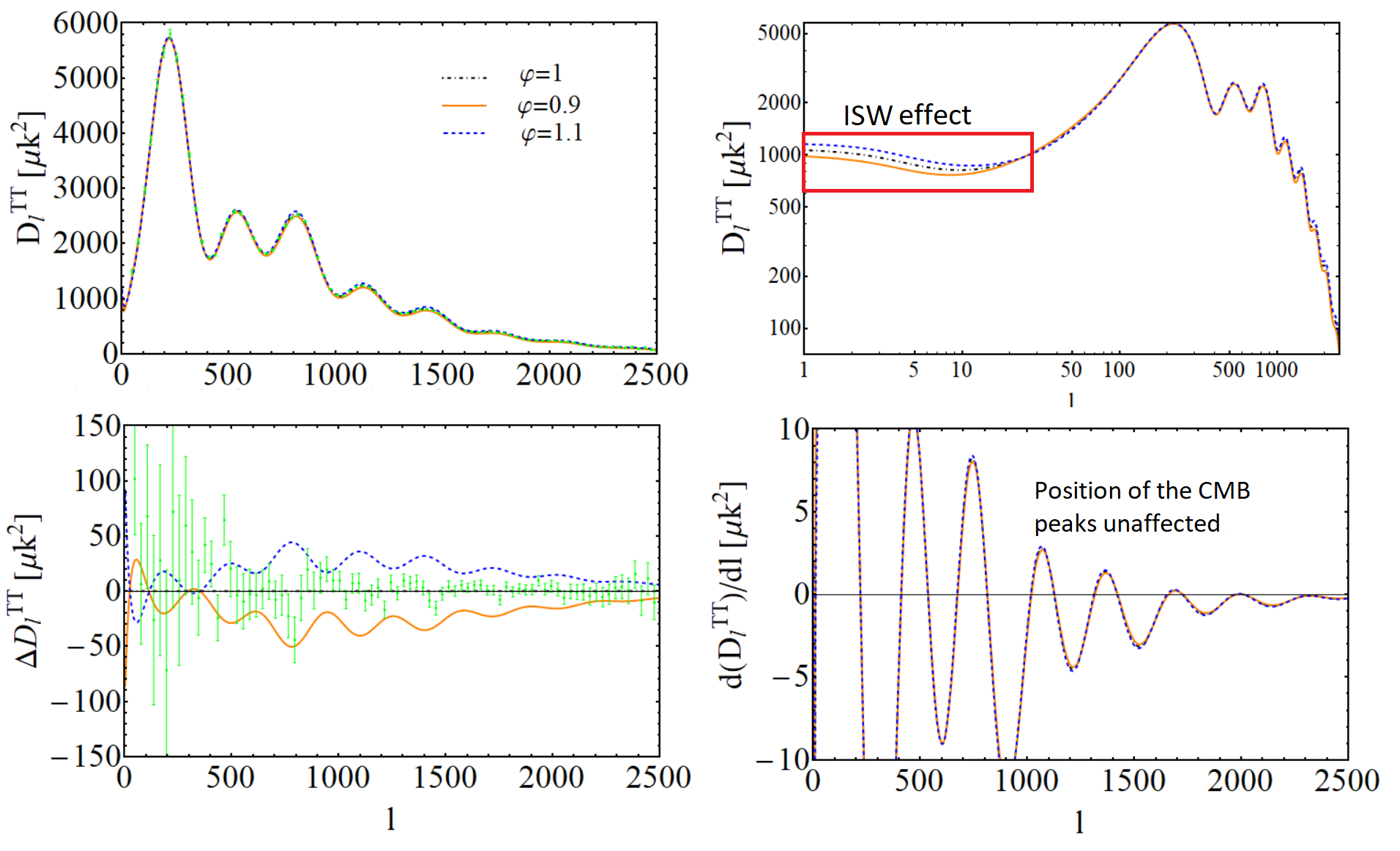}
\caption{\scriptsize {{\it Left-upper plot:} Theoretical curves of the CMB temperature anisotropies obtained by fixing the current energy densities, $\tau$ and the parameters of the primordial power spectrum to the GR-$\Lambda$CDM best-fit values obtained from the analysis of TTTEEE+lowE data by Planck 2018 \cite{Aghanim:2018eyx} (which we will refer to as the $\CC$CDM baseline configuration, denoted by a dot-dashed black line), against the BD-$\Lambda$CDM model  keeping $\eBD=0$ for two different constant values  $\varphi=0.9,1.1$  (orange and dashed blue lines, respectively); {\it Right-upper plot:} The same, but using a logscale in the $x$-axis to better appreciate the integrated Sachs-Wolfe effect at low multipoles; {\it Left-bottom plot:} Absolute differences of the data points and theoretical curves for $\varphi=0.9,1.1$ {\it w.r.t.} the $\varphi=1$ case; {\it Right-bottom plot:} Derivative of the functions plotted in the upper plots, to check the effect of $\varphi$ on the position of the peaks, which corresponds to the location of the zeros in this plot. No shift is observed, as expected (cf. the explanations in the main text).}}
\end{center}
\end{figure}
\newline
\newline
\newline
{\bf CMB temperature anisotropies}
\newline
\newline
\noindent We expect the peak positions of the CMB temperature (TT, in short) power spectrum to remain unaltered under changes of $\varphi$ (when $\eBD=0$), since they are always related with an angle, which is basically a ratio of cosmological distances ({a transverse distance to the line of sight divided by the angular diameter distance}).  If $\varphi=$const.,  such constant cancels  in the ratio, so there is no dependence on $\varphi$. In the right-bottom plot of Fig. 1 we show the derivative of the $\mathcal{D}_l^{TT}$'s for three alternative values of $\varphi$. It is clear that the location of the zeros does not depend on the value of this parameter. Hence $\varphi$ does not shift the peaks of the TT CMB spectrum when we consider it to be a constant throughout the entire cosmic history, as expected. Nevertheless, there are two things that affect its shape and both are due to the impact that $\varphi$ has on the Bardeen potentials. We have seen before that the matter density contrast is not affected by $\varphi$ when it is constant, but taking a look on the Poisson equation in the BD model (cf. \ref{Appendix_B_BD}), we can convince ourselves that $\varphi$ does directly affect the value of $\Psi$ and $\Phi$, since both functions are proportional to $\rho_m\delta_m/\varphi$ at subhorizon scales, see Eqs.\,\eqref{eq:PhiplusPsi}-\eqref{eq:Poisson3}. This dependence modifies two basic CMB observable:
\begin{itemize}
\item The CMB lensing, at low scales {(large multipoles, $500\lesssim l\lesssim 2000$)}. In the left-bottom plot of Fig. 1 we show the difference of the TT CMB spectra {\it w.r.t.} the GR-$\Lambda$CDM model. A variation of $\varphi$ changes the amount of CMB lensing, which in turn modifies the shape of the spectrum mostly from the third peak onwards. {In that multipole range also the Silk damping plays an important role and leaves a signature \cite{Silk1968}.}
\item The integrated Sachs-Wolfe effect \cite{Sachs:1967er}, at large scales {(low multipoles, $l\lesssim 30$). Values of $\varphi<1$ (which, recall, lead to higher values of $H_0$) suppress the power of the $\mathcal{D}_l^{TT}$'s in that range.} This is a welcome feature, since it could help us to explain the low multipole ``anomaly'' that is found in the context of the GR-$\Lambda$CDM model (see e.g. Fig. 1 of \cite{Aghanim:2018eyx}, and \cite{Das:2013sca}). {Later on in the chapter, we further discuss and confirm the alleviation  of this intriguing anomaly in light of the final fitting results, see Sec.\,\ref{sec:LowMultipoles_BD_article} and Fig.\,12. This is obviously an additional bonus of the BD-$\CC$CDM framework.}
\end{itemize}
%
\begin{figure}[t!]
\begin{center}
\label{fig:nsVSvarphi}
\includegraphics[width=6in, height=2.4in]{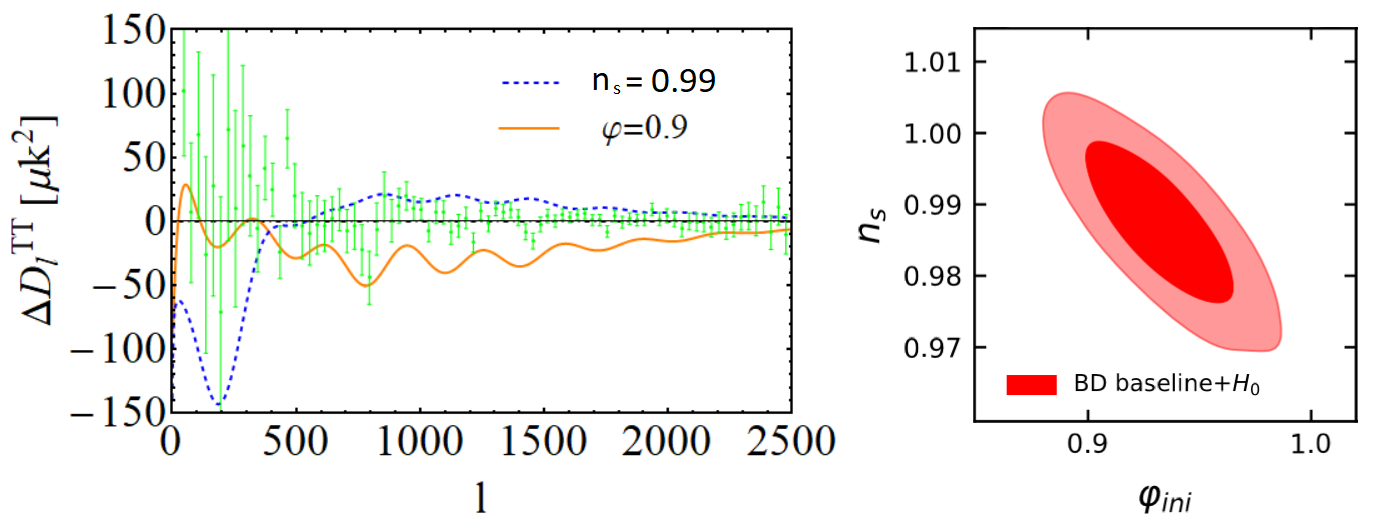}
\caption{\scriptsize {{\it Left plot:} Differences between the CMB temperature spectrum obtained using the original $\CC$CDM baseline configuration (as used in Fig. 1) and those obtained using $\varphi=0.9$ (with the baseline $n_s=0.9649$) and $n_s=0.99$ (with the baseline $\varphi=1$). This is to show that the effects induced by a lowering of $\varphi$ can be compensated by an increase of the spectral index; {\it Right plot:} We also show the $1\sigma$ and $2\sigma$ c.l. regions in the $(\varphi_{ini},n_s)$-plane, obtained using the baseline dataset, together with the Gaussian prior on $H_0$ from \cite{Reid:2019tiq}, see Sec. \ref{sec:MethodData_BD_article} for details. Here one can clearly see the  anticorrelation between these two parameters.}}
\end{center}
\end{figure}
%
Even though in Fig. 1 we are considering large ($\sim 10\%$) relative deviations of $\varphi$ with respect to $1$, the induced deviations of the $\mathcal{D}_l^{TT}$'s are fully contained in the observational error bars at low multipoles, and they are not extremely big at large ones. The latter are only $1-2\sigma$ away from most of the data points.  We emphasize that we are only varying $\varphi$ here, so there is still plenty of room to correct these deviations by modifying the value of other parameters entering the model. To do that it would be great if we could still keep the values of the current energy densities as in the concordance model, since this would ensue the automatic fulfillment of the constraints imposed by the datasets discussed before. But is this possible ? In Fig. 2 we can see that e.g. an increase of $n_s$ can compensate for the decrease of $\varphi$ pretty well. This is why in the BD model we obtain higher best-fit values of the spectral index {\it w.r.t.} the GR-$\Lambda$CDM, and a clear anti-correlation between these two parameters ({cf. Fig. 2, Tables 3-5 and 10, and Sec. \ref{sec:NumericalAnalysis_BD_article} for details}). Small variations in other parameters can also help to improve the description of the data, of course, but the role of $n_s$ seems to be important. In Fig. 3 we can appreciate the change in the matter power spectrum induced by different values of $n_s$. There is a modification in the range of scales that can be observationally accessed to with the analysis of RSD, but these differences are negative at $k\lesssim 0.07\,h{\rm\,Mpc^{-1}}$ and positive at larger values of the wave number (lower scales), so there can be a compensation when $\sigma_8$ is computed through \eqref{eq:sigma8}, leaving the value of the latter stable. Moreover, we will see below that $\eBD$ can also help to decrease the value of $P(k)$ at $k\gtrsim 0.02\,h{\rm\,Mpc^{-1}}$, so the correct shape for the power spectrum is therefore guaranteed.
\newline
\newline
The upshot of this section is worth emphasizing:  as  it turns out,  the sole fact of considering a cosmological Newtonian coupling about $\sim 10\%$ larger than the one measured locally can allow us to fit very well all the cosmological datasets, loosening the $H_0$-tension and keeping standard values of $\sigma_8$. It has become common in the literature to divide the theoretical proposals  able to decrease  the $H_0$-tension into two different classes depending on the stage of the Universe's expansion at which new physics are introduced \cite{Knox:2019rjx}: pre- and post-recombination solutions. The one we are suggesting here (see also the preceding paper \cite{Sola:2019jek}) cannot be identified with any of these two categories, since it modifies the strength of gravity at cosmological scales not only before the decoupling of the CMB photons or the late-time Universe, but during the whole cosmological history, relying on a screening mechanism able to generate $\Geff=G=G_N$ in high density ({non-relativistic}) environments where non-linear processes become important, as e.g. in our own solar system\footnote{The study of these screening mechanisms, see e.g. \cite{Tsujikawa:2008uc,Amendola:2015ksp,Clifton:2011jh,Li:2020uaz} and references therein, can be the subject of  future work, but here we  remark that e.g. chameleon \cite{Khoury:2003aq}, symmetron \cite{Hinterbichler:2010es} or Vainshtein mechanisms (see \cite{Kimura:2011dc} and references therein), do not screen the value of $\varphi$ during the radiation-dominated epoch. This is important to loose the $H_0$-tension in the BD-$\Lambda$CDM framework through the increase of $H(z)$ at both, the early and late Universe.}. That there is indeed a  change of the gravity  strength  throughout the entire cosmological history in our study follows from the fact that $\eBD\neq0$ in the BD framework, and this is exactly the feature that we are going to exploit in the next section, a feature which adds up to the mere change of the global strength of the gravity interaction, which is still possible for $\eBD=0$ in the BD context, and that we have explored in the previous section.

\begin{figure}[t!]
\begin{center}
\label{fig:pkns}
\includegraphics[width=6.5in, height=2.3in]{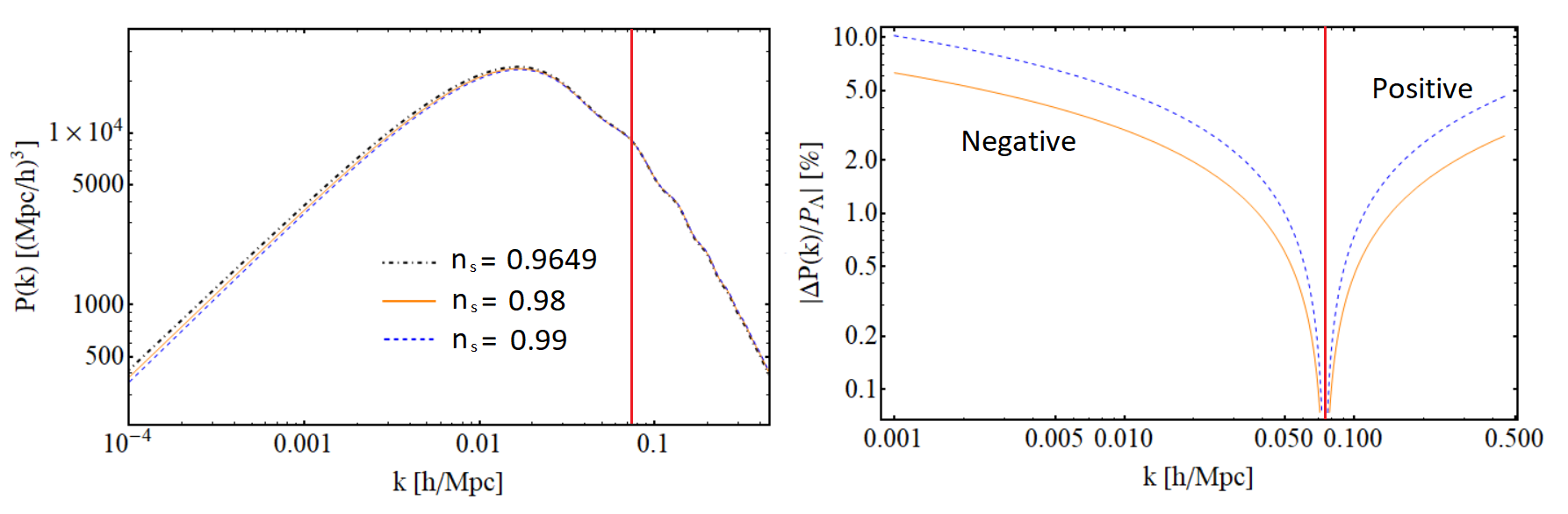}
\caption{\scriptsize {{\it Left plot:} Here we compare the linear matter power spectrum obtained for the $\CC$CDM baseline configuration (dot-dashed black line) and two alternative values of the spectral index $n_s$. The vertical red line indicates the value of the wave number at which there is the break in the right plot; {\it Right plot:} The absolute relative differences in $P(k)$ {\it w.r.t.} the baseline $\CC$CDM model. The ``positive'' (``negative'') region is the one in which $P(k)$ is larger (lower) than in the baseline setup. Both zones are delimited by a vertical red line. See the comments in the text.}}
\end{center}
\end{figure}

\subsubsection{Role of $\eBD$, and the $\sigma_8$-tension}\label{sec:eBDands8_BD_article}
Next we study the effect of $\eBD\neq0$. We know that when we introduce the matter bispectrum information from BOSS \cite{Gil-Marin:2016wya} (which definitely prefers a lower amount of structure in the Universe than the data reported in \cite{Alam:2016hwk}) in our fitting analyses, we find a stronger signal for negative values of  $\eBD$ (cf. Sec. \ref{sec:NumericalAnalysis_BD_article}). When $\eBD\ne 0$ equation \eqref{eq:DC} is not valid anymore, since it was derived under the assumption of $\varphi=$const.  {We start once more from the exact perturbations equation for the matter density contrast  in the linear regime within the BD-gravity, i.e. Eq.\,\eqref{eq:ExactPerturScaleFactor}.  Let us consider its form within the approximation $|\eBD|\ll 1$ (as it is preferred by the data, see e.g. Table 3). We come back to the BD-field equations  \eqref{eq:Friedmann_equation_varphi_BD_article} and \eqref{eq:Pressure_equation_varphi_BD_article} and apply such approximation.  In this way Eq.\,\eqref{eq:ExactPerturScaleFactor} can be expanded linearly in $\eBD$ as follows (primes are still denoting derivatives with respect to the scale factor)}:
\begin{equation}\label{eq:DC2}
\delta_m^{\pp}+\left[\frac{3}{2a}\left(1-w(a)\right)+\frac{\mathcal{F}^\p}{2}-\frac{\varphi^\p}{2\varphi}\right] \delta_m^\p-\frac{3}{2a^2}\frac{\rho_m(a)}{\rho(a)}\delta_m\left(1+\frac{\eBD}{2}-\mathcal{F}\right)=0\,,
\end{equation}
where we have defined
\begin{equation}\label{eq:F}
\mathcal{F}=\mathcal{F}\left(\frac{\varphi^\p}{\varphi}\right)=-a\frac{\varphi^\p}{\varphi}+\frac{\omega_{\rm BD}}{6}a^2\left(\frac{\varphi^\p}{\varphi}\right)^2\,.
\end{equation}
In particular, we have expanded the effective gravitational coupling  \eqref{eq:LocalGNa} linearly as in \eqref{eq:LocalGN2}.
As we can show easily, the expression \eqref{eq:F} can be treated as a perturbation, since it is proportional to $\eBD$. To prove this, let us borrow the solution for the matter-dominated epoch (MDE) derived from the analysis of fixed points presented in Appendix \ref{Appendix_D}. Because the behaviour of the BD-field towards the attractor at the MDE is governed by a power law of the form $\varphi\sim a^{\eBD}$ (cf. Eq.\,\eqref{eq:psiMDE}), we obtain
\begin{eqnarray}\label{eq:ximatter}
a\frac{\varphi^\p}{\varphi} &=& \eBD+\mathcal{O}(\epsilon^2_{\rm BD})\,,\\
\mathcal{F}&=&-\frac{5}{6}\eBD+\mathcal{O}(\epsilon^2_{\rm BD})\simeq -\frac{5}{6}\eBD\,;\,\ \ \ \ \ \ \ \ \mathcal{F}^\p=\mathcal{O}(\epsilon^2_{\rm BD})\simeq 0\,.\label{eq:Fximatter}
\end{eqnarray}
This proves our contention that the function $\mathcal{F}$ in  \eqref{eq:F} is of order $\eBD$ and its effects can be treated as a perturbation to the above formulas.
Incidentally, the relative change of $\varphi$ does not depend on $\varphi$ itself.
%
%
From the definition of $\mathcal{F}$ we can now refine the old Friedmann's equation \eqref{eq:H1} or \eqref{eq:H2} (only valid for $\eBD=0$) as follows. Starting from Eq.\,\eqref{eq:Friedmannequation_BD_chapter}, it is easy to see that it can be cast in the Friedmann-like form:
\begin{equation}\label{eq:FriedmannWithF}
H^2(a)=\frac{8\pi G_N}{3\varphi (a) (1-\mathcal{F}(a))}\,\rho(a).
\end{equation}
Despite $\mathcal{F}(a)$ evolves with the expansion, as shown by \eqref{eq:F}, it is of order $\eBD$ and evolves very slowly. In this sense, Eq.\,\eqref{eq:FriedmannWithF} behaves approximately as an ${\cal O}(\eBD)$ correction to Friedmann's equation \eqref{eq:H1}.
Setting $a=1$, the value of the current Hubble parameter satisfies
\begin{equation}\label{eq:FriedmannWithFAtPresent}
H_0^2=\frac{8\pi G_N}{3\varphi_0 (1-\mathcal{F}_0)} \rho^0,
\end{equation}
where $\varphi_0 \equiv \varphi (a=1)$ and $\mathcal{F}_0 \equiv \mathcal{F}(a=1)$. The above equation implies that $\rho^0=\rho_c^0 \varphi_0 (1-\mathcal{F}_0 )$. We may now rewrite \eqref{eq:FriedmannWithF} in the suggestive form:
\begin{equation}\label{eq:H2withF}
H^2(a)= H_0^2\left[\hat{\Omega}_{m}(a)a^{-3}+\hat{\Omega}_{\gamma}(a)a^{-4}+\hat{\Omega}_\nu(a)+\hat{\Omega}_\CC(a)\right]\,,
\end{equation}
provided we introduce the new `hatted' parameters $\hat{\Omega}_i(a)$, which are actually slowly varying functions of the scale factor:
\begin{figure}[t!]
\begin{center}
\label{fig:varphi_BD_article}
\includegraphics[width=4.2in, height=2.8in]{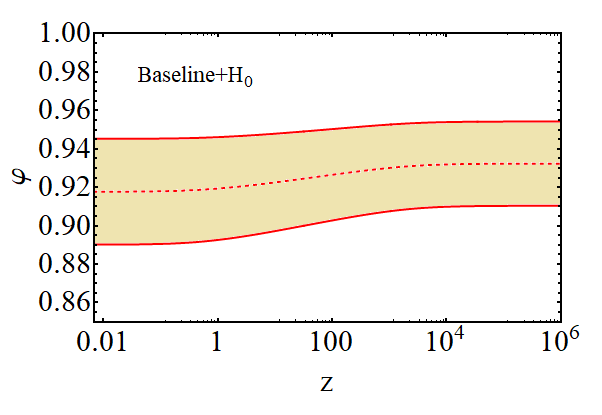}
\caption{\scriptsize{Exact numerical analysis of the evolution of $\varphi$ as a function of the redshift across the entire cosmic history, starting from the radiation-dominated epoch up to our time. We use here the values of the BD-$\CC$CDM Baseline+$H_0$ dataset indicated in the figure itself (cf. Secs. \ref{sec:MethodData_BD_article} and \ref{sec:NumericalAnalysis_BD_article}). In particular, $\eBD=-0.00199^{+0.00142}_{-0.00147}$. The band around the central (dotted) curve shows the computed $1\sigma$ uncertainty from the Markov chains of our statistical analysis. }}
\end{center}
\end{figure}
\begin{equation}\label{eq:hatOmega}
\hat{\Omega}_i (a)=\frac{\Omega_i}{\varphi (a)}\frac{1}{1-\mathcal{F}(a)}\simeq \frac{\Omega_i}{\varphi (a)}\left(1+\mathcal{F}(a)\right) \,,
\end{equation}
with $\Omega_i = \rho^0_i /\rho_c^0$ as previously. These functions also satisfy, exactly, the canonical sum rule at present: %
$\sum_i\hat{\Omega}_i(a=1)=1$.For $\eBD=0$, the hatted parameters reduce to the old tilded ones \eqref{eq:tildeOmegues}, $\hat{\Omega}_i=\tilde{\Omega}_i$, and for typical values of $|\eBD|\sim \mathcal{O}(10^{-3})$ the two sets of parameter differ by $\mathcal{O}(\eBD)$ only:
\begin{equation}\label{eq:hatOmega2}
\hat{\Omega}_i (a)=\frac{\Omega_i}{\varphi}+\mathcal{O}(\eBD)=\tilde{\Omega}_i+\mathcal{O}(\eBD) \,.
\end{equation}
From \eqref{eq:DC2} we obviously recover the previous Eq.\,\eqref{eq:DC} in the limit $\eBD\to 0$, and it is easy to see that for non-null values of $\eBD$ the density contrast acquires a dependence on the ratio $\varphi^\p/\varphi$ and its derivative, so it is sensitive to the relative change of $\varphi$ with the expansion.  Its time evolution is now possible by virtue of the third BD-field equation \eqref{eq:Wave_equation_varphi_BD_article}, which can be expanded linearly in $\eBD$ in a similar way. After some calculations, we find
\begin{equation}\label{eq:FieldeqPsiapprox}
\varphi^{\pp}+\frac{1}{2a}\left(5-3w(a)\right)\varphi^\p=\frac{3\eBD}{2a^2}(1-3w(a))\varphi\,.
\end{equation}
{For $\eBD=0$ we recover the solution $\varphi=$const. In the radiation dominated epoch (RDE), $w\simeq 1/3$,  the \textit{r.h.s.} vanishes and in this case $\varphi$ need not be constant.  It is easy to see that the exact solution of this equation in that epoch is
\begin{equation}\label{eq:varphiRDE}
\varphi(a)=\varphi^{(0)}+\frac{\varphi^{(1)}}{a}\,,
\end{equation}
for arbitrary constants $\varphi^{(i)}$. The variation during the RDE is therefore very small since the dominant solution is a constant and the variation  comes only through a decaying mode $1/a\sim t^{-1/2}$ ($t$ is the cosmic time). For the MDE (for which $w=0$) there is some evolution, once more with a decaying mode but then through a sustained logarithmic term:
\begin{equation}\label{eq:varphiMDE}
\varphi(a)\sim \varphi^{(0)}\left(1+\eBD\ln a\right)+ \varphi^{(1)}a^{-3/2}\rightarrow \varphi^{(0)} \left(1+\eBD\ln a\right) \,,
\end{equation}
where coefficient $\varphi^{(0)}$ is to be adjusted from the boundary conditions between epochs\footnote{Approximate solutions to the BD-field equations for the main cosmological variables  in the different epochs are discussed in Appendix \ref{Appendix_C}.}. The dynamics of $\varphi$ for $\eBD\neq0$  is actually mild in all epochs since $\eBD$  on the \textit{r.h.s.} of Eq.\,\eqref{eq:FieldeqPsiapprox} is small.
However mild it might be, the  dynamics of $\varphi$ modifies both the friction and Poisson terms in Eq. \eqref{eq:DC2}, and it is therefore  of pivotal importance to understand what are the changes that are induced by positive and negative values of $\eBD$ on these terms during the relevant epochs of the structure formation history.  An exact (numerical) solution is displayed in Fig. 4, where we can see that $\varphi$ remains within the approximate interval  $0.918\lesssim\varphi\lesssim 0.932$ for the entire cosmic history (starting from the RDE up to our time).  This plot has been obtained from the overall numerical fit performed to the observational data used in this analysis within one of the BD-$\CC$CDM baseline datasets considered (cf. Sec.\,\ref{sec:NumericalAnalysis_BD_article}). The error band around the main curve includes the $1\sigma$-error computed from our statistical analysis.  Two very important things are to be noted at this point: on the one hand the variation of $\varphi$  is indeed small, and on the other hand $\varphi<1$, and hence $G(\varphi)>G_N$ for the whole cosmic span.}
\begin{figure}[t!]
\begin{center}
\label{fig:terms_deltaEq_BD_article}
\includegraphics[width=6.2in, height=2.3in]{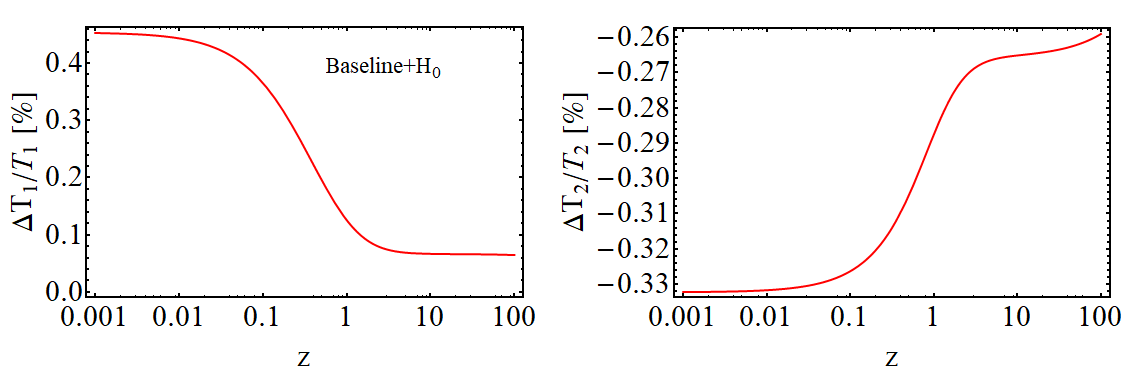}
\caption{\scriptsize{{\it Left plot:} Relative difference between the friction term $T_1$ of Eq. \eqref{eq:DC2}, $\delta_m^\pp+T_1\delta_m^\p-T_2\delta_m=0$, using the best-fit values of the BD-$\CC$CDM model obtained with the Baseline+$H_0$ dataset, i.e. with $\eBD=-0.00199$, and the case with $\eBD=0$ (cf. Table 3 for the values of the other parameters); {\it Right plot:} The same, but for the last term of Eq. \eqref{eq:DC2}, $T_2$ (Poisson term). We can clearly appreciate that negatives values of $\eBD$ produce a higher Hubble friction, i.e. a higher $T_1$, and a lower $T_2$, {\it w.r.t.} the $\eBD=0$ case. Both things lead to a decrease of $\delta_m$. The two plots have been obtained with our modified version of \texttt{CLASS}. See the main text for further details.}}
\end{center}
\end{figure}
\newline
\newline
We can start considering what is the influence of the scalar field dynamics on the perturbations during the pure MDE. Using the relations \eqref{eq:ximatter} and \eqref{eq:Fximatter} in \eqref{eq:DC2} we find ({setting $w=0$, $\rho\simeq \rho_m$ and neglecting $\rL$ in the MDE}):
\begin{equation}\label{eq:DCmatter}
\delta_m^{\pp}+\frac{\delta_m^\p}{2a}(3-\eBD)-\frac{3}{2a^2}\delta_m\left(1+\frac{4\eBD}{3}\right)=0\,.
\end{equation}
{If $\eBD<0$ the Poisson term (the last in the above equation) decreases and, on top of that, the Hubble friction increases {\it w.r.t.} the case with $\varphi=$const. (or the GR-$\CC$CDM model, if we consider the same energy densities). Both effects help to slow down the structure formation in the Universe. Of course, if $\eBD>0$ the opposite happens.   This is confirmed by solving explicitly Eq.\eqref{eq:DCmatter}.  Despite an exact solution to Eq.\,\eqref{eq:DCmatter} can be found, it suffices to quote it at $\mathcal{O}(\eBD)$ and neglect the $\mathcal{O}(\epsilon^2_{\rm BD})$ corrections.  The growing and  decaying modes at leading order read  $\delta^+_m(a)\sim a^{1+\eBD}$ and  $\delta^-_m(a)\sim a^{-\frac12(3+\eBD)}$, respectively.  The latter just fades soon into oblivion  and the former explains why negative values of $\eBD$ are favored by the data on RSD, since $\eBD<0$ obviously slows down the rate of structure formation and hence acts as an effective (positive) contribution to the vacuum energy density\footnote{This feature was already noticed in the preliminary treatment of Ref.\,\cite{Perez:2018qgw} for the BD theory itself, and it was actually pointed out as a general feature of the class of Running Vacuum Models (RVM),  which helps to cure the $\sigma_8$-tension\,\cite{Gomez-Valent:2018nib,Gomez-Valent:2017idt}. This is remarkable, since the RVM's  turn out  to mimic BD-gravity, as first noticed in \cite{Peracaula:2018dkg}. For a summary, see Appendix \ref{sec:RVMconnection_BD_article}.}.  The preference for negative values of $\eBD$ is especially clear when the RSD data include  the matter bispectrum information, which tends to accentuate the slowing down of the growth function, as noted repeatedly in a variety of  previous works\,\cite{Sola:2016ecz , Sola:2017jbl, Sola:2017znb,sola2017first}.  We may clearly appraise this feature also in the present study, see e.g. Tables 3-4 (with spectrum {\it and} bispectrum) and 5 (with spectrum but no bispectrum), where the $\sigma_8$ value is in general well-behaved ($\sigma_8\simeq 0.8$) in the BD-$\CC$CDM framework when $\eBD<0$, but it is clearly reduced (at a level $\sigma_8\simeq 0.78-0.79$)  in the presence of bispectrum data.  And in both cases the value of $H_0$  is in the range of $70-71$ km/s/Mpc. Most models trying to explain both tensions usually increase $\sigma_8$ substantially  ($0.82-0.85$). }
\newline
\newline
We can also study the pure vacuum-dominated epoch (VDE) in the same way. In this case $\varphi\sim a^{2\eBD}$ (cf. Appendix \ref{Appendix_D}), and hence
\begin{equation}\label{eq:xivacuum}
a\frac{\varphi^\p}{\varphi} = 2\eBD+\mathcal{O}(\epsilon^2_{\rm BD})\simeq 2\eBD\,,
\end{equation}
again with $\mathcal{F}^\p=\mathcal{O}(\epsilon^2_{\rm BD})\simeq 0$. The Poisson term can be neglected in this case since $\rho_m\ll\rho\simeq \rL$, and hence,
\begin{equation}\label{eq:DCvacuum}
\delta_m^{\pp}+\frac{\delta_m^\p}{a}(3-\eBD)=0\,.
\end{equation}
When the vacuum energy density rules the expansion of the Universe, there is a stable constant mode solution $\delta_m=$const. and a decaying mode that decreases faster than in the GR scenario if $\eBD<0$, again due to the fact that the friction term is in this case larger than in the standard picture, specifically the latter reads $\delta^-_m(a)\sim a^{-2+\eBD}$ in the ${\cal  O}(\eBD)$ approximation.
\begin{figure}[t!]
\begin{center}
\label{fig:pkepsilon}
\includegraphics[width=6in, height=2.1in]{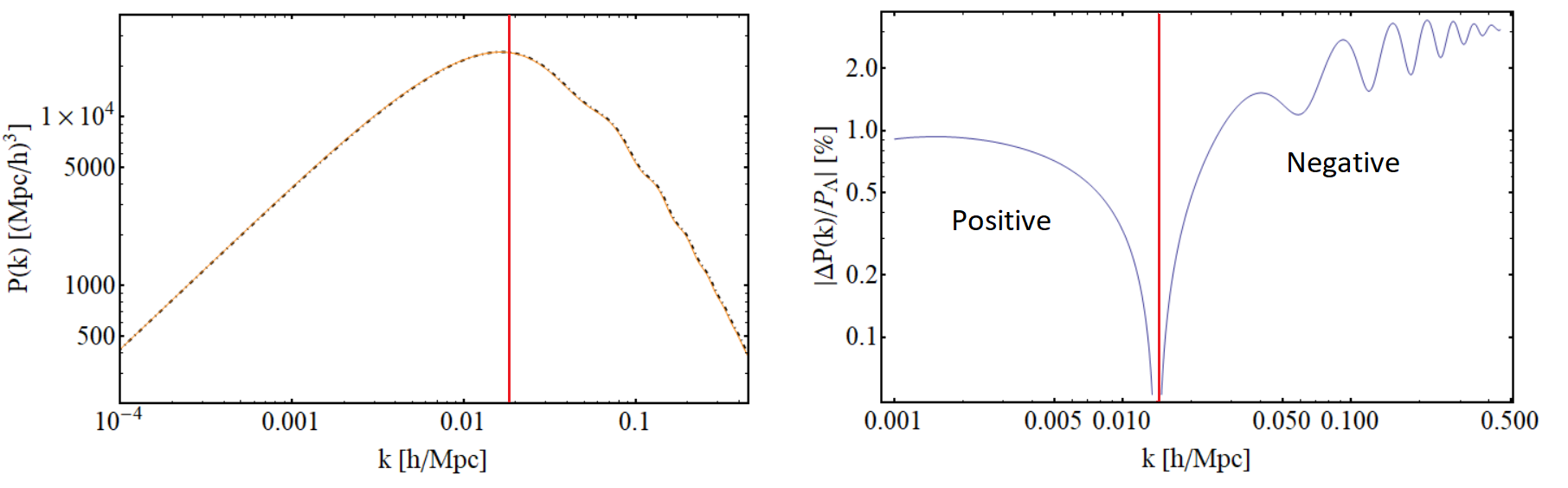}
\caption{\scriptsize{As in Fig. 3, but comparing now the $\CC$CDM baseline configuration defined there with the case $\eBD=-0.003$ (and $\varphi_{ini}=1$) of the BD-$\CC$CDM model.}}
\end{center}
\end{figure}
The analytical study of the transition between the matter and vacuum-dominated epochs is more difficult, but with what we have already seen it is obvious that the amount of structure generated also in this period of the cosmic expansion will be lower than in the $\varphi=$const. case if $\eBD$ takes a negative value. In Fig. 5 we show this explicitly.
\newline
\newline
From this analysis it should be clear that if $\eBD<0$ there is a decrease of the matter density contrast for fixed energy densities when compared with the GR-$\Lambda$CDM scenario, and also with the BD scenario with $\varphi={\rm const}$. In Fig. 6 we can see this feature directly in the matter power spectrum, which is  seen to be suppressed with respect to the case $\eBD=0$ {for those scales that are relevant for the RSD, i.e. within the range of wave numbers  $0.01 h {\rm Mpc^{-1}}\lesssim k\lesssim 0.1 h {\rm Mpc^{-1}}$  (corresponding to distance scales roughly between a few dozen to a few hundred Mpc)}. However, we still don't know whether these negative values of $\eBD$ can also be accommodated by the other datasets.
%
\begin{figure}[t!]
\begin{center}
\label{fig:Cls-epsilon_BD_article}
\includegraphics[width=6in, height=4in]{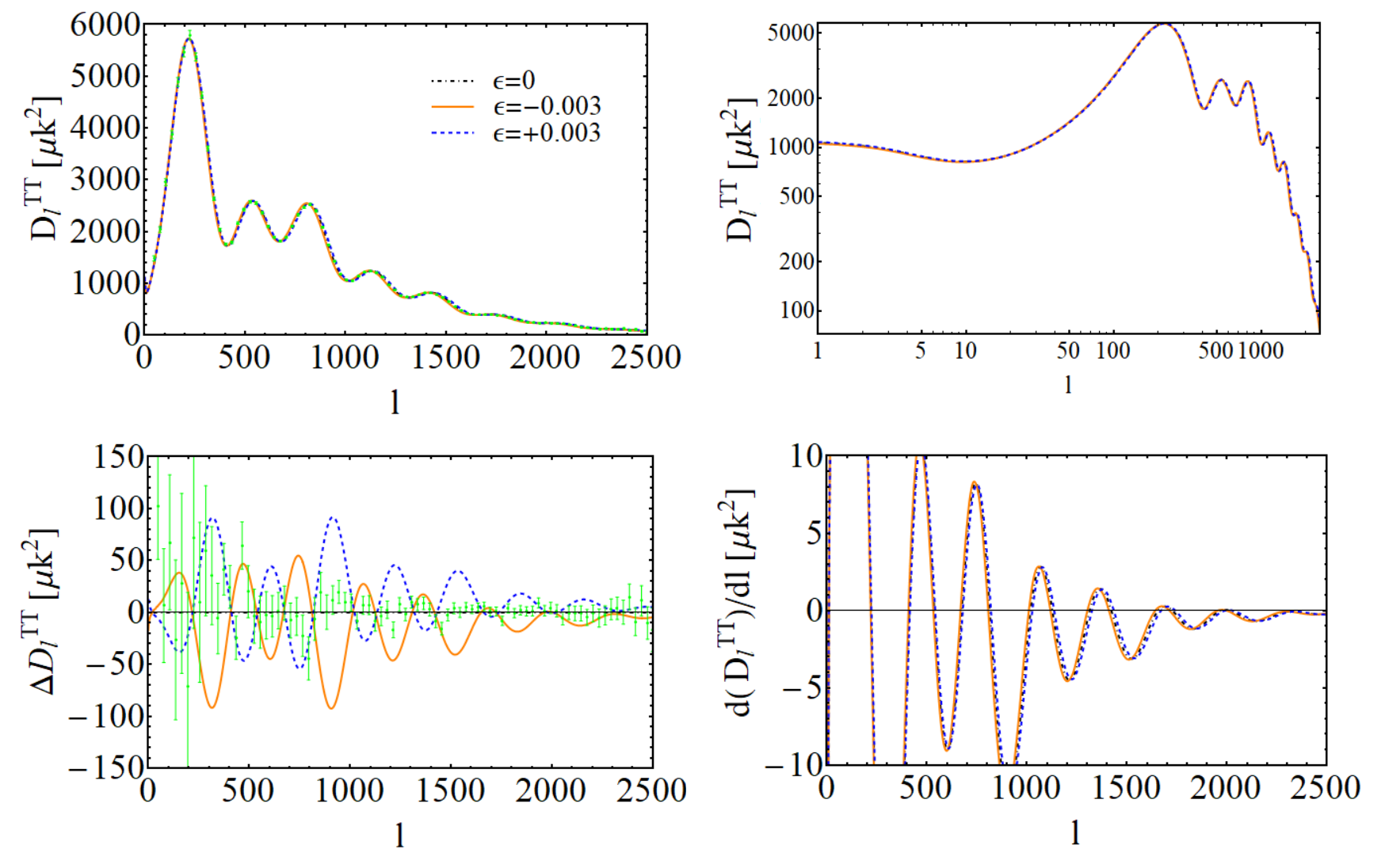}
\caption{\scriptsize{As in Fig. 1, but comparing the GR-$\CC$CDM baseline scenario (equivalent to the BD-$\CC$CDM one for  $\eBD=0$, $\varphi=1$) with the case $\eBD=\pm 0.003$, using as initial condition $\varphi_{ini}=1$.}}
\end{center}
\end{figure}
\newline
\newline
To check this, let us recall from our discussion above (see also Appendices \ref{Appendix_C} and \ref{Appendix_D} for more details) that the evolution of the BD-field takes place basically during the MDE.  In the RDE the scalar field is essentially frozen once the decaying mode becomes irrelevant, and although in the late-time Universe $\varphi$ evolves faster than in the MDE (compare \eqref{eq:ximatter} and \eqref{eq:xivacuum}) it remains almost constant in the redshift range $z\lesssim \mathcal{O}(1)$, in which all the non-CMB data points lie, particularly the LSS data.  {Let us note that the typical values of $\eBD$  fitted from the overall set of data used in our analysis (cf. Sec. \ref{sec:MethodData_BD_article})  place that parameter  in the ballpark  of $|\eBD|\sim \mathcal{O}(10^{-3})$ (cf. Tables 3-5, for example).  Schematically, we can think that the field takes a value $\varphi_{ini}$ during the RDE, then if $\eBD<0$ it decreases an amount $\Delta\varphi_{ini}<0$  with respect to its original value $\varphi_{ini}$, during the MDE, and  finally $\varphi^{0}\approx\varphi_{ini}+\Delta\varphi_{ini}$, with $\varphi(z)\approx\varphi^0$ at $z\lesssim\mathcal{O}(1)$}. Thus, Eq. \eqref{eq:H1} still applies in good approximation during the RDE and the late-time Universe, but with two different values of the cosmological Newton's coupling in the two (widely separated) cosmological eras. Taking these facts into account it is easy to understand why SNIa data are not able to tell us anything about $\eBD$ when used alone, since again $\varphi^{0}$ is fully degenerated with the absolute magnitude parameter $M$ of the supernovae. This does not mean, though, that SNIa data cannot tell us anything about the $H_0$-tension when they are considered together with other datasets that do provide constraints on $H_0$, since the constraints that SNIa impose on $\Om$ help to break degeneracies present in the other datasets and to tighten the allowed region of parameter space. H0LICOW and CCH data, for instance, will allow us to put constraints again on $\Om$ and also on $H_0$ (hence on $\rho^{0}/\varphi^{0}$). BAO will constrain $\rho_b^{0}$, $\Om$, and now also $\Delta\varphi_{ini}/\varphi_{ini}$, which is proportional to $\eBD$. For instance, $r_s H(z)\propto \left(1-\frac{\Delta\varphi_{ini}}{2\varphi_{ini}}\right)$.
\newline
\newline
The BD effect caused on the temperature spectrum of CMB anisotropies is presented in the fourfold plot in Fig. 7. Due to the fact that now we have $\eBD\ne 0$, $\Delta\varphi\ne 0$,  a small shift in the location of the peaks is naturally generated. In the right-bottom plot of such figure one can see that negative values of $\eBD$ move the peaks slightly towards lower multipoles, and the other way around for positive values of this parameter. It is easy to understand why.  To start with, let us remark that we have produced all the curves of this plot using the same initial condition $\varphi_{ini}=1$ and the same $\rho^0_b$, $\rho^0_{cdm}$ and $\rho_\Lambda$ and, hence, fixing in the same way the complete evolution of the energy densities for the different plots in that figure. This means that the differences in the Hubble function can only be due to differences in the evolution of the BD scalar field. The modified expansion histories produce changes in the value of $\theta_*=r_s/D_{A,rec}$ (with $D_{A,rec}$ being the angular diameter distance to the last scattering surface), so also in the location of the peaks. If $\eBD<0$,  $\varphi$ decreases with the expansion, so its value at recombination and at present is lower than when $\eBD=0$, and correspondingly $G(\varphi)$ will be higher. Because of this, the value of the Hubble function will be larger, too, and the cosmological distances lower, so the relation between $\theta_*(\eBD\ne 0)$ and  $\theta_*(\eBD= 0)$ can be written as follows:
\begin{equation}
\theta_*(\eBD\ne 0)=\frac{r_s(\eBD\ne 0)}{D_{A,rec}(\eBD\ne 0)}=\frac{X\cdot r_s(\eBD= 0)}{Y\cdot D_{A,rec}(\eBD=0)}= \frac{X}{Y}\,\theta_*(\eBD=0)\,,
\end{equation}
{where the rescaling factors satisfy  $0<X,Y<1$ for $\eBD<0$. As we have already mentioned before, most of the variation of $\varphi$ occurs during the MDE, so the largest length reduction will be in the cosmic stretch from recombination to the present time, and thereby  $Y<X$}. Thus, if $\eBD< 0$ we find $\theta_*(\eBD< 0)>\theta_*(\eBD=0)$ and the peaks of the TT CMB spectrum shifts towards lower multipoles. Analogously, if $\eBD$ is positive $X,Y>1$, with $Y>X$, so $\theta_*(\eBD> 0)<\theta_*(\eBD=0)$ and the peaks move to larger multipoles. It turns out, however, that these shifts, and also the changes in the amplitude of the peaks, can be compensated by small changes in the baryon and DM energy densities, as we will show in Sec. \ref{sec:NumericalAnalysis_BD_article}.
\newline
At this point we would like to recall why in the GR-$\Lambda$CDM concordance model it is not possible to reconcile the local measurements of $H_0$ with its CMB-inferred value. In the concordance model, which we assume spatially flat,  the current value of the cold dark matter density is basically fixed by the amplitude of the first peak of the CMB temperature anisotropies, and $\rho_b^0$ by the relative amplitude of the second and third peaks with respect to the first one. {As a result, even if the cosmological term plays no role in the early Universe, one finds that  in order to explain the precise location of the CMB peaks the value of $\rL$ obtained from the matching of the predicted  $\theta_*$ with the measured peak positions determines $\rL$ so precisely that it leaves little margin. This causes a problem since such narrow range of values is not in the right range to explain the value of the current Hubble parameter measured with the cosmic distance ladder technique \cite{Riess:2018uxu,Riess:2019cxk,Reid:2019tiq} and the Strong Lensing time delay angular diameter distances from H0LICOW \cite{Wong:2019kwg}. In other words,  despite the concordance model fits in a remarkably successful way the CMB and BAO, and also the SNIa data, there is an irreducible internal discordance in the parameter needs to explain with precision both the physics of the early and of the late-time Universe. This is of course the very expression of the  $H_0$-tension, to which we have to add the $\sigma_8$ one.}
\newline
Cosmographic analyses based on BAO and SNIa data calibrated with the GR-$\Lambda$CDM Planck preferred value of $r_s$ also lead to low estimates of $H_0$. This is the so-called inverse cosmic distance ladder approach, adopted for instance in \cite{Aubourg:2014yra,Bernal:2016gxb,Feeney:2018mkj,Macaulay:2018fxi}. This has motivated cosmologists to look for alternative theoretical scenarios (for instance, the generic class of  EDE  proposals)  able to increase the expansion rate of the Universe before the decoupling of the CMB photons and, hence, to lower $r_s$ down. This, in principle, demands an increase of the Hubble function at present in order not to spoil the good fit to the BAO and CMB observable. {Nevertheless, not all the models passing the BAO and CMB constraints and predicting a larger value of $H_0$ satisfy the `golden rule' mentioned in the Introduction, since they can lead e.g. to a worsening of the $\sigma_8$-tension}.  As an example, we can mention some early DE models, e.g. those discussed in \cite{Poulin:2018cxd}. In these scenarios there is a very relevant DE component which accounts for the $\sim 7\%$ of the total matter-energy content of the Universe at redshifts $\sim 3000-5000$, before recombination. This allows of course to enhance the expansion rate and reduce $r_s$. After such epoch, the DE decays into radiation. In order not to alter the position of the CMB peaks and BAO relative distances, an increase of the DM energy density is needed. According to \cite{Hill:2020osr}, this leads to an excess of density power and an increase of $\sigma_8$ which is not welcome by LSS measurements, including RSD, Weak Lensing and galaxy clustering data.  Another example is the interesting modified gravity model analyzed in \cite{Ballesteros:2020sik}, based on changing the cosmological value of $G$ also in the pre-recombination era, thus mimicking an increase of the effective number of relativistic degrees of freedom in such epoch. The additional component gets eventually diluted at a rate faster than radiation in the MDE and it is not clear if an effect is left at present\footnote{{In stark contrast to the model of \cite{Ballesteros:2020sik}, in BD-$\Lambda$CDM cosmology the behaviour of the effective $\rho_{\rm BD}$  (acting as a kind of additional DE component during the late Universe) mimics pressureless matter during the MDE epoch and modifies the effective EoS of the DE at present, see the next  Sec. \ref{sec:EffectiveEoS_BD_article} for details.}}. This model also fits the CMB and BAO data well and loosens at some extent the $H_0$-tension, but violates the golden rule of the tension solver, as it spoils the structure formation owing to the very large values of  $\sigma_8\sim 0.84-0.85$ that are predicted (see the discussion in  Sec.\,\ref{sec:Discussion_BD_article} for more details).
\newline
\newline
Our study shows that a value of the cosmological gravitational coupling about $\sim 10\%$ larger than $G_N$ can ameliorate in a significant way the $H_0$-tension, while keeping the values of all the current energy densities very similar to those found in the GR-$\Lambda$CDM model. If, apart from that, we also allow for a very slow running (increase) of the cosmological $G$ triggered by negative values of order $\eBD\sim -\mathcal{O}(10^{-3})$, we can  mitigate at the same time the $\sigma_8$-tension when only the CMB TT+lowE anisotropies are considered. When the CMB polarizations and lensing are also included in the analysis, then $\sigma_8$ is kept at the GR-$\Lambda$CDM levels, and the sign of $\eBD$ is not conclusive.  {In all situations we can preserve the golden rule}. We  discuss in detail the numerical  results of our analysis  in Sec. \ref{sec:NumericalAnalysis_BD_article}.
\subsection{Effective equation of state of the dark energy in the BD-$\CC$CDM model}\label{sec:EffectiveEoS_BD_article}
\noindent
{Our aim in this section is to write down the Brans-Dicke cosmological equations \eqref{eq:Friedmann_equation_varphi_BD_article}-\eqref{eq:Wave_equation_varphi_BD_article} in the context of what we may call the ``effective GR-picture''. This means to rewrite them  in such a way that they can be thought of as an effective model within the frame of GR, thus providing a parametrized departure from GR at the background level. We will see that the main outcome of this task, at least qualitatively, is that the BD-$\CC$CDM model  (despite it having a constant vacuum energy density $\rL$) appears as one in the GR class, but with a dynamical DE rather than a CC. The dynamics of such an effective form of DE is a function of the BD-field $\varphi$.  We wish to compute its effective EoS.  In order to proceed, the first step is to  rewrite Eq.\eqref{eq:Friedmann_equation_varphi_BD_article} \`a la  Friedmann:}
\begin{equation}\label{eq:BDFriedmann}
3H^2 = 8\pi{G_N}(\rho + \rvphi)\,,
\end{equation}
where $\rho$ is the total energy density as defined previously (coincident with that of the GR-$\CC$CDM model),  and  $\rvphi$ is the additional ingredient that is needed, which reads
\begin{equation}\label{rhoBD}
\rvphi\equiv \frac{3}{8\pi{G_N}}\left(H^2\dvphi - H\dot{\varphi} + \frac{\oD}{6}\frac{\dot{\varphi}^2}{\varphi}\right).
\end{equation}
Remember the definition $\varphi(t) \equiv G_N\psi(t)$ made in \eqref{eq:definitions_BD_article}, and we have now introduced
\begin{equation}\label{eq:Deltaphi}
\dvphi(t)\equiv 1-\varphi(t)\,,
\end{equation}
which tracks the small departure of $\varphi$ from one and hence of $G(\varphi)$ from $G_N$  (cf. Sec. \ref{sec:BDgravity_BD_article}). Note that $\varphi=\varphi(t)$ evolves in general with the expansion, but very slowly  since $\eBD$ is presumably fairly small.
\newline
From the above  Eq.\,\eqref{rhoBD} it is pretty clear that we have absorbed all the terms beyond the $\Lambda$CDM model into the expression of $\rvphi$. While it is true that we define this quantity as if it were an energy density, it is important to bear in mind that it is not associated to any kind of particle, it is just a way to encapsulate those terms that are not present in the standard model.  This quantity, however, satisfies a local conservation law as if it were a real energy density, as we shall see in a moment.  From the generalized Friedmann equation \eqref{eq:BDFriedmann} and the explicit expression for $\rvphi$ given above we can write down the generalized cosmic sum rule verified by the BD-$\CC$CDM model in the effective GR-picture:
\begin{equation}\label{eq:SumRuleBDexact}
\Omega_m+\Omega_r+\Omega_\CC+\Omega_\varphi=1\,,
\end{equation}
where the $\Omega_i$ are the usual (current) cosmological parameters of the concordance $\CC$CDM, whereas $\Omega_\varphi$ is the additional one that parametrizes the departure of the  BD-$\CC$CDM model from the  GR-$\CC$CDM  in the context of the GR-picture, and reads
\begin{equation}\label{eq:OmegaBD}
\Omega_{\varphi}=\frac{\rho_{\varphi}^0}{\rco}.
\end{equation}
Notice that the above sum rule is exact and it is different from that in Eq.\,\eqref{eq:SumRuleBD} since the latter is only approximate for the case when $\eBD=0$ or very small.  These are two different pictures of the same BD-$\CC$CDM model. The modified cosmological parameters (\ref{eq:tildeOmegues}) depend on $\varphi$ whereas here the $\varphi$-dependence has been fully concentrated on $\Omega_\varphi$.  It is interesting to write down the exact equation \eqref{eq:SumRuleBDexact} in the form
\begin{equation}\label{eq;OMegaBD}
\Omega_m+\Omega_r+\Omega_\CC=1-\Omega_{\varphi}= 1-\Delta\varphi_0+\frac{\dot{\varphi}_0}{H_0}-\frac{\omega_{\rm BD}}{6}\frac{\dot{\varphi}_0^2}{H_0^2\varphi_0}\,.
\end{equation}
in which $\varphi_0=\varphi(z=0)$ and $\dot{\varphi}_0=\dot{\varphi}(z=0)$. For $\eBD\simeq 0$ we know that $\varphi\simeq$const. and we can neglect the time derivative terms and then we find the approximate form $\Omega_m+\Omega_r+\Omega_\CC=1-\Omega_{\varphi}\simeq  1-\Delta\varphi_0$.
This equation suggests that a value of $\Delta\varphi_0\neq 0$ would emulate the presence of  a small  fictitious spatial curvature in the GR-picture. See e.g.\cite{Park:2019emi,Khadka:2020vlh,Cao:2020jgu} and references therein for the study of a variety models explicitly involving spatial curvature.
\newline
\newline
The second step in the process of constructing the GR-picture of the BD theory  is to express \eqref{eq:Pressure_equation_varphi_BD_article} as in the usual pressure equation for GR, and this forces us to define a new pressure quantity $p_{\varphi}$ associated to $\rvphi$. We find
\begin{equation}
2\dot{H} + 3H^2 = -8\pi{G_N}(p + \pvphi),
\end{equation}
with
\begin{equation} \label{pBD}
p_{\varphi}\equiv \frac{1}{8\pi{G_N}}\left(-3H^2\dvphi -2\dot{H}\dvphi  + \ddot{\varphi} + 2H\dot{\varphi} + \frac{\oD}{2}\frac{\dot{\varphi}^2}{\varphi}\right).
\end{equation}
{On the face of the above definitions \eqref{rhoBD} and \eqref{pBD}, we can now interpret the BD theory as an effective theory within the frame of General Relativity, which deviates from it an amount $\dvphi$. Indeed, for $\dvphi=0$ we have $\rvphi=\pvphi=0$ and we recover GR.  Mind that $\dvphi=0$ means not only that $\varphi=$const. (hence that $\eBD=0$, equivalently $\omega_{\rm BD} \to\infty$), but also that that constant is exactly $\varphi=1$.  In such case $\Geff$ is also constant and $\Geff=G_N$ exactly. The price that we have to pay for such a GR-like description of the BD model is the appearance of the fictitious BD-fluid with energy density $\rvphi$ and pressure $\pvphi$, which complies with the following conservation equation throughout the expansion of the Universe}\,\footnote{{The new `fluid' that one has to add to GR to effectively mimic BD plays a momentous role to explain the $H_0$ and $\sigma_8$-tensions. In a way it mimics the effect of the `early DE' models mentioned in the previous section, except that the BD-fluid persists for the entire cosmic history and is instrumental both in the early as well as in the current Universe so as to preserve the golden rule of the tension solver: namely, it either softens the two tensions of GR or improves one of them without detriment of the other. }}:
\begin{equation}
\dot{\rho}_{\varphi}+3H(\rho_{\varphi}+p_{\varphi})= 0.
\end{equation}
{One can check that this equation holds after a straightforward calculation, which makes use of the three  BD-field equations \eqref{eq:Friedmann_equation_varphi_BD_article}-\eqref{eq:Wave_equation_varphi_BD_article}.}
Although at first sight the above conservation equation can be surprising actually it is not, since it is a direct consequence of the Bianchi identity. Let us now assume that the effective BD-fluid can be described by an equation of state like $\pvphi = w_{\varphi}\rvphi$, so
\begin{equation}\label{BDEoS0}
w_{\varphi}=\frac{\pvphi}{\rvphi}=\frac{-3H^2\dvphi -2\dot{H}\dvphi  + \ddot{\varphi} + 2H\dot{\varphi} + \frac{\oD}{2}\frac{\dot{\varphi}^2}{\varphi}}{3H^2\dvphi - 3H\dot{\varphi} + \frac{\oD}{2}\frac{\dot{\varphi}^2}{\varphi}}.
\end{equation}
The contribution from those terms containing derivatives of the BD-field are subdominant for the whole cosmic history.  We have verified this fact numerically, see Fig. 4. While the variation of $\varphi$ between the two opposite ends of the cosmic history is of $\sim 1.5\%$ and is significant for our analysis, the instantaneous variation is actually negligible. Thus,  $H\dot{\varphi}$ and  $\ddot{\varphi}$  are both much smaller than $\dot{H}\dvphi$,  and  in this limit we can approximate \eqref{BDEoS0} very accurately as
\begin{equation}\label{BDEoS}
w_{\varphi}(t) \simeq -1 - \frac{2}{3}\frac{\dot{H}}{H^2}\,\ \ \ \ \ \ ({\rm for}\ H\dot{\varphi}, \ddot{\varphi}\ll \dot{H}\dvphi)\,.
\end{equation}
This EoS turns  out to be the standard total EoS of the $\Lambda$CDM, which boils down to the EoS corresponding to the different epochs of the cosmic evolution (i.e. $w=1/3, 0, -1$ for RDE, MDE and VDE).  This means that the EoS of the  BD-fluid mimics these epochs.   We can go a step further and define not just the BD-fluid but the combined system of the BD-fluid and the vacuum energy density represented by the density $\rL$ associated to the cosmological constant.  {We define the following effective EoS for such combined fluid:}
\begin{figure}[t!]
\begin{center}
\label{fig:weff_varphi_BD_article}
\includegraphics[width=6.8in, height=2.8in]{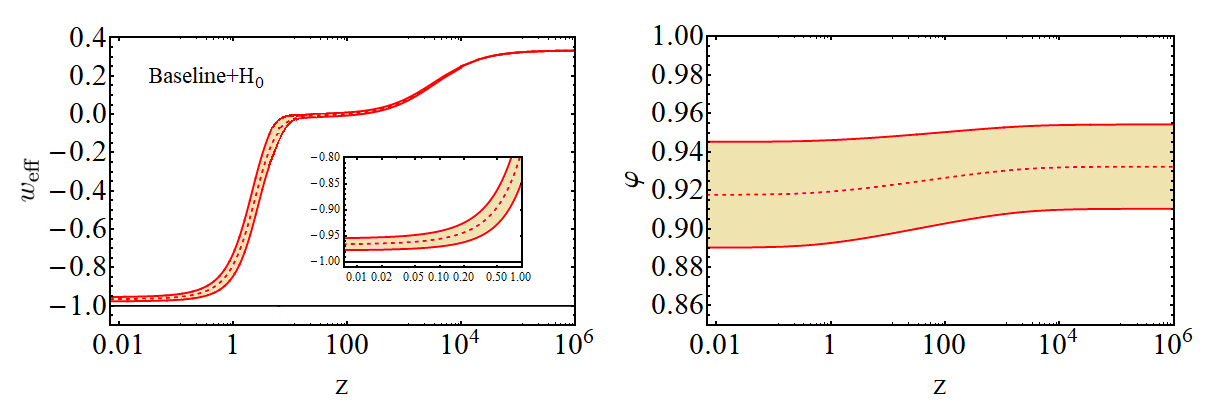}
\caption{\scriptsize{{\it Left plot:} Effective equation of state of the DE in the BD-$\CC$CDM model as a function of the redshift. The inner plot magnifies the region around our time. We can see that the BD model mimics quintessence with a significance of more than $3\sigma$; {\it Right plot:} It shows once more Fig.\, 4 in order to ease the comparison of the EoS evolution, which is associated to the evolution of the BD-field $\varphi$ -- cf. Eq.\eqref{effEoS}. The shadowed bands in these plots correspond to the 1$\sigma$ regions}.}
\end{center}
\end{figure}
%
\begin{equation}\label{effEoS}
\weff\equiv \frac{p_\Lambda + p_{\varphi}}{\rho_\Lambda + \rho_{\varphi}} = -1 + \frac{p_{\varphi} + \rho_{\varphi}}{\rho_\Lambda+ \rho_{\varphi}}=-1+\frac{-2\dot{H}\dvphi+f_1(\varphi,\dot{\varphi},\ddot{\varphi})}{\CC+ 3 H^2\,\dvphi+f_2(\varphi,\dot{\varphi})}\,,
\end{equation}
where the two functions
\begin{equation}\label{eq:f1f2}
  f_1(\varphi,\dot{\varphi},\ddot{\varphi})=\ddot{\varphi}-H\dot{\varphi}+\omega_{\rm BD} \frac{\dot{\varphi}^2}{\varphi}\,,\ \ \ \ \ \ \  f_2(\varphi,\dot{\varphi})= -3H\dot{\varphi}+\frac{\omega_{\rm BD}}{2}  \frac{\dot{\varphi}^2}{\varphi}
\end{equation}
{involve differentiations with respect to the slowly varying field $\varphi$ and as before they are negligible,  in absolute value,  as compared to $\dot{H}\Delta\varphi$ and  ${H^2}\Delta\varphi$. The effective EoS \eqref{effEoS} is a time-evolving quantity which mimics dynamical DE at low redshifts.  At very high redshifts $z\gg1$, well beyond the DE dominated epoch, we can neglect $\CC$ in the denominator of the EoS and the dominant term is $ 3 H^2\,\dvphi$. Similarly, in the numerator the dominant term is always $-2\dot{H}\dvphi$.  Therefore, at high redshifts the effective EoS \eqref{effEoS} behaves as \eqref{BDEoS} since $\dvphi$ cancels out: $\weff(z)\simeq w_{\varphi}(z)\,\ ( z\gg1)$, which means that it just reproduces the standard EoS of the GR-$\CC$CDM.}
The exact EoS \eqref{effEoS}, however,  must be computed numerically, and it is displayed in Fig.\,8, together with the numerical plot of $\varphi$ (which we have already shown in Fig.\,4). We have used the Baseline+$H_0$ dataset defined in Sec. \ref{sec:MethodData_BD_article}.  For a semi-qualitative discussion of the combined EoS it will suffice an analytical approximation, as we did before with $w_{\varphi}$. The most relevant part of $\weff (z)$ as to the possibility of disentangling the dynamical DE effects triggered by the underlying BD model is near the present time ($z<1$). Thus, neglecting the contribution from the functions $f_{1,2}$, but now keeping the $\CC$-term in the denominator of  \eqref{effEoS} we can use the Hubble function of the concordance model and we find the following result at linear order in $\dvphi$:
\begin{equation}\label{BDEoSz0}
\weff(z)\simeq-1-\frac{2\dot{H}\dvphi}{\CC}\simeq -1+\dvphi\,\frac{\Omo}{\OLo}\,(1+z)^3\,,
\end{equation}
where $\Omo$ and $\OLo$ are the current values of the cosmological parameters, which satisfy $\Omo+\OLo=1$  for spatially flat Universe.
{As has been stated before, the previous approximate formula is valid only for $z<1$, but it shows very clearly that for $\dvphi>0$ (resp. $<0$) we meet quintessence-like (resp. phantom-like) behaviour.  As we have repeatedly emphasized, our analysis points to $\eBD<0$ and hence $\varphi$ decreases with the expansion, remaining smaller than one. From Eq.\,\eqref{eq:Deltaphi} this means $\dvphi>0$  and therefore we find that the effective GR behaviour of the BD-$\CC$CDM is quintessence-like.}  We can be more precise at this point.
We have numerically computed the value of the exact function \eqref{effEoS} at $z=0$, taking into account the contribution from all the terms, in particular the slowly varying functions \eqref{eq:f1f2}, see Tables 3-6. The results obtained from three of the most prominent datasets defined in Sec.\,\ref{sec:MethodData_BD_article} read as follows:
\begin{align}
&{\bf Baseline}:\quad &\weff(0)=& -0.983^{+0.015}_{-0.014}\\
&{\bf Baseline+H_0}:\quad &\weff(0)=& -0.966^{+0.012}_{-0.011}\\
&{\bf Baseline+H_0+SL}:\quad &\weff(0) =& -0.962\pm 0.011.
\end{align}
As can be seen, there is a non-negligible departure from the constant EoS value $-1$  of the  GR-$\Lambda$CDM, which reaches the $\sim 3\sigma$ c.l. when the prior on $H_0$ from the local distance ladder measurement by SH0ES \cite{Reid:2019tiq} is included in the analysis, and $\sim 3.5\sigma$ c.l. when also the angular diameter distances from H0LICOW \cite{Wong:2019kwg} are taken into account. The effective quintessence EoS $\weff(0)>-1$ is one of the ingredients that allows the BD-$\CC$CDM model to significantly loosen the $H_0$-tension, since it is a direct consequence of having $\varphi<1$ (or, equivalently, $G>G_N$) (cf. Sec. \ref{sec:preview_BD_article} for details).
\newline
\newline
We have obtained the above results from the equations of motion once a metric was assumed; however, it is possible to obtain all the expressions listed in this section starting from the BD action \eqref{eq:BDaction} itself and then considering the FLRW metric.  {To show this, let us use the dimensionless field $\varphi=G_N\psi$ and the variable $\dvphi$ defined in \eqref{eq:Deltaphi}. }First of all we split the whole action in three pieces
\begin{equation}
S_{\rm BD}[\varphi]=S_{\rm EH}+S_{\rm GR}[\varphi]+S_m,
\end{equation}
where
\begin{equation}
S_{\rm EH}\equiv\int d^4x \sqrt{-g}\left[\frac{R}{16 \pi G_N}-\rho_\Lambda \right],
\end{equation}
is the usual Einstein-Hilbert action, whereas
\begin{equation}
S_{\rm GR}[\varphi] \equiv \int d^4x \sqrt{-g}\frac{1}{16\pi G_N}\left[ -R\dvphi-\frac{\oD}{\varphi}g^{\mu \nu}\partial_\nu \varphi \partial_\mu \varphi \right],
\end{equation}
is the action parametrizing the deviation of the BD theory from the  GR-picture expressed  in terms of the scalar field $\varphi$. As expected, for $\varphi=1$ that action vanishes identically. Finally,
\begin{equation}
S_m\equiv\int d^4x \sqrt{-g}\mathcal{L}_m (\chi_i,g_{\mu \nu})
\end{equation}
is the action for the matter fields. Since there is no interaction involving $\varphi$ with other components,  the BD-field $\varphi$  is covariantly conserved, as remarked in Sec. \ref{sec:BDgravity_BD_article}. In order to compute the energy-momentum tensor and find out the effective density and pressure of the BD-field, we apply the usual definition of that tensor in curved space-time:
\begin{equation} \label{energy-momentum}
T_{\mu \nu}^{\rm BD}=-\frac{2}{\sqrt{g}}\frac{\delta S_{\rm GR}[\varphi]}{\delta g^{\mu \nu}}\,.
\end{equation}
 After some calculations we arrive at
\begin{equation} \label{BDEMT}
\begin{split}
T_{\mu \nu}^{\rm BD}=\frac{R_{\mu \nu}}{8\pi G_N}\dvphi-\frac{\nabla_\nu \nabla_\mu\varphi}{8\pi G_N}&+\frac{g_{\mu \nu}\Box \varphi}{8\pi G_N}+\frac{\oD}{8\pi \varphi}\partial_\nu \varphi \partial_\mu \varphi\\
&-\frac{g_{\mu \nu}}{16\pi G_N}\left(R\dvphi+\frac{\oD}{\varphi}g^{\alpha \beta}\partial_\alpha \varphi  \partial_\beta \varphi \right).
\end {split}
\end{equation}
Since $\varphi$ has no interactions it behaves as any free scalar field, so its energy-momentum tensor must adopt the perfect fluid form at the background level:
\begin{equation}
T_{\mu \nu}^{\rm BD}=\pvphi g_{\mu \nu}+(\rvphi+\pvphi)u_\mu u_\nu.
\end{equation}
Now we can compare this form with \eqref{BDEMT}. It is straightforward to obtain the energy density as well as the corresponding pressure, we only need to compute $\rvphi=T_{00}^{\rm BD}$ and $\pvphi = (T^{\rm BD} + \rvphi)/3$, being $T^{\rm BD} =g^{\mu\nu}T^{\rm BD}_{\mu\nu}$ the trace of the tensor. {Using at this point the spatially flat FLRW metric one can work out  the explicit result for $\rvphi$ and $\rvphi$  and reconfirm that it acquires the form previously indicated in the equations \eqref{rhoBD} and \eqref{pBD}. This  provides perhaps a more formal derivation of these formulas and serves as a cross-check of them.}
\subsection{Structure formation  in the linear regime. Perturbations equations}\label{sec:StructureFormation_BD_article}
In order to perform a complete analysis of the model, we need to study the evolution of the perturbed cosmological quantities in the context of BD theory. For a review of the standard model perturbations equations, see e.g. \cite{Ma:1995ey,liddle_lyth_2000,Lyth:2009zz}.
We assume a FLRW metric written in conformal time, denoted by $\eta$, in which the line element is $ds^2=a^2(\eta)[-d\eta^2+(\delta_{ij}+h_{ij})dx^idx^j]$. Here $h_{ij}$ is a perturbation on the spatial part of the metric which can be expressed in momentum space as follows,
\begin{equation}\label{eq:MainhFourier}
h_{ij}(\eta,\vec{x})=\int d^3k\, e^{-i\vec{k}\cdot\vec{x}}\left[\hat{k}_i\hat{k}_j h(\eta,\vec{k})+\left(\hat{k}_i\hat{k}_j-\frac{\delta_{ij}}{3}\right)6\xi(\eta,\vec {k})\right].
\end{equation}
As we see, in momentum space the trace $h\equiv \delta^{ij}h_{ij}$ decouples from the traceless part of the perturbation, $\xi$. Now, we are going to list the perturbations equations at late stages of expansion in momentum space at deep subhorizon scales, that is, we assume $\mathcal{H}^2\ll k^2$, with  $\mathcal{H}\equiv a^\prime/a$. Although primes were used previously for derivatives with respect to the scale factor, they will henceforth stand for  derivatives with respect to the conformal time: $()^\prime \equiv d()/d \eta$.  For example, it is easy to see that $\mathcal{H}= a H$.
One may work with the standard differential equation for the density contrast at deep subhorizon scales,
\begin{equation} \label{DensityConstrastLCDM}
\delta_m^{\prime \prime}+\mathcal{H}\delta_m^\prime-4\pi{G_N}\bar{\rho}_m a^2 \delta_m=0,
\end{equation}
where $\delta_m \equiv \delta \rho_m / \bar{\rho}_m$ is the density contrast, the bar over $\bar{\rho}_m$ indicates that is a background quantity and the evolution of $\mathcal{H}$ and $\bar{\rho}_m$ is the one expected by the background equations of the BD theory in Section \ref{sec:BDgravity_BD_article}. This expression is just the corresponding one for the $\Lambda$CDM, completely neglecting any possible perturbation in the BD-field, namely $\delta \varphi$. However, it is possible to see that a second order differential equation for the density contrast can be written, even if the perturbation in $\varphi$ is not neglected. This is done in detail in the Appendix  \ref{Appendix_B}. In this section, we present the main perturbations equations in the case of the Synchronous gauge and discuss the interpretation of the result.
\newline
\newline
If  $\vec{v}_m$ is the physical 3-velocity of matter  (which is much smaller than 1 and can be treated as a perturbation), then we can define its divergence, $\theta_m\equiv \vec{\nabla} \cdot (\vec{v}_m)$. At deep subhorizon scales it is possible to see that the equation governing its evolution is
\begin{equation}
\theta_m^\prime+\mathcal{H}\theta_m=0.
\end{equation}
{Since $da^{-1}/d\eta=- \mathcal{H}/a$,  we arrive to a decaying solution  $\theta_m  \propto a^{-1}$. A common assumption is to set $\theta_m \sim 0$ in the last stages of the Universe, which is what we will do in our analysis. This allows us to simplify the equations. Another simplification occurs if we take into account that we are basically interested in computing the matter perturbations only at deep subhorizon scales, namely for $k^2\gg \mathcal{H}^2$, which allows us to neglect some terms as well (cf. Appendix \ref{Appendix_B}). Altogether we are led to the following set of perturbations equations in the synchronous gauge:}
\begin{equation}\label{eq:Mainsimpli0}
\delta_m^\prime=-\frac{h^\prime}{2}\,.
\end{equation}
\begin{equation}\label{eq:Mainsimpli1}
k^2\delta \varphi+\frac{h^\prime}{2}\bar{\varphi}^\prime =\frac{8 \pi G_N}{3+2\oD}a^2 \bar{\rho}_m\delta_m\,,
\end{equation}
\begin{equation}\label{eq:Mainsimpli2}
\bar{\varphi}(\mathcal{H}h^\prime-2\xi k^2)+k^2\delta\varphi+\frac{h^\prime}{2}\bar{\varphi}^\prime=8\pi G_N a^2 \bar{\rho}_m\delta_m\,,
\end{equation}
\begin{equation}\label{eq:Mainsimpli3}
2k^2\delta \varphi+\bar{\varphi}^\prime h^\prime+\bar{\varphi}\left(h^{\prime \prime}+2h^\prime \mathcal{H}-2k^2 \xi\right)=0\,.
\end{equation}
{Combining these four equations simultaneously and  without doing any further approximation, one finally obtains the following compact equation for the matter density contrast of the BD theory at deep subhorizon scales:}
\begin{equation}\label{eq:ExactPerturConfTime}
\delta_m^{\prime\prime}+\mathcal{H}\delta_m^\prime-\frac{4\pi G_N a^2}{\bar{\varphi}}\bar{\rho}_m\delta_m\left(\frac{4+2\oD}{3+2\oD}\right)=0\,.
\end{equation}
{In other words},
\begin{equation}\label{eq:ExactPerturConfTime2}
\delta_m^{\prime\prime}+\mathcal{H}\delta_m^\prime- 4\pi \Geff(\bar{\varphi}) a^2\,\bar{\rho}_m\delta_m=0\,.
\end{equation}
The quantity
\begin{equation}\label{eq:MainGeffective}
 \Geff(\bar{\varphi})=\frac{G_N}{\bar{\varphi}}\left(\frac{4+2\oD}{3+2\oD}\right)=\frac{G_N}{\bar{\varphi}}\left(\frac{2+4\eBD}{2+3\eBD}\right)
\end{equation}
is precisely the effective coupling previously introduced in Eq.\,\eqref{eq:LocalGNa}; it modifies the Poisson term of the perturbations equation with respect to that of the standard model, Eq.\,\eqref{DensityConstrastLCDM}.  There is, in addition, a modification in the friction term between the two models, which is of course associated to the change in ${\cal H}$.
\newline
\newline
The argument of $\Geff$ in \eqref{eq:MainGeffective} is not $\varphi$ but the background value $\bar{\varphi}$ since the latter  is what remains in first order of perturbations from the consistent splitting of the field into the background value and the perturbation:  $\varphi=\bar{\varphi}+\delta\varphi$.  Notice that there is no dependence left on the perturbation $\delta\varphi$.  As we can see from \eqref{eq:MainGeffective}, the very same effective coupling that rules the attraction of two tests masses in BD-gravity is the coupling strength that governs the formation of structure in this theory, as it could be expected.  But this does not necessarily mean that the effective gravitational strength governing the process of structure formation is the same as for two tests masses on Earth. We shall elaborate further on this point in the next section. At the moment we note that if we compare  the above perturbations equation with the standard model one \eqref{DensityConstrastLCDM},  the former reduce to the latter in the limit $\omega_{\rm BD}\to\infty$ (i.e. $\eBD\to 0$) \textit{and} $\bar{\varphi}=1$.
\newline
\newline
The form of \eqref{eq:ExactPerturConfTime2} in terms of the scale factor variable rather than in conformal time was given previously in Sec. \ref{sec:rolesvarphiH0_BD_article}   when we considered a preview of the implications of BD-gravity on structure formation data\footnote{{Recall, however, that prime in Eq.\,\eqref{eq:ExactPerturConfTime2} stands for differentiation  {\it w.r.t.}  conformal time whereas in Eq.\,\eqref{eq:ExactPerturScaleFactor} denotes differentiation  {\it w.r.t.}  the scale factor.  These equations perfectly agree and represent the same perturbations equation for the matter density field in BD-gravity in the respective variables. They are also in accordance with the perturbations equation obeyed by the matter density field within the context of scalar-tensor theories with the general action \eqref{eq:BDactionST}\cite{Boisseau:2000pr} of which  the form (\ref{eq:BDaction2}) is a  particular case. }}. The transformation of derivatives between the two variables can be easily performed with the help of the chain rule $d/d\eta=a{\cal H} d/da$.
\subsection{Different BD scenarios and Mach's Principle}\label{sec:Mach_BD_article}
As previously indicated, the  relation \eqref{eq:LocalGNa}, which appears now in the cosmological context in the manner \eqref{eq:MainGeffective}, follows from the computation of the gravitational field felt by two test point-like (or spherical) masses in interaction in BD-gravity within the weak-field limit\cite{BransDicke1961, dicke1962physical}, see also \cite{Fujii:2003pa} and references therein. Such relation shows in a manifest way the integration of Mach's principle within the BD context, as it postulates a link between the measured local value of the gravitational strength, $G_N$, as measured at the Earth surface, and its cosmological value, $\Geff(\varphi)$, which depends on $\varphi$ and $\omega_{\rm BD}$. In particular, $\varphi$  may be sensitive to the mean energy densities and pressures of all the matter and energy fields that constitute the Universe. {If there is no mechanism screening the BD-field on Earth, $\Geff(\bar{\varphi})(z=0)=G_N$.} {However, one can still fulfill this condition  if  Eq.\,\eqref{eq:MainGeffective} constraints the current value of the cosmological BD-field $\bar{\varphi}$ }
\begin{equation}\label{eq:constraintvarphi}
\bar{\varphi}(z=0)=\frac{4+2\oD}{3+2\oD}=\frac{2+4\eBD}{2+3\eBD}\simeq 1+\frac12\,\eBD+{\cal O}(\eBD^2)\,.
\end{equation}
That is to say, such constraint permits to reconcile $\Geff(\bar{\varphi})(z=0)$ with $G_N$ by still keeping  $\bar{\varphi}(z=0)\ne 1$ and $\eBD\ne 0$. Hence the BD-field can be dynamical and there can be a departure of $G(\bar{\varphi})$ from $G_N$ even at present. This constraint, however, is much weaker than the one following from taking the more radical approach in which $\Geff(\varphi)$ and  $G_N$  are enforced to coincide upon imposing the double condition $\eBD\to0$ (i.e. $\omega_{\rm BD}\to\infty$) \textit{and}  $\bar{\varphi}=1$.  It is this last setup which anchors the BD theory to remain exactly (or very approximately) close to GR at all scales.
\newline
\newline
However, if we take seriously the stringent constraint imposed  by the Cassini probe on the post-Newtonian parameter $\gamma^{\rm PN}$\cite{Bertotti:2003rm}, which leads to a very large  value of $\omega_{\rm BD}\gtrsim 10^4$ (equivalently, a very small value of $\eBD=1/\omega_{\rm BD}$), as we discussed in Sec.\,\ref{sec:BDgravity_BD_article},  $\bar{\varphi}$ must remain almost constant throughout the cosmic expansion, thus essentially equal to $\bar{\varphi}(z=0)$. However, the Cassini limit leaves $\bar{\varphi}(z=0)$ unconstrained, so this constant  is not restricted to be in principle equal to $1$.  In this case the relation \eqref{eq:constraintvarphi} may or may not apply; there is in fact no especial reason for it {(it will depend on the effectiveness of the screening mechanism on Earth)}. If it does, i.e. if there is no screening, $\Geff$ is forced to be very close to $G_N$ {$\forall{z}$};
if it does not, $\bar{\varphi}$ can freely take {(almost constant)} values  which do not push $\Geff$ to stay so glued to $G_N$.  It is  interesting to see  the  extent  to which the cosmological constraints can compete with the local ones given the current status of precision they can both attain.
\newline
So, as it turns out, we find that one of the two interpretations  leads to values of $\Geff$ very close to $G_N$ {$\forall{z}$} on account of the fact that  we are  imposing  very large values of  $\oD$ {and assuming \eqref{eq:constraintvarphi},} 
whereas the other achieves the same aim (viz. $\Geff $  can stay very close to $G_N$) for intermediate values of $\omega_{\rm BD}$ provided they are  linked to $\bar{\varphi}(z=0)$ through the constraint  \eqref{eq:constraintvarphi}. This last option, as indicated, is not likely {since this would imply the existence of a screening at the scales probed by Cassini that may become ineffective on Earth, where the densities are higher}. Finally, we may as well have a situation where the cosmological $\Geff $  remains different from  $G_N$ even if the Cassini limit is enforced. For this to occur we need an (essentially constant) value of  $\bar{\varphi}\neq 1$ (different from the one associated to that constraint) at the cosmological level. This can still be compatible with the local constraints provided $\varphi$ is screened {on Earth} at $z=0$.
\newline
{A more open-minded and general approach, which we are going to study in this chapter, is to take the last mentioned option but without the Cassini limit. This means that $\eBD$ is not forced to be so small and hence  $\varphi$ can still have some appreciable dynamics. We  assume that the pure BD model applies
from the very large cosmological domains to those at which the matter perturbations remain linear.  Equation \eqref{eq:MainGeffective} predicts the cosmological value of $\Geff$ once $\omega_{\rm BD}$ and the initial value $\varphi_{ini}$ are fitted to the data. We can dispense with the Cassini constraint (which affects $\omega_{\rm BD}$ only) because we assume that some kind of screening mechanism
acts at very low (astrophysical) scales, namely in the non-linear domain, without altering the pure BD model at the cosmological level.
To construct a concrete screening mechanism would imply to specify some microscopic interaction properties of $\varphi$ with matter, but these do not affect the analysis at the cosmological level, where there are no place with high densities of material particles. But once such mechanism is constructed (even if not being the primary focus of our work) the value of the BD-field $\varphi$  is ambient dependent, so to speak, since $\varphi$  becomes sensitive to the presence of large densities of matter.   This possibility is well-known in the literature through the chameleon mechanism\,\cite{Khoury:2003aq} and in the case of the BD-field was previously considered in \cite{Avilez:2013dxa} without letting the Brans-Dicke parameter $\oD$ to acquire negative values, and using  datasets which now can be considered a bit obsolete. Here we do, instead,  allow negative values for $\oD$ (we have seen in Sec.\,\ref{sec:preview_BD_article} the considerable advantages involved in this possibility), and moreover we are using a much more complete set of observations from all panels of data taking.  In this scenario we cannot make use of \eqref{eq:MainGeffective} to connect the locally measured value of the gravitational strength $G_N$ with the BD-field at cosmological scales. We just do not need to know how the theory exactly connects these two values. We reiterate once more: we will not focus on the screening mechanism itself here but rather on the properties of the BD-field in the Universe {in the large}, i.e. at the cosmological level. As it is explained in \cite{Avilez:2013dxa} -- see also \cite{Clifton:2011jh} --  many scalar-tensor theories of gravity belonging to the Horndeski class\,\cite{Horndeski:1974wa} could lead to such kind of BD behaviour at cosmological level and, hence, it deserves a dedicated and updated analysis, which is currently lacking in the literature.}
\newline
\newline
{To summarize, the following interpretations of the BD-gravity framework considered here are, in principle, possible in the light of the current observational data:}
\begin{itemize}

\item {\bf BD-Scenario I}: \textit{Rigid Scenario for both the Local and Cosmological domains}. In it, we have $\Geff(\varphi (z))\simeq G(\varphi (z))\simeq G_N$, these three couplings being so close that in practice BD-gravity is indistinguishable from GR.  In this case, the BD-gravity framework is assumed to hold on equal footing with all the scales of the Universe, local and cosmological.  There are no screening effects from matter. In this context, one interprets that the limit from the Cassini probe\,\cite{Bertotti:2003rm}  leads to a very large value of $\omega_{\rm BD}$, which enforces $\varphi$ to become essentially rigid, and one assumes that such constant value is very close to $1$ owing to the relation \eqref{eq:constraintvarphi}.  Such a rigid scenario is, however, unwarranted. It is possible, although  is not necessary since, strictly speaking,  there is no direct connection between the Cassini bound on  $\omega_{\rm BD}$ and the value that $\varphi$ can take. Thus, in this scenario the relation \eqref{eq:constraintvarphi} is just assumed. In point of fact, bounds on  $\omega_{\rm BD}$  can only affect the time evolution of $\varphi$, they do not constraint its value.

\item {\bf BD-Scenario II (Main)}: {\it  Locally  Constrained but Cosmologically Unconstrained Scenario}. It is our main scenario.  It assumes a constrained  situation in the local domain, caused by the presence of chameleonic forces,  but permits an unconstrained picture for the entire cosmological range. In other words,  the Cassini limit that holds for the post-Newtonian parameter $\gamma^{\rm PN}\, $in the local astrophysical level (and hence on $\omega_{\rm BD}$) is assumed to reflect just the presence of screening effects of matter in that non-linear domain. These effects are acting on $\varphi$  and  produce the illusion that  $\omega_{\rm BD}$  has a very large value (as if `dressed' or `renormalized' by the chamaleonic forces). One expects that the `intrusive' effects of matter are only possible in high density (hence non-relativistic) environments, and in their presence we cannot actually know the real (`naked') value of $\omega_{\rm BD}$  through local experiments alone. We assume that the screening disappears as soon as we leave the astrophysical scales and plunge into the cosmological ones; then, and only then, we can measure the naked or ``bare' value of $\omega_{\rm BD}$  (stripped from such effects). We may assume that the screening ceases already at the LSS scales where linear structure formation occurs, {see e.g. \cite{Tsujikawa:2008uc} for examples of potentials which can help to realize this mechanism.} The bare value of $\omega_{\rm BD}$ can then be fitted to the overall data, and in particular to the LSS formation data.  Since $\omega_{\rm BD}$ does no longer appear that big (nor it has any a priori sign) the BD-field $\varphi$ can evolve in an appreciable way at the cosmological level: it increases with the expansion if $\omega_{\rm BD}>0$, and decreases with the expansion if $\omega_{\rm BD}<0$.  In this context, its  initial value, $\varphi_{ini}$, becomes a relevant cosmological parameter, which must be taken into account as a fitting parameter on equal footing with  $\omega_{\rm BD}$ and all of the conventional parameters entering the fit.  Using the large wealth of cosmological data, these parameters can be fixed at the cosmological level without detriment of the observed physics at the local domain, provided there is a screening mechanism insuring that $\Geff(\varphi)(0)$  stays sufficiently  close  to $G_N$ in the local neighbourhood. The numerical results for this important scenario are presented in Tables 3-6 and 9.

\item  {\bf BD-Scenario III}:  {{\it Cassini-constrained Scenario}}. A more restricted version of scenario II  can appear if the `bare value' of $\omega_{\rm BD}$ is as large as in the Cassini bound.  In such case $\omega_{\rm BD}$ is, obviously,  perceived large in both domains,  local  and cosmological. Even so, and despite of the fact that $\varphi$ varies very slowly in this case, one can still exploit the dependence of the fit on the initial value of the BD-field, $\varphi_{ini}$, and use it as a relevant cosmological parameter.  In practice, this situation has only one additional degree of freedom as compared to Scenario I (and one less than in Scenario II), but it is worth exploring -- see our results in Table 10.  As these numerical results show, the Cassini bound still leaves considerable freedom to the BD-$\CC$CDM model for improving the $H_0$ tension (without aggravating the $\sigma_8$ one) since the value of $\varphi$ is still an active degree of freedom, despite its time evolution is now more crippled.

 \item  {\bf BD-Scenario IV}: {\it Variable-$\omega_{\rm BD}(\varphi)$  Scenario}.  Here one admits that the parameter $\omega_{\rm BD}$ is actually a function of the BD-field, $\omega_{\rm BD}=\omega_{\rm BD}(\varphi)$, which can be modeled and adapted to the constraints of the  local and cosmological domains, or even combined with the screening effects of the local Universe.  We have said from the very beginning that we would assume $\omega_{\rm BD}=$const. throughout our analysis, and in fact we shall stick to that hypothesis; so here we mention the variable $\omega_{\rm BD}$ scenario  only for completeness.  In any case, if a function $\omega_{\rm BD}(\varphi)$ exists such that  it takes very large values in the local Universe while it takes much more moderate values in the cosmological scales,  that sort of scenario would be in the main tantamount to Scenario II insofar as concerns its cosmological implications.

\end{itemize}
{In our analysis we basically choose BD- Scenarios II and III (the latter being a particular case of the former), which represent the most tolerant  point of view within the canonical  $\omega_{\rm BD}=$const. option. Scenario II offers the most powerful framework amenable to provide a cure for the tensions afflicting the conventional $\CC$CDM model based on  GR.  Thus, we assume that we can  measure the cosmological  value of the gravity strength in BD theory -- i.e. the value given in Eq.\,\eqref{eq:MainGeffective} -- {by using only cosmological data. We combine the information from the LSS processes involving linear structure formation with the background information obtained from low, intermediate, and very high redshift probes, including BAO, CMB, and the distance ladder measurement of $H_0$.}
The values of  $\varphi_{ini}$ and $\omega_{\rm BD}$ are fitted to the data, and with them we obtain not only $\varphi(z=0)$  but we determine the effective cosmological gravity strength at all epochs  from \eqref{eq:MainGeffective}. The cosmological value of the gravity coupling  can be considered as the `naked' or `bare' value of the gravitational interaction, stripped from screening effects of matter, in the same way as $\omega_{\rm BD}$ measured at cosmological scales is the bare value freed of these effects.  Even though $\Geff$ can be different from $G_N,$  we do not object to that since it can be ascribed to screening forces caused by the huge amounts of clustered matter in the astrophysical environments.  For this reason we do not adopt the local constraints for our cosmological analysis presented in this paper, i.e. we adhere to Scenario II as our main scenario.  Remarkably enough, we shall see that Scenario III still possesses a large fraction of the potentialities inherent to Scenario II, notwithstanding the Cassini bound.  In this sense Scenarios II and III are both extremely interesting. A  smoking gun of such overarching possible picture  is the possible detection of the dynamical dark energy EoS encoded in the BD theory within the GR-picture (cf. Sec.\,\ref{sec:EffectiveEoS_BD_article}), which reveals itself in the form of effective quintessence,  as well as through the large smoothing achieved of  the main tensions afflicting the conventional $\CC$CDM. From here on, we present the bulk of our analysis and detailed results after we have already discussed to a great extent their possible implications.}
\subsection{Data and methodology}\label{sec:MethodData_BD_article}
We fit the BD-$\Lambda$CDM together with the concordance GR-$\Lambda$CDM model and the GR-XCDM (based on the XCDM parametrization of the DE \cite{Turner:1998ex}) to the wealth of cosmological data compiled from distant type Ia supernovae (SNIa), baryonic acoustic oscillations (BAO), a set of measurements of the Hubble function at different redshifts, the large scale structure (LSS) formation data encoded in $f(z_i)\sigma_8(z_i)$, and the CMB temperature and low-$l$ polarization data from the Planck satellite. The joint combination of all these individual datasets will constitute our \textbf{Baseline Data} configuration. Moreover, we also study the repercussion of some alternative data, by adding them to the aforementioned baseline setup. These additional datasets are: a prior on the value of $H_0$ (or alternatively an effective calibration prior on M) provided by the SH0ES collaboration; the CMB high-$l$ polarization and lensing data from Planck; the Strong Lensing (SL) time delay angular diameter distances from H0LICOW; and, finally, Weak Lensing (WL) data from KiDS. Below we provide a brief description of the datasets employed in our analyses, together with the corresponding references.
\newline
\newline
\textbf{SNIa}: We use the full Pantheon likelihood, which incorporates the information from 1048 SNIa \cite{Scolnic:2017caz}. In addition, we also include the 207 SNIa from the DES survey \cite{Abbott:2018wog}. These two SNIa samples are uncorrelated, but the correlations between the points within each sample are non-null and have been duly incorporated in our analyses through the corresponding covariance matrices.
\newline
\newline
\textbf{BAO}: We use data on both, isotropic and anisotropic BAO analyses. We provide the detailed list of data points and corresponding references in Table 1. A few comments are in order about the use of some of the BAO data points considered in this chapter. Regarding the Ly$\alpha$-forest data, we opt to use the auto-correlation information from \cite{Agathe:2019vsu}. Excluding the Ly$\alpha$ cross-correlation data allows us to avoid double counting issues between the latter and the eBOSS data from \cite{Gil-Marin:2018cgo}, due to the partial (although small) overlap in the list of quasars employed in these two analyses. It is also important to remark that we consider in our baseline dataset the BOSS data reported in \cite{Gil-Marin:2016wya}, which contains information from the spectrum (SP) and the bispectrum (BP). The bispectrum information could capture some details otherwise missed when only the spectrum is considered, so it is worth to use it\footnote{See also Ref.\cite{Sola:2018sjf} for additional comments on the significance of the bispectrum data as well as its potential implications on the possible detection of dynamical dark energy.}. Therefore, we study the possible significance of the bispectrum component in the data by carrying out an explicit comparison of the results obtained with the baseline configuration to those obtained by substituting the data points from \cite{Gil-Marin:2016wya} with those from \cite{Alam:2016hwk}, which only incorporate the SP information. The results are provided in Tables 3 and 5, respectively. In Tables 4 and 6-8 we use the SP+BSP combination \cite{Gil-Marin:2016wya}. In Table 10 we employ both SP and SP+BSP.
%
\begin{table}[t!]
\setcounter{table}{0}
\begin{center}
\resizebox{14.5cm}{!}{
\begin{tabular}{| c | c |c | c |c|}
\multicolumn{1}{c}{Survey} &  \multicolumn{1}{c}{$z$} &  \multicolumn{1}{c}{Observable} &\multicolumn{1}{c}{Measurement} & \multicolumn{1}{c}{{\small References}}
\\\hline
6dFGS+SDSS MGS & $0.122$ & $D_V(r_s/r_{s,\rm fid})$[Mpc] & $539\pm17$[Mpc] &\cite{Carter:2018vce}
\\\hline
 WiggleZ & $0.44$ & $D_V(r_s/r_{s,\rm fid})$[Mpc] & $1716.4\pm 83.1$[Mpc] &\cite{Kazin:2014qga} \tabularnewline
\cline{2-4} & $0.60$ & $D_V(r_s/r_{s,\rm fid})$[Mpc] & $2220.8\pm 100.6$[Mpc]&\tabularnewline
\cline{2-4} & $0.73$ & $D_V(r_s/r_{s,\rm fid})$[Mpc] &$2516.1\pm 86.1$[Mpc] &
\\\hline

DR12 BOSS (BSP)& $0.32$ & $Hr_s/(10^{3}\rm km/s)$ & $11.549\pm0.385$   &\cite{Gil-Marin:2016wya}\\ \cline{3-4}
 &  & $D_A/r_s$ & $6.5986\pm0.1337$ &\tabularnewline \cline{3-4}
 \cline{2-2}& $0.57$ & $Hr_s/(10^{3} \rm km/s)$  & $14.021\pm0.225$ &\\ \cline{3-4}
 &  & $D_A/r_s$ & $9.389\pm0.1030$ &\\\hline

DR12 BOSS (SP) & $0.38$ & $D_M(r_s/r_{s,\rm fid})$[Mpc] & $1518\pm22$   &\cite{Alam:2016hwk} \\ \cline{3-4}
 &  & $H(r_{s,\rm fid}/r_s)$[km/s/Mpc] & $81.5\pm1.9$ & \tabularnewline \cline{3-4}
 \cline{2-2}& $0.51$ & $D_M(r_s/r_{s,\rm fid})$[Mpc] & $1977\pm27$ & \\ \cline{3-4}
 &  & $H(r_{s,\rm fid}/r_s)$[km/s/Mpc] & $90.4\pm1.9$ & \\ \cline{3-4}
 \cline{2-2}& $0.61$ & $D_M(r_s/r_{s,\rm fid})$[Mpc]  & $2283\pm32$ & \\ \cline{3-4}
 &  & $H(r_{s,\rm fid}/r_s)$[km/s/Mpc] & $97.3\pm2.1$ & \\\hline

DES & $0.81$ & $D_A/r_s$ & $10.75\pm0.43$ &\cite{Abbott:2017wcz}
\\\hline
eBOSS DR14 & $1.19$ & $Hr_s/(10^{3}\rm km/s)$ & $19.6782\pm1.5867  $ &\cite{Gil-Marin:2018cgo}\\ \cline{3-4}
 &  & $D_A/r_s$ & $12.6621\pm0.9876$ &\tabularnewline \cline{3-4}
 \cline{2-2}& $1.50$ & $Hr_s/(10^{3}\rm km/s)$  & $19.8637\pm2.7187$ &\\ \cline{3-4}
 &  & $D_A/r_s$ & $12.4349\pm1.0429$ &\\ \cline{3-4}
 \cline{2-2}& $1.83$ & $Hr_s/(10^{3}\rm km/s)$  & $26.7928\pm3.5632$ &\\ \cline{3-4}
 &  & $D_A/r_s$ & $13.1305\pm1.0465$ &\\\hline

Ly$\alpha$-F eBOSS DR14 & $2.34$ & $D_H/r_s$ & $8.86\pm0.29$   &\cite{Agathe:2019vsu}
\\ \cline{3-4} &  & $D_M/r_s$ & $37.41\pm 1.86$
&\\\hline
\end{tabular}}
\caption{\scriptsize Published values of BAO data, see the quoted references and text in Sec. \ref{sec:MethodData_BD_article}. Although we include in this table the values of $D_H/r_s=c/(r_sH)$ and $D_M/r_s$ for the Ly$\alpha$-forest auto-correlation data from \cite{Agathe:2019vsu}, we have performed the fitting analysis with the full likelihood. The fiducial values of the comoving sound horizon appearing in the table are $r_{s,{\rm fid}} = 147.5$ Mpc for \cite{Carter:2018vce}, $r_{s,{\rm fid}} = 148.6$ Mpc for \cite{Kazin:2014qga}, and $r_{s,{\rm fid}} = 147.78$ Mpc for \cite{Alam:2016hwk}.}
\end{center}
\end{table}
%
%
\newline
\newline
\textbf{Cosmic Chronometers}: We use the 31 data points on $H(z_i)$, at different redshifts, from \cite{Jimenez:2003iv,Simon:2004tf,Stern:2009ep,Moresco:2012jh,Zhang:2012mp,Moresco:2015cya,Moresco:2016mzx,Ratsimbazafy:2017vga}. All of them have been obtained making use of the differential age technique applied to passively evolving galaxies \cite{Jimenez:2001gg}, which provides cosmology-independent constraints on the Hubble function, but are still subject to systematics coming from the choice of the stellar population synthesis technique, and also the potential contamination of young stellar components in the quiescent galaxies \cite{Lopez-Corredoira:2017zfl,Lopez-Corredoira:2018tmn,Moresco:2018xdr}. For this reason we consider a more conservative dataset that takes into account these additional uncertainties. To be concrete, we use the processed sample presented in Table 2 of \cite{Gomez-Valent:2018gvm}. See therein for further details.
\newline
\newline
\textbf{CMB}: In our baseline dataset we consider the full Planck 2018 TT+lowE likelihood \cite{Aghanim:2018eyx}. In order to study the influence of the CMB high-$l$ polarizations and lensing we consider two alternative (non-baseline) datasets, in which we substitute the TT+lowE likelihood by: (i) the TTTEEE+lowE likelihood, which incorporates the information of high multipole polarizations; (ii) the full TTTEEE+lowE+lensing likelihood, in which we also incorporate the Planck 2018 lensing data. In Tables 6 and 7 these scenarios are denoted as B+$H_0$+pol and B+$H_0$+pol+lens, respectively.
%
%
\begin{table}[t!]
\begin{center}
\resizebox{10cm}{!}{
\begin{tabular}{| c | c |c | c |}
\multicolumn{1}{c}{Survey} &  \multicolumn{1}{c}{$z$} &  \multicolumn{1}{c}{$f(z)\sigma_8(z)$} & \multicolumn{1}{c}{{\small References}}
\\\hline
6dFGS+2MTF & $0.03$ & $0.404^{+0.082}_{-0.081}$ & \cite{Qin:2019axr}
\\\hline
SDSS-DR7 & $0.10$ & $0.376\pm 0.038$ & \cite{Shi:2017qpr}
\\\hline
GAMA & $0.18$ & $0.29\pm 0.10$ & \cite{Simpson:2015yfa}
\\ \cline{2-4}& $0.38$ & $0.44\pm0.06$ & \cite{Blake:2013nif}
\\\hline
DR12 BOSS (BSP)& $0.32$ & $0.427\pm 0.056$  & \cite{Gil-Marin:2016wya}\\ \cline{2-3}
 & $0.57$ & $0.426\pm 0.029$ & \\\hline
 WiggleZ & $0.22$ & $0.42\pm 0.07$ & \cite{Blake:2011rj} \tabularnewline
\cline{2-3} & $0.41$ & $0.45\pm0.04$ & \tabularnewline
\cline{2-3} & $0.60$ & $0.43\pm0.04$ & \tabularnewline
\cline{2-3} & $0.78$ & $0.38\pm0.04$ &
\\\hline

DR12 BOSS (SP) & $0.38$ & $0.497\pm 0.045$ & \cite{Alam:2016hwk}\tabularnewline
\cline{2-3} & $0.51$ & $0.458\pm0.038$ & \tabularnewline
\cline{2-3} & $0.61$ & $0.436\pm0.034$ &
\\\hline

VIPERS & $0.60$ & $0.49\pm 0.12$ & \cite{Mohammad:2018mdy}
\\ \cline{2-3}& $0.86$ & $0.46\pm0.09$ &
\\\hline
VVDS & $0.77$ & $0.49\pm0.18$ & \cite{Guzzo:2008ac},\cite{Song:2008qt}
\\\hline
FastSound & $1.36$ & $0.482\pm0.116$ & \cite{Okumura:2015lvp}
\\\hline
eBOSS DR14 & $1.19$ & $0.4736\pm 0.0992$ & \cite{Gil-Marin:2018cgo} \tabularnewline
\cline{2-3} & $1.50$ & $0.3436\pm0.1104$ & \tabularnewline
\cline{2-3} & $1.83$ & $0.4998\pm0.1111$ &

\\\hline
 \end{tabular}}
\caption{\scriptsize Published values of $f(z)\sigma_8(z)$, see the quoted references and text in Sec. \ref{sec:MethodData_BD_article}.}
\end{center}
\end{table}
%
%
\newline
\newline
\textbf{LSS}: In this chapter the LSS dataset contains the data points on the product of the ordinary growth rate $f(z_i)$ with $\sigma_8(z_i)$ at different effective redshifts. They are all listed in Table 2, together with the references of interest. In order to correct the potential bias introduced by the particular choice of a fiducial model in the original observational analyses we apply the rescaling correction explained in \cite{Macaulay:2013swa}. See also Sec. II.2 of \cite{Nesseris:2017vor}. The internal correlations between the BAO and RSD data from \cite{Gil-Marin:2016wya}, \cite{Alam:2016hwk} and \cite{Gil-Marin:2018cgo} have been duly taken into account through the corresponding covariance matrices provided in these three references.
\newline
\newline
\textbf{Prior on $H_0$}: We include as a prior in almost all the non-baseline datasets the value of the Hubble parameter measured by the SH0ES collaboration, $H_0= (73.5\pm 1.4)$ km/s/Mpc \cite{Reid:2019tiq}. It is obtained with the cosmic distance ladder method using an improved calibration of the Cepheid period-luminosity relation. As a possible alternative, we also consider the case in which we use instead the SH0ES effective calibration prior on the absolute magnitude  $M$ of the SNIa, as provided in \cite{Camarena:2019rmj}:  $M=-19.2191\pm 0.0405$. It is obtained from the calibration of `nearby' SNIa (at $z\lesssim 0.01$) with Cepheids\,\cite{Riess:2019cxk}.
It has been recently argued in \cite{Camarena:2019moy,Camarena:2019rmj} (and later on also in \cite{Dhawan:2020xmp,Benevento:2020fev}) that in cosmological studies it is better to use this SH0ES constraint rather than the direct prior on $H_0$ when combined with low-redshift SNIa data to avoid double counting issues. We show that the results obtained with these two methods are compatible and lead to completely consistent results (see the discussion in Sec. \ref{sec:NumericalAnalysis_BD_article}, and also Tables 3 and 6).
\newline
\newline
\textbf{SL}: In one of the non-baseline datasets we use in combination with the SH0ES prior on $H_0$ the data extracted from the six gravitational lensed quasars of variable luminosity reported by the H0LICOW team \cite{Wong:2019kwg}. In order to know all the details of the methodology to include these data in the analysis see Appendix \ref{Appendix_E}.
\newline
\newline
\textbf{WL}: We include the information provided by the survey The Kilo-Degree Survey (KiDS) \cite{Hildebrandt:2016iqg,Joudaki:2017zdt,Kohlinger:2018sxx,Wright:2020ppw} in one of our non-baseline scenarios. The reason why we prefer to keep away this data from the Baseline combination is because in order to model the non-linear effects it is necessary to choose a particular cosmological model. As a consequence this data is model-dependent. See Appendix \ref{Appendix_E} for more details.  
\newline
\newline
We study the performance of the BD-$\CC$CDM, GR-$\CC$CDM and GR-XCDM models under different datasets. In the following we briefly summarize the composition of each of them:
\begin{itemize}
\item {\bf Baseline (B)}: Here we include the Planck 2018 TT+lowE CMB data, together with SNIa +BAO+$H(z_i)$+LSS (see Table 3 and 8). It is important to remark that for the BOSS BAO+LSS data we consider \cite{Gil-Marin:2016wya}, which includes the information from the spectrum (SP) as well as from the bispectrum (BSP). We construct some other datasets using this baseline configuration as the main building block. See the other items, below.

\item {\bf Baseline{\boldmath+$H_0$}} ({\bf B{\boldmath+$H_0$}}): Here we add the SH0ES prior on the $H_0$ parameter from \cite{Reid:2019tiq} to the baseline dataset (see again Tables 3 and 8).

\item {\bf Baseline{\boldmath+$H_0$+}SL}: The inclusion of the Strong Lensing (SL) data from \cite{Wong:2019kwg} exacerbates  more the $H_0$-tension in the context of the GR-$\CC$CDM model (see e.g. \cite{Verde:2019ivm} and Sec. \ref{sec:Discussion_BD_article}), so it is interesting to also study the ability of the BD-$\CC$CDM to fit the SL data when they are combined with the previous B+$H_0$ dataset, and compare the results with those obtained with the GR-$\CC$CDM. The corresponding fitting results are displayed in Table 4.

\item  {\bf Spectrum}: In this dataset we replace the SP+BSP data from \cite{Gil-Marin:2016wya} used in the Baseline dataset (see the first item, above) by the data from \cite{Alam:2016hwk}, which only contains the spectrum (SP) information (i.e. the usual matter power spectrum).

\item {\bf Spectrum{\boldmath+$H_0$}}: As in the preceding item, but including the $H_0$ prior from SH0ES \cite{Reid:2019tiq}.

\end{itemize}
The aforementioned datasets are all based on the BD-Scenario II (cf. Sec. \ref{sec:Mach_BD_article}) and can be considered as the main ones (cf. Tables 3-5, and 8), nevertheless we also consider a bunch of alternative datasets (also based on the BD-Scenario II). We present the corresponding results in Table 6 and the first five columns of Table 7 for the BD-$\CC$-CDM and GR-$\CC$CDM models, respectively.

\begin{itemize}

\item {\bf B{\boldmath+$M$}}: In this scenario we replace the prior on $H_0$ \cite{Reid:2019tiq} employed in the B+$H_0$ dataset with the effective SH0ES calibration prior on the absolute magnitude of SNIa $M$ provided in \cite{Camarena:2019rmj}.

\item {\bf B{\boldmath+$H_0$+}pol}: Here we add the CMB high-$l$ polarization data from Planck 2018 \cite{Aghanim:2018eyx} to the B+$H_0$ dataset described before, i.e. we consider the Planck 2018 TTTEEE+lowE likelihood for the CMB.

\item {\bf B{\boldmath+$H_0$+}pol+lens}: In addition to the datasets considered in the above case we also include the CMB lensing data from Planck 2018 \cite{Aghanim:2018eyx}, i.e. we use the Planck 2018 TTTEEE+lowE+ lensing likelihood.

\item {\bf B{\boldmath+$H_0$+}WL}: In this alternative case we consider the Weak Lensing (WL) data from KiDS \cite{Hildebrandt:2016iqg,Kohlinger:2018sxx}, together with the B+$H_0$ dataset.

\item  {\bf CMB+BAO+SNIa}: By considering only this data combination, we study the performance of the the BD-$\CC$CDM and the GR-$\CC$CDM models under a more limited dataset, obtained upon the removal of the data that trigger the $H_0$ and $\sigma_8$ tensions, i.e. the prior on $H_0$ from SH0ES and the LSS data. The use of the BAO+SNIa data helps to break the strong degeneracies found in parameter space when only the CMB is considered. Here we use the TT+lowE Planck 2018 likelihood \cite{Aghanim:2018eyx}.

\end{itemize}
Finally, in Table 10 we present the results obtained for the BD-$\CC$CDM in the context of the Cassini-constrained scenario, or Scenario III (see Sec. \ref{sec:Mach_BD_article} for the details). The corresponding results for the GR-$\CC$CDM are shown in the third column of Table 3, and the last three rows of Table 7. In all these datasets we include the Cassini bound\,\cite{Bertotti:2003rm} (see Sec. \ref{sec:BDgravity_BD_article} for details). The main purpose of this scenario is to test the ability of the BD-$\CC$CDM to fit the observational data with $\eBD\simeq 0$ {\it and} $\varphi\neq 1$.
\begin{itemize}

\item {\bf B{\boldmath+$H_0$+}Cassini}: It contains the very same datasets as in the Baseline+$H_0$ case, but here we also include the Cassini constraint.

\item {\bf B{\boldmath+$H_0$+}Cassini (No LSS)}: Here we study the impact of the LSS data in the context of Scenario III, by removing them from the previous B+$H_0$+Cassini dataset.

\item {\bf Dataset \cite{Ballesteros:2020sik}}: To ease the comparison with the results obtained in \cite{Ballesteros:2020sik}, here we use exactly the same dataset as in that reference, namely: the Planck 2018 TTTEEE+lowE+lensing likelihood \cite{Aghanim:2018eyx}, the BAO data from \cite{Beutler:2011hx,Ross:2014qpa,Alam:2016hwk}, and the SH0ES prior from \cite{Riess:2019cxk}, $H_0=74.03\pm 1.42$ km/s/Mpc.

\item {\bf Dataset\cite{Ballesteros:2020sik}+LSS}: Here we consider an extension of the previous scenario by adding the LSS data on top of the data from \cite{Ballesteros:2020sik}.

\end{itemize}
We believe that all the datasets and scenarios studied in this work cover a wide range of possibilities and show in great detail which is the phenomenological performance of the BD-$\CC$CDM, GR-$\CC$CDM and GR-XCDM models.
\newline
The speed of gravitational waves at $z\approx 0$, $c_{gw}$, has been recently constrained to be extremely close to the speed of light, $|c_{gw}/c-1|\lesssim 5\cdot 10^{-16}$ \cite{TheLIGOScientific:2017qsa}. In the BD-$\CC$CDM model $c_{gw}=c\,$ $\forall{z}$, so this constraint is automatically fulfilled. We have also checked that the BD-$\CC$CDM respects the bound on $G(\varphi)$ at the Big Bang Nucleosynthesis (BBN) epoch, $|G(\varphi_{\rm BBN})/G_N-1|\lesssim 0.1$ \cite{Uzan:2010pm}, since $G(\varphi_{\rm BBN})\simeq G(\varphi_{ini})$ and our best-fit values satisfy $G(\varphi_{ini})>0.9 G_N$ regardless of the dataset under consideration, see the fitting results in Tables 3-6 and 10.
\newline
\newline
To obtain the posterior distributions and corresponding constraints for the various dataset combinations described above we have run the Monte Carlo sampler \texttt{MontePython}\footnote{http://baudren.github.io/montepython.html} \cite{Audren:2012wb} together with the Einstein-Boltzmann system solver \texttt{CLASS}\footnote{http://lesgourg.github.io/class\_public/class.html} \cite{Blas:2011rf}. We have duly modified the latter to implement the background and linear perturbations equations of the BD-$\Lambda$CDM model. {We use adiabatic initial conditions for all matter species. Let us note that the initial perturbation of the BD-field and its time derivative can be set to zero, as the DM velocity divergence when the synchronous gauge is employed, see Appendix C.5 of \cite{Sola:2020lba} for a brief discussion.} To get the contour plots and one-dimensional posterior distributions of the parameters entering the models we have used the \texttt{Python} package \texttt{GetDist}\footnote{https://getdist.readthedocs.io/en/latest/} \cite{Lewis:2019xzd}, and to compute the full Bayesian evidences for the different models and dataset combinations, we have employed the code \texttt{MCEvidence}\footnote{https://github.com/yabebalFantaye/MCEvidence} \cite{Heavens:2017afc}. The Deviance Information Criterion (DIC) \cite{DIC} has been computed with our own numerical code. The results are displayed in Tables 3-10, and also in Figs. 9-11. They are discussed in the next section.
\subsection{Numerical analysis. Results}\label{sec:NumericalAnalysis_BD_article}
In the following we put the models under consideration to the test, using the various datasets described in Sec. \ref{sec:MethodData_BD_article}. We perform the statistical analysis of the models in terms of a joint likelihood function, which is the product of the individual likelihoods for each data source and includes the corresponding covariance matrices. For a fairer comparison with the GR-$\CC$CDM  we use standard information criteria in which the presence of extra parameters in a given model is conveniently penalized so as to achieve a balanced comparison with the model having less parameters. More concretely, we employ the full Bayesian evidence to duly quantify the fitting ability of the BD-$\CC$CDM model as compared to its GR analogue. Given a dataset $\mathcal{D}$, the probability of a certain model $M_i$ to be the best one among a given set of models $\{M\}$ reads,
\begin{figure}[t!]
\begin{center}
\label{fig:BayesRatio_BD_article}
\includegraphics[width=4.5in, height=3.5in]{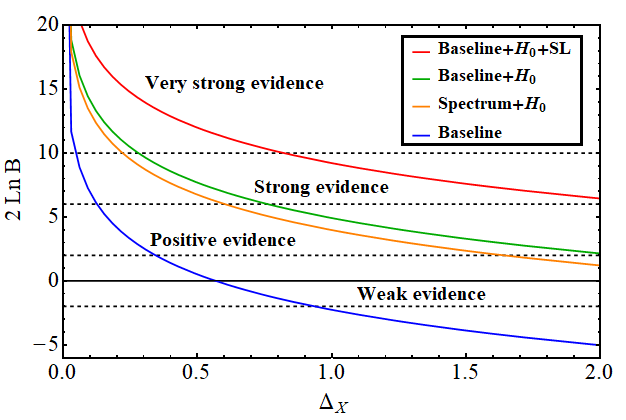}
\caption{\scriptsize{The full Bayesian evidence curves for the BD-$\CC$CDM as compared to the GR-$\CC$CDM, using different datasets and as a function of $\Delta_X=\Delta\eBD/10^{-2}=\Delta\varphi_{ini}/0.2$, with $\Delta\eBD$ and $\Delta\varphi_{ini}$ being the (flat) prior ranges for $\eBD$ and $\varphi_{ini}$, respectively. The curves are computed using the exact evidence formula, Eq. \eqref{eq:evidence}. The marked evidence ranges conform with the conventional Jeffreys' scale, see the main text in Sec. \ref{sec:NumericalAnalysis_BD_article}.}}
\end{center}
\end{figure}
%
\begin{equation}\label{eq:BayesTheorem}
P(M_i|\mathcal{D})=\frac{P(M_i)\mathcal{E}(\mathcal{D}|M_i)}{P(\mathcal{D})}\,,
\end{equation}
{where $P(M_i)$ is the prior probability of the model $M_i$, $P(\mathcal{D})$ the probability of having the dataset $\mathcal{D}$, and the normalization condition $\sum_{j\in\{M\}}P(M_j)=1$ is assumed. The quantity $\mathcal{E}(\mathcal{D}|M_i)$ is the so-called marginal likelihood or evidence\,\cite{Amendola:2015ksp}. If the model $M_i$ has $n$ parameters contained in the vector $\vec{p}^{M_i}=(p^{M_i}_1, p^{M_i}_2,...,p^{M_i}_n)$, the evidence takes the following form:}

\renewcommand{\arraystretch}{1.1}
\begin{table}[t!]
\begin{center}
\resizebox{1\textwidth}{!}{

\begin{tabular}{|c  |c | c |  c | c | c  |}
 \multicolumn{1}{c}{} & \multicolumn{2}{c}{Baseline} & \multicolumn{2}{c}{Baseline+$H_0$}
\\\hline
{\scriptsize Parameter} & {\scriptsize GR-$\Lambda$CDM}  & {\scriptsize BD-$\Lambda$CDM} & {\scriptsize GR-$\Lambda$CDM}  &  {\scriptsize BD-$\Lambda$CDM}
\\\hline
{\scriptsize $H_0$ (km/s/Mpc)}  & {\scriptsize $68.20^{+0.41}_{-0.40}$} & {\scriptsize $68.86^{+1.15}_{-1.24}$} & {\scriptsize $68.57^{+0.36}_{-0.42}$}  & {\scriptsize $70.83^{+0.92}_{-0.95}$}
\\\hline
{\scriptsize$\omega_b$} & {\scriptsize $0.02227^{+0.00019}_{-0.00018}$}  & {\scriptsize $0.02251^{+0.00026}_{-0.00027}$} & {\scriptsize $0.02238\pm 0.00019$}  &  {\scriptsize $0.02275^{+0.00024}_{-0.00026}$}
\\\hline
{\scriptsize$\omega_{cdm}$} & {\scriptsize $0.11763^{+0.00090}_{-0.00092}$}  & {\scriptsize $0.11598^{+0.00159}_{-0.00152}$} & {\scriptsize $0.11699^{+0.00092}_{-0.00083}$}  &  {\scriptsize $0.11574^{+0.00164}_{-0.00158}$}
\\\hline
{\scriptsize$\tau$} & {{\scriptsize$0.050^{+0.004}_{-0.008}$}} & {{\scriptsize$0.052^{+0.006}_{-0.008}$}} & {{\scriptsize$0.051^{+0.005}_{-0.008}$}}  &   {{\scriptsize$0.053^{+0.006}_{-0.008}$}}
\\\hline
{\scriptsize$n_s$} & {{\scriptsize$0.9683^{+0.0039}_{-0.0038}$}}  & {{\scriptsize$0.9775^{+0.0084}_{-0.0086}$}} & {{\scriptsize$0.9703^{+0.0038}_{-0.0036}$}} &   {{\scriptsize$0.9873^{+0.0076}_{-0.0075}$}}
\\\hline
{\scriptsize$\sigma_8$}  & {{\scriptsize$0.797^{+0.005}_{-0.006}$}}  & {{\scriptsize$0.785\pm 0.013$}} & {{\scriptsize$0.796^{+0.006}_{-0.007}$}}  &   {{\scriptsize$0.789\pm 0.013$}}
\\\hline
{\scriptsize$r_s$ (Mpc)}  & {{\scriptsize$147.83^{+0.29}_{-0.30}$}}  & {{\scriptsize$145.89^{+2.26}_{-2.49}$}} & {{\scriptsize$147.88\pm 0.31$}}  &   {{\scriptsize$142.46^{+1.84}_{-1.86}$}}
\\\hline
{\scriptsize $\eBD$} & - & {{\scriptsize $-0.00184^{+0.00140}_{-0.00142}$}} & - &   {{\scriptsize $-0.00199^{+0.00142}_{-0.00147}$}}
\\\hline
{\scriptsize$\varphi_{ini}$} & - & {{\scriptsize $0.974^{+0.027}_{-0.031}$}} & - &    {{\scriptsize $0.932^{+0.022}_{-0.023}$}}
\\\hline
{\scriptsize$\varphi(0)$} & - & {{\scriptsize $0.960^{+0.032}_{-0.037}$}} & - &    {{\scriptsize $0.918^{+0.027}_{-0.029}$}}
\\\hline
{\scriptsize$\weff(0)$} & - & {{\scriptsize $-0.983^{+0.015}_{-0.014}$}} & - &    {{\scriptsize $-0.966^{+0.012}_{-0.011}$}}
\\\hline
{\tiny $\dot{G}(0)/G(0) (10^{-13}yr^{-1})$} & - & {{\scriptsize $2.022^{+1.585}_{-1.518}$}} & - &    {{\scriptsize $2.256^{+1.658}_{-1.621}$}}
\\\hline
{\scriptsize$\chi^2_{min}$} & {\scriptsize 2271.98}  & {\scriptsize 2271.82} & {\scriptsize 2285.50}  &  {\scriptsize 2276.04}
\\\hline
{\scriptsize$2\ln B$} & {\scriptsize -}  & {\scriptsize -2.26} & {\scriptsize -}  &  {\scriptsize +4.92}
\\\hline
{\scriptsize$\Delta {\rm DIC}$} & {\scriptsize -}  & {\scriptsize -0.54} & {\scriptsize -}  &  {\scriptsize +4.90}
\\\hline
\end{tabular}}
\end{center}
\label{tableFit3}
\caption{\scriptsize The mean fit values and $68.3\%$ confidence limits for the considered models using our baseline dataset in the first block, i.e. SNIa+$H(z)$+BAO+LSS+CMB TT data, and baseline+$H_0$ in the second one (cf. Sec. \ref{sec:MethodData_BD_article} for details). These results have been obtained within our main BD scenario (Scenario II of Sec.\,\ref{sec:Mach_BD_article}). In all cases a massive neutrino of $0.06$ eV has been included. First we display the fitting results for the six conventional parameters, namely: $H_0$, the reduced density parameters for baryons ($\omega_{b}=\Omega_b h^2$) and CDM ($\omega_{cdm}=\Omega_{cdm} h^2$), the reionization optical depth $\tau$, the spectral index $n_s$ of the primordial power-law power spectrum, and, for convenience, instead of the amplitude $A_s$ of such spectrum we list the values of $\sigma_8$. We also include the sound horizon at the baryon drag epoch, $r_s$, obtained as a derived parameter. Right after we list the values of the free parameters that characterize the BD model: $\eBD$ \eqref{eq:definitions_BD_article} and the initial condition for the BD-field, $\varphi_{ini}$. We also include the values of the BD-field, the (exact) effective EoS parameter \eqref{effEoS}, and the ratio between the derivative and the value of Newton's coupling, all computed at $z=0$. Finally, we report the values of the minimum of the $\chi^2$-function, $\chi^2_{min}$, the exact Bayes ratios (computed under the conditions explained in the main text of Sec. \ref{sec:NumericalAnalysis_BD_article}), and the DIC. It is also worth to remark that the baseline dataset employed here includes the contribution not only of the spectrum, but also the bispectrum information from BOSS \cite{Gil-Marin:2016wya}, see Sec. \ref{sec:MethodData_BD_article} for details.}
\end{table}
\begin{equation}\label{eq:evidence}
\mathcal{E}(\mathcal{D}|M_i)=\int \mathcal{L}(\mathcal{D}|\vec{p}^{M_i},M_i)\pi(\vec{p}^{M_i}) d^np^{M_i}\,,
\end{equation}
{with $\mathcal{L}(\mathcal{D}|\vec{p}^{M_i},M_i)$ the likelihood and $\pi(\vec{p}^{M_i})$ the prior of the parameters entering the model $M_i$. The evidence is larger for those models that have more overlapping volume between the likelihood and the prior distributions, but penalizes the use of additional parameters having a non-null impact on the likelihood. Hence, the evidence constitutes a good way of quantifying the performance of the model by implementing in practice the Occam razor principle. We can compare the fitting performance of BD-$\CC$CDM and GR-$\Lambda$CDM models by assuming equal prior probability for both of them, i.e. $P({\rm BD-}\CC{\rm CDM})=P({\rm GR-}\CC{\rm CDM})$ (``Principle of Insufficient Reason''). The ratio of their associated probabilities can then be directly written as the ratio of their corresponding evidences, i.e.}
%
%
\renewcommand{\arraystretch}{1.1}
\begin{table}[t!]
\begin{center}
\resizebox{0.65\textwidth}{!}{

\begin{tabular}{|c  |c | c |  c |}
 \multicolumn{1}{c}{} & \multicolumn{2}{c}{Baseline+$H_0$+SL}
\\\hline
{\scriptsize Parameter} & {\scriptsize GR-$\CC$CDM}  & {\scriptsize BD-$\CC$CDM}
\\\hline
{\scriptsize $H_0$ (km/s/Mpc)}  & {\scriptsize $68.74^{+0.37}_{-0.40}$} & {\scriptsize $71.30^{+0.80}_{-0.84}$}
\\\hline
{\scriptsize$\omega_b$} & {\scriptsize $0.02242^{+0.00018}_{-0.00019}$}  & {\scriptsize $0.02281\pm 0.00025$}
\\\hline
{\scriptsize$\omega_{cdm}$} & {\scriptsize $0.11666^{+0.00087}_{-0.00086}$}  & {\scriptsize $0.11560^{+0.00158}_{-0.00169}$}
\\\hline
{\scriptsize$\tau$} & {{\scriptsize$0.051^{+0.005}_{-0.008}$}} & {{\scriptsize$0.053^{+0.006}_{-0.008}$}}
\\\hline
{\scriptsize$n_s$} & {{\scriptsize$0.9708^{+0.0036}_{-0.0038}$}}  & {{\scriptsize$0.9901^{+0.0075}_{-0.0070}$}}
\\\hline
{\scriptsize$\sigma_8$}  & {{\scriptsize$0.795^{+0.006}_{-0.007}$}}  & {{\scriptsize$0.789\pm 0.013$}}
\\\hline
{\scriptsize$r_s$ (Mpc)}  & {{\scriptsize$147.93^{+0.30}_{-0.31}$}}  & {{\scriptsize$141.68^{+1.69}_{-1.73}$}}
\\\hline
{\scriptsize $\eBD$}  & {\scriptsize -}   & {{\scriptsize$-0.00208^{+0.00151}_{-0.00140}$}}
\\\hline
{\scriptsize$\varphi_{ini}$}  & {\scriptsize -}   & {{\scriptsize$0.923^{+0.019}_{-0.021}$}}
\\\hline
{\scriptsize$\varphi(0)$}  & {\scriptsize -}   & {{\scriptsize$0.908^{+0.026}_{-0.028}$}}
\\\hline
{\scriptsize$\weff(0)$}  & {\scriptsize -}   & {{\scriptsize$-0.962\pm0.011$}}
\\\hline
{\tiny $\dot{G}(0)/G(0) (10^{-13}yr^{-1})$} & {\scriptsize -}   & {{\scriptsize$2.375^{+1.612}_{-1.721}$}}
\\\hline
{\scriptsize$\chi^2_{min}$} & {\scriptsize 2320.40}  & {\scriptsize 2305.80}
\\\hline
{\scriptsize$2\ln B$} & {\scriptsize -}  & {\scriptsize $+9.22$}
\\\hline
{\scriptsize$\Delta {\rm DIC}$} & {\scriptsize -}  & {\scriptsize $+9.15$}
\\\hline
\end{tabular}}
\end{center}
\label{tableFit4}
\caption{\scriptsize Fitting results as in Table 3, but adding the Strong Lensing data, i.e. we use here the dataset Baseline+$H_0$+SL, for both the GR-$\CC$CDM and the BD-$\CC$CDM.}
\end{table}
%
%
%
\renewcommand{\arraystretch}{1.1}
\begin{table}[t!]
\begin{center}
\resizebox{1\textwidth}{!}{

\begin{tabular}{|c  |c | c |  c | c | c  |}
 \multicolumn{1}{c}{} & \multicolumn{2}{c}{Spectrum} & \multicolumn{2}{c}{Spectrum+$H_0$}
\\\hline
{\scriptsize Parameter} & {\scriptsize GR-$\Lambda$CDM}  & {\scriptsize BD-$\Lambda$CDM} & {\scriptsize GR-$\Lambda$CDM}  &  {\scriptsize BD-$\Lambda$CDM}
\\\hline
{\scriptsize $H_0$ (km/s/Mpc)}  & {\scriptsize $68.00^{+0.47}_{-0.48}$} & {\scriptsize $68.86^{+1.27}_{-1.34}$} & {\scriptsize $68.61^{+0.46}_{-0.49}$}  & {\scriptsize $70.94^{+1.00}_{-0.98}$}
\\\hline
{\scriptsize$\omega_b$} & {\scriptsize $0.02223^{+0.00020}_{-0.00021}$}  & {\scriptsize $0.02241\pm 0.00027$} & {\scriptsize $0.02239^{+0.00020}_{-0.00019}$}  &  {\scriptsize $0.02264^{+0.00026}_{-0.00025}$}
\\\hline
{\scriptsize$\omega_{cdm}$} & {\scriptsize $0.11809^{+0.00112}_{-0.00095}$}  & {\scriptsize $0.11743^{+0.00168}_{-0.00170}$} & {\scriptsize $0.11695\pm 0.00104$}  &  {\scriptsize $0.11702^{+0.00169}_{-0.00167}$}
\\\hline
{\scriptsize$\tau$} & {{\scriptsize$0.051^{+0.005}_{-0.008}$}} & {{\scriptsize$0.053^{+0.006}_{-0.008}$}} & {{\scriptsize$0.053^{+0.006}_{-0.008}$}}  &   {{\scriptsize$0.054^{+0.007}_{-0.008}$}}
\\\hline
{\scriptsize$n_s$} & {{\scriptsize$0.9673^{+0.0039}_{-0.0044}$}}  & {{\scriptsize$0.9742^{+0.0086}_{-0.0090}$}} & {{\scriptsize$0.9705^{+0.0040}_{-0.0041}$}} &   {{\scriptsize$0.9845^{+0.0076}_{-0.0077}$}}
\\\hline
{\scriptsize$\sigma_8$}  & {{\scriptsize$0.800^{+0.006}_{-0.007}$}}  & {{\scriptsize$0.798\pm 0.014$}} & {{\scriptsize$0.798\pm 0.007$}}  &   {{\scriptsize$0.801^{+0.015}_{-0.014}$}}
\\\hline
{\scriptsize$r_s$ (Mpc)}  & {{\scriptsize$147.75^{+0.31}_{-0.35}$}}  & {{\scriptsize$145.82^{+2.33}_{-2.52}$}} & {{\scriptsize$147.88^{+0.33}_{-0.32}$}}  &   {{\scriptsize$142.55^{+1.71}_{-1.96}$}}
\\\hline
{\scriptsize $\eBD$} & - & {{\scriptsize $-0.00079^{+0.00158}_{-0.00157}$}} & - &   {{\scriptsize $-0.00081^{+0.00162}_{-0.00165}$}}
\\\hline
{\scriptsize$\varphi_{ini}$} & - & {{\scriptsize $0.976^{+0.028}_{-0.032}$}} & - &    {{\scriptsize $0.935^{+0.020}_{-0.024}$}}
\\\hline
{\scriptsize$\varphi(0)$} & - & {{\scriptsize $0.970^{+0.034}_{-0.038}$}} & - &    {{\scriptsize $0.929^{+0.028}_{-0.030}$}}
\\\hline
{\scriptsize$\weff(0)$} & - & {{\scriptsize $-0.987^{+0.016}_{-0.014}$}} & - &    {{\scriptsize $-0.971^{+0.013}_{-0.011}$}}
\\\hline
{\tiny $\dot{G}(0)/G(0) (10^{-13}yr^{-1})$} & - & {{\scriptsize $0.864^{+1.711}_{-1.734}$}} & - &    {{\scriptsize $0.913^{+1.895}_{-1.791}$}}
\\\hline
{\scriptsize$\chi^2_{min}$} & {\scriptsize 2269.04}  & {\scriptsize 2268.28} & {\scriptsize 2283.66}  &  {\scriptsize 2274.64}
\\\hline
{\scriptsize$2\ln B$} & {\scriptsize -}  & {\scriptsize $-2.94$} & {\scriptsize -}  &  {\scriptsize $+3.98$}
\\\hline
{\scriptsize$\Delta {\rm DIC}$} & {\scriptsize -}  & {\scriptsize $-3.36$} & {\scriptsize -}  &  {\scriptsize $+4.76$}
\\\hline
\end{tabular}}
\end{center}
\label{tableFit5}
\caption{\scriptsize  As in Table 3, but replacing the BOSS BAO+LSS data from \cite{Gil-Marin:2016wya} with those from \cite{Alam:2016hwk}, which only includes the spectrum information. See the discussion of the results in Sec. \ref{sec:NumericalAnalysis_BD_article}.}
\end{table}
%
%
%
\begin{equation}\label{eq:BayesRatio}
\frac{P({\rm BD-}\CC{\rm CDM}|\mathcal{D})}{P({\rm GR-}\CC{\rm CDM}|\mathcal{D})}=\frac{\mathcal{E}(\mathcal{D}|{\rm BD-}\CC{\rm CDM})}{\mathcal{E}(\mathcal{D}|{\rm GR-}\CC{\rm CDM})} \equiv B\,,
\end{equation}
where $B$ is the so-called Bayes ratio (or Bayes factor) and is the quantity we are interested in. Notice that when $B>1$ this means that data prefer the BD-$\CC$CDM model over the GR version, but of course depending on how much larger than 1 it is we will have different levels of statistical significance for such preference. It is common to adopt in the literature the so-called Jeffreys' scale to categorize the level of evidence that one can infer from the computed value of the Bayes ratio. Jeffrey's scale actually is usually written not directly in terms of $B$, but in terms of $2\ln B$. The latter is sometimes estimated with a simple Schwarz (or Bayesian) information criterion $\Delta$BIC \cite{Schwarz1978,KassRaftery1995}, although   $2\ln B$ is a much more rigorous, sophisticated (and difficult to compute) statistics than just the usual  $\Delta$BIC estimates based on using the minimum value of $\chi^2$, the number of points and the number  of independent fitting parameters. If   $2\ln B$ lies below $2$ in absolute value, then we can conclude that the evidence in favor of BD-$\CC$CDM (against GR-$\CC$CDM)  is at most only {\it weak}, and in all cases {\it not conclusive}; if $2<2\ln B<6$ the evidence is said to be {\it positive}; if, instead, $6<2\ln B<10$, then it is considered to be {\it strong}, whereas if $2\ln B>10$ one is entitled to speak of {\it very strong} evidence in favor of the BD-$\CC$CDM over the GR-$\CC$CDM model. For more technical details related with the evidence and the Bayes ratio we refer the reader to \cite{KassRaftery1995,Burnham2002,Amendola:2015ksp}. Notice that the computation of \eqref{eq:BayesRatio} is not easy in general; in fact, it can be rather cumbersome since we usually work with models with a high number of parameters, so the multiple integrals that we need to compute become quite demanding from the computational point of view. We have calculated the evidences numerically, of course, processing the Markov chains obtained from the Monte Carlo analyses carried out with \texttt{CLASS}+\texttt{MontePython} \cite{Blas:2011rf,Audren:2012wb} with the numerical code \texttt{MCEvidence} \cite{Heavens:2017afc}, which is publicly available (cf. Sec. \ref{sec:MethodData_BD_article}). We report the values obtained for $2\ln B$ \eqref{eq:BayesRatio} in Tables 3-6, 8, and 10. Table 3 contains the fitting results for the BD- and GR$-\CC$CDM models obtained with the Baseline and Baseline+$H_0$ datasets. In Table 4 we present the results for the Baseline+$H_0$+SL dataset. In Table 5 we show the output of the fitting analysis for the same models and using the same data as in Table 3, but changing the BOSS data from \cite{Gil-Marin:2016wya}, which contain both  the mater spectrum and bispectrum information, by the BOSS data from \cite{Alam:2016hwk}, which only incorporate the spectrum part (i.e. the usual matter power spectrum). In Table 6 we plug the results obtained for the BD-$\CC$CDM with the alternative datasets described in Sec. \ref{sec:MethodData_BD_article}, and in Table 7 we show the corresponding results for the GR-$\CC$CDM. Next, in Table 8 we present the results with the Baseline and Baseline+$H_0$ data configurations obtained using the GR-XCDM parametrization. In Table 9 we display the values of the parameters $\sigma_8$ and $S_8$ for the GR- and BD-$\CC$CDM models, as well as the parameter $\tilde{S}_8 = S_8/\sqrt{\varphi(0)}$ for the BD. Finally, in Table 10 we present the fitting results for the BD-$\CC$CDM model, considering the (Cassini-constrained) Scenario III described in Sec. \ref{sec:Mach_BD_article}.
\renewcommand{\arraystretch}{1.1}
\begin{table}[t!]
\begin{center}
\resizebox{1\textwidth}{!}{
\begin{tabular}{|c  |c | c |  c | c | c | c | c | c| c|}
 \multicolumn{1}{c}{} & \multicolumn{2}{c}{} & \multicolumn{1}{c}{} & \multicolumn{1}{c}{} & \multicolumn{1}{c}{} & \multicolumn{1}{c}{} & \multicolumn{1}{c}{}
\\\hline
Datasets & {\small $H_0$}  & {\small $\omega_m$} & {\small $\sigma_8$ } & {\small $r_s$ (Mpc) }  & {\small $\eBD\cdot 10^{3}$ }  & {\small $\weff(0)$}  & {\small $2\ln B$}
\\\hline
{\small B+$M$}  & {\small $71.19^{+0.92}_{-1.02}$} & {\small $0.1390^{+0.0014}_{-0.0015}$} & {\small $0.788^{+0.012}_{-0.013}$} & {\small $141.87^{+2.06}_{-1.81}$} & {\small $-2.16^{+1.42}_{-1.36}$} & {\small $-0.963^{+0.012}_{-0.011}$} & {\small $+10.38$}
\\\hline
{\small B+$H_0$+pol}  & {\small $69.85^{+0.81}_{-0.85}$} & {\small $0.1409^{+0.0012}_{-0.0011}$} & {\small $0.801\pm 0.011$} & {\small $144.72^{+1.51}_{-1.83}$} & {\small $-0.30^{+1.20}_{-1.23}$} &  {\small $-0.985^{+0.012}_{-0.009}$} & {\small $-1.44$ }
\\\hline
{\small B+$H_0$+pol+lens}  & {\small $69.74^{+0.82}_{-0.77}$} & {\small $0.1416^{+0.0011}_{-0.0010}$} & {\small $0.808\pm 0.009$} & {\small $144.66^{+1.56}_{-1.61}$} & {\small $0.00^{+1.11}_{-1.07} $} &  {\small $-0.986\pm 0.010$} & {\small $-1.98$ }
\\\hline
{\small B+$H_0$+WL}  & {\small $70.69^{+0.91}_{-0.90}$} & {\small $0.1398^{+0.0015}_{-0.0013}$} & {\small $0.794^{+0.011}_{-0.012}$} & {\small $142.76^{+1.79}_{-1.86}$} & {\small $-1.42^{+1.29}_{-1.37}$} & {\small $-0.970^{+0.011}_{-0.010}$} & {\small $+4.34$ }
\\\hline
{\small CMB+BAO+SNIa}  & {\small $68.63^{+1.44}_{-1.50}$} & {\small $0.1425\pm 0.0019$} & {\small $0.818^{+0.017}_{-0.018}$} & {\small $146.62^{+2.92}_{-2.93}$} & {\small $1.14^{+1.84}_{-1.68}$} &  {\small $-0.999^{+0.020}_{-0.017}$} & {\small $-3.00$ }
\\\hline
\end{tabular}}
\end{center}
\label{tableFit6}
\caption{\scriptsize Fitting results for the BD-$\Lambda$CDM model obtained with some alternative datasets and in all cases within the main BD Scenario II. Due to the lack of space, we employ some abbreviations, namely: {\bf B} for the Baseline dataset described in Sec. \ref{sec:MethodData_BD_article}; {\bf pol} for the Planck 2018 (TE+EE) high-$l$  polarization data; and {\bf lens} for the CMB lensing. {The $\omega_m$ parameter contains the contribution of baryons and dark matter.  In the last row, CMB refers to the TT+lowE Planck 2018 likelihood (cf.  Sec. \ref{sec:MethodData_BD_article} for details on the data).   $H_0$ is given in km/s/Mpc. For a discussion of the results, see Sec. \ref{sec:NumericalAnalysis_BD_article}}.}
\end{table}
%
%
The evidence \eqref{eq:evidence} clearly depends on the priors for the parameters entering the model. For the 6 parameters in common in the BD- and GR-$\Lambda$CDM models, namely, $(\omega_b=\Omega_b h^2,\omega_{cdm}=\Omega_{cdm}h^2,H_0,\tau,A_s,n_s)$, if we use the same flat priors in both models they cancel exactly in the computation of the Bayes ratio \eqref{eq:BayesRatio}. Thus, the latter does not depend on the priors for these parameters if their ranges are big enough so as to not alter the shape of the likelihood severely. The Bayes ratio is, though, sensitive to the priors for the two additional parameters introduced in the BD-$\CC$CDM model in our Scenario II (cf. Sec. \ref{sec:Mach_BD_article}), i.e. $\varphi_{ini}$ and $\eBD$, since they are not canceled in \eqref{eq:BayesRatio}. We study the dependence of the evidence on these priors in Fig.\,9, where we plot $2\ln B$ obtained for the BD-$\CC$CDM model from different datasets, and as a function of a quantity that we call $\Delta_X$, defined as
\begin{equation}\label{eq:DeltaX}
\Delta_X\equiv \frac{\Delta\eBD}{10^{-2}}=\frac{\Delta\varphi_{ini}}{0.2}\,,
\end{equation}
with $\Delta\eBD$ and $\Delta\varphi_{ini}$ being the flat prior ranges of $\eBD$ and $\varphi_{ini}$, centered at $\eBD=0$ and $\varphi_{ini}=1$, respectively. $\Delta_X$ will be equal to one when $\Delta\eBD=10^{-2}$ and $\Delta\varphi_{ini}=0.2$, which are natural values for these prior ranges. The former implies $\oD>100$ and the latter could be associated to the range $0.9<\varphi_{ini}<1.1$. We do not expect $\oD\lesssim 100$ since this would imply an exceedingly large departure from GR, even at cosmological scales, where this lower bound was already set using the first releases of WMAP CMB data almost twenty years ago, see e.g. \cite{Nagata:2003qn,Acquaviva:2004ti}. Regarding the prior range $0.9<\varphi_{ini}<1.1$, it is also quite natural, since this is necessary to satisfy the BBN bounds \cite{Uzan:2010pm}. In all tables we report the values of $2\ln B$ obtained by setting the natural value $\Delta_X=1$, and in Fig.\,9, as mentioned before, we also show how this quantity changes with the prior width, in terms of the variable $\Delta_X$ \eqref{eq:DeltaX}.
\newline
In the Cassini-constrained Scenario III (cf. again Sec. \ref{sec:Mach_BD_article}), we also allow variations of $\varphi_{ini}$ and $\eBD$ in our Monte Carlo runs, of course, but the natural prior range for $\eBD$ is now much smaller than in Scenario II, since now we expect it to be more constrained by the local observations. It is more natural to take in this case a prior range $\Delta\eBD=5\cdot 10^{-5}$ (still larger than Cassini's bound), and this is what we do in all the analyses of this scenario. See the comments in Sec. \ref{sec:Discussion_BD_article}, and Table 10.
\newline
In Tables 3-5 apart from the Bayes ratio, we also include the Deviance Information Criterion \cite{DIC}, which is strictly speaking an approximation of the exact Bayesian approach that works fine when the posterior distributions are sufficiently Gaussian. The DIC is defined as
\begin{equation}
{\rm DIC}=\chi^2(\hat{\theta})+2p_D\,.
\end{equation}
Here $p_D=\overline{\chi^2}-\chi^2(\hat{\theta})$ is the effective number of parameters of the model and $\overline{\chi^2}$ the mean of the overall $\chi^2$ distribution. DIC is particularly suitable for us, since we can easily compute all the quantities involved directly from the Markov chains generated with \texttt{MontePython}. To compare the ability of the BD- and GR-$\Lambda$CDM models to fit the data, one has to compute the respective differences of DIC values between the first and second models. They are denoted $\Delta$DIC in our tables, and this quantity is the analogous to $2\ln B$.
\subsubsection{Comparing with the  XCDM parametrization}
\renewcommand{\arraystretch}{1.1}
\begin{table}[t!]
\begin{center}
\resizebox{1\textwidth}{!}{
\begin{tabular}{|c  |c | c |  c | c | c |}
 \multicolumn{1}{c}{} & \multicolumn{2}{c}{} & \multicolumn{1}{c}{} & \multicolumn{1}{c}{}
\\\hline
{\scriptsize Datasets} & {\small $H_0$}  & {\small $\omega_m$} & {\small $\sigma_8$ } & {\small $r_s$ (Mpc) }
\\\hline
{\small B+$M$} & {\small $68.64^{+0.39}_{-0.38}$} & {\small $0.1398\pm 0.0009$} & {\small $0.796^{+0.005}_{-0.006}$} & {\small $147.87^{+0.29}_{-0.30}$}
\\\hline
{\small B+$H_0$+pol}  & {\small $68.50^{+0.33}_{-0.36}$} & {\small $0.1408^{+0.0007}_{-0.0008}$} & {\small $0.799\pm 0.006$} & {\small $147.53\pm 0.022$}
\\\hline
{\small B+$H_0$+pol+lens}  & {\small $68.38^{+0.35}_{-0.33}$} & {\small $0.1411\pm 0.0007$} & {\small $0.803\pm0.006$} & {\small $147.44^{+0.22}_{-0.21}$}
\\\hline
{\small B+$H_0$+WL}  & {\small $68.64\pm 0.37$} & {\small $0.1399^{+0.0008}_{-0.0009}$} & {\small $0.795^{+0.005}_{-0.007}$} & {\small $147.89^{+0.31}_{-0.29}$}
\\\hline
{\small CMB+BAO+SNIa}  & {\small $67.91^{+0.39}_{-0.41}$} & {\small $0.1413\pm 0.0009$} & {\small $0.805^{+0.006}_{-0.007}$} & {\small $147.66^{+0.31}_{-0.29}$}
\\\hline\hline
{\small B+$H_0$ (No LSS)}  & {\small $68.38^{+0.42}_{-0.37}$} & {\small $0.1405\pm 0.0009$} & {\small $0.802^{+0.007}_{-0.008}$} & {\small $147.76\pm 0.30$}
\\\hline
{\small Dataset \cite{Ballesteros:2020sik}}  & {\small $68.17^{+0.43}_{-0.44}$} & {\small $0.1416\pm 0.0009$} & {\small $0.810^{+0.006}_{-0.007}$} & {\small $147.35^{+0.22}_{-0.24}$}
\\\hline
{\small Dataset \cite{Ballesteros:2020sik} + LSS (SP)}  & {\small $68.36\pm 0.42$}  & {\small $0.1412\pm 0.0009$} & {\small $0.806\pm 0.006$} & {\small $147.43\pm $0.23}
\\\hline
\end{tabular}}
\end{center}
\label{tableFit7}
\caption{\scriptsize Different fitting results for the GR-$\Lambda$CDM model. The first five rows correspond to the different non-baseline datasets explored for the  BD-$\CC$CDM  in Table 6. The last three rows correspond  to other scenarios tested with the BD-$\CC$CDM model in Table 10, see  Sec.\,\ref{sec:Discussion_BD_article} for more details. $H_0$ is given in km/s/Mpc.}
\end{table}
%
As mentioned, in our numerical analysis of the data we also wish to consider the effect of a simple but powerful DDE parametrization, which is the traditional XCDM\cite{Turner:1998ex}. In this very simple framework, the DE is self-conserved and is associated to some unspecified entity or fluid (called X) which exists together with ordinary baryonic and cold dark matter, but it does not have any interaction with them.  The energy density of X is simply given by $\rho_X(a) = \rho_{X0}a^{-3(1+w_0)}$,   $\rho_{X0}=\rho_\Lambda$ being the current DE density value and $w_0$ the (constant) EoS parameter of such fluid. More complex parametrizations of the EoS can be considered, for instance the CPL one\,\cite{Chevallier:2000qy,Linder:2002et}, in which there is a  time evolution of the EoS itself. However, we have previously shown its incapability to improve the XCDM performance in solving the two tensions, see\,\,\cite{Sola:2018sjf}. Thus, in this work we prefer to stay as closer as possible to the standard cosmological model and we will limit ourselves to analyze the XCDM only. By setting $w_0=-1$ we retrieve the $\Lambda$CDM model with constant $\rho_\Lambda$. For $w_0 \gtrsim-1$ the XCDM mimics quintessence, whereas for $w_0 \lesssim -1$ it mimics phantom DE. The fitting results generated from the XCDM on our datasets are used in our analysis as a figure of merit or benchmark to compare with the corresponding fitting efficiency of both the BD-$\CC$CDM and the GR-$\CC$CDM models. In the next section, we comment on the comparison.
%



\setlength{\tabcolsep}{0.7em}
\renewcommand{\arraystretch}{1.1}
\begin{table}[t!]
\begin{center}
\resizebox{0.7\textwidth}{!}{

\begin{tabular}{|c  |c | c |  c |}
 \multicolumn{1}{c}{} & \multicolumn{2}{c}{GR-XCDM}
\\\hline
{\normalsize Parameter} & {\normalsize Baseline}  & {\normalsize Baseline+$H_0$}
\\\hline
{\normalsize  $H_0$ (km/s/Mpc) }  & {\normalsize  $67.34^{+0.63}_{-0.66}$ } & {\normalsize $68.40^{+0.60}_{-0.62}$}
\\\hline
{\normalsize$\omega_b$} & {\normalsize   $0.02235^{+0.00021}_{-0.00020}$}  & {\normalsize $0.02239^{+0.00019}_{-0.00020}$}
\\\hline
{\normalsize$\omega_{cdm}$} & {\normalsize $0.11649^{+0.00108}_{-0.00111}$}  & {\normalsize $0.11671^{+0.00117}_{-0.00109}$}
\\\hline
{\normalsize$\tau$} & {{\normalsize$0.053^{+0.006}_{-0.008}$}} & {{\normalsize$0.051^{+0.005}_{-0.008}$}}
\\\hline
{\normalsize$n_s$} & {{\normalsize$0.9709\pm 0.0043$}}  & {{\normalsize$0.9707^{+0.0042}_{-0.0043}$}}
\\\hline
{\normalsize$\sigma_8$}  & {{\normalsize$0.782^{+0.011}_{-0.010}$}}  & {{\normalsize$0.792\pm 0.011$}}
\\\hline
{\normalsize$r_s$ (Mpc)}  & {{\normalsize$148.05^{+0.32}_{-0.34}$}}  & {{\normalsize$147.95^{+0.33}_{-0.34}$}}
\\\hline
{\normalsize$w_0$}  & {{\normalsize$-0.956\pm0.026$}}  & {{\normalsize$-0.991^{+0.026}_{-0.024}$}}
\\\hline
{\normalsize$\chi^2_{min}$} & {\normalsize 2269.88}  & {\normalsize 2285.22}
\\\hline
{\normalsize$2\ln B$} & {\normalsize $-2.23$}  & {\normalsize $-5.21$}
\\\hline
\end{tabular}}
\end{center}
\label{tableFit8}
\caption{\scriptsize As in Table 3, but for the XCDM parametrization (within GR). Motivated by previous works (see e.g. \cite{Sola:2018sjf}), we have used the (flat) prior $-1.1<w_0<-0.9$.}
\end{table}

\vspace{0.3cm}

\subsection{Discussion and extended considerations}\label{sec:Discussion_BD_article}
In this chapter, we have dealt with Brans-Dicke  (BD) theory in  extenso. Our main goal was to assess if BD-gravity can help to smooth out the main two tensions besetting the usual concordance $\CC$CDM model (based on GR): i) the $H_0$-tension (the most acute existing discordance at present), and ii) the $\sigma_8$-tension, which despite not being so sharp it often occurs that the (many) models in the market dealing with the former tend to seriously  aggravate the latter.  As we have explained at the beginning,  the  `golden rule' to be preserved by the tension solver should be to find a clue on how to tackle the main discrepancy (on the local $H_0$ parameter)  while at the same time to curb the $\sigma_8$ one, or at least not to worsen it. We have found that BD-gravity could be a key paradigm capable of such achievement.  Specifically, we have considered the original BD model with the only addition of a cosmological constant (CC), and we have performed a comprehensive analysis  in the light of a rich and updated set of observations. These involve a large variety of experimental inputs of various kinds, such as  the long chain SNIa+$H(z)$+BAO+LSS+CMB of data sources, which we have considered  at different levels and combinations; and tested with the inclusion of other potentially important factors such as the influence of the bispectrum component in the structure formation data (apart from the ordinary power spectrum); and also assessed the impact of  gravitational lensing data of different sorts (Weak and Strong Lensing).
\newline
Although  BD-gravity is fundamentally different from GR, we have found very useful to try to pick out possible measurable signs of the new gravitational paradigm by considering the two frameworks in the (spatially flat) FLRW metric and compare the  versions of the $\CC$CDM model resulting in each case, which we have called BD-$\CC$CDM and GR-$\CC$CDM, respectively. We have parametrized the departure of the former from the latter at the background level  (cf. Sec.\,\ref{sec:EffectiveEoS_BD_article}) and we have seen that BD-$\CC$CDM can appear in the form of a dynamical dark energy (DDE) version of the GR-$\CC$CDM, in which the vacuum energy density is evolving through a non-trivial EoS  (cf. Fig.\,8).  We have called it  the `GR-picture' of the BD theory.  The resulting effective behaviour is  $\CC$CDM-like with, however, a mild time-evolving  quasi-vacuum component. In fact, such behaviour is not `pure vacuum' -- which is why we call it quasi-vacuum --  despite of the fact that the original BD-$\CC$CDM theory possesses a rigid cosmological constant.  Specifically, using the numerical fitting results of our analysis we find that such EoS  shows up in effective quintessence-like form at more than $3\sigma$ c.l. (this is perfectly appreciable at naked eye in Fig. 8). Our fit to the data demonstrates that such an effective  representation of BD-gravity can be competitive with the concordance model with a rigid $\CC$-term. It may actually create the fiction that the DE is dynamical when viewed within the GR framework, whilst it is  just a rigid CC in the underlying  BD action.  The practical outcome is that the BD approach with a CC definitely  helps to smooth out some of the tensions afflicting the $\CC$CDM in a manner very similar to the Running Vacuum Model, see e.g.\,\cite{Sola:2016hnq,Sola:2016ecz,Sola:2017znb,Sola:2017jbl,Gomez-Valent:2017idt,Gomez-Valent:2018nib}, and this success might ultimately reveal the signs of the  BD theory. We conclude that finding  traces of vacuum dynamics, accompanied with apparent deviations from the standard matter conservation law\,\cite{Fritzsch:2012qc} could be the `smoking gun' pointing to the possibility that the gravity theory sitting behind these effects is not GR but BD.

\renewcommand{\arraystretch}{1.1}
\begin{table}[t!]
\begin{center}
\resizebox{1\textwidth}{!}{

\begin{tabular}{|c  |c | c |  c |c|c|c|}
 \multicolumn{1}{c}{} & \multicolumn{2}{c}{}
\\\hline
{\normalsize Scenarios} & {\normalsize ${\sigma}_{8(\rm GR)}$} & {\normalsize ${\sigma}_{8(\rm BD)}$} & {\normalsize ${S}_{8(\rm GR)}$} & {\normalsize ${S}_{8(\rm BD)}$}  & {\normalsize $\tilde{S}_{8(\rm BD)}$}
\\\hline
{\normalsize Baseline } & {\normalsize $0.797^{+0.005}_{-0.006}$} & {\normalsize $0.785\pm 0.013$} & {\normalsize $0.800^{+0.010}_{-0.011}$} & {\normalsize $0.777^{+0.019}_{-0.020}$} & {\normalsize $0.793\pm 0.012$}
\\\hline
{\normalsize Baseline+$H_0$} & {\normalsize $0.796^{+0.006}_{-0.007}$} & {\normalsize $0.789\pm 0.013$} & {\normalsize $0.793^{+0.011}_{-0.010}$} & {\normalsize $0.758\pm 0.018$} & {\normalsize $0.792\pm 0.013$}
\\\hline
{\normalsize Baseline+$H_0$+SL} & {\normalsize $0.795^{+0.006}_{-0.007}$} & {\normalsize $0.789\pm 0.013$}& {{\normalsize$0.789^{+0.011}_{-0.010}$}} & {\normalsize $0.753^{+0.017}_{-0.018}$}  & {{\normalsize$0.791^{+0.012}_{-0.013}$}}
\\\hline
{\normalsize Spectrum}  & {\normalsize $0.800^{+0.006}_{-0.007}$}& {\normalsize $0.798\pm 0.014$} &{\normalsize $0.807\pm 0.013$} & {\normalsize $0.793^{+0.021}_{-0.022}$} & {\normalsize $0.805\pm 0.014$}
\\\hline
{\normalsize Spectrum+$H_0$} & {\normalsize $0.798\pm 0.007$} & {\normalsize $0.801^{+0.015}_{-0.014}$} & {{\normalsize$0.794^{+0.012}_{-0.013}$}} & {\normalsize $0.773^{+0.019}_{-0.021}$} & {{\normalsize$0.802^{+0.013}_{-0.014}$}}
\\\hline
\end{tabular}}
\end{center}
\label{tableFit9}
\caption{\scriptsize Fitted values for the $\sigma_{8(M)}$ (here M stands for GR or BD) obtained under different dataset configurations. We also include the derived values of $S_{8(M)}$ = $\sigma_8\sqrt{\Omega_m/0.3}$ and the renormalized $\tilde{S}_8$ = $\sigma_8\sqrt{\tilde{\Omega}_m/0.3}$, with $\tilde{\Omega}_m$ defined as in Sec.\,\ref{sec:rolesvarphiH0_BD_article}. These results correspond to the main Scenario II of the BD-$\CC$CDM model.}
\end{table}

%
\subsubsection{Alleviating the $H_0$-tension}\label{sec:H0-tension_BD_article}
Our analysis of the BD theory with a CC, taken at face value,  suggests that the reason for the enhancement of $H_0$ in the BD model is because the effective gravitational coupling acting at cosmological scales, $\Geff\sim G_N/\varphi$, is higher than the one measured on Earth (see Fig.\,4). This possibility allows the best-fit current energy densities of all the species to remain compatible at $\lesssim 1\sigma$ c.l. with the ones obtained in the GR-$\CC$CDM model (cf. e.g. Tables 3-5).  
Thus, since $\varphi<1$  we find that the increase of the Hubble parameter is basically due to the increase of  the effective $G$, and there is no need for a strong modification of the energy densities of the various species filling the Universe. This is a welcome feature since  the measurable cosmological mass parameter in the BD-$\CC$CDM model is, for sufficiently small $\eBD$,  not the usual $\Omega_m$, but precisely the tilded one, related to it through $\tilde{\Omega}_m=\Omega_m/\varphi$.  The latter   is about   $\sim  8-9\%$  bigger than its standard model counterpart  ($\tilde{\Omega}_m> \Omega_m$)   as it follows from  Fig. \,4, where we can read off the current value  $\varphi(z=0)\simeq 0.918$.  Now, because at the background level  it is possible to write an approximate  Friedmann's equation \eqref{eq:H2}  in terms of the tilded parameters, these are indeed the ones that are actually  measured from SNIa and BAO observations in the BD context\footnote{Recall that for $\eBD\neq0$ the tilded  parameters $\tilde{\Omega}_i$  (which were originally defined for $\eBD=0$) receive a correction and become the hatted parameters $\hat{\Omega}_i$ introduced in Eq.\,\eqref{eq:hatOmega}.  However, the difference between them is of ${\cal O}(\eBD)$, see Eq.\,\eqref{eq:hatOmega2}, and since $|\eBD|\sim \mathcal{O}(10^{-3})$ it can be ignored.}. The differences, however, as we have just pointed out, are not to be attributed to a change in the physical energy content of matter but to the fact that $\varphi<1$  throughout the entire cosmic history, as clearly shown in Fig.\, 4.  Obviously, the measurement of the parameters $\tilde{\Omega}_i$  can be performed  through the very same data  and procedures well accounted for  in the context of the GR-$\CC$CDM framework.  This explanation is perfectly consistent with the fact that when the Friedmann's law is expressed in terms of the effective $G$, as indicated  in Eq.\,\eqref{eq:H1}, the local value of the Hubble parameter $H_0$ becomes bigger owing to $\Geff=G_N/\varphi$ being bigger than $G_N$.  Thus,  when we compare the early and local measurements of $H_0$ we do not meet  any anomaly in this approach.
\newline
We also recall at this point that there is no correction from $\omega_{\rm BD}$  on the effective coupling $\Geff$, Eq.\,\eqref{eq:MainGeffective},  in the local domain. This is because in our context  $\omega_{\rm BD}$ appears as being very large owing to the assumed screening of the BD-field caused by the clustered matter (cf. Sec.\,\ref{sec:Mach_BD_article}).  From Fig.\,4 and Table 3 we find that the BD model leads to a value of  $H_0$ a factor  $\Geff^{1/2}(z=0)/G_N^{1/2}\sim 1/\varphi^{1/2}(z=0)=1/\sqrt{0.918}$, i.e. $\sim 4.5\%$, bigger than the one inferred from the CMB in the GR-$\CC$CDM model, in which $\Geff=G_N\,(\forall{z})$. It is reassuring to realize  that such a `renormalization factor'  can  enhance the low Planck 2018  CMB measurement of the  Hubble parameter (viz. $H_0=67.4\pm 0.5$ km/s/Mpc\,\cite{Aghanim:2018eyx})  up to the range of  $70-71$km/s/Mpc (cf. e.g. Tables 3-6), hence  much closer to the local measurements. For example,   SH0ES  yields $ H_0= (73.5\pm 1.4)$ km/s/Mpc\,\cite{Reid:2019tiq}; and when the latter is combined  with Strong Lensing data from the H$0$LICOW collab.\cite{Wong:2019kwg}  it leads to  $ H_0= (73.42\pm 1.09)$ km/s/Mpc.  This combined  value  is $5\sigma$ at odds with the Planck 2018 measurement, a serious tension.

\renewcommand{\arraystretch}{1.1}
\begin{table}[t!]
\begin{center}
\resizebox{1\textwidth}{!}{

\begin{tabular}{|c  |c | c |  c | c | c  |}
 \multicolumn{1}{c}{} & \multicolumn{4}{c}{BD-$\Lambda$CDM (Scenario III: Cassini-constrained)}
\\\hline
{\scriptsize Parameter} & {\scriptsize B+$H_0$ (No LSS)}  & {\scriptsize B+$H_0$ } & {\scriptsize Dataset \cite{Ballesteros:2020sik}}  &  {\scriptsize Dataset \cite{Ballesteros:2020sik} + LSS (SP)}
\\\hline
{\scriptsize $H_0$ (km/s/Mpc)}  & {\scriptsize $70.99^{+0.94}_{-0.97}$} & {\scriptsize $70.80^{+0.81}_{-0.91}$} & {\scriptsize $70.01^{+0.86}_{-0.92}$}  & {\scriptsize $70.03^{+0.90}_{-0.88}$}
\\\hline
{\scriptsize$\omega_b$} & {\scriptsize $0.02257\pm 0.00021$}  & {\scriptsize $0.02256^{+0.00019}_{-0.00020}$} & {\scriptsize $0.02271\pm 0.00016$}  &  {\scriptsize $0.02272^{+0.00015}_{-0.00016}$}
\\\hline
{\scriptsize$\omega_{cdm}$} & {\scriptsize $0.11839^{+0.00093}_{-0.00094}$}  & {\scriptsize $0.11748\pm 0.00089$} & {\scriptsize $0.11885^{+0.00092}_{-0.00095}$}  &  {\scriptsize $0.11827^{+0.00089}_{-0.00093}$}
\\\hline
{\scriptsize$\tau$} & {\scriptsize $0.057^{+0.007}_{-0.008}$} & {\scriptsize$0.050^{+0.004}_{-0.008}$} & {\scriptsize$0.061^{+0.007}_{-0.008}$}  &   {\scriptsize$0.058^{+0.006}_{-0.008}$}
\\\hline
{\scriptsize$n_s$} & {\scriptsize $0.9824^{+0.0057}_{-0.0058}$}  & {\scriptsize$0.9811^{+0.0051}_{-0.0052}$} & {\scriptsize$0.9783^{+0.0052}_{-0.0059}$} &   {\scriptsize$0.9701^{+0.0056}_{-0.0054}$}
\\\hline
{\scriptsize$\sigma_8$}  & {\scriptsize $0.815^{+0.008}_{-0.009}$}  & {\scriptsize$0.804^{+0.006}_{-0.007}$} & {\scriptsize$0.817\pm 0.007$}  &   {\scriptsize$0.812^{+0.006}_{-0.007}$}
\\\hline
{\scriptsize$r_s$ (Mpc)}  & {\scriptsize $142.14^{+1.91}_{-1.72}$}  & {\scriptsize$143.31^{+1.72}_{-1.63}$} & {\scriptsize$143.58^{+1.62}_{-1.55}$}  &   {\scriptsize$144.10^{+1.62}_{-1.52}$}
\\\hline
{\scriptsize $\eBD$} & {\scriptsize $-0.00002\pm 0.00002$} & {\scriptsize $-0.00002\pm 0.00002$} & {\scriptsize $-0.00002\pm 0.00002$} &  {\scriptsize $-0.00002\pm 0.00002$}
\\\hline
{\scriptsize$\varphi_{ini}$} & {\scriptsize $0.933\pm 0.021$} & {\scriptsize $0.944\pm 0.020$} &{\scriptsize $0.955^{+0.018}_{-0.019}$}&    {\scriptsize $0.960^{+0.020}_{-0.018}$}
\\\hline
{\scriptsize$\varphi(0)$} & {\scriptsize $0.933^{+0.020}_{-0.021}$} & {\scriptsize $0.944^{+0.019}_{-0.020}$} &{\scriptsize $0.955^{+0.018}_{-0.019}$}     &{\scriptsize $0.960^{+0.020}_{-0.017}$}
\\\hline
{\scriptsize$\weff(0)$} & {\scriptsize $-0.972\pm 0.009$} & {\scriptsize $-0.977\pm 0.008$} & {\scriptsize $-0.981^{+0.008}_{-0.007}$} &    {\scriptsize $-0.983\pm 0.008$}
\\\hline
{\tiny $\dot{G}(0)/G(0) (10^{-13}yr^{-1})$} & {\scriptsize $0.025^{+0.025}_{-0.026}$} & {\scriptsize $0.026^{+0.027}_{-0.028}$} & {\scriptsize $0.023^{+0.026}_{-0.027}$} &   {\scriptsize $0.020\pm 0.026$}
\\\hline
{\scriptsize$\chi^2_{min}$} & {\scriptsize 2256.14}  & {\scriptsize 2278.34} & {\scriptsize 2797.44}  &  {\scriptsize 2812.68}
\\\hline
{\scriptsize$2\ln B$} & {\scriptsize +9.03}  & {\scriptsize $+5.21$} & {\scriptsize +3.45}  &  {\scriptsize +2.21}
\\\hline
\end{tabular}}
\end{center}
\label{tableFit10}
\caption{\scriptsize Fitting results for the BD-$\CC$CDM, in the context of the BD-Scenario III explained in Sec. \ref{sec:Mach_BD_article} under different datasets.  As characteristic of Scenario III, in all of these datasets the Cassini constraint on the post-Newtonian parameter $\gamma^{\rm PN}$ has been imposed \cite{Bertotti:2003rm}. In the first two fitting columns we use the Baseline+$H_0$ dataset described in Sec. \ref{sec:MethodData_BD_article}. However, we exclude the LSS data in the first column while it is kept in the second. In the third and fourth fitting columns we report on the results obtained using the very same dataset as in Ref.\,\cite{Ballesteros:2020sik}, just to ease the comparison between the BD-$\CC$CDM and the variable $G$ model studied in that reference (cf. their Table 1). This dataset includes the Planck 2018 TTTEEE+lensing likelihood \cite{Aghanim:2018eyx}, BAO data from \cite{Beutler:2011hx,Ross:2014qpa,Alam:2016hwk} and the SH0ES prior from \cite{Riess:2019cxk}. In the last fitting column, however, we add the LSS data to the previous set but with no bispectrum (cf. Table 2 and Sec. \ref{sec:MethodData_BD_article}). The corresponding results for the  GR-$\CC$CDM can be found in Tables 3 and 7.}
\end{table}


\begin{figure}[t!]
\begin{center}
\label{fig:triangular_BD_article}
\includegraphics[width=6.5in, height=6.5in]{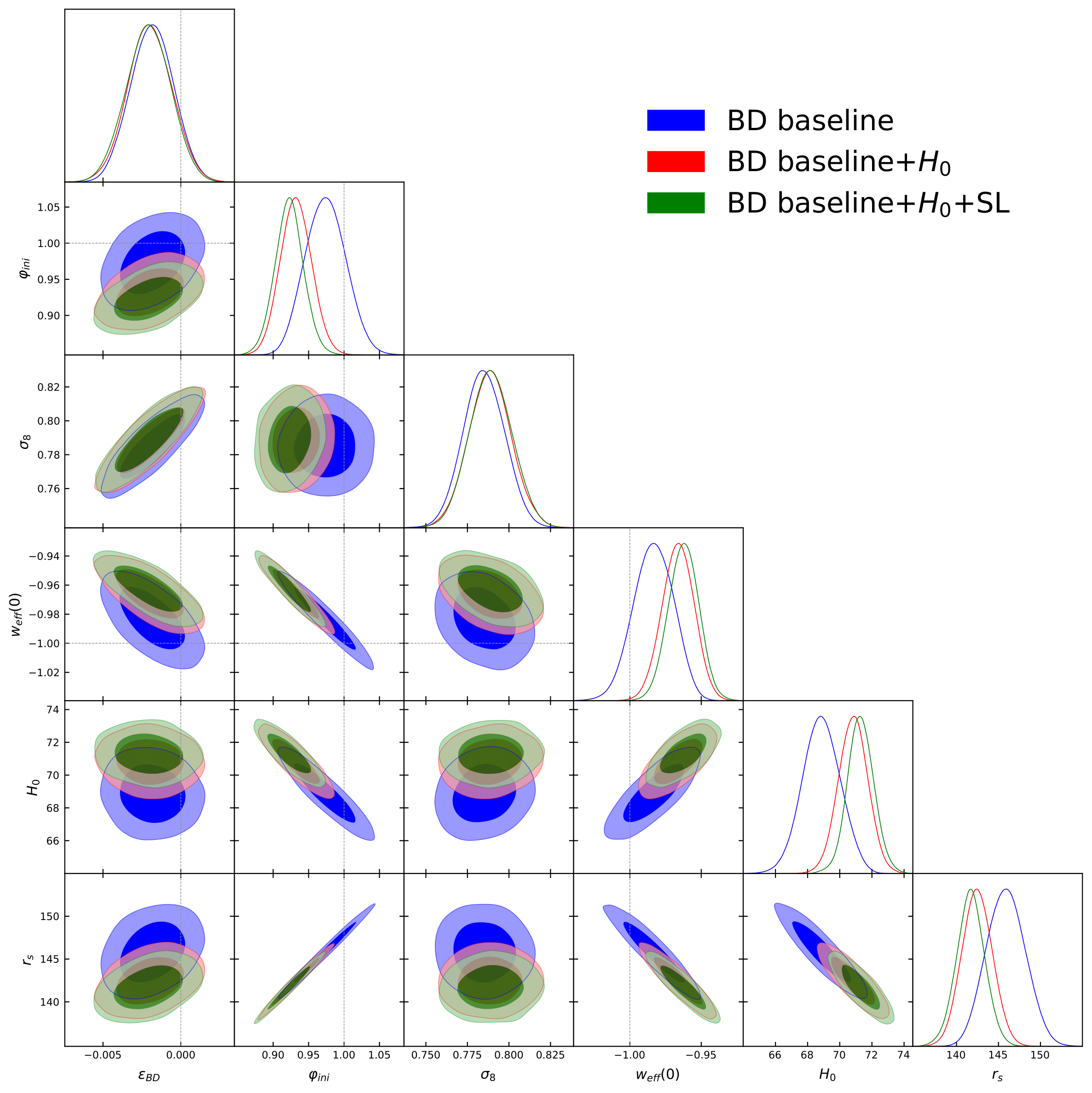}
\caption{\scriptsize{Triangular matrix containing the two-dimensional marginalized distributions for some relevant combinations of parameters in the BD-$\CC$CDM model (at $1\sigma$ and $2\sigma$ c.l.), together with the corresponding one-dimensional marginalized likelihoods for each of them. $H_0$ is expressed in km/s/Mpc, and $r_s$ in Mpc. See Tables 3 and 4 for the numerical fitting results.}}
\end{center}
\end{figure}
%
On the other hand, if we compare e.g. the fitting value predicted within the BD-$\CC$CDM model from Table 3 (namely $H_0=70.83^{+0.92}_{-0.95}$ km/s/Mpc) with the aforementioned  SH0ES determination, we can see that the difference is of only $1.58\sigma$. If we next compare our fitting result from Table\,4  ($H_0= 71.30^{+0.80}_{-0.84}$ km/s/Mpc), which incorporates the H$0$LICOW Strong Lensing data in the fit as well, with the combined SH0ES  and H$0$LICOW result (viz. the one which is in $5\sigma$ tension with the CMB value)  we obtain once more an inconspicuous tension of only  $1.55\sigma$.  In either case  it is far away from any perturbing discrepancy.  In fact, no discrepancy which is not reaching a significance of at least  $3\sigma$ can be considered sufficiently worrisome.
\subsubsection{Alleviating the $\sigma_8$-tension}\label{sec:s8-tension_BD_article}
Furthermore, the smoothing of the tension applies to the $\sigma_8$ parameter as well, with the result that it essentially disappears within a similar level of inconspicuousness. In fact, values such as $\sigma_8=0.789\pm 0.013$ and $\tilde{S}_8=0.792\pm 0.013$ (obtained within the Baseline$+H_0$ dataset, see Table 9) are in good agreement with weak gravitational lensing observations derived from shear data (cf. the WL data block mentioned in Sec. \ref{sec:MethodData_BD_article}). Let us take  the value by Joudaki et al. 2018 of the combined observable $S_8 = 0.742\pm 0.035$, for example,  obtained by KiDS-450, 2dFLenS and BOSS collaborations from a joint analysis of weak gravitational lensing tomography and overlapping redshift-space galaxy clustering\,\cite{Joudaki:2017zdt}.  These observations can be compared  with our prediction for ${S}_8=\sigma_8\sqrt{\Omega_m/0.3}$  and  with the `renormalized' form of that quantity within the BD-$\CC$CDM model, namely $\tilde{S}_8=\sigma_8\sqrt{\tilde{\Omega}_m/0.3}$,  which depends on the modified cosmological parameter $\tilde{\Omega}_m$, which is slightly higher,  recall  our Eq.\,\eqref{eq:tildeOmegues}\footnote{Although we could use the hatted parameter $\hat{S}_8=\sigma_8\sqrt{\hat{\Omega}_m/0.3}$, instead of  $\tilde{S}_8$, we have already pointed out  that the difference between $\hat{\Omega}_m$ and $\tilde{\Omega}_m$ is negligible for $|\eBD|\sim \mathcal{O}(10^{-3})$, and so is between $\hat{S}_8$ and  $\tilde{S}_8$.}. Both ${S}_8$ and $\tilde{S}_8$ are displayed together in Table 9 for the main scenarios, also in company with $\sigma_8$ values for the GR and BD models.  Differences of the mentioned experimental measurements with respect to e.g. our prediction for the Baseline$+H_0$ dataset,   are at the level of  $0.5\sigma-1.3\sigma$ depending on whether we use $S_8$ or $\tilde{S}_8$, whence statistically irrelevant in any case. More details can be appraised on some of  these observable and their correlation with $H_0$ in Figs. 10 and 11, on which we shall further comment later on.

\begin{figure}[t!]
\begin{center}
\label{fig:H0S8_BD_article}
\includegraphics[width=4.3in, height=3.8in]{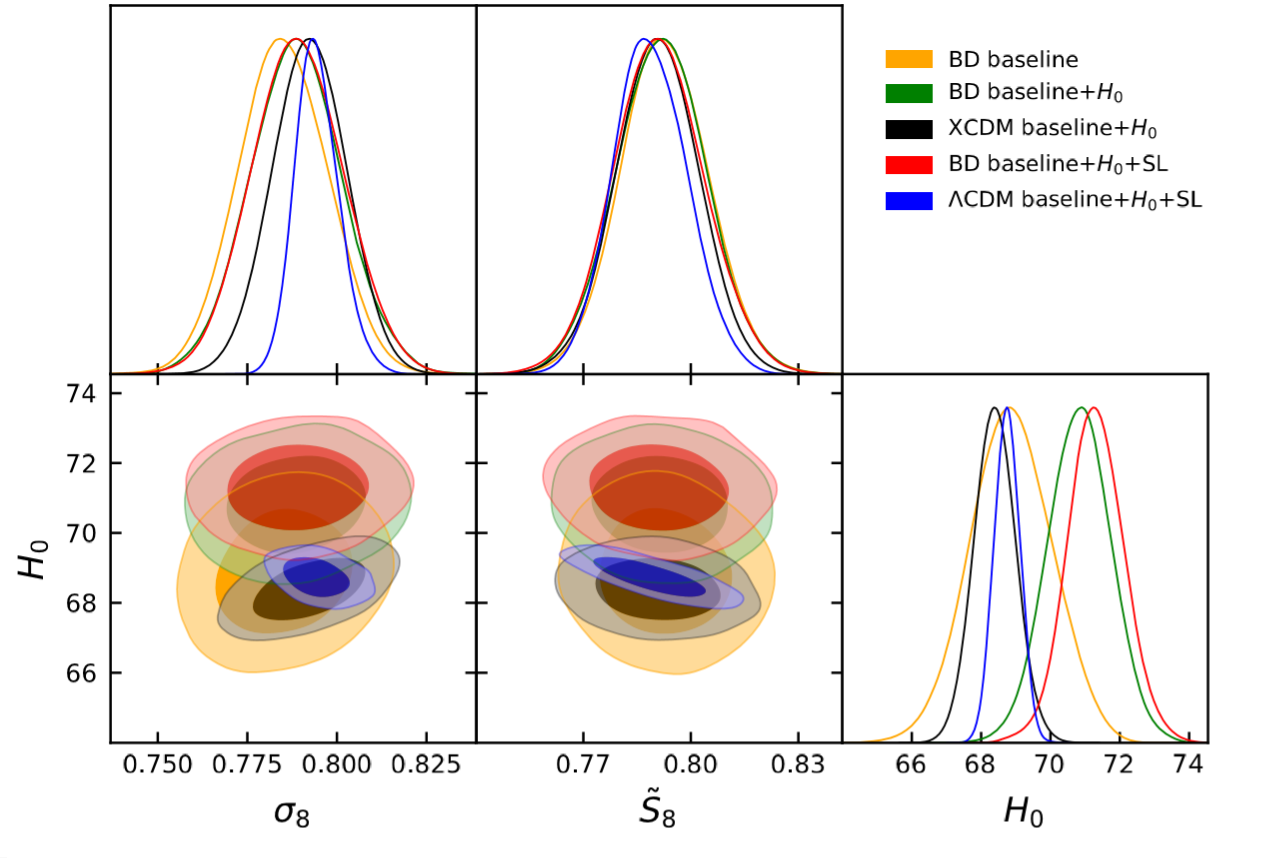}
\caption{\scriptsize {Constraints obtained for $\sigma_8$ and $\tilde{S}_8$ versus $H_0$ (in km/s/Mpc) from the fitting analyses of the GR- and BD-$\CC$CDM models, and the GR-XCDM parametrization. We show both, the contour lines in the corresponding planes of parameter space, and the associated marginalized one-dimensional posteriors. The centering of the parameters in the ranges $\sigma_8<0.80$ and $H_0\gtrsim71$ km/s/Mpc is a clear sign of the smoothening of the $\sigma_8$-tension and, more conspicuously, of the $H_0$ one within the BD-$\CC$CDM model. We can also see that while a simple XCDM parametrization for the DE can help to diminish $\sigma_8$ as compared to the concordance model, it is however unable to improve the $H_0$ tension, which is kept at a similar level as within the concordance model.}}
\end{center}
\end{figure}
%
\subsubsection{Comparing different scenarios}\label{sec:Others_BD_article}
We have also tested the performance of the BD- and GR-$\CC$CDM models when we include the CMB high-$l$ (TE+EE) polarization data from Planck 2018, with and without the CMB lensing, in combination with the baseline dataset and the SH0ES prior on $H_0$ (cf. Tables 6 and 7). The values of the Hubble parameter  in these cases are a little bit lower than when we consider only the temperature and low-$l$ polarization (TT+lowE) CMB data, but the tension is nevertheless significantly reduced, being now of only $\sim 2.2\sigma$ c.l., whereas in the GR-$\CC$CDM model it is kept at the $\sim 3.5\sigma$ level. The values of $\sigma_8$ are still low, $\sim 0.80-0.81$. The information criteria in these cases, though, have no preference for any of the two models, they are not conclusive.
\newline
We also examine the results that are obtained when we do not include in our fitting analyses any of the data sources that trigger the tensions. We consider here the CMB, BAO and SNIa datasets (denoted as CMB+BAO+SNIa in Tables 6 and 7), but exclude the use of the SH0ES prior on $H_0$ and the LSS data.  As expected,  the evidence for the BD model decreases since now we do not give to it the chance of showing its power.  Even though the description of the data improves,  it is not enough to compensate for the penalty received owing to the use of the two additional parameters $(\eBD,\varphi_{ini})$,  and in this case we read  $2\ln B=-3.00$ (from Table 6). Thus, there is here a marginal preference for the GR scenario, but as previously mentioned, this is completely normal, since we are removing precisely the data sources whose correct description demands for new physics. Even so, the $H_0$-tension is again remarkably reduced from  $3.8\sigma$  in  GR-$\CC$CDM to only  $2.4\sigma$  in the BD-$\CC$CDM. The respective values of $\sigma_8$ remain compatible with $0.80$  within  $\sim 1\sigma$.
\newline
It is also interesting to compare the results that we have obtained within the BD framework  with other approaches in the literature,  in which the variation of $G$ is dealt with as a small departure from GR, namely in a context where the action still contains a large mass scale $M$ near the Planck mass  $m_{\rm Pl}$, together with some scalar field which parametrizes the deviations from it. This is of course fundamentally different from the BD paradigm but it bears relation owing to the variation of the effective $G$, and it has also been used to try to smooth out the tensions. However, as already mentioned at the beginning, and also in Sec. \ref{sec:preview_BD_article}, it is not easy at all for a given model to fulfill the `golden rule', i.e. to loosen the two tensions at a time, or just to alleviate one of them without worsening the other.  Different proposals have appeared in the market trying to curb the $H_0$-tension, e.g. the so-called early dark energy models\,\cite{Poulin:2018cxd,Chudaykin:2020acu,Braglia:2020bym}, and the model with variable $G$ recently considered in \cite{Ballesteros:2020sik,Braglia:2020iik,Ballardini:2020iws}. Although the physical mechanism of the EDE and the aforementioned models with variable $G$ is of course very different, their aim is pretty similar. They reduce the sound horizon $r_s$ at recombination in order to force the increase of the Hubble function in the late Universe. This allows them to generate larger values of $H_0$ in order to keep a good fit to the CMB and BAO data, but this happens only at the expense of increasing the tension in $\sigma_8$, since they do not have any compensation mechanism able to keep the structure formation in the late Universe at low enough levels.  Some of these models  appear not to be particularly disfavored notwithstanding. But this is simply because they did not use LSS data in their fits, i.e. they did not put their models to the test of structure formation and for this reason they have more margin to adjust the remaining  observable without getting any statistical punishment.  So the fact that the significant increase of $\sigma_8$  that they find is not statistically penalized is precisely because they do not use  LSS data,  such as e.g. those on the observable $f(z)\sigma_8(z)$ displayed in our Table 2.  In this respect, EDE cosmologies are an example; they seem to be unable to alleviate the $H_0$-tension when LSS data are taken into account, as shown in \cite{Hill:2020osr}.
\subsubsection{Imposing the Cassini constraint}\label{sec:Cassini_BD_article}
To further illustrate the capability of the BD-$\CC$CDM model to fit the data under more severe conditions, in  Table 10 we consider four possible settings to fit our Cassini-constrained  BD-Scenario III defined in Sec.\,\ref{sec:Mach_BD_article}. Recall that this BD scenario involves the stringent Cassini bound on the post-Newtonian parameter $\gamma^{\rm PN}$ \cite{Bertotti:2003rm}, which we have discussed in Sec.\,\ref{sec:BDgravity_BD_article}. The first two fitting columns of Table 10 correspond to our usual Baseline$+H_0$ dataset,  in one case (first fitting column in that table) we omit the LSS data, whereas in the second column we restore it. In this way we can check the effect of the structure formation data on the goodness of the fit. The comparison between the results presented in these two columns shows, first and foremost, that the Cassini bound does not have a drastically nullifying effect, namely it does not render the BD-$\CC$CDM model irrelevant to the extent of  making it indistinguishable from GR-$\CC$CDM, not at all, since the quality of the fits is still fairly high (confer the Bayes factors in the last row). In truth, the fit quality is still comparable to that of the Baseline$+H_0$ scenario (cf. Table 3). However, the description of the LSS data is naturally poorer since $\eBD$ is smaller and the model cannot handle so well  the features of the structure formation epoch, thus yielding slightly higher values of $\sigma_8$. Second, the fact that the scenario without LSS furnishes a higher Bayes factor  just exemplifies the aforementioned circumstance that when cosmological models are tested without using this kind of data the results may in fact not be sufficiently reliable. When the LSS data enter the fit (third fitting column in that table), we observe, interestingly enough, that the BD-$\CC$CDM model is still able to keep $H_0$  in the safe range, it does improve the value of $\sigma_8$ as well  (i.e. it becomes lower) and, on top of that, it carries a (`smoking gun')  signal of almost $2.9\sigma$ c.l. --  encoded in the value of $\weff$ --  pointing to quintessence-like behaviour. Overall it is quite encouraging since it shows that the Cassini bound does not exceedingly hamper the BD-$\CC$CDM model capabilities. Such bound constraints the time evolution of $\varphi$ (because $|\eBD|$ is forced to be much smaller) but it does not preclude $\varphi$ from choosing a suitable value in compensation (cf. BD-Scenario III in Sec.\,\ref{sec:Mach_BD_article}).
\newline
In the last two columns of Table 10, we can further  check what are the changes in the previous fitting results when we use a  more restricted dataset,  e.g. the one used in Ref.\,\cite{Ballesteros:2020sik}, in which the Cassini bound is also implemented. These authors study a model which represents a modification of GR through an effective $G\sim 1/M^2$, with $M$ a mass very near the Planck mass, $m_{\rm Pl}$, which is allowed to change slowly through a scalar field $\phi$ as follows: $M^2\to M^2+\beta\phi ^2$, where $\beta$ is a small (dimensionless) parameter. The authors assume that the Cassini bound on the post-Newtonian parameter $\gamma^{\rm PN}$ \cite{Bertotti:2003rm} is in force (see Sec. \ref{sec:MethodData_BD_article} for details). However, they do not consider LSS data (only CMB and BAO). We may compare the results they obtain within that variable $G$ model (cf. their Table 1) with those we obtain within the BD-Scenario III under the very same dataset as these authors. The results are displayed in the third fitting column of Table 10.  We obtain  $H_0=(70.01^{+0.86}_{-0.92})$ km/s/Mpc and $\sigma_8=0.817\pm 0.007$, whereas they obtain $H_0=(69.2^{+0.62}_{-0.75})$ km/s/Mpc and $\sigma_8=0.843^{+0.015}_{-0.024}$. Clearly, the BD-$\CC$CDM is able to produce larger central values of $H_0$ and lower values of $\sigma_8$, even under the Cassini bound, although the differences are  within errors. The value of $2\ln\,B$ lies around $+3.5$ and hence points to a mild positive evidence in favor of the BD model. This is consistent with the associated deviation we find  of $G(0)$ from $G_N$ at $2.43\sigma$ c.l., and with a signal of effective quintessence at $2.53\sigma$ c.l. within the GR-picture.
\newline
Let us  now consider what is obtained if we add up the LSS data to this same BD-Scenario III, still with the restricted dataset of Ref.\,\cite{Ballesteros:2020sik}. As expected, the inclusion of the structure formation data
pushes the value of $\sigma_8=0.812^{+0.006}_{-0.007}$ down as compared to their absence ($\sigma_8=0.817\pm 0.007$). This is the most remarkable difference between the two cases, as one cannot appreciate significant changes in the other parameters. Something very similar happens when we compare the values of $\sigma_8$ of the first two fitting columns of Table 10, in which we consider the B+$H_0$ dataset without LSS data (first fitting column) and with LSS data (second fitting column). The relative improvement {\it w.r.t.} the model of \cite{Ballesteros:2020sik} is therefore greater in the presence of LSS data, whose use has been omitted in that reference. In that variable $G$ model one finds a larger value of $H(z)$ at recombination thanks to the larger values of $G$ in that epoch, but at present $G$ is forced to be almost equal to $G_N$ (being the differences not relevant for cosmology). This means that: (i) in that model $G$ decreases with time, which leads to a kind of effective $\eBD>0$; (ii) the model cannot increase $H_0$ with a large value of $G(z=0)$. Both facts do limit significantly  the effectiveness of the model in loosening the tensions. In the BD-$\CC$CDM model under consideration, instead, we find that $G$ has to be $\sim 8-9\%$ larger than $G_N$ not only in the pre-recombination Universe, but also at present, and this allows to reduce significantly the $H_0$-tension. Moreover, we find that a mild increase of the cosmological $G$ with the expansion leads also to an alleviation of the $\sigma_8$-tension.
\newline
Under all of the datasets studied in Table 10 we obtain central values of $|\eBD|\sim \mathcal{O}(10^{-5})$, which are compatible with 0 at $1\sigma$. Notice that this value is just of order of the Cassini bound on $\eBD$, as could be expected. Notwithstanding, and remarkably enough, the stringent bound imposed by the Cassini constraint, which enforces $\eBD$ to remain two orders of magnitude lower than in the main Baseline  scenarios, is nevertheless insufficient to wipe out the  positive effects from the BD-$\CC$CDM model. They are still capable to emerge with a sizeable part of the genuine BD signal. This is, as anticipated in Sec.\,\ref{sec:Mach_BD_article}, mainly due to the fact that the Cassini bound cannot restrict the value of the BD field  $\varphi$, only its time evolution.
\subsubsection{More on alleviating simultaneously the two tensions}\label{sec:TwoTensions_BD_article}
A few additional comments on our results concerning the parameters $H_0$  and $\sigma_8$ are now in order. Their overall impact can be better assessed by examining  the triangular matrix of fitted contours involving all the main parameters, as shown in Fig.\,10,
in which we offer the numerical results of several superimposed analyses based on different datasets, all of them within Baseline scenarios. We project the contour lines in the corresponding planes of parameter space, and  show the associated marginalized one-dimensional posteriors.   The fitted value of the EoS parameter at $z=0$  shown there, $\weff(0)$, is to be understood, of course, as a derived parameter from the prime ones of the fit, but we include this information along with the other parameters in order to further display the significance of the obtained signal:  $\gtrsim 3\sigma$  quintessence-like behaviour.  Such signal, therefore, mimics  `GR+ DDE' and  hints at something beyond pure GR-$\CC$CDM. What we find in our study is that such time-evolving DE behaviour is actually of quasi-vacuum type and appears as a kind of  signature of the underlying BD theory\footnote{As we recall in Appendix \ref{Appendix_C}, such  kind of  dynamical behaviour of the vacuum is characteristic of the Running Vacuum Model (RVM),  a version of the   $\CC$CDM in which the vacuum energy density is not just a constant  but involves also a dynamical term $\sim H^2$.  The description of the BD-$\CC$CDM model in the GR-picture mimics a  behaviour of this sort \,\cite{Peracaula:2018dkg,Perez:2018qgw}.}.
\newline
Figure\,10 provides  a truly panoramic and graphical view of our main fitting results,  and  from where we can comfortably  judge the impact of the BD framework for describing the overall cosmological data. It is fair to say that it appears at a level highly competitive with GR --  in fact, superior to it.  In all datasets involving the local $H_0$ input in the fit analyses, the improvement is substantial and manifest. Let us stand out only three of the entries in that graphical matrix:  i) for the parameters $(\sigma_8,H_0)$,  all the contours in the main dataset scenarios are centered around values of $\sigma_8<0.80$ and $H_0\gtrsim71$ km/s/Mpc, which are the coveted ranges for every model aiming at smoothing the two tensions at a time;  ii) for the pair $(H_0,r_s)$, the contours are centered around the same range of relevant  $H_0$ values as before, and also around values of the comoving sound horizon (at the baryon drag epoch) $r_s\lesssim142$ Mpc, these being significantly smaller  than those of the concordance $\CC$CDM  (cf. Table 7) and hence consistent with larger values of the expansion rate at that epoch;  iii) and for  $(\weff(0), H_0)$,  the relevant  range $H_0\gtrsim71$ km/s/Mpc is once more picked out, together with an effective quintessence signal $\weff(0)>-1$ at more than $3\sigma$ (specifically $3.45\sigma$ for the scenario of Table 4, in which strong lensing data are included in the fit).
\newline
It is also interesting to focus once more our attention on  Fig. 11, where we provide devoted contours involving  both  $\tilde{S}_8$ and $\sigma_8$  versus $H_0$.  On top of the observations we have previously  made on these observable, we  can compare here our basic dataset scenarios for the BD-$\CC$CDM model with the yield of a simple  XCDM parametrization of the DDE. In previous studies we had already shown that such parametrization can help to deal with the $\sigma_8$ tension\,\cite{Sola:2018sjf}. Nonetheless, as we can see here, it proves completely impotent for solving  or minimally helping to alleviate the $H_0$-tension since the values predicted for this parameter stay as low as in the concordance model. This shows, once more,  that in order to address a possible solution to the two tensions simultaneously, it is not enough to have just some form of dynamics in the DE sector; one really needs a truly specific one,  e.g. the one provided (in an effective way) by the BD-$\CC$CDM model.
\subsubsection{Predicted relative variation of the effective gravitational strength}\label{sec:VariationG_BD_article}
In the context of the BD framework it is imperative, in fact mandatory,  to discuss the current values of the relative variation of the effective gravitational strength, viz. of $\dot{G}(0)/G(0)$, which follow from our fitting analyses (see the main Tables 3-5). The possible time evolution of that quantity hinges directly on $\eBD$, of course, since the latter is the parameter that controls the (cosmological) evolution of the gravitational coupling in the BD theory. It is easy to see from Eq.\,\eqref{eq:definitions_BD_article} that $\dot{G}(0)/G(0)=-\dot{\varphi}(0)/\varphi(0)\simeq -\eBD H_0$, where we use the fact that $\varphi\sim a^{\eBD}$ in the matter-dominated epoch (cf. Appendix \ref{Appendix_C}). Recalling that $H_0\simeq 7\times 10^{-11}$ yr$^{-1}$ (for $h\simeq 0.70$), we find  $\dot{G}(0)/G(0)\simeq -\eBD\cdot 10^{-10}\,yr^{-1}$. Under our main BD-Scenario II (cf.\,Sec.\,\ref{sec:Mach_BD_article}) we obtain values for $\dot{G}(0)/G(0)$ of order $\mathcal{O}(10^{-13})\,yr^{-1}$ (and positive), just because $\eBD\sim \mathcal{O}(10^{-3})$ (and negative). Being  $\dot{G}(0)/G(0)>0$ it means that the effective gravitational coupling obtained by our global cosmological fit increases with the expansion, and hence it was smaller in the past. This suggests that the sign $\eBD<0$, which is directly picked out by the data, prefers a kind of asymptotically free behaviour for the gravitational coupling since the epochs in the past are more energetic,  in fact characterized by larger values of $H$ (with natural dimension of energy).  The central values show a mild time variation at present, at a level of
 $1.3\sigma$,  both for $\eBD$ and $\dot{G}(0)/G(0)$, when the bispectrum data from BOSS is also included in the analysis (cf. Tables 3 and 4). Such departure goes below $1\sigma$ level when only the spectrum is considered (see Table 5). In the context of the BD-Scenario III, in which $\eBD$ is very tightly constrained by the Cassini bound \cite{Bertotti:2003rm}, namely at a level of $\mathcal{O}(10^{-5})$, we find $\dot{G}(0)/G(0)\sim 10^{-15}\,yr^{-1}$, which is compatible with 0 at $1\sigma$. All that said, we should emphasize once more  that the fitting values that we obtain for $\dot{G}(0)/G(0)$ refer to the cosmological time variation of $G$ and, therefore, cannot be directly compared with constraints existing in the literature based on strict local gravity measurements, such as e.g. those from the lunar laser ranging experiment -- $\dot{G}(0)/G(0)=(2\pm 7)\cdot 10^{-13}yr^{-1}$ \cite{Muller:2007zzb} -- (see e.g. the review \cite{Uzan:2010pm} for a detailed presentation of many other local constraints on $\dot{G}(0)/G(0)$). Even though this bound turns out to be preserved within our analysis, it is not in force at the cosmological level provided  an screening mechanism acting at these scales is assumed, as in our case. Thus, the local measurements have no bearing a priori on the BD-$\CC$CDM cosmology. The opposite may not be true, for despite  the fact that the values reported in our tables are model-dependent, they prove to be quite efficient and show that the cosmological observations can compete in precision with the local measurements.

\begin{figure}[t!]
\begin{center}
\label{fig:Cls-ISW_BD_article}
\includegraphics[width=4.in, height=3in]{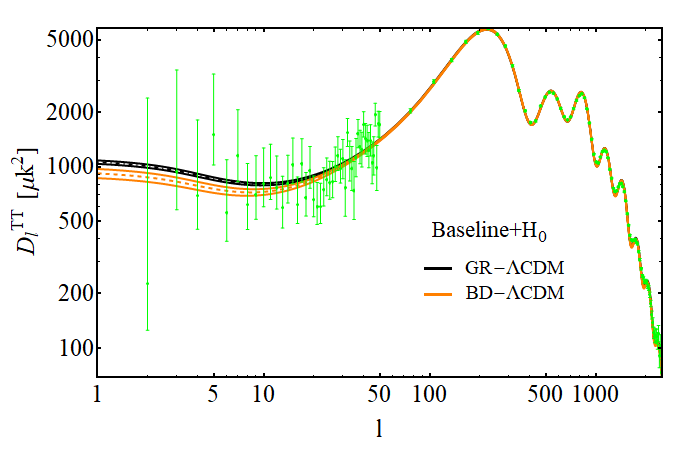}
\caption{\scriptsize{CMB temperature power spectrum for the GR-$\CC$CDM (in black) and BD-$\CC$CDM (in orange), obtained from the fitting results within the Baseline+$H_0$ dataset (cf. Table 3 and Sec. \ref{sec:MethodData_BD_article}). We plot the central curves together with the corresponding $1\sigma$ bands. In the inner plot we zoom in the multipole range $l\in [0,30]$ and include the Planck 2018 \cite{Aghanim:2018eyx} data error bars (in green). At low multipoles ($l\lesssim 30$) the BD-$\CC$CDM model produces less power than the concordance GR-$\CC$CDM model owing to the suppression of the Integrated Sachs-Wolfe effect that we have discussed in Sec. \ref{sec:preview_BD_article}. The differences are $\gtrsim 1\sigma$, and allow to soften a well-known low-multipole CMB anomaly. See the main text for further discussion.}}
\end{center}
\end{figure}

 \begin{table}[t!]
\begin{center}
\begin{tabular}{|c  |c | c |  c | c | c  |}
 \multicolumn{1}{c}{} & \multicolumn{2}{c}{Baseline} & \multicolumn{2}{c}{Baseline+$H_0$}
\\\hline
{\scriptsize Parameter} & {\scriptsize GR-$\Lambda$CDM}  & {\scriptsize BD-$\Lambda$CDM} & {\scriptsize GR-$\Lambda$CDM}  &  {\scriptsize BD-$\Lambda$CDM}
\\\hline
{\scriptsize $H_0$ (km/s/Mpc)}  & {\scriptsize $67.75^{+0.46}_{-0.48}$} & {\scriptsize $68.86^{+1.26}_{-1.22}$} & {\scriptsize $68.35^{+0.49}_{-0.46}$}  & {\scriptsize $70.81^{+0.95}_{-0.92}$}
\\\hline
{\scriptsize$\omega_b$} & {\scriptsize $0.02231^{+0.00018}_{-0.00020}$}  & {\scriptsize $0.02214\pm 0.00031$} & {\scriptsize $0.02239\pm 0.00019$}  &  {\scriptsize $0.02239^{+0.00029}_{-0.00032}$}
\\\hline
{\scriptsize$\omega_{cdm}$} & {\scriptsize $0.11655^{+0.00110}_{-0.00111}$}  & {\scriptsize $0.11860^{+0.00194}_{-0.00197}$} & {\scriptsize $0.11625^{+0.00099}_{-0.00106}$}  &  {\scriptsize $0.11821^{+0.00214}_{-0.00200}$}
\\\hline
{\scriptsize$\tau$} & {{\scriptsize$0.053^{+0.006}_{-0.008}$}} & {{\scriptsize$0.054^{+0.006}_{-0.008}$}} & {{\scriptsize$0.053^{+0.006}_{-0.008}$}}  &   {{\scriptsize$0.054^{+0.006}_{-0.009}$}}
\\\hline
{\scriptsize$n_s$} & {{\scriptsize$0.9706^{+0.0041}_{-0.0040}$}}  & {{\scriptsize$0.9691^{+0.0104}_{-0.0093}$}} & {{\scriptsize$0.9717^{+0.0041}_{-0.0042}$}} &   {{\scriptsize$0.9791^{+0.0097}_{-0.0090}$}}
\\\hline
{\scriptsize$\sigma_8$}  & {{\scriptsize$0.770^{+0.016}_{-0.018}$}}  & {{\scriptsize$0.759^{+0.0184}_{-0.0163}$}} & {{\scriptsize$0.780^{+0.018}_{-0.015}$}}  &   {{\scriptsize$0.762^{+0.021}_{-0.018}$}}
\\\hline
{\scriptsize$r_s$ (Mpc)}  & {{\scriptsize$148.03\pm0.33$}}  & {{\scriptsize$146.16^{+2.40}_{-2.58}$}} & {{\scriptsize$148.04^{+0.32}_{-0.34}$}}  &   {{\scriptsize$142.81^{+1.91}_{-2.03}$}}
\\\hline
{\scriptsize$m_{\nu}$ (eV)}  & {{\scriptsize$0.161^{+0.058}_{-0.059}$}}  & {{\scriptsize$0.409^{+0.139}_{-0.198}$}} & {{\scriptsize$0.118^{+0.053}_{-0.058}$}}  &   {{\scriptsize$0.409^{+0.183}_{-0.228}$}}
\\\hline
{\scriptsize $\eBD$} & - & {{\scriptsize $0.00459^{+0.00316}_{-0.00319}$}} & - &   {{\scriptsize $0.00433^{+0.00395}_{-0.00336}$}}
\\\hline
{\scriptsize$\varphi_{ini}$} & - & {{\scriptsize $0.979^{+0.028}_{-0.032}$}} & - &    {{\scriptsize $0.938^{+0.023}_{-0.024}$}}
\\\hline
{\scriptsize$\varphi(0)$} & - & {{\scriptsize $1.016^{+0.041}_{-0.049}$}} & - &    {{\scriptsize $0.972^{+0.042}_{-0.043}$}}
\\\hline
{\scriptsize$w_{\rm eff}(0)$} & - & {{\scriptsize $-1.005^{+0.021}_{-0.017}$}} & - &    {{\scriptsize $-0.986^{+0.017}_{-0.016}$}}
\\\hline
{\tiny $\dot{G}(0)/G(0) (10^{-13}yr^{-1})$} & - & {{\scriptsize $-5.048^{+3.485}_{-3.434}$}} & - &    {{\scriptsize $-4.915^{+3.749}_{-4.501}$}}
\\\hline
{\scriptsize$\chi^2_{min}$} & {\scriptsize 2270.54}  & {\scriptsize 2268.44} & {\scriptsize 2285.02}  &  {\scriptsize 2274.76}
\\\hline
{\scriptsize$2\ln B$} & {\scriptsize -}  & {\scriptsize -0.12} & {\scriptsize -}  &  {\scriptsize +9.34}
\\\hline
{\scriptsize$\Delta {\rm DIC}$} & {\scriptsize -}  & {\scriptsize -0.81} & {\scriptsize -}  &  {\scriptsize +8.22}
\\\hline
\end{tabular}
\caption{{\scriptsize As in Table 3, but here we allow the variation of the mass of the massive neutrino ($m_\nu$) in the Monte Carlo routine, instead of setting it to $0.06$ eV. The other two neutrinos remain massless. We have used the same conservative prior range for $m_\nu\in [0,1]$ eV in both, the GR- and BD-$\Lambda$CDM models.}}
\end{center}
\label{tableFit3}
\end{table}

\subsubsection{One more bonus: suppressing the power at low multipoles }\label{sec:LowMultipoles_BD_article}
An additional bonus from the  BD cosmology is worth mentioning before we close this lengthy study. It is found in the description of the CMB temperature anisotropies. As we have discussed in Sec. \ref{sec:preview_BD_article}, the BD-$\CC$CDM model is, in principle, able to suppress the power at low multipoles ($l\lesssim 30$), thereby softening one of the so-called CMB anomalies that are encountered in the context of the GR-$\CC$CDM model. This is basically due to the low values of $\varphi<1$ preferred by the data, which in turn produce a suppression  of the Integrated Sachs-Wolfe (ISW) effect \cite{Sachs:1967er,Das:2013sca}. We have confirmed that this suppression actually occurs for the best-fit values of the parameters in our analysis, cf. Fig. 12. The aforementioned anomaly is not very severe, since the  power at low multipoles is affected by a large cosmic variance and cannot be measured very precisely. Nevertheless, it is a subtle anomaly which has been there unaccounted for a long time and could not be improved in a consistent way: that is to say, usually models ameliorating the low tail of the spectrum do spoil the high part of it. However, here  the suppression of power with respect to the GR-$\CC$CDM at $l\lesssim 30$ is fully consistent and is another very welcome feature of the BD-$\CC$CDM model, which is not easy at all to attain.  It is interesting to mention that the ISW effect can be probed by cross correlating the CMB temperature maps with the LSS data, e.g. with foreground galaxies number counts, especially if using  future surveys which should have much smaller uncertainties. This can be a useful probe for DE and a possible distinctive signature for DDE theories\,\cite{Pogosian:2004wa,Zucca:2019ohv}\,\footnote{We thank L. Pogosian for interesting comments along these lines.}.  The upshot of our investigation is that once more  the `golden rule' mentioned at the beginning is preserved here and the curing effects from the BD-$\CC$CDM stay aligned: the  three tensions of the  GR-$\CC$CDM ($H_0$, $\sigma_8$ and  the exceeding CMB power at low multipoles) can be improved at a time.
\subsubsection{The effect of massive neutrinos }\label{sec:MassiveNeutrino_BD_article}
{Finally, it is worth assessing the impact of leaving the sum of the three neutrino masses ($M_\nu\equiv\sum m_\nu$) as a free parameter, rather than fixing it. We model this scenario in two different ways:}

\begin{itemize}
    \item {Considering one massive ($m_\nu$) and two massless neutrinos, so $M_\nu=m_\nu$. Here we try to mimic the physics encountered when the neutrino masses follow the normal {hierarchy}, in which one of the neutrinos is much heavier than the other two, for reasonable values of $M_\nu$.}
    \item {Considering three massive neutrinos of equal mass $m_\nu$, such that $M_\nu=3m_\nu$. This is the degenerated case, utilized also in the analysis by Planck 2018 \cite{Aghanim:2018eyx} (cf. Sec. 7.5.1 therein).}
\end{itemize}
{In both cases, the BD-$\CC$CDM has one additional parameter {\it w.r.t.} the same model with fixed mass $M_\nu=m_\nu=0.06$ eV that we have discussed in the previous sections, and three more than the vanilla $\Lambda$CDM model: $(\eBD,\varphi_{ini}, M_\nu)$. Upon taking into account the constraints obtained from neutrino oscillation experiments on the mass-squared splittings through the corresponding likelihoods we would perhaps find more precise bounds on the sum of the neutrino masses\,\cite{Loureiro:2018pdz}, but here we want to carry out a more qualitative analysis to study how massive neutrinos may impact on our results, considering only cosmological data. The results of the first mass scenario are shown in Table 11. Those for the second scenario are not tabulated, but are commented below.}
\newline
{As can be seen from Table 11, the effect of having a neutrino with an adjustable mass picked out by the fitting process is non-negligible. It produces a significant lowering of the value of $\sigma_8$ while preserving the value of $H_0$ at a level comparable to the previous tables in which the neutrino had a fixed mass of $0.06$ eV, therefore preserving what we have called the `golden rule'.  Let us note, however, that the sign of $\eBD$ has changed now with respect to the situation with a fixed light mass, it is no longer negative but positive and implies a value of the BD parameter of $\omega_{\rm BD}\simeq230$.  Last but not least, the fitted value of the neutrino mass is $m_\nu=0.409$ eV, which is significantly higher than the upper bound placed by Planck 2018 under the combination TTTEEE+lowE+lensing+BAO: $M_\nu\equiv\sum m_\nu<0.120$ eV ($95\%$ c.l.)\,\cite{Aghanim:2018eyx}. }
\newline
{Similar results are obtained with the second mass scenario mentioned above.  Let us summarize them. For the Baseline+$H_0$ dataset, the common fitted mass value obtained for the three neutrinos is $m_\nu=0.120^{+0.054}_{-0.068}$ eV, which means that $M_\nu\simeq 0.360$ eV, slightly lower than in the first scenario. The corresponding BD parameter reads comparable, $\omega_{\rm BD}\simeq 261$, and the values for $\sigma_8$ and $H_0$ are also very close to the previous case, so the two scenarios share similar advantages. The fact that the Planck 2018 upper limit for $M_\nu$ is overshooted in both neutrino mass scenarios does not necessarily exclude them, as the limits on  $M_\nu$ are model-dependent, see e.g. \cite{Loureiro:2018pdz} and \cite{Ballardini:2020iws}. In particular, Planck 2018 obviously used  GR-$\CC$CDM. We can check in  Table 11 that the neutrino mass values for BD-$\CC$CDM are substantially different. The results that we have obtain for GR-$\CC$CDM are fully compatible with those from Planck 2018.}
{We conclude that the influence of neutrino masses on the BD-$\CC$CDM fitting results is potentially significant, but it cannot be settled at this point. The subject obviously deserves further consideration in the future. }
\subsection{Conclusions}\label{sec:Conclusions_BD_article}
To summarize, we have presented a rather comprehensive work on the current status of the Brans-Dicke theory with a cosmological constant in the light of the modern observations. Such framework constitutes a new version of the concordance $\CC$CDM model in the context of a distinct gravity paradigm, in which the gravitational constant is no longer a fundamental constant of Nature but a dynamical field. We have called this framework BD-$\CC$CDM model to distinguish it from the conventional one, the GR-$\CC$CDM model, based on General Relativity.  Our work is a highly expanded and fully updated analysis of our previous and much shorter presentation\,\cite{Sola:2019jek}, in which we have  replaced the Planck 2015 data by the Planck 2018 data,  and we have included additional sets of modern cosmological observations. We reconfirm the results of \,\cite{Sola:2019jek} and provide now a bunch of new results which fit in with the conclusions of our previous work and reinforce its theses.  To wit:  in the light of the figures and tables that we have presented in the current study we may assert that the BD-$\CC$CDM model fares better not only as compared to the GR-$\CC$CDM  with a rigid cosmological constant (cf. Tables 3-7 and 10) but also when the CC term is replaced with a dynamical parametrization of the DE, such as the traditional XCDM, which acts as a benchmark (cf. Table 8).  We find that the GR-XCDM is completely unable to enhance the value of $H_0$ beyond that of the concordance model.  In particular, in Tables 3-4  we can see that the information criteria (Bayes factor and Deviance Information Criterion) do favor  significantly and consistently the BD-$\CC$CDM model as compared to GR-$\CC$CDM and  GR-XCDM. There is a very good resonance between the Bayesian evidence criterion and the DIC differences, which definitely uphold the BD framework at a level of $+5$ units for the Baseline+$H_0$ dataset scenario, meaning that the degree of support of BD versus GR is in between \textit{positive} to \textit{strong} (cf. Sec.\,\ref{sec:NumericalAnalysis_BD_article}). This support is further enhanced up to more than $+9$ units, hence  in between \textit{strong} to \textit{very strong}, for the case  when we include the Strong Lensing data in the fit (see Table 4). The exact Bayesian evidence curves computed in Fig.\,9 reconfirm these results in a graphical way. The pure baseline dataset scenario (in which the local $H_0$ value is not included) shows weak evidence; however, as soon as the local $H_0$ value is fitted along with the remaining parameters the evidence increases rapidly and steadily,  reaching the status of positive, strong and almost very strong depending on the datasets.
\newline
{Another dataset scenario which is particularly favored in our analysis} is the one based on considering the effective calibration prior on the absolute magnitude $M$ of the nearer SNIa data in the distance ladder (as defined in Sec.\,\ref{sec:MethodData_BD_article}), instead of the local value $H_0$ from SH0ES. The results for the Baseline dataset in combination with $M$ (denoted B+$M$) can be read off from Table 6 (first row).  We can see it yields a tantalizing overall output, with values of $H_0$ and  $\sigma_8$ in the correct ranges for solving the two tensions, and fully compatible with the results obtained using the prior on $H_0$, as expected. In addition, the corresponding Bayes factor for this scenario points to a remarkably high value $2\ln B>+10$, thereby carrying a very strong Bayesian evidence,  in fact comparable to the Baseline+$H_0$+SL scenario of Table 4.  In all of the mentioned cases in our summary the information criteria  definitely endorse the BD-cosmology versus the GR one.
\newline
{Finally, we have also assessed the influence of the neutrino masses in the context of the BD-$\CC$CDM model, see Table 11. We have found that massive neutrinos can help to further reduce the predicted value of $\sigma_8$ to the level of what is precisely needed to describe the weak gravitational lensing observations derived from direct shear data. This would completely dissolve the $\sigma_8$-tension without detriment of the positive results obtained to loosen the $H_0$-tension, i.e. by preserving the `golden rule'.  Such a conclusion is, however, provisional as it requires a devoted study of the neutrino sector (extending the analysis of Sec.\,\ref{sec:MassiveNeutrino_BD_article}) which is beyond the scope of the current work.}
\newline
Overall, the statistical support in favor of the BD-$\CC$CDM model against the concordance  GR-$\CC$CDM model is rather significant. It is not only that the $H_0$ and $\sigma_8$ tensions are simultaneously dwarfed to a level where they are both rendered inessential ($\lesssim 1.5\sigma$), but also that all tested BD scenarios involving the local $H_0$ value provide a much better global fit than the concordance model on the basis  of a rich and updated set of modern observations from all the main cosmological data sources available at present. If we take into account that the BD-$\CC$CDM framework is not just some \textit{ad hoc} phenomenological toy-model, or some last-minute smart parametrization just concocted to solve or mitigate the two tensions, but the next-to-leading fundamental theory candidate directly competing with GR, it may give us a sense of the potential significance of these results.
\newpage
\setcounter{secnumdepth}{-1}
\section{Conclusions} 
Throughout this thesis a wide variety of models, beyond the standard one, have been studied. The corresponding analyses have been carried out by studying in detail the theoretical predictions at the background and perturbation level, with the purpose of testing them with the large amount of cosmological data to which we currently have access. The ultimate goal is to see if we can detect signs of new physics that help to alleviate some of the tensions that affect the $\Lambda$CDM. As we have stated before, the concordance model, has remained robust and unbeaten for a long time since it is roughly consistent with a large body of cosmological data. Because of this fact, it is not reasonable to look for models with a very different behaviour than the $\Lambda$CDM, but to study models that exhibit small departures with respect to the standard model in key aspects. 
\newline
We have studied the Running Vacuum Models (RVM's) in depth. They are characterized by having a time-evolving vacuum energy density, whose functional expression is motivated in the context of quantum field theory (QFT) in curved space-time. Neglecting the terms that are not relevant for the late-time Universe, the vacuum energy density, can be written as:
\begin{equation}\nonumber
\rho_{\rm vac}(H,\dot{H}) = \frac{3}{8\pi{G_N}}(c_0 + \nu{H^2} + \alpha{\dot{H}}). 
\end{equation}
The constant term $c_0$ turns out to be a key ingredient of the above expression. Without it, it is not possible to generate the expected transition from a decelerated to an accelerated Universe and, what is more, the fit of the structure formation data is ruined. This is in total agreement with what we have said above, since the small dimensionless parameters $\nu$ and $\alpha$ introduce a mild evolution of the vacuum energy density with respect to the standard behaviour, represented by the constant term $c_0$. 
\newline
\newline
We have also studied the Peebles \& Ratra model, which is a particularly successful scalar field model $\phi$CDM for which the potential takes the form $V(\phi)\sim \phi^{-\alpha}$. The dimensionless parameter $\alpha$ encodes the extra degree of freedom that this model has with respect to the standard model. It is found to be small and positive, therefore $V(\phi)$ can mimic and approximate cosmological constant that is decreasing slowly with time. In the late Universe the contribution of the scalar field, $\phi$, surpasses the matter density, thus becoming the dominant component. 
\newline
\newline
Not all the models studied are motivated within a theoretical framework. For instance we have analyzed the performance of a couple of dark energy parameterizations, namely, the XCDM and the CPL. They were introduced as the simplest way to track a possible dynamics for dark energy. The rigid cosmological constant is replaced by an unspecified quantity $X$ whose energy density coincides with $\rho^0_\Lambda$ at present. For the XCDM parameterization we have a constant equation of state (EoS) parameter, but different from $-1$, whereas CPL is characterized by having a time-evolving EoS. 
\newline
We have also contrasted the performance of the RVM against two phenomenological models, denoted by: $Q_{dm}$ and $Q_\Lambda$, respectively. For both a totally {\it ad-hoc} expression for $\dot{\rho}_{\rm vac}$ is chosen, and then in order to allow the time evolution of the vacuum energy density an interaction with dark matter (DM) is considered. 
\newline
\newline
Last but not least, at the end of the thesis the Brans \& Dicke (BD) gravity model was studied in detail. The main feature of this model is that the Newtonian constant coupling $G_N$ is replaced by a dynamical scalar field $G(t) = 1/\psi(t)$, coupled to the curvature. As a consequence the gravitational interaction is not only mediated by the metric field, as in the General Relativity (GR) case but also for the aforementioned scalar field $\psi$. The theory contains a dimensionless parameter, named BD-parameter and denoted by $\omega_{\rm BD}$. Unlike in the original formulation of the model we do consider the presence of the constant $\rho_\Lambda$ in the action. Defined in this way, the model contains two extra {\it d.o.f.} with respect to the standard model, to wit: the before mentioned parameter $\omega_{\rm BD}$ and the initial value of the BD scalar field. Therefore, in order to recover the $\Lambda$CDM we have to enforce both limits, $\omega_{\rm BD}\rightarrow\infty$ {\it and} $\psi\rightarrow{G_N}$. As we have seen the fact of having these two degrees of freedom turns out to be fundamental to solve the $H_0$-tension and $\sigma_8$-tension at a time. 
\newline
\newline
Let us summarize the main conclusions obtained in this dissertation:
\begin{itemize}
\item In the light of the results obtained we can confirm that the general class of RVM appears as a serious candidate for the description of the current state of the Universe in accelerated expansion. Very clear signs of dynamical vacuum energy have been found, being the evidence in the range 3-4$\sigma$ depending on the dataset employed. The dimensionless parameter $\nu$, which can be related with the $\beta$-function of the running, takes a value of order $\sim \mathcal{O}(10^{-3})$, which is in complete agreement with the theoretical predictions. A decreasing vacuum energy density (due to $\nu >0$) (which is more natural from a thermodynamic point of view) throughout the cosmic history is favoured by the data, meaning that the amount of vacuum energy grows in the past. As a result the repulsive force exerted by this component is greater in the past that it is in the $\Lambda$CDM, therefore there is less structure formation predicted in this model. This important feature is revealed as essential to loosen the $\sigma_8$-tension that affects the standard model of cosmology. 
\item By looking at the results obtained for the $\phi$CDM, with the Peebles \& Ratra potential, we can reconfirm that the hypothesis $\Lambda={\rm const.}$ is once again, disfavoured. As in the case of the RVM's the level of significance achieved is between 3-4$\sigma$ and the dimensionless parameter $\alpha$, which is responsible for the slow evolution of the scalar field, takes a value of order $\sim \mathcal{O}(0.1)$. It is reassuring the fact of finding the same level of evidence in favor of dynamical DE even when two different models like $\phi$CDM and the general class of RVM's, are test against the very same dataset. Both type of models seem to point out in the same direction, namely, that the DE is decreasing with the expansion and therefore that it behaves effectively as quintessence. 
\item Even for the XCDM and CPL, which are simple DE parameterizations, it is possible to detect significant signs of evolving dark energy. This may be pointing out that perhaps the signature of such dynamical signature is sitting in the cosmological data and therefore is not exclusive of a particular model. On the other hand the level of significance does indeed vary from one model to another. In fact, this is what happens if we compare the performance of the phenomenological models $Q_{dm}$ and $Q_\Lambda$ with the performance of the RVM. Neither of the first two models reaches the level of significance achieved by the Running Vacuum Model. 
\item The quality fits obtained in the different chapters of this thesis, that point out the preference of the cosmological data for the option $\Lambda \neq {\rm const.}$ have been reassessed with the help of the time-honored Akaike and Bayesian information criteria. These two tests, conveniently penalize the fact of having a greater number of degrees of freedom. According to the statistical standards, the evidence in favour of the dynamical dark energy goes from ``positive evidence" to ``very strong evidence" depending on the model under consideration. This supports the conclusion that there exists an improvement of the description of the cosmological data, when we consider a dynamical DE model, in comparison to the $\Lambda$CDM. 
\item By studying the effect that each one of the datasets has on the results, we have identified which are the most important data ingredients responsible for the dynamical dark energy signal. When the triad BAO+LSS+CMB is not included in the analysis the signal is completely lost. The impact of the $H(z)$ and SNIa data turns out to be more moderate, however they are still important to consolidate the results. We have also verified that the inclusion of the Gravitational Lensing data, either the Weak Lensing one or the Strong Lensing data, has a significant impact on the results. 
\item While the dynamical vacuum energy models and also the quasi-vacuum models do a great job of reducing the $\sigma_8$-tension, they are not able to alleviate by any means the $H_0$-tension. When the LSS data is included in the dataset the values obtained for the $H_0$ are in full agreement with the one obtained by the Planck collaboration, but in tension with the value reported by the SH0ES team obtained from the application of the cosmic distance ladder method. 
\item In regards to the BD model, or BD-$\Lambda$CDM as we have called it in Chapter \ref{BD_gravity_chapter}, the results clearly show that even with a $G_{\rm eff}$ being higher than in the GR-$\Lambda$CDM, the model is capable to fit the wealth of the cosmological data employed in the analysis. The main consequence of having higher value of $G_{\rm eff}$ is that the value of the $H_0$ parameter can be in the range 70-72 km/s/Mpc (depending on the dataset utilized), alleviating then the aforementioned $H_0$-tension. What is more, due to the role played by the $\omega_{\rm BD}$ parameter in helping to suppress the structure formation processes in the Universe, the model is also able to loosen the $\sigma_8$-tension. Therefore, the main outcome that can be extracted from the comprehensive analysis of the BD-$\Lambda$CDM model is that it is able to alleviate the main two tensions affecting the GR-$\Lambda$CDM at a time. 
\end{itemize}
As it can be appreciated the results obtained in this thesis, after years of studying the different cosmological models in depth, provide very encouraging signs in favor of dynamical dark energy. It is very remarkable the fact that this signal has been perceived even if the models are based on totally different assumptions which clearly reinforces the conclusions collected.
Such promising results may be reconfirmed thanks to the incoming new generation of cosmological data. Perhaps, in a very short time we will obtain more clues that help us to decipher some of the existing mysteries in cosmology, either from the observational side or from the theoretical perspective. We are entering in an era that promises to be fascinating for cosmology !

\newpage

\setcounter{secnumdepth}{+1}
\appendix  

\section{Sign conventions in General Relativity}\label{Appendix_A}
Due to the existence of different sign conventions in cosmology we deem it is necessary to dedicate this appendix to see how the different elements that we work with are affected by these conventions. \footnote{This appendix is based on the notes that the supervisor of this thesis, the Professor Joan Sol\`a Peracaula, prepared for his students.}
\newline
\newline
Let us define the following four signs, $S1$ (metric), $S2$ (Riemann), $S3$ (Einstein) and $S_4$ (Ricci). Only three of them are independent. We arbitrarily take to be the first three the independent ones, and we call the sign sequence
\begin{equation}\label{eq:MTW1}
(S1,S2,S3)
\end{equation}
the ``MTW sign convention'' for a given author. The name comes from  the time-honored truly encyclopedic GR book by Misner-Thorn-Wheeler \cite{Misner:1974qy}.  Each GR book uses a given one of the 8 possible conventions. For this book the convention is, of course, $(+,+,+)$.
These three fundamental signs are defined as follows:
\begin{equation}\label{eq:S1}
g_{\mu\nu}=\Su \left(-,+,+,+\right)\,,
\end{equation}
\begin{equation}\label{eq:S2}
R^{\alpha}_{\mu\beta\nu}=\Sd \left(\partial_\beta\Gamma_{\mu\nu}^\alpha- \partial_\nu\Gamma^{\alpha}_{\mu\beta} + \Gamma^{\alpha}_{\sigma\beta}\Gamma^{\sigma}_{\nu\mu} - \Gamma^{\alpha}_{\sigma\nu}\Gamma^{\sigma}_{\beta\mu} \right),
\end{equation}
\begin{equation}\label{eq:S3}
G_{\mu\nu}=\St 8\pi G_{N} T_{\mu\nu}\,.
\end{equation}
The fourth sign is for the Ricci tensor, but is no longer independent if the first three ones are given:
\begin{equation}\label{eq:S4}
R_{\mu\nu}=\Sq R^{\alpha}_{\mu\alpha\nu}=\Sd \St R^{\alpha}_{\mu\alpha\nu}=\St \left(\partial_\alpha\Gamma_{\mu\nu}^\alpha-...\right)\,.
\end{equation}
There is no need to show the rest of the terms in the Riemann tensor since it all depends on the starting ones and then the rest is unambiguously fixed.
\newline
\newline
The line element squared for a metric widely employed in this thesis, the FLRW one, therefore reads
\begin{equation}\label{eq;ds2}
ds^2 = \Su\left(-dt^2 + a^2(t)\delta_{ij}dx^{i}dx^{j}\right).
\end{equation}
Sometimes the MTW convention is expressed as
\begin{equation}\label{eq:MTW2}
(S1,S2,S4)\,.
\end{equation}
In that case $[S3]$ is fixed from $[S3]=\Sd [S4]$.
\newline
\newline
The following observations should be bear in mind for the rest of the appendix:
\begin{itemize}
\item  The Christoffel symbols do not change under metric conventions since the metric appears twice, derivately and non-derivatively.
\item  $ R_{\mu\nu}$  does not flip sign under metric flip changes. The reason is that the Riemann tensor is made of Christoffel symbols and hence does not flip sign under metric flips, and then $R_{\mu\nu}$ involves one more  bilinear dependence on the metric that leads to a $\delta$-Kronecker, and hence no additional sign flip:
\begin{equation}\label{eq:Rmunusign}
R_{\mu\nu}=\Sq R^{\alpha}_{\mu\alpha\nu}=\Sq g^{\alpha\beta} R_{\alpha\mu\beta\nu}= \Sq g^{\alpha\beta} g_{\alpha\sigma} R^{\sigma}_{\mu\beta\nu}=\Sq \delta^{\beta}_{\alpha} R^{\alpha}_{\mu\beta\nu}.
\end{equation}
\item  In contrast, the sign of the Ricci scalar
\begin{equation}\label{eq:Rdef}
R=g^{\mu\nu} R_{\mu\nu}\,,
\end{equation}
does flip with the metric  since $ R_{\mu\nu}$ does \textit{not}.
\item It follows that $g_{\mu\nu} R$ does not flip with the change of the metric signature and hence  the entire Einstein tensor $G_{\mu\nu}=R_{\mu \nu }-\frac{1}{2}g_{\mu \nu }R$ does not flip either.
\item Let us repeat here that if we take the first three signs to be independent, then the Ricci sign (i.e. the sign $[S4]$ on the \textit{r.h.s.} of Einstein equations, see Eq.\,(\ref{eq:S4})) is fixed as follows:
\begin{equation}\label{eq:S4fixed}
  [S4]=\Sd [S3]\,,
\end{equation}
as it can be seen from (\ref{eq:Rmunusign}) and (\ref{eq:S3}) and  the definition of $G_{\mu\nu}$.  This can be shown by substituting the last piece of the \textit{r.h.s.} of (\ref{eq:S4}), which already used Eq.(\ref{eq:S4fixed}), into  $G_{\mu\nu}$  on the \textit{l.h.s.} of (\ref{eq:S3}) and observe that all signs cancel out in Einstein's equations, once these are expressed in terms of $\partial_\alpha\Gamma_{\mu\nu}^\alpha-...$, as these terms are no longer dependent on any sign, not even on $[S1]$.  Conversely, if we use the alternative signs (\ref{eq:MTW2}), then the Einstein sign is fixed through $ [S3]=\Sd [S4].$
\end{itemize}
Let us now derive the signs of Einstein's equations from the signs in the EH-action, which in any MTW convention must have the form
\begin{eqnarray}
S_{\rm EH} &=& \frac{1}{16\pi\,G_{N}}\,\int d^4
x\sqrt{-g}\, \left(\Su \St\,R - 2\,\CC\,\right)\nonumber\\
&=&\ \ \ \int d^4
x\sqrt{-g}\,\left(\frac{\Su [S3]}{16\pi\,G_N} \,R-\rL\right)\,. \ \ 
\label{eq:EHgeneral}
\end{eqnarray}
Notice that the sign in front  of $\Lambda$ or $\rL$ must \emph{always} be negative since for $\Lambda>0$ it must give a positive contribution to the potential energy (density) of the Lagrangian. As for the sign of  $R$, recall it flips sign with the metric, hence $\Su R$ does not, and finally there must be $[S3]$ because this controls the sign of $G_{\mu\nu}$ and hence of $R_{\mu\nu}$, see Eq.\,(\ref{eq:S4}), which appears in the variation of $R$. Indeed, let us check that. Recall the formulae:
\begin{equation}\label{eq:deltaR}
\delta\sqrt{-g}=  -\frac12\,\sqrt{-g}\,g_{\mu\nu}\delta g^{\mu\nu}\,,\ \ \ \ \ \  \ \delta \left(\sqrt{-g}\,R\right)=\sqrt{-g}\,G_{\mu\nu}\,\delta g^{\mu\nu}\,.
\end{equation}
Now we must variate the total action $S=S_{\rm EH}+S_m$, where $S_m$ is the matter action. The latter is related to the energy-momentum tensor as follows:
\begin{equation}\label{eq:deltaTmunu}
  T^{\mu\nu}=+\Su\frac{2}{\sqrt{-g}}\frac{\delta S_m}{\delta g_{\mu\nu}}\ \ \Longleftrightarrow \ \ \delta S_m=+\frac12\,\Su \int d^4 x\sqrt{-g}\,T^{\mu\nu}\,\delta g_{\mu\nu}\,.
\end{equation}
From these formulas it is now straightforward to compute the total variation of the action $\delta S=\delta S_{\rm EH}+\delta S_m$ and equate it to zero:
\begin{eqnarray}\label{eq:deltaStotal}
 \delta S&=&\int d^4x\,\sqrt{-g}\,\left\{\frac{1}{16\pi\,G_N}\left(\Su\St G_{\mu\nu} +\Lambda g_{\mu\nu}\right)  -\frac12\Su T_{\mu\nu} \right\}\delta g^{\mu\nu}=0\nonumber.\\
\end{eqnarray}
Equating the braced expression to zero and multiplying it all with $16\pi G_N$ we get
\begin{equation}\label{eq:EE2}
\Su\St G_{\mu\nu}+\CC g_{\mu\nu}=\Su  8\pi G_N T_{\mu\nu}.
\end{equation}
Multiplying this equation by $\Su[S3]$ we find Einstein's equations in all MTW conventions:
\begin{equation}\label{eq:EE}
G_{\mu\nu}+\Su\St \CC\,g_{\mu\nu}=\St 8\pi G_N\ {T}_{\mu\nu}.     
\end{equation}
So far we have been dealing with the geometric part of Einstein's equations, but now it is time to consider some of the particular forms that the energy-momentum tensor can take and how they change under the sign conventions. Let us consider two of the most common cases the energy-momentum tensor of a scalar field and the energy-momentum tensor of a perfect fluid. 
\subsection{Signs for the energy-momentum tensor of a scalar field}
Notice that since the coordinates are defined carrying upper indices, the metric tensor to which they couple is $g_{\mu\nu}$ rather than its inverse $g^{\mu\nu}$.  So the primary gravitational field is $g_{\mu\nu}$, not $g^{\mu\nu}$, and with respect to the primary one we have a $+$ sign in the definition of $ T^{\mu\nu}$ -- see (\ref{eq:deltaTmunu}).  However, owing to $\delta g^{\mu\nu}=- g^{\mu\alpha}g^{\beta\nu}\delta g_{\alpha\beta}$ an extra sign appears, which comes from the fact that  $g^{\mu\alpha}g_{\alpha\nu}=\delta^\mu_\nu$.  Incidentally, the one-dimensional analog is  $\delta x^{-1}=-x^{-2}\delta x$, which appears from differentiating $x^{-1}\,x=1$.
This allows to extend the formulae (\ref{eq:deltaR}) as follows:
\begin{equation}\label{eq:deltasqrtgandR2b}
\delta\sqrt{-g}=  -\frac12\,\sqrt{-g}\,g_{\mu\nu}\delta g^{\mu\nu}=+\frac12\,\sqrt{-g}\,g^{\mu\nu}\delta g_{\mu\nu}\,,\ \ \ \ \delta \left(\sqrt{-g}\,R\right)=\sqrt{-g}\,G_{\mu\nu}\,\delta g^{\mu\nu}=-\sqrt{-g}\,G^{\mu\nu}\,\delta g_{\mu\nu}\,.
\end{equation}
The sign difference between using $\delta g^{\mu\nu}$ and  $\delta g_{\mu\nu}$ is transferred to the corresponding formula for  $T_{\mu\nu}$ with lower indices:
\begin{equation}\label{eq:deltaTmunu2}
  T_{\mu\nu}=-\Su\frac{2}{\sqrt{-g}}\frac{\delta S_m}{\delta g^{\mu\nu}}\ \ \Longleftrightarrow \ \ \delta S_m=-\frac12\,\Su \int d^4 x\sqrt{-g}\,T_{\mu\nu}\,\delta g^{\mu\nu}\,.
\end{equation}
We are going to work out the expression for the energy-momentum tensor of a scalar field from the corresponding action
\begin{eqnarray}
S_{\Phi}=\int d^4x\,\sqrt{-g}\,\left[-\Su\frac12\,g^{\mu\nu}\,\partial_{\mu}\Phi\,\partial_{\nu}\Phi-U(\Phi)\right]\,,
\label{ScalarGeneric}
\end{eqnarray}
we obtain the generic form for $T^{\mu\nu}$
\begin{equation}\label{eq:TmunuScalarField1}
  T^{\mu\nu}=\partial^{\mu}\Phi\, \partial^{\nu}\Phi+\Su g^{\mu\nu}\left[-\Su \frac12 g^{\alpha\beta}\partial_{\alpha}\Phi \,\partial_{\beta}\Phi -U(\Phi)\right]=\partial^{\mu}\Phi\, \partial^{\nu}\Phi+\Su g^{\mu\nu}{\cal L}
\end{equation}
or
\begin{equation}\label{eq:TmunuScalarField2}
  T_{\mu\nu}=\partial_{\mu}\Phi\, \partial_{\nu}\Phi+ \Su g_{\mu\nu}\left[-\Su \frac12 g^{\alpha\beta}\partial_{\alpha}\Phi \,\partial_{\beta}\Phi -U(\Phi)\right]=\partial_{\mu}\Phi\, \partial_{\nu}\Phi+\Su g_{\mu\nu}{\cal L}\,.
\end{equation}
The previous calculations follow from the standard rule for functional differentiation, but it is possible to just use normal partial differentiation for any action of the form
\begin{equation}\label{eq:ActionLagrangian}
S_\Phi=\int d^4x\,\sqrt{-g}\,{\cal L}(\Phi, g_{\mu\nu},\partial\Phi)
\end{equation}
in which the Lagrangian (density) ${\cal L}$ does not depend on the derivatives of the metric, as in (\ref{ScalarGeneric}).  Using (\ref{eq:deltaTmunu}) we have (see below):
\begin{equation}\label{eq:Tmunupartialdiff}
{T^{\mu\nu}=\Su\left[ -2\frac{\partial\cal L}{\partial g_{\mu\nu}}+g^{\mu\nu}{\cal L}\right]}.    
\end{equation}
The same formula is valid with lower indices and no sign change anywhere, as can be easily checked.  Using these formulas we can easily recover equations (\ref{eq:TmunuScalarField1}) and (\ref{eq:TmunuScalarField2})  using normal partial differentiation.
\newline
\newline
To check (\ref{eq:Tmunupartialdiff}) we can either use variational calculus or functional differentiation. Let us start with the latter. Let us  recall that from the point of view of functional differentiation the formulas (\ref{eq:deltasqrtgandR2b}) must be interpreted conveniently. For instance,
\begin{equation}\label{eq:deltasqrtFunctional}
\frac{\delta\sqrt{-g(x)}}{\delta g^{\mu\nu}(x^{\prime})}=  -\frac12\,\sqrt{-g(x)}\,g_{\mu\nu}\,\delta^{(4)}(x-x^{\prime})\,,\ \ \ \ \frac{\delta\sqrt{-g}}{\delta g_{\mu\nu}(x^{\prime})}=  \frac12\,\sqrt{-g}\,g^{\mu\nu}\,\delta^{(4)}(x-x^{\prime})\,.
\end{equation}
Similarly, from $\delta g^{\mu\nu}=- g^{\mu\alpha}g^{\beta\nu}\delta g_{\alpha\beta}$  we obtain
\begin{equation}\label{eq:deltagmunuFunctional}
\frac{\delta g^{\mu\nu}(x)}{\delta g_{\alpha\beta}(x^{\prime})}=  -g^{\mu\alpha} g^{\beta\nu}\delta^{(4)}(x-x^{\prime})\,.
\end{equation}
Using these formulas we may functionally differentiate (\ref{eq:ActionLagrangian}):
\begin{equation}\label{eq:FunctDiffSphi}
\frac{\delta S_\Phi}{\delta g_{\mu\nu}(x^{\prime})}=\int d^4x\left(\frac{\delta\sqrt{-g(x)}}{\delta g_{\mu\nu}(x^{\prime})}\,{\cal L} +\frac{\delta{\cal L}}{\delta g_{\mu\nu}(x^{\prime})}\sqrt{-g(x)}\right)
\end{equation}
where the first term in the parenthesis is given by the second in (\ref{eq:deltasqrtFunctional}). And the second in the parenthesis  is computed with the help of (\ref{eq:deltagmunuFunctional}), rendering:
$$
  \frac{\delta{\cal L}}{\delta g_{\mu\nu}(x^{\prime})}=\frac{\delta}{\delta g_{\mu\nu}(x^{\prime})}\left[-\Su\frac12\,g^{\alpha\beta}\,\partial_{\alpha}\Phi\,\partial_{\beta}\Phi-U(\Phi)\right]
  $$
 \begin{equation}\label{eq:deltacal L}
 = -\Su\frac12\,\frac{\delta g^{\alpha\beta}(x)}{\delta g_{\mu\nu}(x^{\prime})}\,\partial_{\alpha}\Phi\,\partial_{\beta}\Phi=+\Su\frac12\,\partial^{\mu}\Phi\,\partial^{\nu}\Phi\,\delta^{(4)}(x-x^{\prime}).
\end{equation}
Integrating out the $\delta$-functions, and trivially noting that $[S1]^2=1$, we find from (\ref{eq:deltaTmunu})  and the previous results:
\begin{equation}\label{eq:Tmunuupperphi}
   T^{\mu\nu}=+\Su\frac{2}{\sqrt{-g}}\frac{\delta S_\Phi}{\delta g_{\mu\nu}}=\partial^{\mu}\Phi\,\partial^{\nu}\Phi+\Su g^{\mu\nu}{\cal L}\,.
\end{equation}
This result is exactly equivalent to (\ref{eq:Tmunupartialdiff}) because the second term is already matched, whereas the first term is just also right so:
\begin{equation}\label{eq:equiv1}
\Su\left[ -2\frac{\partial\cal L}{\partial g_{\mu\nu}}\right]=\Su\left[-2\Su\frac12\frac{\partial g^{\alpha\beta}}{\partial g_{\mu\nu}}\,\partial_\alpha\Phi\partial_\beta\Phi\right]=\partial^{\mu}\Phi\,\partial^{\nu}\Phi\,,
\end{equation}
where from $g^{\alpha\beta}=g^{\alpha\mu}g^{\beta\nu}g_{\mu\nu}$ we get $\partial g^{\alpha\beta}/\partial g_{\mu\nu}=g^{\alpha\mu}g^{\beta\nu}$. Notice that upper and lower indices variables are treated as independent variables in the differentiation process.  We could equally proceed differentiating with respect to $g^{\mu\nu}$ (i.e. with upper indices) and compute
\begin{equation}\label{eq:FunctDiffSphi2}
\frac{\delta S_\Phi}{\delta g^{\mu\nu}(x^{\prime})}=\int d^4x\left(\frac{\delta\sqrt{-g(x)}}{\delta g^{\mu\nu}(x^{\prime})}\,{\cal L} +\frac{\delta{\cal L}}{\delta g^{\mu\nu}(x^{\prime})}\sqrt{-g(x)}\right)
\end{equation}
where the first term in the parenthesis is given by the first of (\ref{eq:deltasqrtFunctional}) and has now a minus sign, and  the second is computed from
$$
  \frac{\delta{\cal L}}{\delta g^{\mu\nu}(x^{\prime})}=\frac{\delta}{\delta g^{\mu\nu}(x^{\prime})}\left[-\Su\frac12\,g^{\alpha\beta}\,\partial_{\alpha}\Phi\,\partial_{\beta}\Phi-U(\Phi)\right]
  $$
\begin{equation}\label{eq:deltacal L2}
 = -\Su\frac12\,\frac{\delta g^{\alpha\beta}(x)}{\delta g^{\mu\nu}(x^{\prime})}\,\partial_{\alpha}\Phi\,\partial_{\beta}\Phi=-\Su\frac12\,\partial_{\mu}\Phi\,\partial_{\nu}\Phi\,\delta^{(4)}(x-x^{\prime})
\end{equation}
where we have used
\begin{equation}\label{eq:deltagmunuFunctionalallup}
\frac{\delta g^{\mu\nu}(x)}{\delta g^{\alpha\beta}(x^{\prime})}=  +\delta^{\mu}_{\alpha} \delta^{\beta}_\nu\delta^{(4)}(x-x^{\prime})\,.
\end{equation}
Note the following: since all metric indices are in upper positions, the previous relation obviously does not introduce an extra sign --  in contrast to (\ref{eq:deltagmunuFunctional}).  These sign changes in each term are now compensated with the extra minus sign that the definition of $T_{\mu\nu}$ (with lower indices)  involves, and we obtain the desired result:
\begin{equation}\label{eq:Tmunudown}
T_{\mu\nu}=-\Su\frac{2}{\sqrt{-g}}\frac{\delta S_\Phi}{\delta g^{\mu\nu}}=\partial_\mu\Phi\partial_\nu\Phi+\Su g_{\mu\nu}{\cal L}\,.
\end{equation}
This result is consistent with (\ref{eq:Tmunuupperphi}) because they must be related of course through
\begin{equation}\label{eq:tensorRelation}
  T^{\mu\nu}=g^{\mu\alpha}\,g^{\nu\beta}\,T_{\alpha\beta}\,.
\end{equation}
One can entirely avoid using functional differentiation and just use variational calculus. In this case one has to vary the action and then identify $T^{\mu\nu}$ using the equivalent expression on the \textit{r.h.s.} of Eqs.\,(\ref{eq:deltaTmunu}) or (\ref{eq:deltaTmunu2}) etc.  The results are of course the same.  For example, let us variate (\ref{eq:ActionLagrangian}) with respect to the metric with \emph{upper} indices, $\delta g^{\mu\nu}$:
$$
\delta S_\Phi=\int d^4x\left[ \delta\sqrt{-g}\,{\cal L}+\sqrt{-g}\left(-\Su\frac12\delta g^{\mu\nu}\partial_{\mu}\Phi\partial_\nu\Phi\right)\right]
$$
\begin{equation}\label{eq:variation1}
=\int d^4x \sqrt{-g}\left[-\frac12\,g_{\mu\nu}{\cal L}-\Su\frac12\partial_\mu\Phi\partial_\nu\Phi\right]\delta g^{\mu\nu}\,.
\end{equation}
Comparing with the second expression of (\ref{eq:deltaTmunu2})   we must have
\begin{equation}\label{eq:comparison1}
  -\frac12\Su T_{\mu\nu}=-\frac12\,g_{\mu\nu}{\cal L}-\Su \frac12\partial_\mu\Phi\partial_\nu\Phi.
\end{equation}
Multiplying both sides for $-2[S1]$ we find the desired result:
\begin{equation}\label{eq:comparison3}
T_{\mu\nu}=\partial_{\mu}\Phi\,\partial_{\nu}\Phi+\Su g_{\mu\nu}{\cal L}\,.
\end{equation}
Had we varied the action (\ref{eq:ActionLagrangian}) with respect to the metric with \emph{lower} indices, $\delta g_{\mu\nu}$, we would have started again from
\begin{equation}\label{eq:repeat1}
\delta S_\Phi=\int d^4x \delta\sqrt{-g}\,{\cal L}+\sqrt{-g}\left[-\Su\frac12\delta g^{\mu\nu}\partial_{\mu}\Phi\partial_\nu\Phi\right]\,,
\end{equation}
but now we use $\delta\sqrt{-g}=  +\frac12\,\sqrt{-g}\,g^{\mu\nu}\delta g_{\mu\nu}$ for the first term, whereas we use
\begin{equation}\label{eq:notice1}
\delta g^{\mu\nu}\partial_{\mu}\Phi\,\partial_\nu\Phi=-\,g^{\mu\alpha}\,g^{\nu\beta}\delta g_{\alpha\beta}\,\partial_{\mu}\Phi\,\partial_\nu\Phi=-\partial^\alpha\Phi\,\partial^\beta\Phi\,\delta g_{\alpha\beta}=
-\partial^\mu\Phi\,\partial^\nu\Phi\,\delta g_{\mu\nu}\,,
\end{equation}
for the second term.
Therefore (\ref{eq:repeat1}) becomes
\begin{equation}\label{eq:variation2}
\delta S_\Phi=\int d^4x \sqrt{-g}\left[+\frac12\,g^{\mu\nu}{\cal L}+\Su\frac12\partial^\mu\Phi\partial^\nu\Phi\right]\delta g_{\mu\nu}\,.
\end{equation}
Comparing with the second expression of  (\ref{eq:deltaTmunu})  we must have
\begin{equation}\label{eq:comparison4}
  +\frac12\Su T_{\mu\nu}=+\frac12\,g_{\mu\nu}{\cal L}+\Su \frac12\partial^\mu\Phi\partial^\nu\Phi.
\end{equation}
Multiplying both sides for $2[S1]$ we find
\begin{equation}\label{eq:Tmunuupper}
T^{\mu\nu}=\partial^{\mu}\Phi\,\partial^{\nu}\Phi+\Su g^{\mu\nu}{\cal L}\,,
\end{equation}
which is consistent once more with the previous form (\ref{eq:comparison3}) because both must be related through Eq.\,(\ref{eq:tensorRelation}).
It is perhaps faster with variational calculus, but functional differentiation is a very useful tool in many other situations and it is very convenient to practice  it as well. It is actually the most rigorous procedure, see below.
\newline
\newline
Let us notice that while the functional differentiation was unambiguous all the time and provided the right answer whether differentiating with respect to $g_{\mu\nu}$ or $g^{\mu\nu}$, the variational calculation was ambiguous at the point (\ref{eq:repeat1}). In fact, we could have written the kinetic Lagrangian as  ${\cal L}=-\Su (1/2)g_{\mu\nu} \partial^\mu\Phi\,\partial_\nu\Phi$ where now the metric indices are down and the partial derivative indices up. The variation with respect to the metric could have been interpreted simply as  as $\delta{\cal L}=-\Su (1/2)\delta g_{\mu\nu} \partial^\mu\Phi\,\partial_\nu\Phi$, with no sign change as compared to (\ref{eq:notice1}).  If so, the sign of the first term of (\ref{eq:Tmunuupper}) would be negative , and hence wrong because it would very evidently violate the fundamental tensor relation (\ref{eq:Tmunuupper}).
\newline
\newline
\subsection{Signs for the energy-momentum tensor of a perfect fluid}
Let us also note that the energy-momentum tensor for a perfect fluid is given by
\begin{equation}\label{eq:TmunuPerfectFluid}
 T_{\mu\nu}=\Su p g_{\mu\nu} +(\rho+p) U_{\mu}U_{\nu}
\end{equation}
where we generically consider a fluid characterized by its energy density $\rho$ and the corresponding pressure $p$. The square of the 4-velocity vector satisfies the closure relation
\begin{equation}\label{eq:4velocitysquare}
 U_{\mu}U^{\mu} = g_{\mu\nu}U^{\mu}U^{\nu}=-[S1].
\end{equation}
The above relation is fulfilled in any frame since it is an invariant. Let us now consider the components of the energy-momentum tensor in  the proper frame, where we simply have $U^\mu=(-1,0,0,0)$. It is important to note that the components have different form depending on whether we use the mixed tensor $T^\mu_\nu$, with upper and lower indices, or the original tensor $T_{\mu\nu}$ given in (\ref{eq:TmunuPerfectFluid}) with all indices in the same position. As we shall see, the first is easy to operate because the $\delta$-Kronecker appears in it;
\begin{equation}\label{eq:TmunuPerfectFluidMixed}
  T^{\mu}_\nu=\Su p \delta^{\mu}_\nu +(\rho+p) U^{\mu}U_{\nu}\,,
\end{equation}
but on the other hand the components of this tensor do depend on the signature of the metric, i.e. on $[S1]$, whereas  $T_{\mu\nu}$ does not. This is expected since $T_{00}=T^{00}$ gives the sign of the energy density in \emph{all} conventions and therefore it cannot depend on the convention, whereas $T_0^0$ is convention-dependent.  
\newline
\newline
Let us check it.  From the definition of $U^\mu$ given above and the property $U_{\mu}U^{\mu} = -[S1]$ we have 
\begin{equation}
T^0_0 = [S1]\times{p} - (\rho + p)\times[S1], \quad  T^i_i=\Su p   
\end{equation}
consequently
\begin{equation}\label{eq:Tupmudownnu}
   T^{\mu}_\nu=
   \Su {\rm diag}(-\rho, p, p, p)\,.
\end{equation}
From here we can compute the trace:
\begin{equation}\label{eq:Trace Tmunu}
   T^{\mu}_\mu=
   \Su (3 p-\rho)\,,
\end{equation}
which does indeed depend on the metric signature since it is not an observable.  Notice, however,  from (\ref{eq:S1})  and (\ref{eq:Tupmudownnu}) that
$$
T^{00}=g^{\mu 0} T^0_\mu=g^{00} T^0_0=\Su\left(-T^0_0\right)=\Su\Su\rho =\rho \,,
$$
and similarly
$$
T^{ii}=g^{\mu i}T^i_{\mu}=g^{ii}T^i_i=\Su T^i_i=\Su\Su p=p\,.
$$
Therefore the energy-momentum tensor with upper indices in the proper frame reads
\begin{equation}\label{eq:Tupmuupnu}
   T^{\mu\nu}=
    {\rm diag}(\rho, p, p, p)\,,
\end{equation}
which is fully independent of the metric convention. These are the observable quantities. Similarly
\begin{equation}\label{eq:Tdownmudownnu}
   T_{\mu\nu}=
    {\rm diag}(\rho, p, p, p)\,.
\end{equation}
owing to the tensor property Eq.\,(\ref{eq:tensorRelation}), which involves twice the metric tensor.
\newpage
\newpage\phantom{123}
\newpage\phantom{123}

\section{Cosmological perturbations}\label{Appendix_B}
It is a fact that our Universe is really far from being completely homogeneous and isotropic. Only on large scales ($\sim$100 Mpc) the description of the expanding Universe in terms of quantities that only depend on time, remains valid. As we have explained in the Introduction the aforementioned scenario is well described in terms of the Friedmann-Lema\^\i tre-Robertson-Walker (FLRW) metric and the corresponding Friedmann's and pressure equations. However if we want to go a step further, the previous description is no longer valid and we need to employ the perturbation theory in order to explain the observed structures in the Universe. So, how were these structures formed ? Let us answer this question with a brief summary, without going into the mathematical details, of this fascinating process based on a simple concept, the gravitational instability. It is commonly believed that after the Big Bang, the Universe undergone a phase of rapidly expansion called inflation, where the space expanded itself at an approximately exponential rate. While it is true that we do not have a solid evidence of this theory, the absence of alternatives and the fact that it provides beautiful solutions for some worrisome problems, like the observed flatness, the homogeneity and isotropy of the current Universe plus the absence of exotic relics, makes it very appealing for cosmologists. One of the most important predictions of inflation is the generation of quantum-mechanical fluctuations when the relevant scales were causally connected. These tiny fluctuations were amplified by inflation and seeds the structure formation that we can observe. The end of inflation is called reheating. This process is poorly understood, nevertheless the most accepted theory stands that along this period the particle responsible for the inflationary period (whatever it is) decays into a hot thermal plasma of other particles. Therefore, at this stage the Universe was made up of a dense plasma of photons, electrons, baryons and dark matter (DM). Because of the radiation pressure, structure formation in the baryonic sector was inhibited at that time, nevertheless since DM does not interact except gravitationally  remained in the center of the overdensities. The electromagnetic interaction forced baryons and photons to move together outward from those overdense regions, creating spherical sound waves which are the origin of the baryonic acoustic oscillations (BAO). As the Universe expanded it cooled, so at some point the temperature was cool enough to allow protons to capture electrons and form neutral hydrogen atoms. This is known as the recombination time and as can be imagined it was of tremendous importance in the cosmic history. The interaction between photons and baryons ceased and the first ones started to stream freely making up, what we know as, the cosmic microwave background radiation. Baryons fell into the potential wells, previously created by DM, forming the first structures in the Universe. The Universe continued to expand and cool and eventually (well within the matter dominated epoch) the gravitational attraction exerted by baryons and DM allows the structure formation started to grow. This is the starting point where we will focus on this appendix. 
\newline
\newline

\subsection{Cosmological perturbation equations for dark energy models}
In this section we present the basic steps that must be followed to obtain the set of perturbation equations at first order. We will keep the FLRW metric as the background metric (it will be denoted as $\bar{g}_{\mu\nu}$) and for the rest of the chapter we stick to the convention (+,+,+) explained in the previous appendix. We introduce a tiny perturbation in the metric in the following way:
\begin{equation}\label{general_perturbed_metric_Appendix_B}
ds^2 = g_{\mu\nu}dx^{\mu}dx^{\nu} = (\bar{g}_{\mu\nu} + \delta{g_{\mu\nu}})dx^{\mu}dx^{\nu} = a^2(\eta_{\mu\nu} + {h_{\mu\nu}})dx^{\mu}dx^{\nu},
\end{equation}
where $\eta_{\mu\nu} = {\rm diag}(-1,+1,+1,+1)$ is the Minkowski metric. We have to bear in mind that the elements of the perturbed metric $\delta{g_{\mu\nu}}$ (or equivalently ${a^2}h_{\mu\nu}$) are small in comparison with the background components $\bar{g}_{\mu\nu}$. The metric field $g_{\mu\nu}$ must be symmetric, therefore $\delta{g_{\mu\nu}}$ also has to be. This means that the total number of {\it d.o.f.} is reduced from 16 to 10. The General Relativity (GR) field equations are invariant under a general coordinate change, as a consequence, we can fix 4 of the remaining {\it d.o.f.} by fixing the gauge. It can be shown that in a completely general way the perturbed metric can be decomposed as 
\begin{equation}
ds^2 = a^2\left[ -(1+2\Phi)d\eta^2 + \omega_i{d\eta{dx^{i}}} + (\delta_{ij} + h_{ij})dx^i{dx^{j}}\right],
\end{equation}
where $\Phi$ is a scalar function, $\omega_i$ is a 3-vector and $h_{ij}$ is a traceless tensor. Remember that $\eta$ represents the conformal time and it can be related with the cosmic time through the expression $\eta = \int^t_0 dt^{\prime}/a(t^\prime)$. There are three types of perturbation, to wit: scalar, vector and tensor. Here we are only interested in the first type. The vector perturbations are suppressed in the early Universe whereas the tensor ones contribute as a gravitational waves to the B-mode of the CMB polarizations. The scalar perturbations are the only ones coupled to the matter perturbations and this is why here we only focus on this particular type. The perturbed metric can be rewritten as:
\begin{equation}
ds^2 = a^2\left[ -(1+2\Phi)d\eta^2 + \partial_{i}E{d\eta{dx^{i}}} + ((1 + 2\Psi)\delta_{ij} + D_{ij}B)dx^i{dx^{j}}  \right]
\end{equation}
and now $\Phi$, $E$, $\Psi$ and $B$ are all scalar functions. We can now fix the gauge by imposing some conditions over these functions. For instance, if we want to work in the synchronous gauge we must set up $\Phi$ = $E$ = 0. On the other hand if we want to work in the conformal Newtonian gauge then we impose $E$ = $B$ = 0. Let us focus on the last one, for which the final form of the perturbed metric can be written as:
\begin{equation}\label{PerturbedMetric_Appendix_B}
ds^2 = a^2\left[ -(1+2\Phi)d\eta^2  + (1 + 2\Psi)\delta_{ij}dx^i{dx^{j}}  \right].
\end{equation}
Once the perturbed metric, which is a key element in the perturbation theory, is clearly determined we can start the process that will lead us to obtain the perturbed cosmological equations. 
\newline
\newline
The following comment about the notation employed in this appendix is in order. When we write a quantity (could be a scalar, a vector or a tensor) generally denoted as $X$ and we do not add nor a bar neither a $\delta$ right in front of the term, we are referring to a quantity which is the sum of a background part $\bar{X}$ plus a perturbed part $\delta{X}$. For instance in the case of the metric we have (as it can be observed from \eqref{general_perturbed_metric_Appendix_B}) $g_{\mu\nu} = \bar{g}_{\mu\nu} + \delta{g_{\mu\nu}}$.
\newline
\newline
To derive the first-order Einstein field equations we need to introduce a perturbation not only in the Einstein tensor $G_{\mu\nu}$ but also in the energy-momentum tensor $T_{\mu\nu}$. We end up with the set of equations 
\begin{equation}
\bar{G}_{\mu\nu} + \delta{G}_{\mu\nu} = 8\pi{G_N}\left(\bar{T}_{\mu\nu} + \delta{T_{\mu\nu}}\right).
\end{equation}
From now on, we are going to drop the pure background terms since in this part we are only interested in the perturbation equations. It turns out really convenient to write down the covariant elements of the perturbed metric and the corresponding contravariant elements: 
\begin{equation}
\begin{split}
g_{00}=-a^2 (1+2\Phi)\qquad g_{ij}=a^2 (1+2\Psi)\delta_{ij} \quad \quad g^{00}=-\frac{1}{a^2}(1-2\Phi) \qquad g^{ij}=\frac{1}{a^2}(1-2\Psi)\delta^{ij}.
\end{split}
\end{equation}
Once we have the entries of the perturbed metric we can move forward and compute the perturbed Christoffel symbols $\delta\Gamma^{\mu}_{\nu\lambda}$ by using the expression
\begin{equation}  
\delta\Gamma^{\mu}_{\nu\lambda} = \frac{1}{2}\delta{g^{\mu\alpha}}(   \partial_{\lambda}\bar{g}_{\alpha\nu}  + \partial_{\nu}\bar{g}_{\alpha\lambda} - \partial_{\alpha}\bar{g}_{\nu\lambda}   ) + \frac{1}{2}\bar{g}^{\mu\alpha}( \partial_{\lambda}\delta g_{\alpha\nu}  + \partial_{\nu}\delta g_{\alpha\lambda} - \partial_{\alpha}\delta g_{\nu\lambda}  ).
\end{equation}
The Christoffel symbols different from zero, in the Newtonian gauge, are: 
\begin{equation}
\Gamma^0_{00}=\mathcal{H}+\Phi^\prime \qquad \Gamma^0_{0i} = \Gamma^i_{00}=\partial_i\Phi \qquad \Gamma^0_{ij}=\delta_{ij}[\mathcal{H}(1+2\Psi-2\Phi)+\Psi^\prime]
\end{equation}
$$\Gamma^i_{j0}=\delta^i_j (\mathcal{H}+\Psi^\prime) \qquad \Gamma^i_{jl}=\delta^i_j\partial_l\Psi+\delta^i_l\partial_j\Psi-\delta_{jl}\partial_i\Psi .$$
Where $\mathcal{H} = \frac{1}{a}\frac{da}{d\eta}$ represents the Hubble function in conformal time and the primes denote derivatives with respect to that variable. We can repeat exactly the same procedure to obtain from 
\begin{equation}
\delta{R_{\mu\nu}} = \partial_{\alpha}\delta{\Gamma^{\alpha}_{\mu\nu}} - \partial_{\nu}\delta{\Gamma^{\alpha}_{\mu\alpha}} + \delta\Gamma^{\alpha}_{\mu\nu}\bar{\Gamma}^{\beta}_{\alpha\beta} + \bar{\Gamma}^{\alpha}_{\mu\nu}\delta\Gamma^{\beta}_{\alpha\beta} - \delta\Gamma^{\alpha}_{\mu\beta}\bar{\Gamma}^{\beta}_{\alpha\nu} - \bar{\Gamma}^{\alpha}_{\mu\beta}\delta\Gamma^{\beta}_{\alpha\nu}
\end{equation}
the non-null components of the Ricci tensor: 
\begin{equation}
R_{00}=-3\mathcal{H}^\prime+\nabla^2\Phi-3\Psi^{\prime\prime}+3\mathcal{H}(\Phi^\prime-\Psi^\prime), \qquad R_{0i}=-2\partial_i\Psi^\prime+2\mathcal{H}\partial_i\Phi,
\end{equation}
$$R_{ij}=-\partial_i\partial_j(\Psi+\Phi)+\delta_{ij}\left[(2\mathcal{H}^2+\mathcal{H}^\prime)(1+2\Psi-2\Phi)-\nabla^2\Psi+\Psi^{\prime\prime}+5\mathcal{H}\Psi^\prime-\mathcal{H}\Phi^\prime\right].$$
Contracting the indices of the previous tensor we are able to compute the Ricci scalar
\begin{equation}
{a^2}R=6(\mathcal{H}^2+\mathcal{H}^\prime)(1-2\Phi)-2\nabla^2(\Phi+2\Psi)+6\Psi^{\prime\prime}-6\mathcal{H}\Phi^\prime+18\mathcal{H}\Psi^\prime.
\end{equation}
Finally we are in a position to calculate the components of Einstein's tensor 
\begin{equation}\label{perturbed_Einstein_Tensor_Appendix_B}
G_{00}=3\mathcal{H}^2+6\mathcal{H}\Psi^\prime-2\nabla^2\Psi,
\end{equation}
$$G_{ij}=-\partial_i\partial_j(\Psi+\Phi)+\delta_{ij}\left[-(\mathcal{H}^2+2\mathcal{H}^\prime)(1+2\Psi-2\Phi)+2\mathcal{H}(\Phi^\prime-2\Psi^\prime)+\nabla^2(\Psi+\Phi)-2\Psi^{\prime\prime}\right],$$
$$G_{0i}=-2\partial_i\Psi^\prime+2\mathcal{H}\partial_i\Phi.$$
It is time now to obtain the perturbed expression for the energy-momentum tensor of a perfect fluid, which contains the contribution of the different species considered. The starting point is the expression valid at the background level
\begin{equation}
\bar{T}_{\mu\nu} = (\bar{\rho} + \bar{p})\bar{U}_{\mu}\bar{U}_\nu + \bar{p}{\bar{g}_{\mu\nu}},
\end{equation}
by plugging the corresponding perturbations we can get
\begin{equation}
\delta{T}_{\mu\nu} = (\bar{\rho} + \bar{p})(\delta{U}_{\mu}\bar{U}_\nu + \bar{U}_\mu\delta{U}_{\nu} ) + (\delta\rho + \delta{p})\bar{U}_{\mu}\bar{U}_\nu + \bar{p}\delta{g}_{\mu\nu} + \delta{p}\bar{g}_{\mu\nu}.
\end{equation}
Where we denote the total energy density and pressure, at the background level, by $\bar{\rho}$ and $\bar{p}$ respectively. We need to obtain the components of the perturbed 4-velocity. To do so, we only require the definition of the total 4-velocity $U^\mu = dx^{\mu}/d\tau$, where $\tau$ is the proper time and the closure relation $U^{\mu}U_{\mu} = -1$
\begin{equation}
g_{\mu\nu}\frac{dx^\mu}{d\tau}\frac{dx^\nu}{d\tau} = \left(\frac{d\eta}{d\tau}\right)^2\left[g_{00} + g_{ij}\frac{dx^i}{d\eta}\frac{dx^j}{d\eta}\right] = -1, 
\end{equation}
where $v^i\equiv dx^i/d\eta$ is defined as the peculiar velocity with respect to the comoving frame. If we assume that the peculiar velocities are small quantities, we can get the perturbed expression for the first component of the 4-velocity vector
\begin{equation}
\frac{d\eta}{d\tau} = \frac{1}{a}(1-\Phi) + \mathcal{O}(2). 
\end{equation}
Regarding the spatial components we only need to employ the chain rule together with the aforementioned expression for the peculiar velocities
\begin{equation}
\frac{dx^i}{d\tau} = \frac{dx^i}{d\eta}\frac{d\eta}{d\tau} = \frac{v^i}{a} + \mathcal{O}(2). 
\end{equation}
Once we have computed the different components we can write the expression for the 4-velocity, at first order, in the covariant and contravariant form:
\begin{equation}
U^{\mu} = \left[\frac{1}{a}(1-\Phi), \frac{v^i}{a}\right] \quad U_\mu = [-a(1 + \Phi), av^i].
\end{equation}
For convenience we are going to rewrite the 3-vector of the peculiar velocities using the Helmholtz decomposition $\vec{v} = \vec{v}_\parallel + \vec{v}_\perp$, where $\vec{\nabla} \times \vec{v}_{\parallel} = 0 $ and $\vec{\nabla}\cdot\vec{v}_{\perp} = 0$. As we stated at the beginning of this section we are only interested in scalar perturbations, therefore, only $\vec{v}_\parallel = \vec{\nabla}{v}$ plays a role in the calculations. For the sake of convenience we list the non-null components of the total energy-momentum tensor:
\begin{align}
T_{00} &= a^2\sum_n\left[(1+2\Phi)\bar{\rho}_n + \delta\rho_n\right]\label{eq:T_00_Appendix_B} \\ 
T_{ij} &= a^2\sum_n\left[(1+2\Psi)\bar{p}_n + \delta{p}_n\right]\delta_{ij}\label{eq:T_ij_Appendix_B} \\ 
T_{0i} &= -a^2\partial_{i}\sum_n (\bar{\rho}_n + \bar{p}_n)v_n\label{eq:T_0i_Appendix_B},
\end{align}
where the subindex $n$ runs for all components considered. Once we have computed all the required elements we can write down the perturbation equations for the different DE models under study. Before doing so, we are going to perform a Fourier expansion of the perturbed quantities, following the transformation between the position-space and the momentum-space: 
\begin{equation}
f(\eta,\vec{x}) = \frac{1}{(2\pi)^3}\int f_k(\eta,\vec{k})e^{i\vec{k}\cdot\vec{x}}d^3k.
\end{equation}
The above expression means that we decompose all of the perturbed quantities considered in the equations, as a sum of plane waves. Due to the fact that we have restricted the analysis at first order, we do not have to worry about the mixing between the different modes. They all $f_k(\eta, \vec{k})$ evolve independently. 
\newline
\newline
It is finally time to write down the perturbed equations in momentum space derived from Einstein's field equations:
\begin{itemize}
\item $\delta{G_{00}} = 8\pi{G_N}\delta{T_{00}}$
\begin{equation}
k^2\Psi + 3\mathcal{H}\left(\Psi^{\prime}-\mathcal{H}\Phi\right) = 4\pi{G_N}a^2\sum_n \delta\rho_n
\end{equation}
\item $\delta{G_{0i}} = 8\pi{G_N}\delta{T_{0i}}$ 
\begin{equation}
\Psi^{\prime}-\mathcal{H}\Phi = 4\pi{G_N}a^2\sum_n\left(\bar{\rho}_n + \bar{p}_n\right)v_n
\end{equation} 
\item $\delta{G_{ij}} = 8\pi{G_N}\delta{T_{ij}}$ \quad ($i\neq j$)
\begin{equation}\label{Psi_and_Phi_Appendix_B}
\Psi = -\Phi     
\end{equation}
\item $\delta{G_{ij}} = 8\pi{G_N}\delta{T_{ij}}$\quad  ($i=j$)
\begin{equation}
\Psi^{\prime\prime} - \Phi\left(\mathcal{H}^2 + 2\mathcal{H}^{\prime}\right) - \mathcal{H}\left(\Phi^\prime - 2\Psi^\prime\right) = -4\pi{G_N}a^2\sum_n \delta{p}_n
\end{equation}
\item $\nabla^{\mu}T_{\mu{0}} = 0$
\begin{equation}\label{tensor_conservation_0_Appendix_B}
\sum_n \left[ \delta\rho^{\prime}_n + 3\mathcal{H}\left(\delta\rho_n + \delta{p_n}\right) + \left(\bar{\rho}_n + \bar{p}_n\right)\left(  3\Psi^{\prime} -k^{2}v_n\right)\right] = 0     
\end{equation}
\item $\nabla^{\mu}T_{\mu{i}} = 0$
\begin{equation}\label{tensor_conservation_i_Appendix_B}
\sum_n \left[ \left(\bar{\rho}^{\prime}_n + \bar{p}^{\prime}_n \right)v_n + \left(\bar{\rho}_n + \bar{p}_n \right)\left(4\mathcal{H}v_n + v^{\prime}_n + \Phi\right)  + \delta{p_n}\right] = 0.   
\end{equation}
\end{itemize}
Although not all the equations listed are independent, they are all useful in one way or another. If we look at \eqref{Psi_and_Phi_Appendix_B} we can see that it is only valid if we consider that the energy budget is composed just by perfect fluids. We do know that this is not accurate if we consider, either the interaction between photons and baryons or we deal with the full treatment of the massive neutrinos. Taking into account that we are considering equations valid at the matter dominated epoch, well after the decoupling time when the interaction between photons and baryons was relevant, as a good approximation we can work with the assumption that both of them evolve independently. As far as neutrinos are concerned, as a first approximation we can consider that they behave as if they were a perfect fluid. As a consequence the relation $\Psi+\Phi\simeq0$ remains valid.
\newline
\newline
Unfortunately it is not possible to find an analytical solution for this set of differential equations, nevertheless, we can find solutions under some approximations. In this section we are going to consider physical modes much more shorter than the Hubble radius, $1/\mathcal{H}$, since they are the only ones that can be observed. What is more, since the structure formation started within the matter dominated epoch, well after the recombination time, we drop the radiation contributions, namely we make the approximation $\bar{\rho}_r/\bar{\rho}_m\simeq 0 $. So, considering the before mentioned conditions for the modes, $k^2\gg\mathcal{H}^2$, we can find the following set of equations:
\begin{align}
&k^2\Psi -4\pi{G_N}a^2\left(\delta\rho_m + \delta\rho_\Lambda\right)  = 0 \\
&\sum_{i=m,\Lambda} \left[ \delta\rho^{\prime}_i + 3\mathcal{H}\left(\delta\rho_i + \delta{p_i}\right) + \left(\bar{\rho}_i + \bar{p}_i\right)\left(  3\Psi^{\prime} -k^{2}v_i\right)\right] = 0 \\
&\sum_{i = m,\Lambda} \left[ \left(\bar{\rho}^{\prime}_i + \bar{p}^{\prime}_i \right)v_i + \left(\bar{\rho}_i + \bar{p}_i \right)\left(4\mathcal{H}v_i + v^{\prime}_i + \Phi\right)  + \delta{p_i}\right] = 0. \\
\end{align}
As usual $\bar{\rho}_m = \bar{\rho}_b + \bar{\rho}_{dm} + \bar{\rho}^{\rm NR}_{\nu}$ (and of course $\delta\rho_m = \delta\rho_b + \delta\rho_{dm} + \delta\rho^{\rm NR}_\nu$), consequently we get $v_m = \left(\bar{\rho}_b{v_b} + \bar{\rho}_{dm}v_{dm} + \bar{\rho}^{\rm NR}_{\nu}v^{\rm NR}_{\nu}\right)/\bar{\rho}_m $. Where $\rho^{\rm NR}_\nu$ stands for the contribution of massive neutrinos in the non-relativistic regime. It is worthwhile to note that we make no assumptions about DE, since we do not consider any particular value for $w_\Lambda = \bar{p}_\Lambda/\bar{\rho}_\Lambda$ and we do not neglect the contribution of $\delta\rho_\Lambda$. In order to compare the theoretical predictions for a given model with the large scale structure (LSS) observational data it turns out essential to compute the total matter density contrast, defined as:
\begin{equation}\label{total_matter_density_contrast_Appendix_B}
\delta_m = \frac{\delta\rho_b + \delta\rho_{dm} + \delta\rho^{\rm NR}_{\nu}}{\bar{\rho}_m}.    
\end{equation}
It goes without saying that the particularities of the models under study, like the equation of sate (EoS) parameter considered or some sort of possible interactions between the different species, will affect the value of \eqref{total_matter_density_contrast_Appendix_B}. Therefore when the background data is not enough to distinguish between two given models, the large scale structure (LSS) data can help by pointing out departures in the theoretical predictions of the models at the perturbation level. 
\newline
\newline
\subsection{Perturbation equations for the scalar field models}\label{Appendix_B_scalar_fields}
In this section our aim is to write down the most important perturbation equations for the non-interactive scalar field models. Throughout this thesis we have denoted the 1-dimensional scalar field as $\Phi$, however, due to the fact that here we are going to deal with the perturbed metric, which in the Newtonian gauge already includes the term $\Phi$, we are forced to change slightly the notation. Thus we refer to the scalar field and the corresponding perturbation as: 
\begin{equation}\label{perturbed_scalar_field_Appendix_B}
\Phi_S = \bar{\Phi}_S + \delta\Phi_S.     
\end{equation}
Consequently the action at the background level must be rewritten as
\begin{equation}\label{SFM_action_perturbations_Appendix_B}
S[\bar{\Phi}_S] = -\int d^4x\sqrt{-g}\left(\frac{1}{2}\bar{g}^{\alpha\beta}\partial_\alpha\bar{\Phi}_S\partial_\beta\bar{\Phi}_S + U(\bar{\Phi}_S)\right). 
\end{equation}
Considering \eqref{SFM_action_perturbations_Appendix_B}, employing \eqref{perturbed_scalar_field_Appendix_B} and taking into account the perturbations of the metric $g_{\mu\nu} = \bar{g}_{\mu\nu} + \delta{g}_{\mu\nu}$, we can work out the expression for the symmetric energy-momentum tensor of a scalar field: 
\begin{equation}
T^{\Phi_S}_{\mu\nu} = \bar{T}^{{\Phi}_S}_{\mu\nu} + \delta{T}^{\Phi_S}_{\mu\nu} 
\end{equation}
where 
\begin{align}
\bar{T}^{{\Phi}_S}_{\mu\nu} &= \partial_\mu\bar{\Phi}_S\partial_\nu\bar{\Phi}_S + \bar{g}_{\mu\nu}\left[\frac{\bar{\Phi}^{\prime{2}}_S}{2a^2} - U(\bar{\Phi}_S)\right]\\
\delta{T}^{\Phi_S}_{\mu\nu} &= \partial_\mu\bar{\Phi}_S\partial_\nu\delta\Phi_S + \partial_\mu\delta\Phi_S\partial_\nu\bar{\Phi}_S + \bar{g}_{\mu\nu}\left[\frac{\bar{\Phi}^{\prime}_{S}\delta\Phi^{\prime}_S}{a^2} - \frac{\Psi\bar{\Phi}^{\prime{2}}_{S}}{a^2} - \frac{\partial{U(\bar{\Phi}_S)}}{\partial\bar{\Phi}_S}\delta\Phi_S\right]\\
&+ \delta{g}_{\mu\nu}\left[\frac{\bar{\Phi}^{\prime{2}}_S}{2a^2} - U(\bar{\Phi}_S)\right]\nonumber.
\end{align}
It will be useful to write down the components of the total energy-momentum tensor $T^{\Phi_S}_{\mu\nu}$:
\begin{align}
T^{\Phi_S}_{00} &=   \frac{\bar{\Phi}^{\prime{2}}_S}{2} + a^2{U(\bar{\Psi}_S)} + \bar{\Phi}^{\prime}_{S}\delta\Phi^{\prime}_S +  a^2\frac{\partial{U(\bar{\Phi}_S)}}{\partial\bar{\Phi}_S}\delta\Phi_S + 2a^2\Psi{U(\bar{\Psi}_S)}\\
T^{\Phi_S}_{ij} &= \left[\frac{\bar{\Phi}^{\prime{2}}_S}{2} - a^2{U(\bar{\Psi}_S)} + \bar{\Phi}^{\prime}_{S}\delta\Phi^{\prime}_S +  a^2\frac{\partial{U(\bar{\Phi}_S)}}{\partial\bar{\Phi}_S}\delta\Phi_S\right.\\
&\left.+ (\Psi-\Phi)\bar{\Phi}^{\prime{2}}_{S} -a^2\frac{\partial{U(\bar{\Phi}_S)}}{\partial\bar{\Phi}_S}\delta\Phi_S   - 2a^2\Psi{U(\bar{\Psi}_S)} \right]\delta_{ij}\nonumber\\
T^{\Phi_S}_{0i} &= \partial_i\left(\bar{\Phi}^{\prime}_{S}\delta\Phi^{\prime}_S \right).
\end{align}
Once we have the non-null components of the energy-momentum tensor in addition to the components of the Einstein tensor \eqref{perturbed_Einstein_Tensor_Appendix_B} and the matter energy momentum tensor $T^{m}_{\mu\nu}$ \eqref{eq:T_00_Appendix_B}-\eqref{eq:T_0i_Appendix_B} (considering that here the sum over the different components only contains the contribution of the non-relativistic matter) we can list the Einstein equations in momentum space:
\begin{itemize}
\item $\delta{G}_{00} = 8\pi{G_N}\left(\delta T^{m}_{00} + \delta{T^{\Phi_S}_{00}} \right)$
\begin{equation}
k^2\Psi + 3\mathcal{H}\left(\Psi^{\prime}-\mathcal{H}\Phi\right) = 4\pi{G_N}a^2\left[\delta\rho_m + \frac{\bar{\Phi}^{\prime}_S\delta\Phi^{\prime}_{S}}{a^2} + \frac{\partial{U(\bar{\Phi}_S)}}{\partial\bar{\Phi}_{S}}\delta\Phi_S - \frac{\bar{\Phi}^{\prime{2}}_{S}\Phi}{a^2}  \right] 
\end{equation}
\item $\delta{G}_{0i} = 8\pi{G_N}\left(\delta T^{m}_{0i} + \delta{T^{\Phi_S}_{0i}} \right)$
\begin{equation}
\Psi^{\prime} - \mathcal{H}\Phi = 4\pi{G_N}a^2\left((\bar{\rho}_m + \bar{p}_m)v_m -\frac{\bar{\Phi}^{\prime}_S\delta\Phi_S}{a^2}\right)   
\end{equation}
\item $\delta{G}_{ij} = 8\pi{G_N}\left(\delta T^{m}_{ij} + \delta{T^{\Phi_S}_{ij}} \right)$ \quad ($i\neq j$)
\begin{equation}
\Psi = -\Phi     
\end{equation}
\item $\delta{G}_{ij} = 8\pi{G_N}\left(\delta T^{m}_{ij} + \delta{T^{\Phi_S}_{ij}} \right)$ \quad ($i= j$)
\begin{equation}
\Psi^{\prime\prime} - \Phi\left(\mathcal{H}^2 + 2\mathcal{H}^{\prime}\right) - \mathcal{H}\left(\Phi^\prime - 2\Psi^\prime\right) = -4\pi{G_N}a^2\left(\delta{p_m} +  \frac{\bar{\Phi}^{\prime}_S\delta\Phi^{\prime}_{S}}{a^2} - \frac{\partial{U(\bar{\Phi}_S)}}{\partial\bar{\Phi}_{S}}\delta\Phi_S - \frac{\bar{\Phi}^{\prime{2}}_{S}\Phi}{a^2}  \right)    
\end{equation}
\item $\nabla^{\mu}{T^{\Phi_S}}_{\mu{0}}=0$
\begin{align}
\frac{\bar{\Phi}^{\prime{2}}_{S}}{a^2}\left(3\Psi^{\prime}-\Phi^{\prime} -4\mathcal{H}\Phi\right)  &+ \frac{\bar{\Phi}^{\prime}_{S}}{a^2}\left(4\mathcal{H}\delta\Phi^{\prime}_S -2\Phi\bar{\Phi}^{\prime\prime}_S + k^2\delta\Phi_S + \delta\Phi^{\prime\prime}_S\right)\label{00_conservation_SFM}\\   + &\frac{d}{d\eta}\left(\frac{\partial{U(\bar{\Phi}_S)}}{\partial\bar{\Phi}_S}\right)\delta\Phi_S + \delta\Phi^{\prime}_S\left(\frac{\partial{U(\bar{\Phi}_S)}}{\partial\bar{\Phi}_S} + \frac{\bar{\Phi}^{\prime\prime}_{S}}{a^2}\right) = 0. \nonumber
\end{align}
\end{itemize}
It is important to note that at scales deep inside the horizon, where remember the condition $k^2\gg \mathcal{H}^2$ is fulfilled, the perturbation of the scalar field is strongly suppressed, $k^2\delta\Phi_S\simeq0$ as it can be deduced from  \eqref{00_conservation_SFM}. As a consequence the differential equation for $\delta_m$ at subhorizon scales turns out to be the same as in the $\Lambda$CDM case, with of course the replacement $\rho^0_\Lambda\rightarrow \rho_{\Phi_S}$. 
\newline
\newline
\subsection{Perturbation equations for the Brans-Dicke model with cosmological constant}\label{Appendix_B_BD}
We are now interested in obtaining the perturbation equations for the Brans-Dicke model when a cosmological constant is included in the action. As it can be easily understood by taking a quick look at the equations \eqref{eq:Friedmann_equation_varphi_BD_article} and \eqref{eq:Pressure_equation_varphi_BD_article} the process will be a little longer than the previous cases, however, the philosophy remains untouched, namely: we introduce the usual perturbations $\delta{g}_{\mu\nu}$, $\delta\rho_n$, $\delta{p}_n$ in addition to the $\delta\varphi$, which corresponds to the perturbation of the BD scalar field, in the field equations and we obtain the perturbed equations up to the first order. First of all, we provide the formulas of some expressions depending on $\varphi$ that will be useful in subsequent computations:
\begin{align}
a^2\Box\varphi &=  -\bar{\varphi}^{\prime\prime}-2\mathcal{H}\bar{\varphi}^\prime-\delta\varphi^{\prime\prime}+2\bar{\varphi}^{\prime\prime}\Phi+\nabla^2\delta\varphi-2\mathcal{H}\delta\varphi^\prime+\bar{\varphi}^\prime(\Phi^\prime-3\Psi^\prime)+4\mathcal{H}\Phi\bar{\varphi}^\prime, \\
\partial_\alpha\varphi\partial^\alpha\varphi &=-\frac{(\bar{\varphi}^{\prime})^2}{a^2}(1-2\Phi)-\frac{2}{a^2}\bar{\varphi}^\prime\delta\varphi^\prime,\\
\nabla_0\nabla_0\varphi &= \bar{\varphi}^{\prime\prime}+\delta\varphi^{\prime\prime}-\bar{\varphi}^{\prime}(\mathcal{H}+\Phi^\prime)-\mathcal{H}\delta\varphi^\prime,\\
\nabla_i\nabla_j\varphi &=-k_i{k}_j\delta\varphi-\bar{\varphi}^\prime\delta_{ij}[\mathcal{H}(1+2\Psi-2\Phi)+\Psi^\prime]-\mathcal{H}\delta\varphi^\prime\delta_{ij},\\
\nabla_i\nabla_0\varphi &=i{k_i}(\delta\varphi^\prime-\bar{\varphi}^\prime\Phi-\mathcal{H}\delta\varphi).
\end{align}
Since the local energy conservation equation \eqref{eq:FullConservationLaw_BD_chapter} has exactly the same form than the one of the standard model, the equations $\nabla^{\mu}T_{\mu{0}}=0$ and $\nabla^{\mu}T_{\mu{i}}=0$, do not change with respect to the ones displayed in \eqref{tensor_conservation_0_Appendix_B} and \eqref{tensor_conservation_i_Appendix_B}, respectively. The field equations, written in the momentum space, that have been modified because of the extra {\it d.o.f.}, are: 
\begin{itemize}
\item For $\mu=\nu=0$
\begin{equation}
\begin{split}
\bar{\varphi}(6\mathcal{H}\Psi^\prime+2k^2\Psi)+3\mathcal{H}^2\delta\varphi+&k^2\delta\varphi+3\mathcal{H}\delta\varphi^\prime+3\Psi^\prime\bar{\varphi}^\prime\\
  {}&+\frac{\oD}{2\bar{\varphi}}\left[\frac{\delta\varphi}{\bar{\varphi}}(\bar{\varphi}^\prime)^2-2\bar{\varphi}^\prime\delta\varphi^\prime\right]=8\pi G_N a^2\sum_n(2\Phi\bar{\rho}_n+\delta\rho_n)\,.\label{eq:gField00Mom}
\end{split}
\end{equation}
\item $\mu = i$ and $\nu = j$  \quad ($i\neq j$)
\begin{equation}\label{eq:gFieldij1Mom}
\Psi+\Phi=-\frac{\delta\varphi}{\bar{\varphi}}.    
\end{equation}
\item $\mu = i$ and $\nu = j$  \quad ($i = j$)
\begin{equation}
\begin{split}
&(\Phi^\prime-2\Psi^\prime)(2\mathcal{H}\bar{\varphi}+\bar{\varphi}^\prime)+(\Phi-\Psi)\left[2\bar{\varphi}(\mathcal{H}^2+2\mathcal{H}^\prime)+2\bar{\varphi}^{\prime\prime}+2\mathcal{H}\bar{\varphi}^\prime+\frac{\oD}{\bar{\varphi}}(\bar{\varphi}^\prime)^2\right] \\
  {}&-2\Psi^{\prime\prime}\bar{\varphi}-\delta\varphi^{\prime\prime}-\mathcal{H}\delta\varphi^\prime-\delta\varphi(\mathcal{H}^2+2\mathcal{H}^\prime)+\frac{\oD}{2\bar{\varphi}}\left[\frac{\delta\varphi}{\bar{\varphi}}(\bar{\varphi}^\prime)^2-2\bar{\varphi}^\prime\delta\varphi^\prime\right]= 8\pi G_N a^2\sum_n(2\bar{p}_n\Psi+\delta p_n)\,. \label{eq:gFieldij2Mom}
\end{split}
\end{equation}    
\item The perturbed Klein-Gordon equation
\begin{equation}\label{eq:BDfield2Mom}
-\delta\varphi^{\prime\prime}+2\bar{\varphi}^{\prime\prime}\Phi-k^2\delta\varphi-2\mathcal{H}\delta\varphi^\prime+\bar{\varphi}^\prime(\Phi^\prime-3\Psi^\prime)+4\mathcal{H}\Phi\bar{\varphi}^\prime=\frac{8\pi G_N}{3+2\oD}a^2\sum_n(3\delta p_n-\delta\rho_n).
\end{equation}
\end{itemize}
As has been done for the other models, we study now the evolution of matter perturbations at deep subhorizon scales, i.e. $k^2\gg \mathcal{H}^2$. In this limit, Eq. \eqref{tensor_conservation_0_Appendix_B} boils down to
\begin{equation}
\delta_m^\prime+\theta_m+3\Psi^\prime=0\,,
\end{equation}
and \eqref{tensor_conservation_i_Appendix_B} can be written as
\begin{equation}
\theta^\prime_m +\mathcal{H} \theta_m-k^2\Phi=0,
\end{equation}
where we have used the definition $\theta_m \equiv \partial_i{v^i_m}$.
These equations can be easily combined to make disappear the dependence on $\theta_m$. If we do that, we obtain the following approximate second order differential equation for the matter density contrast,
\begin{equation}\label{eq:perturbk2Phi}
\delta_m^{\prime \prime}+\mathcal{H}\delta_m^\prime+k^2 \Phi=0.
\end{equation}
The problem is now reduced to find an expression for $k^2\Phi$ in terms of background quantities and $\delta_m$. Collecting \eqref{eq:gField00Mom}, \eqref{eq:gFieldij1Mom} and \eqref{eq:BDfield2Mom},
\begin{equation}\label{eq:PhiplusPsi}
-\frac{\delta \varphi}{\bar{\varphi}}=\Psi+\Phi,
\end{equation}
\begin{equation}\label{eq:Poisson1}
2k^2 \Psi+k^2 \frac{\delta \varphi}{\bar{\varphi}}=\frac{8\pi G_N a^2}{\bar{\varphi}} \bar{\rho}_m \delta_m,
\end{equation}
\begin{equation}\label{eq:Poisson2}
k^2 \delta \varphi=\frac{8\pi G_N a^2}{3+2\omega_{\rm BD}}\bar{\rho}_m \delta_m,
\end{equation}
respectively. We can see, as expected, that for $\omega_{\rm BD}\to\infty$ \textit{and} $\bar{\varphi}=1$ the first equation above gives $\Phi=-\Psi$ (no anisotropy stress) and the third one renders a trivial equality ($0=0$), whereas the second equation yields Poisson equation $k^2\Psi=-k^2\Phi=4\pi G_N a^2\sum_n\delta\rho_n$.
In the general case, by combining the above equations one finds:
\begin{equation}\label{eq:Poisson3}
k^2\Phi=-\frac{4\pi G_N a^2 \bar{\rho}_m \delta_m}{\bar{\varphi}}\left( \frac{4+2\omega_{\rm BD}}{3+2\omega_{\rm BD}} \right)\,.
\end{equation}
So, finally, inserting the previous relation in \eqref{eq:perturbk2Phi} we are led to the desired equation for the density contrast at deep subhorizon scales:
\begin{equation}
\delta_m^{\prime\prime}+\mathcal{H}\delta_m^\prime-\frac{4\pi G_N}{\bar{\varphi}} a^2\bar{\rho}_m\delta_m \left(\frac{4+2\omega_{\rm BD}}{3+2\omega_{\rm BD}}\right)=0\,,
\end{equation}
or, alternatively, in terms of the cosmic time $t$,
\begin{equation}\label{eq:dcEq}
\ddot{\delta}_m+2H\dot{\delta}_m-\frac{4\pi G_N}{\bar{\varphi}}\bar{\rho}_m\delta_m\left(\frac{4+2\omega_{\rm BD}}{3+2\omega_{\rm BD}}\right)=0\,,
\end{equation}
with the dots denoting derivatives with respect to $t$. As expected, we can recover the standard $\Lambda$CDM result for  $\omega_{\rm BD}\to \infty$ {\it and} $\bar{\varphi}\to 1$.
\newline
As already mentioned above, because of \eqref{eq:PhiplusPsi} non-null scalar field perturbations induce a deviation of the anisotropic stress, $-\Psi/\Phi$, from $1$, i.e. the GR-$\Lambda$CDM (see the notation employed in Chapter \ref{BD_gravity_chapter}) value. At scales well below the horizon,
\begin{equation}
-\frac{\Psi}{\Phi} = \frac{1+\oD}{2+\oD}=1-\epsilon_{\rm BD}+\mathcal{O}(\epsilon^2_{\rm BD})\,,
\end{equation}
so constraints on the anisotropic stress directly translate into constraints on $\epsilon_{\rm BD} = 1/\oD$, and a deviation of this quantity from $1$ at the linear regime would be a clear signature of non-standard gravitational physics. A model-independent reconstruction of the anisotropic stress from observations has been recently done in \cite{Pinho:2018unz}. Unfortunately, the error bars are still of order $\mathcal{O}(1)$, so these model-independent results cannot put tight constraints on $\epsilon_{\rm BD}$ (yet).
\newpage


%
%

\section{Semi-analytical solutions for the Brans-Dicke model in different epochs}\label{Appendix_C}
Our aim in this appendix is to provide semi-analytical solutions for the Brans-Dicke (BD) equations in the various epochs of the cosmic history, by using a perturbative approach. The nomenclature and the notation employed here are the same that can be found in Chapter \ref{BD_gravity_chapter}. We express the BD-field \eqref{eq:definitions_BD_article} and the scale factor up to linear order in $\eBD$ as follows,
\begin{align}
&\varphi(t)=\varphi^{(0)}+\epsilon_{\rm BD}\varphi^{(1)} (t)+\mathcal{O}(\epsilon_{\rm BD}^2)\label{ExpansionOfvarphi} \\
& a(t)=a^{(0)}(t)+\epsilon_{\rm BD} a^{(1)}(t)+\mathcal{O}(\epsilon_{\rm BD}^2)\,, \label{ExpansionOfa}
\end{align}
with the functions with superscript $(0)$ denoting the solutions of the background equations in standard GR with a constant Newtonian coupling that can be in general different from $G_N$, and the functions with superscript $(1)$ denoting the first-order corrections induced by a non-null $\epsilon_{\rm BD}$. Neglecting the higher-order terms is a very good approximation for all the relevant epochs of the expansion history due to the small values of $\epsilon_{\rm BD}$ allowed by the data. Plugging these expressions in Eqs. \eqref{eq:Friedmann_equation_varphi_BD_article}-\eqref{eq:local_conservation_equation_BD_article} we can solve the system and obtain the dominant energy density at each epoch and the Hubble function, which of course can also be written as
\begin{align}
&\rho_N(t) = \rho_N^{(0)}(t)+\epsilon_{\rm BD} \rho_N^{(1)}(t)+\mathcal{O}(\epsilon_{\rm BD}^2)\,\label{ExpansionOfrho}\\
& H(t) = H^{(0)}(t)+\epsilon_{\rm BD} H^{(1)}(t)   +\mathcal{O}(\epsilon_{\rm BD}^2)\,,\label{ExpansionOfH}
\end{align}
respectively, where $N=R,M,\Lambda$ denotes the solution at the radiation, matter, and $\Lambda$-dominated epochs\footnote{For the sake of simplicity and to ease the obtaining of analytical expressions, in this appendix we consider three massless neutrinos.}. We make use of the following relations,
\begin{equation}
\frac{1}{2\oD+3}=\frac{\epsilon_{\rm BD}}{2}+\mathcal{O}(\epsilon_{\rm BD}^2)\,, \qquad
\frac{\dot{\varphi}}{\varphi}=\epsilon_{\rm BD} \frac{\dot{\varphi}^{(1)}}{\varphi^{(0)}}+\mathcal{O}(\epsilon_{\rm BD}^2)\,,\qquad
\frac{\oD}{2}\left(\frac{\dot{\varphi}}{\varphi} \right)^2=\frac{\epsilon_{\rm BD}}{2}\left(\frac{\dot{\varphi}^{(1)}}{\varphi^{(0)}}\right)^2+\mathcal{O}(\epsilon_{\rm BD}^2)\,.
\end{equation}
The Klein-Gordon equation is already of first order in $\epsilon_{\rm BD}$,
\begin{equation}
\ddot{\varphi}^{(1)}+3H^{(0)}\dot{\varphi}^{(1)}= 4\pi G_N(\rho^{(0)}_N-3p^{(0)}_N)\,, \label{eq:KGFirstOrder}
\end{equation}
and this allows us to find $\varphi^{(1)}$ without knowing $H^{(1)}$ nor the linear corrections for the energy densities and pressures. The Friedmann equation leads to
\begin{equation}
H^{(0)}=\left(\frac{8\pi{G_N}}{3\varphi^{(0)}}\rho_N^{(0)}\right)^{1/2}\,,\label{eq:FriedmannZeroOrder}
\end{equation}
at zeroth order, and
\begin{equation}
6H^{(0)}H^{(1)}+3H^{(0)}\frac{\dot{\varphi}^{(1)}}{\varphi^{(0)}}-\frac{1}{2}\left(\frac{\dot{\varphi}^{(1)}}{\varphi^{(0)}}\right)^2=3\left(H^{(0)}\right)^2 \left(\frac{\rho^{(1)}_N}{\rho_N^{(0)}} - \frac{\varphi^{(1)}}{\varphi^{(0)}}\right)\,, \label{eq:FriedmannFirstOrder}
\end{equation}
at first order, which let us to compute $a^{(1)}(t)$ once we get $\rho^{(1)}_N(t,a^{(1)}(t))$ from the conservation equation and substitute it in the \textit{r.h.s.} $H^{(1)}$ is then trivially obtained using the computed scale factor. Alternatively, one can also combine \eqref{eq:FriedmannFirstOrder} with the pressure equation to obtain a differential equation for $H^{(1)}$ and directly solve it. Proceeding in this way one obtains, though, an additional integration constant that must be fixed using the Friedmann and conservation equations.

\subsection{Radiation dominated epoch (RDE)}
In the RDE the trace of the energy-momentum tensor is negligible when compared with the total energy density in the Universe, so we can set the right-hand side of \eqref{eq:KGFirstOrder} to zero. The leading order of the scale factor and the Hubble function take the following form, respectively: $a^{(0)}(t)=A t^{1/2}$,  $H^{(0)}(t)=1/(2t)$, with $A \equiv (32\pi{G_N}\rho^{0}_r/3\varphi^{(0)})^{1/2}$ and $\rho^{0}_{r}$ the current value of the radiation energy density. Using these relations we can find the BD-field as well as the scale factor at first order. They read,
\begin{equation}
\varphi (t) =\varphi^{(0)}+\epsilon_{\rm BD}\left(C_{R1} + C_{R2}\,t^{-1/2}\right)+\mathcal{O}(\epsilon_{\rm BD}^2)\,,
\end{equation}
and
\begin{equation}
a(t)= At^{1/2}\left( 1  + \epsilon_{\rm BD}\left[\frac{C_{R2}^2 \ln(t)}{24(\varphi^{(0)})^2{t}}-\frac{C_{R1}}{4\varphi^{(0)}}+\frac{C_{R3}}{t} \right] +\mathcal{O}(\epsilon_{\rm BD}^2) \right)\,,
\end{equation}
where $C_{R1}$, $C_{R2}$ and $C_{R3}$ are integration constants. {Let us note that in the RDE the evolution of the BD-field $\varphi$ is essentially frozen, since there is no growing mode. The term evolving as $\sim t^{-1/2}$ is the decaying mode, and after some time we are eventually left with a constant contribution}.
\newline
\newline
Finally, it is easy to find the corresponding Hubble function
\begin{equation}
H(t) = \frac{1}{2t}\left(1  + \epsilon_{\rm BD}\left[ -\frac{C_{R2}^2}{12(\varphi^{(0)})^2}\frac{\ln(t)}{t}+\frac{1}{t}\left(\frac{C_{R2}^2}{12(\varphi^{(0)})^2}-2C_{R3} \right) \right] +\mathcal{O}(\epsilon_{\rm BD}^2)\right)\,.
\end{equation}
At late enough times it is natural to consider that the decaying mode is already negligible, and this allows us to simplify a lot the expressions, by setting $C_{R2}=0$. The scalar field remains then constant in very good approximation when radiation rules the expansion of the Universe, and the other cosmological functions are the same as in GR ($a\sim t^{1/2}$,  $H\sim \frac{1}{2t}$), but with an effective gravitational coupling $G=G_N/\varphi$.

\subsection{Matter dominated epoch (MDE)}

When non-relativistic matter is dominant in the Universe the scalar field evolves as
\begin{equation}\label{eq:varphiMDEApp}
\varphi(t) = \varphi^{(0)} + \epsilon_{\rm BD}\left[\frac{C_{M1}}{t}+\frac{2\varphi^{(0)}}{3}\ln(t) + C_{M2}\right] +\mathcal{O}(\epsilon_{\rm BD}^2)\,,
\end{equation}
where $C_{M1}$ and $C_{M2}$ are integration constants. At leading order the scale factor and the Hubble function take the following form, $a^{(0)}(t) = Bt^{2/3}$ and $H^{(0)}(t) = 2/3t$ respectively, with $B \equiv \left(6\pi{G_N}\rho^0_m/\varphi^{(0)}\right)^{1/3}$ and $\rho^0_m$ the current value of the matter energy density. If we neglect the decaying mode in \eqref{eq:varphiMDEApp} we find $\varphi(a)=\varphi^{(0)}(1+\epsilon_{\rm BD}\ln a+\mathcal{O}(\epsilon^2_{\rm BD}))$. This solution is also found from the analysis of fixed points at leading order in $\epsilon_{\rm BD}$ that will be presented in Appendix \ref{Appendix_D}. The scale factor reads in this case,
\begin{equation}
a(t) = Bt^{2/3}\left(1 + \epsilon_{\rm BD}\left[ -\left(\frac{C_{M1}}{\varphi^{(0)}}\right)^2\frac{1}{8t^2}-\frac{C_{M2}}{3\varphi^{(0)}}-\frac{1}{18}-\frac{2}{9}\ln(t)+\frac{C_{M3}}{t} \right] +\mathcal{O}(\epsilon_{\rm BD}^2) \right)\,,
\end{equation}
where $C_{M3}$ is another integration factor, and the Hubble function,
\begin{equation}
H(t) = \frac{2}{3t}\left(1 +\epsilon_{\rm BD}\left[\left(\frac{C_{M1}}{\varphi^{(0)}}\right)^2\frac{3}{8t^2}-\frac{1}{3}-\frac{3C_{M3}}{2t} \right]+\mathcal{O}(\epsilon_{\rm BD}^2)\right)\,.
\end{equation}
{Once more, after we neglect the contribution from the decaying modes, the usual cosmological functions are as in GR in the MDE ($a\sim t^{2/3}$,  $H\sim \frac{2}{3t}$). However, in contrast to the RDE there is some mild evolution (a logarithmic one with the cosmic time) of the BD-field. Since $\varphi^{(0)}$ is obviously positive, it follows from  Eq.\,\eqref{eq:varphiMDEApp} that the sign of such evolution (implying growing or decreasing behaviour) is entirely defined by the sign of the BD-parameter $\eBD$.  The fit to the overall data clearly shows that $\epsilon_{\rm BD}<0$ and hence $\varphi$ decreases with the expansion during the MDE, which means that the effective gravitation coupling $G=G_N/\varphi$ increases with the expansion.}
\subsection{$\Lambda$-dominated or VDE}
Here we assume that the energy density of the Universe is completely dominated by a vacuum fluid, with constant energy density $\rho_\Lambda$.  {This period occurs in the very early Universe during inflation, and in the very late one when matter has diluted significantly and a new period of inflation occurs}. The usual solution for the Hubble function in that epoch is $H^{(0)}(t)=H_\Lambda$, where $H_\Lambda$ is a constant, fixed by \eqref{eq:FriedmannZeroOrder}. Taking this into account we find
\begin{equation}
\varphi(t)=\varphi^{(0)}+\epsilon_{\rm BD}\left(2H_\Lambda \varphi^{(0)} t+C_{\Lambda 1} e^{-3H_\Lambda t}+C_{\Lambda 2} \right)+\mathcal{O}(\epsilon_{\rm BD}^2), \label{eq:varphi(t)Inflation}
\end{equation}
and for the scale factor
\begin{equation}
\begin{split}
&a(t)=C_{\Lambda 4}e^{H_\Lambda t}\left(1+\epsilon_{\rm BD}\left[-\frac{H_\Lambda^2 t^2}{2}-\left(\frac{C_{\Lambda 1}}{\varphi^{(0)}}\right)^2\frac{e^{-6H_\Lambda t}}{8}+C_{\Lambda 3}H_\Lambda t\left(\frac{C_{\Lambda 2}}{2\varphi^{(0)}}+\frac{2}{3} \right) \right]+\mathcal{O}(\epsilon_{\rm BD}^2)\right),
\end{split}
\end{equation}
where $C_{\Lambda 1},$ $C_{\Lambda 2}$, $C_{\Lambda 3}$, and $C_{\Lambda 4}$ are integration constants. Note that by performing the limit $\epsilon_{\rm BD} \rightarrow 0$ we recover the usual GR expressions, as in all the previous formulas. The Hubble function at first order takes the form
\begin{equation}
\begin{split}
&H(t)=H_\Lambda\left(1 +\epsilon_{\rm BD}\left[-\frac{2}3+\frac{3}{4}\left(\frac{C_{\Lambda 1}}{\varphi^{(0)}}\right)^2e^{-6H_\Lambda t}-H_\Lambda t-\frac{C_{\Lambda 2}}{2\varphi^{(0)}} \right]+\mathcal{O}(\epsilon_{\rm BD}^2)\right).
\end{split}
\label{H(t)Inflation}
\end{equation}
The term accompanied by the constant $C_{\Lambda 1}$ in \eqref{eq:varphi(t)Inflation} is a decaying mode, which we considered to be negligible already in the RDE. Thus, we can safely remove it, and $C_{\Lambda 2}$ can be fixed by the condition $\varphi(t_*)=\varphi_*$, at some $t_*$ deeply in the VDE. The scalar field evolves then as $\varphi(t)\sim 2\epsilon_{\rm BD}H_\Lambda t\sim 2\epsilon_{\rm BD}\ln a$, which is the behaviour we obtain also from the analysis of fixed points (cf. Appendix \ref{Appendix_D}). {We can also see that during this epoch the BD-field decreases with the expansion because $\epsilon_{\rm BD}<0$, as indicated before.}

\subsection{Mixture of matter and vacuum energy}

In a Universe with a non-negligible amount of vacuum and matter energy densities, it is also possible to obtain an analytical expression for the BD scalar field at leading order in $\epsilon_{\rm BD}$. Unfortunately, this is not the case for the scale factor and the Hubble function, so we will present here only the formula for $\varphi$. In the $\Lambda$CDM the scale factor is given by
\begin{equation}
a^{(0)}(t)=\left(\frac{\tilde{\Omega}_{m}}{\tilde{\Omega}_{\Lambda}}\right)^{1/3} \sinh^{2/3} \left(\frac{3}{2}\sqrt{\tilde{\Omega}_{\Lambda}}H_0 t \right)\,,
\end{equation}
so
\begin{equation}
H^{(0)}(t)=H_0\sqrt{\tilde{\Omega}_\CC} \coth \left(\frac{3}{2}\sqrt{\tilde{\Omega}_{\Lambda}}H_0 t\right)\,.
\end{equation}
Solving the Klein-Gordon equation, we obtain
\begin{equation}\label{eq:varphi1VDE}
\varphi^{(1)}(t)=\sqrt{\tilde{\Omega}_\Lambda}H_0 t  \coth \left(\frac{3}{2}\sqrt{\tilde{\Omega}_\Lambda}H_0 t \right)+\frac{2}{3}\ln\left( \sinh \left(\frac{3}{2}\sqrt{\tilde{\Omega}_\Lambda}H_0 t \right) \right)\,.
\end{equation}
One can easily check that in the limits $H_0t\ll 1$ and $H_0 t\gg 1$ we recover the behaviour that we have found in previous sections for the matter and $\Lambda$-dominated Universes, respectively. {When we substitute the previous expression in Eq.\,\eqref{ExpansionOfvarphi}, we confirm once more that $\varphi$ decreases with the expansion (since $\eBD<0$ and $\dot{\varphi}^{(1)}(t)>0$ $\forall{t}$). This is the period when the Universe is composed of a  mixture of matter and vacuum energy at comparable proportions, and corresponds to the current Universe.  Thus  $G=G_N/\varphi$ increases with the expansion in the present Universe as it was also the case in the preceding MDE period, which is of course an important feature that helps to solve the $H_0$-tension, as explained in different parts of the Chapter \ref{BD_gravity_chapter}}.

%
\subsection{Connection of the BD-$\CC$CDM model with the Running Vacuum Model}\label{sec:RVMconnection_BD_article}
As we have just seen it is possible to search for  approximate solutions in the different epochs, which can help to better understand the numerical results and the qualitative behaviour of the BD-$\CC$CDM model.  Actually, a first attempt in this direction trying to show that BD-$\CC$CDM  can mimic the Running Vacuum Model (RVM) was done in \cite{Peracaula:2018dkg,Perez:2018qgw} and we refer the reader to these references for details. Here we just summarize the results and adapt them to the notation employed in this thesis. It is based on searching for  solutions in the MDE in the form of a power-law ansatz in which the BD-field $\varphi$ evolves very slowly:
\begin{equation}
\varphi(a) = \varphi_0\,a^{-\epsilon}\  \qquad (|\epsilon|\ll1)\,.\label{powerlaw}
\end{equation}
Obviously $\epsilon$ must be a very small parameter in absolute value since $G(a)\equiv G(\varphi(a))$ cannot depart too much from $G_N$.  On comparing with the analysis of fixed points given in Appendix \ref{Appendix_D} -- cf. Eq.\,\eqref{eq:psiMDE} -- we can anticipate that $\epsilon\propto -\epsilon_{\rm BD}$, although we do not expect perfect identification since \eqref{powerlaw} is a mere ansatz solution in the MDE whereas \eqref{eq:psiMDE} is an exact phase trajectory in that epoch.  For $\epsilon>0$ (hence $\epsilon_{\rm BD}<0$), the effective  coupling increases with the expansion and therefore is asymptotically free since  $G(a)$  is smaller in the past, which is the epoch when the
Hubble rate (with natural dimension of energy) is bigger. For  $\epsilon<0$ ($\epsilon_{\rm BD}>0$), instead, $G(a)$  decreases with the expansion. Using the power-law ansatz (\ref{powerlaw}) we find:
\begin{equation}
 \frac{\dot\varphi}{\varphi}= -\epsilon {H}\,,\ \ \ \ \ \ \ \ \ \ \ \
 \frac{\ddot\varphi}{\varphi} = -\epsilon\dot{H} + \epsilon^2{H^2}. \label{derivatives}
\end{equation}
Plugging these  relations into the system of equations  \eqref{eq:Friedmann_equation_varphi_BD_article}-\eqref{eq:Wave_equation_varphi_BD_article} and after some calculation it is possible to arrive at the following pair of Friedmann-like equations to ${\cal O}(\epsilon)$\cite{Peracaula:2018dkg,Perez:2018qgw}:
\begin{equation}\label{eq:effective Friedmann}
   H^2=\frac{8\pi G}{3}\left(\rho^0_{m} a^{-3+\epsilon}+\rDE(H)\right)
\end{equation}
and
\begin{equation}\label{eq:currentacceleration}
\frac{\ddot{a}}{a}=-\frac{4\pi G}{3}\,\left(\rho_m^0 a^{-3+\epsilon}+\rDE(H)+3p_{\Lambda}\right)\,,
\end{equation}
with $G=G_N/\varphi_0$. The first equation emulates an effective Friedmann's equation with time-evolving cosmological term, in which the DE appears as dynamical:
\begin{equation}\label{eq:rLeff}
  \rDE(H)=\rL+\frac{3\,\nu_{\rm eff}}{8\pi G} H^2\,.
\end{equation}
Here
\begin{equation}
\ \nu_{\rm eff}\equiv\epsilon\left(1+\frac16\,\oD\epsilon\right) \label{nueff}
\end{equation}
is the coefficient controlling the dynamical character of the dark energy \eqref{eq:rLeff}. The structure of this dynamical dark energy  (DDE) is reminiscent of the Running Vacuum Model (RVM), see \cite{Sola:2013gha,sola2015cosmology,Gomez-Valent:2017tkh} and references therein. { In the language used in Chapter \ref{BD_gravity_chapter}, the analogue of the above RVM form of the Friedmann equation can be derived from Eq.\,\eqref{eq:FriedmannWithF} upon taking into account that the function ${\cal F}$ is of ${\cal O}(\eBD)$. In fact,  ${\cal F}$ is the precise analogue of $\nu_{\rm eff}$, for if we set $\epsilon\to -\eBD$  in \eqref{nueff} it boils down to the value quoted in  Eq.\eqref{eq:Fximatter}. The two languages are similar, but not identical, for the reasons explained above.} Notice from \eqref{eq:effective Friedmann} that,  to ${\cal O}(\epsilon)$:
\begin{equation}\label{eq:SumRule}
\Omega_m +\Omega_\Lambda =1-\nu_{\rm eff}\,,
\end{equation}
so the usual sum rule of GR is slightly violated by the BD model when parametrized as a deviation with respect to GR. Only for $\epsilon=0$ we have $\nu_{\rm eff}=0$ and then  we recover the usual cosmic sum rule. The parameter $\nu_{\rm eff}$ becomes associated to the dynamics  of the DE. Worth noticing, the above expression adopts  the form of the RVM, see  \cite{Sola:2013gha,sola2015cosmology,Gomez-Valent:2017tkh} and references therein, in particular \cite{Shapiro:2000dz,Sola:2007sv,Shapiro:2009dh} -- where the running parameter is usually denoted $\nu$ and is associated to the $\beta$-function of the running vacuum.  Recently, the parameter $\nu$ (and in general the structure of the RVM) has been elucidated from direct calculations in QFT in curved space-time within GR\,\cite{Moreno-Pulido:2020anb}. For additional discussions on the running of the CC term, see e.g. \cite{Babic:2004ev,Ward:2010qs,Antipin:2017pbt}. The RVM has been shown to be phenomenologically promising to alleviate some of the existing tensions within the $\CC$CDM, particularly the  $\sigma_8$-tension\,\,\cite{Sola:2016ecz , Sola:2017jbl , Gomez-Valent:2018nib ,Gomez-Valent:2017idt,Sola:2017znb , sola2017first , Sola:2016hnq , Sola:2015wwa , Geng:2017apd,Rezaei:2019xwo,Geng:2020mga}. It is therefore not surprising that the mimicking of the RVM by the BD-$\CC$CDM model enjoys of the same virtues. In actual fact, the particular RVM form obtained in BD-gravity (we may call it ``BD-RVM'' for short)  is even more successful since it can cure both tensions, the $H_0$ and $\sigma_8$ one.  The reason why the BD-RVM can cure also the $H_0$-tension is because we need the evolution of the effective gravitational coupling $\Geff$ to achieve that, as we have seen in the preview Sec.\,\ref{sec:preview_BD_article}, whereas the $\sigma_8$-tension can be cured with $\nueff$, which is associated to $\epsilon\propto-\eBD$  (the second ingredient characteristic of BD-gravity), and hence the two key elements are there to make a successful phenomenological performance.
\newline
\newline
On the other hand, from \eqref{eq:currentacceleration} it follows that the EoS for the effective DDE is
\begin{equation}\label{eq:EffEoS}
\weff(z)=\frac{p_{\Lambda}}{\rDE(H)}\simeq -1+\frac{3\nueff}{8\pi G \rL}\,H^2(z)=-1+\frac{\nueff}{\Omega_\CC}\,\frac{H^2(z)}{H_0^2}\,,
\end{equation}
where use has been made of (\ref{eq:rLeff}). It follows that the BD-RVM, in contrast to the original RVM, does not describe a  DE of pure vacuum form ($p_{\Lambda}=-\rL$) but a DE whose EoS departs slightly from the pure vacuum. In fact, for $\epsilon>0\ (\epsilon<0) $  we have $\nu_{\rm eff}>0\  (\nu_{\rm eff}<0)$ and the effective DDE behaves quintessence (phantom)-like.  For $\epsilon\to 0$ (hence  $\nu_{\rm eff}\to 0$) we have  $\weff\to -1$ ($\CC$CDM).   As could be expected, Eq.\eqref{eq:EffEoS} is the BD-RVM version of the effective EoS that we obtained in Sec.\,\ref{sec:EffectiveEoS_BD_article}  -- see Eq.\,\eqref{BDEoSz0}.  The two languages are consistent. Indeed, by comparison we see that $\nueff$ here plays the role of $\dvphi$ there. We know that $\dvphi=1-\varphi>0$, i.e. $\varphi<1$,  for $\eBD<0$, as we have shown previously, which is consistent with the fact that $\nueff\propto\epsilon\propto -\eBD>0$.  Finally, since $\weff$  approaches $-1$ from above it corresponds to an effective quintessence behaviour, which is more pronounced the more we explore the EoS into our past.

%
\newpage

\section{Fixed Points in BD-$\CC$CDM  cosmology}\label{Appendix_D}
In order to study the fixed points of the BD theory we must define new variables such that the system of differential equations that characterize the model becomes of first order in the derivatives when it is rewritten in terms of the new variables. It is useful though to firstly carry out the change $t\to N\equiv \ln(a)$. This preliminary step will help us to identify in an easier way how we must define the new variables. When written in terms of $N$ the system takes the following form\footnote{{Primes in this appendix stand for derivatives {\it w.r.t.}  to the variable $N=\ln a$, i.e.  $()^\prime \equiv d()/d N$. We consider, as in Appendix \ref{Appendix_C}, three massless neutrinos.}}:
\begin{equation}
\frac{\psi^{\prime\prime}}{\psi}+\frac{H^\prime}{H}\frac{\psi^\prime}{\psi}+3\frac{\psi^\prime}{\psi}=\frac{8\pi}{3+2\oD}\left(\frac{\rho_m+4\rho_\Lambda}{\psi H^2}\right)\,,
\end{equation}
\begin{equation}
3+3\frac{\psi^\prime}{\psi}-\frac{\oD}{2}\left(\frac{\psi^\prime}{\psi}\right)^2=\frac{8\pi}{\psi H^2}(\rho_r+\rho_m+\rho_\Lambda)\,,
\end{equation}
\begin{equation}
3+\frac{H^\prime}{H}\left(2+\frac{\psi^\prime}{\psi}\right)+\frac{\psi^{\prime\prime}}{\psi}+2\frac{\psi^\prime}{\psi}+\frac{\oD}{2}\left(\frac{\psi^\prime}{\psi}\right)^2=\frac{8\pi}{\psi H^2}\left(\rho_\Lambda-\frac{\rho_r}{3}\right)\,,
\end{equation}
\begin{equation}
\rho^\prime_r+4\rho_r=0\,,
\end{equation}
\begin{equation}
\rho^\prime_m+3\rho_m=0\,.
\end{equation}
\noindent
Now one can define the following quantities:
\begin{equation}
\xp\equiv \frac{\psi^\prime}{\psi}\quad ; \quad x_i^2\equiv\frac{8\pi \rho_i}{H^2\psi}\,,
\end{equation}
\noindent
where $i=r,m,\Lambda$. In terms of these variables the system of equations can be easily written as follows:
\begin{equation}\label{eq:eq1}
\xp^\prime+\xp^2+\frac{H^\prime}{H}\xp+3\xp=\frac{\xm^2+4\xl^2}{3+2\oD}\,,
\end{equation}
\begin{equation}\label{eq:eq2}
3+3\xp-\frac{\oD}{2}\xp^2=\xr^2+\xm^2+\xl^2\,,
\end{equation}
\begin{equation}\label{eq:eq3}
3+\frac{H^\prime}{H}\left(2+\xp\right)+\xp^\prime+\xp^2+2\xp+\frac{\oD}{2}\xp^2=\xl^2-\frac{\xr^2}{3}\,,
\end{equation}
\begin{equation}\label{eq:eq4}
\xr^\prime=-\xr\left(\frac{H^\prime}{H}+2+\frac{\xp}{2}\right)\,,
\end{equation}
\begin{equation}\label{eq:eq5}
\xm^\prime=-\xm\left(\frac{H^\prime}{H}+\frac{3}{2}+\frac{\xp}{2}\right)\,.
\end{equation}
\noindent
This system is of first order, as wanted. We have five equations and five unknowns, namely: $\xr,\xm,\xp,\xl,H^\prime/H$. We can reduce significantly the complexity of the system if we just isolate $H^\prime/H$ from \eqref{eq:eq1} and $\xl$ from \eqref{eq:eq2},
\begin{equation}
\frac{H^\prime}{H}=\frac{1}{\xp}\left[\frac{\xm^2+4\xl^2}{3+2\oD}-\xp^\prime-3\xp-\xp^2\right]\,,
\end{equation}
\begin{equation}
\xl^2=3+3\xp-\frac{\oD}{2}\xp^2-\xr^2-\xm^2\,,
\end{equation}
\noindent
and substitute the resulting expressions in the other equations, i.e. in \eqref{eq:eq3}, \eqref{eq:eq4}, and \eqref{eq:eq5}. Doing this, and after a little bit of algebra, one finally obtains three equations written only in terms of $\xr,\xm,\xp$:
\begin{equation}\label{DynSysPsi}
\xp^\prime=-\xp\left[3+3\xp-\frac{1}{2}\oD\xp^2-\frac{2}{3}\xr^2-\frac{\xm^2}{2}-\left(\frac{1}{\xp}+\frac{1}{2}\right)\left(\frac{12+12\xp-2\oD\xp^2-4\xr^2-3\xm^2}{3+2\oD}\right)\right]\,,
\end{equation}
\begin{equation}\label{DynSysR}
\xr^\prime=-\xr\left[2+\frac{5}{2}\xp-\frac{\oD}{2}\xp^2-\frac{2}{3}\xr^2-\frac{\xm^2}{2}-\frac{1}{2}\left(\frac{12+12\xp-2\oD\xp^2-4\xr^2-3\xm^2}{3+2\oD}\right)\right]\,,
\end{equation}
\begin{equation}\label{DynSysM}
\xm^\prime=-\xm\left[\frac{3}{2}+\frac{5}{2}\xp-\frac{\oD}{2}\xp^2-\frac{2}{3}\xr^2-\frac{\xm^2}{2}-\frac{1}{2}\left(\frac{12+12\xp-2\oD\xp^2-4\xr^2-3\xm^2}{3+2\oD}\right)\right]\,.
\end{equation}
\noindent
They allow us to search for the fixed points of the system. There is an important restriction produced by \eqref{eq:eq4} and \eqref{eq:eq5}. Supposing that $x_r \neq 0$ and $x_m\neq 0$, we see that the mentioned equations impose that
\begin{equation}
\frac{x_r^\prime}{x_r}=-\frac{x_m^\prime}{x_m}+\frac{1}{2}.
\end{equation}
This equation is not compatible with the conditions of fixed point, so that we should assume that $x_r=0$, $x_m=0$ or both conditions at the same time. The fixed points are:
\newline
\newline
\textbf{RDE}
\begin{equation}
(\xr,\xm,\xl,\xp)_{\rm RD} = \left(\sqrt{3},0,0,0\right)\,,
\end{equation}
the Jacobian of the non-linear system \eqref{DynSysPsi}, \eqref{DynSysR}, \eqref{DynSysM} has eigenvalues $\lambda^{\rm RD}_1=-1$, $\lambda^{\rm RD}_2 =1/2$, $\lambda^{\rm RD}_3=4$ so that is an unstable point.
\newline
\newline
\textbf{MDE}
\begin{equation}
(\xr,\xm,\xl,\xp)_{\rm MD} = \left(0,\frac{\sqrt{12+17\oD+6\oD^2}}{\sqrt{2}|1+\oD|},0,\frac{1}{1+\oD}\right)\,,
\end{equation}
the Jacobian of the non-linear system \eqref{DynSysPsi}, \eqref{DynSysR}, \eqref{DynSysM} has eigenvalues $\lambda^{\rm MD}_1=-1/2,$ $\lambda^{\rm MD}_2\simeq-3/2,$ $\lambda^{\rm MD}_3\simeq 3$ so that is also an unstable point.
\newline
\newline
\textbf{$\Lambda$-dominated or VDE}
\begin{equation}
(\xr,\xm,\xl,\xp)_{\rm \Lambda D} = \left(0,0,\frac{\sqrt{15+28\oD+12\oD^2}}{|1+2\oD|},\frac{4}{1+2\oD}\right)\,
\end{equation}
the Jacobian of the non-linear system \eqref{DynSysPsi}, \eqref{DynSysR}, \eqref{DynSysM} has eigenvalues $\lambda^{\Lambda}_1=-2,$ $\lambda^{\Lambda}_2 =-3/2,$ $\lambda^{\Lambda}_3 \simeq -3$ so that is a stable point.
\newline
\newline
The first and second fixed points are unstable, whereas the latter is stable. We have assumed that $|\oD|\gg1$ for approximating the eigenvalues. The first one is very well-known, since regardless of the initial conditions for the BD scalar field we already know that the velocity of the scalar field decays during the RDE, and the solution tends to the attractor with $\psi^\prime=0$ ($\xp=0$) and full domination of radiation, i.e. $\xr=\sqrt{3}$. The Hubble function and BD scalar field are:
\begin{equation}
\psi_{\rm RD}(a)=\psi_r\,,
\end{equation}
\begin{equation}
H^2_{\rm RD}(a)=\frac{8\pi}{3\psi_r}\rho^{0}_{r}a^{-4}\,,
\end{equation}
where $\psi_r$ is an arbitrary constant and $\rho^{0}_{r}$ is the value of the radiation energy density at present.
This fixed point is unstable because at some moment non-relativistic matter starts to dominate the expansion. When this happens the solution starts to look for the new attractor, the one of the MDE. We stress  that this is an exact solution. The BD scalar field and the Hubble function during the MDE take the following form:
\begin{equation}\label{eq:psiMDE}
\psi_{\rm MD}(a)=Ca^{\frac{1}{1+\oD}}\,,
\end{equation}
\begin{equation}
H^2_{\rm MD}(a)=\frac{16\pi \rho_m^{0}(1+\oD)^2a^{-\left(\frac{4+3\oD}{1+\oD}\right)}}{C(12+17\oD+6\oD^2)}\,,
\end{equation}
\noindent
where $C$ is an arbitrary constant and $\rho_m^{0}$ is the value of the matter energy density at present. The MD fixed point is, again, unstable, because the MDE finishes and the VDE starts. The solution searches now for the last fixed point, which is stable. In this last case, we have obtained the following Hubble function and BD field,
\begin{equation}
\psi_\Lambda (a)= D a^{4/(1+2\oD)}\,,
\end{equation}
\begin{equation}
H^2_\Lambda (a)=\frac{8\pi \rho_\Lambda }{D}\frac{(1+2\oD)^2}{15+28\oD +12 \oD^2}a^{-4/(1+2\oD)}\,,
\end{equation}
where $D$ is an arbitrary constant and $\rho_\Lambda$ is the constant value of vacuum energy.
\newline
\newline
One can easily check that the solutions computed in Appendix \ref{Appendix_C} of first order in $\epsilon_{\rm BD}=1/\oD$, coincide (once the decaying modes become irrelevant) with the ones presented here when the latter are Taylor-expanded up to first order in this parameter as well. The results we have found here are consistent with previous studies on fixed points in cosmological dynamical systems, see e.g. \cite{Bahamonde:2017ize} and references therein.
\newpage
\thispagestyle{empty}
\mbox{}
\newpage

\section{Cosmological data and statistical analysis}\label{Appendix_E}
If one desires to carry out a serious study of a given cosmological model it is fundamental to go beyond the theoretical predictions, and put them in the light of the increasing amount of observational data at our disposal, allowing then to test their performance and make a model comparison, with the ultimate purpose of discerning which is the most favoured model. In this appendix we explicitly show the theoretical definitions for the various observable used in the analysis, necessary to compare with the experimental data. Additionally, we also collect some basic (but technical) information about the fitting procedure that has been employed to get the results displayed in Chapters \ref{First_evidence_chapter}-\ref{BD_gravity_chapter}.
\newline
\newline
Let us start by defining the $\chi^2$ function which represents the building block of the fitting procedure:

\begin{equation}\label{eq:chi_square_function_appendix_E}
\chi^2_{\rm tot}(\vec{p})= \sum_s[\vec{x}_s(\vec{p})-\vec{d}_s]^{T}C_s^{-1}[\vec{x}_s(\vec{p})-\vec{d}_s]\,,
\end{equation}
where the subindex $s$ runs over all the data sets considered in the analysis, for instance the CMB, BAO or SNIa \footnote{We must remark that actually the form of \eqref{eq:chi_square_function_appendix_E} does not apply for the Strong Lensing data, as it will be explained in the corresponding section. }. Later on, we will go into the details of the different data considered. The $\vec{x}_s$ are the vectors containing the theoretical predictions extracted from the different models under study, $\vec{p}$ is the vector containing the free parameters, they represent the different {\it d.o.f.} that characterize a certain model. Finally $C_s$ is the covariance matrix for each data set $s$, which can be constructed from the correlation matrix $\rho_s$ and the $1\sigma$ uncertainties as follows: $C_{s,ij}=\rho_{s,ij}\sigma_{s,i}\sigma_{s,j}$.
\newline
\newline
As we shall see, it is fundamental to properly build the function $\chi^2_{\rm tot}(\vec{p})$. One must be very careful to only pick up data for which either the correlations with other data are well under control or there are no correlations at all. Another issue that must be considered it is the possibility of the double counting. Throughout the last years, many data releases from the different surveys have appeared and as it can be understood they are not independent since they have been obtained considering at least an overlapping sample. Let us specify how to construct the $\chi^2$ function by treating all the data sets employed in this thesis one by one. 
%
%
%
\subsection{Cosmological data}
In the following we comment on the most important details of the observational data employed in the different chapters, as well as, the cosmological observable that allow us to contrast them with the theoretical predictions. While it is true that we do not take into consideration all the possible sources of cosmological data, we do use, to the best of our knowledge, the most reliable ones. It is important to remark that the extraction of data, through cosmological observations, is a task far from be a piece of cake. A great effort has been made, over the years, to perfect the employed techniques. This is the reason why we prefer to be conservative when adding new data to the dataset. 
\subsubsection{SNIa}
As we have mentioned in the introductory chapter supernovae are extremely luminous explosions that can be divided in different types according to their spectrum. They played a fundamental role in the confirmation of the accelerated expansion of the Universe and they are one of the most consolidated sources of cosmological data at our disposal. For the SNIa sector, the fitted quantity is the distance modulus, defined as:
\begin{equation}
\mu(z,\vec{p})=5\log{D_{L}(z,\vec{p})}+M    
\end{equation}
where $M$ is a normalization parameter and $D_{L}(z,\vec{p})$ is the luminosity distance:
\begin{equation}\label{LumDist}
D_{L}(z,\vec{p})={c}{(1+z)} \int_{0}^{z} \frac{{ d}z^\prime}{H(z^\prime,\vec{p})}\;,
\end{equation}
with $c$ being the speed of light (now included explicitly in this formula for better clarity). The previous formula \eqref{LumDist} for the luminosity distance applies only for a spatially flat Universe, which is something that we have assumed throughout this thesis.
\newline
\newline
There are two different options to deal with $M$ usually referred to as nuisance parameter: it can be included as if it were just another free parameter in the fitting procedure or it can be integrated out through the corresponding marginalization of the SNIa likelihood. This integration can be performed analytically, so this also helps to improve the computational efficiency by reducing in one dimension the fitting parameter space. Let us provide the details of the analytical integration.
\newline
\newline
After this marginalization, the resulting effective $\chi^2$ function for this data set reads,
\begin{equation}\label{chi2SNIa}
\chi^2_{\rm SNIa}(\vec{p})=\vec{y}(\vec{p})^{T}J\vec{y}(\vec{p})-\frac{[\sum_{i,j=1}^{N} J_{ij}y_{i}(\vec{p})]^2}{\sum_{i,j=1}^{N} J_{ij}}\,,
\end{equation}
where $J\equiv C_{\rm SNIa}^{-1}$, $y_i(\vec{p})\equiv 5\log{D_{L}(z_i,\vec{p})}-{\mu}_{\rm obs,i}$ and $N$ is the number of data points for each case. \eqref{chi2SNIa} is the direct generalization of formula (13.16) of \cite{Amendola:2015ksp}, since we are now considering a non-diagonal covariance matrix, instead of a diagonal one. Let us prove \eqref{chi2SNIa}. The original SNIa likelihood can be written as follows,
\begin{equation}
\mathcal{L}_{\rm SNIa}(\vec{p},M)=\mathcal{L}_0\,e^{-\frac{1}{2}[y_i(\vec{p})+M]J_{ij}[y_j(\vec{p})+M]}\,,
\end{equation}
where we have assumed the Einstein summation notation in order to write the result more concisely and $\mathcal{L}_0$ is the normalization constant. Let us now marginalize this distribution over the parameter $M$,
\begin{equation}
\tilde{\mathcal{L}}(\vec{p})\equiv \int_{-\infty}^{+\infty}\mathcal{L}_{\rm SNIa}(\vec{p},M)\,dM\,.
\end{equation}
This can be expressed like
\begin{equation}\label{demo0}
\tilde{\mathcal{L}}(\vec{p})={\mathcal{L}}_0 e^{-\frac{1}{2}B_0(\vec{p})}F(\vec{p})\,,
\end{equation}
with
\begin{equation}
F(\vec{p})\equiv \int_{-\infty}^{+\infty}e^{-\frac{B_1}{2}M^2-MB_2(\vec{p})}\,dM\,,
\end{equation}
and
\begin{equation}
\begin{array}{rcl}
B_0 (\vec{p}) &\equiv &  \sum_{i,j=1}^{31}y_{i}(\vec{p})J_{ij}y_{j}(\vec{p})\,,\\
B_1 & \equiv & \sum_{i,j=1}^{31}J_{ij}\,, \\
B_2(\vec{p}) & \equiv & \sum_{i,j=1}^{31}J_{ij}y_{i}(\vec{p})\,.
\end{array}
\end{equation}
The problem is basically reduced to compute the function $F(\vec{p})$ appearing in \eqref{demo0}. Let us perform the derivative of the latter with respect to $M$ and the i$_{th}$ component of the fitting vector $\vec{p}$,
\begin{equation}\label{demo1}
\frac{\partial F}{\partial M}=0=-\int_{-\infty}^{+\infty}\left[MB_1+B_2({\vec{p}})\right]e^{-\frac{B_1}{2}M^2-MB_2(\vec{p})}\,dM\,,
\end{equation}
\begin{equation}\label{demo2}
\frac{\partial F}{\partial p_i}= -\frac{\partial B_2(\vec{p})}{\partial p_i}\int_{-\infty}^{+\infty}M\,e^{-\frac{B_1}{2}M^2-MB_2(\vec{p})}\,dM\,.
\end{equation}
Combining these two expressions we can obtain a differential equation for $F(\vec{p})$, which is very easy to solve,
$$\frac{1}{F(\vec{p})}\frac{\partial F}{\partial p_i}=\frac{\partial B_2}{\partial p_i}\frac{B_2(\vec{p})}{B_1}\,,$$
\begin{equation}
F(\vec{p})\propto e^{\frac{B_2^2(\vec{p})}{2B_1}}\,.
\end{equation}
The integration constant is irrelevant since it can be absorbed by the normalization factor $\mathcal{L}_0$ in \eqref{demo0}, giving rise to a new one, $\tilde{\mathcal{L}}_0$. Thus, the marginalized likelihood reads,
\begin{equation}
\tilde{\mathcal{L}}(\vec{p})=\tilde{\mathcal{L}}_0\,e^{-\frac{\chi^2_{\rm SNIa}(\vec{p})}{2}}\,,
\end{equation}
with $\chi^2_{\rm SNIa}(\vec{p})$ given by \eqref{chi2SNIa}.

\subsubsection{BAO}
Galaxy surveys measure angular and redshift distributions of galaxies. Using these data, they are able to obtain the matter power spectrum $P_m(z,\vec{k})$ in redshift space upon modeling the bias factor and transforming these angles and redshifts into distances.
However, the distance of two sources along the line of sight, being
given by the difference of redshift $\Delta z$, depends on the cosmological model. Similarly, the distance of two sources in the direction perpendicular to the line of sight is given by the angular separation, $\Delta\theta$, and also depends on the cosmological model. Therefore, it is convenient to first define a fiducial model to which one refers the predictions of the new models being studied. Usually such model is the $\CC$CDM with appropriately chosen values of the free parameters. One also has to disentangle the effect of the redshift space distortions that are due to the peculiar velocities of galaxies falling into the gravitational potential wells. The comoving wavevector $\vec{k}$ is usually decomposed in a parallel to the line of sight component, $k_{\parallel}$, and another one which is perpendicular to it, $k_{\perp}$. These two directions determine the two-dimensional (2D) space in which the distribution of galaxies is studied. The analysis of the BAO signal in the 2D power spectrum carries precious information about the angular and longitudinal size of the BAO standard ruler at the measured redshift $z$, namely
\begin{equation}\label{eq:BAO1}
\Delta\theta_s (z)=\frac{a r_s(a_{d})}{D_A(a)}=\frac{r_s(z_{d})}{(1+z)D_A(z)}\,,
\end{equation}
where $a=1/(1+z)$ is the scale factor, and
\begin{equation}\label{eq:BAO2}
\Delta z_s (z)= \frac{r_s(z_{d})H(z)}{c}\,,
\end{equation}
$r_s(z_d)$ being the sound horizon at the redshift of the drag epoch (see below), with $D_A$ the proper diameter angular distance,
\begin{equation}\label{SomDist}
D_A(z,\vec{p}) \equiv \frac{c}{(1+z)}\int^{z}_{0}\frac{dz^\prime}{H(z^\prime,\vec{p})}\,.
\end{equation}
In principle, galaxy surveys are able to extract anisotropic BAO information from their analyses. That is to say, they can grant constraints on the quantities $r_s(z_d)/D_A(z)$ and $H(z)r_s(z_d)$ through the perpendicular and parallel dilation scale factors
\begin{eqnarray}\label{DilaScales}
\alpha_\perp &\equiv & \frac{r^{\rm fid}_s(z_d)/D^{\rm fid}_A(z)}{r_s(z_d)/D_A(z)}=\frac{D_A(z)r^{\rm fid}_s(z_d)}{D^{\rm fid}_A(z)r_s(z_d)}\nonumber\\ \alpha_\parallel &\equiv& \frac{H^{\rm fid}(z)r^{\rm fid}_s(z_d)}{H(z)r_s(z_d)}\,,
\end{eqnarray}
where $r^{\rm fid}_s(z_d)$, $D^{\rm fid}_A(z)$ and $H^{\rm fid}(z)$ are obtained in the fiducial cosmology that the survey has used to convert redshift and angles into distances. The quantity $r_s(z_d)/D_A(z)$ is the sound horizon at the drag epoch measured in units of the angular diameter distance at redshift $z$, and similarly $r^{\rm fid}_s(z_d)/D^{\rm fid}_A(z)$ is the corresponding value in the fiducial model. These ratios are just the comoving angular sizes (\ref{eq:BAO1}), i.e. the angular sizes divided by the scale factor.  Similarly, $H(z)r_s(z_d)$ and $H^{\rm fid}(z)r^{\rm fid}_s(z_d)$ are, respectively, the sound horizon in units of the Hubble horizon in the given model and in the fiducial model.
\newline
\newline
Sometimes the surveys are not equipped with enough BAO data to obtain \eqref{eq:BAO1} and \eqref{eq:BAO2} separately. In this case they limit themselves to compute the 
volume-averaged spectrum in a conveniently defined volume $V$ and obtain a constraint which is usually encapsulated in the isotropic BAO estimator $r_s(z_d)/D_V(z)$ (sometimes denoted $d_z$\,\cite{Blake:2011en}). Here $D_V(z)$ is an effective distance or dilation scale\,\cite{Eisenstein:2005su} obtained from the cubic root of the volume $V$ (whose value is defined from the square of the transverse dilation scale times the radial dilation scale):
\begin{equation}\label{eq:DV}
D_V(z,\vec{p}) = \left[zD_H(z,\vec{p})D^{2}_M(z,\vec{p})\right]^{1/3}\,.
\end{equation}
In this expression, $D_M=(1+z) D_A$ is the comoving angular diameter distance (playing the role of the transverse dilation scale at redshift $z$)
and $D_H$ is the Hubble radius
\begin{equation}\label{SomDist2}
D_H(z,\vec{p}) \equiv \frac{c}{H(z,\vec{p})},
\end{equation}
with $zD_H$ acting as the radial dilation scale at redshift $z$. Therefore, the distilled BAO estimator $d_z$  measures the sound horizon distance at the drag epoch in units of the effective dilation scale \eqref{eq:DV}.
\newline
\newline
Again a comparison with a fiducial value is necessary. The isotropic BAO parameter relating the fiducial model value of $D_V$ with the actual value of a given model is defined in \cite{Beutler:2011hx} as
\begin{equation}\label{CombDilaScales0}
\alpha\equiv\frac{D_V(z)}{D^{\rm fid}_V(z)}\,.
\end{equation}
In other cases $\alpha$ is defined in terms of $d_z$,
\begin{equation}\label{CombDilaScales}
\alpha\equiv \frac{r^{\rm fid}_s(z_d)/D^{\rm fid}_V(z)}{r_s(z_d)/D_V(z)}= \frac{D_V(z)r^{\rm fid}_s(z_d)}{D^{\rm fid}_V(z)r_s(z_d)}\,,
\end{equation}
as in \cite{Ross:2014qpa,Kazin:2014qga}. These isotropic BAO estimators (\ref{CombDilaScales0} and \ref{CombDilaScales}) are akin to the Alcock-Paczynski (AP) test \cite{Alcock:1979mp}, in which the ratio of the observed radial/redshift to the angular size at different redshifts, $\Delta z_s/\Delta\theta_s$, is used to obtain cosmological constraints on the product $H(z)D_M(z)$ and hence of $H(z)D_A(z)$. Both for the AP test and the isotropic BAO measurement, the value of the Hubble parameter $H(z)$ cannot be disentangled from $D_A(z)$, there is a degeneracy. This is in contradistinction to the situation with the anisotropic BAO parameters (\ref{DilaScales}), in which a measurement of both $\alpha_\perp$ and $\alpha_\parallel$ permits to extract the individual values of  $H(z)$ and $D_A(z)$.
\newline
\newline
In our fitting analysis, carried out in the different chapters of this thesis, we use data of both types of BAO estimators: isotropic and anisotropic ones. The anisotropic BAO data contains more information than the isotropic one because the aforementioned degeneracy between $H(z)$ and $D_A(z)$ has been broken. Thus, anisotropic BAO is richer and yields stronger cosmological constraints \cite{Shoji:2008xn,Anderson:2013zyy}.
\newline
\newline
To compute the theoretical predictions for the above BAO estimators, we need some extra formulas. For instance, the sound horizon at redshift $z$, i.e. $r_s(z)$, is given by the expression
\begin{equation}\label{comdist}
r_s(z) = \int^{\infty}_{z}\frac{c_s(z^\prime)dz^\prime}{H(z^\prime)}\,,
\end{equation}
where
\begin{equation}\label{soundSpeed}
c_s(z) = \frac{c}{\sqrt{3(1+\mathcal{R}(z)})}\,
\end{equation}
is the sound speed in the photo-baryon plasma and $\mathcal{R}(z)=\frac{\delta\rho_b(z)}{\delta\rho_\gamma(z)}$. If the expressions for the baryon and photon energy densities take the usual form, then we have  $\mathcal{R}(z) = (3\Omega_b/4\Omega_\gamma)/(1+z)$, being $\Omega_b$ and $\Omega_\gamma$ the baryon and photon density parameter at present time, respectively. The redshift at the drag epoch is $z_d\sim \mathcal{O}(10^{3})$. Below such redshift value the baryon perturbations effectively decouple from the photon ones and start to grow with dark matter perturbations. The precise value of $z_d$ depends in a complicated way on the different cosmological parameters. The fitted formula obtained in \cite{Eisenstein:1997ik}, can help us get an idea of how $z_d$ depends on the standard parameters:
\begin{align}\label{zdrag}
&z_d = \frac{1291\,\omega_m^{0.251}}{1 + 0.659\,{\omega_m}^{0.828}}\left[1 + \beta_1{\omega_b}^{\beta_2}\right]\nonumber\\
&\beta_1 = 0.313\,{\omega_m}^{-0.419}\left[1 + 0.607\,{\omega_m}^{0.674}\right]\\
&\beta_2 = 0.238\,{\omega_m}^{0.223}\nonumber\,.
\end{align}
Let us point out that many experimental groups do not make use of these fitted formulas (denoted with a subindex $ff$ below), but of the complete set of Boltzmann (Boltz) equations, which must be solved numerically. Both approaches can give rise to differences that are around $2-3\%$. Consequently, if the observational values are computed with a Boltzmann code and one is interested in using the fitted formulas to get the theoretical observable, then it is necessary to apply a re-scaling in order to perform the comparison with our theoretical predictions. In these cases we follow the same procedure applied in \cite{Kazin:2014qga} upon re-scaling the observational data by $f_{r}\equiv r_s(z_d)^{\rm fid}_{ff}/r_s(z_d)^{\rm fid}_{{\rm Boltz}}$. The two quantities involved in this ratio are computed using the same fiducial $\Lambda$CDM cosmology chosen by the observational teams, but while the first is obtained with the fitted formulas \eqref{comdist}-\eqref{zdrag}, the second is obtained with a Boltzmann code. One can also compute the latter using the existing approximated formulas for the sound horizon at $z_d$ that are obtained by fitting the data extracted from e.g. the CAMB code, since they are very accurate, with errors less than the 0.1\% \cite{Anderson:2013zyy},
\begin{equation}\label{rsBoltz}
r_s(z_d)_{\rm Boltz}=\frac{55.234(1+\Omega_\nu h^2)^{-0.3794}\,{\rm Mpc}}{(\Omega_{dm}h^2+\Omega_b h^2)^{0.2538}(\Omega_b h^2)^{0.1278}}\,,
\end{equation}
where $\Omega_\nu h^2$ is the neutrino reduced density parameter and depends, of course, on the effective number of neutrino species and the photon CMB temperature, which is taken from \cite{Fixsen:2009ug}.
\newline
\newline
We point out  that the aforementioned re-scaling does not change the values of the dilation scale factors \eqref{DilaScales}, \eqref{CombDilaScales0} and \eqref{CombDilaScales} furnished by the experimental teams. If they, for instance, hand over their results through the product $\alpha_\perp(D^{\rm fid}_A/r_s(z_d)^{\rm fid}_{{\rm Boltz}})$, then we just multiply it  by $f_r^{-1}$ in order to keep our fitting procedure consistent.  For the same reason, we re-scale $(H^{\rm fid}r_s(z_d)^{\rm fid}_{{\rm Boltz}})/\alpha_\parallel$ multiplying it by $f_r$. But we insist that the dilation scale factors, which are the fundamental outputs extracted from the analysis of the BAO signal, are not modified by this re-scaling.
\newline
\newline
In the different analysis considered in this thesis the BAO data has been updated and corrected (when necessary) as the experimental teams have made it public. Take a look at the Chapters \ref{First_evidence_chapter}-\ref{BD_gravity_chapter} in order to know the particular data points employed in each case.

\subsubsection{Cosmic Chronometers} 
Data on the Hubble parameter at different redshift points. We use $31$ (in Chapters \ref{First_evidence_chapter}-\ref{Possible_signals_chapter} we employ 30 but when the data point from \cite{Ratsimbazafy:2017vga} was released we incorporated it into our table) data points on $H(z_i)$ at different redshifts. We use only $H(z_i)$ values obtained by the so-called differential-age (or cosmic chronometer) techniques applied to passively evolving galaxies. That is to say, we use data where one estimates $H(z)$ from  $(1+z)H(z)=-dz/dt$, where $dz/dt$ is extracted from a sample of passive galaxies (i.e. with essentially no active star formation) of different redshifts and ages. See the corresponding references for more details. The important point to remark here is that these $H(z_i)$ values, although relying on the theory of spectral evolution of galaxies, are  uncorrelated with the BAO data points. \newline
\newline
The covariance matrix for the $H(z_i)$ data is diagonal and therefore, $C_{H,ij}=\sigma_{H,i}^2\delta_{ij}$. The possibility of having non-zero off-diagonal terms in the correlation matrix is not excluded, but these coefficients are not found in the literature. Thus, the corresponding $\chi^2$ function adopts the following simple form,
\begin{equation}\label{chi2H}
\chi^{2}_{ H}(\vec{p})=\sum_{i=1}^{31} \left[ \frac{ H(z_{i},\vec{p})-H_{\rm obs}(z_{i})}
{\sigma_{H,i}} \right]^{2}\,.
\end{equation}
Where $H(z_{i},\vec{p})$ represents the theoretical value, for a given model which depends on the parameters encapsulated in $\vec{p}$, evaluated at each redshift $z_i$ and $H_{\rm obs}(z_i)$ are the observational values.
\subsubsection{Large Scale Structure}
By observing the existing anisotropies in the clustering of galaxies (in the redshift space) produced by the peculiar velocities, it is possible to obtain measurements that will allow us to test the growth rate of structure of a given model. An observable that turns out to be particularly useful, since it does not contain the bias factor which is model dependent, is the combination $f(a,\vec{k})\sigma_8(a)$. The term $f(a,\vec{k})$ is known as the growth rate and it can be computed from the following expression:
\begin{equation}
f(a,\vec{k}) =  \frac{a}{\delta_m(a,\vec{k})}\frac{d\delta_m(a,\vec{k})}{da},
\end{equation}
where $\delta_m = \delta_m/\rho_m$ is the density contrast for non-relativistic matter. If we are well within the horizon ($k\gg aH$) and it is valid the linear regime it is possible to express the matter density contrast as:
\begin{equation}
\delta_m(a,\vec{k}) = D(a)F(\vec{k})    
\end{equation}
where as it can be appreciated the dependence on $\vec{k}$ has been factored out. The properties of $F(\vec{k})$ are determined by the initial conditions and $D(a)$ is called the growth function. The above relation is going to be useful due to the fact that we want to make use of the matter power spectrum (which is related with the matter density contrast as we are about to see) to obtain the value of $f(a)$. The relation between the matter power spectrum and the density contrast reads as follows:
\begin{equation}\label{Power_spectrum_density_contrast}
P_m(a,\vec{k}) = C\langle \delta_m(a,\vec{k})\delta^{*}_m(a,\vec{k}) \rangle = CD^2(a)\langle F(\vec{k})F^{*}(\vec{k}) \rangle \equiv D^2(a)P_0(\vec{k})    
\end{equation}
where $C$ is a constant and $P_0(\vec{k}) = C\langle F(\vec{k})F^{*}(\vec{k}) \rangle$ is the primordial power spectrum which is determined from the theory of inflation. Thanks to \eqref{Power_spectrum_density_contrast} we can compute $f(a)$:
\begin{equation}
f(a) =  \frac{a}{\delta_m(a,\vec{k})}\frac{d\delta_m(a,\vec{k})}{da}  = \frac{a}{D(a)}\frac{dD(a)}{da} =  \frac{a}{2P_m(a,\vec{k})}\frac{dP_m(a,\vec{k})}{da}.     
\end{equation}
We have used $D(a) \sim \sqrt{P_m(a,\vec{k})}$. The observable $f(a)\sigma_8(a)$ is computed at subhorizon scales and for most of the cosmological models this leads to a second order differential equation for the density contrast, which can be generally expressed as: 
\begin{equation}
\delta^{\prime\prime}_m + F(a,\vec{p})\delta^{\prime}_m + G(a,\vec{p})\delta_m = 0    
\end{equation}
where the dependence on $\vec{k}$ has been removed and the primes represent derivatives with respect to the scale factor. The functions $F(a,\vec{p})$ and $G(a,\vec{p})$ can vary from one model to another and they depend on the free parameters embodied in the vector $\vec{p}$. 
\newline
\newline
In order to compute $\sigma_8(a)$, which is the root mean square of matter fluctuations on spheres of radius $R_8 = 8{h^{-1}}$ Mpc at the given value of the scale factor, we need to employ the following expression:
\begin{equation}
\sigma^2_8(a) = \frac{1}{2\pi^2}\int^{\infty}_0 d{k}{k}^{2}P_m(a,\vec{k})W^2({k}R_8)     
\end{equation}
where ${k}\equiv |\vec{k}|$ and $W({k}R)$ is the top hat smoothing function: 
\begin{equation}
W({k}R) = \frac{3\left({\rm sin}({k}R) - {k}R{\rm cos}({k}R)  \right)}{({k}R)^2}    
\end{equation}
Usually the data is given in terms of the redshift instead of the scale factor and without depending on $\vec{k}$, namely: $f(z)\sigma_8(z)$. 
\newline
\newline
We want to stress the fact that the LSS data turn out to be crucial to differentiate between models that have very similar behaviour at the background level but not at the perturbation level. It is important to employ a set of $f(z)\sigma_8(z)$ points totally free from correlations or double counting in order to test the models in the fairest way possible. In Chapters \ref{First_evidence_chapter}-\ref{BD_gravity_chapter} this idea has been followed carefully and when the correlations between the data points have been provided, by the experimental teams, they have been duly taken into account. To conclude this part we want to note that part of the information, embodied in the data points, comes from such small scales that non-linear effects come into play and therefore they need to be modeled. Great improvements have been made in this area, turning the redshift-space distortions data in a robust and reliable source of data. 
\subsubsection{Gravitational Lensing}
General Relativity tells us that the presence of matter creates a local curvature in space-time. As a consequence when a ray of light is affected by this curvature its path is bent. This phenomenon causes a process known as Gravitational Lensing. The light coming from distant galaxies is bent by the gravitational effect exerted by the massive bodies present in the path of the photon towards us. Therefore the real image of the distant galaxy, which is behind the massive bodies but in the same line of sight, appears to be distorted. In the recent years the improvements in the techniques to measure this effect, carried out by the experimental teams \cite{Joudaki:2017zdt,Hildebrandt:2016iqg,Wong:2019kwg,Birrer:2020tax}, have turned this field into a very promising one and the information extracted can be used to constraint cosmological models. 
\newline
\newline
Depending on the amount of bending produced by the intervening masses we consider two different regimes: Weak Lensing (WL) or Strong Lensing (SL). 
\newline
\newline
{\bf Weak Lensing}
\newline
\newline
The experimental teams like KiDS-450, DES or the HSC are able to measure the Weak Lensing statistical distortion of angles and shapes of galaxy images caused by the presence of inhomogeneities in the low-redshift Universe \cite{Hildebrandt:2016iqg,Joudaki:2017zdt,Kohlinger:2018sxx,Wright:2020ppw,Hikage:2018qbn,Troxel:2017xyo}. The two-point correlation functions of these angle distortions are related to the power spectrum of matter density fluctuations, and from it, it is possible to obtain constraints on the parameter combination $S_8=\sigma_8\sqrt{\Omega_m/0.3}$. However, one have to be careful when including this data in the analysis, because of the following reason: at the computational level, non-linear effects for small angular scales are calculated with the Halofit module \cite{Takahashi:2012em} (in case \texttt{CLASS} is used) which only works fine for the $\Lambda$CDM and minimal extensions of it, as XCDM \cite{Turner:1998ex} or CPL parametrization of the dark energy EoS parameter\, \cite{Chevallier:2000qy,Linder:2002et}. Thus, it is may not be able to model accurately the potential low-scale particularities of the model under study. Indeed it is possible to remove these low angular scales from the analysis, in order to avoid the operation of Halofit, but in that case the loss of information is very big, and $S_8$ remains basically unconstrained. \footnote{See http://kids.strw.leidenuniv.nl/sciencedata.php for more details.}
\newline
\newline
{\bf Strong Lensing}
\newline
\newline
In Chapter \ref{BD_gravity_chapter} we use the data extracted from the six gravitational lensed quasars of variable luminosity reported by the H0LICOW team. They measure the time delay produced by the deflection of the light rays due to the presence of an intervening lensing mass. After modeling the gravitational lens it is possible to compute the so-called time delay distance $D_{\Delta{t}}$ (cf. \cite{Wong:2019kwg} and references therein). The fact of being absolute distances (instead of relative, as in the SNIa and BAO datasets) {allows them to directly constrain the Hubble parameter in the context of the $\Lambda$CDM as follows:  $H_0=73.3^{+1.7}_{-1.8}$ km/s/Mpc} . It turns out that for the three sources B1608+656, RX51131-1231 and HE0435-1223, the posterior distribution of $D_{\Delta t}$ can be well approximated by the analytical expression of the skewed log-normal distribution,
\begin{equation}\label{eq:skewed}
\mathcal{L}(D_{\Delta t}) = \frac{1}{\sqrt{2\pi}(D_{\Delta t} - \lambda_D)\sigma_D}\,{\rm exp}\left[-\frac{(\ln(D_{\Delta t} -\lambda_D) - \mu_D)^2}{2\sigma^2_D}\right]\,,
\end{equation}
where the corresponding values for the fitted parameters $\mu_D$, $\sigma_D$ and $\lambda_D$ are reported in Table 3 of \cite{Wong:2019kwg}. On the other hand, the former procedure cannot be applied to the three remaining lenses, i.e. SDSS 1206+4332, WFI 2033-4723 and PG 1115+080. From the corresponding Markov chains provided by H0LICOW\footnote{http://shsuyu.github.io/H0LiCOW/site/} we have instead constructed the associated analytical posterior distributions of the time delay angular diameter distances for each of them. Taking advantage of the fact that the number of points in each bin is proportional to $\mathcal{L}(D_{\Delta t})$ evaluated at the average $D_{\Delta t}$ for each bin, we can get the values for $-\ln(\mathcal{L}/\mathcal{L}_{max})$ and fit the output to obtain its analytical expression as a function of $D_{\Delta t}$. The fitting function can be as accurate as wanted, e.g. using a polynomial of order as high as needed. The outcome of this procedure is used instead of \eqref{eq:skewed} for the three aforesaid lenses.
Interestingly, the non-detection of Strong Lensing time delay variations can be used to put an upper bound on $\dot{G}/G$ at the redshift and location of the lens \cite{Giani:2020fpz}. 
\newline
\newline
\subsubsection{Prior on $H_0$}
In Chapters \ref{H0_tension_chapter} and \ref{BD_gravity_chapter} we include, as a prior, the value of the Hubble parameter measured by the SH0ES collaboration, $H_0= (73.24\pm 1.74)$ km/s/Mpc \cite{Riess:2016jrr} and $H_0= (73.5\pm 1.4)$ km/s/Mpc \cite{Reid:2019tiq} respectively. Both measures have been obtained by applying the cosmic distance ladder method, based on chaining different techniques, to measure distance to celestial objects, to finally obtain the desired final distance. The second measurement includes some refinements in the method which includes an improved calibration of the Cepheid period-luminosity relation. It is based on distances obtained from detached eclipsing binaries located in the Large Magellanic Cloud, masers in the galaxy NGC $4258$ and Milky Way parallaxes. This measurement is in $4.1\sigma$ tension with the value obtained by the Planck 2018 team under the TTTEEE+lowE+lensing dataset, and using the $\Lambda$CDM model, $H_0= (67.36\pm 0.54)$ km/s/Mpc. \cite{Aghanim:2018eyx}. This represents the most acute tension affecting the standard model of cosmology and as cannot be otherwise has triggered a search for models able to accommodate the SH0ES measurement while the CMB data is included in the analysis.  
\newline
\newline
\subsubsection{CMB}
Without doubt the cosmic microwave background is one of the most important tools at our disposal to test cosmological models. It provides us with a unique opportunity to obtain information from the early Universe, namely at redshifts of order $z\sim \mathcal{O}(1000)$, and then if it is combined with the late-time cosmological data  we obtain a robust data set to tighten the constrains on free parameters. In order to obtain the perturbation equations, that contain the physics behind the CMB, one must deal with the coupled differential equations expressing the contribution of each component which includes, among other things, the interaction between photons and baryons. See \cite{Amendola:2015ksp} and \cite{Dodelson:2003ft} for a detailed description. 
\newline
\newline
The question now is, how is this temperature field described today ? The deviations with respect to the mean value of the CMB temperature can be expressed as $\delta{T} = \delta{T}(z,\vec{x}, \hat{n})$, where $\hat{n}$ is a unitary vector that indicates the direction of the incoming photons. The CMB can only be observed here in the Earth (actually near to the Earth thanks to the satellites) and at present time, $z = 0$. Taking this into account, while the temperature field depends on $z$, $\vec{x}$ and $\hat{n}$, in practise, we are only able to measure the dependence on the direction, or in other words the anisotropies. It turns out to be convenient expand $\delta{T}$ in terms of spherical harmonics, which can be understood as a kind of Fourier transformation in 2-dimensions:
\begin{equation}\label{TemperatureExpansion}
\frac{\delta{T}(z,\vec{x}, \hat{n})}{T(z)} = \sum^{\infty}_{\ell=1}\sum^{\ell}_{m=-\ell}a_{{\ell}m}(z,\vec{x})Y(\hat{n})    
\end{equation}
being $Y(\hat{n})$ the spherical harmonics containing the angular dependence. Upon inverting \eqref{TemperatureExpansion} we can get the expression for the $a_{\ell{m}}$ coefficients
\begin{equation}
a_{\ell{m}} = \int d\Omega Y^{*}(\hat{n})\frac{\delta{T}(z,\vec{x}, \hat{n})}{T(z)}. 
\end{equation}
They are characterized by having zero mean value but non-null variance, whose expression can be obtained from: 
\begin{equation}
C_\ell = \frac{1}{2\ell +1}\sum^{\ell}_{m=-\ell}|a_{{\ell}m}|^2. 
\end{equation}
This is the observable used to compare the theoretical predictions with the observed power spectrum of the CMB. $\ell$ is known as the multipole moment and is related to the angle $\theta$ through $\theta = \pi/\ell$. This last relation means that the low-$\ell$ multipoles contain the information from the large scale whereas the high-$\ell$ multipoles the small scale one. It is not possible to have an analytical expression for $C_\ell$ since they involve quantities that can only be obtained by getting the numerical solutions of a coupled system of differential equations. So, they must be obtained numerically. In Chapters \ref{Signs_chapter}-\ref{BD_gravity_chapter} we have used the Einstein-Boltzmann system solver \texttt{CLASS} \cite{Blas:2011rf} to do so. 
\newline
\newline
It is possible to compress, in a very efficient way, the CMB information in what is called as CMB distance priors, namely $R$ (shift parameter), $\ell_a$ (acoustic length) and their correlations  with the reduced baryon mass parameter and the spectral index  $(\omega_b,n_s)$. It has been shown in \cite{Mukherjee:2008kd} that although this compression loses information compared to the full CMB likelihood, such information loss becomes negligible when more data is added. The theoretical expressions for $R$ and $\ell_a$ are: 
\begin{align}
R(\vec{p})&=\sqrt{\Omega_{m}}\int_{0}^{z_{*}} \frac{dz}{E(z)}\\ \ell_a(\vec{p})&=\pi(1+z_*)\frac{D_A(z_*)}{r_s(z_*)}    
\end{align}
where $z_*$ is the redshift at decoupling. Its precise value depends weakly on the parameters, and it is obtained from the fitting formula\,\cite{Hu:1994uz}:
\begin{equation}\label{zlastscatt}
z_{*}=1048\,\left(1+0.00124\,\omega_b^{-0.738}\right)\left(1+g_1 \omega_m^{g_2}\right)\,,
\end{equation}
with
\begin{align}\label{zlastscatt1}
&g_1 = \frac{0.0783\,\omega_b^{-0.238}}{1+39.5\,\omega_b^{0.763}}\nonumber\\
&g_2 = \frac{0.560}{1+21.1\omega_b^{1.81}}\,.
\end{align}
\newline
\newline
\subsection{Statistical analysis}
As we have stated, statistics is a fundamental part of the analysis of a given cosmological model. Here we are going to present the fundamental concepts that we must consider if we want to constraint the parameter space of the models under study. Furthermore, we will provide the statistical tools that allow us to compare the performance of the different models, even if they have different number of degrees of freedom. 
\subsubsection{Model fitting}
It is time to properly introduce the methods that have been used in this thesis in order to get the results displayed in the corresponding tables and figures that can be found in Chapters \ref{First_evidence_chapter}-\ref{BD_gravity_chapter}. First of all, we should bear in mind, the ultimate goal of the model fitting procedure, namely: extracting information on the cosmological parameters, that characterize the studied model, from the considered observational data. The more precise data, covering a wide range of redshift we have, the better constrained the free parameters of a given model will be. As a clear example of this statement one can take the evolution of the constrains on the $\Omega_\Lambda$ parameter over the years, since as mentioned in the introductory chapter, $\Omega_\Lambda$ was found to be different from zero at $\sim 5\sigma$ c.l. employing only distant supernovae, while by considering the recent CMB data from the Planck team this evidence rises until the astonishing level of $\sim 94\sigma$ c.l. As it is well-known, the $\Lambda$CDM model, has proven to be very successful in passing the cosmological observational tests over the last years. What we want to know now is whether models that allow small deviations with respect to $\Lambda$CDM's predictions are favoured by the data.  
\newline
\newline
Two approaches can be taken in order to extract the desired information from the cosmological data, the frequentist and the bayesian. Determining the conceptual differences among them is beyond the scope of this thesis, so, we are just going to focus on explaining the bayesian approach since it has been proven to be the most convenient one to study cosmological models. It is based on the famous Bayes theorem, which can be expressed as:
\begin{eqnarray}\label{eq:BayesTheorem_appendix_E}
p(\theta|x,M)=\frac{p(x|\theta,M) p(\theta | M)}{p(x|M)}\,.
\end{eqnarray}
In words, it says that the posterior probability of $\theta$ (which represents the free parameters), namely $p(\theta|x,M)$, given the model $M$ and the data $x$ is equal to the probability of the data given the parameters of the model (i.e. the likelihood  $p(x|\theta,M)$) times the prior probability of  $\theta$,  $p(\theta | M)$, divided by the probability of the data $x$  (usually in the form of a probability distribution function (PDF)).
\newline
\newline
What we are looking for, is the expression of the posterior probability function, which contains the information on how the characterizing parameters of a given model are distributed once the data $x$ have been considered in addition to some previous knowledge on the parameters encoded in the priors. The function $p(x|\theta,M)$ is usually referred as the likelihood and is related to the $\chi^2(\theta)$ (as long as the data can be described by a Gaussian function) function defined in \eqref{eq:chi_square_function_appendix_E} in the following way:
\begin{equation}
\mathcal{L} = \mathcal{L}_0e^{-\frac{\chi^2(\theta)}{2}}.    
\end{equation}
Where $\mathcal{L}_0$ is just a normalization constant. Note that while in the previous section we have employed $\vec{p}$ to denote the vector of free parameters, here for convenience we use $\theta$. The likelihood function gives us information about the parameters, $\theta$, given the data. In particular, it says how likely a given data would be for a specific set of values of the parameters $\theta$. It is important to remark that it is not a probability distribution function of the parameters. In the next section we will provide more information about the function $p(x|M)$ due to the fact that it plays an important role in the model selection procedure. For the purpose of this section what is important is the relation 
\begin{equation}\label{eq:Posterior_proportionality_appendix_E}
p(\theta|x,M) \propto p(x|\theta,M) p(\theta | M) 
\end{equation}
since it allows us to sample the posterior distribution function once we have established the shape of the priors and the form of the likelihood function. This is tantamount to say that we will be working with the kernel of the posterior instead of with the full probability distribution function. The kernel is the part of the probability distribution function that remains when all the parts that do not depend on the parameters are removed. This concept can be easily understood with a simple example. Consider a random variable $X$ which is distributed according to the expression of a Gaussian function $p(X|\sigma,\mu) = e^{-(X-\mu)^{2}/2\sigma^2}/\sqrt{2\pi\sigma^2}$, then the kernel would be in this case $e^{-(X-\mu)^{2}/2\sigma^2}$.
\newline
\newline
Let us briefly discuss the general features of some of the most common model fitting procedures. The first one is the method known as the grid method, which is a quite straight-forward approach. Basically it consists in building a grid for the involved parameters in the parameter space and compute the value of the likelihood function at each point of the grid. Once all the calculations are done and thanks to the relation \eqref{eq:Posterior_proportionality_appendix_E} it is possible to estimate not only those values of the parameters that maximizes the likelihood function (or minimizes the $\chi^2$ function) but also the errors associated to the free parameters that represent the {\it d.o.f.} of the model under consideration. This method is very robust as it does not make any assumption about the shape of the posterior distribution, only relies on the fact that the probability distribution functions of the experimental measurements are correct. However it has an important drawback, even if we consider that we are able to make a reasonable guess about where is the maximum of the likelihood function in order to be sure that it is  into the grid that we build. The main problem is, as could not be otherwise, the computational time. When our model contains more than 3-4 parameters, and note that this is perfectly reasonable, this method requires an enormous amount of time to compute the value of the likelihood function at all the desired point. 
\newline
\newline
A much more efficient approach is the so-called Fisher matrix method, which is based on the central limit theorem that claims that any well behaved distribution function can be approximated by a Gaussian distribution near its maximum. In other words, even if we do not know the shape of the posterior distribution, we can obtain at least an approximation, considering that the likelihood is well-described by a multivariate Gaussian near the maximum, and once again, the relation \eqref{eq:Posterior_proportionality_appendix_E}. Let us write down the approximation taken into account:
\begin{equation}\label{eq:likelihood_appendix_E}
\mathcal{L}(\theta) \simeq \mathcal{L}(\hat{\theta})e^{-\frac{1}{2}(\theta_i -\hat{\theta}_i)F_{ij}(\theta_j -\hat{\theta}_j)}    
\end{equation}
where by $\hat{\theta}$ we mean the vector of best fit values, which maximizes the likelihood. The term $F_{ij}$ is called the Fisher matrix and its expression will be clarified in a moment. Note that the above expression implies that the likelihood is not only a Gaussian function of the data but also of the parameters. Depending on the shape of the priors the posterior distribution is also a multivariate-Gaussian distribution. Taking the logarithm on both sides of \eqref{eq:likelihood_appendix_E} and expanding at second order near $\hat{\theta}$ we can obtain the expression: 
\begin{equation}
\ln(\mathcal{L}(\theta)) \simeq \ln(\mathcal{L}(\hat{\theta})) + \frac{1}{2}\frac{\partial^2\ln{\mathcal{L}(\theta)}}{\partial\theta_i\partial\theta_j}\Bigg|_{\theta = \hat{\theta}}(\theta_i -\hat{\theta}_i)(\theta_j -\hat{\theta}_j).
\end{equation}
We have not included the first derivative of the logarithm of the likelihood function because the expression is evaluated at $\hat{\theta}$ which by definition is the vector of parameters that extremizes (in this case maximizes) the likelihood. Now we are able to identify the Fisher matrix with its expression
\begin{equation}
F_{ij} \equiv \frac{\partial^2\ln{\mathcal{L}(\theta)}}{\partial\theta_i\partial\theta_j}\Bigg|_{\theta = \hat{\theta}}.     
\end{equation}
Even if it is not possible to compute analytically the Fisher matrix, very often we can easily compute it numerically due to the smooth behaviour of the function around the maximum. What makes this method extremely useful is the fact that once the Fisher matrix is computed it is straight-forward to get the errors of the parameters and the two-dimensional contour plots which are going to be ellipses. In order to marginalize over the parameters we are not interested in, we only need to remove the corresponding rows and columns from the inverse of the Fisher matrix. On the other hand, to obtain the length of the semiaxes of the ellipsoidal contours we need to take the square root of the eigenvalues of the inverse of the Fisher matrix. Some numerical problems could arise from the procedure to get the inverse of the Fisher matrix, so we must be careful with that. As it can be understood this method implies a considerable reduction in the computational time in comparison to the grid method even if we study models with a large number of free parameters. Nevertheless, it is important to remark that this is only accurate, as long as, the likelihood function and consequently the posterior distribution can be well-approximated by a multivariate Gaussian function, otherwise the deviations from the normal distribution can infer biased estimation for the errors. In order to obtain the results presented in Chapters \ref{First_evidence_chapter}-\ref{H0_tension_chapter} the Fisher matrix method has been employed. However, we have carefully studied possible deviations from the Gaussian behaviour. See the before mentioned chapters to know all the details.
\newline
\newline
We will finish this brief summary by talking about the well-known Monte Carlo Markov Chains (MCMC) method. We remember that the main disadvantage of the grid  method is the huge computationally time that is required to explore the desired piece of the parameter space. The solution provided to this issue by the MCMC method is to explore the landscape of the space parameter under study with random jumps. In other words, using the likelihood function and the selected priors, according to the relation \eqref{eq:Posterior_proportionality_appendix_E} the MCMC techniques allow us to explore the parameter space and keep those points that are accepted by the selection algorithm employed (see for instance \cite{Metropolis:1953am,Hastings:1970aa} ). Once this is done we end up with a collection of n-dimensional points labeled by the values of the free parameters. We call this the Markov Chains, and from them we can get the corresponding confidence level for the involved parameters, the correlations among them and the contour plots. As the reader may have noticed, we have not mentioned that the likelihood function must have a Gaussian shape. This can be considered the main advantage of the MCMC method over the Fisher matrix approach. In other words, we do not need to make any assumption about the form of the likelihood function which allows us to study cases where the non-gaussianities may have an important role. On the contrary, clearly the MCMC method is going to take more time than the Fisher matrix method.  
A natural question at this point could be, how long should the Markov Chains be ? A convergence criterion must be applied at this point, see for instance \cite{Gelman:1992zz}. 
\newline
\newline 
As it can be appreciated all the methods have advantages and disadvantages which makes recommendable to bear in mind all of them in order to use the most suitable one in each case. As stated at the beginning of this appendix the model fitting is unavoidable if one wants to carry out a proper study of a given model. No matter how well-looking are the theoretical predictions of some particular model, if at the end of the day, the model is not able to pass the experimental tests it must be ruled out, at least until new experimental data come out.

\subsubsection{Model selection}\label{sec:Model_Selection_Appendix_E}
So far, we have been dealing with the problem of comparing the theoretical predictions for a given model, with the observational data considered and how we can narrowing down the parameter space through different methods like the Fisher matrix or like the Monte Carlo Markov Chains techniques. When all this is done is time for the next natural question: does, the model studied in the first place, reproduce the data better than any other one ? Answering this question is known as the model selection problem. 
\newline
\newline
Among the different options, in this appendix, we focus on the selection procedure called the Bayesian evidence or marginal likelihood. 
The significance of the results obtained can be further appraised by computing the Bayesian evidence, which is based on evaluating the posterior probability of a model $M$  given the data $x$ and the priors, see e.g. \cite{Amendola:2015ksp}. The relevant quantity we are looking for is the so-called Bayes factor, which does not depend on the arbitrarily assigned prior probability to the given model $M$. This factor can be obtained by making use of the Bayes theorem previously mentioned \eqref{eq:BayesTheorem_appendix_E}. As can be appreciated by taking a look of that expression, the quantity $p(x|M)$ does not depend on the values of $\theta$. Therefore, if we normalize the posterior probability to one and  integrate over all the parameters $\theta$ on both sides of Eq.\,(\ref{eq:BayesTheorem_appendix_E}), the corresponding integral in the numerator on the \textit{r.h.s.}  must be equal to $p(x|M)$:
\begin{eqnarray}\label{eq:evidence_appendix_E}
 p(x|M) = \int d\theta\, p(x|\theta,M)\,p(\theta|M)\,.
\end{eqnarray}
This likelihood integral over all the values that can take  the parameters $\theta$ is called the marginal likelihood or evidence\,\cite{Amendola:2015ksp}.
\newline
\newline
An analogous formula to (\ref{eq:BayesTheorem_appendix_E}) can also be applied to estimate the posterior probability that a model $M$ is true given a measured data set $x$. Thus, following the same scheme, the posterior probability of the model given the data, $p(M|x)$, must be equal to the probability of the data given the model (irrespective of the values of $\theta$) -- i.e. the marginal likelihood (\ref{eq:evidence_appendix_E}) --- times the prior probability of the model, divided by the PDF of the data. Writing this relation for two cosmological models $M_i$  and  $M_j$ that are being compared, we find that the ratio of posterior probabilities of these models is equal to the ratio of model priors times a factor:
\begin{equation}
\frac{p(M_i|x)}{p(M_j|x)} =
\frac{p(M_i)}{
p(M_j)}\, B_{ij}\,.
\end{equation}
Such factor, $B_{ij} = p(x| M_i)/p(x| M_j)$,  is the so-called  Bayes factor of the model $M_i$ with respect to model $M_j$. It gives the ratio of marginal likelihoods (i.e. of evidences) of the two models. Note that it coincides with the ratio of posterior probabilities of the models only if the prior probabilities of these models are the same. This is generally assumed (``Principle of Insufficient Reason'') and hence the comparison of the two  models $M_i$ and $M_j$ is usually performed directly  in terms of $B_{ij}$.
\newline
\newline
In  the literature it has been customary to  define the Bayes information criterion (BIC) through the parameter ${\rm BIC}=\chi^2_{\rm min}+n\,\ln N$, where $\chi^2_{\rm min}$ is the minimum of the $\chi^2$ function, $n$ is the number of independent fitting parameters and $N$ is the total number of data points. One can show that the Bayes factor can be estimated through $B_{ij}= e^{\Delta{\rm BIC}/2}$ where $\Delta {\rm BIC}={\rm BIC}_j-{\rm BIC}_i$ is the difference between the values of BIC for models $M_j$ and $M_i$\,\cite{KassRaftery1995,Burnham2002}, i.e.
\begin{equation}\label{eq:DeltaBIC_E_appendix}
\Delta {\rm BIC}=\Delta\chi^2_{\rm min}-\Delta n\,\ln N\,,
\end{equation}
%
\begin{table}[t!]
\begin{center}
\setcounter{table}{0}
\begin{tabular}{cc}
\hline
$\Delta {\rm BIC}=2\ln B_{ij}$ &$\ \ \ \ \ \ $ Bayesian evidence of model ${M}_i$ versus $M_j$ $\ \ $ \\ \hline
$0<\Delta {\rm BIC} < 2$ & {\rm weak evidence (consistency between both models)} \\
$2<\Delta {\rm BIC}< 6$ & {\rm  positive evidence} \\
$6<\Delta {\rm BIC}< 10$ & {\rm strong evidence} \\
$\Delta {\rm BIC} \geq 10$ & {\rm very strong evidence} \\
$\Delta {\rm BIC} < 0$ & {\rm counter-evidence against model $M_i$} \\
\hline
\end{tabular}
\caption{\scriptsize Conventional ranges of values of $\Delta {\rm BIC} $  used to judge the observational support for a given model $M_i$ with respect to the reference one $M_j$. }
\label{tab:Delta BIC_E_appendix}
\end{center}
\end{table}
%
\newline
where $ \Delta\chi^2_{\rm min}=\left(\chi^2_{\rm min}\right)_j-\left(\chi^2_{\rm min}\right)_i $ is the difference between the minimum values of  $\chi^2$ for each model,  and $\Delta n=n_i-n_j$ is the difference in the number of independent fitting parameters of $M_i$ and $M_j$, both describing the same $N$ data points. The added term to $\Delta\chi^2_{\rm min}$, i.e. $\Delta n\,\ln N$, represents the penalty assigned to the model having the largest number of parameters ($\Delta n>0$). Indeed, if $M_j$ has less parameters than $M_i$  (i.e. $n_i>n_j$) we expect that $\left(\chi^2_{\rm min}\right)_i< \left(\chi^2_{\rm min}\right)_j$ and hence $\Delta\chi^2_{\rm min}>0$. However, since $\Delta n>0$ the last term of \eqref{eq:DeltaBIC_E_appendix} indeed penalizes the model with more parameters and compensates in part for its  (presumably smaller) value of $\chi^2_{\rm min}$. Defined in this way, a positive value of the expression  \eqref{eq:DeltaBIC_E_appendix} denotes that the model $M_i$ is better than model $M_j$, and the larger is $\Delta {\rm BIC}$  the better is  $M_i$ as compared to $M_j$.  Clearly, the Bayesian criterion allows a  quantitatively implementation of Occam's razor.
\newline
\newline
Even though the simple and very economic formula \eqref{eq:DeltaBIC_E_appendix} is useful and has been applied in many places of the literature, see e.g. \,\cite{Sola:2016jky,Sola:2017lxc,Sola:2016ecz,Sola:2017jbl} and references therein, it is only an approximate formula.  The exact value of $\Delta {\rm BIC}$ associated to the full Bayes factor requires to evaluate
\begin{equation}\label{eq:DeltaBICExact_appendix_E}
\Delta {\rm BIC}=2\,\ln B_{ij}=2\,\ln\frac{\int d\theta_i\, p(x|\theta_i,M_i)\,p(\theta_i|M_i)}{\int d\theta_j\, p(x|\theta_j,M_j)\,p(\theta_j|M_j)}\,,
\end{equation}
where $\theta_i$ and $\theta_j$ are the two sets of free parameters integrated over for each model. We refer to this (exact) value of  $\Delta {\rm BIC}$ as the full ``Bayesian evidence'' of model $M_i$ versus model $M_j$. It represents the optimal implementation of Occam's razor. Formula \eqref{eq:DeltaBIC_E_appendix} provides an estimate (sometimes good, sometimes rough)  of the cumbersome expression \eqref{eq:DeltaBICExact_appendix_E}. The evaluation of the latter is carried out in practice using the MCMC's for the statistical analysis of the model and it can be a formidable numerical task. Fortunately it can be speeded up with the help of the recent numerical code \texttt{MCEvidence} \cite{Heavens:2017hkr}. In Table 1 we collect the  ranges of values of $\Delta {\rm BIC} $ that are conventionally used in the literature to quantify the observational support of a given model $M_i$ with respect to some reference one $M_j$\,\cite{KassRaftery1995,Burnham2002}. Consistently with the approximate formula discussed above, positive values of $\Delta {\rm BIC}$ in the indicated intervals favor the model $M_i$ over  model $M_j$. Usually $M_j$ is the $\CC$CDM and $M_i$ is any of the considered beyond the standard model (BSM). Small positive values near one just denote that the two models under competition are comparable  (i.e. that there is no marked preference of $M_i$ over $M_j$), but larger and positive values of $\Delta {\rm BIC} $ increase the support of $M_i$ (BSM) versus $M_j$ ($\CC$CDM). Negative values of $\Delta {\rm BIC}$, in contrast,  would lead to the opposite conclusion, i.e. that the BSM is penalized as compared to the $\CC$CDM  and hence that the latter is preferred.
\newline
\newline
\newpage
\thispagestyle{empty}
\mbox{}
\newpage

\bibliographystyle{ieeetr}
\bibliography{references}

\end{document}